\documentclass{ULBreport}

\usepackage{amsmath} 
\usepackage{amssymb} 
\usepackage{xspace} 
\usepackage[backref=false, sorting=none, bibstyle=ieee, citestyle=numeric-comp, minbibnames=1, maxbibnames=3]{biblatex} 
\usepackage[export]{adjustbox} 
\usepackage[justification=centering]{subfig} 
\usepackage{graphicx} 
\usepackage{tikz} 
\usepackage{siunitx} 
\usepackage{mathtools} 
\usepackage{booktabs}
\usepackage{tocbibind} 
\usepackage{epigraph} 
\usepackage{multicol}

\usepackage{xfrac} 
\usepackage{braket} 
\usepackage{rotating,threeparttable} 

\usepackage{array}
\usepackage{xparse}
\usepackage{multirow} 
\usepackage{placeins} 
\usepackage{emptypage} 
\usepackage{anyfontsize} 
\usepackage{afterpage} 
\usepackage{float} 

\usepackage{etoolbox}
\appto\TPTdoTablenotes{\footnotesize}

\usepackage{xurl} 
\usepackage[%
	hidelinks,
	linktoc=all, 
	bookmarksnumbered, 
	pdfauthor={Martin Habedank},
	pdftitle={Leaving No Matter Unturned - Analysing existing LHC measurements and events with jets and missing transverse energy measured by the ATLAS Experiment in search of Dark Matter},
	pdfsubject={Dissertation},
	pdfkeywords={Dark Matter, Dunkle Materie, LHC, ATLAS Experiment, missing transverse energy, fehlende transversale Energie, jets, 2HDM+a, Contur, measurement, Messung, search, Suche, unfolding},
	pdflang = {en-GB},
	]{hyperref} 
\usepackage{bookmark} 
\usepackage[acronym,toc,nonumberlist,nomain,shortcuts]{glossaries}[=v4.46] 
\usepackage{enumitem} 
\usepackage[a-1b]{pdfx} 

\usepackage{thesis-bib} 
\usepackage{acronyms} 
\usepackage{thesis-defs} 
\usetikzlibrary{positioning} 
\usetikzlibrary{shapes.geometric} 
\usetikzlibrary{fadings, decorations.markings} 

\makeatletter
\AtBeginDocument{\let\mathaccentV\AMS@mathaccentV}
\makeatother

\begin{document}
\emergencystretch 3em 
\newcommand{\abstractText}{%
Various astrophysical observations point towards an as-of-yet unexplained, mainly gravitationally interacting type of matter.
If this matter, called Dark Matter, is an elementary particle, it could be produced in particle collisions at the Large Hadron Collider.
Given its weak interaction with ordinary matter, however, it would not be directly observable with the general-purpose detectors at the Large Hadron Collider.
Its production would therefore manifest as events in which detector-visible objects recoil against the detector-invisible Dark Matter, giving rise to missing transverse energy.
This thesis focuses on final states in which these visible objects are jets.

A measurement of the final state of large missing transverse energy and at least one jet in \SI{139}{\ifb} of proton--proton collisions at \SI{13}{TeV} recorded with the ATLAS detector at the Large Hadron Collider is performed in this thesis.
Good agreement between measured data and Standard-Model prediction is found in a statistical fit, corresponding to a reduced chi-square of \fixedRmissChiTwoNdF.
The measurement is corrected for detector effects to facilitate later reinterpretation.
Measurements prepared in such a way can, for example, be exploited by the \Contur toolkit to set constraints on new theories.
Both, the results of the measurement and the \Contur toolkit making use of existing measurements at the Large Hadron Collider, are employed to set exclusion limits on a model able to explain Dark Matter, the two-Higgs-doublet model with a pseudoscalar mediator to Dark Matter.
At $\tanB=1$, masses of the pseudoscalar~$A$ up to \limitmamAmAmin and larger than \limitmamAmAmax are excluded at \SI{95}{\%} confidence level.
At $\mAeqmHeqmHpm=\SI{600}{GeV}$, masses of the pseudoscalar~$a$ up to \limitmatbma and values of \tanB up to \limitmatbtbmin as well as larger than \limitmatbtbmax are excluded at \SI{95}{\%} confidence level.%
}

\begin{titlepage}
	\begin{center}
		\scalebox{0.8}{
			\fontfamily{txr}\selectfont
			\begin{tikzpicture}
				\node (L) {\textcolor{titlenumbercolor}{\fontsize{100}{0}$\mathcal{L}$}};
				\node [xshift=98pt, yshift=5pt] (eaving) at (L) {{\textcolor{titlenumbercolor}{\fontsize{40}{0}\selectfont \bfseries\textit{eaving}}}};
				\node [yshift=-32pt, xshift=65pt] (no) at (L) {{\textcolor{titlenumbercolor}{\fontsize{40}{0}\selectfont \bfseries\textit{No}}}};
				\node [xshift=83pt] (matter) at (no) {{\textcolor{titlenumbercolor}{\fontsize{40}{0}\selectfont \bfseries\textit{Matter}}}};
				\node [yshift=-75pt, xshift=30pt] (U) at (L) {{\textcolor{titlenumbercolor}{\fontsize{100}{0}\selectfont \bfseries$\mathcal{U}$}}};
				\node [xshift=102pt, yshift=5pt] (nturned) at (U) {{\textcolor{titlenumbercolor}{\fontsize{40}{0}\selectfont \bfseries\textit{nturned}}}};
			\end{tikzpicture}
		}

		\vspace{5pt}
		{\Large		
		Analysing existing LHC measurements and events with jets and missing transverse energy measured by the ATLAS Experiment in search of Dark Matter}
		
		\vspace{15pt}
		\includegraphics[width=0.7\textwidth]{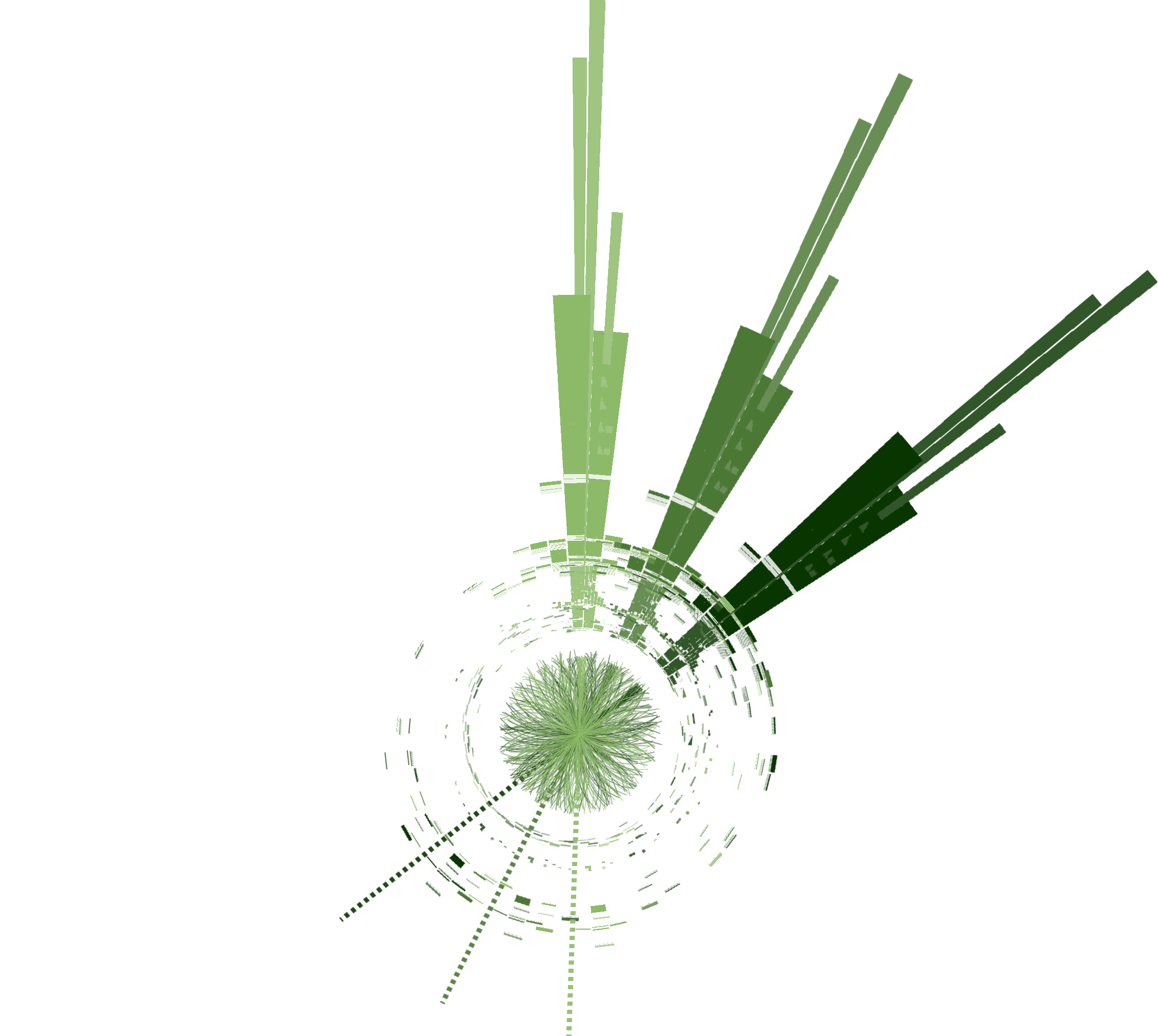}
		
		{\Large Dissertation}\\
		{\footnotesize
		zur Erlangung des akademischen Grades \textit{Doctor Rerum Naturalium} (Dr. rer. nat.)\\
		im Fach \textit{Physik}, Spezialisierung \textit{Experimentalphysik}\\
		eingereicht an der Mathematisch-Naturwissenschaftlichen Fakultät\\
		der Humboldt-Universität zu Berlin von\\}
		\textbf{\Large M. Sc. Martin Habedank}\\

	\end{center}

	\footnotesize
	\noindent
	Präsidentin der Humboldt-Universität zu Berlin: \textit{Prof. Dr. Julia von Blumenthal}\\
	Dekanin der Mathematisch-Naturwissenschaftlichen Fakultät: \textit{Prof. Dr. Caren Tischendorf}

	\noindent
	Gutachter*innen:
	\vspace{-12pt}
	\begin{multicols}{2}
		\begin{enumerate}[noitemsep]
			\item Dr. Priscilla Pani
			\item Prof. Dr. David Berge
			\item Prof. Dr. Thomas Kuhr
		\end{enumerate}
	
		\columnbreak

		\noindent weiteres Mitglied: Prof. Dr. Thorsten Kamps\\
		Vorsitz: Prof. Dr. Christophe Grojean\\[21pt]
		Tag der mündlichen Prüfung: 13. November 2023
	\end{multicols}
\end{titlepage}

\thispagestyle{empty}%
\vspace*{\fill}
\noindent
The illustration on the front page is an artistic adaptation (threefold overlay and rotation) of \refcite{ATLAS:2021mje}.
See \secref{sec:metJets_strategy} for an explanation of the illustration.

\tableofcontents

\ChapterStarSimple{Declaration of independent work}

I declare that I have completed the thesis independently, using only the aids and tools specified.
I have not applied for a doctor’s degree in the doctoral subject elsewhere and do not hold a corresponding doctor’s degree.
I have taken due note of the Faculty of Mathematics and Natural Sciences PhD Regulations, published in the Official Gazette of Humboldt-Universität zu Berlin no. 42/2018 on 11th July 2018.

\vspace{30pt}
\noindent
Berlin, 4th August 2023\\[10pt]
\noindent
Martin Habedank

\ChapterStarSimple{Abstract}
\abstractText 

\ChapterStarSimple{Zusammenfassung}

Diverse astrophysikalische Beobachtungen weisen auf bislang unerklärte Materie hin, die hauptsächlich gravitativ interagiert.
Wenn diese Materie, bezeichnet als Dunkle Materie, ein Elementarteilchen ist, könnte sie in Teilchenkollisionen am Large Hadron Collider produziert werden.
Aufgrund ihrer schwachen Wechselwirkung mit normaler Materie könnte sie jedoch nicht direkt mit den Vielzweckdetektoren am Large Hadron Collider beobachtet werden.
Ihre Produktion würde sich stattdessen in Ereignissen zeigen, in denen ein Rückstoß von Detektor-sichtbaren Objekten gegen die Detektor-unsicht\-bare Dunkle Materie besteht, was fehlende transversale Energie verursacht.
Diese Dissertation beschäftigt sich mit Endzuständen, in denen diese sichtbaren Objekte so\-ge\-nann\-te Jets sind.

In dieser Dissertation wird eine Messung des Endzustands aus großer fehlender transversaler Energie und mindestens einem Jet in \SI{139}{\ifb} an Proton--Proton-Kollisionen bei \SI{13}{TeV}, aufgezeichnet mit dem ATLAS-Detektor am Large Hadron Collider, durchgeführt.
Zwischen den gemessenen Daten und der Standard-Modell-Vorhersage wird in einer statistischen Optimierung gute Übereinstimmung gefunden.
Diese entspricht einem reduzierten Chi-Quadrat von \fixedRmissChiTwoNdFGerman.
Die Messung wird von Detektor-Effekten bereinigt, um spätere Reinterpretation zu vereinfachen.
Messungen, die auf diese Art aufbereitet wurden, können beispielsweise von der \Contur-Software benutzt werden, um Pa\-ra\-me\-tergrenzen für neue Theorien festzustellen.
Sowohl die Ergebnisse der Messung, als auch die \Contur-Software, welche auf existierende Messungen am Large Hadron Collider zurückgreift, werden verwendet, um Ausschlussgrenzen für ein Modell, das Dunkle Materie erklären kann, festzustellen.
Betrachtet wird das Modell von zwei Higgs-Dubletts und einem pseudoskalaren Botenteilchen zu Dunkler Materie.
Unter Annahme von\linebreak $\tanB=1$ werden Massen des Pseudoskalars~$A$ bis zu \limitmamAmAmin und größer als \limitmamAmAmax bei einer Konfidenz von \SI{95}{\%} ausgeschlossen.
Unter Annahme von $\mAeqmHeqmHpm=\SI{600}{GeV}$ werden Massen des Pseudoskalars~$a$ bis zu \limitmatbma und Werte von \tanB bis zu \limitmatbtbminGerman sowie größer als \limitmatbtbmax bei einer Konfidenz von \SI{95}{\%} ausgeschlossen.


\Chapter{Getting on track}{Introduction}{%
	Ja, ich puzzle gern und puzzle lang und puzzle viel.\\
	Ich bin beinahe fertig und vom Glück beseelt,\\
	da entdecke ich mit Schrecken: Das letzte Stück fehlt.%
}{Bodo Wartke}{BodoWartke:2009dls}
\label{sec:introduction}

Particle physics is at a crossroads.
For decades, the Standard Model of particle\linebreak physics~(\SM) has served as an immensely successful guiding theory, indicating which experimental paths to follow to discover missing particles and consolidate the theoretical framework.
With the discovery of a Higgs boson in 2012~\cite{CMS:2012qbp,ATLAS:2012yve}, however, the Standard Model is complete -- and yet open questions remain: numerous conceptional
and experimental shortcomings
demand novel theories.

Concretely, the nature of Dark Matter (\DM) poses one of the most profound open questions.
On the one hand, there is clear observational evidence for a type of matter that interacts gravitationally, but only very weakly with other known particles and is therefore difficult to observe in usual detectors.
On the other hand, there is no suitable candidate for this kind of particle in the Standard Model.

A wealth of models going \textit{beyond the Standard Model} (\BSM) has been proposed to address the open questions.
Each of them has its own parameters, resulting in a plethora of experimental signatures.
Given the lack of observational clues, however, it remains completely unclear if one of these models is realised in nature.
This comes with large freedom in research where a multitude of avenues can be equitably explored.
Apart from this, it also causes the need for new approaches in physics analyses.
Physics analyses are commonly designed with a specific model in mind.
The likelihood for this exact model to prove correct in the future, however, is small.
Physics analyses should therefore be reinterpreted with respect to other models as well to cover the broadest possible phase space for as many models as possible.
For this, it is imperative to routinely preserve physics analyses in a way that is easy to reinterpret.

\bigskip
This work shows how a physics analysis at a particle collider can be designed to give maximum impact also beyond the exact scope considered in the analysis itself.
This is done in a measurement targetting events with large momentum imbalance and at least one jet ($E_\text{T}^\text{miss}$+jets) using the ATLAS detector at the Large Hadron Collider (\LHC).
This measured final state is particularly sensitive to \DM production.
Therefore, this work also demonstrates how preserved physics analyses can be used to vet \BSM models.
For this, the two-Higgs-doublet model with a pseudoscalar mediator to Dark Matter (\THDMa) is used.
This model is the simplest gauge-invariant and renormalisable extension of the Standard Model with a pseudoscalar mediator to Dark Matter consistent with existing constraints on properties of the Higgs boson~\cite{LHCDarkMatterWorkingGroup:2018ufk}.
Pseudoscalar mediators to Dark Matter are particularly interesting to explore at particle colliders because their spin-dependent couplings are difficult to probe at other facilities, \eg direct-detection experiments.

Both, measurement and interpretation, will also be summarised in a publication of the ATLAS collaboration currently in preparation~\cite{ATLAS:2023mjt}.

Further, this thesis illustrates how a plethora of existing physics analyses from different experiments at the \LHC can be employed at once using the \Contur toolkit~\cite{Buckley:2021neu} to cover a broad range of models and parameter spaces in search of a new guiding theory.
This is performed for the \THDMa connected to a previously published study~\cite{Butterworth:2020vnb}.

\subsubsection{Levels of physics representation}
\label{sec:level_interpretation}

The \METjets measurement covered in this thesis studies how predictions from a physics model relate to data measured with the ATLAS detector.
A translation from physics model to detector signals has to take place to allow for a one-to-one comparison.
This is performed in different processing steps.
\figref{fig:detectorCorrection} gives a schematic overview of these steps and the intermediate representation levels.
Corrections to data and theory are marked by blue and red arrows, respectively.
In general, it can be understood that the more complex the corrections to data become, the more trivial the corrections for the theory prediction get and vice versa.

\begin{myfigure}{
		Schematic illustrating at which levels collision data and theory can be represented.
		The blue arrows indicate which corrections need to be applied to data to compare to theory at a higher level, increasing the model dependence.
		The red arrows give the steps necessary to compare a theory prediction to measured data at a chosen level.
		Steps along dashed arrows are not covered in this thesis.
		The levels which require high-level object (re)construction are indicated in green.
	}{fig:detectorCorrection}
	\begin{tikzpicture}[node distance=71pt]
		\definecolor{theoryred}{RGB}{228,26,28}
		\definecolor{experimentalblue}{RGB}{55,126,184}
		\definecolor{boxfillcolour}{rgb}{0.9, 0.9, 0.9}
		\newlength{\xdiff}
		\setlength{\xdiff}{6pt}
		\newlength{\xoffset}
		\setlength{\xoffset}{120pt}
		\newlength{\minwidth}
		\setlength{\minwidth}{130pt}
		\newlength{\minheight}
		\setlength{\minheight}{33pt}
		\tikzset{
			rectnode/.style={
				draw=none,
				fill=boxfillcolour,
				rounded corners=1mm,
				very thick,
				align=center,
				minimum width=\minwidth,
				minimum height=\minheight,
			},
			trapnode/.style={
				trapezium,
				draw=none,
				fill=boxfillcolour,
				very thick,
				align=center,
				minimum height=\minheight,
			},
			%
			arrowStyle/.style={
				-latex,
				midway,
				line width=5,
			},
			uparrowtext/.style={
				align=right,
				left,
				xshift=-5pt,
			},
			downarrowtext/.style={
				align=left,
				right,
				xshift=5pt,
			},
			%
			chapnodeStyle/.style={
				draw=black!70,
				thick,
				fill=subtitlecolor,
			},
			pics/uparrow/.style args={#1,#2,#3}{
				code={
					\path[arrowStyle] ([xshift=-\xdiff, yshift=5pt]#1.north) edge [draw=experimentalblue, path fading=south] node [uparrowtext] {#3} ([xshift=-\xdiff, yshift=-5pt]#2.south);
				}
			},
			pics/uparrowLabel/.style args={#1,#2,#3,#4,#5}{
				code={
					\path[arrowStyle] ([xshift=-\xdiff, yshift=5pt]#1.north) edge [draw=experimentalblue, path fading=south] ([xshift=-\xdiff, yshift=-5pt]#2.south);
					\path[arrowStyle] ([xshift=-\xdiff, yshift=5pt]#1.north) edge [draw=none] node [uparrowtext] {#3} node [chapnodeStyle,xshift=#5] {\ref{#4}} ([xshift=-\xdiff, yshift=-5pt]#2.south);
				}
			},
			pics/downarrowGeneral/.style args={#1,#2,#3,#4,#5,#6}{
				code={
					\path[arrowStyle] ([xshift=\xdiff, yshift=#5]#1.south) edge [draw=theoryred, path fading=north] ([xshift=\xdiff, yshift=#6]#2.north);
					\path[arrowStyle] ([xshift=\xdiff, yshift=#5]#1.south) edge [draw=none] node [downarrowtext] {#3} node [chapnodeStyle] {\ref{#4}} ([xshift=\xdiff, yshift=#6]#2.north);
				}
			},
			pics/downarrow/.style args={#1,#2,#3,#4}{
				code={
					\pic{downarrowGeneral={#1,#2,#3,#4,-5pt,5pt}};
				}
			},
			%
			pics/chapnode/.style args={#1,#2,#3,#4}{
				code={
					\path[midway] ([yshift=-5pt, xshift=#4]#1.south) edge [draw=none] node [chapnodeStyle] {\ref{#3}} ([yshift=5pt, xshift=#4]#2.north);
				}
			},
			%
			chapbox/.style={
				chapnodeStyle,
				align=center,
				minimum width=10pt,
				minimum height=10pt,
				yshift=-5pt,
			},
		}
		
		\node [trapnode, trapezium angle=-50, inner xsep=0pt] (SM) {(B)\SM model};
		\node [rectnode, yshift=-\minheight-2pt] (model) at (SM) {model parameters};
		\node [below=of model, rectnode] (incFS) {inclusive final state};
		\node [below=of incFS, yshift=30pt] (partKinCuts) {};
		\node [below=of incFS, rectnode, fill=titlebackcolor!90, draw=none] (partLevel) {fiducial final state:\\"particle level"};
		\node [below=of partLevel, rectnode, fill=titlebackcolor!90, draw=none] (detLevel) {reconstructed objects:\\"detector level"};
		\node [below=of detLevel, yshift=50pt] (detKinCuts) {};
		\node [below=of detLevel, rectnode] (readout) {detector signals};
		\node [trapnode, yshift=-\minheight-2pt, inner xsep=-15pt, trapezium angle=50] (detector) at (readout) {collider \& detector};
		
		\path[arrowStyle] ([xshift=\xdiff, yshift=5pt]readout.north) edge [draw=theoryred, path fading=south] node {} ([xshift=\xdiff, yshift=-5pt]detKinCuts.south);
		\path[arrowStyle] ([xshift=\xdiff, yshift=-10pt]detKinCuts.north) edge [draw=theoryred, path fading=south] ([xshift=\xdiff, yshift=-5pt]detLevel.south);
		\pic{uparrowLabel={readout,detKinCuts,{high-level\\reconstruction},sec:objReco,\xdiff}};
		\path[arrowStyle] ([xshift=-\xdiff, yshift=-10pt] detKinCuts.north) edge [draw=experimentalblue, path fading=south] ([xshift=-\xdiff, yshift=-5pt]detLevel.south);
		\path[arrowStyle] ([xshift=-\xdiff, yshift=-10pt] detKinCuts.north) edge [draw=none] node [uparrowtext] {kinematic cuts} node [chapnodeStyle,xshift=\xdiff] {\ref{sec:metJets}} ([xshift=-\xdiff, yshift=-5pt]detLevel.south);
		\pic{uparrowLabel={detLevel,partLevel,correction for\\detector effects,sec:metJets_detectorCorrection,0pt}};
		\path[arrowStyle] ([xshift=-\xdiff, yshift=5pt]partLevel.north) edge [dashed, draw=experimentalblue!70, path fading=south] ([xshift=-\xdiff, yshift=-5pt]incFS.south);
		\path[arrowStyle] ([xshift=-\xdiff, yshift=5pt]partLevel.north) edge [draw=none] node [uparrowtext, black!70] {extrapolation\\of phase space} ([xshift=-\xdiff, yshift=-5pt]incFS.south);
		\path[arrowStyle] ([xshift=-\xdiff, yshift=5pt]incFS.north) edge [dashed, draw=experimentalblue!70, path fading=south] ([xshift=-\xdiff, yshift=-5pt]model.south);
		\path[arrowStyle] ([xshift=-\xdiff, yshift=5pt]incFS.north) edge [draw=none] node [uparrowtext, black!70] {determination\\of parameters} ([xshift=-\xdiff, yshift=-5pt]model.south);
		
		\pic{downarrow={model,incFS,calculation\\of observables,sec:MC}};
		\pic{downarrow={incFS,partKinCuts,construction,sec:objReco}};
		\pic{downarrowGeneral={partKinCuts,partLevel,kinematic cuts,sec:metJets,10pt,5pt}};
		\path[arrowStyle] ([xshift=\xdiff, yshift=-5pt]partLevel.south) edge [dashed, draw=theoryred!70, path fading=north] ([xshift=\xdiff, yshift=5pt]detLevel.north);
		\path[arrowStyle] ([xshift=\xdiff, yshift=-5pt]partLevel.south) edge [draw=none] node [downarrowtext, black!70] {smearing} ([xshift=\xdiff, yshift=5pt]detLevel.north);
		\path[arrowStyle] ([xshift=5pt]incFS.east) edge [bend left=50,draw=theoryred, path fading=north] ([xshift=5pt]readout.east);
		\path[arrowStyle] ([xshift=5pt]incFS.east) edge [bend left=50,draw=none] node [downarrowtext] {detector\\simulation} node [chapnodeStyle, xshift=1pt] {\ref{sec:MC}} ([xshift=5pt]readout.east);
		
		\path[arrowStyle, black!70] ([xshift=-\xoffset]readout.center) edge [path fading=south] ([xshift=-\xoffset]partLevel.center);
		\path[arrowStyle, black!70] ([xshift=-\xoffset]readout.center) edge [draw=none] node [sloped, below] {detector model} ([xshift=-\xoffset]partLevel.center);
		\path[arrowStyle, black!70] ([xshift=-\xoffset-2\xdiff,yshift=-100pt]partLevel.center) edge [path fading=south] ([xshift=-\xoffset-2\xdiff]model.center);
		\path[arrowStyle, black!70] ([xshift=-\xoffset-2\xdiff]partLevel.center) edge [draw=none] node [sloped, below, path fading=south] {(B)\SM model} ([xshift=-\xoffset-2\xdiff]model.center);
		\path[midway,line width=5, dotted, black!20] ([xshift=-\xoffset-2\xdiff]readout.center) edge [path fading=south] ([xshift=-\xoffset-2\xdiff]partLevel.center);
		\path[arrowStyle, black!70] ([xshift=-\xoffset-2\xdiff]readout.center) edge [draw=none] node [sloped, above, path fading=south] {dependence on} ([xshift=-\xoffset-2\xdiff]model.center);
		
		\node[chapbox, xshift=46pt] at (SM) {\ref{sec:theory}};
		\node[chapbox, xshift=46pt] at (partLevel) {\ref{sec:interpretation}};
		\node[chapbox, xshift=55pt] at (detector) {\ref{sec:experiment}};
		
		\path[arrowStyle] ([xshift=-5pt]incFS.west) edge [bend right=50,draw=none] node [uparrowtext] {\rule{45pt}{0pt}} ([xshift=-5pt]readout.west);
		
		\node[chapbox, xshift=-100pt, yshift=-25pt] at (detector) {X};
		\node[xshift=-94pt, yshift=-30pt,right] at (detector) {: discussed in Chapter X};
	\end{tikzpicture}
\end{myfigure}

\bigskip
Starting at the bottom of the figure, the particle collider and used detector provide the fundament for data taking.
The representation of the data that is the closest to the detector and still meaningful is calibrated \textit{detector signals} without interpreting those with respect to actual physics objects.
Examples are energy depositions and trajectories of particle--detector interactions.
At this level, the smallest possible model assumptions are applied to experimental data.
This, however, makes even data-to-data comparisons difficult because the exact experimental conditions are run- and time-dependent.

Following the data corrections (blue arrows), these conditions are taken into account when reconstructing high-level physics objects, like electrons or photons.
Afterwards, kinematic cuts can be applied to select events fulfilling specific criteria.
This gives observables in \textit{detector-level} representation, \eg event counts as a function of momentum intervals.

The next step is to correct for various detector effects, which comprise misidentified objects, detector resolutions and reconstruction efficiencies.
All of these detector effects can affect every single event, but it is a posteriori impossible to know to which amount a specific event was influenced.
Contrary to calibrations which adjust directly the properties of physics objects, detector corrections are therefore applied statistically to ensembles of events.
After this step, a specific final state has been selected that is \textit{fiducial} as it is still restricted to selected kinematic regimes within the detector acceptance.
The fiducial final state represents physical observables at the so-called \textit{particle level}, \ie corresponding to the quantities of detector-stable particles originating from Monte-Carlo event-generation.
This can for example be a differential cross section as a function of momentum intervals.

Subsequently, extrapolation from the fiducial to the inclusive phase space can take place to remove the kinematic constraints, \eg in a variable in which the detector acceptance is limited, like the solid angle. In the example used before, the total cross section for a physics process could be obtained.

The last step is to determine specific parameters of the investigated model from the measured data.
This allows comparing measurement and prediction directly, without any further corrections having to be applied to the theory.
An example of this would be the extraction of the mass of a particle from the data which can be directly compared to its theoretical value.

The examined quantities become more model dependent with each of these processing steps: going from detector signals over detector level to particle level, increasingly better understanding of the detector has to be assumed, \eg of electronics, subsystems and efficiencies.
Until the particle level, only mostly unspecific assumptions about the physics model have to be made.
These assumptions can for example concern conservation laws or the interaction of particles with the detector.
This changes going from particle level over the inclusive final state to actual model parameters, where more and more assumptions about the physics model have to be built in.
This can for example be about fixed model parameters or the relation of observables in the observed to the unobserved phase space.

\bigskip
Starting at the top of \figref{fig:detectorCorrection} with a specific physics model, more and more processing steps for the theoretical model need to be performed for comparisons of data and prediction at a given level.
Conversely, less and less processing steps for the detector signals are needed.
Following the corrections for the prediction (red arrows), the first step is to calculate physics observables from the chosen model.
This gives processes in the inclusive final state.
To proceed towards the fiducial final state at particle level, physics objects have to be constructed, and kinematic cuts need to be applied.

Simulated detector signals can be obtained by employing detector simulation to the inclusive final state~\cite{GEANT4:2002zbu}.
Following this, high-level reconstruction and kinematic cuts can be applied to obtain a detector-level representation.
Alternatively, the detector-level representation can be reached by smearing the particle-level observables~\cite{deFavereau:2013fsa,Bierlich:2019rhm}. 

\bigskip

\figref{fig:detectorCorrection} can be used as an outline for the first parts of this thesis, starting at the outermost layers (top and bottom) and successively moving towards the inside (centre of the figure).
\chapref{sec:theory} presents the theoretical framework of modern particle physics, the Standard Model of particle physics, motivates the search for Dark Matter and introduces the \THDMa.
The collider used to produce proton--proton collisions and the detector measuring them in this thesis, the \LHC and the ATLAS Experiment, respectively, are described in \chapref{sec:experiment}.
\chapref{sec:MC} details how these collisions can be simulated and observables calculated.
Explanations on how to (re)construct and calibrate objects from detector signals and particle simulation are given in \chapref{sec:objReco}.
\chapref{sec:analysisPreservation} motivates comparing data and prediction at particle level. 
In \chapref{sec:metJets}, the \METjets final state is explored at detector level.
Detector effects in this final state are corrected for in \chapref{sec:metJets_detectorCorrection}, giving measurement results in the particle-level representation.
The contributions of the \THDMa to the \METjets final state at particle level are investigated in \chapref{sec:2HDMa_metJetsMeasurement}.
In \chapref{sec:interpretation}, the results of the measurement are interpreted with regard to the agreement between data and generated Standard-Model prediction as well as \THDMa.

Measurements that are corrected for detector effects and preserved in a suitable way allow for fast and easy reinterpretation.
This is demonstrated with the large amount of existing collider analyses for the \THDMa in \chapref{sec:Contur}.
The results obtained in this thesis are compared to existing results in \chapref{sec:comparison}.
Finally, \chapref{sec:conclusion} provides conclusions and an outlook into the future.

\Chapter[1]{Setting the stage}{To the Standard Model and beyond}{%
	Muse}{United States of Eurasia~\cite{Muse:2009use}}{verse 2, lines 3-4}
\label{sec:theory}



Physics strives to explain the universe at a fundamental level and unveil its guiding principles. \textit{Particle} physics consequently must be recognised as the field following this goal to the limit because it aspires to explain nature as being formed by elementary building blocks, \textit{matter}, that exact forces, \textit{interactions}, upon another. In this approach, all larger entities can be understood as composites constructed from these fundamentals.

Over the last decades, particle physics was thereby able to erect a finely detailed framework of matter particles and their interactions -- the Standard Model of particle physics (\SM). This framework has constantly proved astonishingly accurate: the Standard Model was able to successfully predict the existence of new particles, starting with the $Z$ boson~\cite{Glashow:1959wxa,Salam:1968rm,Weinberg:1967tq,UA1:1983mne,UA2:1983mlz} over the top quark~\cite{Kobayashi:1973fv,CDF:1995wbb,D0:1995jca} to the most recently discovered Higgs boson~\cite{Englert:1964et,Higgs:1964pj,Guralnik:1964eu,CMS:2012qbp,ATLAS:2012yve},
to name only a few. In addition, it also correctly predicts physics observables spanning 15 orders of magnitudes with unparalleled precision, as can be seen for production cross sections in \figref{fig:SM_crossSections}.


\begin{myfigure}{Production cross sections for various processes as predicted by the Standard Model (bars) compared to their measured values (symbols). Figure taken from \refcite{ATL-PHYS-PUB-2022-009}.}{fig:SM_crossSections}
	\includegraphics[width=0.9\textwidth]{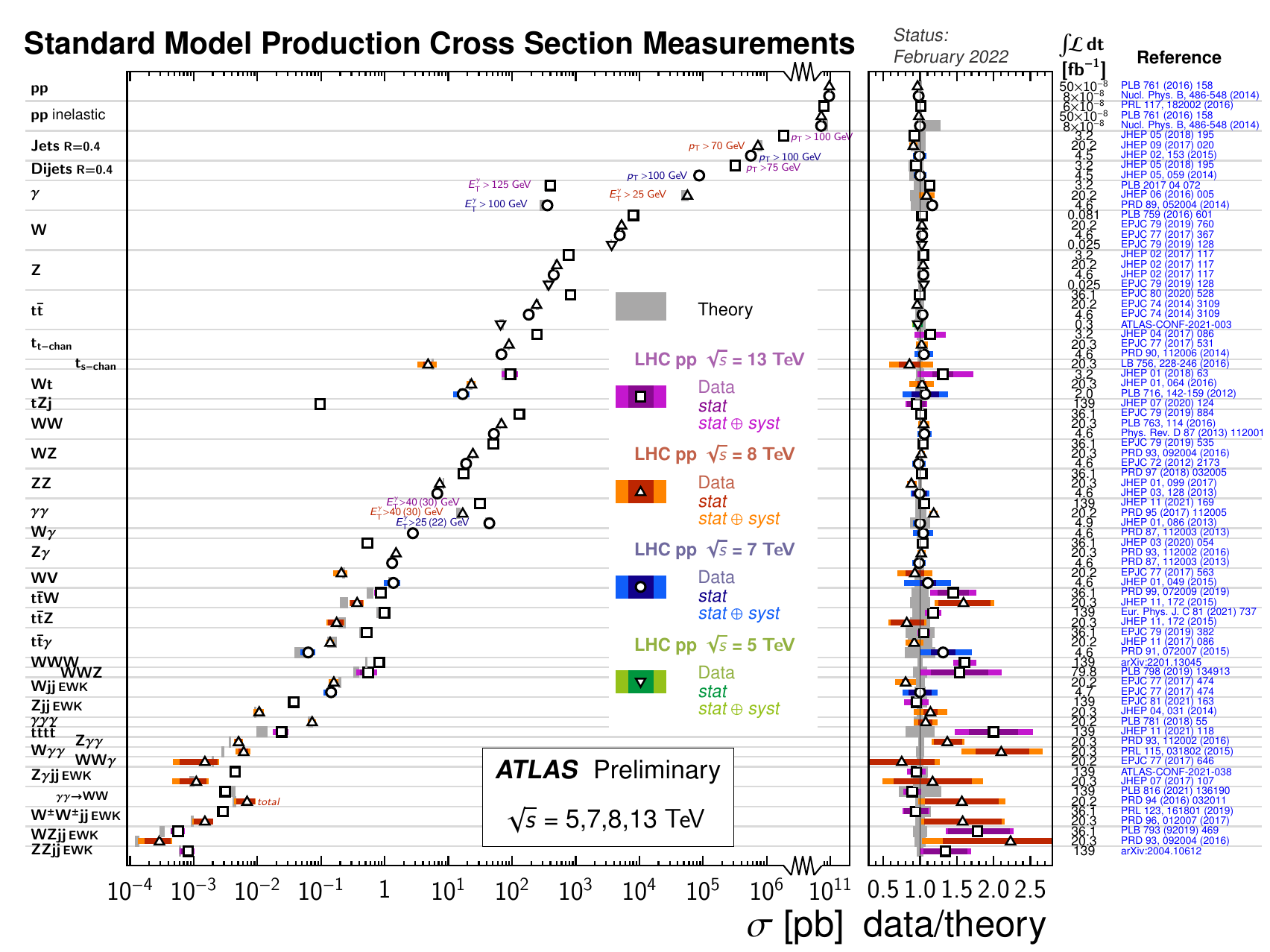}
\end{myfigure}

Despite this tremendous success, however, phenomena are observed that are unaccounted for in the Standard Model and require physics \textit{beyond the Standard Model} (\BSM) to explain them.
This extension of the framework has been attempted in plentiful ways, answering one or even multiple of the open questions. 

\secref{sec:SM} is dedicated to outlining the basic principles of the Standard Model, \secref{sec:BSM} to its shortcomings that are clear signs of \BSM physics.
Particular emphasis is spent in \secref{sec:DM} on the topic of Dark Matter, and in \secref{sec:2HDMa} on one of many attempts for a theory incorporating it, the two-Higgs-doublet model with a pseudoscalar mediator to Dark Matter~(\THDMa).
Both of these are focal points of this thesis.

\markright{The Standard Model of particle physics} 
\section{The Standard Model of particle physics (\SM)}
\markright{The Standard Model of particle physics}
\label{sec:SM}

As mentioned before, the Standard Model has two fundamental building blocks: matter particles and interactions.
For structural simplification, they are discussed separately in the following

\subsection{Matter particles}

In the Standard Model, all particles constituting matter are fermions, particles with\linebreak spin~$\sfrac{1}{2}$. This connection between spin and matter is in fact not accidental: traditionally, matter has been understood as anything that has a mass and a volume, \ie occupies space. While clear intuitively, this definition proves cumbersome when moving from the original extended, composite objects to point-like, elementary particles. Consequently, only particles following the Pauli exclusion principle, \ie fermions, can occupy space despite a point-like nature.

The fermions in the Standard Model can be subdivided by their interactions: \textit{quarks} are those fermions that posses colour charge and consequently interact strongly, \textit{leptons} those that do not~\cite{Griffiths:1987tj}. Both categories are then further distinguished by their electric charge.
There are leptons with electric charge $-1$, and for each a corresponding electrically uncharged lepton, a neutrino. For quarks, there are those with electric charge $+\sfrac{2}{3}$ and $-\sfrac{1}{3}$; up- and down-type quarks, respectively. 
Furthermore, the fermions are subdivided into three \textit{generations} with increasing mass, leading to 12 matter particles in total.
Generally speaking, all charged particles from the second and third generation are unstable because they decay to first-generation fermions on short timescales due to the difference in mass.

In addition to all the already mentioned fermions, there is an antiparticle with the same mass but opposite charges, \ie additive quantum numbers, for each of them. \figref{fig:SM_particles} shows an overview of the \SM particle-content, including the mentioned matter particles in the three left-most columns.

\begin{myfigure}{Tabular representation of the particle content of the Standard Model. Figure adapted from \refcite{Burgard:2016bug}.}{fig:SM_particles}
	\includegraphics[width=\textwidth]{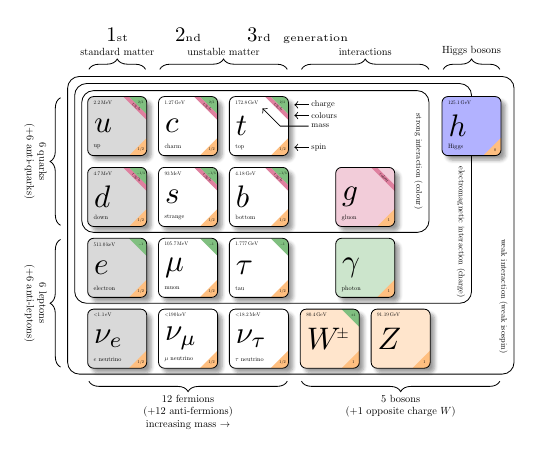}
\end{myfigure}

\subsection{Interactions}
The Lagrangian $\Lagr_{\SM}$ describing the behaviour of the quantum fields in the Standard Model can be conceptionally broken down into two components, the electroweak~(\EW) interaction and the strong interaction related to quantum chromodynamics~(\QCD):
\begin{equation*}
	\Lagr_{\SM} = \Lagr_{\EW}+\Lagr_{\QCD}.
\end{equation*}

The electroweak interaction is a unification of electromagnetism described by quantum electrodynamics (\QED) and weak interaction. Gravity, the fourth fundamental interaction, was not successfully integrated into the \SM framework yet although there has been considerable research on this topic~\cite{Polchinski:1998rq,Polchinski:1998rr,Ashtekar:2004eh}.
At particle colliders, however, gravitational interactions are weak enough to be negligible due to the low particle masses and particle densities.

\subsubsection{Electroweak theory}
\label{sec:SM_EW_theory}

Only a short overview of the electroweak theory is attempted here, for more complete reviews see \eg\refscite{Burgess:2006hbd,Nechansky:2021dxn}.
In the electroweak theory, two different fields are defined: $B$ follows the symmetry group \UOY and couples to particles according to their hypercharge
\begin{equation*}
	Y\coloneqq2\left(Q_\textnormal{em}-I_3\right).
\end{equation*}
Hereby, $Q_\textnormal{em}$ is a particle's electric charge and $I_3$ the third component of its weak isospin.
As fermions with left-handed chirality have weak isospins $I=\sfrac{1}{2}$ and right-handed fermions $I=0$, the coupling of $B$ to fermions therefore depends on their handedness.

Secondly, there are the fields $W_i,\conditionGap i\in\{1,2,3\}$ which couple exclusively to left-handed fermions and follow the symmetry group $SU(2)_L$. Quark flavours can mix through this interaction as
\begin{equation}
	\label{eq:SM_CKM}
	\begin{pmatrix}
		d'\\
		s'\\
		b'
	\end{pmatrix}
	=
	\begin{pmatrix}
		V_{ud}& V_{us}&	V_{ub}\\
		V_{cd}& V_{cs}&	V_{cb}\\
		V_{td}&	V_{ts}&	V_{tb}
	\end{pmatrix}
	\begin{pmatrix}
		d\\
		s\\
		b
	\end{pmatrix}
\end{equation}
which is called the Cabibbo--Kobayashi--Maskawa (\CKM) matrix~\cite{Cabibbo:1963yz,Kobayashi:1973fv}.
The mixing is formulated for down-type quarks here but can equivalently be defined for up-type quarks.
In \eqref{eq:SM_CKM}, $f\in\left\{d,s,b\right\}$ denote the mass eigenstates, and $f'$ the weak eigenstates that couple to up-type quarks.
$\abs{V_{ij}}^2$ is proportional to the transition probability of a quark of flavour $j$ to a quark of flavour $i$. As the \CKM matrix is unitary, its entries can be described by three real mixing angles and one complex phase. The latter has an important consequence: while in most interactions charge--parity (\CP) is conserved, \ie processes are identical if a particle is exchanged for its antiparticle ($C$ conjugation) while simultaneously inverting its spatial coordinates ($P$~transformation), in weak interactions it can be violated. For neutrinos, there exists a mixing mechanism corresponding to the \CKM matrix, the Pontecorvo--Maki--Nakagawa--Sakata (\PMNS) matrix~\cite{Pontecorvo:1957qd,Maki:1962mu}.

The $B$ and $W$ fields mix through electroweak symmetry breaking~\cite{Higgs:1964pj,Englert:1964et,Guralnik:1964eu} according to
\begin{equation*}
	\begin{pmatrix}
		W^+\\
		W^-\\
		Z\\
		\gamma
	\end{pmatrix}
	=
	\begin{pmatrix}
		\frac{1}{\sqrt{2}}& -\frac{i}{\sqrt{2}}&	0&		0\\
		\frac{1}{\sqrt{2}}& \frac{i}{\sqrt{2}}&		0&		0\\
		0&					0&						\cosW&	-\sinW\\
		0&					0&						\sinW&	\cosW
	\end{pmatrix}
	\begin{pmatrix}
		W_1\\
		W_2\\
		W_3\\
		B
	\end{pmatrix}
\end{equation*}
and the $W^\pm$, $Z$ and $\gamma$ bosons are observed as the actual mass and charge eigenstates. Hereby, \thetaW is the electroweak mixing angle which relates the masses of $W^\pm$ and $Z$ bosons as
\begin{equation*}
	m_W=m_Z\cosW.
\end{equation*}

While the combination of many measurements for $m_W$ has been in very good agreement with the \SM predictions so far~\cite{Zyla:2020zbs}, a recent result~\cite{CDF:2022hxs} has put significant strain on this compatibility.

Electroweak symmetry breaking occurs because when a complex weak isodoublet~$\Phi$, the Higgs field, is introduced, an additional potential
\begin{equation}
	\label{eq:SM_EW_Vpot}
	V\left(\Phi^\dagger\Phi\right)=-\mu^2\Phi^\dagger\Phi+\lambda\left(\Phi^\dagger\Phi\right)^2
\end{equation}
with $\lambda,\mu^2\in\mathbb{R},\conditionGap\lambda>0$ is obtained. This potential is invariant (\textit{symmetric}) under rotations in the phase of $\Phi$, resulting in a degenerate minimum at
\begin{equation}
	\label{eq:SM_EW_minimiseV}
	\Phi^\dagger_0\Phi_0=\frac{\mu^2}{2\lambda}
\end{equation}
if $\mu^2>0$.
Any specifically chosen ground state $\Phi_0$, however, is not invariant under this transformation (\textit{spontaneously breaks the symmetry}). As an example the choice
\begin{equation}
	\label{eq:SM_Higgs_groundState}
	\Phi_0 =
	\begin{pmatrix}
		0\\
		\frac{1}{\sqrt{2}}\left(v+h(x)\right)
	\end{pmatrix}
\end{equation}
can be made, where $h(x)$ is a real field and $v\coloneqq\frac{\mu^2}{\lambda}$ a real constant satisfying the minimising condition \eqrefNoLabel{eq:SM_EW_minimiseV}.
The latter is referred to as \textit{vacuum expectation value} and takes the value\footnote{%
	Natural units are used throughout this thesis.
} $v\approx\SI{246}{GeV}$.

Inserting the potential \eqrefNoLabel{eq:SM_EW_Vpot} with the chosen ground state \eqrefNoLabel{eq:SM_Higgs_groundState} into the Lagrangian for the electroweak interaction obtains the aforementioned masses for $W^\pm$ and $Z$ boson as well as that the mass for the photon~$\gamma$ vanishes.
Furthermore, an additional term
\begin{equation*}
	\Lagr\supset-\lambda v^2h^2
\end{equation*}
corresponding to a new particle arises: the Higgs boson with mass $\mh\coloneqq\sqrt{2\lambda}v$. In a last step, couplings between the fermionic fields $\psi$ and the Higgs field $\Phi$ are introduced. This gives a contribution
\begin{equation*}
	\Lagr\supset y_f\overline{\psi}\Phi\psi=\frac{y_fv}{\sqrt{2}}\overline{\psi}\psi+\frac{y_f}{\sqrt{2}}\overline{\psi}h\psi
\end{equation*}
for each species of fermions $f$. The fermion mass is identified as $m_f\coloneqq\frac{y_fv}{\sqrt{2}}$, giving
\begin{equation*}
	\Lagr\supset m_f\overline{\psi}\psi+\frac{m_f}{v}\overline{\psi}h\psi.
\end{equation*}
A coupling of a fermionic with a scalar field as seen in the second term is referred to as \textit{Yukawa coupling}~\cite{Yukawa:1935xg}. Most importantly, the coupling of fermions to the Higgs boson is therefore proportional to their mass.

\subsubsection{Quantum chromodynamics}
\label{sec:SM_QCD}

The strong interaction in the Standard Model is described by the field of quantum chromodynamics (\QCD).
The strong interaction follows the symmetry group $SU(3)_C$ and is mediated by \textit{gluons} which couple to the colour charge of particles. There are three different colour charges for particles: red, green and blue. Joining all three as well as colour--anticolour combinations result in colour-neutral states.

In the Standard Model, only quarks and the gluons themselves are colour charged.
The colour charge of gluons has an immediate impact on the strong coupling constant~$\alpha_s$:
while in \QED the interaction strength decreases with increasing distance between the particles, in \QCD the leading order contribution follows~\cite{Skands:2012ts}
\begin{equation}
	\label{eq:SM_QCD_scale}
	\alphas\left(Q^2\right)=\frac{12\pi}{\left(11N_c-2N_f\right)\ln\left(Q^2/\Lambda_{\QCD}^2\right)}
\end{equation}
which is also depicted in \figref{fig:alpha_s_coupling}. Hereby, $Q$ is the momentum transfer, $N_c=3$ the number of colours and $N_f\leq6$ the number of quark flavours with $2 m_f<\abs{Q}$. $\Lambda_{\QCD}=\order{\SI{0.3}{GeV}}$ is the \textit{\QCD scale}, the Landau pole at which the perturbatively derived theory breaks down. From \eqref{eq:SM_QCD_scale} it can be seen that for shorter distances (\ie $Q\rightarrow\infty$) the coupling decreases, $\alphas\rightarrow0$, which is called \textit{asymptotic freedom}.
For larger distances (\ie $Q\rightarrow\Lambda$), however, $\alphas\rightarrow\infty$ is obtained.
This means that as two colour-charged particles are separated, their interaction strength increases due to the self-coupling of gluons.
At one point, the invested energy is large enough to form intermediate \qqbar pairs.

\begin{myfigure}{
		Overview of measurements of \alphas as a function of the momentum transfer $Q$. The markers indicate the type of measurement and degree of \QCD perturbation theory used to extract the coupling. The black line corresponds to the theory prediction extrapolated from the world average of \alphas. Figure adapted from \refscite{Zyla:2020zbs,Skands:2012ts}.
	}{fig:alpha_s_coupling}
	\begin{tabular}{m{0.2\linewidth}m{0.5\linewidth}m{0.2\linewidth}}
		\begin{tikzpicture}[node distance=2.1cm, font=\scriptsize\linespread{0.8}, text centered]
			\node (A) {};
			\node [left=of A] (B) {};
			
			\draw[latex-,line width=2pt] (B) -- node [above=1pt] {Landau pole} (A);
			\path (B) -- node [below=1pt, text width=2.1cm] {\& confinement} (A);
		\end{tikzpicture} &
		\includegraphics[width=0.5\textwidth]{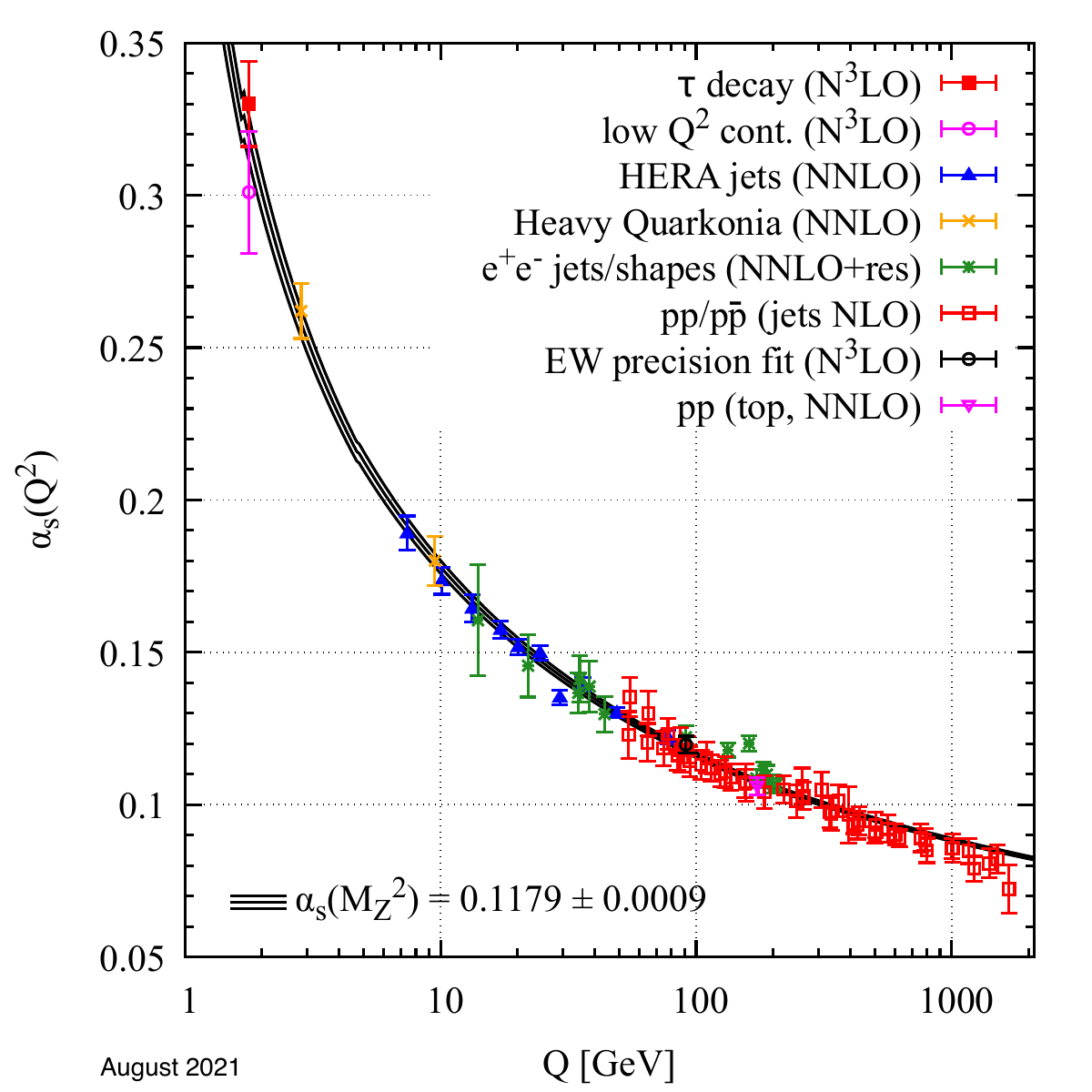} &
		\begin{tikzpicture}[node distance=2.1cm, font=\scriptsize\linespread{0.8}, text centered]
			\node (A) {};
			\node [right=of A] (B) {};
			
			\draw[-latex,line width=2pt] (A) -- node [above=1pt,text width=2.1cm] {asymptotic freedom} (B);
			\path (A) -- node [below=1pt, text width=2.1cm] {\& Grand Unified Theory?} (B);
		\end{tikzpicture}
	\end{tabular}
\end{myfigure}

This cycle can only be broken by forming colour-neutral objects, which is called \textit{confinement} of quarks to colour-neutral states. At large separation, only specific combinations of quarks are therefore found which have net-zero colour from the outside.
On the one hand, there are \textit{mesons} consisting of an equal number of valence quarks and antiquarks, with matched colour--anticolour pairs.
On the other hand, there are \textit{baryons} containing an odd number of (and at least three) valence quarks. The latter take advantage of the fact that combinations of all three colours red, green and blue are also colour-neutral. The hypernym of mesons and baryons is \textit{hadrons}. Historically, only \qqbar and $qqq$ states had been observed as mesons and baryons, respectively, but recently there have also been discoveries of tetra-~\cite{BESIII:2013ris,Belle:2013yex,LHCb:2021uow} and pentaquarks~\cite{LHCb:2015yax,LHCb:2019kea}.

In high-energy collisions, confinement comes with another consequence: while parts of the collision itself may be described as the interaction of bare quarks, when moving away from the collision point they start to form colour-neutral, composite particles, \ie hadrons, which is called \textit{hadronisation}.
Therefore, bare quarks cannot be observed in a particle detector but instead manifest as conical sprays of particles, termed \textit{jets}.
A more detailed discussion of hadronisation and jets follows in \secsref{sec:MCEG_hadronisation}{sec:objReco_jets}, respectively.

\subsection{Summary}

Combining electroweak theory and strong interaction,
\begin{equation*}
	SU(3)_C \otimes  SU(2)_L \otimes\UOY
\end{equation*}
is obtained as the symmetry group for the Standard Model. Following Noether's theorem~\cite{Noether1918} that for each symmetry there is a corresponding conservation law, conserved quantities in the Standard Model can be directly read off: colour charge $C$, weak isospin and weak hypercharge $Y$.

The particle content of the Standard Model is shown in \figref{fig:SM_particles}. Although there is considerable interplay between the various \SM parameters, not all of them can be imposed from theory. In total, there are 26 free parameters that have to be determined in experiments~\cite{Thomson:2013zua}, \eg the masses of the twelve fermions $m_f$, the Higgs mass $m_h$, the four mixing parameters of \CKM and \PMNS matrix each, the coupling strengths of the three considered interactions as well as the vacuum expectation value $v$ and \QCD vacuum angle. The latter is needed to account for the fact that no \CP violation has been observed in the strong interaction~\cite{Kim:2008hd}.

\section{Going beyond the Standard Model}
\label{sec:BSM}

As mentioned earlier, despite its astonishingly accurate predictions, the Standard Model provides an incomplete description of nature, as there are questions it lacks an answer for.
In general, the challenges of the Standard Model can be grouped into two categories: conceptional shortcomings, which concern the elegance and com\-pre\-hen\-sive\-ness of the \SM framework, and experimental shortcomings, which concern the consistency with observations.
The topics mentioned in the following are by far not exhaustive.

\subsubsection{Conceptional shortcomings}
\begin{itemize}
	\item In principle, the Standard Model could account for \CP violation also in the strong sector. The fact that there has been no \CP violation observed in \QCD to date is ad-hoc fixed in the Standard Model by setting the \QCD vacuum angle to zero.
	There might, however, be an underlying mechanism that the Standard Model simply does not take into consideration~\cite{Peccei:1977hh}.
	\item The electromagnetic and weak interactions have been successfully unified into the electroweak interaction. Naturally, the question arises whether a further unification of electroweak and strong interaction at high energy scales is realised~\cite{Georgi:1974sy,Buras:1977yy,Croon:2019kpe}.
	Models covering this approach are called \textit{Grand Unified Theories}.
	\item The Standard Model does not account for gravity. At the latest when considering the Planck scale -- the energy scale at which quantum effects of gravity become significant, which is at \order{\SI{e19}{GeV}} -- this needs to be addressed~\cite{Sevrin:2001jwl,Ashtekar:2004vs}.
	\item The Higgs field couples to particles according to their mass.
	In consequence, any (so far unobserved) particle with mass larger than of known \SM particles would yield a significant contribution to the Higgs mass \mh.
	This hypothetical particle's mass can, however, be anywhere up to the Planck scale.
	The Higgs mass \mh can therefore only be as close to the mass of the other \SM particles as it is if there is either considerable fine-tuning of the \SM parameters, a mechanism cancelling this other particle's contribution to the Higgs mass~\cite{Martin:1997ns} or a mechanism to reduce the Planck scale~\cite{Arkani-Hamed:1998jmv}.
\end{itemize}

\subsubsection{Experimental shortcomings}

\begin{itemize}
	\item The observable universe is made up almost exclusively of matter and not of antimatter.
	However, equal amounts of both should have been produced after the Big Bang in \CP conserving processes.
	While, as mentioned before, the Standard Model allows for \CP violating processes, these are \order{10^{10}} too small to explain the observed matter--antimatter asymmetry~\cite{Hou:2008xd}.

	\item In the Standard Model, neutrinos do not have mass.
	By now, however, it is firmly established that neutrinos have non-vanishing masses~\cite{Aitchison:2003tq,Mertens:2016ihw} although the exact figure and even mass ordering remains unclear~\cite{Super-Kamiokande:2010orq,KATRIN:2022ayy,Planck:2018vyg}.

	\item There are numerous discrepancies between \SM predictions and observations,\linebreak among others concerning lepton flavour universality in $B$ meson decays~\cite{BaBar:2012obs,Belle:2016dyj,LHCb:2023cjr}, the mass of the $W$ boson~\cite{CDF:2022hxs} and the anomalous magnetic dipole moment of muons~\cite{Muong-2:2021ojo}.
	
	\item There is tremendous evidence for Dark Matter (\DM), which the Standard Model has no suitable particle candidate for.
	As this can be considered the most concrete and thereby most pressing experimental shortcoming of the Standard Model, there has been considerable research on this topic in the last decades.
	Also this work focuses on shedding light on the \DM problem.
	A more detailed introduction to Dark Matter is given in the next section.
\end{itemize}

\markright{Dark Matter} 
\section{Dark Matter (\DM)}
\markright{Dark Matter}
\label{sec:DM}

The compelling evidence for Dark Matter is reviewed in \secref{sec:DM_evidence}.
Search strategies as well as possible theoretical models for it are discussed in \secsref{sec:DM_search_strategies}{sec:DM_models}, respectively.
Lastly, the signatures of \DM models at particle colliders are described in \secref{sec:DM_signature}.

\subsection{Observational evidence, nature and origin of Dark Matter}
\label{sec:DM_evidence}

There is a tremendous amount of as-of-yet unexplained observations for non-luminous matter in the universe. A few examples of these shall be given in the following:

\begin{itemize}
	\item Galaxies, being held together exclusively by gravitational pull, should follow Kepler's third law and thus the rotation velocity $v$ of matter in them roughly correspond to
	\begin{equation*}
		v(r)\approx\sqrt{G\frac{m(r)}{r}},
	\end{equation*}
	where $G$ is the gravitational constant, $r$ the distance of the considered object from the spin axis and $m(r)$ the mass enclosed between spin axis and object. Observations clearly indicate that the rotation velocities in galaxies do not follow this law if only the luminous components in the galaxies are considered~\cite{Rubin:1980zd,Begeman:1991iy} as shown in \figref{fig:GalaxyRotationCurve}.

	\item Galaxy clusters in a mechanically stationary state are expected to adhere to the virial theorem relating average kinetic energy $\braket{T}$ of galaxies to their average potential energies $\braket{U}$ as
	\begin{equation*}
		\braket{T}=-\frac{1}{2}\braket{U}.
	\end{equation*}
	Inferring back to the mass of the galaxies and the galaxy cluster from this relation clearly indicates that the observed luminous matter is too small to constitute the whole mass content~\cite{Zwicky:1933gu,Faber:1976sn}.
	
\begin{myfigure}{
		Rotation velocity $V_\mathrm{cir}$ of components in galaxy NGC~6503 as a function of their distance from the spin axis. Predictions from visible components (dashed line) and gas (dotted line) are drawn in. A good fit to the measured data (markers) is only achieved when a dark halo contribution (dash-dot line) is assumed, giving the total yield (solid line). Figure taken from \refcite{Begeman:1991iy}.
	}{fig:GalaxyRotationCurve}
	\includegraphics[width=0.7\textwidth]{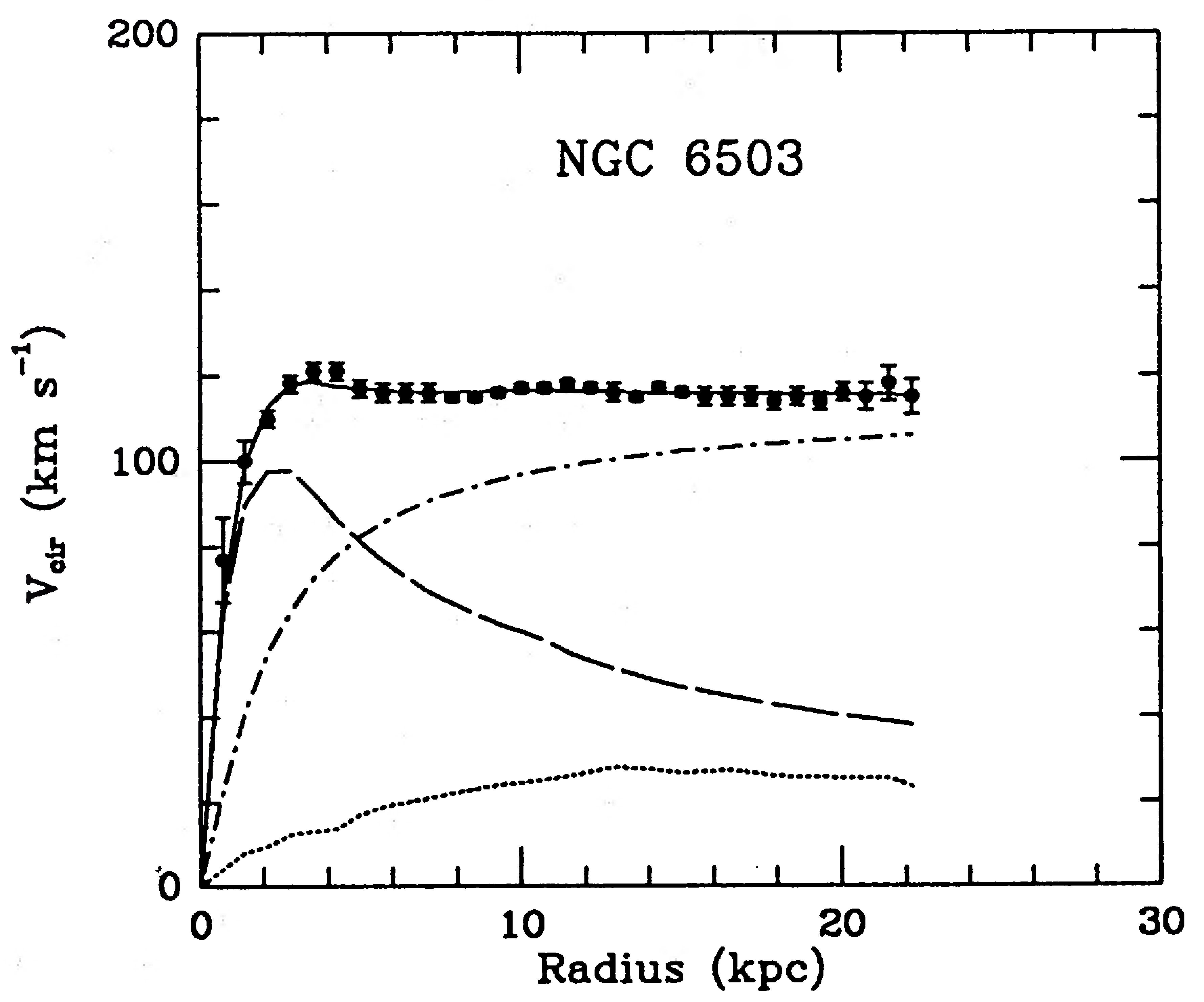}
\end{myfigure}

	\item Being massive, all matter bends spacetime and alters the trajectory of photons travelling from a distant source to the solar system.
	This effect is called \textit{gravitational lensing}. Measurements of the size and origin of these distortions are sensitive to the mass density located between the radiating objects and the observer, giving indications for non-luminous matter in the universe~\cite{vanWaerbeke:2000rm}.

	\item There are different methods to determine the mass of galaxy clusters: from the distribution of radial velocities of galaxies in the cluster, from the radiation of the gas in the cluster assuming a balance between gas pressure and gravity as well as from gravitational lensing. These clearly indicate the presence of non-luminous matter~\cite{Allen:2011zs}.

	\item Comparing the optical and X-ray spectrum of some galaxies or galaxy clusters with the mass distribution obtained from gravitational lensing effects shows clear separations or differences in orientation.
	This is the case \eg for the Bullet Cluster~\cite{Markevitch:2003at}, the Cosmic Horseshoe~\cite{Schuldt:2019vza} or the galaxy cluster MACSJ0025.4-1222~\cite{Bradac:2008eu}. The mass density and spectrum of electromagnetic radiation for the latter is shown in \figref{fig:MACSJ0025.4-1222}. Every single of these cases demonstrates that the mass density follows the motion of the luminous matter only to some degree and large fractions can in particular traverse collisionless.

\begin{myfigure}{
		Image of the galaxy cluster MACSJ0025.4-1222 in the optical regime. Overlaid are the contours for the X-ray brightness (yellow), I band (white) and mass density (red). The peak positions of the total mass and their errors are indicated by the cyan crosses. Figure taken from \refcite{Bradac:2008eu}.
	}{fig:MACSJ0025.4-1222}
	\includegraphics[width=0.7\textwidth]{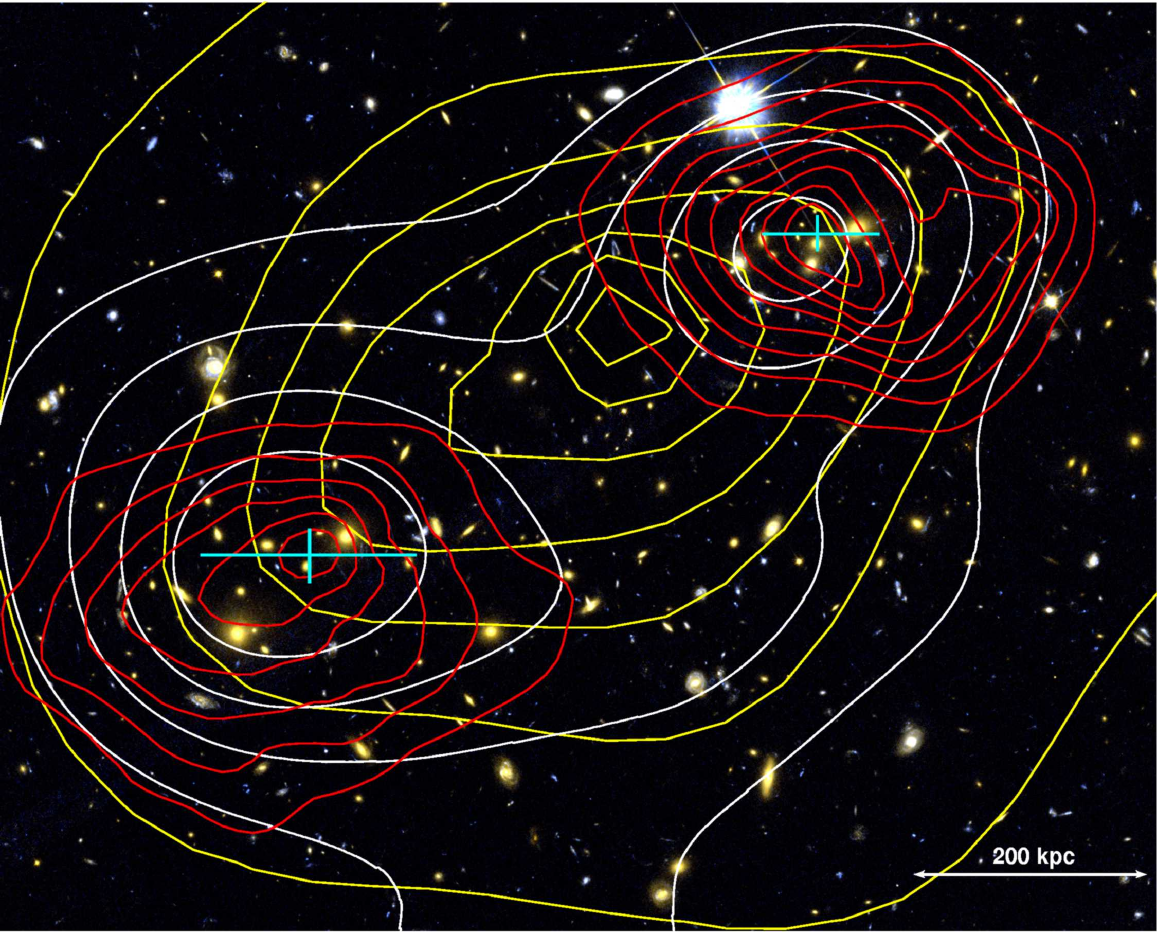}
\end{myfigure}

	\item In the early universe, photons and baryons were coupled in what can be modelled as a photon--baryon fluid.
	This fluid was subject to acoustic oscillations in gravitational wells.
	About \SI{380,000}{years} after the Big Bang, the universe had cooled through expansion so far that electrons and protons recombined to electromagnetically neutral atoms and photons decoupled~\cite{Brandenberger:1995qk}.
	This gives rise to the cosmic microwave background that can be measured today.
	Its anisotropy power spectrum is sensitive to the amount of matter that was and the amount that was not coupled to photons at the moment of their decoupling because of the acoustic oscillations.
	The anisotropy power spectrum strongly indicates the existence of massive matter particles not coupling to photons~\cite{Planck:2018vyg}.
\end{itemize}


All these observations point towards a similar kind of matter, which is given the name "Dark Matter" (\DM).
This matter must not have electric charge, thereby not interacting with photons and becoming non-luminous.
Further, it has to be weakly interacting in general, thereby becoming mostly collision-free.
The facts that it still exists in large amounts today and no signs of decays can be found from regions with high \DM concentration~\cite{Palomares-Ruiz:2007egs} indicate that it is stable on cosmological timescales~\cite{Covi:2009xn}.

It is the amount of independent observations that all indicate non-luminous matter with identical properties that makes the hypothesis about the existence of a \DM particle so convincing. The Standard Model does not offer a suitable candidate explaining all these observations~\cite{White:1983fcs,Ibarra:2015wfa,Gross:2018ivp,MACHO:2000qbb}.

One possible class of \DM candidates beyond the Standard Model are weakly-interact\-ing massive particles (\WIMPs) with masses \order{\textnormal{GeV}-\textnormal{TeV}}.
In the following, the term "Dark Matter" always refers to \WIMPs as these shall be the focus of this thesis.
There exist also other promising candidates~\cite{Zyla:2020zbs}, however, \eg axion-like particles~\cite{Bauer:2017ris}, dark photons~\cite{An:2014twa} and sterile neutrinos~\cite{Dodelson:1993je}.

In the early universe, \WIMPs could have been involved in perpetual production and annihilation processes. The expansion and gradual cooling of the universe would then lead to a decoupling of \WIMPs from the thermal bath when production processes become kinematically inaccessible as well as annihilations rare due to reduced density and overall small cross sections. This is called \textit{freeze out}. The \DM density measured today~\cite{Planck:2018vyg} as
\begin{equation*}
	\Omega_ch^2 = 0.1198 \pm 0.0012
\end{equation*}
would then be a relic of the early universe.
This value can be obtained when assuming \WIMP masses and couplings similar to the masses and couplings commonly observed at the weak scale, \order{\SI{100}{GeV}} and \order{1}, respectively~\cite{Kane:2008gb}.
This agreement is a compelling argument for the hypothesis of \WIMPs as Dark Matter. 

\subsection{Searching for Dark Matter}
\label{sec:DM_search_strategies}

In the hunt for Dark Matter, three general strategies can be distinguished, as is also depicted in \figref{fig:DM_search_strategies}.

In \textit{indirect detection}, indications of \DM existence can be found in \DM--\DM annihilations producing \SM particles.
These can be for example neutrinos, photons produced via quark-loops or antimatter like positrons and antiprotons.
They would dominantly originate from regions where the \DM density is high, \ie Earth, sun or galactic centre.
Experiments searching for Dark Matter with this strategy are for example ANTARES and IceCube~\cite{ANTARES:2020leh}, HESS~\cite{HESS:2018kom} or AMS~\cite{Kopp:2013eka}.

\begin{myfigure}{Interactions between \DM and \SM particles and their corresponding detection channels.}{fig:DM_search_strategies}
	\includegraphics[width=0.6\textwidth]{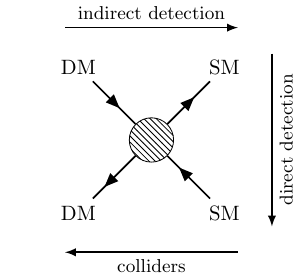}
\end{myfigure}

The gravitational binding of Dark Matter and ordinary matter would lead to the existence of \DM halos around galaxies.
This implies that Dark Matter could be found everywhere in the Milky Way and there is a steady stream of \DM particles through every volume on Earth due to the revolution of the Earth around the sun and of the sun around the galactic centre.
This is taken advantage of in \textit{direct detection}:
the scattering of baryonic matter off Dark Matter in dedicated experimental setups would lead to measurable recoil energies.
Examples for direct detection experiments are XENON1T~\cite{XENON:2018voc} and SuperCDMS~\cite{SuperCDMS:2014cds}.
The recoil energies are enhanced for higher masses of baryonic matter, rendering heavy atomic nuclei, like xenon or germanium, ideal targets.

The last search strategy is to produce Dark Matter directly in the collision of \SM particles at \textit{colliders}.
This strategy allows for very controlled and reproducible conditions and is the focus of this thesis.

\subsection{Theoretical models for Dark Matter}
\label{sec:DM_models}

The thermal properties of the \DM relic, \eg density and velocity distribution, are important for indirect and direct detection of Dark Matter.
At particle colliders, Dark Matter could be newly produced.
Therefore, the production mechanism is an important aspect when studying Dark Matter at colliders because it defines the kinematics of the Dark Matter and accompanying particles.
The production mechanism can vary depending on the theoretical model for Dark Matter that is assumed.

Models for Dark Matter come in variable complexity as sketched in \figref{fig:DM_model_classes}.
In principle, this classification of models can be performed for various topics.
Here, the focus shall be exclusively on how the corresponding model classes could explain Dark Matter.

\bigskip
The simplest approach are effective field theories (\EFT{}s).
\EFT{}s explaining Dark Matter only introduce a new \DM particle in addition to the \SM particle content and treat all interactions with the \DM particle as four-point contact interactions.
An example for this kind of interaction is shown in \subfigref{fig:DM_Feynman_diagrams}{a}.
A detailed explanation of why the radiation of an additional \SM particle is needed is given in \secref{sec:DM_signature}.
These models, like \DM-\EFT~\cite{Cao:2009uw}, are the most model-independent.
However, they are only valid for momentum transfers $Q$ up to a cut-off scale $\Lambda$.
Inaccuracies arise as soon as $Q\ll\Lambda$ is violated and the particles mediating the interaction between \DM and \SM particles can be resonantly produced.
This is the case at the large centre-of-mass energies at the \LHC~\cite{Busoni:2013lha,Busoni:2014sya,Busoni:2014haa}.

\begin{myfigure}{
		Overview of different classes of \DM models.
		The circles denote Dark-Matter particle (\DM, blue), mediator to Dark Matter ($a$, red) and other particles of the model ($X$ and $Y$, green).
		The examples given for the different classes of \DM models are Dark-Matter effective field theory (\DM-\EFT), a simplified model with a pseudoscalar mediator to Dark Matter (\DMP model), a two-Higgs-doublet model with a pseudoscalar mediator to Dark Matter (\THDMa) and supersymmetric theories.
	}{fig:DM_model_classes}
	\includegraphics[width=\textwidth]{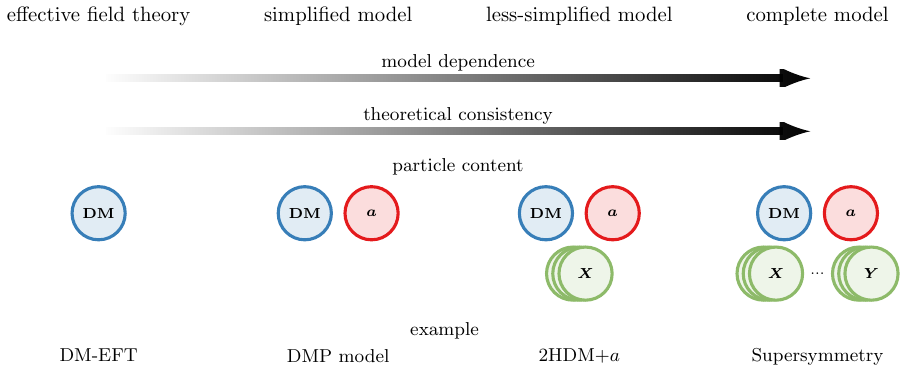}
\end{myfigure}

\myTwoFeynmanFigure{%
		Feynman diagrams for the production of \DM particles$~\chi$ in (a) an effective field theory and (b) in a model with a mediator~$a$ to Dark Matter.
	}{fig:DM_Feynman_diagrams}{%
		EFT%
	}{%
		DMP/SMSM_to_a_to_xx+ISRphoton%
}

The other end of the scale is marked by \textit{complete} models.
They are valid to arbitrarily high energies.
Their phenomenology is considerably more complex as they often come with an extensive number of new particles and parameters that need to be constrained.
They can commonly solve multiple of the shortcomings of the Standard Model.
If they provide \DM candidates, this is a natural consequence of the theory and not by construction.
Supersymmetry~\cite{Martin:1997ns} is a famous example for a class of complete models.

In between these two extremes lie simplified and less-simplified models.
The former introduce exactly the mediator to Dark Matter in addition to the actual \DM particle to overcome the problems of \EFT{}s.
An example for this kind of interaction is shown in \subfigref{fig:DM_Feynman_diagrams}{b}.
Simplified models do, however, violate gauge invariance if the centre-of-mass energy of an interaction becomes large~\cite{LHCDarkMatterWorkingGroup:2018ufk}, leading to divergences.
A simplified model could for example only introduce a pseudoscalar mediator and the actual \DM particle (\DMP model)~\cite{Abercrombie:2015wmb}.

Less-simplified models strive to have the minimally necessary theoretical framework to be gauge invariant and valid to arbitrarily high energies in addition to offering an explanation for Dark Matter.
Their theoretical framework is therefore more complex than for simplified models but has a reduced amount of particles and parameters compared to complete models.
As such, they form a compromise between theoretical consistency and model complexity.
The two-Higgs-doublet model with a pseudoscalar mediator to Dark Matter (\THDMa) that is the focus of this thesis is such a less-simplified model.
The \THDMa is detailed in \secref{sec:2HDMa}.

\subsection{Signatures of Dark-Matter models at colliders}
\label{sec:DM_signature}

The nominal signature of \DM models at colliders if the \DM particle itself is produced in the collision is shown in \subfigref{fig:DM_MET}{a}.
\DM particles, however, have a low probability to interact with the material of the detectors at the collider and can leave the detector completely unobserved.
In principle, momentum conservation can be taken advantage of to infer the presence of \DM particles in that case.
No indication can be observed that any collision happened at all, however, if only invisible particles are in the event.
This is demonstrated in \subfigref{fig:DM_MET}{c} on the example of the production of a mediator~$a$ decaying to \DM particles (\cf\subfigref{fig:DM_MET}{a}).

It is therefore necessary that the invisible particles recoil against some visible final-state object, like a photon or Higgs boson, as in \subfigref{fig:DM_MET}{b}.
If the visible particles have a total momentum $\vv p$, the total momentum of the invisible particles is known to be $\vv{p}^\mathrm{miss}\coloneqq-\vv p$ (\cf\subfigref{fig:DM_MET}{d}).
More details on this quantity are given in \secref{sec:objReco_MET}.

\begin{myfigure}{
		Top: Feynman diagrams for the production of an invisibly decaying mediator~$a$ (a) without and (b) with radiation of a \SM particle.
		Bottom: Schematic of the corresponding final state which (c) does not produce a visible signal in a detector at all and (d) is a \SM particle recoiling against missing momentum $\vv{p}^\mathrm{miss}$, respectively.
		Solid (dashed) black arrows are visible (invisible) to the detector, red arrows represent observables.
	}{fig:DM_MET}
	\begin{tabular}{cc}
		\subfloat[]{\includegraphics[width=\twoFeynmanWidth]{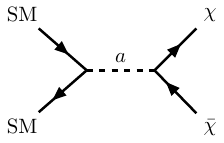}\hspace{10pt}} &
		\subfloat[]{\includegraphics[width=\twoFeynmanWidth]{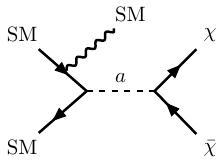}}\\
		\subfloat[]{\includegraphics[width=0.49\textwidth]{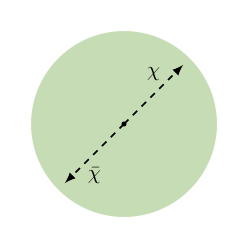}} &
		\subfloat[]{\includegraphics[width=0.49\textwidth]{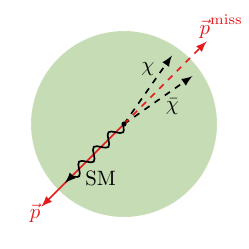}}
	\end{tabular}
\end{myfigure}

\myTwoFeynmanFigure{%
		Feynman diagrams for a \BSM particle decaying to (a) \SM particles and (b) a \SM particle and another \BSM particle.
	}{fig:DM_Feynman_diagrams_SM}{%
		DMP/SMSM_to_a_to_SMSM%
	}{%
		2HDMa/SMSM_to_A_to_ah%
}

\bigskip
In addition, \DM models that introduce more particles than only the \DM candidate, \ie simplified, less-simplified and complete models, give also rise to other important final states:
the resonant production of the mediator with subsequent decays of it back to \SM particles (\cf\subfigref{fig:DM_Feynman_diagrams_SM}{a}) and decays of it to other particles introduced by the theory (\cf\subfigref{fig:DM_Feynman_diagrams_SM}{b}).
These final states are then also sensitive to the \DM model.

\section{The Two-Higgs-doublet model with a pseudoscalar mediator to Dark Matter (\THDMa)}
\markright{The Two-Higgs-doublet model with a pseudoscalar mediator to Dark Matter}
\label{sec:2HDMa}
Dark Matter by definition has to be a massive particle and as such couple to the Higgs sector.
Measurements of the decay of the \SM Higgs $h$, however, show that the branching fraction of $h$ decaying invisibly, \ie to particles that escape detectors unobserved, $\brf{h\to\text{invisible}}<0.11$~\cite{ATLAS-CONF-2020-052}. Similarly, the branching fraction of undetected $h$ decay channels, which also takes into account processes analyses are currently not sensitive to, like decays to light quarks, $\brf{h\to\text{undetected}}<0.12$~\cite{ATLAS:2022vkf}. This means that the necessary \DM abundance cannot be achieved if there is a direct coupling between Dark Matter and \SM Higgs boson~\cite{Bauer:2017ota}. The Higgs sector has to be expanded.

The option to extend the Higgs sector chosen for this work, two-Higgs-doublet models, is introduced in \secref{sec:2HDMs}.
An explanation how these can be extended to solve the problem of Dark Matter is given in \secref{sec:2HDMa_addA}.

\subsection{Two-Higgs-doublet models (\THDMs)}
\label{sec:2HDMs}

In \secref{sec:SM_EW_theory}, the simplest possible choice for the Higgs sector was made for the Standard Model, consisting of a single scalar doublet and accordingly a potential following \eqref{eq:SM_EW_Vpot}.
Given current knowledge there is freedom regarding the exact choice of multiplicity and kind of Higgs multiplets, however.
Two-Higgs-doublet models (\THDMs) are some of the simplest extensions to the \SM Higgs-sector, introducing only a single additional Higgs doublet.
They are of particular interest because they appear also in different complete theories, like supersymmetrical or Peccei--Quinn models~\cite{Branco:2011iw}.
In principle, also more complex extensions are imaginable, however, for example involving one or more Higgs singlets or triplets.

Adding a second Higgs doublet generally can give rise to flavour-changing neutral currents, \eg $\bar{s}+d\to h \to s+\bar{d}$, at tree level which would have been detected.
In consequence, only one of the Higgs doublets is allowed to couple to fermions of a given charge~\cite{Glashow:1976nt}.
Different \textit{types} of \THDMs are distinguished depending on which doublet couples to up-type quarks, down-type quarks and leptons.
Most importantly there are type-I, type-II, lepton-specific and flipped \THDMs~\cite{Branco:2011iw}.
The most common choice -- referred to as \textit{type II} -- is to couple the first Higgs doublet $\Phi_1$ to down-type quarks as well as leptons and the second Higgs doublet $\Phi_2$ only to up-type quarks because this is the structure that also emerges in supersymmetric models.
This is also the choice adopted in this thesis.
The insights obtained here for the type-II \THDM can, however, given sufficient scrutiny, also be transferred to other types of \THDMs.

\subsubsection{Particles and parameters for the \THDM}
\label{sec:2HDM_parameters}

\begin{myfigure}{
		Tabular representation of the particle content of the \THDMa.
		The masses of all particles are free parameters of the model (to be determined, t.b.d.).
		The sole exception is the \SM-like Higgs boson~$h$ which takes the mass of the \SM Higgs boson in the alignment limit.
	}{fig:2HDMa_particles}
	\includegraphics[width=0.6\textwidth]{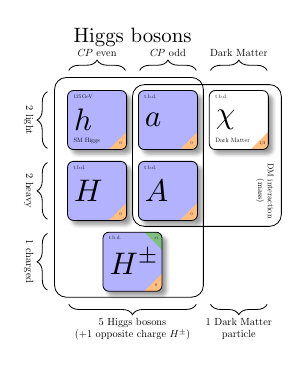}
\end{myfigure}

Four physical particles are obtained when changing to the mass eigenbasis in this model: a light and a heavy \CP-even Higgs boson ($h$, $H$), a \CP-odd Higgs boson ($A$) as well as a charged Higgs boson \Hpm.
These are also shown in \figref{fig:2HDMa_particles}.
In this basis, the parameters steering the model are

\begin{itemize}
	\item the masses of these four particles, \mh, \mH, \mA and \mHpm.
	\item the vacuum expectation values $v_1$ and $v_2$ of the Higgs doublet $\Phi_1$ and $\Phi_2$, respectively.
	\item the mixing angle $\alpha$ of the \CP-even weak eigenstates.
	\item three Higgs quartic couplings \lT, \lPO and \lPT.
\end{itemize}

In a more common notation, the electroweak vacuum expectation value $v$ is identified as $v=\sqrt{v_1^2+v_2^2}=\SI{246}{GeV}$ and $\tanB\coloneqq\frac{v_1}{v_2}$ is defined as the ratio of the vacuum expectation values. The degree of freedom for $\alpha$ can then be expressed as \cosBA.
The default choices for the parameters are summarised in the upper part of \tabref{tab:LHCDMWG_params}.

Fits to the measured rates of the \SM Higgs boson production and decay exclude most of the parameter space for $\cosBA\neq0$~\cite{ATLAS-CONF-2021-053,Haller:2018nnx}.
In this work, the \textit{alignment limit} $\cosBA\equiv0$ is therefore adopted which corresponds to the lightest \CP-even Higgs boson $h$ having \SM-like couplings and mass $\mh=\SI{125}{GeV}$.

The widths for the decays of the scalar~$H$ and the pseudoscalar~$A$ become large if $\abs{\lT-\lPO}\gg0$ or $\abs{\lT-\lPT}\gg0$~\cite{Bauer:2017ota}.
This would mean that the production of a particle cannot be separated from its decay which complicates computations (see \secref{sec:MCEG_hadronisation}).
For simplification, $\lambda'\coloneqq\lTeqlPOeqlPT$ is therefore adopted in this work.
Furthermore, perturbativity enforces $\lT<4\pi$.
At the same time, bounding the scalar potential from below requires $\lT\gg0$~\cite{Gunion:2002zf}, \eg $\lambda>2$ has to be fulfilled to allow $\mH>\SI{1}{TeV}$~\cite{LHCDarkMatterWorkingGroup:2018ufk}.
As a compromise, $\lambda'=\lT=3$ is used in this work.

Electroweak precision constraints, among others of the mass of the $W$ boson, require that either $\mHeqmHpm$ or $\mA\equiv\mHpm$~\cite{Haber:1992py,Gerard:2007kn,Haber:2010bw,Haller:2018nnx}.
For simplification, \mAeqmHeqmHpm is adopted in this thesis.
This has the additional advantage that the conclusions drawn in this work using \mAeqmHeqmHpm can be applied to different supersymmetric theories~\cite{Djouadi:2005gj}.
Measurements of $b\to s\gamma$ decays exclude $\mHpm<\SI{570}{GeV}$ independent of \tanB~\cite{Hermann:2012fc,Misiak:2015xwa,Misiak:2017bgg}.
In this work, $\mAeqmHeqmHpm=\SI{600}{GeV}$ is therefore used as default value.
The whole range of $\SI{200}{GeV}<\mAeqmHeqmHpm<\SI{2000}{GeV}$, however, is explored in different scans.

Decays of hadrons $B_s\to\mu^+\mu^-$ exclude extreme values of \tanB, \ie $\tanB\ll1$ as well as $\tanB\gg10$~\cite{Logan:2000iv,LHCb:2017rmj}.
In addition, perturbativity requires $\tanB>0.3$~\cite{Branco:2011iw}.
In this work, $\tanB=1$ is therefore used as a default value.
This has the additional advantage that the conclusions drawn in this work using $\tanB=1$ can be applied to different types of \THDMs~\cite{LHCDarkMatterWorkingGroup:2018ufk}.
The whole range of $0.5<\tanB<40$, however, is explored in different scans.

\begin{table}
	\centering
	\THDM parameters\\
	\begin{tabular}{cccc}%
		\toprule
		\mAeqmHeqmHpm & \tanB & \cosBA & $\lambda'\coloneqq\lTeqlPOeqlPT$\\
		\midrule
		\SI{600}{GeV}& 1& 0&	3\\ 
		\bottomrule
	\end{tabular}%
	\vspace{10pt}\\
	additional \THDMa parameters\\
	\begin{tabular}{cccc}%
		\toprule
		\ma & \mX & \sinP & \yX\\
		\midrule
		\SI{250}{GeV} & \SI{10}{GeV}& 0.35& 1\\ 
		\bottomrule
	\end{tabular}%
	\caption{%
		Default parameter settings used in the \THDMa model.
		The upper part gives the parameters for the \THDM, the lower part the additional parameters when adding a pseudoscalar mediator to Dark Matter.
	}
	\label{tab:LHCDMWG_params}
\end{table}

\subsection{Adding a pseudoscalar mediator to Dark Matter}
\label{sec:2HDMa_addA}

\THDMs can be extended in numerous ways to incorporate \DM candidates.
The \DM particle and mediator introduced by a theoretical model can be systematically classified by their spin and \CP properties.
Possible options are reviewed in the following.

For the \DM particle, real or complex scalars (spin~0), Dirac or Majorana fermions (spin~$\sfrac{1}{2}$) and neutral vectors (spin~1) are commonly considered~\cite{Abdallah:2015ter}.
For fermionic Dark Matter, it can be further distinguished whether there is a distinct antiparticle (Dirac fermion) or it is its own antiparticle (Majorana fermion).
While for direct and indirect detection the phenomenology can drastically change because of the possible spin-dependence of interactions, none of these classifications plays a role to first order at particle colliders~\cite{Abdallah:2015ter,Abercrombie:2015wmb}.
In this work, the \DM particle is chosen to be a Dirac fermion.

For the mediator, scalars (spin~0, \CP even), pseudoscalars (spin~0, \CP odd), vectors (spin~1, \CP even) and axial-vectors (spin~1, \CP odd) are commonly considered~\cite{Abercrombie:2015wmb}.
To avoid flavour constraints~\cite{Buras:2000dm}, spin-0 mediators are commonly assumed to have Yukawa couplings, spin-1 mediators to have flavour-universal couplings.
The spin of the mediator does therefore impact the production cross section at particle colliders and all options have to be investigated systematically.
In this work, the focus shall be on a spin-0 mediator.
The \CP properties of the mediator do not matter significantly at particle colliders.
This is different to direct-detection experiments where the \SM--\DM coupling becomes spin-dependent and the scattering effectively only occurs off unpaired nucleons for \CP-odd mediators~\cite{Goodman:1984dc}.
Direct detection therefore in general is very sensitive to spin-independent couplings, \ie\CP-even mediators, and considerably less for spin-dependent couplings, \ie\CP-odd mediators.
Particle-collider results are therefore more important for models with \CP-odd mediators.
This is therefore the choice made for this work.
In total, the focus shall be on \CP-odd mediators with spin~0, \ie pseudoscalar mediators.

\bigskip

In summary, the choice is to extend a two-Higgs-doublet model with a pseudoscalar mediator to a fermionic Dirac \DM particle.
This model is named \THDMa.
Dark Matter couples to \SM particles via mixing of the pseudoscalar mediator with the \CP-odd Higgs boson~$A$ introduced by the \THDM framework.

\subsubsection{Particles and parameters for the \THDMa}
\label{sec:2HDMa_parameters}

The \THDMa adds two additional particles to the \THDM.
In the mass eigenbasis, they are another \CP-odd Higgs boson ($a$) and a \SM-singlet, stable fermion ($\chi$).
These are also shown in \figref{fig:2HDMa_particles}.
In the mass eigenbasis, the additional parameters steering the model are

\begin{itemize}
	\item the masses of the two additional particles, \ma and \mX.
	\item the mixing angle~$\theta$ of the \CP-odd weak eigenstates.
	\item the coupling of the mediator to Dark Matter, \yX.
\end{itemize}

Bounding the scalar potential from below requires $\sinP\ll1$~\cite{Gunion:2002zf}, \eg $\sinP<\frac{\sqrt{2}}{2}\approx0.7$ has to be fulfilled to allow $\mH>\SI{1}{TeV}$~\cite{LHCDarkMatterWorkingGroup:2018ufk}.
At the same time, the production cross section for \DM particles vanishes for $\sinP\to0$.
As a compromise, $\sinP=0.35$ is adopted in this work.

Very small masses of the pseudoscalar~$a$ allow decays of the \SM Higgs boson $h\to aa$.
This can lead to a completely new phenomenology and in general be in conflict with the observed branching fraction for invisible and undetected decays of the Higgs boson, as mentioned before.
There are dedicated searches for these Higgs boson decays~\cite{ATLAS:2021ykg}.
At the same time, the decays $H\to aZ$ and $A\to ah$ are phenomenologically interesting and require $\ma\lessapprox \mA-\mh$.
In this work, $\ma=\SI{250}{GeV}$ is therefore used as a default value.
The whole range of $\SI{100}{GeV}<\ma<\SI{800}{GeV}$, however, is explored in different scans.

Sizeable branching fractions \brf{a\to\xx} and correspondingly production cross sec\-tions for Dark Matter for all considered values of \ma are obtained as long as $\mX<\frac{\ma}{2}$.
Apart from this, the exact choice of the mass of the \DM particle~\mX or its coupling to the mediator~\yX do not matter significantly in collider phenomenology~\cite{LHCDarkMatterWorkingGroup:2018ufk}.
In this work, $\mX=\SI{10}{GeV}$ and $\yX=1$ are adopted.

The default choices for the parameters added by extending the \THDM with a pseudoscalar mediator to Dark Matter are given in the lower part of \tabref{tab:LHCDMWG_params}.
The \LHC Dark-Matter working group (\LHCDMWG)~\cite{LHCDarkMatterWorkingGroup:2018ufk} is a symposium of experimentalists from the ATLAS and CMS collaborations as well as theorists striving to formulate the minimal basis of \DM models to influence the design of searches for Dark Matter at the \LHC~\cite{Abercrombie:2015wmb}.
The parameter choices outlined here and in \secref{sec:2HDM_parameters} are in line with the recommendations of the \LHCDMWG~\cite{LHCDarkMatterWorkingGroup:2018ufk}.

\subsection{Couplings and decays in the \THDMa}
The most relevant scalar--fermion interactions in the \THDMa can be expressed by~\cite{Bauer:2017ota}
\begin{equation}
	\begin{aligned}
		\label{eq:2HDMa_Lagrangian}
		\Lagr\supset
		&-\frac{m_t}{v}\bar{t}\left[h+\xi_tH+i\xi_t\left(\cosP\,A-\sinP\,a\right)\gamma_5\right]t\\
		&-\sum_{f=b,\tau}\frac{m_f}{v}\bar{f}\left[h+\xi_fH+i\xi_f\left(\cosP\,A-\sinP\,a\right)\gamma_5\right]f\\
		&+\frac{V_{tb}}{v}H^+\left(m_b\xi_b\bar{t}_Lb_R-m_t\xi_t\bar{t}_Rb_L\right)+\mathrm{h.c.}\\
		&-i\yX\left(\sinP\,A+\cosP\,a\right)\bar{\chi}\gamma_5\chi.
	\end{aligned}
\end{equation}
Hereby, $V_{tb}$ is an entry of the \CKM matrix as introduced in \secref{sec:SM_EW_theory}.
The masses of fermion fields $f'\in\left\{t,b,\tau\right\}$ are denoted $m_{f'}$.
In a type-II \THDM, $\xi_{f'}$ depends on the type of fermion:
\begin{equation}
	\label{eq:2HDMa_xi}
	\xi_t\coloneqq-\cotB\hspace{40pt}\xi_b\equiv\xi_\tau\coloneqq\tanB.
\end{equation}
Most importantly, the couplings of the uncharged new bosons $H$, $a$ and $A$ to bottom quarks and taus are therefore proportional to \tanB while their coupling to top quarks is inversely proportional to it.
For the charged boson \Hpm, the coupling is to bottom and top quarks simultaneously and there are terms proportional as well as inversely proportional to \tanB.
The couplings of the Higgs bosons to other \SM fermions are negligible because the coupling strength is always proportional to the fermion mass, which is largest for top quarks, bottom quarks and taus.

In the following, a brief overview for the most important couplings and decay widths of the new Higgs bosons is given.
These are helpful to understand the model and the phenomenology in \chapsref{sec:2HDMa_metJetsMeasurement}{sec:Contur}.
All formulas are taken from \refcite{Bauer:2017ota}.

\subsubsection{Light pseudoscalar $a$}
The field $a$ decays at tree level most importantly only to pairs of \DM particles or fermions:
\begin{align}
	\notag
	\width[a\to\xx]=&\frac{\yX^2}{8\pi}\ma\beta_{\chi/a}\cosPSq\\
	\label{eq:2HDMa_Gamma_a_ff}
	\width[a\to f\bar{f}]=&\frac{N_c^f\xi_f^2}{8\pi}\frac{m_f^2}{v^2}\ma\beta_{f/a}\sinPSq
\end{align}
where
\begin{equation*}
	\beta_{i/a}\coloneqq\sqrt{1-\tau_{i/a}}
\end{equation*}
is the velocity of the particle $i$ in the rest frame of the final-state pair and
\begin{equation*}
	\tau_{i/a}\coloneqq4\frac{m_i^2}{\ma^2}.
\end{equation*}
$N_c^f$ denotes the number of relevant colour factors, which is 3 for quarks and 1 for leptons.
In practise, the pseudoscalar~$a$ almost exclusively decays to \xx as long as $\ma>2\mX$, as can be seen in \figref{fig:2HDMa_BR_a}.
There are contributions from $a\to\ttbar$ as well as from cascade decays of the pseudoscalar~$a$ to one of the other \BSM bosons and a \SM boson if kinematically allowed.

\newcommand{\myBRFigure}[2]
{%
	\begin{myfigure}{
			The largest branching fractions~$\mathcal{B}_f$ for decays of the #1 as a function of \ma for different values of \mAeqmHeqmHpm.
			All parameters are set according to \tabref{tab:LHCDMWG_params} if they are not varied.
			The branching fractions are obtained using the generation setup described in \secref{sec:MC_2HDMa_Contur}.
		}{fig:2HDMa_BR_#2}
		\includegraphics[width=0.8\textwidth]{figures/Contur/xsecBR/mA_sinP0.35/BR_#2.pdf}
	\end{myfigure}
}
\myBRFigure{pseudoscalar~$a$}{a}
\myBRFigure{pseudoscalar~$A$}{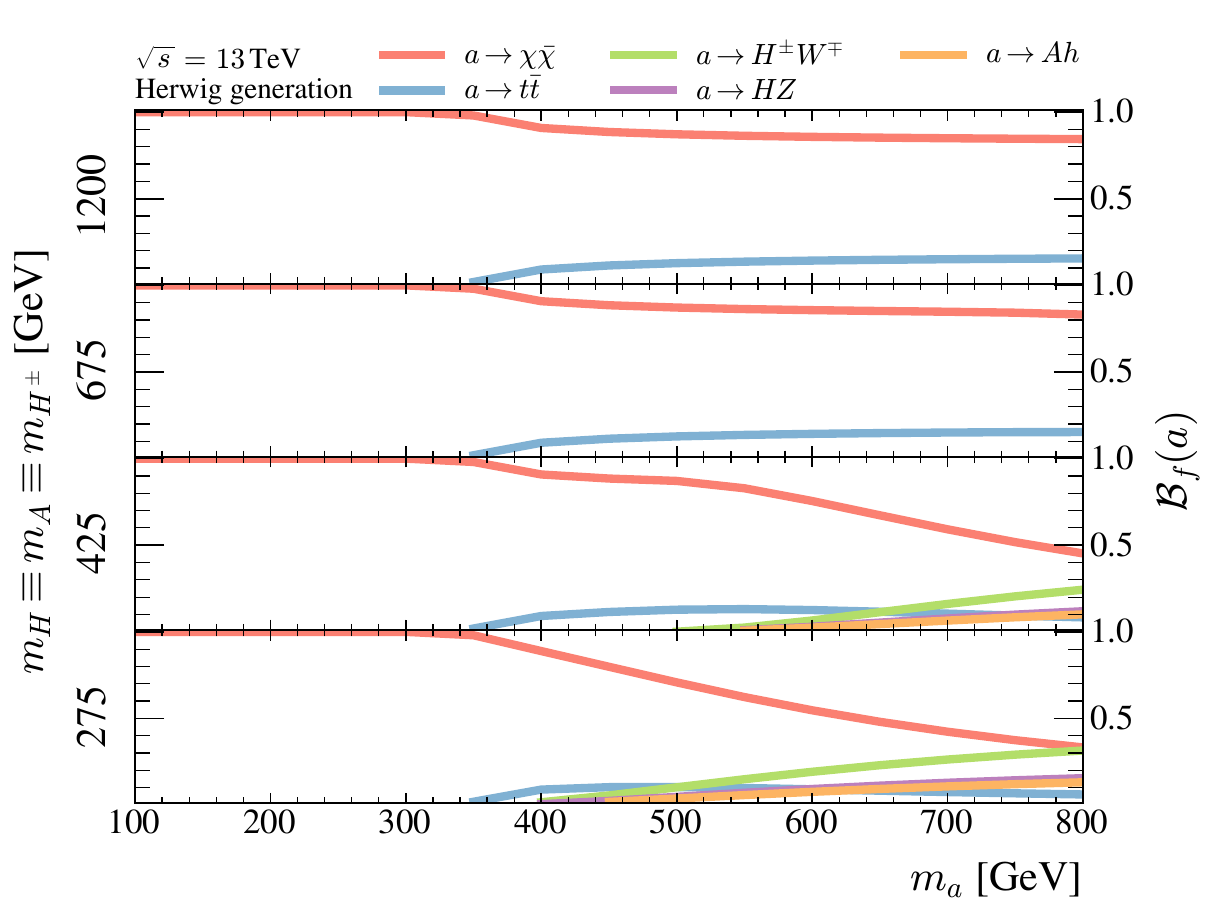}

The functional dependence of the production cross section can be estimated using $\sigma(AB\to C)\propto\width[C\to AB]$~\cite{Griffiths:1987tj} as
\begin{equation}
	\label{eq:2HDMa_sigma_a}
	\begin{aligned}
		\sigma\left(f\bar{f}\to a\right)&\propto\xi_f^2\frac{m_f^2}{v^2}\ma\beta_{f/a}\sinPSq\\
		\sigma\left(gg\to a\right)&\propto\frac{\alpha_s^2}{v^2}\ma^3\abs{\sum_{q=t,b,c}\xi_qf\hspace{-1pt}\left(\tau_{q/a}\right)}^2\sinPSq
	\end{aligned}
\end{equation}
with
\begin{equation*}
	f\hspace{-1pt}\left(\tau\right)\coloneqq\tau\arctan^2\left(\frac{1}{\sqrt{\tau-1}}\right).
\end{equation*}
In consequence, the most important production modes of the field $a$ in proton--proton collisions taking into account the particle content of protons (parton distributions functions, see \secref{sec:MCEG_hardProcess}) and \eqref{eq:2HDMa_xi} are $gg\to a$ for small values of \tanB and $\bbbar\to a$ for large values of \tanB.

\subsubsection{Heavy pseudoscalar $A$}

Production and decay for the heavy pseudoscalar field $A$ are similar to those of the light pseudoscalar field $a$:
\begin{align}
	\notag
	\width[A\to\xx]&=\frac{\yX^2}{8\pi}\mA\beta_{\chi/A}\sinPSq\\
	\label{eq:2HDMa_Gamma_A_ff}
	\width[A\to f\bar{f}]&=\frac{N_c^f\xi_f^2}{8\pi}\frac{m_f^2}{v^2}\mA\beta_{f/A}\cosPSq.
\end{align}
The important difference is the swapped proportionality to \sinPSq and \cosPSq (\cf\eqref{eq:2HDMa_Gamma_a_ff}).
At \sinPLow, it results in a smaller branching fraction for \brf{A\to\xx} than for\linebreak \brf{a\to\xx} for identical masses \mA and \ma (\cf\figsref{fig:2HDMa_BR_a}{fig:2HDMa_BR_A}).
Instead, the branching fraction for \brf{A\to f\bar{f}} is larger than for \brf{a\to f\bar{f}}.
Most importantly, also the production cross section for the pseudoscalar~$A$ from processes involving fermions is larger than for the pseudoscalar~$a$ at identical masses.

In addition, the pseudoscalar~$A$ can decay to $ah$. The decay width for this process is
\begin{equation}
	\label{eq:2HDMa_Gamma_A_ah}
	\width[A\to ah]=\frac{1}{16\pi}\frac{\kappa^{1/2}\left(\mA, \ma, \mh\right)}{\mA}g_{Aah}^2
\end{equation}
which is sizeable as long as $\mA>\ma+\mh$. Hereby, using \lTeqlPOeqlPT,
\begin{equation}
	\label{eq:2HDMa_g_Aah}
	g_{Aah}=\frac{1}{\mA v}\left[\mh^2-2\mH^2-\mA^2+4\mHpm^2-\ma^2\right]\sinP\cosP
\end{equation}
and $\kappa$ is defined as
\begin{equation*}
	\kappa\left(m_1, m_2, m_3\right)\coloneqq\left(m_1^2-m_2^2-m_3^2\right)^2-4m_2^2m_3^2.
\end{equation*}

\subsubsection{Heavy scalar $H$}

The decay widths for the heavy scalar field $H$ are most importantly
\begin{align}
	\label{eq:2HDMa_Gamma_H_ff}
	\width[H\to f\bar{f}]&=\frac{N_c^f\xi_f^2}{8\pi}\frac{m_f^2}{v^2}\mH\beta_{f/H}^3\\
	\label{eq:2HDMa_Gamma_H_aa}
	\width[H\to aa]&=\frac{1}{32\pi}g_{Haa}^2\mH\beta_{a/H}\\
	\label{eq:2HDMa_Gamma_H_aZ}
	\width[H\to aZ] &=\frac{1}{16\pi}\frac{\kappa^{3/2}\left(\mH, \ma, m_Z\right)}{\mH^3v^2}\sinPSq.
\end{align}
Hereby, using $\lambda'\coloneqq\lTeqlPOeqlPT$ and \mAeqmHeqmHpm,
\begin{equation}
	\label{eq:2HDMa_g_Haa}
	g_{Haa} = \frac{2\sinPSq}{\mH v}\cot\left(2\beta\right)\left(\mh^2-\lambda'v^2\right).
\end{equation}
As shown in \figref{fig:2HDMa_BR_H}, the decays $H\to\ttbar$ and $H\to aZ$ dominate if they are kinematically allowed.

\myBRFigure{scalar~$H$}{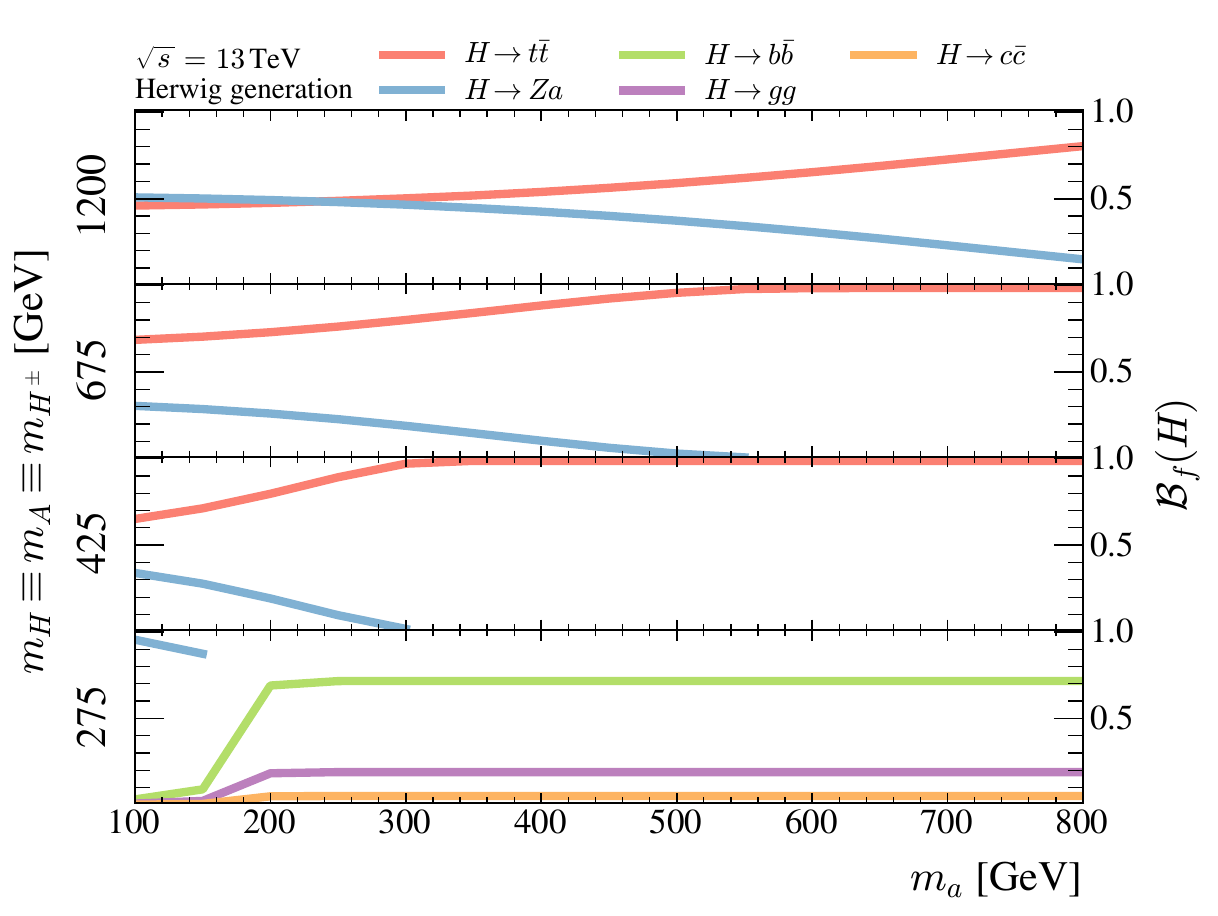}

\subsubsection{Light scalar $h$}

In the alignment limit, the light scalar field $h$ in general has \SM-like couplings.
Large decay widths, however, are also
\begin{align*}
	\width[h\to aa]&=\frac{1}{32}g_{haa}^2\mh\beta_{a/h}\\
	\width[h\to a\xx]&=\frac{\yX^2}{32\pi^3}g_{haa}^2\mh\beta_{\chi/a}k\hspace{-1pt}\left(\tau_{a/h}\right)\cosPSq
\end{align*}
with, using $\lambda'\coloneqq\lTeqlPOeqlPT$ and \mAeqmHeqmHpm,
\begin{equation}
	\label{eq:2HDMa_g_haa}
	g_{haa} = \frac{1}{\mh v}\left[
	\left(2\left(\mA^2-\ma^2\right)+\mh^2\right)\sinPSq
	-2\lambda'v^2
	\right].
\end{equation}
Further, $k$ is defined as
\begin{equation*}
	k\hspace{-1pt}\left(\tau\right)\coloneqq\frac{1}{8}\left(1-\tau\right)\left[4-\ln\left(\frac{\tau}{4}\right)\right]-\frac{5\tau-4}{4\sqrt{\tau-1}}\left[\arctan\left(\frac{\tau-2}{2\sqrt{\tau-1}}\right)-\arctan\left(\frac{1}{\sqrt{\tau-1}}\right)\right].
\end{equation*}
The high branching fraction for invisible $a$ decays, $a\to\xx$, renders also $h\to aa$ decays mostly invisible. Invisible $h$ decays are still allowed by experiments because the current limit is $\brf{h\to\text{invisible}}<0.11$~\cite{ATLAS-CONF-2020-052}, as mentioned before. Nonetheless, these limits set stringent constraints on the \THDMa parameter space for small masses of the pseudoscalar~$a$.

\subsubsection{Charged scalar \Hpm}
The charged scalar \Hpm is dominantly produced in association with a top quark%
\footnote{%
	To simplify notation, the bar indicating an antifermion ( $\bar{}$ ) or the sign indicating the charge of a lepton is omitted sometimes when \Hpm bosons, $W^\pm$ bosons or taus are involved.
	These cases are to be understood as using (anti)particles where needed such that charge is always conserved.%
}
, $pp\to t\Hpm$\cite{Pani:2017qyd}.
It decays most importantly into a bottom and a top quark, $\Hpm\to \bar{b}t$, or into another \BSM boson and a $W$ boson, $\Hpm\to H'W^\pm$ where \mbox{$H'\in\left\{a, A, H\right\}$}, if these decays are kinematically allowed~\cite{Bauer:2017ota}.
This is shown for different masses~\ma and \mAeqmHeqmHpm in \figref{fig:2HDMa_BR_H+}.

\myBRFigure{charged scalar~\Hpm}{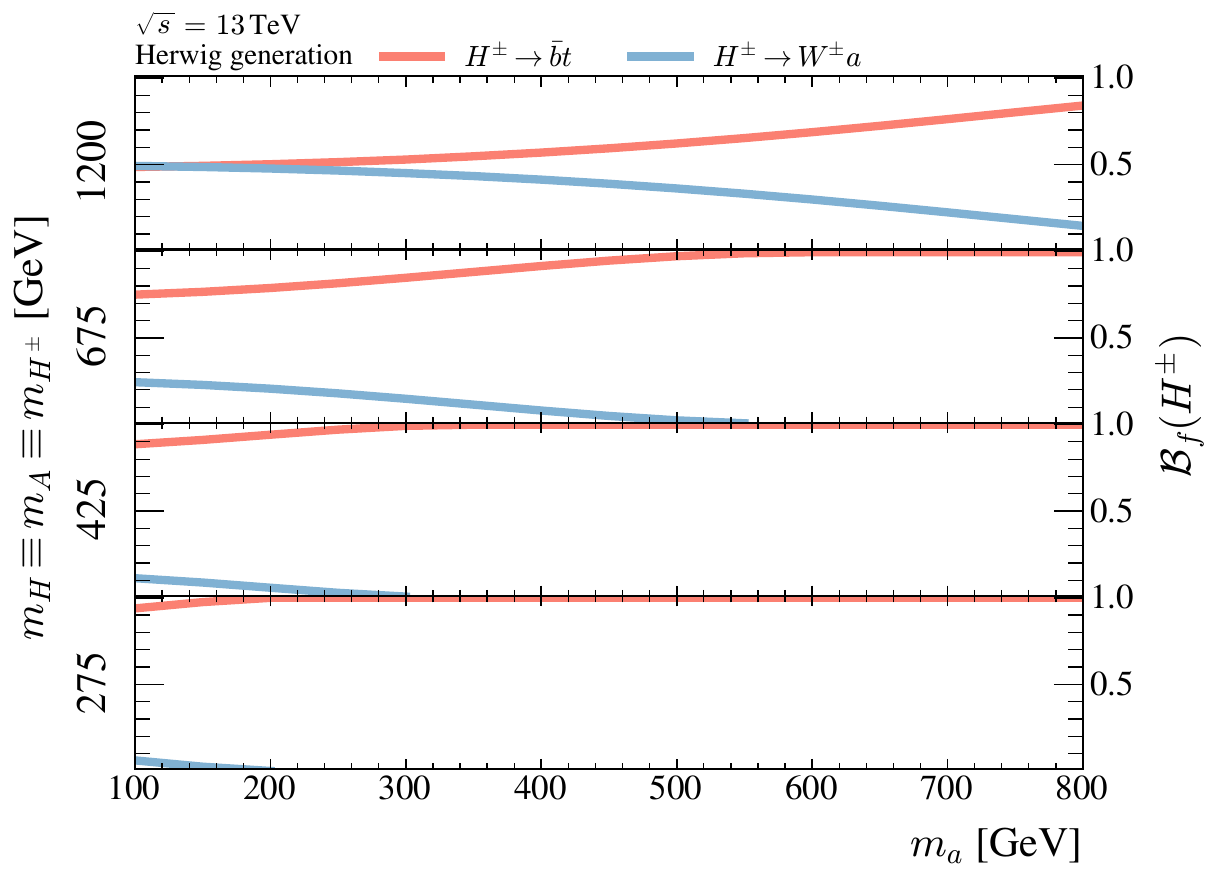}

\subsection{Investigated parameter planes}
\label{sec:2HDMa_parameterPlanes}

Following from the discussion above, two main parameter planes are studied in which the parameters deviate from the choices in \tabref{tab:LHCDMWG_params}.
On the one hand, the \mamA plane is investigated, where the masses \ma and \mAeqmHeqmHpm of all \BSM bosons are varied.
Different decay channels of the \BSM bosons are kinematically open with changing masses \ma and \mAeqmHeqmHpm.
In this plane, $\tanB=1$ and according to \eqref{eq:2HDMa_xi}, $\abs{\xi_{f'}}=1$ where $f'$ is any charged fermion.
The absolute value of the coupling of uncharged \BSM bosons to charged fermions therefore only differs by the mass of the fermion~$m_{f'}$.
In consequence, the largest coupling of the uncharged \BSM bosons is to top quarks (\cf\eqref{eq:2HDMa_Lagrangian}).
On the other hand, the \matanB plane at $\mAeqmHeqmHpm=\SI{600}{GeV}$ is investigated, where the mass~\ma and \tanB are varied.
As can be seen from \eqsref{eq:2HDMa_Lagrangian}{eq:2HDMa_xi}, the coupling of the charged \BSM bosons to top quarks decreases and the one to bottom quarks and taus increases with increasing \tanB so different production channels become important.
\Chapter[1]{Performing experiments}{The \LHC and the ATLAS Experiment}{%
	Linkin Park}{Until It Breaks~\cite{LinkinPark:2012uib}}{verse 1, line 1}
\label{sec:experiment}

Particle colliders provide a promising approach to observe physics beyond the Standard Model (\BSM, \cf\secref{sec:BSM}) because of their very controlled and reproducible conditions, as discussed in \secref{sec:DM_search_strategies}.
One of the theoretically simplest possibilities why physics beyond the Standard Model has not been observed yet is if it involves particles at masses higher than those explored so far.
For this reason, proton--proton collisions from the world's most powerful particle accelerator, the Large Hadron Collider (\LHC)~\cite{Evans:2008zzb}, are studied in this thesis.
Details on the \LHC are given in \secref{sec:LHC}.

The ATLAS and CMS Experiments are detectors designed to explore the complete spectrum of collisions produced at the \LHC.
Data from the ATLAS Experiment is used for the \METjets measurement described in \chapsref{sec:metJets}{sec:interpretation}.
The ATLAS detector is introduced in detail in \secref{sec:ATLAS}.
Publicly available physics analyses from the CMS Collaboration are used in \secref{sec:Contur} together with those from the ATLAS Collaboration to study the general sensitivity of the \LHC experiments to the \THDMa.
Differences between the CMS and ATLAS detectors are pointed out in \secref{sec:CMS}.
An overview of the other experiments located at the \LHC is provided in \secref{sec:LHC_furtherExperiments}.

\section{The Large Hadron Collider (\LHC)}
\markright{The Large Hadron Collider}
\label{sec:LHC}

The \LHC is located in the \SI{26.7}{km} long tunnel that was originally occupied by the Large Electron--Positron Collider~(\LEP) 45 to \SI{170}{m} below CERN, near Geneva on the border between France and Switzerland~\cite{Evans:2008zzb}.
Its construction started in the year 2000, and it has been in operation since 2008.
The \LHC consists of two rings hosting counter-rotating beams of protons or heavy ions. 
Particles are pre-accelerated by a chain of other accelerators, the CERN accelerator complex, which is shown in \figref{fig:exp_acceleratorComplex}.
The acceleration chain starts with the linear accelerator Linac4, continues with the Proton Synchrotron Booster and the Proton Synchrotron (\PS) and finishes with the Super Proton Synchrotron (\SPS) which injects particles into the \LHC with an energy of \SI{450}{GeV}.
For protons, the \LHC is designed to reach a centre-of-mass energy $\sqrt{s}$ for the collisions of \SI{14}{TeV} by accelerating the particles in radio-frequency cavities. During its first operational period, \textit{Run 1} from 2009 to 2013, it delivered $\sqrt{s}=7-\SI{8}{TeV}$ for protons during the main data-taking period. Run 2 from 2015 to 2018 was conducted with $\sqrt{s}=\SI{13}{TeV}$. In 2022, Run 3 started with $\sqrt{s}=\SI{13.6}{TeV}$.

\begin{myfigure}{
		Schematic of the \LHC accelerator complex at CERN. The locations of the four large experiments are marked with green dots. Figure adapted from \refcite{Horvath:2022arb}.
	}{fig:exp_acceleratorComplex}
	\includegraphics[width=0.8\textwidth]{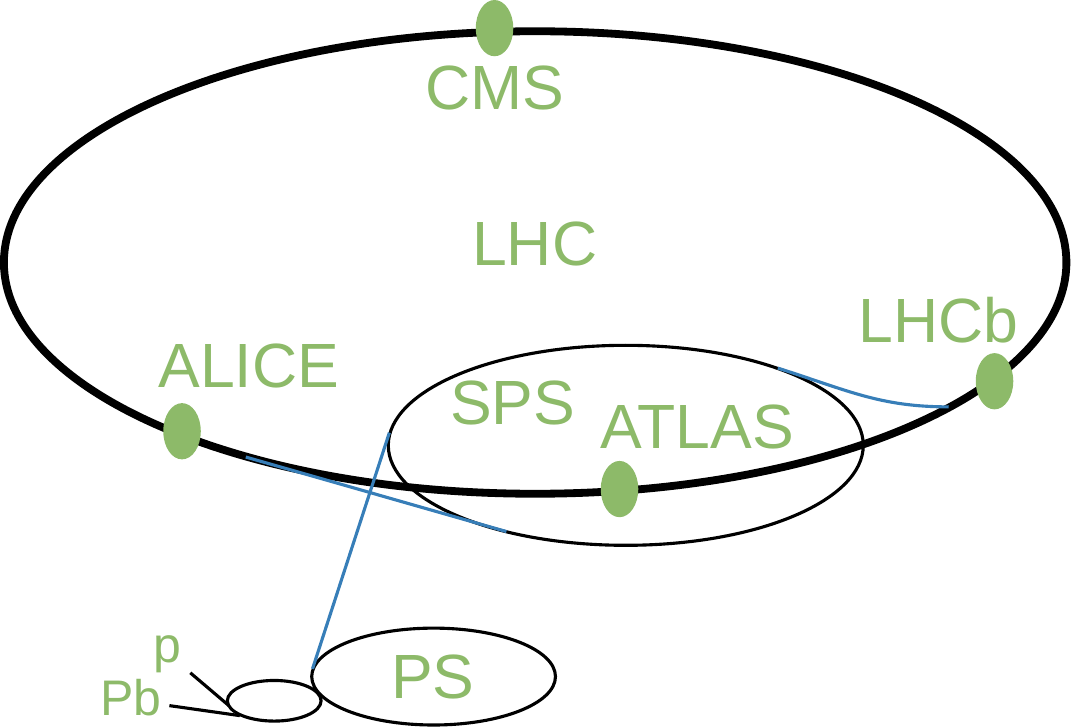}
\end{myfigure}

There are eight crossing points of the two beams, four of which host large experiments, which are marked by green dots in \figref{fig:exp_acceleratorComplex}: ALICE, ATLAS, CMS and LHCb.
Particles are collected in \textit{bunches} of more than \SI{e11}{protons} each to achieve high collision rates at the interaction points.
During Run 2, $N_b=\order{2400}$ bunches per beam were simultaneously stored in the \LHC rings~\cite{ATLAS-CONF-2019-021}.
The nominal bunch spacing is \SI{25}{ns}, giving a revolution frequency $f_r$ of \SI{40}{MHz}.


\subsection{Luminosity}
\label{sec:experiment_luminosity}

The event rate $\frac{\mathrm{d}N}{\mathrm{d}t}$ for a physics process with cross section $\sigma$ is
\begin{equation*}
	\frac{\mathrm{d}N}{\mathrm{d}t} \eqqcolon L\sigma,
\end{equation*}
where the proportionality factor $L$ is the \textit{instantaneous luminosity} delivered by the collider.
The instantaneous luminosity can be calculated as
\begin{equation*}
	L = f_rN_b\frac{n_1n_2}{2\pi\Sigma_x\Sigma_y}.
\end{equation*}
Hereby, $n_i$ is the number of particles in bunch $i$ and $\Sigma_x$ ($\Sigma_y$) are the convolved beam sizes in the horizontal (vertical) plane. In Run 2, instantaneous luminosities of 5 to \SI{21e33}{cm^{-2}s^{-1}} were achieved in the \LHC~\cite{ATLAS-CONF-2019-021,ATLAS:2022lum}. The \textit{integrated luminosity}~\intLumi~ -- and correspondingly the total number of events -- is obtained from the instantaneous luminosity by integrating in time,
\begin{equation*}
	\intLumi\coloneqq\int L\mathrm{d}t.
\end{equation*}

\subfigref{fig:lumiAndPileup}{a} shows the integrated luminosity available to the ATLAS Experiment during Run 2 of the \LHC as a function of time. In the end, the \LHC delivered a total of $\intLumi=$\SI{156}{fb^{-1}} of proton--proton collisions for the ATLAS Experiment, of which \SI{147}{fb^{-1}} were recorded and \SI{139}{fb^{-1}} are considered to be usable for physics analyses because of good efficiencies and data quality~\cite{ATLAS:2022lum}.

\begin{myfigure}{
		(a) Integrated luminosity of proton--proton collisions as a function of time during Run 2 delivered by the \LHC to the ATLAS Experiment (green), recorded by the ATLAS Experiment (yellow) and usable for physics analyses (blue). (b) Integrated luminosity recorded by the ATLAS Experiment during Run 2 as a function of the average number of interactions per bunch crossing. The colours indicate the distributions for the different years of data taking. Figures taken from \refcite{ATLAS:2022lum}.
	}{fig:lumiAndPileup}
	\subfloat[]{\includegraphics[width=0.49\textwidth]{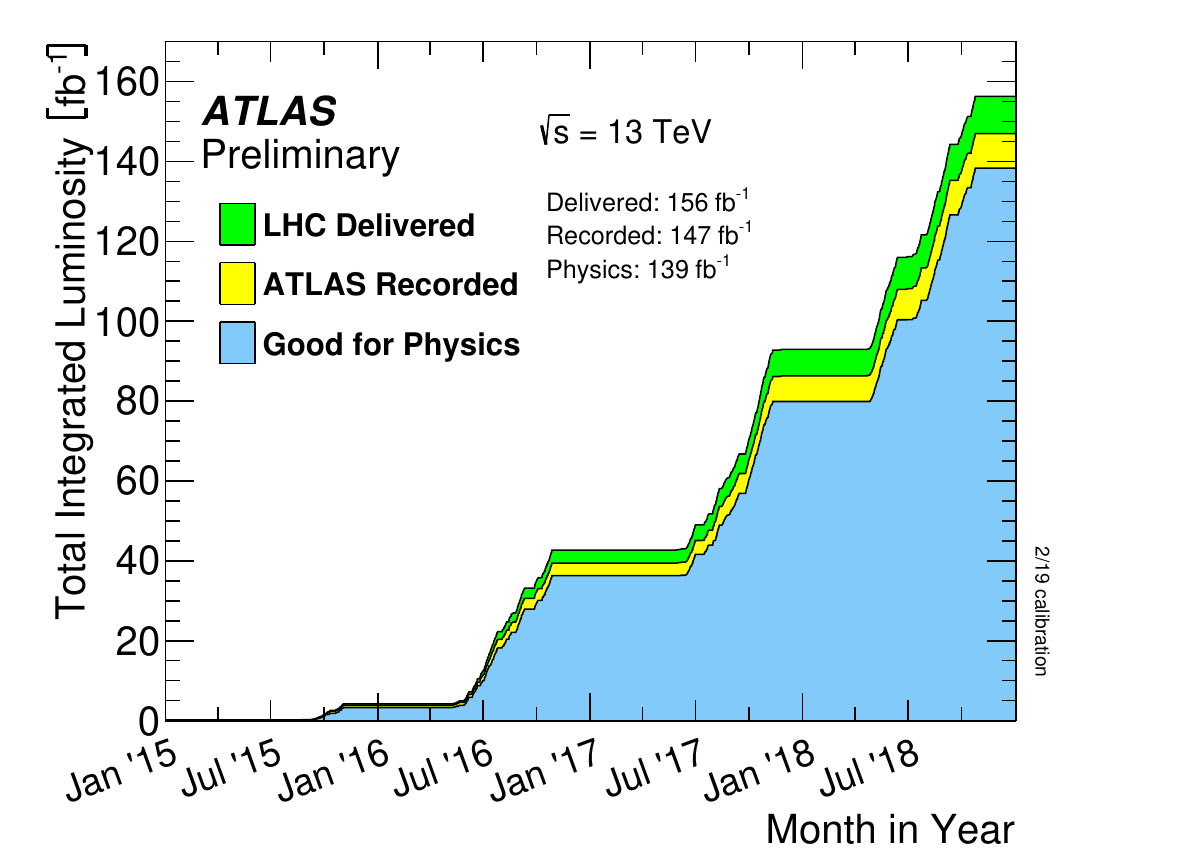}}
	\subfloat[]{\includegraphics[width=0.49\textwidth]{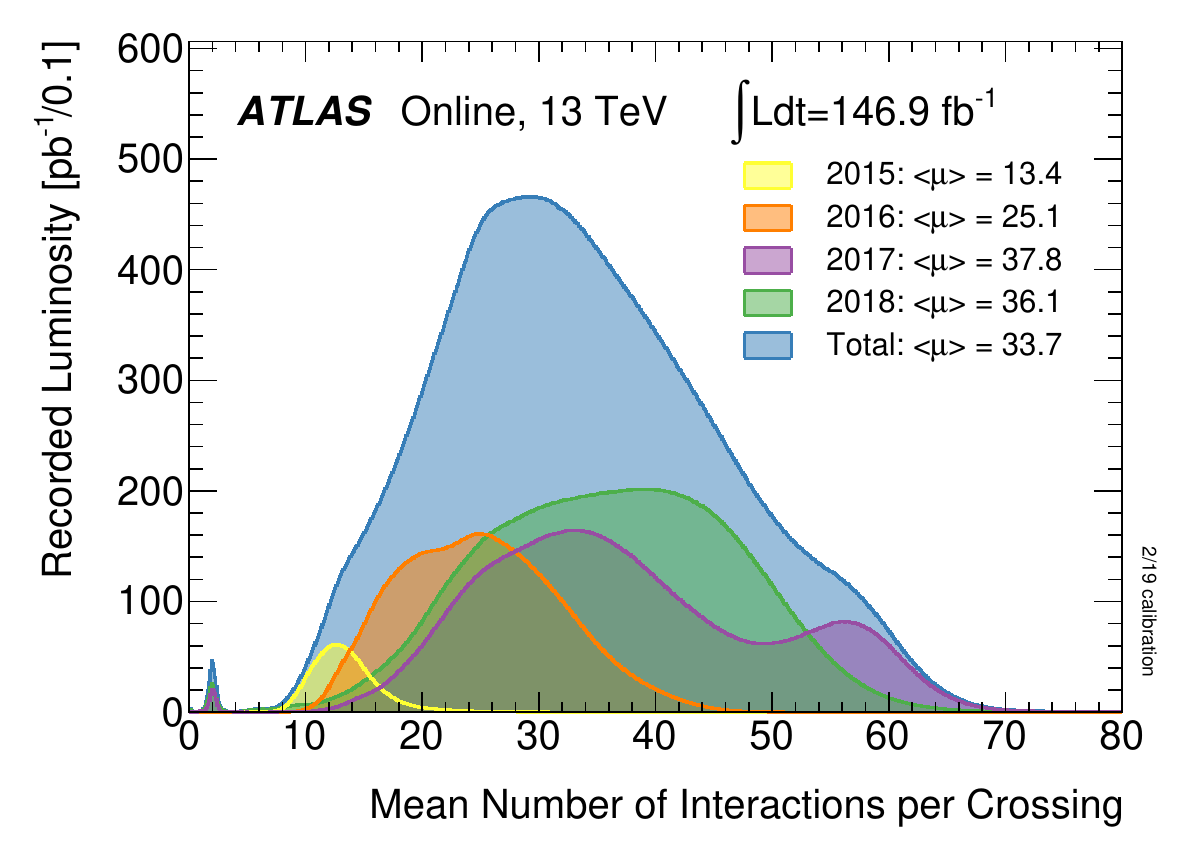}}\\
\end{myfigure}

\subsection{Pileup}
The collision of interest in each event is often the interaction with the highest sum of absolute transverse momenta, the \textit{hard-scatter} interaction~\cite{ATLAS:2015ull}.
Additional collisions not coming from the collision of interest are called \textit{pileup}.
In-time pileup refers to additional collisions in the same bunch crossing, out-of-time pileup to remnants from neighbouring bunch crossings~\cite{ATLAS:2015ull}.
Other contributions to pileup come from cavern background, beam-gas and beam-halo events~\cite{Marshall:2014mza}.

Following the considerations regarding luminosity, a higher number of particles per bunch leads to larger luminosity and therefore to a higher number of events for a physics process of interest.
A higher number of particles per bunch does, however, also mean that more than one interaction can happen in each bunch crossing, contributing to the pileup. \subfigref{fig:lumiAndPileup}{b} shows the recorded integrated luminosity as a function of the mean number of interactions per bunch crossing for Run 2 in the ATLAS detector. In total, the average number of interactions per bunch crossing was 33.7.


\section{The ATLAS Experiment}
\label{sec:ATLAS}

The ATLAS Experiment~\cite{ATLAS:2008xda} is one of the two general-purpose particle-detectors at the \LHC.
The experiment aims to measure all kinds of \SM parameters, conduct precision tests of the Standard Model -- in particular regarding \QCD, electroweak and flavour physics -- and search for \BSM phenomena up to the TeV scale. In consequence, it is designed to detect all elementary particles predicted by the Standard Model.
Electrons, muons and photons can be detected directly, quarks are observed as jets (see \secref{sec:objReco_jets}).
Almost all other \SM particles, \ie taus and bosons, can be reconstructed from their decay products. The only \SM particles that cannot be detected directly are neutrinos which traverse the detector volume without interacting. Their presence has to be inferred from a momentum imbalance in the collision, \MET (see \secref{sec:objReco_MET}).
Depending on their nature, \BSM particles can also either interact with the detector directly, be reconstructed from their decay products or inferred from \MET.

The \LHC is designed for high energies and high luminosity such that also rare processes can be observed.
In consequence, the ATLAS detector is built to be fast and radiation hard, and provide a high granularity to be able to dissolve multiple interactions due to pileup.
At the same time, it is meant to give the best-possible resolution of the measured physics quantities, reconstruction efficiency and background rejection in a wide geometric coverage.

To meet all these design and physics goals, the ATLAS Experiment is constructed as a forward--backward symmetric cylindrical detector of \SI{25}{m} height and \SI{44}{m} length.
\figref{fig:ATLAS_detector} shows a cut-away view of the ATLAS detector and its subdetector systems:
The part of the detector closest to the interaction point is the Inner Detector, followed by calorimeters. Muon spectrometers make up the outermost part of the detector.
The coordinate system and observables used by the ATLAS Experiment are introduced in \secref{sec:ATLAS_observables}.
General features of the design and geometry of the ATLAS detector are given in \secref{sec:ATLAS_generalDesign}.
All subdetector systems are described in more detail in \secsref{sec:ATLAS_ID}{sec:ATLAS_calo}{sec:ATLAS_muonSpec}, respectively.

\begin{myfigure}{Schematic of the ATLAS detector and its subdetector systems. Figure taken from \refcite{ATLAS:2008xda}.}{fig:ATLAS_detector}
	\includegraphics[width=\textwidth]{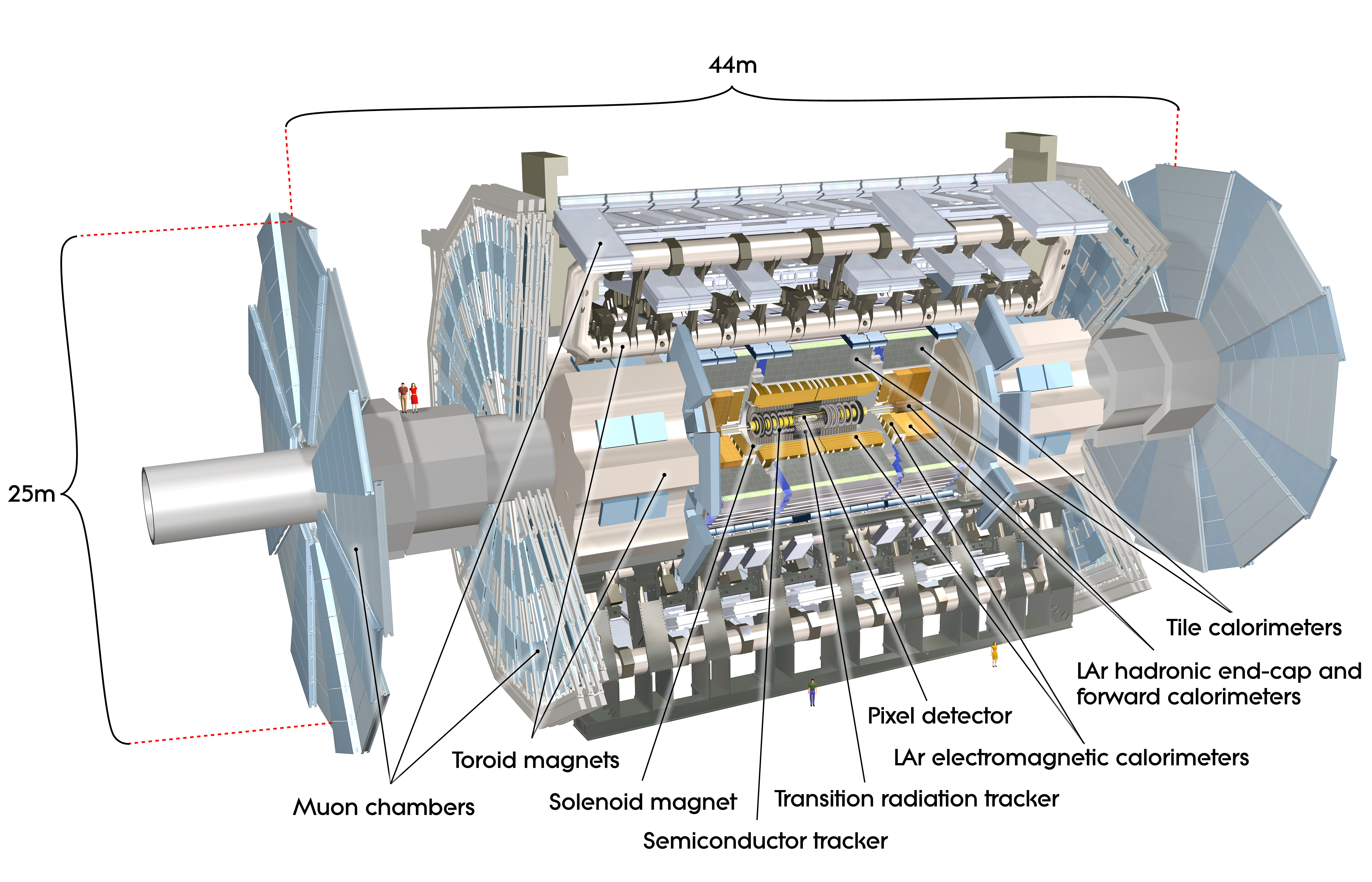}
\end{myfigure}

The subdetector systems are supplemented by a collection of smaller detectors located tangentially to the beamline at large distances from the collision point. These are briefly introduced in \secref{sec:ATLAS_forwardDetectors}.

The extraordinarily high collision rate at the \LHC renders recording the information of every single collision (\textit{event}) unfeasible.
Therefore, the ATLAS Experiment makes use of a triggering system that marks events deemed interesting for permanent storage. This is described in more detail in \secref{sec:ATLAS_trigger}.

\subsection{Coordinates and observables}
\label{sec:ATLAS_observables}

\begin{myfigure}{Schematic of the coordinate system and common observables used in the ATLAS Experiment. Figure adapted from \refcite{Neutelings:2021neu}.}{fig:ATLAS_coordinateSystem}
	\includegraphics[width=\textwidth]{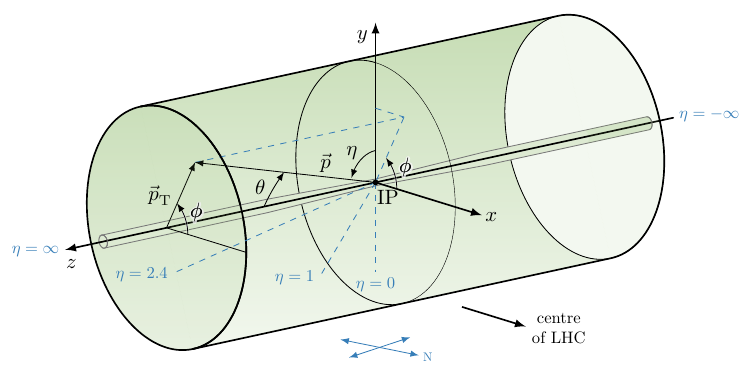}
\end{myfigure}

\figref{fig:ATLAS_coordinateSystem} gives an overview of the coordinate system and common observables used by the ATLAS Experiment. The origin of the right-handed coordinate system lies at the nominal interaction point (\IP). The $x$-axis points from the interaction point to the centre of the \LHC ring, the $y$-axis upwards and the $z$-axis in the direction of the beam. In spherical coordinates, the azimuthal angle around the beam axis is denoted with $\phi$, the polar angle from the beam axis $\theta$ and the distance from the beam axis $r$.

The \textit{rapidity} $y$ of a particle with energy $E$ and momentum $p_z$ in $z$-direction is defined as
\begin{equation*}
	y \coloneqq\frac{1}{2}\ln\frac{E+p_z}{E-p_z}.
\end{equation*}
For massless particles or in the limit of large momentum, this converges against the \textit{pseudorapidity}
\begin{equation*}
	\eta\coloneqq-\ln\tan\frac{\theta}{2}=\frac{1}{2}\ln\frac{p+p_z}{p-p_z},
\end{equation*}
where $p\equiv\abs{\vv p}$ is the total particle momentum. These observables are useful because differences in rapidity are Lorentz-invariant under boosts along the $z$-axis. Distances are commonly expressed as
\begin{equation*}
	\Delta R\coloneqq\sqrt{\left(\Delta\eta\right)^2+\left(\Delta\phi\right)^2}.
\end{equation*}
or
\begin{equation*}
\Delta R_y\coloneqq\sqrt{\left(\Delta y\right)^2+\left(\Delta\phi\right)^2}.
\end{equation*}

The ATLAS detector is fully hermetic along $\phi$ but not completely along $\theta$ because of the beam pipe.
That is why often instead of the total momentum $p$ the momentum in the \tuple{x}{y} plane, the \textit{transverse momentum} $\pT\coloneqq\sqrt{p_x^2+p_y^2}$, is considered.

\subsection{General design and geometry features of the ATLAS detector}
\label{sec:ATLAS_generalDesign}
Hadron collisions at high luminosity and for a long period of time lead to stringent requirements on radiation hardness for detector systems at the \LHC.
Two design guidelines follow from this for the ATLAS detector.
On the one hand, different hardware is used for the same functionality at different distances from the beam axis because lower radiation levels can be expected at greater distances.
The total ionising dose at $\eta=0$ is for example
$\order{\SI{e3}{\gray\per\ifb}}$
at $r=\SI{10}{cm}$ and decreases to
$\order{\SI{e1}{\gray\per\ifb}}$
at $r=\SI{100}{cm}$~\cite{Dawson:2021xku}.
On the other hand, many interactions from hadrons in colliding bunches only lead to a small deflection of particles from the beam axis.
In consequence, the forward and backward parts of the ATLAS detector are irradiated considerably more than the cylinder barrel.
The total ionising dose at a distance of \SI{1}{m} from the collision points is for example
$\order{\SI{e1}{\gray\per\ifb}}$
at $\eta=0$ but
$\order{\SI{7e3}{\gray\per\ifb}}$
at $\absEta=4$~\cite{Dawson:2021xku}.
To mitigate imbalanced detector degradation, the ATLAS subsystems therefore consist of two parts: the barrel subdetector systems are formed by concentric cylinders; those in the end-caps by disks perpendicular to the beam axis which use a different, more radiation-resistant hardware.

There are two magnetic systems in the ATLAS detector that bend the trajectories of charged particles (\textit{tracks}) and correspondingly allow determining their momentum:
The Inner Detector is immersed into a \SI{2}{T} magnetic-field from superconducting solenoids.
Superconducting toroids with a magnetic field strength of \SI{0.5}{T} (\SI{1}{T}) surround the calorimeters in the barrel (end-caps).
They consist of radial coils, eight of which are symmetrically distributed around the beam axis for the barrel and each of the two end-caps.

\subsection{Inner Detector (\ID)}
\label{sec:ATLAS_ID}

The ATLAS Inner Detector~\cite{CERN-LHCC-97-016,Haywood:331064,ATLASIBL:2018gqd} provides data points for tracking, momentum and vertex measurements as well as for pattern recognition that benefits particle identification.
The Inner Detector is the detector closest to the interaction point and has a very high granularity to allow for excellent resolution of \textit{vertices}, the space points where the interactions occurred.
The Inner Detector extends from $r=\SI{23.5}{mm}$ to $r=\SI{1.25}{m}$~\cite{ATLASIBL:2018gqd} and is \SI{5.3}{m} in length.
Its innermost part is made up by the silicon Pixel detector consisting of pixels as small as $\SI{250}{\mu m}\times\SI{50}{\mu m}$.
Four layers in the barrel region and three disks in the end-caps are formed to obtain multiple interactions with particles along their trajectory.
This is followed by the Semiconductor tracker (\SCT) made of four layers of silicon microstrips in the barrel region and up to nine disks in the end-caps.
In each layer and disk, pairs of microstrips are glued back-to-back with a small relative rotation to allow for two-dimensional space-point resolution.
Pixel detector and \SCT extend up to $\absEta<2.5$ and even $\absEta<3.0$ for the innermost layer of the Pixel detector. Charged particles passing the subdetector material create electron--hole pairs which are successively collected and measured.

The part of the Inner Detector at larger radii is the Transition radiation tracker (\TRT).
The \TRT is made of drift straw tubes interleaved with material with varying refractive indices and covers the region $\absEta<2.0$.
The straw tubes are filled with a xenon-based gas mixture that is ionised by passing charged particles and transition-radiation photons.
The latter yield much larger signal amplitudes and allow for improved electron identification.
Electric signals are collected from anode wires in the tubes.

\subsection{Calorimeters}
\label{sec:ATLAS_calo}

The energy and position of particles produced in collisions is measured by stopping them in the calorimeters if possible.
The ATLAS Experiment uses sampling calorimeters which consist of a high-density absorber material with maximised stopping power layered with an active material to determine the deposited energy.
The calorimeters are designed to fully absorb the particles' energy to avoid spurious missing energy.

Electrons and photons require little material to be stopped due to their low mass. Their interaction with the detector material gives rise to cascades of electromagnetic interactions, \textit{showers}. The inner part of the calorimeters therefore consists of the so-called electromagnetic (\EM) calorimeters~\cite{CERN-LHCC-96-041} covering $\absEta<3.2$.
They are made of lead absorber plates and use liquid argon (\LAr) as active material.
This gives a small radiation length $X_0$ which keeps the \EM showers narrow and well-contained.
In total, the \EM calorimeters have a thickness of more than 22 radiation lengths $X_0$.
The typical cell size is $\Delta\eta\times\Delta\phi=0.025\times0.025$, allowing the accurate measurement of the shower position and determination of the deposited energy as precise as $\frac{\sigma_E}{E}=\frac{\SI{400}{MeV}}{E}\oplus\frac{\SI{8}{\%}}{\sqrt{E}}\oplus\SI{0.7}{\%}$.
Readout takes place through accordion-shaped Kapton electrodes. In the barrel, the \EM calorimeters extend up to $\absEta<1.475$, in the end-caps they cover from $1.375<\absEta<3.2$.

Particles passing the electromagnetic calorimeters reach the hadronic calorimeters.
In the barrel region, $\absEta<1.7$, these use steel as the absorber and scintillating tiles as the active material with a segmentation of $\Delta\eta\times\Delta\phi=0.1\times0.1$~\cite{CERN-LHCC-96-042}.
A resolution of $\frac{\sigma_E}{E}=\frac{\SI{50}{\%}}{\sqrt{E}}\oplus\SI{3}{\%}$ is achieved.
In the end-caps, $1.5<\absEta<3.2$, copper is used as absorber and liquid argon as active material~\cite{CERN-LHCC-96-041}.
The high density and extent of the hadronic calorimeters ensure that hadronic showers are fully contained: in total, the hadronic calorimeters have a thickness of more than 9.7 interaction lengths ($\lambda$).
This improves the energy resolution and limits the punch through to the Muon spectrometer which would deteriorate its precision. 

The forward region, $3.1<\absEta<4.9$, uses copper (tungsten) as absorber for the electromagnetic (hadronic) calorimeters and liquid argon as active material.

\subsection{Muon spectrometer (\MS)}
\label{sec:ATLAS_muonSpec}

Muons are minimum-ionising particles and can therefore traverse the ATLAS detector without being stopped. After passing the calorimeters, they are deflected by the toroidal magnetic field which allows for precision measurements of their track coordinates.
In the barrel region, $\absEta<2.7$ $\left(\absEta<2\text{ for the innermost muon chambers}\right)$, this is conducted by proportional drift tubes called "Monitored drift tubes" (\MDT{}s) with a diameter of \SI{30}{mm}~\cite{CERN-LHCC-97-022}.
In the end-caps, $2.0<\absEta<2.7$, multi-wire proportional chambers with cathodes segmented into strips, called "Cathode strip chambers" (\CSC{}s), are used.
Their anode--anode as well as anode--cathode spacing is \SI{2.54}{mm}.

A second set of muon chambers is aimed at fast readout instead of highest-possible precision to allow triggering events (see \secref{sec:ATLAS_trigger}) containing muons.
At the same time, these chambers contribute information on the muon coordinates orthogonal to the precision tracking chambers.
In the barrel region, $\absEta<1.05$, these chambers are made of gaseous parallel electrode-plates with a gas gap of \SI{12}{mm} called "Resistive plate chambers" (\RPC{}s).
In the end-caps, multi-wire proportional chambers with particularly small anode--cathode distances of \SI{1.4}{mm} for shorter drift times, "Thin gap chambers" (\TGC{}s), are used. They cover the region $1.05<\absEta<2.7$ but only the region $<2.4$ is used for triggering.

The Muon spectrometer~\cite{CERN-LHCC-97-022} is arranged in three layers in the barrel and four disks in the end-caps, each composed of a pair of precision and fast-readout chambers.
Overall, the Muon spectrometer constitutes the outermost layer of the ATLAS detector, covering the region $\SI{5}{m}<r<\SI{12.5}{m}$ in the barrel region and $\SI{7.4}{m}<z<\SI{22}{m}$ in the end-caps.
The Muon spectrometer achieves a resolution of the muon transverse momentum of $2-\SI{3}{\%}$.

\subsection{Forward detectors}
\label{sec:ATLAS_forwardDetectors}
The aforementioned main detector components are supplemented by a number of smaller forward detectors for specific purposes.
Their data are not explicitly used in this work, and they are therefore only briefly described in the following:

\begin{itemize}
	\item LUCID-2~\cite{Avoni:2018iuv} is an array of several small Cherenkov detectors located at $z=\pm\SI{17}{m}$ for luminosity measurements.
	\item The Zero degree calorimeter for ATLAS~\cite{ATLAS:2007aa} consists of quartz rods and tungsten plates to determine the impact parameter between the centres of colliding ions in heavy-ion collisions.
	The experiment is located at $z=\pm\SI{140}{m}$, measuring neutral particles with $\absEta\geq8.2$.
	\item ALFA consists of scintillating fibre trackers inside Roman pots located at $z=\pm\SI{240}{m}$ and reaching down to $r=\SI{1}{mm}$ to measure particles at very low deflection angles.
	\item AFP aims at tagging very forward protons and measuring their momentum and emission angle at $\abs{z}\approx 210-\SI{420}{m}$. These protons stem from elastic or diffractive scatterings in which one or both protons remain intact. Their tracking is conducted by a silicon-pixel system and their time of flight measured by a quartz detector exploiting Cherenkov radiation.
\end{itemize}

\subsection{Trigger}
\label{sec:ATLAS_trigger}

The rate of bunch crossings at the \LHC is \SI{40}{MHz}~\cite{ATLAS:2021tnq}.
At an instantaneous luminosity of $L=\SI{e34}{cm^{-2}s^{-1}}$, proton--proton interactions happen at a rate of \order{\SI{1}{GHz}} because multiple proton--proton interactions can take place in each bunch crossing.
As it is unfeasible to store all these data, the ATLAS Experiment employs a trigger system that selects which events are saved.

The first step is a hardware-based trigger, called Level 1 (\LOT)~\cite{CERN-LHCC-98-014,ATLAS:2013lic,ATLAS:2021tnq}.
The \LOT trigger uses information from the subdetectors specialised on triggering, \RPC{}s and \TGC{}s, as well as low-granularity information from the calorimeters.
The \LOT trigger searches for events with charged leptons, photons or jets with high \pT, a large momentum imbalance \MET or high total energy in the transverse plane.
If an event is selected, a region of interest is defined that specifies the \tuple{\eta}{\phi} coordinates and nature of the identified feature.
This information is passed to the next stage.
By this approach, the event rate is reduced to about \SI{100}{kHz}.

The second step is a software-based trigger, called High-level trigger (\HLT)~\cite{Jenni:616089,ATLAS:2016wtr}.
In addition to the information passed on by the \LOT trigger, it draws on the full granularity and precision of all ATLAS subdetector systems.
The HLT performs a fast reconstruction of physics objects and trajectories of particles as well as tags jets for whether they presumably originate from bottom quarks.
Events are selected according to predefined algorithms.
Thresholds on the minimum transverse momentum of objects leading to an event being selected need to be imposed as the bandwidth for storing events is limited.
This impacts the selection criteria for physics analyses, \eg for the \METjets measurement (see \secref{sec:metJets_regions}).
The event rate is reduced by the high-level trigger to approximately \SI{1}{kHz}.

The ATLAS data acquisition system (\DAQ)~\cite{Jenni:616089} receives the data from the trigger levels. If an event is triggered on by the \LOT trigger, it is temporarily stored by the ATLAS \DAQ. If also the \HLT criteria are met, the event is stored permanently.
Reconstruction of physics objects from the stored data is described in \chapref{sec:objReco}.


\section{The CMS Experiment}
\label{sec:CMS}

The CMS Experiment~\cite{CMS:2008xjf} is the other general-purpose particle-detector at the \LHC. Its physics goals are similar to that of the ATLAS Experiment, leading to comparable detector architectures. A major design difference is that the solenoidal magnetic field of the CMS Experiment encompasses not only the subdetector systems used for tracking, as in the ATLAS detector, but also the calorimeters.
The CMS solenoidal magnetic field has a field strength of \SI{3.8}{T}.
Iron yokes are used for magnetic flux return, resulting in homogeneous a magnetic field of up to \SI{2}{T} outside the calorimeters~\cite{CMS:2013vyz,CMS:2009moq}.

The tracking closest to the interaction point in the CMS Experiment is conducted by silicon-pixel detectors.
This is followed by silicon-strip detectors for tracking in a larger volume.
In the \EM calorimeters, lead-tungstate (PbWO$_4$) crystals are used. The hadronic calorimeters consist of plastic scintillators as active and brass as the absorber material. To achieve a sufficient thickness in terms of interaction lengths $\lambda$, tail-catchers using iron as absorber material are placed in addition outside the solenoid in the barrel region.
In the end-caps, sampling calorimeters with iron as absorber and quartz-fibres as active material provide coverage up to $\absEta=5.0$.
The muon system consists of drift tubes and \CSC{}s for precision measurements in the barrel and end-cap regions, respectively, as well as \RPC{}s for triggering. In total, the CMS detector is \SI{21.6}{m} in length and \SI{14.6}{m} in height. \figref{fig:CMS_detector} shows a schematic view of the CMS detector and its subdetector systems.

Like the ATLAS Experiment, the CMS Experiment has a number of smaller forward detectors: CASTOR~\cite{CMS:2020ldm} and the Zero degree calorimeter for CMS~\cite{Grachov:2006ke}.
For triggering, the CMS Experiment maintains a system based on a Level-1 and a High-level trigger similar to the ATLAS Experiment.

\begin{myfigure}{Schematic of the CMS detector and its subdetector systems. Figure taken from \refcite{Sakuma:2013jqa}.}{fig:CMS_detector}
	\includegraphics[width=0.9\textwidth]{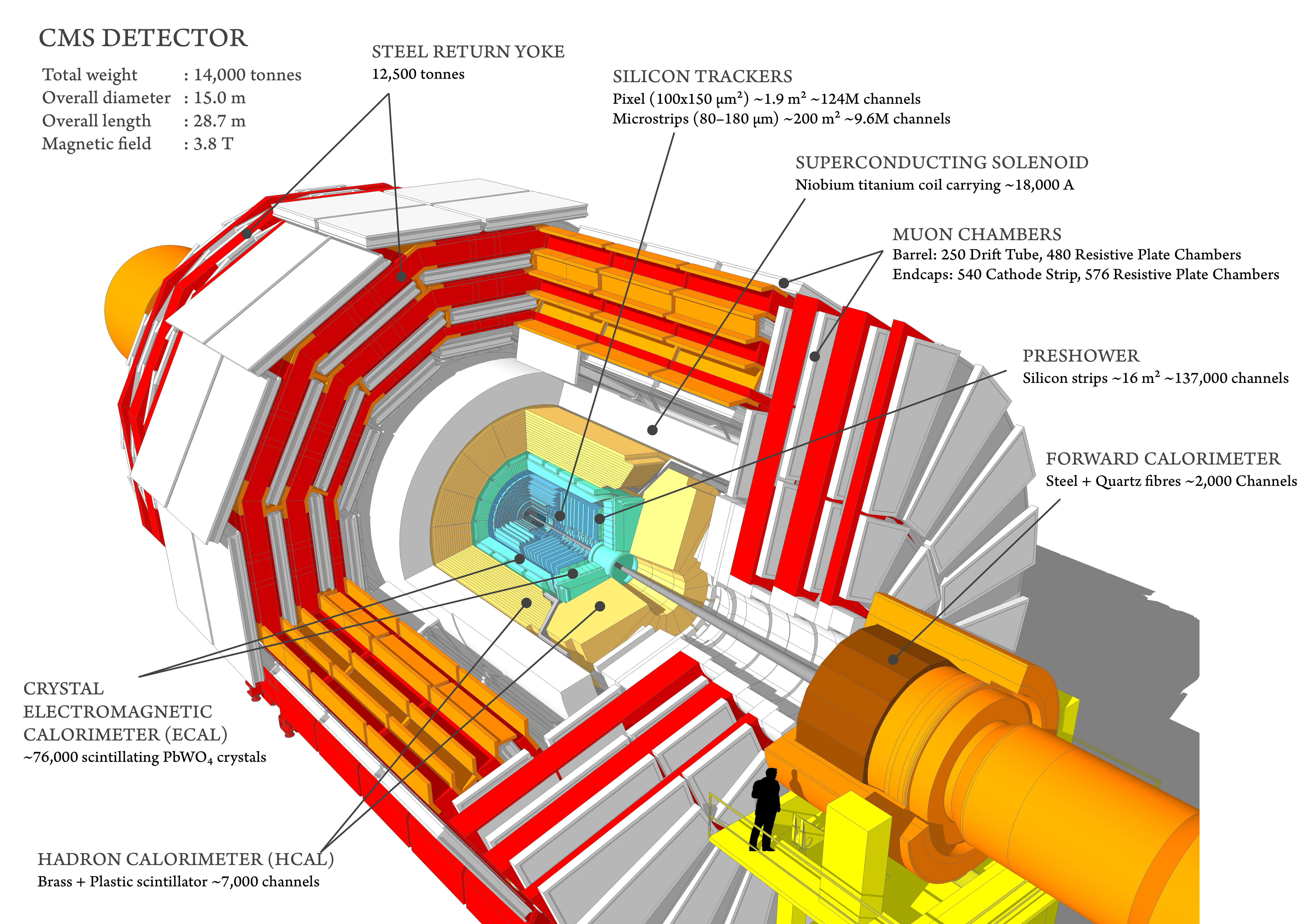}
\end{myfigure}


\subsection{Performance differences between the ATLAS and CMS detectors}

The detector layouts of the ATLAS and CMS Experiments are very similar in many aspects because they pursue the same physics goals, as outlined above.
The order of the subdetector systems is predetermined by obtaining the maximum physics information of all objects.
Large angular coverage is needed to constrain undetected energy.
Full coverage in \phi is reached; the coverage in \absEta is similar for both experiments: up to $\absEta=4.9$ for the ATLAS Experiment~\cite{ATLAS:2008xda} and up to $\absEta=5.0$ for the CMS Experiment~\cite{CMS:2008xjf}.
Finally, a triggering mechanism is required to select the interesting events out of the multitude of collisions.

The two experiments have different design foci, however.
The emphasis of the CMS Experiment is on excellent resolution for tracks, electrons and photons~\cite{Seiden:2012zz}.
The focus of the ATLAS Experiment is to obtain more than the minimally required amount of measurements for each object by using high granularity and many detector layers.
This enables cross checks between the subdetector systems of the ATLAS detector and ultimately excellent background rejection.
The different design foci of ATLAS and CMS Experiment lead to distinct choices of architecture and correspondingly in physics performance.
The main items shall be outlined in the following.

The ATLAS Experiment employs a \SI{2}{T} magnetic field in the Inner Detector to bend the trajectories of charged particles.
The magnet is optimised to enable the best momentum resolution while minimising the material in front of the \EM calorimeters.
The CMS Experiment, in contrast, uses an extraordinarily powerful, \SI{3.8}{T} magnetic field.
This gives better momentum resolution for charged particles, at the cost of worse energy resolution in the hadronic calorimeters due to the more voluminous magnets in front of the tail-catchers.

The CMS trackers are made entirely of silicon, allowing for precise position measurement in a large volume.
The ATLAS Experiment instead makes use of straw tubes in the outer regions of the Inner Detector.
They yield a worse track resolution but are more radiation-hard, resulting in a slower degradation of the detector over time.
In addition, they allow for improved electron identification by detecting transition radiation.

Two different detector technologies for the \EM calorimeters are explored for the ATLAS and CMS Experiments, which increases the safety against technology failure through diversification.
The CMS Experiment uses a homogeneous calorimeter, where the entire volume is active and absorber material at once, made of lead-tungstate crystals which are located inside the so\-le\-noi\-dal coil.
The ATLAS Experiment uses a sampling calorimeter with lead absorber plates and liquid argon as active material.
This is a more radiation-hard technology.
They are placed outside the solenoidal coil of the ATLAS detector, which diminishes resolution of electrons and photons because it renders multiple scattering in the magnet possible.

Overall, the CMS Experiment has a better energy and momentum re\-so\-lu\-tion for tracks, e\-lec\-trons and photons as shown in the first two rows of \tabref{tab:exp_ATLAS_CMS_resolution}.

\begin{mytable}{Typical values for the resolution of the transverse momentum for objects with $\pT=\SI{100}{GeV}$ resulting from the design choices of the ATLAS and CMS Experiments. Table adapted from \refscite{Seiden:2012zz, Ragusa:2007zz}.}{tab:exp_ATLAS_CMS_resolution}{llcc}
	Subdetector system& physics objects& \multicolumn{2}{c}{\pT resolution [\%] for}\\
	&& ATLAS& CMS\\
	\midrule
	Inner tracker& charged particles & 4& 1\\
	\EM calorimeter& electrons, photons& 1.1& 0.5\\
	Hadronic calorimeter& jets& 6& 11\\
	Muon system& muons& 1& 1\\
\end{mytable}

The drawback of placing the calorimeters within the magnetic solenoid is a reduced resolution for jet observables in the hadronic calorimeters for the CMS detector: due to limited space and the magnetic field, the CMS hadronic calorimeters have to be very compact and non-magnetic. The CMS Experiment therefore makes use of brass as absorber and plastic scintillators as active material.
The setup still necessitates tail-catchers outside the solenoid for sufficient thickness in terms of interaction lengths $\lambda$. Multiple scattering within the solenoid deteriorates the jet resolution and correspondingly the measurement for transverse momentum imbalances.
The ATLAS hadronic calorimeters are not as constrained in space and magnetic properties, resulting in a larger subdetector volume and more complete hermeticity.
In the barrel region, steel is used as absorber as well as scintillating tiles as active material.
Overall, this leads to better resolutions for jet energy and momentum imbalance as shown in the third row of \tabref{tab:exp_ATLAS_CMS_resolution}.

For the muon system, the ATLAS Experiment focuses on very high precision by employing four different detector technologies. Separate toroids not only in the barrel but also in the forward and backward region enable good resolution also at large \absEta. The magnetic field in the muon system for the CMS Experiment is generated by the iron return yokes. The muon chambers are placed in-between these.
Both, the ATLAS and the CMS Experiment, combine measurements in the Muon spectrometer with those in the tracker to improve the resolution of muon momenta.
Details for the approach of the ATLAS Experiment for this are given in \secref{sec:objReco_muons}.
At low muon momentum, the curvature radius is paramount for the momentum resolution, leading to better results for the CMS Experiment due to the stronger magnetic solenoid.
At high momentum, the tracker volume is too small for adequate curvature measurement and the performance of the Muon spectrometer becomes more important.
This leads to better muon momentum resolution at high momenta for the ATLAS Experiment.
At transverse momenta of \SI{100}{GeV}, both experiments achieve similar resolutions as shown in the last row of \tabref{tab:exp_ATLAS_CMS_resolution}.

\section{Further experiments at the Large Hadron Collider}
\label{sec:LHC_furtherExperiments}

Apart from the two mentioned, general-purpose detectors -- ATLAS and CMS~-- there are also two other large experiments at the \LHC:
\begin{itemize}
	\item The LHCb Experiment~\cite{LHCb:2008vvz} is focused on heavy flavour physics: it is optimised to look for rare decays of $B$ and charm hadrons -- hadrons with a valence bottom- or charm-quark, respectively -- and new physics in \CP violation. The emphasis on heavy flavour physics allows the LHCb Experiment to run at lower luminosity, resulting in less pileup, lower overall occupancy and less radiation damage while still providing enough statistics for \bbbar processes. As \bbbar pairs are predominantly produced in the same forward or backward cone, the LHCb detector is constructed as a single-arm spectrometer covering 10 to \SI{300}{mrad}.
	\item The ALICE Experiment~\cite{ALICE:2008ngc} is a general-purpose detector with a focus on heavy-ion collisions. The nominal emphasis is on lead--lead collisions at $\sqrt{s}<\SI{7}{TeV}$, but also symmetric collisions of lighter ions (including protons at $\sqrt{s}\geq\SI{7}{TeV}$) and asymmetric proton--nucleus collisions are studied. These conditions allow investigating \QCD effects as well as quark--gluon plasmas at extreme values of energy density and temperature.
\end{itemize}

There are also six smaller, sophisticated experiments at the \LHC that each use the same collision points as one of the larger experiments:
FASER~\cite{FASER:2019aik},
\LHC{}f~\cite{LHCf:2008lfy},
MATHUSLA~\cite{MATHUSLA:2018bqv,MATHUSLA:2020uve},
milliQan~\cite{Haas:2014dda},
MoEDAL~\cite{MoEDAL:2009jwa,MoEDAL:2014ttp} and
TOTEM~\cite{TOTEM:2008lue}.
\Chapter[1]{Simulating collisions}{Monte-Carlo event-generators}{%
	Muse}{Algorithm~\cite{Muse:2018alg}}{verse 1, lines 3-5}
\label{sec:MC}

Theoretical predictions can be compared to measurements at particle colliders to make statistical statements about the theory.
Many properties of events from particle collisions are, however, not calculable analytically. In consequence, it is necessary to run simulations based on Monte-Carlo~(\MC) integration methods and Markov chains in the form of random walks to obtain reliable predictions. Further, these \MC simulations can be used to extract \SM parameters, design and tune new experiments as well as analyses and extract possible \BSM signals from the \SM background.

In \secref{sec:MC_overview}, it is described how particle-collision events are generated to model physics processes of interest.
This gives events in the inclusive final state (\cf\secref{sec:level_interpretation}).
How the detector response to generated events is simulated is discussed in \secref{sec:MCEG_detectorSimulation}.
This step gives events in the detector-signal representation.
The setups for event generation and detector simulation employed in this work are described in \secref{sec:MC_thesis}.

\section{Event generation}
\label{sec:MC_overview}

There exists a variety of different Monte-Carlo event-generators (\MCEGs) which broadly adopt the same procedure.
The following gives a short overview about the most important \MC techniques, concentrating on the items required to make an informed choice regarding the \MCEG to use from the experiment's point of view.
The focus is on the simulation of hadron--hadron collisions as this thesis concerns the usage of \LHC data.
Most information, however, is also applicable in the context of other collision particles.
The section follows the points made in \refcite{Buckley:2011ms}, \qv for a complete review of the topic.

\bigskip
\MC generation relies heavily on the concept of factorisation, \ie, in this context, to separate the treatment of the highest-energy scales from those at low energy.
At high energies, the quarks and gluons, collectively referred to as \textit{partons}, forming the incoming hadrons produce highly energetic outgoing partons, leptons or gauge bosons when colliding.
This \textit{hard process} is perturbatively computable and commonly the main interest in the collision.
It is crucial to also account for the lower-energy scales to obtain an accurate description of the complete multi-particle final states, however.

At scales with low momentum transfer, typically \order{\SI{1}{GeV}}, the coupling constant of the strong interaction~\alphas becomes large. This results in incoming partons being confined and outgoing partons interacting non-per\-tur\-ba\-tive\-ly, forming hadrons. \textit{Hadronisation} cannot be calculated from first principles as of now but has to be modelled in a way that is inspired by, but not derived from, \QCD.

The transition from high to low scales is described by an evolutionary process called \textit{parton showering}.
Parton showering can in principle be calculated by perturbative \linebreak\QCD~\cite{Buckley:2011ms}.
In practise, it is, however, commonly determined using Markov processes because of their better computational scaling.

\bigskip
The calculation of the hard process and parton showering are discussed in \linebreak\secsref{sec:MCEG_hardProcess}{sec:MC_partonShower}, respectively.
Overlap between these two steps is handled by matching and merging techniques, which are detailed in \secref{sec:MCEG_matchMerge}.
Hadronisation and decay of particles in generated events is examined in \secref{sec:MCEG_hadronisation}.
The discussion up to that point focuses on the generation of events according to the Standard Model.
Specific considerations that have to be taken into account when generating events for \BSM models are detailed in \secref{sec:MCEG_BSM}.
An overview of the \MCEGs most important for this thesis is given in \secref{sec:MCEG_overview}.

\subsection{Hard process}
\label{sec:MCEG_hardProcess}

The main interest in physics analyses typically lies in the properties of hard processes, \ie processes with large momentum transfer $Q^2$, because they produce particles with large masses or high transverse momenta.
Their calculation makes use of the fact that \QCD quanta are asymptotically free at these high energy scales because \alphas becomes small, and properties can be calculated from perturbative \QCD (\cf\secref{sec:SM_QCD}).
The starting point is to choose a hard process and generate it according to its matrix element (\ME) and phase space.
The hard process is typically calculated at leading order (\LO), next-to-leading order (\NLO) or single-loop level.
Higher orders provide higher accuracy but come with immensely increased computational complexity.
This limits the choice for the available and commonly used order of the calculation.

Regarding the generation of the hard process in detail, the cross section $\sigma_{AB\to n}$ for incoming partons $A$ and $B$ forming $n$ outgoing partons in a collision has a functional dependence~\cite{Buckley:2011ms}
\begin{equation}
	\label{eq:MCEG_xs}
	\sigma_{AB\to n}= f\left(\PDF^{h_1}_A\left(x_A, \mu_F\right), \PDF^{h_2}_B\left(x_B, \mu_F\right), \Phi_n, \mathcal{M}_{AB\to n}\right).
\end{equation}
Hereby, $\PDF^{h_i}_Y\left(x_Y, \mu_F\right)$ gives the probability to find a parton of species $Y$ with momentum fraction $x_Y$ in hadron $h_i$ if probed at the factorisation scale $\mu_F$ and is called \textit{parton distribution function} (\PDF)~\cite{Bjorken:1969ja}.
A parton distribution function has to follow the equation
\begin{equation*}
	\sum_Y \int_x\PDF^{h_i}_Y\left(x_Y, \mu_F\right) \mathrm{d}x_Y\equiv1,
\end{equation*}
summing over all hadron constituents $Y$ and integrating over all momentum fractions~$x_Y$. \PDFs depend on the hadron wave function, which is non-perturbative, and can consequently not be derived from first principles but have been measured~\cite{Bailey:2020ooq,Ball:2012cx,NNPDF:2014otw,Harland-Lang:2014zoa,Butterworth:2015oua,Dulat:2015mca}.
Example \PDFs for the proton are shown in \figref{fig:MCEG_PDFs}.
For typical processes at the \EW scale, $Q=\order{\SI{100}{GeV}}$ and correspondingly $Q^2=\order{\SI{e4}{GeV^2}}$.
At a centre-of-mass energy of \SI{13}{TeV}, the probed minimum momentum fraction $x$ is small and in a region where gluons dominate the \PDF (see \figref{fig:MCEG_PDFs}, right).
Therefore, integrating over the momentum fraction, the most important contributions inducing \EW processes come from gluons and to a lesser extent also valence up- and down-quarks.

\begin{myfigure}{
		68\% confidence-level uncertainty bands for the proton \PDFs $f\left(x, \mu_F=Q^2\right)$ at next-to-next-to-leading order (\NNLO) for $Q^2=\SI{10}{GeV^2}$ (left) and $Q^2=\SI{e4}{GeV^2}$ (right). Figures taken from \refcite{Bailey:2020ooq}.
	}{fig:MCEG_PDFs}
	\begin{tabular}{rr}
		\includegraphics[width=0.46\textwidth]{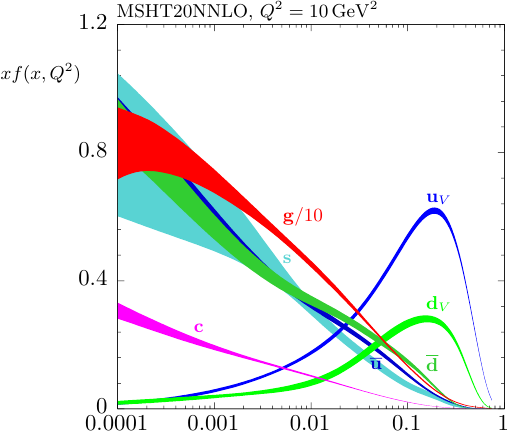}&
		\includegraphics[width=0.46\textwidth]{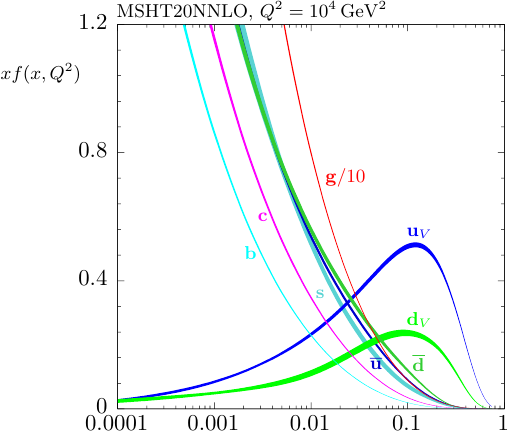}\\
		\scriptsize{$x$}\hspace*{10pt}& \scriptsize{$x$}\hspace*{10pt}
	\end{tabular}
\end{myfigure}

In \eqref{eq:MCEG_xs}, $\mathcal{M}_{AB\to n}$ is the matrix element for the process $AB\to n$ which can be calculated as the sum over the contributing Feynman diagrams. In general, it depends on the phase space $\Phi_n$ for the $n$-particle final state, the factorisation scale $\mu_F$ and the renormalisation scale $\mu_R$. These scales $Q^2\coloneqq\mu_F, \mu_R$ typically are chosen corresponding to the $s$-channel resonance mass, $Q^2=M^2$, or the transverse momentum of a massless particle, $Q^2=\pT^2$. Its importance in calculating the cross section is the reason why the hard process is often also referred to as "matrix element".\\
The computational load for calculating matrix elements as well as phase spaces analytically increases drastically with the number of final state particles. Phase spaces are therefore integrated using \MC sampling techniques.
Matrix elements are calculated using either dedicated \MCEGs, like \MadGraph~\cite{Alwall:2014hca,Frederix:2018nkq}, or limiting the number of final state particles.
The latter is the approach taken for example by \Herwig~\cite{Bahr:2008pv,Bellm:2015jjp,,Bellm:2019zci} which takes into account only \twoToTwo and $s$-channel processes for \BSM models when no external \ME generators are employed.

\subsection{Parton showering}
\label{sec:MC_partonShower}

The hard process gives a good description of the energetic particles in a collision at a fixed order.
This, however, is not sufficient for representing the observables of a full process because for example jet substructures depend heavily on lower-scale emissions.
Parton showering therefore generates the effect of all higher orders to emissions below the hard scale defined by $Q^2$ by connecting it through an evolutionary chain to the outgoing partons and hadronisation scale at $Q_0^2$. In this evolution, coloured particles can radiate gluons which in turn radiate other gluons or produce quark--antiquark pairs. This can be simulated using Markov chains, taking into account the probability to radiate a new from a given parton in each step. This procedure is thereby in principle independent of the hard process and multiplicative to its total cross section.
The process history can therefore be neglected in subsequent steps.

The iterative splitting does, however, require an ordering scale that simultaneously approaches two separate but interconnected limits:
\begin{itemize}
	\item In the \textit{collinear limit} the angle $\theta$ between emitting and emitted parton becomes increasingly small.
	\item In the \textit{infrared limit} the transverse momentum \kT of the emitted parton relative to the emitting parton decreases.
\end{itemize}
In both limits the partons become indistinguishable by measurement. As
\begin{equation*}
	\frac{\mathrm{d}\theta^2}{\theta}=\frac{\mathrm{d}\kT^2}{\kT},
\end{equation*}
a common natural cut-off scale $Q_0^2$ for resolvable partons is defined below which no parton evolution is needed. At the same time, the running coupling \alphas increases when successively lowering the ordering scale, enhancing the splitting probability.
This means that at the \QCD scale $\Lambda_{\QCD}$, where $\alphas\gtrsim1$, parton showering becomes computationally unaffordable. Conveniently, $\order{\SI{1}{GeV}}=Q_0^2\gg\Lambda_{\QCD}=\order{\SI{0.3}{GeV}}$ is found~\cite{Buckley:2011ms}.

There are two established choices regarding the ordering scale:
\begin{itemize}
	\item \textit{Angular ordering} is the natural ordering following from the collinear limit. The infrared limit is simultaneously taken into account because low-energy gluons at wide angles are not emitted by individual partons but due to interference by the process as a whole. This can be seen when inspecting the corresponding matrix elements. Mathematically, this is equivalent to having low-energy, wide-angle emissions first before more collinear emissions take place. As such, they can be included into angular-ordered parton showering. This is the approach taken for example by the \Herwig generator.
	It has to be pointed out, however, that this is not physically equivalent as the shower history is altered.
	\item \textit{\kT ordering} is the natural ordering following from the infrared limit. Here, the event is decomposed into colour flows where the connecting lines can be interpreted as colour dipoles. Emissions from dipoles then influence the event globally. The shower evolution starts with the highest-\kT emission and proceeds towards successively lower \kT. This readily approaches the infrared limit but can also incorporate the collinear limit~\cite{Gustafson:1987rq, Hamilton:2020rcu}. This is the approach followed for example by \Pythia~\cite{Sjostrand:2014zea} and \Sherpa~\cite{Sherpa:2019gpd}.
\end{itemize}

The description above has been focused on partons coming out of the hard process, called final-state radiation (\FSR).
The partons causing the hard process can, however, analogously emit other partons before the collision.
Parton showering for this initial-state radiation (\ISR) can, in principle, work similarly to that for \FSR.
Special attention, however, has to be paid to keeping only those partons in the final generation output that are indeed involved in a hard scatter.
All others have to be folded back into the proton remnant as they do not interact, which is computationally inefficient.
In practise, parton showering for initial-state radiation is therefore performed by a backward evolution from the hard process~\cite{Sjostrand:1985xi,Marchesini:1987cf}.
In this calculation, the probability for a parton to come from another parton with a higher momentum fraction is taken into account.

\subsection{Matching and merging}
\label{sec:MCEG_matchMerge}

The need for starting event generation with the hard-process calculation for high precision at high scales as well as parton showering to simulate the process evolution to low scales has been outlined in the previous sections.
In principle, there arises double counting between partons from the hard process and partons from the parton shower when combining multiple matrix elements.
This is sketched in \figref{fig:MCEG_MatchingMerging}: when generating for example events with at least one \ISR parton, processes with exactly two \ISR partons are added by parton showering events with one \ISR parton in the matrix element (first row) as well as by the matrix element directly (second row). Different methods have been established to handle this potential double counting.

\begin{myfigure}{
		Schematic overview of the Feynman diagrams added by the matrix element and by the parton shower for a hadron collision producing at least one \ISR parton without matching or merging applied.
	}{fig:MCEG_MatchingMerging}
	\includegraphics[width=\textwidth]{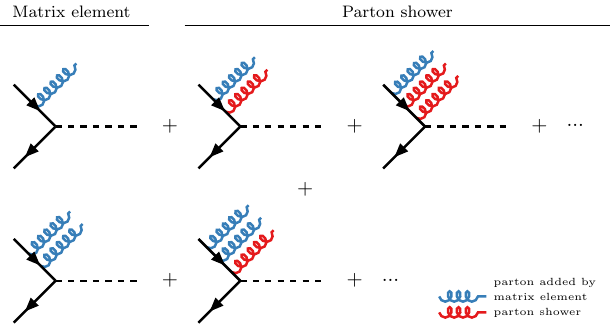}
\end{myfigure}

In \textit{matching} techniques~\cite{Bengtsson:1986hr}, processes at a chosen order from the matrix element are directly matched to their parton shower equivalents and excluded from additional calculation in the parton shower.
This gives exact generation results up to the chosen order. \LO matching has been common for a long time, but by now also \NLO matching is firmly established, \eg by \MCatNLO~\cite{Frixione:2010wd} and \POWHEG~\cite{Nason:2004rx,Frixione:2007vw}.

The second approach is \textit{merging}~\cite{Catani:2001cc,Lonnblad:2001iq,Hoeche:2009rj,Hamilton:2009ne}, in which a cut-off scale is introduced above which partons are taken from the matrix element and below which they come from the parton shower.
The result is then not accurate to a specific order but a mixture of the order used by the matrix element and all orders considered by the parton shower.
There exist numerous algorithms to combine \LO \ME-calculations with parton showers, \eg CKKW~\cite{Catani:2001cc}, CKKW-L~\cite{Lonnblad:2001iq} and MLM~\cite{Mangano:2006rw}, and fewer for merging \NLO \ME-calculations with parton showers, \eg\MePsAtNLO~\cite{Hoeche:2012yf}.

\subsection{Hadronisation and decays}
\label{sec:MCEG_hadronisation}
The interaction between partons becomes strong when the shower evolution approaches $\Lambda_{\QCD}$. The consequence is a non-perturbative confinement of partons into colourless bound states, hadrons. The simulation of the hadronisation process is not directly derived from \QCD but merely inspired by it. Similar to the parton shower it is insensitive to the exact production history of the partons.
This allows developing common hadronisation models independent of the kind and energy of the hard process or shower history.
There are two models for hadronisation in use.

In the \textit{string model}~\cite{Andersson:1983ia,Andersson:1997xwk}, quarks are connected by colour strings and gluons are represented by colour--anticolour kinks in the strings. The potential energy stored in the string increases the further the partons move apart until the string breaks, forming a quark--antiquark pair between the original partons. The system becomes separated into two colour-singlet, \ie non-interacting, parts. For each of those parts the procedure is repeated recursively until the potential energy becomes too low to form new quark--antiquark pairs. Colour-connected quark--antiquark pairs are then considered mesons. Baryons can be produced in this model by either matching multiple quark--antiquark pairs or by allowing not only splitting of the string into quark--antiquark but also diquark--antidiquark pairs.
It can be shown that neither collinear nor low-energetic partons impact the outcome of this hadronisation significantly~\cite{Buckley:2011ms} which renders it insensitive to the exact cut-off scale $Q_0^2$ of the parton shower.
However, there are many other parameters related to flavour properties that have to be determined from measurement.
The string model is the hadronisation scheme most prominently employed by \Pythia.

The second approach is the \textit{cluster model}~\cite{Amati:1979fg,Wolfram:1980gg}.
Here, colour singlet combinations of partons, "clusters", are formed in a pre-confinement step. These have asymptotically universal mass distributions which only depend on $Q_0$ and $\Lambda_{\QCD}$ but not on $Q$ as long as $Q\gg Q_0$~\cite{Amati:1979fg}. In general, this model leads to a less good description of event properties after hadronisation than the string model~\cite{Buckley:2011ms}, but it has considerably fewer parameters.
The cluster model is employed by \Sherpa~\cite{Winter:2003tt} and \Herwig~\cite{Webber:1983if}.


\bigskip

Closely intertwined with hadronisation is the decay of the hadrons into other lighter particles. In general, \MCEGs work in the narrow-width approximation, \ie the decay of particles is factorised from their production.
This holds for $\width\ll M$ where $M$ is the mass of the decaying particle and \width its decay width because in this case interference effects between production and decay can be neglected.

Particle decays can either be accounted for in the \MCEG itself or by usage of a sophisticated external package like \EvtGen~\cite{Lange:2001uf}.
There are two choices each decay software has to make: Firstly, which decay channels to consider; because there is an overwhelming number of options and non-negligible uncertainties connected to these. Secondly, each specific decay channel can either be calculated from the matrix element or simulated. 

After hadronisation and decay, the generated events are in the representation of the inclusive final state (\cf\secref{sec:level_interpretation}).
Constructing physics objects and applying kinematic cuts brings them into particle-level representation.
Performing a detector simulation on the events converts them into detector signals (see \secref{sec:MCEG_detectorSimulation}).

\subsection{Event generation for \BSM models}
\label{sec:MCEG_BSM}

The discussion up to this point has focused on generating events for \SM processes.
Generating \BSM events follows in general the exact same principles outlined in Sections~\ref{sec:MCEG_hardProcess} to~\ref{sec:MCEG_hadronisation}.
It does, however, require the implementation of new physics processes.
Many \MCEGs commonly have built-in support for a smaller number of chosen \BSM models, but further effort is needed for \BSM models beyond those.
\FeynRules~\cite{Alloul:2013bka} is a package that calculates the underlying Feynman rules for an arbitrary \BSM model.
The package is commonly interfaced to various \MCEGs using the Universal \FeynRules Output (\UFO,~\cite{Degrande:2011ua}). The \MCEGs then automatically calculate the matrix element from this input. Thereby, two different approaches are chosen: \Herwig, for example, keeps the calculations computationally lightweight by limiting them to \twoToTwo processes and the production of \BSM particles in the $s$-channel.
These are followed by cascade decays of the heavy particles. \MadGraph and \Sherpa generate any $2\to N$ process to treat unstable intermediate particles correctly and to be able to take into account correlation effects~\cite{Buckley:2011ms}.

For \BSM models, usually only the hard process has to be generated differently from \SM processes.
This is the case for the \THDMa.
There are also models impacting the complete simulation down to parton shower and hadronisation which have to be treated in a more sophisticated manner, however.

\subsection{Overview of Monte-Carlo event-generators}
\label{sec:MCEG_overview}

A summary of the most important features of the \MCEGs used in this thesis following from the outline above is given in \tabref{tab:MCEGs}:
\begin{itemize}
	 \item \Herwig~\cite{Bahr:2008pv,Bellm:2015jjp,,Bellm:2019zci} and \Sherpa~\cite{Sherpa:2019gpd} are general-purpose \MCEGs that can simulate the complete physics from the hard process over the parton shower to hadronisation and decay.
	 \item \Pythia~\cite{Sjostrand:2014zea} is also a general-purpose \MCEG but with a clear focus on parton shower and hadronisation:
	 the matrix element is limited for \QCD calculations to $2\to N$, $N\leq3$ processes in the Standard Model and no automatic generation of code for \BSM models exists.
	 For particle decays, \Pythia is often interfaced with \EvtGen~\cite{Lange:2001uf}, a sophisticated external package.
	 \item \MadGraphAMCatNLO\cite{Alwall:2014hca,Frederix:2018nkq} is an \MCEG specialised on the calculation of the hard process.
	 It offers automatic code generation for \BSM models, \eg from \UFO{}s~\cite{Degrande:2011ua}.
	 For parton shower and hadronisation, it has to be interfaced with one of the general-purpose \MCEGs to which it supplies the information for matching matrix element to parton shower following the \MCatNLO~\cite{Frixione:2010wd} approach.
	 \item \POWHEGBOX~\cite{Alioli:2010xd} is a software framework specialised on the calculation of the hard process.
	 It can, however, only handle \BSM processes if these are implemented by the user themself.
	 For parton shower and hadronisation, it has to be interfaced with one of the general-purpose \MCEGs to which it supplies the information for matching matrix element to parton shower following the \POWHEG~\cite{Nason:2004rx,Frixione:2007vw} approach.
\end{itemize}

\begin{sidewaystable}
	\centering
	\begin{threeparttable}
		\begin{tabular}{lccccc}
			\toprule
			\textbf{\MCEG}& \Herwig& \Pythia& \Sherpa& \MadGraphAMCatNLO& \POWHEGBOX\\
			Major version number& 7.2& 8.2& 2.2& 2.7& 2\\
			Purpose& general& general& general& hard process& hard process\\
			\midrule
			\textbf{Hard process}\\
			\SM& all& all\tnote{c}& all& all& all\\
			Built-in \BSM& yes& yes& yes& yes& no\tnote{d}\\
			Automatic \BSM code& yes\tnote{a}& no& yes& yes& no\\
			\BSM \UFO& in part\tnote{b}& no& yes& yes& no\\
			\midrule
			\textbf{Matching and merging}& various& various& various& matching (\MCatNLO)& matching (\POWHEG)\\
			\midrule
			\textbf{Parton shower}& yes& yes& yes& no& no\\
			Parton shower ordering& angular& \kT& \kT\\
			\cline{1-4}
			\textbf{Hadronisation}& cluster & string& cluster\\
			\textbf{Decays}& yes& yes& yes\\
			\midrule
			\textbf{References}& \cite{Bahr:2008pv},\cite{Bellm:2015jjp},\cite{Bellm:2019zci}& \cite{Sjostrand:2014zea}& \cite{Sherpa:2019gpd}& \cite{Alwall:2014hca},\cite{Frederix:2018nkq}& \cite{Nason:2004rx},\cite{Frixione:2007vw},\cite{Alioli:2010xd}\\
			\bottomrule
		\end{tabular}
		
		\begin{tablenotes}
			\item[a]but no interference effects with other decay chains
			\item[b]matrix element for \twoToTwo and $s$-channel processes
			\item[c]only $2\to N$, $N\leq3$, for \QCD
			\item[d]but can be implemented by user
		\end{tablenotes}
	
		\caption{Overview of the important features of the \MCEGs and software packages mainly used in this thesis.}
		\label{tab:MCEGs}
	\end{threeparttable}
\end{sidewaystable}

\section{Detector simulation}
\label{sec:MCEG_detectorSimulation}
So far, the description of the \MC generation has been independent of the detector intended to measure the generated events.
However, it is essential to simulate the whole detector system and its interaction with the hadrons and leptons coming from the primary collision to allow for direct comparisons of generated to real events.
Effects that need to be considered are
\begin{itemize}
	\item Detector geometry and used materials.
	\item Decay of long-lived particles and interaction of particles with the detector material, producing secondary particles if applicable.
	\item Tracking of primary and secondary particles through the detector material and electromagnetic fields.
	\item The response of the detector sensors to the particle--material interactions.
\end{itemize}

The ATLAS, CMS and LHCb Experiments rely on \GeantFour~\cite{GEANT4:2002zbu,Allison:2006ve} for these simulations~\cite{ATLAS:2010arf,Hildreth:2015kps,Clemencic:2011zza}.
Due to the high amount of material in the detectors and their intricate working, detector simulation is computationally very costly: it can take up to minutes to process a single event. 
This can be mitigated by replacing the steps in the simulation that are particularly slow, \eg the interactions in the calorimeters, with pre-simulated showers~\cite{ATLAS:2010arf}.

When the detector simulation is completed, the signals the detector would record are digitised, \ie for each readout channel it is estimated whether the signal exceeds a predefined threshold in a specified time-window. If it does, the sensor response is recorded. Pileup is overlaid either before digitisation or afterwards.

After these steps, the simulated data is in detector-signal representation (\cf\secref{sec:level_interpretation}).
They have the same format as the real data coming from the data acquisi-\linebreak{}tion~(\DAQ) and can be treated using the same workflow as data, \ie be subsequently processed by high-level trigger~(\HLT) and reconstruction software in the case of the ATLAS detector.
This is described in \chapref{sec:objReco}.
Applying additional kinematic cuts brings the observables for reconstructed objects into detector-level representation.
The sole but important technical difference between real and simulated data is that the latter is enriched with the particle-level information of the physics processes.

\bigskip
One caveat with detector simulation is that the information to accurately emulate the detector response and efficiencies might not necessarily be public.
Even if it is, simulating the interaction between the generated particles and the detector is very costly computationally, as mentioned above.
For this reason, procedures exist to smear particle-level representations and parametrise the detector response to obtain detector-level representations~\cite{deFavereau:2013fsa}.
This, however, comes with additional inaccuracies.

Either point of computational cost, missing public detector information as well as inaccuracies from smearing renders it effortful to reinterpret real data published in detector-level representation.
This is also discussed in \chapref{sec:analysisPreservation}.
The caveats of detector simulation for reinterpretation can be avoided by publishing data directly in particle representation.
It is shown in \chapref{sec:metJets_detectorCorrection} how detector-level representations can be converted into particle-level representations by correcting for detector effects.

\section{Monte-Carlo setups in this work}
\label{sec:MC_thesis}

In the following, the \MC setups used in this work are detailed.
In \secref{sec:metJets_MC}, the setup employed for \SM processes in the \METjets measurement in \chapsrefAnd{sec:metJets}{sec:metJets_detectorCorrection} is described.
The setup employed for generating \SM processes for the interpretation of the measurement in \chapref{sec:interpretation} is given in \secref{sec:interpretation_SM_MC}.
In \secref{sec:interpretation_2HDMa_MC}, the setup for interpreting the measurement with respect to the \THDMa in \chapref{sec:2HDMa_metJetsMeasurement}~and~\secref{sec:interpretation_2HDMa} is given.
The setup for generating events to determine the sensitivity of existing \LHC measurements to the \THDMa with the \Contur toolkit in \secref{sec:contur_2HDMa} is described in \secref{sec:MC_2HDMa_Contur}.
Of all the mentioned \MC setups, only the one in \secref{sec:metJets_MC} needs a description of events in the detector-signal representation and employs a detector simulation.
All other setups exclusively employ event generation and provide events in the inclusive final state.

\subsection{\SM processes for the \METjets measurement}
\label{sec:metJets_MC}

The \MC setup employed in the \METjets measurement described in \chapsrefAnd{sec:metJets}{sec:metJets_detectorCorrection} is detailed in the following.
An overview is given in \tabref{tab:metJets_SM_MC}.
The detector simulation is carried out with \GeantFour~\cite{GEANT4:2002zbu,Allison:2006ve}, after which simulated events can be reconstructed using the same workflow as data.
For all simulated \SM processes, pileup (\cf\secref{sec:experiment_luminosity}) is overlaid over the hard-scatter event based on soft \QCD processes simulated with \toolVersion{\Pythia}{8.186}~\cite{Sjostrand:2014zea} using the \NNPDFTwoThreeLO \PDF set~\cite{Ball:2012cx} and the A3 set of tuned parameters~\cite{ATLAS:2016puo}.
Simulated events are then weighted such that the average number of interactions per bunch crossing in simulation matches the one observed in data.
Uncertainties related to the \MC generation are discussed in \secref{sec:metJets_theoSystUnc}.

\begin{mytable}{%
		Generation setup used for the different considered \SM processes in the \METjets measurement in detector-level representation with the considered \PDF set, corresponding hard-process as well as parton-shower generators and tune for the underlying event.
	}{tab:metJets_SM_MC}{cccccc}
	Physics process & \PDF set& Hard process & Parton shower & Tune\\
	\midrule
	\Vjets & \NNPDFNNLO & \multicolumn{2}{c}{\toolVersion{\Sherpa}{2.2.1}} & \Sherpa default\\
	\EW\Vjj & \NNPDFNNLO & \multicolumn{2}{c}{\toolVersion{\Sherpa}{2.2.11}} & \Sherpa default\\
	di-/tribosons & \NNPDFNNLO & \toolVersion{\Sherpa}{2.2.1/2.2.2} & \toolVersion{\Sherpa}{2.2.2} & \Sherpa default\\
	top quarks & \NNPDFNLO & \toolVersion{\POWHEGBOX}{2} & \toolVersion{\Pythia}{8.230} & A14\\
\end{mytable}

\subsubsection{\Vjets processes}
\label{sec:metJets_MC_Vjets}
The generation of events containing a single vector boson $V\in\left\{W, Z\right\}$ in association with jets, \Vjets, follows the procedure described in \refcite{ATL-PHYS-PUB-2017-006}.
They are generated with \toolVersion{\Sherpa}{2.2.1}~\cite{Sherpa:2019gpd} using matrix elements at \NLO accuracy for up to two jets and at \LO accuracy for up to four jets calculated with \Comix~\cite{Gleisberg:2008fv} and \OpenLoops~\cite{Cascioli:2011va,Denner:2016kdg}.
A dedicated set of tuned parameters developed by the \Sherpa authors for this version based on the \NNPDFNNLO~\cite{NNPDF:2014otw} \PDF set is used and the default \Sherpa parton showering employed.
Matching uses a colour-exact version of \MCatNLO~\cite{Frixione:2010wd}, merging the CKKW algorithm~\cite{Catani:2001cc} with a merging cut of $Q=\SI{20}{GeV}$ extended to \NLO using \MePsAtNLO~\cite{Hoeche:2012yf}.

\NNLO predictions in \QCD are available and would lead to a $5-\SI{10}{\%}$ increase in cross section that is mostly constant as a function of the transverse momentum of the vector boson~\cite{Anastasiou:2003ds}.
For technical reasons, however, predictions at \NLO are used consistently for \QCD and \EW production.

For \QCD \Vjets production, the spectrum of the invariant mass of dijets, \mjj, is severely mismodelled in \toolVersion{\Sherpa}{2.2.1} and older~\cite{Aad:2014dta,D0:2013gro,CMS:2016sun,ATLAS:2017nei}.
In generated events, in general larger values of \mjj are obtained.
In the \METjets measurement in \chapsref{sec:metJets}{sec:interpretation}, the event yield at $\mjj>\SI{3}{TeV}$ is for example overestimated by up to a factor of 5.
This is corrected for in a data-driven approach similar to the one taken in \refscite{Aad:2014dta,ATLAS:2017nei} whereby simulated events are assigned weights such that the distributions for simulated \QCD \Vjets processes and data less simulated non-\QCD \Vjets processes in detector-level representation agree in the \VBF subregion.
These modifications to the weights are indispensable because excessively large discrepancies between simulated and measured data can significantly impact the correction for detector effects (see \secref{sec:detCorr_unfUncertainties}).

\bigskip
Events with electroweak production of a vector boson $V$ in association with two jets (\EW\Vjj), \ie mainly vector-boson fusion (\VBF), are generated with \toolVersion{\Sherpa}{2.2.11} allowing up to one additional parton.
The matrix element is calculated at \LO and the \Sherpa default used for parton showering. Matching follows the \MePsAtLO prescription.

Diagrams of the production of two vector bosons with one of the bosons decaying to hadrons lead to the identical final state. These are removed gauge-invariantly by inhibiting colour exchange between the incoming partons.

\subsubsection{Di- and tribosons processes}
\label{sec:metJets_MC_DiTriboson}
Events with two (three) vector bosons are called \textit{diboson} (\textit{triboson}) events and generated following the procedure described in \refcite{ATL-PHYS-PUB-2017-005}.
The matrix elements for triboson events and diboson events in which both vector bosons decay to leptons are calculated with \toolVersion{\Sherpa}{2.2.2} at \NLO. Effects of virtual \QCD processes are provided by \OpenLoops. \NNPDFNNLO is used as the \PDF set. Parton showering follows the \MePsAtNLO procedure with the default parton-shower parameters from \Sherpa.

Diboson events in which exactly one of the vector bosons decays to leptons and the other to hadrons are generated with an almost identical setup, albeit using \toolVersion{\Sherpa}{2.2.1}.
Diboson events in which both vector bosons decay to hadrons do not lead to \MET, which is an important selection requirement of the \METjets measurement (see \secref{sec:metJets_regions}), and are therefore not considered.

\subsubsection{Top-quark processes}
\label{sec:metJets_MC_top}
Matrix elements for single-top-quark processes, including those where a top quark is produced in association with a $W$ boson ($tW$), and \ttbar processes are generated at \NLO with \toolVersion{\POWHEGBOX}{2}~\cite{Nason:2004rx,Frixione:2007vw,Alioli:2010xd}. The parameter controlling the transverse momentum of the first additional emission beyond the leading-order Feynman diagram, $h_\text{damp}$, is set to twice the mass of the top quark. As \PDF set, \NNPDFNLO is used. Parton showering is done by \toolVersion{\Pythia}{8.230} with the A14 set of tuned parameters~\cite{ATLAS:2014rfk} and the \NNPDFTwoThreeLO \PDF set.

Single-top events are corrected to the theory prediction calculated at \NLO with\linebreak \toolVersion{\HatHor}{2.1}~\cite{Aliev:2010zk,Kant:2014oha}.
\ttbar events are corrected to the \NNLO theory prediction including soft-gluon terms at next-to-next-to-leading logarithmic order using \toolVersion{\TopPP}{2.0}~\cite{Beneke:2011mq,Cacciari:2011hy,Barnreuther:2012wtj,Czakon:2012zr,Czakon:2012pz,Czakon:2013goa,Czakon:2011xx}.

The interference between \ttbar and $tW$ processes is taken into account by generating two distinct $tW$ predictions with the above setup, one employing the diagram-subtraction, the other the diagram-removal scheme~\cite{Frixione:2008yi}. The former is used as the nominal prediction as it provides a better description of the data, the latter is used to derive the systematic uncertainty for the treatment of this interference (see \secref{sec:metJets_theoSystUnc}).

\subsection{\SM processes for the interpretation of the \METjets measurement}
\label{sec:interpretation_SM_MC}

The simulated \MC events employed for the selection studies in \chapref{sec:metJets} and the unfolding in \chapref{sec:metJets_detectorCorrection} use the setup described in the previous section.
During the period of work on these steps, improved \SM predictions became available.
It suffices to use these improved predictions in the interpretation in \chapref{sec:interpretation} to allow for a state-of-the-art comparison to the measured data in particle-level representation.

Most of the \SM predictions are generated similar to the approach described in the previous section.
They are summarised in \tabref{tab:interpretation_SM_MC}.
Most importantly newer versions of the \Sherpa generator are used and \Sherpa is also employed for the generation of \ttbar processes.
The improvements in detail in the \MC generation are:
\begin{itemize}
	\item For \Vjets, matrix elements and parton showering are generated with\linebreak \toolVersion{\Sherpa}{2.2.11}.
	A dedicated set of tuned parameters developed by the \Sherpa authors for this version based on the \PDFfLHC~\cite{Butterworth:2015oua} \PDF set is used.
	It is supplemented with \QED effects from \LUXqed~\cite{Manohar:2016nzj,Manohar:2017eqh}.
	\Vjets events are weighted to \NLO in the electroweak coupling according to \refcite{Lindert:2017olm}.
	
	All other settings are kept unchanged.
	Most importantly, matrix elements are calculated at \NLO accuracy for up to two jets and at \LO accuracy for up to four jets with \Comix and \OpenLoops.
	Matching uses a colour-exact version of \MCatNLO, merging the CKKW algorithm with a merging cut of $Q=\SI{20}{GeV}$ extended to \NLO using \MePsAtNLO.
	
	As mentioned in \secref{sec:metJets_MC_Vjets}, \NNLO predictions in \QCD are available and would lead to a $5-\SI{10}{\%}$ increase in cross section that is mostly constant as a function of the transverse momentum of the vector boson~\cite{Anastasiou:2003ds}.
	For technical reasons, however, predictions at \NLO are used consistently for \QCD and \EW production.
	
	\item Electroweak \Vjj events are weighted to \NLO in the electroweak coupling according to~\refcite{Lindert:2022ejn}.
	
	\item For all di- and triboson events, matrix elements and parton showering are generated with \toolVersion{\Sherpa}{2.2.12} instead of \toolVersion{\Sherpa}{2.2.1/2.2.2}.
	All other settings are kept unchanged.
	
	\item \ttbar processes are generated with \toolVersion{\Sherpa}{2.2.12} using matrix elements at \NLO accuracy matrix elements for up to one additional jet and at \LO accuracy for up to four additional jets.
	Effects of virtual \QCD processes are provided by \OpenLoops.
	\NNPDFNNLO is used as the \PDF set.
	Parton showering follows the \MePsAtNLO procedure with the default parton-shower parameters from \Sherpa.
\end{itemize}
All other settings are kept unchanged.

\begin{mytable}{Generation setup used for the different considered \SM processes in particle-level representation with the considered \PDF set, corresponding hard-process as well as parton-shower generators and tune for the underlying event.}{tab:interpretation_SM_MC}{cccccc}
	Physics process & \PDF set& Hard process & Parton shower & Tune\\
	\midrule
	\Vjets & \PDFfLHC & \multicolumn{2}{c}{\toolVersion{\Sherpa}{2.2.11}} & \Sherpa default\\
	\EW\Vjj & \NNPDFNNLO & \multicolumn{2}{c}{\toolVersion{\Sherpa}{2.2.11}} & \Sherpa default\\
	di-/tribosons & \NNPDFNNLO & \multicolumn{2}{c}{\toolVersion{\Sherpa}{2.2.12}} & \Sherpa default\\
	top quarks (\ttbar) & \NNPDFNNLO & \multicolumn{2}{c}{\toolVersion{\Sherpa}{2.2.12}} & \Sherpa default\\
	top quarks (other) & \NNPDFNLO & \toolVersion{\POWHEGBOX}{2} & \toolVersion{\Pythia}{8.230} & A14\\
\end{mytable}

\subsection{\THDMa processes for the interpretation of the \METjets measurement}
\label{sec:interpretation_2HDMa_MC}

The phase space selected by the \METjets measurement requires large \MET (\cf\secref{sec:metJets_regions}).
In the \THDMa, this can mainly be obtained by the production of Dark Matter in association with other particles.
Different processes, like the decay of the pseudoscalar~$A$ to top-quark pairs, can give rise to small amounts of \MET as well.
These contributions are negligible, however.
Therefore, only processes involving the production of at least one \DM pair are generated for the interpretation of the \METjets measurement with respect to the \THDMa in \chapref{sec:2HDMa_metJetsMeasurement}~and~\secref{sec:interpretation_2HDMa}.
Exclusion limits derived with this generation setup have to be considered approximate as contributions from events without \DM pairs are neglected.

The implementation of the \THDMa physics in a corresponding Universal FeynRules Output (\UFO) format is used~\cite{Haisch:2017ufo}.
\THDMa contributions to the \METjets final state involving the production of one or two \DM pairs are generated with \toolVersion{\MadGraphAMCatNLO}{2.7.3}~\cite{Alwall:2014hca,Frederix:2018nkq} at tree and one-loop level using the\linebreak \NNPDFNLO~\cite{NNPDF:2014otw} \PDF set.
This is interfaced with \toolVersion{\Pythia}{8.245}~\cite{Sjostrand:2014zea} in the A14 tune~\cite{ATLAS:2014rfk} employing the \NNPDFTwoThreeLO \PDF set for parton showering.
For the sake of simplicity, production and decay processes are factorised using the narrow-width approximation (\cf\secref{sec:MCEG_hadronisation}).
Particle decays are handled by \toolVersion{\EvtGen}{1.7.0}~\cite{Lange:2001uf}.
The generation setup is summarised in \tabref{tab:interpretation_2HDMa_MC}.

\begin{mytable}{Generation setup used for the considered \THDMa processes in the \METjets interpretation with the employed \PDF set, corresponding hard-process as well as parton-shower generators and tune for the underlying event.}{tab:interpretation_2HDMa_MC}{cccccc}
	Physics process & \PDF set& Hard process & Parton shower & Tune\\
	\midrule
	\THDMa, $\xx+X$ & \NNPDFNLO & \toolVersion{\MadGraphAMCatNLO}{2.7.3} & \toolVersion{\Pythia}{8.245} & A14\\
\end{mytable}


The uncertainty on the \THDMa predictions from scale variations are assessed by independently taking one half, one or two times the nominal value for the factorisation scale.
The uncertainty due to the choice for renormalisation scale is estimated as a relative uncertainty of \SI{10}{\%} that is constant as a function of \METconst.
The envelope of the distributions for factorisation and renormalisation scale is taken as uncertainty.
Uncertainties related to the choice of \PDF set for the process generation are estimated by taking the envelope from the alternative \PDF sets \CTFTNLO~\cite{Dulat:2015mca} and \MMHT~\cite{Harland-Lang:2014zoa} as well as the statistical uncertainty from \NNPDFNLO.
Uncertainties related to the parton showering are estimated by taking the sum in quadrature from variations concerning underlying events, jet structures and production of additional jets according to \refcite{ATLAS:2014rfk}.

\subsection{\THDMa processes for the \Contur approach}
\label{sec:MC_2HDMa_Contur}

For determining the existing sensitivity of \LHC measurements to the \THDMa with the \Contur toolkit~\cite{contur_zenodo,Buckley:2021neu} in \secref{sec:contur_2HDMa}, events are generated with \toolVersion{\Herwig}{7.2.2} employing \toolVersion{\ThePEG}{2.2.2} based on the \THDMa \UFO.
\Herwig is set to generate all \twoToTwo processes at leading order in which at least one of the outgoing legs is a \BSM particle.
This includes $s$- as well as $t$-channel contributions.
In addition, processes at leading order in which both outgoing legs are \SM particles coming from a \BSM $s$-channel resonance are considered.
\Herwig cannot generate processes in which both outgoing legs are \SM particles interacting via a \BSM particle in the $t$-channel~\cite{Bahr:2008pv}.
\Herwig is used for generating the hard process as well as the parton shower in the default \Herwig tune.
\CTFTLO~\cite{Dulat:2015mca} is employed as \PDF.
\tabref{tab:contur_2HDMa_MC} summarises the generation setup.

\begin{mytable}{
		Generation setup used for all \THDMa processes in the \Contur approach with the considered \PDF set, corresponding hard-process as well as parton-shower generator and tune for the underlying event.
		The generation includes all \twoToTwo processes at leading order in which at least one of the outgoing legs is a \BSM particle or in which both outgoing legs are \SM particles coming from a \BSM $s$-channel resonance.
	}{tab:contur_2HDMa_MC}{cccccc}
	Physics process & \PDF set& Hard process & Parton shower & Tune\\
	\midrule
	\THDMa, $pp\to XY$ & \CTFTLO & \multicolumn{2}{c}{\toolVersion{\Herwig}{7.2.2}} & \Herwig default\\
\end{mytable}

The \Contur toolkit reaches its full physics potential only if the complete breadth of measurements in the repository is exploited.
This requires that the generated final states are as inclusive as possible.
This is why for this study \Herwig is used as \MCEG, as opposed to \MadGraph in the \METjets interpretation, because it can automatically generate all \twoToTwo and $s$-channel processes.
The results obtained with \Herwig were extensively validated with \MadGraph, among others by ensuring that the cross sections and momentum distributions for individual processes agree within uncertainties.
Furthermore, it was ascertained with \MadGraph that the contribution of $2\to3$ processes is negligible~\cite{Butterworth:2020vnb}.

In event generation for \BSM models, \Herwig by default uses merging techniques to prevent double counting between quarks and gluons generated as part of the hard process and those stemming from decays of particles in the parton shower (\cf\secref{sec:MCEG_matchMerge}).
A merging cut-off of $\kT<\SI{50}{GeV}$ for jets from parton showering is used to separate hard process and parton shower.
For the sake of simplicity, production and decay processes are factorised using the narrow-width approximation (\cf\secref{sec:MCEG_hadronisation}).
\Chapter{Interpreting detector signals}{Object reconstruction in ATLAS}{%
	\textnormal{[...]} sie sortieren ihre Stifte\\
	und markieren immer alles in den Büchern, die sie lesen.%
}{Jakob Wich}{ALG:2009fdu}
\label{sec:objReco}


Particles traversing a detector after being produced in a collision at a particle collider can interact with the active material of the detector, \eg ionising it or depositing energy.
These signs of interactions are converted into electronic signals in the ATLAS Experiment and digitised for further processing.
\textit{Object reconstruction} aims to infer backwards from these digital electronic signals to the kind and properties of the passing particle.
Reconstructed objects are used to select events of interest for the physics analysis.
In this chapter, it is discussed as an example how reconstruction of particles and derived physics objects, like jets, is performed in the ATLAS Experiment during Run 2~\cite{ATLAS:2021ath} as used by the \METjets measurement described in \chapsref{sec:metJets}{sec:interpretation}.
It is only described how events are reconstructed from permanently stored signals after data taking.
A simplified approach of the reconstruction methods detailed below is used for the high-level trigger (\cf\secref{sec:ATLAS_trigger}) during data-taking.

\bigskip
\secref{sec:objReco_partSignatures} gives a general overview of the signals different particles leave in the ATLAS detector.
\secref{sec:objReco_tracks} describes tracks, the signals of trajectories of charged particles in the Inner Detector and Muon spectrometer.
The reconstruction of electrons and photons is detailed in \secref{sec:objReco_elePhot}, of muons in \secref{sec:objReco_muons}.
This is followed by a description of the reconstruction of jets and hadronically decaying taus, which are derived from jets, in \secsref{sec:objReco_jets}{sec:objReco_taus}, respectively. Multiple counting of signals is prevented by a procedure called \textit{overlap removal}, which is detailed in \secref{sec:objReco_OLR}. If all objects in an event have been reconstructed, the momentum imbalance can be assessed as described in \secref{sec:objReco_MET}.

These discussions focus on reconstructing and combining objects starting from the detector signals, \ie in the process towards a detector-level representation (\cf\figref{fig:detectorCorrection}).
Object construction and combination coming from the inclusive final state, \ie in the process towards a particle-level representation, is discussed in \secref{sec:objReco_particleLevel}.

\section{Particle signatures}
\label{sec:objReco_partSignatures}
Different kinds of particles produced in a collision give rise to different sets of detector responses, \textit{signatures}, in the ATLAS detector when traversing it, as depicted in \figref{fig:objReco_particleTracks}.
Particles with electric charge ionise active material in the Inner Detector (\cf\secref{sec:ATLAS_ID}), which is reconstructed as tracks.
Electromagnetically interacting particles, \ie particles with electric charge and photons, interact with the detector material in the electromagnetic calorimeters (\cf\secref{sec:ATLAS_calo}), giving rise to electromagnetic showers. Electrons and photons deposit all their energy in the electromagnetic calorimeters. They are distinguished by the existence of tracks in the Inner Detector: electrons leave tracks but photons do not.

Quarks and gluons hadronise before reaching the subdetector systems, as described in \secref{sec:SM_QCD}.
Electromagnetically charged and neutral hadrons, \eg protons and neutrons, respectively, interact with the nuclei in the hadronic calorimeters (\cf\secref{sec:ATLAS_calo}) due to the high density of these subdetectors.
Energy depositions in the electromagnetic or hadronic calorimeters with or without tracks in the Inner Detector are used to reconstruct composite objects, jets.

Muons, as minimum-ionising particles, can traverse the whole detector and leave only small traces in Inner Detector and calorimeters.
They are the only charged \SM particles that commonly reach the Muon spectrometer (\cf\secref{sec:ATLAS_muonSpec}). 
There, they ionise the active material which is converted into electronic signals.

Particles that decay before reaching the ATLAS subdetectors can be determined from their decay products.
Stable particles that are subject to neither electromagnetic nor strong interaction, \ie neutrinos in the Standard Model (\cf\secref{sec:SM}) but also hypothetically Dark Matter (\cf\secref{sec:DM}), interact too weakly to give rise to any distinguishable detector signals.
Their presence can only be deduced from observing a momentum imbalance in the event as a whole, \MET (see \secref{sec:objReco_MET}).

\begin{myfigure}{
		Illustration of the signatures different particles leave in the ATLAS detector.
		Solid lines represent trajectories giving rise electronic signals in the detector, dashed lines are trajectories invisible to the detector.
		Figure adapted from \refcite{Pequenao:1505342}.
	}{fig:objReco_particleTracks}
	\includegraphics[width=0.85\textwidth]{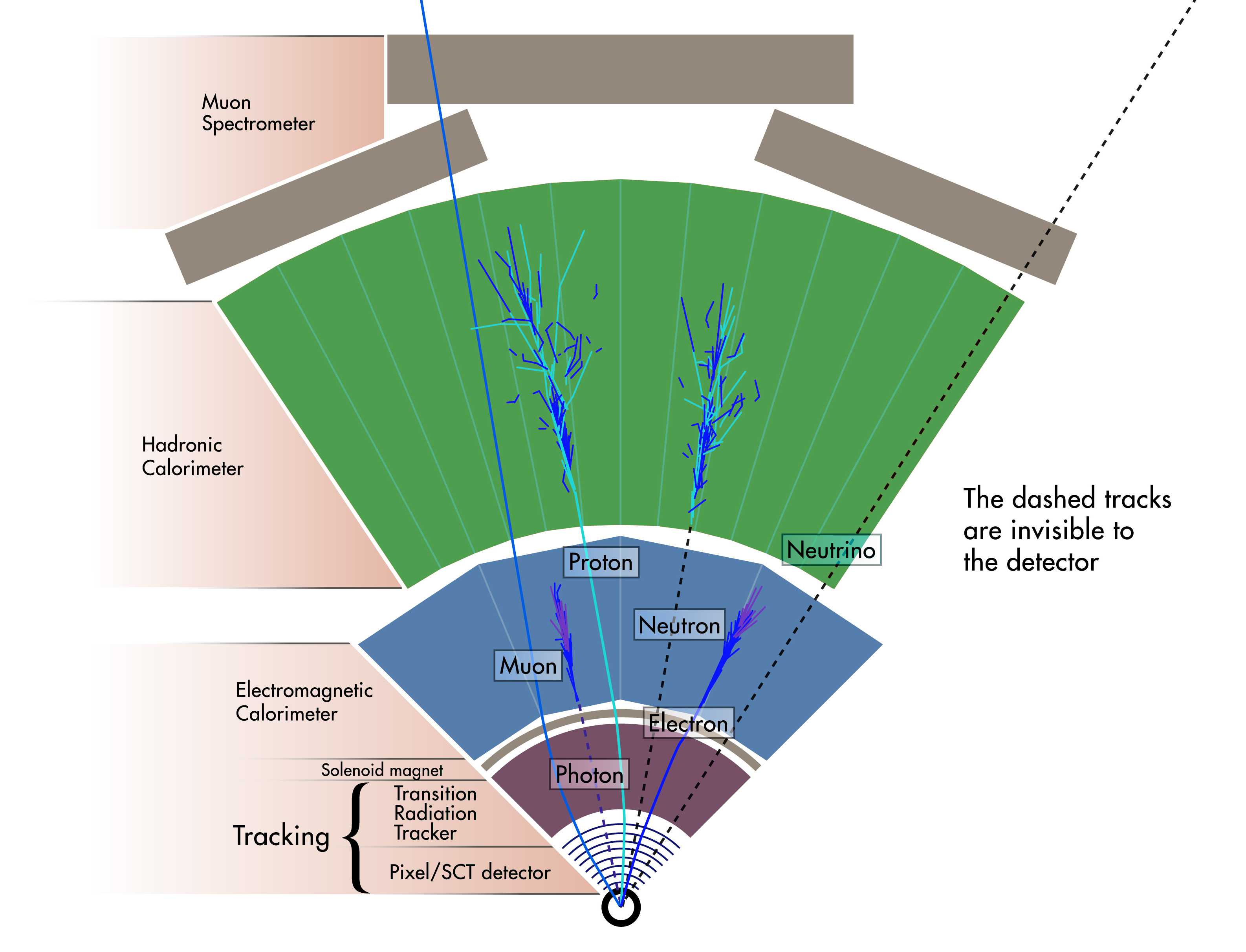}
\end{myfigure}

\section{Tracks in the Inner Detector}
\label{sec:objReco_tracks}
Ionisation in the active material of the Inner Detector caused by a traversing charged particle is translated into a signal by the detector electronics.
If a predefined threshold is passed, this interaction is considered a \textit{hit}.
Hits in the Inner Detector (\cf\secref{sec:ATLAS_ID}) are used for track reconstruction~\cite{Cornelissen:2007vba,ATLAS:2017kyn}.
Those in adjacent cells are grouped into \textit{clusters} by taking their geometric centre, as particles can cross multiple neighbouring cells in a layer.
Clusters in the Pixel detector provide three-dimensional information about the coordinates of the interaction.
Only two-dimensional information on the strip position but not on the coordinate along the strip is known for \SCT clusters.
Therefore, making use of the small angle between them, both sides of a strip layer are used to obtain three-dimensional position information for these.

From sets of three clusters in different layers (\cf\secref{sec:ATLAS_ID}), track seeds are formed.
These are extended to form track candidates by successively adding other clusters.\linebreak
Charged particles follow helical trajectories as a consequence of the solenoidal magnetic field.
Only those clusters adding a chi-square \chiSq contribution below a predefined threshold when assuming a helical trajectory are considered for extending the track candidates.
Clusters shared by multiple track candidates, track candidates with missing clusters, \ie with an intersection with sensitive detector elements without a cluster, and track candidates faked by combinatoric association of unrelated clusters lead to ambiguities in the track reconstruction.
These are treated by ranking the track candidates according to a likelihood metric, iteratively removing shared clusters from the track candidates with lower likelihood and finally discarding track candidates with a likelihood below a predefined threshold.
Track candidates are then required to fulfil a number of quality criteria, among others $\pT>\SI{400}{MeV}$, $\absEta<2.5$ and a minimum cluster multiplicity.

\TRT hits are only taken into account in a last step because the subdetector is limited to $\absEta<2.0$.
By extrapolating track candidates to the \TRT, suitable \TRT hits are added to form the final reconstructed objects, \textit{tracks}.
Subsequently, tracks in the Inner Detector can be combined with tracks in the Muon spectrometer to form muon tracks if appropriate (see \secref{sec:objReco_muons}).

\subsection{Vertices}
\label{sec:objReco_vertices}

A vertex is the space point where a proton--proton interaction occurred.
In the ATLAS Experiment, the reconstruction of vertices~\cite{ATLAS:2016nnj} starts by creating a vertex seed, a space point at the beam spot to which all tracks that have not been previously matched to a vertex are associated.
The $x$- and $y$-coordinates of the vertex seed are placed directly at the beam-spot centre, the $z$-coordinate is the mode of the $z$-coordinates of the associated tracks at their point of closest approach to the beam-spot centre.
The best-estimate vertex position is then obtained by fits to all associated tracks of a vertex seed, successively decreasing the weights of those that prove incompatible.
After the last iteration, the most incompatible tracks are rejected and considered as input for additional vertex seeds.
This is repeated until there are no unassociated tracks, or no additional vertex with at least two associated tracks can be found.

\textit{Primary vertices} are those directly at the beamline where proton--proton interactions happened. Commonly, there is more than one primary vertex per bunch crossing.
The \textit{hard-scatter vertex} is the primary vertex with the highest sum of squared transverse momenta of the associated tracks.
\textit{Secondary vertices} stem from particles with a long lifetime, \eg $B$ hadrons, that travel for a short time before decaying, causing an offset between vertex position and beamline.

The impact parameters of a collision describe the coordinates of closest approach of a track to the beamline, measured relative to the primary vertex of interest.
The coordinate in the plane perpendicular to the beam axis is called transverse impact parameter $d_0$.
The \textit{$d_0$ significance} is the ratio of the absolute value of $d_0$ to its uncertainty, \dZsig.
The longitudinal impact parameter $z_0$ denotes the coordinate along the beam axis of the space point at which $d_0$ is determined.

\section{Electrons and photons}
\label{sec:objReco_elePhot}
Generally speaking, electrons $e^-$ and photons $\gamma$ are reconstructed in the ATLAS Experiment from clusters of energy depositions in the \EM calorimeters (\cf\secref{sec:ATLAS_calo}) with ($e^-$) or without ($\gamma$) a matched track in the Inner Detector~\cite{ATLAS:2019qmc}.
As many of the reconstruction steps are conducted jointly or analogously, they are discussed together here. Only the reconstruction procedure for objects within $\absEta<2.5$ is described.

\bigskip

First, so-called proto-clusters are formed which group cells (see \secref{sec:ATLAS_calo}) in which the deposited energy with regard to the expected cell noise is above a predefined threshold. 
Neighbouring cells are added to these in an iterative procedure if their deposited energy exceeds a relaxed threshold. If after this step the cells of proto-clusters overlap, the proto-clusters are merged. Next, surrounding cells are added independently of the energy deposited in them. Proto-clusters that exhibit more than one local maximum of deposited energy above a certain threshold are subdivided.

Topological clusters are formed from proto-clusters of topologically connected \EM- and hadronic-calorimeter cells. These clusters are dynamic in size which allows recovering energy from bremsstrahlung photons as well as from photons converted into elec\-tron--positron pairs by interaction with the detector material, $\gamma\to e^+e^-$. Topological clusters are matched, if possible, to tracks in the Inner Detector. Tracks are thereby refitted to account for possible bremsstrahlung effects.
It is also attempted to reconstruct vertices in which a photon conversion has taken place, \eg from two opposite-charge tracks.

Topological clusters with matched tracks are considered electrons, with matched conversion vertices are considered converted photons and with neither unconverted photons.

After reconstruction is complete, the energy scale and resolution of the objects is calibrated by comparing $Z\to e^+e^-$ events in simulation and measurement.
Corrections applied to data eradicate discrepancies in the energy scale, corrections applied to simulation discrepancies in the energy resolution.

\bigskip

Two more steps are taken to improve the discrimination between real electrons or photons and background~\cite{ATLAS:2019qmc}:
\begin{itemize}
	\itembf{Identification}
	This steps helps to differentiate against detector signals from other sources, like noise or pions, by taking advantage of intrinsic properties of the reconstructed objects.
	For electrons and photons, a likelihood discriminant is used to improve the purity of the selected objects. Hereby, different \textit{working points} are defined, selection criteria on the likelihood discriminant corresponding to a specific selection efficiency.
	These also determine associated anticorrelated levels of background rejection.
	"Loose", "Medium" and "Tight" working points are available, corresponding to an average selection efficiency of \SI{93}{\%}, \SI{88}{\%} and \SI{80}{\%}, respectively.
	The exact selection efficiency, however, depends on \eta and the energy of the object considering only the momentum in the transverse plane, \ET (\cf\subfigref{fig:objReco_electrons}{a}).
	\itembf{Isolation}
	Partons, \ie quarks and gluons, produced in a collision can repeatedly radiate gluons and gluons can break up into quark--antiquark pairs (see\linebreak \secsref{sec:MC_partonShower}{sec:objReco_jets}).
	This can mimic the signature of electrons or photons or even lead to the radiation of real photons from quarks.
	Furthermore, electrons and photons can be produced in the decay of heavy-flavour hadrons.
	They are then called \textit{non-prompt}.\\
	There is generally more activity close to misidentified and non-prompt electrons and photons due to the other processes happening in the parton showering, fragmentation and decay than for real prompt electrons and photons.
	Imposing \textit{isolation} criteria on the amount of activity close to the reconstructed object helps to discriminate against these.
	Activity can be defined in terms of tracks of nearby charged particles, nearby energy depositions in the calorimeters, or both.
	The efficiencies for different isolation working points are compared in \subfigref{fig:objReco_electrons}{b}.\\
	In the \METjets measurement performed in \chapsref{sec:metJets}{sec:interpretation}, the "HighPtCaloOnly" isolation is used for electrons.
	This isolation working point only makes use of the total \ET in a cone with $R=0.2$ around the object and provides the highest rejection power in the large-\ET region, \ie $\ET>\SI{100}{GeV}$.
	Photons are not specifically selected in the measurement.
\end{itemize}

\begin{myfigure}{
		(a) Efficiency for the electron identification at different working points.
		The inner (outer) uncertainty bars give the statistical (total) uncertainties.
		(b) Efficiency of different isolation working points for electrons fulfilling the selection criteria of the "Medium" identification.
		The uncertainty bars give the total uncertainties.
		For both plots, the bottom panel shows ratio of data to Monte-Carlo (\MC) simulation.
		Figures taken from \refcite{ATLAS:2019qmc}.
	}{fig:objReco_electrons}
	\subfloat[]{\includegraphics[width=0.49\textwidth]{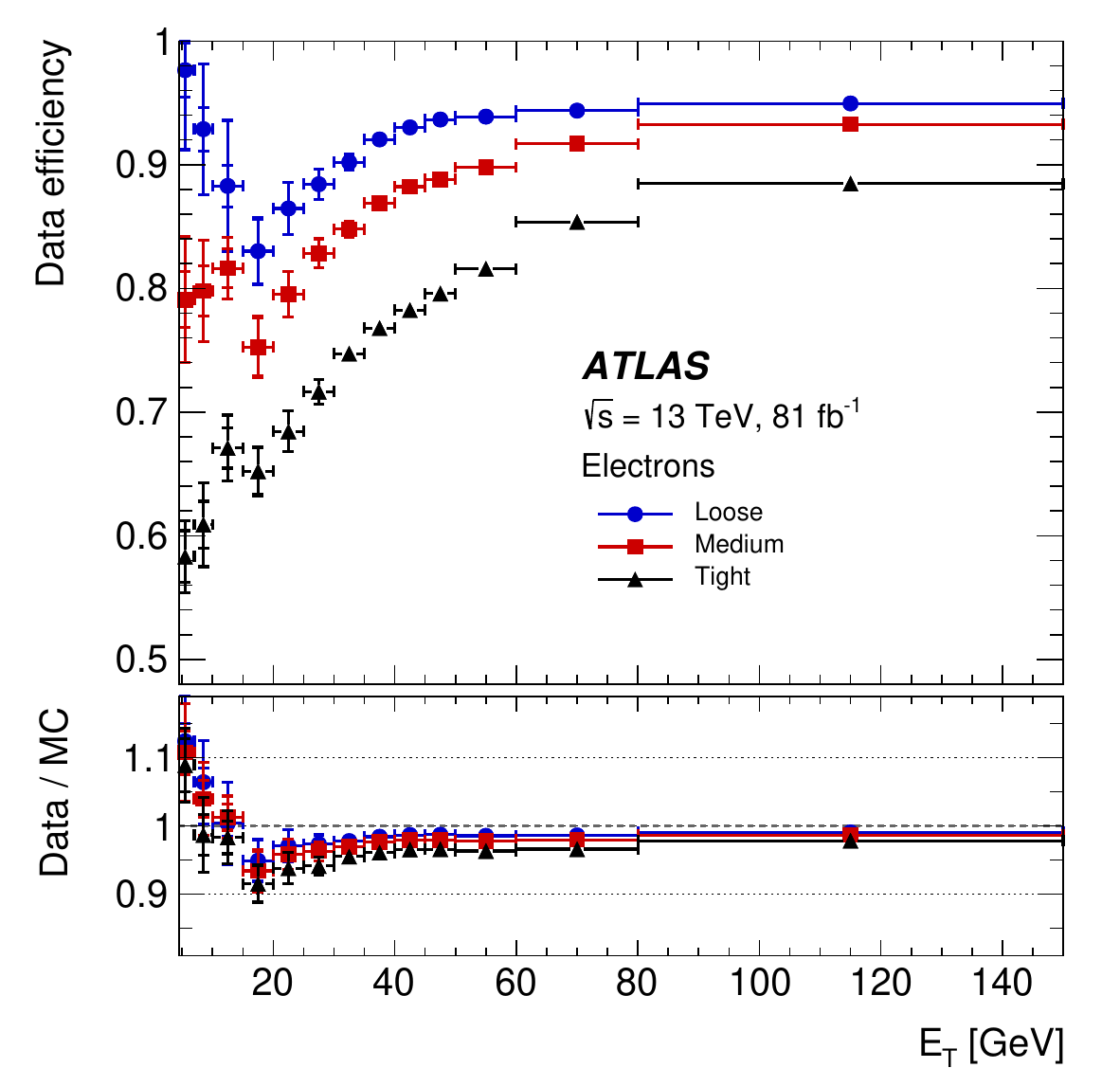}}
	\subfloat[]{\includegraphics[width=0.49\textwidth]{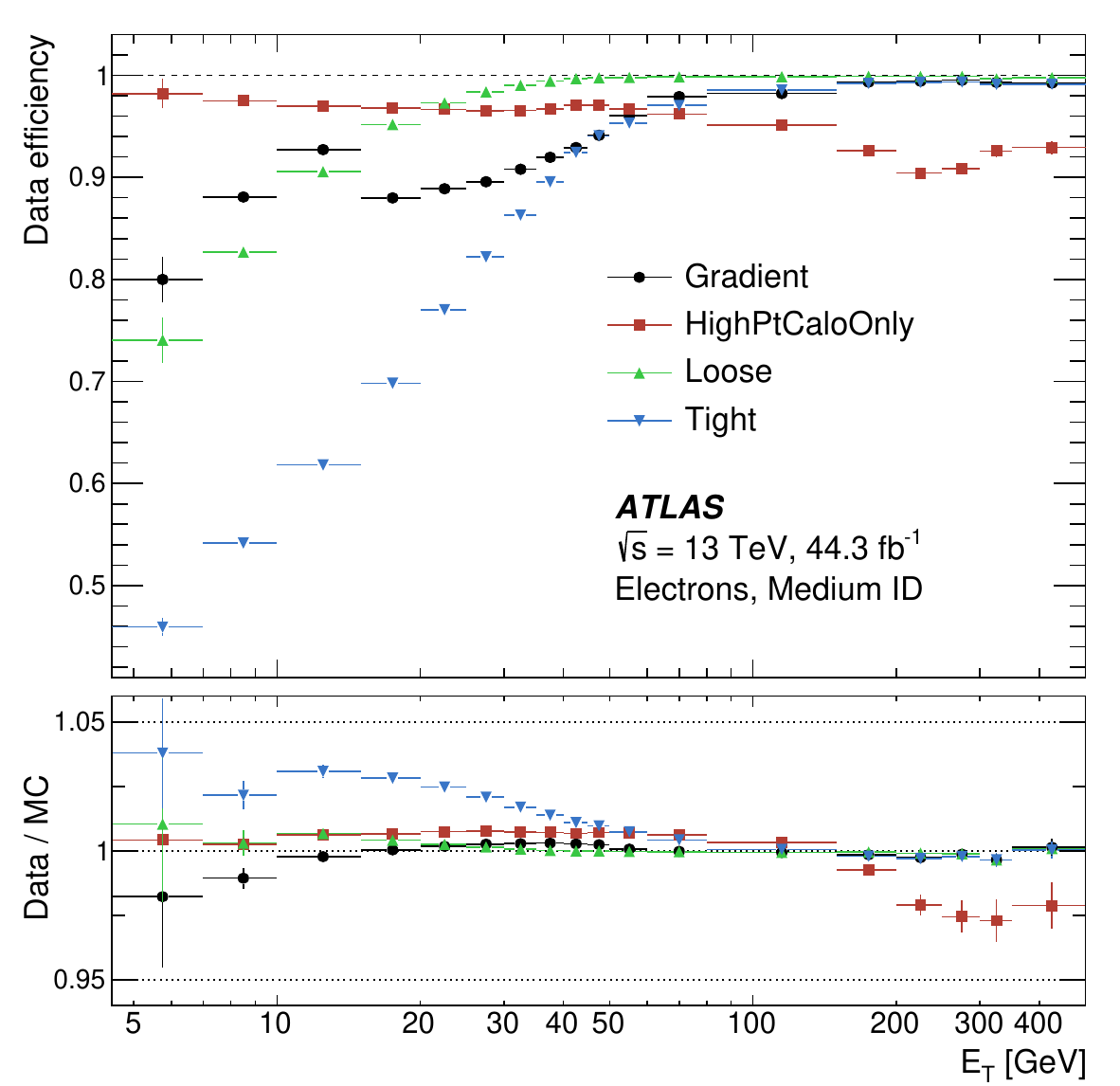}}
\end{myfigure}

Different responses in reconstruction, identification and isolation of objects between data and simulation can lead to discrepancies in the reconstruction efficiencies. This is investigated in $Z\to e^+e^-$ as well as $J/\psi\to e^+e^-$ events and corrected by applying different event weights (\textit{scale factors}) to simulated events. For photon reconstruction efficiencies, radiative $Z$ decays and inclusive photon events are used.

\section{Muons}
\label{sec:objReco_muons}

As minimum-ionising particles, the signature of muons in the ATLAS detector is tracks in the Muon spectrometer (\MS) and characteristic energy depositions in the calorimeters (\cf\secsref{sec:ATLAS_muonSpec}{sec:ATLAS_calo}, respectively). These are used in interplay with tracks in the Inner Detector, as described in \secref{sec:objReco_tracks}, to reconstruct muons~\cite{ATLAS:2020auj}.

Track reconstruction in the Muon spectrometer begins with track segments from hits in the individual \MS layers that form straight lines. Track segments from different layers are then combined into tracks if they fulfil a constraint of approximately originating from the interaction point.
A parabolic trajectory of the tracks is assumed, comprising an approximation to the muons bending in the toroidal magnetic field~\cite{ATLAS:2020auj}.
Afterwards, a \chiSq fit of the muon trajectory is performed, which takes into account possible misalignment effects and interactions with detector material. In this fit, outliers are removed and formerly unassociated hits along the trajectory added. Hits shared with other tracks are removed if the other tracks are of higher quality. Lastly, the tracks are extrapolated to the beamline. Hereby, energy depositions in the calorimeters are taken into account. The transverse momentum cited for a \MS track then corresponds to the \pT at the interaction point.

Muons are reconstructed from tracks in the Inner Detector and Muon spectrometer as well as energy depositions in the calorimeters. Hereby, various reconstruction strategies are used which differ in which information from the subdetector systems is used and how they are combined. Two of these strategies shall be highlighted:
In the "combined" strategy, \ID and \MS tracks are matched and energy depositions from the calorimeters are taken into account.
In the "segment-tagged" strategy, \ID tracks are extrapolated to the Muon spectrometer and angular-matched to \MS track-segments. The track parameters are taken directly from the track fit in the Inner Detector.

Two more steps are taken to improve the discrimination between real muons and background~\cite{ATLAS:2020auj}, just like for electrons and photons:
\begin{itemize}
	\itembf{Identification} Different requirements can be placed on the number of hits in the \ID and \MS subdetector systems, the fit properties of the tracks as well as the compatibility of the momentum and charge measurement between Inner Detector and Muon spectrometer.
	Specific sets of requirements are pooled as working points that correspond to a certain selection efficiency and purity.
	Among others, "Loose", "Medium" and "Tight" working points are defined.
	The "Loose" working point aims for analyses with high muon multiplicities which allow lowering the back\-ground-rejection conditions for the individual muons.
	The "Medium" working point is suitable for all analyses without special requirements on muons.
	The\linebreak "Tight" working point exhibits the highest purity and background rejection and is therefore used by analyses limited by background from non-prompt muons.
	The selection efficiencies for these three working points are  $90 - \SI{99}{\%}$, $70-\SI{97}{\%}$ and $36-\SI{93}{\%}$, respectively, whereby the exact efficiency depends on the muon \pT (\cf\subfigref{fig:objReco_muons}{a}).
	\itembf{Isolation} Like for electrons and photons, the activity close to the reconstructed muons is taken into account from the Inner Detector, from the calorimeters, or from both.
	In the measurement performed in \chapsref{sec:metJets}{sec:interpretation}, the "Loose" isolation is used for muons.
	This isolation working point is particularly useful for analyses in which a good prompt-muon efficiency has a higher priority than the rejection of non-prompt muons.
	Its selection efficiency as a function of the transverse momentum of the muon is shown in \subfigref{fig:objReco_muons}{b}.
\end{itemize}

\begin{myfigure}{
		(a) Efficiency for the muon identification at different working points.
		The efficiencies in data (\MC simulation) are denoted by filled (open) markers.
		(b) Efficiency of the "Loose" isolation working point for muons.
		For both plots, the bottom panel shows ratio of data to \MC simulation.
		Figures taken from \refcite{ATLAS:2020auj}.
	}{fig:objReco_muons}
	\subfloat[]{\includegraphics[width=0.49\textwidth]{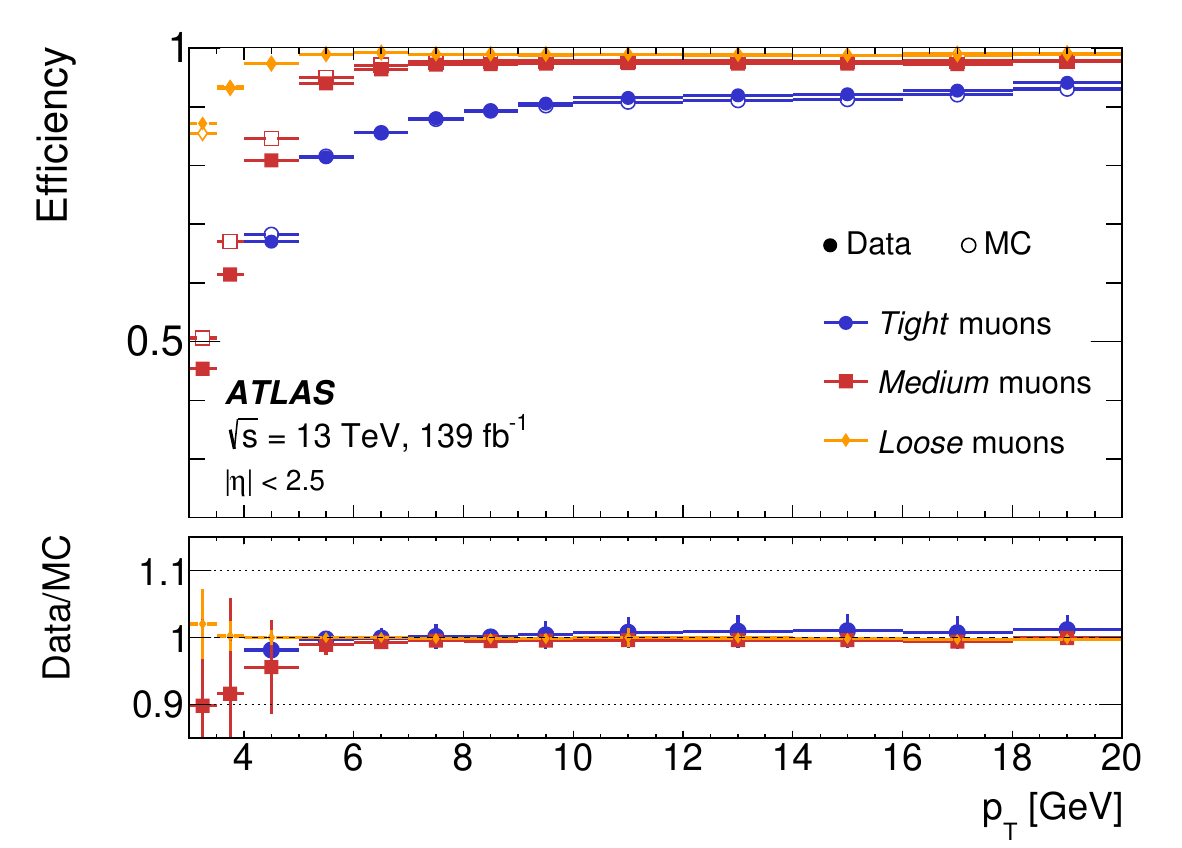}}
	\subfloat[]{\includegraphics[width=0.49\textwidth]{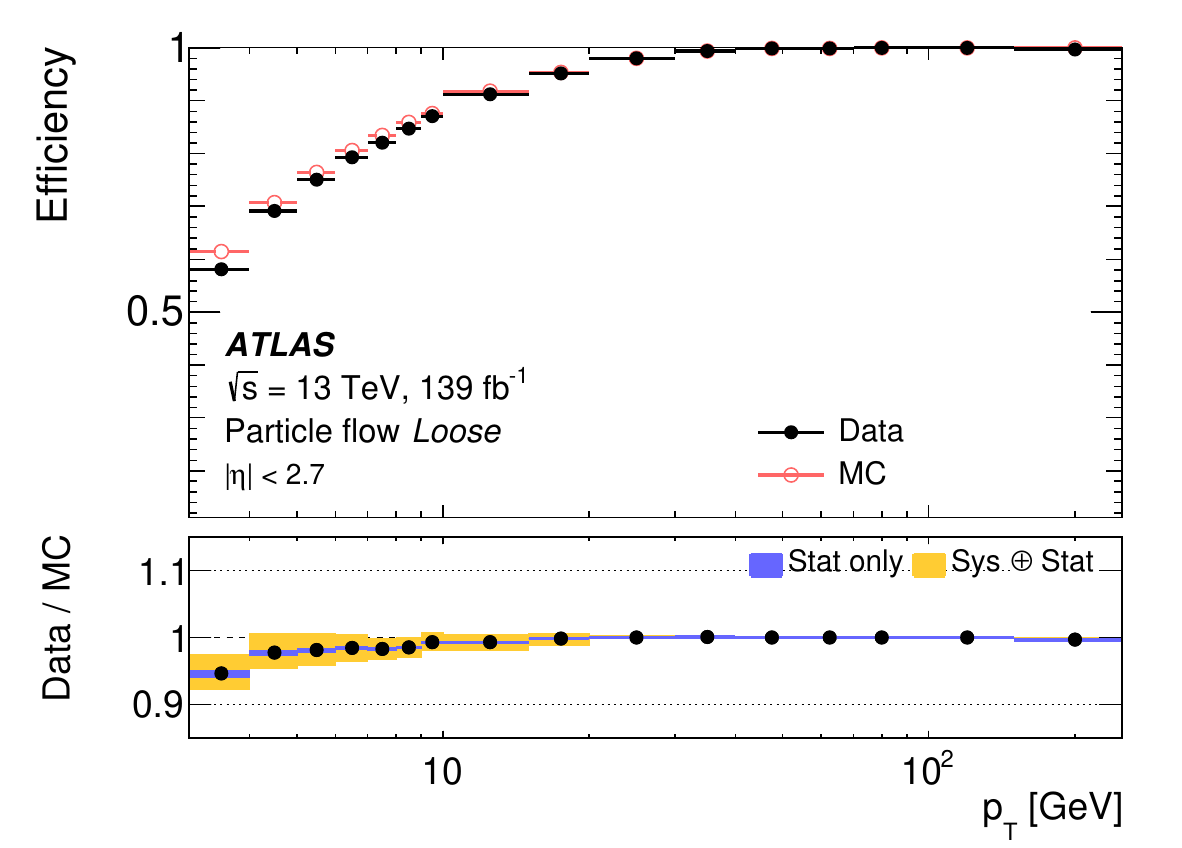}}
\end{myfigure}

Similar as for electrons and photons, different reconstruction efficiencies between data and simulation are investigated in $Z\to \mu^+\mu^-$ and $J/\psi\to \mu^+\mu^-$ events and corrected by applying different event weights to simulated events.

\section{Jets}
\label{sec:objReco_jets}

Partons, \ie quarks and gluons, coming from a collision cannot be observed directly as discussed in \secref{sec:SM_QCD}: they form colourless bound states, hadrons, due to confinement.
Moving away from the collision point, partons can repeatedly radiate gluons and gluons can break up into quark--antiquark pairs.
These parton showers constitute therefore not single hadrons but conical sprays of particles, which are called \textit{jets}. 
\figref{fig:objReco_jets} shows a sketch of the steps involved in jet reconstruction: Partons coming from a proton--proton collision form collinear bundles of hadrons. These can be clustered into jets (\secref{sec:objReco_jets_clustering}) and interact with the detector material, allowing for the jet reconstruction (\secref{sec:objReco_jets_reconstruction}).
Both of these steps aim to collect as many of constituents into a jet object as possible without introducing a bias, \eg from overlapping other objects.

\begin{myfigure}{
		Illustration of how partons coming from a proton--proton collision hadronise and form jets which are reconstructed by a detector. Figure taken from \refcite{Carli:2015qta}.
	}{fig:objReco_jets}
	\includegraphics[width=0.8\textwidth]{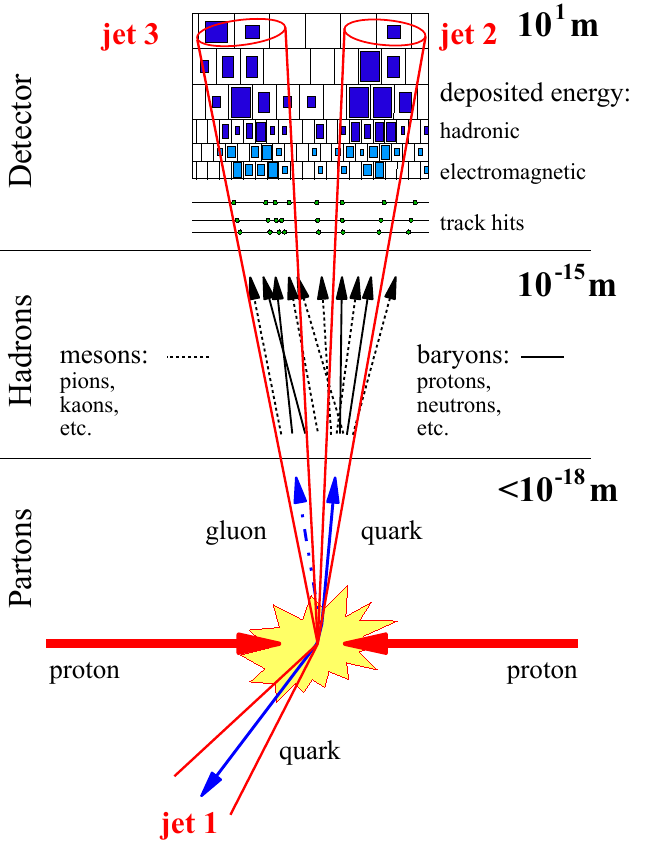}
\end{myfigure}

\subsection{Jet clustering}
\label{sec:objReco_jets_clustering}

Jets are a prominent feature of particle collisions, but in contrast to the objects described so far in this chapter, they are not fundamental particles.
Therefore, jet-clustering algorithms are needed to decide after hadronisation (see \secref{sec:MCEG_hadronisation}) which hadrons belong to a jet.
As the emissions of partons in the shower history are dominantly either collinear or low-energetic, the clustering algorithm needs to be collinear-safe and infrared-safe, \ie replacing any parton with collinear partons of the same total momentum or adding low-energetic partons should lead to identical results. This is related to parton showering needing to simultaneously fulfil the collinear and infrared limit as described in \secref{sec:MC_partonShower}.

The ATLAS Experiment makes use of the anti-$k_t$ algorithm~\cite{Cacciari:2008gp} for jet clustering. Hereby, the measure
\begin{equation*}
	d_{ij}\coloneqq\min\tuple{\frac{1}{p_{\text{T},i}^2}}{\frac{1}{p_{\text{T},j}^2}} \frac{\left(\Delta R_y\right)^2}{R^2}
\end{equation*}
is used for the distance between entities, \ie particles or pseudojets, $i$ and $j$.  $R$ is a radius parameter. 
\begin{equation*}
	d_{iB}\coloneqq\frac{1}{p_{\text{T},i}^2}
\end{equation*}
is used for the distance between entities and the beam. For the jet clustering, the smallest distance for all combinations $i,j,B$ is selected, then $i$ and $j$ are combined into a pseudojet if this distance is of type $d_{ij}$. The entity is labelled as a jet and removed from further considerations instead if the distance is of type $d_{iB}$. Successively, distances are recomputed and the steps repeated until there are no entities left.

The anti-$k_t$ algorithm is collinear-safe because entities with a small angular separation $\Delta R_y$ are clustered first.
The algorithm is also infrared-safe because the distance $d_{ij}$ between any two low-energetic entities is in general large because of the inverse dependence on the transverse momenta.
In consequence, low-energetic entities are clustered first to high-energetic entities.

If high-energetic entities have $\Delta R_y>2R$, they are reconstructed as two individual, conic jets. For $R<\Delta R_y<2R$, the overlap is divided between the two jets. Entities with $\Delta R_y<R$ are merged into a single high-energetic jet. The \METjets measurement described in \chapsref{sec:metJets}{sec:interpretation} uses a radius parameter of $R=0.4$ for its jets.

\subsection{Jet reconstruction}
\label{sec:objReco_jets_reconstruction}

The entities needed for jet clustering in the ATLAS Experiment~\cite{ATLAS:2020cli} at detector level are obtained from energy depositions in the calorimeters. These are clustered making use of the signal-to-noise ratio as is done in electron and photon reconstruction described in \secref{sec:objReco_elePhot}. Event-by-event corrections are applied to take into account that jets are expected to originate from primary vertices.

For the jets used in the measurement in \chapsref{sec:metJets}{sec:interpretation}, the information from calorimeters and Inner Detector is combined by the particle-flow algorithm~\cite{ATLAS:2017ghe} for jets with $\pT\leq\SI{100}{GeV}$.
The particle-flow algorithm approximates the jet constituents as individual particles.
Energy depositions in the calorimeters are replaced in the calculation by the momenta from matched tracks.
For $\pT>\SI{100}{GeV}$, no tracking information is used due to the excellent calorimeter performance at high energies.
In a last step, a scaling is applied to compensate the different response in the Inner Detector and calorimeters.

Reconstructed jets can be tagged as originating from bottom or charm quarks by sophisticated algorithms~\cite{ATLAS:2022qxm}.
These exploit the long lifetime of the order of \SI{1.5}{ps} for $B$ hadrons (\SI{0.5}{ps} for $C$ hadrons)~\cite{Zyla:2020zbs}.
The long lifetime leads to significant mean flight lengths for energetic particles and correspondingly large offsets between primary and secondary vertices.
These tagged jets are particularly useful for selecting events involving heavy-flavour quarks, \eg top quarks decaying to $W$ bosons and bottom quarks~\cite{ATLAS:2019bwq}.
For the \METjets measurement described in \chapsref{sec:metJets}{sec:interpretation}, heavy-flavour quarks do not play an accentuated role and flavour tagging of jets is not performed.

\bigskip
The construction of jets at particle level is described in \secref{sec:objReco_particleLevel}.
Matching particle- and detector-level jets within $\Delta R<0.3$ in dijet events allows the particle- and detector-level jet energy ($E_\textnormal{part}$ and $E_\textnormal{det}$, respectively) to be put in relation~\cite{ATLAS:2020cli}.
The central value of the $E_\textnormal{det}/E_\textnormal{part}$ distribution is called the \textit{jet energy scale} (\JES), its width the \textit{jet energy resolution} (\JER).
They are shown in \subfigsref{fig:objReco_jets_scales}{a}{b}, respectively.


\begin{myfigure}{
		(a) Jet energy scale as a function of reconstructed jet energy at detector level~$E_\textnormal{det}$ and (b) jet energy resolution as a function of transverse momentum of the jet.
		Figures adapted from \refcite{ATLAS:2020cli}.
	}{fig:objReco_jets_scales}
	\subfloat[]{\includegraphics[width=0.48\textwidth]{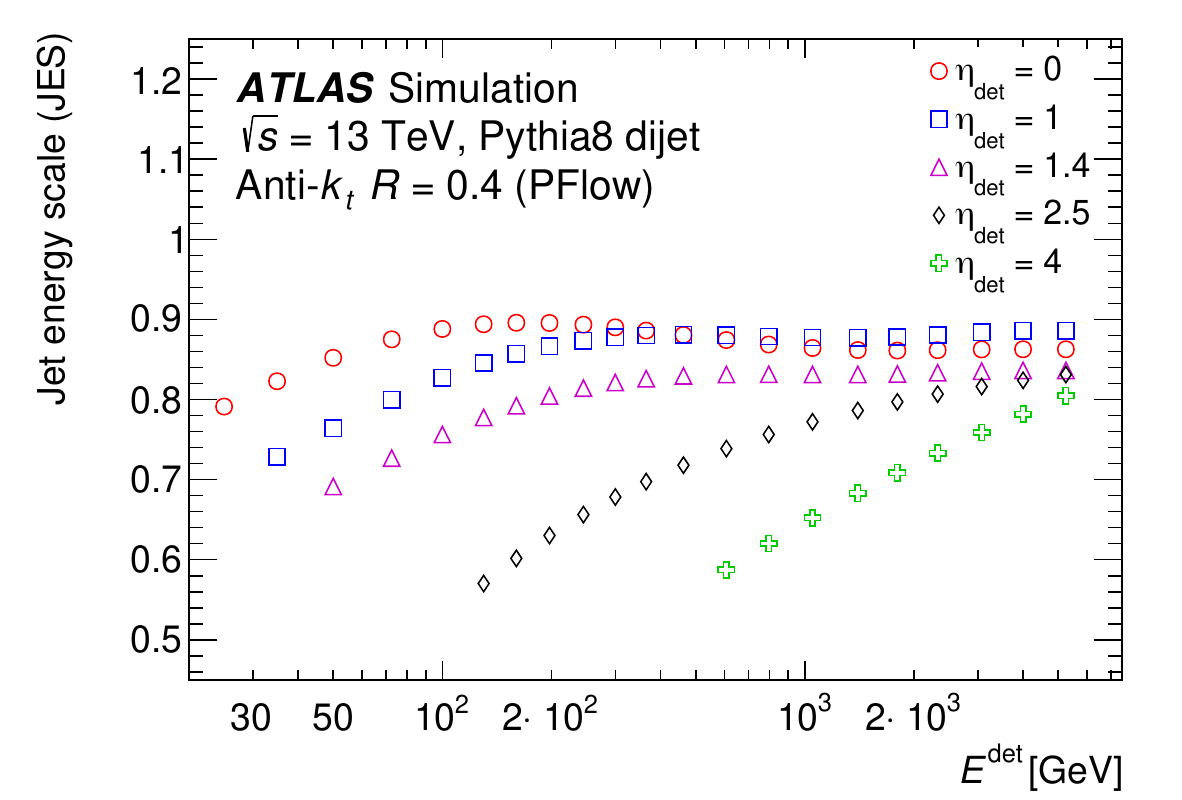}}
	\hspace{10pt}
	\subfloat[]{\includegraphics[width=0.48\textwidth]{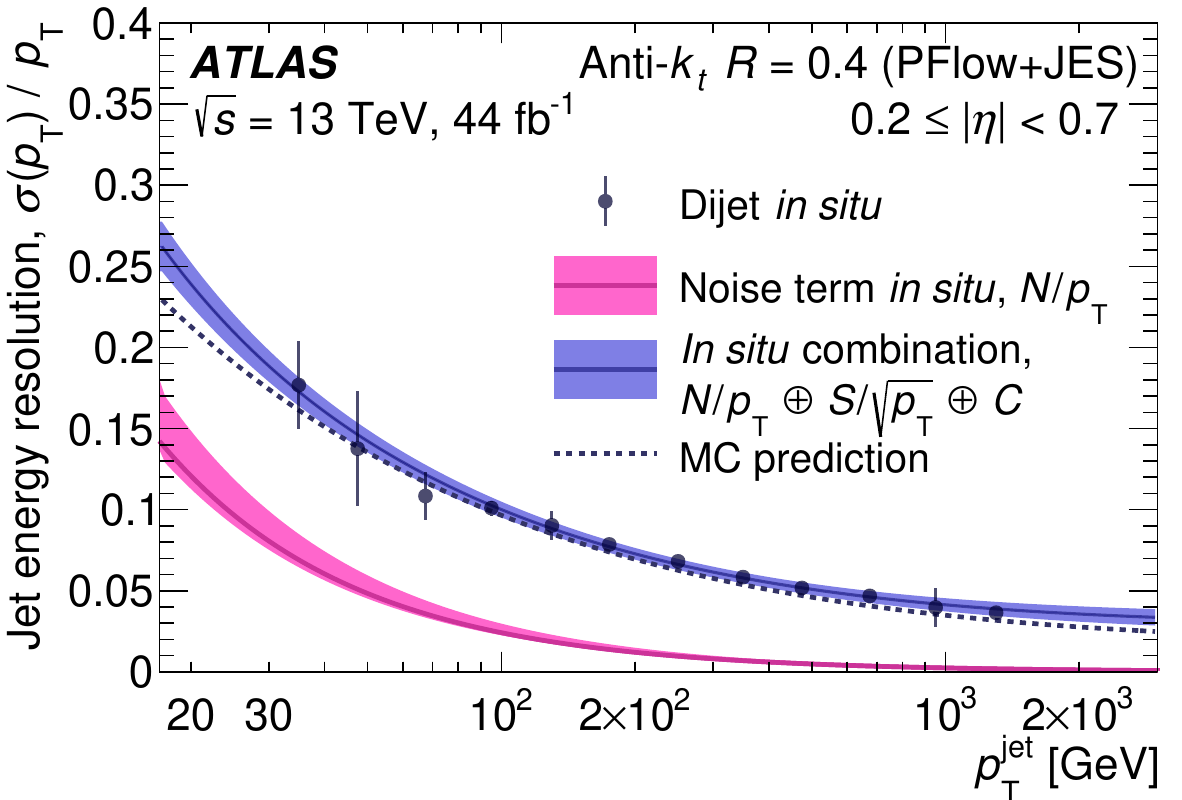}}
\end{myfigure}

\subsubsection{Jet energy scale (\JES) and resolution (\JER) calibration}
\label{sec:objReco_jets_jetCalibration}
Successively, multiple calibration steps are carried out to correct differences in the jet energy between detector- and particle level:
Firstly, pileup contributions are removed.
Secondly, the absolute energy and direction of detector-level jets is matched to that of particle-level jets.
This takes into account calorimeter response, energy losses in passive material, out-of-cone effects as well as biases due the transition between different detector technologies.
Next, the jet resolution is improved by considering the dependence of the properties of reconstructed jets on observables from the subdetector systems.
These dependences stem for example from differences between quark- and gluon-initiated jets.

In a last step, the remaining difference to simulation is corrected in data for the jet energy scale.
For the jet energy resolution, the remaining difference to data is corrected in simulation by smearing the resolution.

The resulting jet energy resolution is determined by three different components:
\begin{itemize}
	\item Contributions from detector noise and pileup, which dominate for $\pT<\SI{30}{GeV}$.
	\item Contributions from statistical fluctuations in the amount of deposited energy, which dominate in the region $\SI{30}{GeV}<\pT<\SI{400}{GeV}$.
	\item Contributions from energy depositions in passive material, the starting point of the hadronic showers and non-uniform calorimeter responses. These dominate for $\pT>\SI{400}{GeV}$.
\end{itemize}

Combined, these effects result in a jet energy resolution of $0.25-0.04$ after applying all \JES corrections, decreasing with the jet \pT (\cf\subfigref{fig:objReco_jets_scales}{b}).

\subsubsection{\JES and \JER uncertainties}
For the jet energy scale, the "category reduction" scheme~\cite{ATLAS:2020cli} is used for assessing the systematic uncertainties in the \METjets measurement in \chapsref{sec:metJets}{sec:interpretation}.
This scheme consists of 30 different components, covering the effect of either detector, modelling or limited statistics, or a mixture of these.
The total uncertainty is $1-\SI{5}{\%}$ as a function of jet~\pT (\cf\subfigref{fig:objReco_jets_uncertainties}{a}) but depends in general also on the jet \eta.

\begin{myfigure}{
		Components of the (a) \JES and (b) \JER systematic uncertainty (coloured lines).
		The total uncertainty is shown as a filled region topped by a solid black line.
		Figures taken from \refcite{ATLAS:2020cli}.
	}{fig:objReco_jets_uncertainties}
	\subfloat[]{\includegraphics[width=0.48\textwidth]{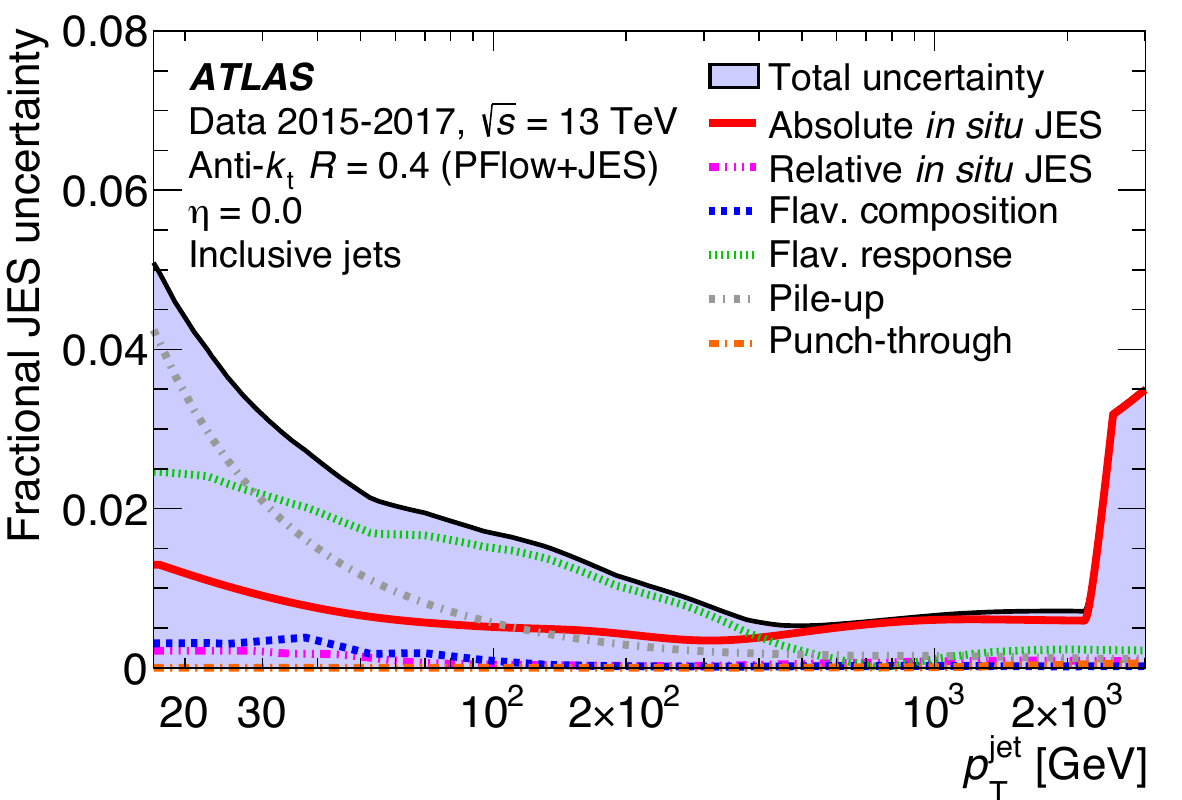}}
	\hspace{10pt}
	\subfloat[]{\includegraphics[width=0.48\textwidth]{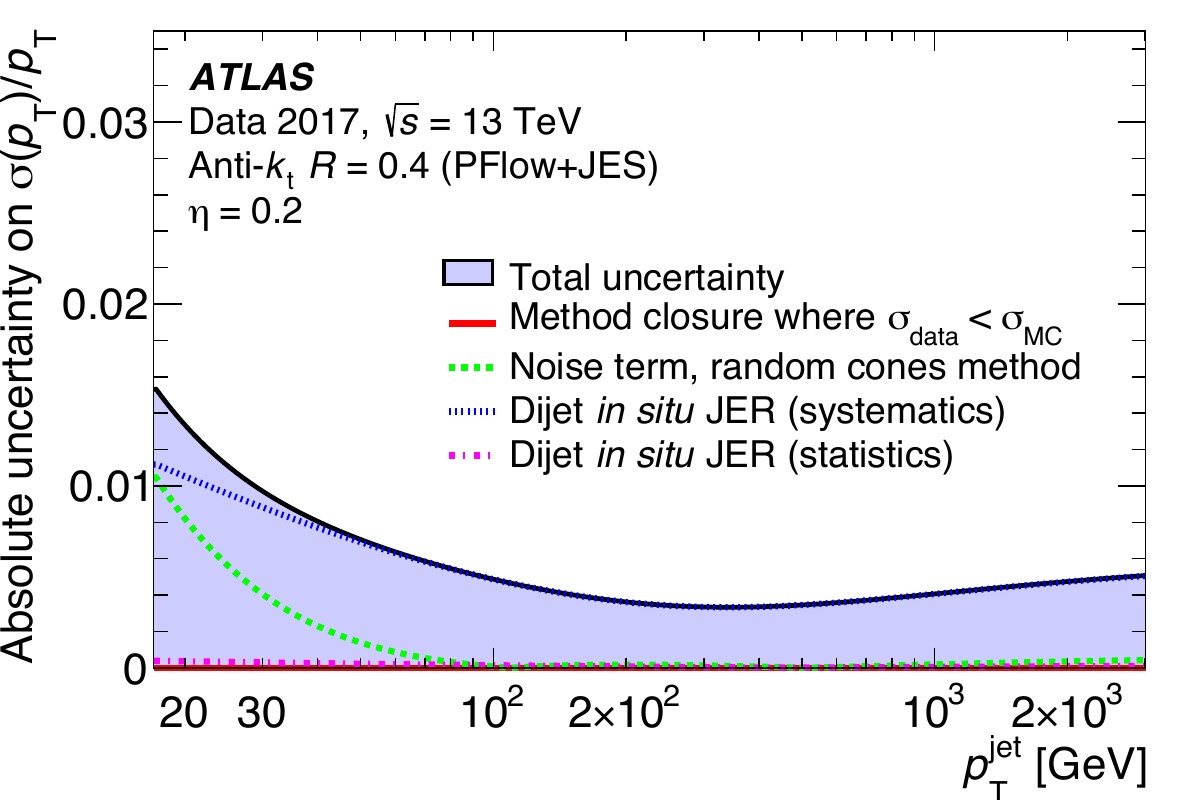}}
\end{myfigure}

For the jet energy resolution, 26 different components for the systematic uncertainties are considered.
They are dominated by propagated \JES uncertainties but take into account also other effects, like estimation of the detector noise and non-closure if the observed resolution in data is better than the one expected from simulation.
The total absolute uncertainty is $0.005-0.015$ as a function of jet \pT (\cf\subfigref{fig:objReco_jets_uncertainties}{b}) but depends in general also on the jet \eta~\cite{ATLAS:2020cli}.

\section{Hadronically decaying taus}
\label{sec:objReco_taus}
Reconstructed taus are only used to reject specific events in the \METjets measurement in \chapsref{sec:metJets}{sec:interpretation} and therefore of reduced importance for this work.
They are only discussed very briefly here.

Taus decay typically within \SI{90}{\micro\metre} of their production vertex~\cite{Zyla:2020zbs}. This decay is either to a $\tau$ neutrino, a lighter lepton and the corresponding light-lepton neutrino or to a $\tau$ neutrino and hadrons. The former decay channel is indistinguishable from the prompt production of electrons or muons apart from a small momentum imbalance in the event. This is why the specific reconstruction of leptonically decaying taus is usually not attempted.
Hadronically decaying taus, on the other hand, can be reconstructed from anti-$k_t$ jets with a radius parameter of $R=0.4$ exhibiting \ID tracks within $\Delta R<0.2$ and can be identified using recurrent neural networks~\cite{ATLAS:2015xbi,ATL-PHYS-PUB-2022-044}.

\section{Overlap removal}
\label{sec:objReco_OLR}

The object-reconstruction steps are so far -- except for electron and photon reconstruction -- completely agnostic of one another. As such, it is possible to doubly count detector signals by assigning them to two different objects. In addition, related objects might be erroneously reconstructed as being independent.
This can for example be muons that are produced in the decay of hadrons inside a jet or calorimeter clusters that are reconstructed as two electrons although originating from the same track.

All of this is prevented by regarding the reconstructed objects in a specific sequence and rejecting them in favour of other objects if certain criteria are fulfilled that indicate they are overlapping. \tabref{tab:objReco_overlapRemoval} shows the steps taken for the \METjets measurement described in \chapsref{sec:metJets}{sec:interpretation}.

\begin{table}[H] 
	\centering
	\begin{tabular}{cccccc}
		\toprule
		Step & Reject	& Against	& \multicolumn{2}{c}{Overlap criteria} \\
		&		&			& $\Delta{R_y}$ less than & Further criteria\\
		\midrule
		1	& tau			& electron	& 0.2\\
		2	& tau			& muon		& 0.2\\
		3	& electron	& muon		& $-$	& shared \ID track\\
		4	& jet			& electron	& 0.2\\
		5	& electron	& jet		& 0.4\\
		6	& jet			& muon		& 0.4	& number of tracks $<3$\\
		7	& muon		& jet		& 0.4\\
		8	& jet			& tau		& 0.2\\
		\bottomrule
	\end{tabular}
	\caption{
		Ordered sequence of the procedure for overlap removal used in the \METjets measurement in \chapsref{sec:metJets}{sec:interpretation}. The objects in the second column are rejected in favour of the objects in the third column if the criteria in the last two columns are fulfilled. The sequence goes from the top to the bottom. Objects removed in previous steps are not considered in the following decisions.
	}
	\label{tab:objReco_overlapRemoval}
\end{table}

\section{Missing transverse energy (\MET)}
\label{sec:objReco_MET}

Particles that have a low probability to interact with the detector material can traverse and leave the detector completely unobserved.
In the Standard Model, this is the case for neutrinos. \BSM models can introduce new particles for which this is also the case, the most prominent example being Dark Matter (see \secref{sec:DM}).
In general terms, also particles outside the detector acceptance have to be counted towards these unobserved particles.

In principle, momentum conservation can be taken advantage of to infer the presence of these effects. 
As no indication can be observed that any collision happened at all if only invisible particles are in the event, the recoil of the invisible particles against visible particles is exploited (\cf\secref{sec:DM_signature}).
The ATLAS detector is hermetic in the \tuple{x}{y} plane but not in $\theta$ due to the beam pipe (\cf\secref{sec:ATLAS_observables}).
Concisely, the momentum imbalance is calculated in the ATLAS Experiment~\cite{ATLAS:2018txj} as:


\begin{equation}
	\label{eq:objReco_METdefinition}
	p_{x(y)}^\text{miss} \coloneqq -\sum_\text{electrons} p_{x(y)}^e -\sum_\text{muons} p_{x(y)}^\mu-\sum_\text{taus} p_{x(y)}^\tau-\sum_\text{photons} p_{x(y)}^\gamma-\sum_\text{jets} p_{x(y)}^j-\sum_{\overset{\text{\scriptsize unused}}{\text{tracks}}} p_{x(y)}^\text{track}.
\end{equation}

Hereby, $p_{x(y)}^\alpha$ gives the momentum in $x$-($y$-)direction of objects of kind $\alpha$ after overlap removal, where $\alpha$ corresponds to electrons~$e$, muons~$\mu$, taus~$\tau$, photons~$\gamma$ and jets~$j$.
The sums iterate over all reconstructed and selected objects of kind $\alpha$.
The last summand in \eqref{eq:objReco_METdefinition} is a track-based soft term that takes into account the momentum in $x$-($y$-)direction of all tracks associated with the hard-scatter vertex but not matched to any selected object. In particular, if objects fail analysis selection-criteria, \eg on \pT or \eta, they contribute to the soft term.
$p_{x(y)}^\text{miss}$ is then the negative sum in $x$-($y$-)direction of all these contributions.

The definition of the momentum imbalance is only in terms of the transverse plane, $\pTmiss\equiv\left(p_{x}^\text{miss}, p_{y}^\text{miss}, 0\right)$, because the \textit{visible} momentum along the $z$-axis is not conserved as the detector is not completely hermetic in that direction and particles may disappear in the beam pipe. The magnitude of the missing transverse momentum is the \textit{missing transverse energy} $\MET\coloneqq \abs{\pTmiss}$.

The correct determination of \MET is challenging as it requires the reconstruction of all other objects and precise measurements of the properties in all detector subsystems, leading to a complete description of the hard scatter of interest.
The constructed value for \MET is therefore strongly influenced by the excellent, but still limited, resolution of the detector and sensitive to pileup.
The \MET resolution for the ATLAS Experiment is shown in \subfigref{fig:objReco_MET_plots}{a}.
The trigger efficiency (\cf\secref{sec:ATLAS_trigger}) is shown in \subfigref{fig:objReco_MET_plots}{b}.

\begin{myfigure}{
		(a) The root-mean-square (\RMS) width of $p_{x(y)}^\text{miss}$ as a function of the scalar sum of all transverse momenta, $\sum\ET$, for data and \MC simulation.
		The bottom panel shows a ratio of data to \MC simulation.
		Figure adapted from \refcite{ATLAS:2018txj}.
		(b) Trigger efficiencies for \MET triggers for each year during the Run-2 data taking as a function of the transverse momentum of a dimuon system.
		Muons are not considered by the \MET triggers.
		Figure taken from \refcite{ATLAS:2020atr}.
	}{fig:objReco_MET_plots}
	\subfloat[]{
		\includegraphics[height=140pt,valign=c]{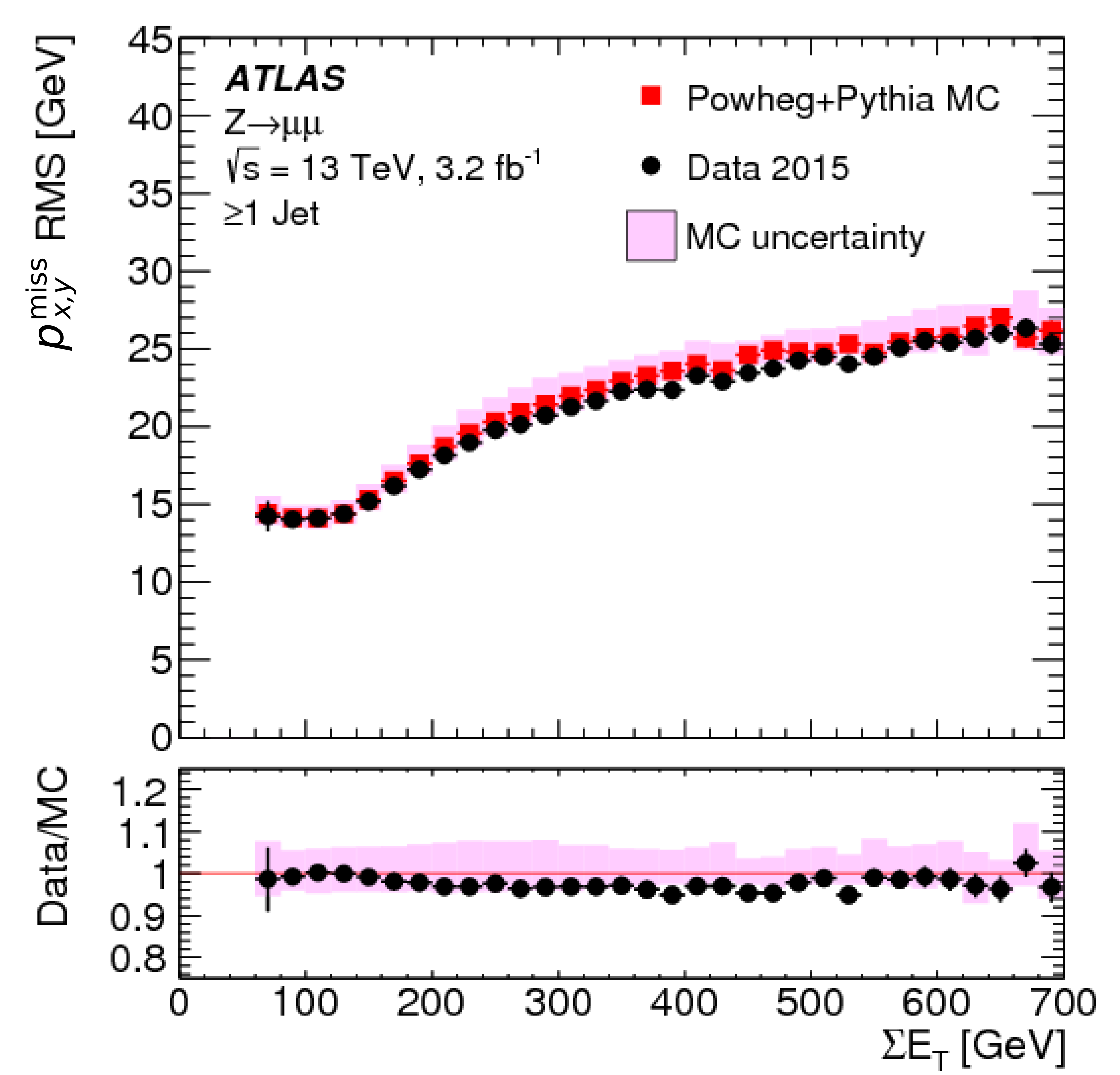}
		\vphantom{\includegraphics[height=150pt,valign=c]{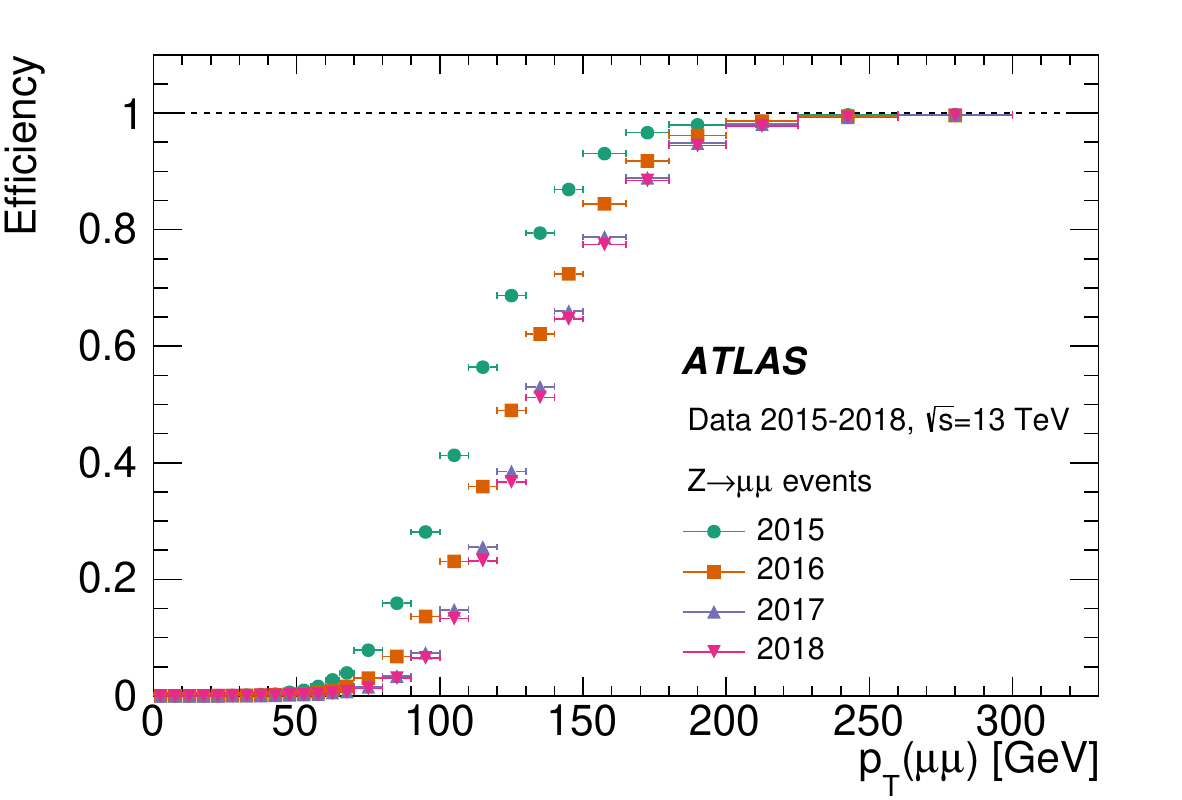}}
	}
	\hspace{10pt}
	\subfloat[]{\includegraphics[height=150pt,valign=c]{figures/reconstruction/MET/MET_trigger_pT}}
\end{myfigure}

\section{Particle-level definitions}
\label{sec:objReco_particleLevel}
The discussion so far has focused on reconstructing objects towards a detector-level representation.
However, it is also necessary to define these objects at particle level (see \chapref{sec:introduction}) in simulation to improve the understanding of detector effects and allow for meaningful comparisons between data and simulation.
The definitions used in the \METjets measurement in \chapsref{sec:metJets}{sec:interpretation} are discussed in the following.

Tracks are intrinsically related to detector signals and have to be neglected at particle level.

Electrons, muons and photons are detector-stable elementary particles and therefore directly retrieved from the simulation history (see \chapref{sec:MC}).
They are required to be prompt, \ie not originate from a hadron or heavier-lepton decay.
To mimic the detector response to photon radiation, photons within $\Delta R<0.1$ of a lepton are added to its four-momentum.
Correspondingly, photons are required to be isolated in accordance with the isolation requirement chosen for detector-level photons.
In particular, the magnitude of the transverse momentum not originating from the photon in a cone of $R=0.4$ around it is required to follow $\ET^\textnormal{cone}<0.044 p_{\textnormal{T},\gamma}+\SI{2.45}{GeV}$ where $p_{\textnormal{T},\gamma}$ is the transverse momentum of the photon.
This corrects for the activity of the underlying event and an energy offset.

\bigskip
Jets, taus and \MET are not fundamental entities  or not detector-stable and have to be constructed at particle level similar to detector level.

For jets, stable final-state particles are used as the input entities for the jet clustering~\cite{ATLAS:2017ghe}.
Neutrinos, other invisible particles like hypothetically Dark Matter as well as muons are excluded from the clustering to mimic the response of the ATLAS calorimeter, in which neither of these leaves significant energy depositions.
Further, contributions from pileup are neglected.
On the remaining particles, the same jet-clustering algorithm as at detector level is applied.

Any jets containing a hadron from a tau decay are classified as hadronically decaying
taus.
Overlap removal is performed at particle level as similar to detector level as possible.
This means requiring the same criteria in $\Delta R_y$ but neglecting all conditions with regard to tracks in the last column of \tabref{tab:objReco_overlapRemoval}.

\MET can be defined in different ways at particle level.
In the \METjets measurement in \chapsref{sec:metJets}{sec:interpretation} it is defined as the negative magnitude of the vector which is the sum of the momenta in the transverse plane of all visible final-state particles.
Only particles with $\absEta<4.9$ are taken into account as the range of the ATLAS detector in \eta is limited. Muons are required to have $\absEta<2.7$ due to the smaller extension of the Muon spectrometer (\cf\secref{sec:ATLAS_muonSpec}).

\Chapter{Providing for the future}{Analysis preservation}{%
	The future is a monster, and now it's turning.\\
	I wanna be future-proof.%
}{Conor Mason}{NothingButThieves:2021ftp}
\label{sec:analysisPreservation}



Physics analyses performed by detector collaborations at particle colliders usually result in measured data being published alongside an \textit{interpretation} of the data with regard to the Standard Model.
Sometimes additionally an interpretation with regard to \BSM models is published because there are questions left unanswered by the Standard Model, like the origin of Dark Matter, as discussed in \chapref{sec:SM}.
Naturally, only a limited number of \BSM models can be exercised in this approach.
Despite best efforts it is unlikely that the exact \BSM model that will prove correct in the future is among this small sample, however.
The importance of certain models may increase in the future and even models inconceived as of now could emerge.
Considering this, analysis results should foresightfully be provided in a format that allows straightforward \textit{reinterpretation} with regard to models other than the already considered.
Only then the physics potential of each analysis can be exploited to its fullest.

This is in line with the currently most important guidelines for Open Science, the FAIR Data Principles for findable, accessible, interoperable, and reusable digital\linebreak assets~\cite{Wilkinson:2016fai}.
Further, analysis preservation was recognised by the European Strategy for Particle Physics~\cite{Akesson:2006we,EuropeanStrategyforParticlePhysicsPreparatoryGroup:2013fia,EuropeanStrategyGroup:2020pow} and the U.S. Community Study on the Future of Particle Physics (Snowmass)~\cite{Butler:2023glv} as good scientific practise.

Mainly two criteria help in simplifying the reinterpretation of a physics analysis and should be adopted in the analysis preservation:
an appropriate choice of representation level as well as standardisation.
They are discussed in \secsref{sec:analysisPreservation_choiceOfLevel}{sec:analysisPreservation_standardisation}, respectively.
\secref{sec:analysisPreservation_metJets} draws conclusion from this for the \METjets measurement that is described in \chapsref{sec:metJets}{sec:interpretation}.

\section{Choice of representation level}
\label{sec:analysisPreservation_choiceOfLevel}

Different levels of representation are available at which measured data and theory prediction can be compared, as discussed in \chapref{sec:introduction} and shown in \figref{fig:detectorCorrection}: Detector signals, detector level, particle level, inclusive final state and model parameters.
Preparing data and prediction at one of these levels comes with different processing steps for either. Each physics analysis has to make a choice at which level to compare data and prediction as well as publish results that allow reinterpretation.

\begin{itemize}

\itembf{Detector signals} This level has the advantage of requiring minimal processing of the measured data but at the cost of being unintelligible to humans and excessively dependent on detector conditions.
Nonetheless, first studies of directly using the detector signals for model interpretation exist~\cite{Dort:2022pga}.

\itembf{Detector level} This representation is a little more complex, but still straightforward for experiments to achieve: it is the lowest level at which the data is human-intelligible, and to obtain a theory prediction at detector level only a detector simulation has to be run on the inputs from the inclusive final state.
After this the theory predictions can be treated identically to measured data.
While setting up the detector simulation is not trivial, it is often required for the experiments in any case because it is relied upon for tuning of detector and algorithms.

\itembf{Particle level} Experiments can only cover a finite range of models in their publications.
Later reinterpretation of the results at detector level with regard to another model poses a much greater problem.
On the one hand, complete detector simulation is very costly computationally and can take several minutes for a single event~\cite{ATLAS:2010arf} (\cf\secref{sec:MCEG_detectorSimulation}).
On the other hand, the information to accurately emulate the detector response and efficiencies might not necessarily be public.
Both points render it effortful to reinterpret data published at detector level.
In contrast, comparisons at particle level allow for easier reinterpretation for researchers outside the analysis: No detector simulation is required any more as detector response and efficiencies are corrected for by the analysers.

\itembf{Inclusive final state} Comparisons at this level require extrapolations of the phase space of the measured data into unobserved regions. This procedure is very model-dependent and error-prone, leading to large uncertainties at best, and is therefore discouraged~\cite{Bierlich:2019rhm}. 

\itembf{Model parameters} At this level, reinterpretation with regard to other models is very difficult which is why this level is generally disfavoured.
\end{itemize}

It is therefore recommended for experiments to publish their data in particle-level representation where possible~\cite{Bierlich:2019rhm}.
This is the level that compromises ideally between dependence on (B)\SM models while at the same time completely removing the need for understanding the detector from the theory side.
Apart from the facilitation of \BSM reinterpretation, results at particle level also allow directly incorporating improvements in the theory predictions when they become available.
This comes of course with additional efforts for the experiments to accurately assess and correct for detector effects.
They, however, possess by far the largest expertise regarding their detectors and can improve the scientific value of their results in this way.

There are also cases in which publishing results at particle level is not reasonable, however:
Unfolding (see~\chapref{sec:metJets_detectorCorrection}), for example, requires the availability of large event counts as it is a purely statistical procedure.
Typically, at least 20 events per interval of a physics quantity are required for stable algorithm performance.
Moving to a more inclusive phase space with higher statistics can reduce the sensitivity in particular in searches for \BSM physics where effects are expected to be rare or are prominent only in the tail of distributions.

Also models which predict a complex interaction of the \BSM particles with the detector cannot be compared to measured data at particle level, \ie independent of detector effects. The prime example for these are models that predict long-lived particles that decay within the detector volume.
They´ require sophisticated reconstruction techniques to take advantage of the distinct kinematic features of the long-lived particles.

\section{Standardisation}
\label{sec:analysisPreservation_standardisation}
The second measure that simplifies reinterpretation is standardisation. Reinterpreters are not as familiar with the subtleties of an analysis as the original authors. They might even want to reinterpret multiple analyses in short order, limiting the reasonable amount of effort spent on a single analysis. A standardised interface for providing analysis results ensures long-term usability of the results and opens up the possibility of automatising reinterpretations, as will be seen in \chapref{sec:Contur}. Regarding distributions and model parameters, \HEPData~\cite{Maguire:2017ypu} has commendably been established as a standard for publication.
It is encouraged to publish not only the measured data, but also theoretical predictions, correlations including systematic uncertainties, likelihoods and metadata preserving the complete analysis logic~\cite{LHCReinterpretationForum:2020xtr}.

Preserving the complete analysis logic in a standardised manner is conducted within the ATLAS Experiment commonly using one of the following tools:
\begin{itemize}
	\item \RECAST~\cite{Cranmer:2010hk} is a tool that preserves the analysis logic at detector level. Reinterpretations with regard to a new model therefore require to run a complete detector simulation to be able to compare data and prediction. More than 90 ATLAS analyses have been preserved in this way~\cite{ATLAS:2022rec}.
	\item \SimpleAnalysis~\cite{ATL-PHYS-PUB-2022-017} is a tool that preserves the analysis logic at particle level, albeit employing either a fast detector simulation or approximate estimate of the detector efficiencies.
	More than 60 ATLAS analyses have been preserved in this way~\cite{ATLAS:2022sia}.
	The tool, however, is internal to the ATLAS collaboration and not accessible to the public, limiting its use.
	\item \Rivet~\cite{Bierlich:2019rhm} is another tool that preserves the analysis logic at particle level.
	This tool can also employ smearing according to approximate estimates of the detector efficiencies if a representation at detector level is desired.
	Given its accessibility to the public, versatility and the general arguments in favour of providing results in particle-level representation given in the previous section, this tool is recommendable for many analyses~\cite{LHCReinterpretationForum:2020xtr}.
	More than 180 ATLAS analyses have been preserved in this way~\cite{Rivet:2022ric}.
\end{itemize}

\section{Analysis preservation for the \METjets measurement}
\label{sec:analysisPreservation_metJets}

Following the discussion above, the \METjets measurement described in \chapsref{sec:metJets}{sec:interpretation} is corrected for detector effects and will be published in a particle-level representation alongside a \Rivet routine~\cite{ATLAS:2023mjt}.
It is important to make these choices in advance to performing the actual measurement because they influence the analysis strategy, \eg with respect to kinematic cuts or the required minimum number of events in a given observable interval.

\figref{fig:analysisPreservation_RivetvsNominal} gives a comparison of events selected by the nominal analysis (blue) and \Rivet routine (red) of the \METjets measurement described in \chapsref{sec:metJets}{sec:interpretation}.
Shown are the yields in the signal region and \Mono subregion, see \secref{sec:metJets_phaseSpaces_observables} for the region definitions and \secref{sec:metJets_detLevelResults} for a detailed description of the distribution.
Generated\linebreak $Z\left(\to\nu\bar{\nu}\right)$+jj events (see \secref{sec:metJets_MC}) are used for comparing the selection.

The bottom panel gives the ratio to the \Rivet yields.
They exhibit very good agreement between the \Rivet and nominal selection.

\begin{myfigure}{
		Yield in generated $Z\left(\to\nu\bar{\nu}\right)$+jj events for the nominal analysis and \Rivet routine for the \METjets measurement described in \chapsref{sec:metJets}{sec:interpretation} in the signal region~(\SR) and \Mono subregion.
		The bottom panel shows the ratio to the \Rivet yield.
	}{fig:analysisPreservation_RivetvsNominal}
	\includegraphics[width=0.7\textwidth]{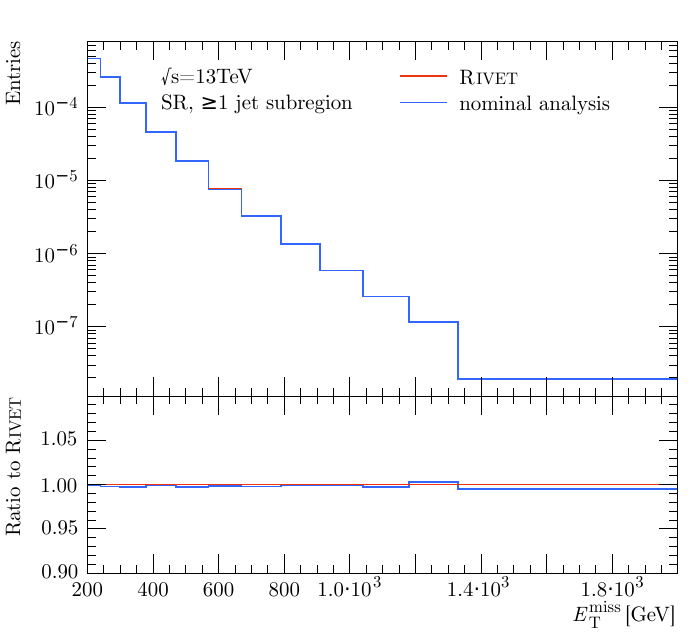}
\end{myfigure}

\ChapterGeneral{Assembling a measurement}{Selection of the \METjets final state\linebreak{}with the ATLAS detector}{Selection of the \METjets final state}{%
	Drei Festplatten voll, viele Ordner angelegt,\\
	analysiert, Nummer notiert und dann gewählt.%
}{Kolja Podkowik \& Jakob Wich}{AntilopenGang:2014enk}
\label{sec:metJets}

The Standard Model does not provide a suitable candidate for Dark Matter, as discussed in \secref{sec:DM}.
This poses one of its most concrete shortcomings.
Investigating whether Dark Matter is produced in events at particle colliders offers an opportunity to unravel the nature of Dark Matter, or at least set constraints on its production.
In the search process, the understanding of \SM processes and their modelling in the same topology is improved.

In detectors at particle colliders, Dark Matter cannot be measured directly:
due to its inherently low interaction cross section with the detector material it leaves the detector unobserved.
It is necessary for invisible particles to recoil against visible objects to give rise to a detectable final state with missing transverse energy, \MET (\cf\secsref{sec:DM_signature}{sec:objReco_MET}).
One of the most generic final states is obtained if the invisible particles recoil against at least one jet ($E_\text{T}^\text{miss}$+jets), as depicted in \subfigref{fig:metJets_Feynman_metJets}{a}.
This final state was investigated at the \LHC under different conditions before.
This is discussed in \chapref{sec:comparison}.


\myTwoFeynmanFigure{%
		Feynman diagrams for the production of the \METjets final state (a) where the invisible particle is Dark Matter, like in the \THDMa, and (b) in the Standard Model.%
	}{fig:metJets_Feynman_metJets}{%
		2HDMa/gg_to_loop_to_a_to_xx+ISR%
	}{%
		SM/qq_to_Z_to_nunu+ISR%
}

In the following chapters (\chapsref{sec:metJets}{sec:interpretation}), a measurement of the differential cross section of the \METjets final state is performed.
The measurement makes use of proton--proton collisions recorded with the ATLAS detector at the \LHC during Run 2 at a centre-of-mass energy of $\sqrt{s}=\SI{13}{TeV}$, corresponding to a total integrated luminosity of \SI{139}{fb^{-1}}.
Given the good agreement of the measured data with \SM predictions after a statistical fit, the results are used to constrain a model for Dark Matter, the \THDMa (\cf\secref{sec:2HDMa}).

In this chapter, \chapref{sec:metJets}, the general analysis setup as well as selection is described.
Comparisons are made between data and simulated \SM predictions in detector-level representation.
Following the discussion in \chapref{sec:analysisPreservation}, the measurement is designed such that it can be easily corrected for detector effects to achieve a representation at the particle level.
This is performed in \chapref{sec:metJets_detectorCorrection}.
In \chapref{sec:2HDMa_metJetsMeasurement}, the contributions the \THDMa  would make to the \METjets phase space are investigated.
In \chapref{sec:interpretation}, the measured data is evaluated with respect to their agreement to generated \SM and \THDMa predictions.

This chapter is structured as follows:
\secref{sec:metJets_strategy} gives an overview of the analysis strategy of the measurement.
The used data, simulation and triggers are described in \secref{sec:metJets_dataMC}.
\secsref{sec:metJets_objDefinitions}{sec:metJets_kinVariables} detail the definitions of objects and variables for the kinematic selection used in the measurement, respectively.
Regions investigated by the measurement are defined in \secref{sec:metJets_phaseSpaces_observables}, uncertainties estimated in \secref{sec:metJets_uncertainties}.
Comparisons between measured data and simulated \SM prediction in detector-level representation are made in \secref{sec:metJets_detLevelResults}.

\section{Analysis strategy}
\label{sec:metJets_strategy}

A typical \METjets event is shown in \figref{fig:metJets_eventDisplay}.
It prominently features large energy depositions in the calorimeters (yellow and green bars) that are reconstructed as a jet.
There are no significant other objects in the event balancing the transverse momentum of the jet, giving rise to large missing transverse momentum \pTmiss (dashed red line).
\MET denotes the absolute value of the missing transverse momentum, the \textit{missing transverse energy}.

\begin{myfigure}{
		A $\MET+$jet event collected in the 2017 with the ATLAS detector, shown (left) in the \tuple{x}{y} plane as well as (right) in a diagonal view.
		Green (yellow) bars correspond to the energy depositions in the electromagnetic (hadronic) calorimeters.
		The prominent energy depositions towards the top are reconstructed as a jet.
		Solid orange lines correspond to reconstructed tracks.
		The missing transverse momentum \pTmiss is displayed as a dashed red line.
		Figure taken from \refcite{ATLAS:2021mje}.
	}{fig:metJets_eventDisplay}
	\includegraphics[width=0.8\textwidth]{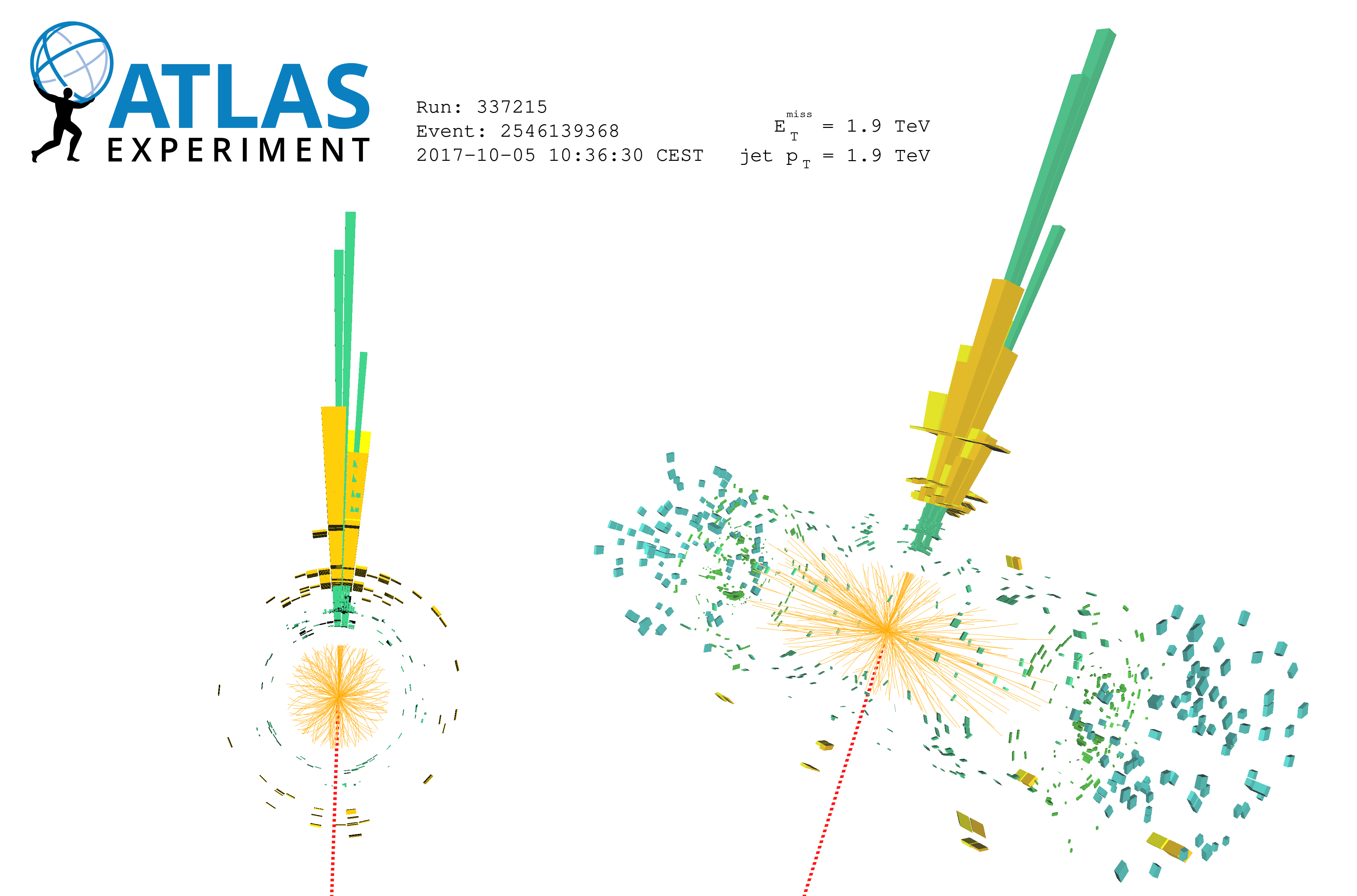}
\end{myfigure}

The phase space selected by the measurement is designed to be inclusive to any \METjets contributions.
Inclusive phase-space definitions are recommended~\cite{LHCReinterpretationForum:2020xtr} because they minimise built-in theoretical assumptions that complicate later reinterpretation.
The intended inclusive selection of the \METjets phase space is sensitive to \SM processes as well as to hypothetical \DM production, while remaining independent of any specific \DM model.
Presence of \BSM physics could be indicated by a significant excess of the measured data over the \SM prediction.

A region focused on the process of interest, the so-called \textit{signal region}~(\SR), is defined to select events in the \METjets final state.
The dominant \SM contribution in this phase space stems from \Znunujets events (\cf\subfigref{fig:metJets_Feynman_metJets}{b}) where the $Z$ boson decaying to neutrinos remains invisible to the detector and can only be inferred from the recoil of the jet or jets as \MET.
The most important subdominant \SM contributions arise from $W\to\ell\nu$ decays in association with jets (\cf\subfigref{fig:metJets_Feynman_AMs}{a}), where the charged lepton $\ell$ is undetected.

One goal of the measurement is to compare whether measured data and simulated \SM prediction agree within their respective uncertainties.
This makes excellent control of the systematic uncertainties and potential mismodellings imperative.
To this end, auxiliary measurements (\AMs) are defined in phase spaces similar to the signal region.
This similarity causes the auxiliary measurements to be influenced by the same uncertainties as well as potential mismodellings.
In a simultaneous fit, the auxiliary measurements then constrain these common uncertainties and mismodellings.
The dominant systematic uncertainties that arise in the signal region and shall be constrained in the auxiliary measurements are caused by jet reconstruction and calibration (\cf\secref{sec:objReco_jets}) as well as theoretical predictions for \Vjets processes with $V\in\left\{Z,W\right\}$.
They are discussed in detail in \secref{sec:metJets_uncertainties}.

\Zll decays, as shown in \subfigref{fig:metJets_Feynman_AMs}{b}, are kinematically almost identical to \Znunu decays.
Their decay products, charged leptons, are visible to the detector, however, allowing \Zll decays to be reconstructed with high precision.
This is different from \Znunu decays where the presence of neutrinos gives rise to \MET.
Calculating \MET requires the reconstruction of all other objects in the event, causing larger uncertainties.
Further negligible differences between \Zll and \Znunu events arise because charged leptons have larger masses and are electrically charged, resulting in a different detector response and radiative corrections~\cite{Novikov:1999af}.
Two auxiliary measurements, \TwoEJetsAM and \TwoMuJetsAM, probe the phase space where the $Z$ boson decays to electrons (\Zee) and muons (\Zmumu), respectively, instead of neutrinos (\Znunu).
The magnitude of the combined transverse momenta of the leptons in \Zll events, \ET, serves as a proxy for \MET in \Znunu events.

\myTwoFeynmanFigure{%
		Feynman diagrams for the dominant \SM contributions to the auxiliary measurements: (a) \Wlnujets for \OneLJetsAMs and (b) \Zlljets for \TwoLJetsAMs.
	}{fig:metJets_Feynman_AMs}{%
		SM/qq_to_W_to_lnu+ISR%
	}{%
		SM/qq_to_Z_to_ll+ISR%
}

\Wlnu decays (\cf\subfigref{fig:metJets_Feynman_AMs}{a}) in which the charged leptons remain undetected can also mimic the \METjets signature and enter the signal region.
\OneEJetsAM (\OneMuJetsAM) are auxiliary measurements for the case where the electron (muon) coming from the decay of the $W$ boson \textit{is} detected.

Taus decay, before reaching the detector, into neutrinos and either hadrons or a lighter charged lepton~\cite{Zyla:2020zbs}.
\textit{Leptonically} decaying taus are difficult to distinguish from prompt electrons or muons.
Their signature differs only by a larger amount of \MET from the two neutrinos in the $\tau\to\ell\nu_\ell\nu_\tau$ decay.
They are therefore already selected by the other auxiliary measurements.
\textit{Hadronically} decaying taus introduce new systematic uncertainties.
Therefore, no specific auxiliary measurement is designed to select taus, and they are only used to reject events.

\bigskip
With this setup, the experimental and theoretical uncertainties exhibit strong correlations for signal region as well as auxiliary measurements.
There is correlation for experimental uncertainties related to luminosity and jets between all regions.
Uncertainties related to electrons (muons) exhibit correlations between \OneEJetsAM and \TwoEJetsAM (\OneMuJetsAM and \TwoMuJetsAM).
For theoretical uncertainties, not only the \Zjets uncertainties exhibit correlations between the different measurements, but there is also a correlation between \Zjets and \Wjets uncertainties~\cite{Lindert:2017olm}.
A simultaneous fit to all these measurements can therefore exploit the correlations.
This improves the statistical statement whether potential discrepancies between measured data and simulated \SM prediction can be accounted for simultaneously in all measurements given the systematic uncertainties.
This can be exploited further by defining
\begin{equation}
	\label{eq:metJets_Rmiss}
	\Rmiss\left(\SR/\AM_X\right)\coloneqq\frac{\sigma_\text{fid}\left(\SR\right)}{\sigma_\text{fid}\left(\AM_X\right)}
\end{equation}
as observable.
It is the ratio between the differential cross sections  $\sigma_\text{fid}$ in the signal region and a chosen auxiliary measurement $\AM_X$.
In \Rmiss, potential mismodellings that appear in both numerator and denominator are partly cancelled.
The same cancellation takes place for common theoretical and experimental uncertainties.

\bigskip
Due to the inclusive definition of the phase spaces of the measurement, \BSM models can not only give significant contributions to the signal region, but also to the auxiliary measurements.
This is demonstrated for the \THDMa in \chapref{sec:2HDMa_metJetsMeasurement}.
\section{Data, simulation and triggers}
\label{sec:metJets_dataMC}

The data used in this measurement correspond to an integrated luminosity of \SI{139}{\ifb}, collected during the \LHC Run 2 in 2015 to 2018 with the ATLAS detector.
\SM processes are simulated using the setup described in \secref{sec:metJets_MC}.

For data and simulation, events have to be selected by \MET~\cite{ATLAS:2020atr}, single-electron or dielectron~\cite{ATLAS:2019dpa} triggers.
The exact list of used triggers is given in \appref{app:metJets_triggers}.
These triggers are chosen because they offer the highest selection efficiency for the signal region, \OneEJetsAM as well as \TwoEJetsAM, respectively.
No additional triggers are required for muons, \ie for \OneMuJetsAM and \TwoMuJetsAM, because the calculation of \MET in the high-level trigger in the ATLAS Experiment does not use information from the Muon spectrometer~\cite{ATLAS:2020atr}.
Therefore, all events with sufficient transverse momentum of the muon are already selected by the \MET triggers.
This approach avoids additional uncertainties from muon triggers.


\section{Object definitions}
\label{sec:metJets_objDefinitions}

Detector-level objects are required to comply to criteria detailed in the following to ensure a high quality of the selected objects.
Among others, the selections reflect the acceptance of the ATLAS detector with the highest resolutions and reconstruction efficiencies.
Particle-level objects are required to comply to the same criteria with respect to transverse momentum and rapidity.

For most objects, relaxed \textit{baseline} and stricter \textit{signal} selections are specified.
Baseline definitions impose the minimal set of requirements such that the type of the object can be assumed.
Signal definitions tighten the baseline criteria to allow for a higher degree of precision, at the cost of lower reconstruction efficiencies, such that a phase-space selection can be reliably based on these objects.
The signal and auxiliary measurements are then differentiated by requiring a specific multiplicity of reconstructed signal objects and no other baseline leptons.
Objects erroneously failing the signal definitions given in this section therefore in general lead to the event being assigned to one of the measurements with a lower multiplicity of signal objects.
In addition, they contribute to \MET as described in \secref{sec:objReco_MET}.

All baseline (and with that signal) objects have to pass the procedure for overlap removal described in \secref{sec:objReco_OLR}.

\subsection{Electrons}
\label{sec:metJets_objDefinitions_electrons}
Baseline electrons, reconstructed as described in \secref{sec:objReco_elePhot},  have to exhibit $\pT>\SI{7}{GeV}$ and \crackVeto. The latter represents the acceptance of the ATLAS \EM calorimeters, excluding the transition region between barrel and end-caps where energy resolution is low~\cite{ATLAS:2008xda}.
The "Loose" identification working point is used but no requirement on the isolation (\cf\secref{sec:objReco_elePhot}) set.

For \OneEJetsAM and \TwoEJetsAM, signal electrons have to fulfil tighter selection criteria in addition to those for baseline electrons to ensure that only well-measured electrons are considered.
Signal electrons are required to be matched to the hard-scatter vertex within $\abs{z_0 \sinP}<\SI{0.5}{mm}$ and $\dZsig<5$, as defined in \secref{sec:objReco_vertices}.
The "HighPtCaloOnly" isolation working point is used, which is optimised for electrons with high transverse momenta.

The main source of misidentified prompt electrons stems from jets misidentified as electrons.
In \OneEJetsAM, signal electrons are required to comply with the stricter "Tight" identification working point to reduce these contributions.
In \TwoEJetsAM, the "Medium" identification working point is demanded. This poses a tighter restriction compared to baseline electrons to ensure higher-quality electrons but relaxed requirements compared to signal electrons in \OneEJetsAM to maintain a high selection efficiency despite requiring two signal electrons in the event.

\subsection{Muons}
Baseline muons have to be reconstructed following the "combined" or "segment-tagged" strategy (\cf\secref{sec:objReco_muons}), exhibit $\pT>\SI{7}{GeV}$ and be within the acceptance of the Muon spectrometer, \muonAcceptance.
The "Loose" identification working point but no isolation criterion is used.

For \OneMuJetsAM and \TwoMuJetsAM, signal muons have to be matched to the hard-scatter vertex within $\abs{z_0 \sinP}<\SI{0.5}{mm}$ and $\dZsig<3$ in addition to the selection criteria for baseline muons. Further, signal muons have to meet the requirements of the "Medium" identification and "Loose" isolation working points. This is advantageous because the required muon multiplicities are not very high, as pointed out in \secref{sec:objReco_muons}.

The only relevant source of misidentified prompt muons is decays of heavy-flavour hadrons involving muons that are then non-prompt.
With the conditions above, the fraction of misidentified prompt muons is small.
Therefore, no tighter working points are used for signal muons in \OneMuJetsAM, contrary to signal electrons in \OneEJetsAM.

\subsection{Taus}
Only hadronically decaying taus are considered because events with leptonically decaying taus differ from events with electrons or muons only in the neutrino multiplicity, as discussed in \secsref{sec:objReco_taus}{sec:metJets_strategy}.
Baseline taus have to be within \crackVeto, similar to electrons.
They are required to have $\pT>\SI{20}{GeV}$, a unit electric charge and fulfil the "Loose" criterion by a recursive neural-network for identification of hadronically decaying taus~\cite{ATL-PHYS-PUB-2022-044}.
Taus are only used to reject events and therefore no signal taus are defined.


\subsection{Jets}
\label{sec:metJets_objDefinitions_jets}
Jets clustered with the anti-\kT algorithm with a radius parameter $R=0.4$ and reconstructed with the particle-flow algorithm are used in this measurement, as mentioned in \secref{sec:objReco_jets}.
They must be within the detector acceptance, $\absY<4.4$, and have $\pT>\SI{30}{GeV}$.
Furthermore, they have to fulfil "Tight" criteria for jet suppression from pileup for $\absEta<2.5$~\cite{ATLAS-CONF-2014-018} and in the forward region~\cite{ATL-PHYS-PUB-2019-026}. No distinction is made between baseline and signal jets.

\subsection{Missing transverse energy (\MET)}
\label{sec:metJets_objDefinitions_MET}

The reconstruction of the missing transverse energy follows exactly the procedure outlined in \secref{sec:objReco_MET}.
This \textit{measured} \MET is called \METmeas.
The production of Dark Matter tends to give rise to large \METmeas.

In the \textit{constructed} \MET, \METconst, all neutral and charged signal leptons are treated as invisible in the \MET calculation.
This is different to \METmeas in which only neutral leptons, \ie neutrinos, are treated as invisible.
If no signal leptons are present, \METmeas and \METconst are identical.
\METconst in general corresponds to the transverse momentum of the vector boson in the dominant \SM processes.
\METconst thereby renders the auxiliary measurements similar to the signal region when imposing selection criteria on \METconst: the dominant \SM contributions in each region now correspond to the cases where the $Z$ boson decays to a pair of neutrinos instead of leptons (\TwoLJetsAMs) or the lepton coming from the decay of the $W$ boson (\OneLJetsAMs) is not in the detector acceptance.
In events where the largest recoil stems from a single jet, \METconst and the leading-jet \pT are similar.
\section{Variables for kinematic selection}
\label{sec:metJets_kinVariables}

Different variables are used to define the kinematic selections of the \METjets measurement. They are described in the following.

The main kinematic variables used for object and consequently event selection are the pseudorapidity \eta and the transverse momentum \pT as introduced in \secref{sec:ATLAS_observables}.
Objects of the same type in an event are ordered by their transverse momentum. The objects with the highest and second highest \pT are denoted the \textit{leading} and \textit{subleading} object, respectively.

The missing transverse energy captures the effect of particles leaving the detector unobserved, \eg neutrinos or Dark Matter.
\METconst is defined such that it is as similar as possible in all measurements, \ie marking signal leptons invisible in its calculation as explained in the previous section.
For \Znunujets processes, \METconst gives an estimate of the transverse momentum of the vector boson.
Selection criteria on \eta, \pT and \METconst are motivated by the acceptance of detector and triggers.

\pTmissMeas and its absolute value \METmeas represent the measured momentum imbalance before treating any reconstructed charged signal lepton as invisible. 
If no signal leptons are present, \METmeas and \METconst are identical.
For \Wlnujets processes, \pTmissMeas and \METmeas give an estimate of the transverse momentum of the neutrino and its absolute value, respectively.

The transverse mass of the system of lepton and neutrino in \Wlnujets events is
\begin{equation*}
	\mT\coloneqq\sqrt{2\hspace{2pt}\pT^\ell\hspace{2pt}\METmeas \left(1 - \cos\left(\Delta\phi\tuple{\vv{p}_\text{T}^\ell}{\vv{p}_\text{T,meas}^\text{miss}}\right) \right)}
\end{equation*}
whereby $\vv{p}_\text{T}^\ell$ denotes the momentum of the charged lepton in the transverse plane.
The transverse mass is used to place constraints on the reconstructed mass of a $W$ boson.
Both, \METmeas and \mT, are used to improve the selection efficiency of \Wlnujets events in \OneEJetsAM.
This helps to constrain the \Wjets contribution in the signal region and its corresponding systematic uncertainties.

The invariant mass of the dilepton system, \mll, is used to place constraints on the reconstructed mass of a $Z$ boson for improving the selection efficiency of \Zlljets events in \TwoLJetsAMs.
This helps to constrain the \Zjets contribution in the signal region and its corresponding systematic uncertainties.

If at least two jets are in the event, the invariant mass of the system of the two highest-momenta jets, \mjj, and their rapidity difference, \DeltaYjj, are used to define a dedicated selection of events which exhibit the fusion of two vector bosons (\VBF events).

Mismeasurements of jet momenta can lead to a momentum imbalance, \ie\pTmiss, in events. In these cases, the momentum of the mismeasured jet and \pTmiss would align. These events are suppressed to an almost negligible size by requirements on the minimum azimuthal distance between the four highest-\pT jets and \pTmissConst,
\begin{equation*}
	\DeltaPhiMin\coloneqq\min_{1\leq i\leq 4}\Delta\phi\left(\vv{p}_{\text{jet}_i},\pTmissConst\right).
\end{equation*}
\section{Measurement regions}
\label{sec:metJets_phaseSpaces_observables}

The \SR and auxiliary measurements used in the \METjets measurement correspond to different fiducial phase-space \textit{regions} defined by requirements on the multiplicity and kinematics of the final-state particles.
These requirements have to be met in detector-level representation and are detailed in \secref{sec:metJets_regions}.
Simulated predictions in particle-level representation have to meet the same selection criteria apart from trigger requirements.
For each region, two phase-space \textit{subregions} are defined to enhance the sensitivity of the selection to different \SM topologies.
They are detailed in \secref{sec:metJets_subregions}.
A schematic representation of the phase spaces selected by the regions and subregions is shown in \figref{fig:metJets_subRegions}.

\begin{myfigure}{
		Schematic of the selection cuts of the \Mono (red) and \VBF (blue) subregions in leading-jet \pT and number of jets.
		Both subregions overlap for leading-jet $\pT>\SI{120}{GeV}$ and number of jets $\geq2$ (purple).
		The five measurement regions differ by the number of selected signal leptons.
	}{fig:metJets_subRegions}
	\includegraphics[width=\textwidth]{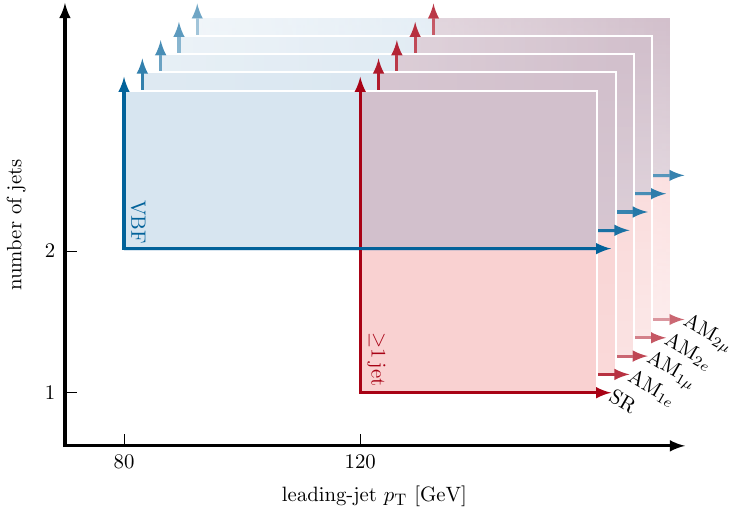}
\end{myfigure}

\subsection{Regions}
\label{sec:metJets_regions}

Five regions -- one signal region and four auxiliary measurements -- are defined.
Their selection criteria are motivated by the acceptance of detector and trigger as well as background rejection and are detailed in \tabref{tab:metJets_regions}.

{
	\renewcommand{\arraystretch}{1.3} 
	\newlength{\columnwidthHere}
	\setlength{\columnwidthHere}{65pt}
	\begin{mytable}{
			Selections defining the five principal phase-space regions of the measurement.
			The top part of the table applies to all regions.
			Only signal leptons are considered for the bottom part of the table.
		}{tab:metJets_regions}{L{115pt}C{\columnwidthHere}C{\columnwidthHere}C{\columnwidthHere}C{\columnwidthHere}C{\columnwidthHere}C{\columnwidthHere}}
		
		& \METjetsSR & \OneEJetsAM & \OneMuJetsAM & \TwoEJetsAM & \TwoMuJetsAM\\
		\midrule
		\METconst [GeV] & \multicolumn{5}{c}{$>200$}\\
		jet multiplicity & \multicolumn{5}{c}{$\geq1$}\\
		add. baseline leptons & \multicolumn{5}{c}{none}\\
		\DeltaPhiMin & \multicolumn{5}{c}{$>0.4$}\\
		\midrule
		trigger & \METmeas & electron & \METmeas & electron & \METmeas\\
		leading-lepton \pT [GeV] & -- & $>30$ & $>7$ & \multicolumn{2}{c}{$>80$}\\
		sublead.-lepton \pT [GeV] & -- & -- & -- & \multicolumn{2}{c}{$>7$}\\
		dilepton mass [GeV] & -- & -- & -- & \multicolumn{2}{c}{$66<\mll<116$}\\
		\METmeas [GeV] & $> 200$& $> 60$  & -- & -- & --\\
		transverse mass [GeV] & -- & $30<\mT<100$ & -- & -- & --\\
	\end{mytable}
}

\subsubsection{Signal region (\SR)}
The focal point of the analysis is the signal region (\SR) which requires events to have large \METconst and at least one jet recoiling against it.
The signal region further rejects events with baseline leptons.
Therefore, $\METconst\equiv\METmeas$ in the signal region.
In addition, events in the signal region have to be selected by a \METmeas trigger. 

$\METconst>\SI{200}{GeV}$ is required to allow for a trigger efficiency of more than \SI{97}{\%} (\cf\subfigref{fig:objReco_MET_plots}{b}).
The full list of triggers is given in \appref{app:metJets_triggers}.
Mismeasurements of jet momenta, in particular from \QCD multijet events, are suppressed by requiring $\DeltaPhiMin>0.4$.

These requirements allow selecting primarily \Znunujets processes in the Standard Model.
In \BSM models, processes producing detector-invisible particles, like Dark Matter, recoiling against at least one jet are selected.

\subsubsection{Auxiliary measurements (\AM{}s)}
The signal region, as stated in the previous paragraph, is dominated by \Znunujets processes.
Auxiliary measurements enriched in other, but similar \Vjets processes are defined to constrain correlated systematic uncertainties and potential common mismodellings.
Correlated systematic uncertainties originate for example from jet reconstruction and calibration as well as theoretical uncertainties related to \Vjets simulation.

In general, the auxiliary measurements are defined as similar to the signal region as possible.
Concretely, they have the same requirements with respect to jet multiplicity, \DeltaPhiMin and \METconst.
In addition, a specific multiplicity and flavour of signal leptons is required for each auxiliary measurement.
As explained in \secref{sec:metJets_objDefinitions_MET}, all signal leptons are treated as invisible in the \METconst calculation to increase the similarity between signal region and auxiliary measurements.
Similar to the signal region, the auxiliary measurements reject events that have any number of baseline leptons in addition to the signal leptons.

The detailed selection criteria for the four auxiliary measurements are as follows:

\begin{itemize}
	\itembf{\OneEJetsAM}
	The \OneEJets auxiliary measurement (\OneEJetsAM) is tailored towards\linebreak \Wenujets processes and selects events passing an electron trigger with exactly one signal electron.
	The requirement on the transverse momentum of the lepton is tightened with respect to the object definition (\cf\secref{sec:metJets_objDefinitions_electrons}) to $\pT>\SI{30}{GeV}$ to ensure a high trigger efficiency.
	Events in which jets have been misidentified as electrons are suppressed by imposing two additional criteria:
	\METmeas, giving an estimate of the transverse momentum of the electron-neutrino, is required to be larger than \SI{60}{GeV}.
	The transverse mass has to be compatible with the mass of a $W$ boson, $\SI{30}{GeV}<\mT<\SI{100}{GeV}$.

	\itembf{\OneMuJetsAM}
	The \OneMuJets auxiliary measurement (\OneMuJetsAM) is tailored towards\linebreak \Wmunujets processes and selects events with exactly one signal muon with $\pT>\SI{7}{GeV}$.
	Events are required to be selected by a \METmeas trigger, as in the signal region.
	The usage of a sophisticated muon trigger can be avoided because the calculation of \METmeas in the high-level trigger in the ATLAS Experiment does not use information from the Muon spectrometer~\cite{ATLAS:2020atr}, as pointed out in \secref{sec:metJets_dataMC}.
	This approach allows lowering the requirement on the transverse momentum below the threshold for triggering single muons of \SI{26}{GeV}~\cite{ATLAS:2020gty}.
	The contamination of the selected phase space by events with jets misidentified as leptons is smaller in \OneMuJetsAM than in \OneEJetsAM.
	Therefore, no additional selection criteria regarding \METmeas and \mT are applied in \OneMuJetsAM.

	\itembf{\TwoEJetsAM}
	The \TwoEJets auxiliary measurement (\TwoEJetsAM) is tailored towards\linebreak \Zeejets processes and selects events passing an electron trigger with exactly two signal electrons of opposite charge.
	Looser electron-identification criteria than for \OneEJetsAM are used ("Medium" instead of "Tight", \cf\secref{sec:metJets_objDefinitions_electrons}) because the two electrons originating from the decay of a $Z$ boson result in a more distinct process signature in \TwoEJetsAM.
	The triggers for "Medium" electrons have a higher \pT threshold with respect to the triggers for "Tight" electrons to provide an acceptable triggering rate despite relaxed identification criteria (\cf\appref{app:metJets_triggers}).
	In consequence, a stricter requirement of $\pT>\SI{80}{GeV}$ for the leading electron in \TwoEJetsAM with respect to \OneEJetsAM is imposed to ensure a high trigger efficiency.
	For the subleading electron, the criterion of the object definition, \ie $\pT>\SI{7}{GeV}$, is sufficient.
	The invariant mass of the dilepton system, \mll, has to be compatible with the mass of a $Z$ boson: $\SI{60}{GeV}<\mll<\SI{116}{GeV}$.

	\itembf{\TwoMuJetsAM}
	The \TwoMuJets auxiliary measurement (\TwoMuJetsAM) is tailored towards\linebreak \Zmumujets processes and selects events passing a \METmeas trigger with exactly two signal muons of opposite charge.
	The same criteria as for \TwoEJetsAM on leading and subleading lepton \pT as well as \mll are imposed.
\end{itemize}

\subsection{Subregions}
\label{sec:metJets_subregions}

Two different phase-space subregions are defined depending on the kinematics of the jets in the event to increase the sensitivity of the selection to different \SM topologies.
The requirements are applied consistently to all regions and listed in \tabref{tab:metJets_subregions}.
Like this, each region obtains two subregions.
These subregions are defined inclusively and are therefore not statistically independent, in contrast to the regions defined previously.
A schematic representation of the phase spaces selected by the regions and subregions is shown in \figref{fig:metJets_subRegions}.

\begin{mytable}{Selections defining the two phase-space subregions of the measurement.}{tab:metJets_subregions}{lC{35pt}C{35pt}}
	& \Mono  & \VBF\\ 
	\midrule
	leading-jet \pT [GeV] & $>120$ & $>80$\\
	leading-jet \absEta & $<2.4$ & --\\
	number of jets & $\geq1$ & $\geq2$\\
	subleading-jet \pT [GeV]& -- & $>50$\\
	\mjj [GeV] & -- & $>200$\\
	$\abs{\DeltaYjj}$ & -- & $>1$\\
	in-gap jets & -- & none\\
\end{mytable}

\myTwoFeynmanFigure{%
		Feynman diagrams for the targetted processes of the (a)~\Mono and (b)~\VBF subregion.
	}{fig:metJets_Feynman_subregions}{%
		SM/qq_to_Z_to_nunu+ISR%
	}{%
		SM/qq_to_Z_to_nunu_VBF%
}

\subsubsection{\Mono subregion}
The \Mono subregion is as inclusive as possible and requires only the presence of at least one jet in a region of the detector with good resolutions, $\absEta<2.4$.
The Feynman diagram for an example process selected in this phase space is shown in \subfigref{fig:metJets_Feynman_subregions}{a}.
The transverse momentum of the leading jet has to exceed \SI{120}{GeV}.
This prevents large imbalances between \METconst and jet \pT in single-jet events.
At the same time, the difference to the simultaneous requirement of $\METconst>\SI{200}{GeV}$ is large enough that the selection efficiency is not excessively decreased.

\subsubsection{\VBF subregion}
The \VBF subregion has kinematic requirements that explicitly select processes involving the fusion of two vector bosons (vector-boson fusion, \VBF).
An example process is shown in \subfigref{fig:metJets_Feynman_subregions}{b}.
\VBF events typically exhibit two high-momenta jets in opposite hemispheres of the detector that are more forward than jets from other processes.
This leads to large values of the invariant mass of the system, \mjj, and the absolute rapidity difference, \absDeltaYjj, of these jets~\cite{ATLAS-CONF-2020-008}.
Little hadronic activity giving rise to \textit{in-gap} jets is expected between the two jets due to the colourless exchange.

The \VBF subregion relaxes the criterion of the leading-jet \pT to \SI{80}{GeV} to avoid an excessive decrease in selection efficiency despite requiring a subleading jet with $\pT>\SI{50}{GeV}$.
No requirements on the rapidity of the jets beyond that of the jet definition described in \secref{sec:metJets_objDefinitions_jets}, $\absY<4.4$, are imposed.
Conversely, $\mjj>\SI{200}{GeV}$ and $\absDeltaYjj>1$ are required.
This enhances contributions from \VBF processes with respect to top and diboson processes with a hadronically decaying vector boson.
No further in-gap jets with $\pT>\SI{30}{GeV}$ are allowed to suppress in particular contributions from \QCD multijet processes.

\section{Uncertainties}
\label{sec:metJets_uncertainties}

Various uncertainties of statistical and systematic nature are considered for the measurements. They are described in the following.

\subsection{Statistical uncertainties}
\label{sec:metJets_statUnc}

The statistical uncertainties of the measurements in data and simulation are obtained using the bootstrap method~\cite{ATL-PHYS-PUB-2021-011}: replica datasets are created by assigning each event in the nominal dataset a unique weight sampled from a Poisson distribution with unit mean. They correspond to considering alternative datasets to the nominal one collected under the same conditions. The statistical covariance between any two observables $x$ and $y$ is then given by
\begin{equation}
	\label{eq:metJets_covariance}
	\Cov\tuple{x}{y}\coloneqq\frac{1}{N_\text{rep}}\sum_{i=1}^{N_\text{rep}}\left(x_i-\bar{x}\right)\left(y_i-\bar{y}\right).
\end{equation}
$N_\text{rep}$ is the number of replicas, $a_i$ the measured value of the observable $a\in\left\{x,y\right\}$ in replica $i$ and $\bar{a}$ the mean value of the observable. The statistical uncertainty for an observable $a$ is then given by
\begin{equation}
	\label{eq:statUncertainty}
	\sigma_a=\sqrt{\Cov\tuple{a}{a}}.
\end{equation}
The correlation between measurements of $x$ and $y$ is obtained as
\begin{equation*}
	\rho\tuple{x}{y}=\frac{\Cov\tuple{x}{y}}{\sigma_x\sigma_y}.
\end{equation*}

Thus, the bootstrap method allows determining the nominal statistical uncertainty and makes the correlation between measurements readily available.
Furthermore, the statistical uncertainty can be propagated through involved processing steps of the datasets by applying the processing to all replicas: the statistical uncertainty of a distribution after the unfolding (\cf\chapref{sec:metJets_detectorCorrection}) can be assessed by unfolding the replicas of the distribution and employing \eqref{eq:statUncertainty}.


\subsection{Experimental systematic uncertainties}
\label{sec:metJets_expSystUnc}

The systematic uncertainty on the integrated luminosity is \SI{1.7}{\%}~\cite{ATLAS-CONF-2019-021}.
Systematic uncertainties related to jets arise from the calibration of the jet energy scale and resolution~\cite{ATLAS:2020cli}, as detailed in \secref{sec:objReco_jets_reconstruction}.
Systematic uncertainties related to electrons~\cite{ATLAS:2019qmc}, muons~\cite{ATLAS:2020auj} as well as taus~\cite{ATL-PHYS-PUB-2022-044} arise related to their reconstruction, as detailed in \secsref{sec:objReco_elePhot}{sec:objReco_muons}{sec:objReco_taus}, respectively.
Systematic uncertainties to \METmeas arise related to the term taking into account unused tracks in \eqref{eq:objReco_METdefinition} and are estimated by varying the resolution parallel and perpendicular to the transverse momentum of the recoiling objects~\cite{ATLAS:2018txj}.

Objects that are misidentified can enter the region selections because of misidentification or mismeasurement.
These \textit{fake} contributions considered in the measurement are fake \METmeas from jet mismeasurement in the signal region, fake electrons from jet misidentification in \OneEJetsAM and \TwoEJetsAM as well as non-prompt muons mistaken to be prompt in \OneMuJetsAM and \TwoMuJetsAM.
Modelling these misidentifications in simulation is in general difficult because of their rarity and close relation to the detector workings. As such, variants of the data-driven matrix method~\cite{ATLAS:2022swp,Buttinger:2018abc} are used to estimate the number of events with misidentified objects, \eg by inverting selection criteria on identification or isolation, or varying jet momenta.
The fake contribution to the respective regions is less than \SI{2}{\%} for fake \METmeas and fake muons as well as less than \SI{3}{\%} for fake electrons.
Uncertainties related to the fake contributions are assessed by varying the estimation techniques. They can reach up to \SI{100}{\%} of the fake yield.

Further experimental systematic uncertainties arise related to the modelling of the average number of interactions per bunch crossing observed in data, called \textit{pileup reweighting}.

Uncertainties related to luminosity, jets, rejecting taus and pileup reweighting exhibit strong correlations between all regions.
Uncertainties related to electrons (muons) exhibit correlations between \OneEJetsAM and \TwoEJetsAM (\OneMuJetsAM and \TwoMuJetsAM).
The uncertainties related to electron identification exhibit correlations for \OneEJetsAM and \TwoEJetsAM despite invoking a loser working point for the latter because they originate from the same sources and are derived in a correlated manner.

\bigskip
\figsref{fig:metJets_expSystematics_Mono}{fig:metJets_expSystematics_VBF} show the experimental systematic uncertainties for the five measurement regions in the \Mono and \VBF subregions respectively.
Jet-related uncertainties, \ie jet energy scale (red) and resolution (orange), and the uncertainty related to the luminosity determination (purple) give large contributions in all regions.
In the lepton auxiliary measurements, uncertainties related to the reconstruction of the respective lepton (electron: green, muon: blue) or their fakes contribute significantly.
Smaller contributions can come from the uncertainty due to pileup reweighting and the fake \METmeas estimate.

\begin{myfigure}{		
		Experimental systematic uncertainties in detector-level representation for the five measurement regions in the \Mono subregion.
	}{fig:metJets_expSystematics_Mono}
	\subfloat[]{\includegraphics[width=0.49\textwidth]{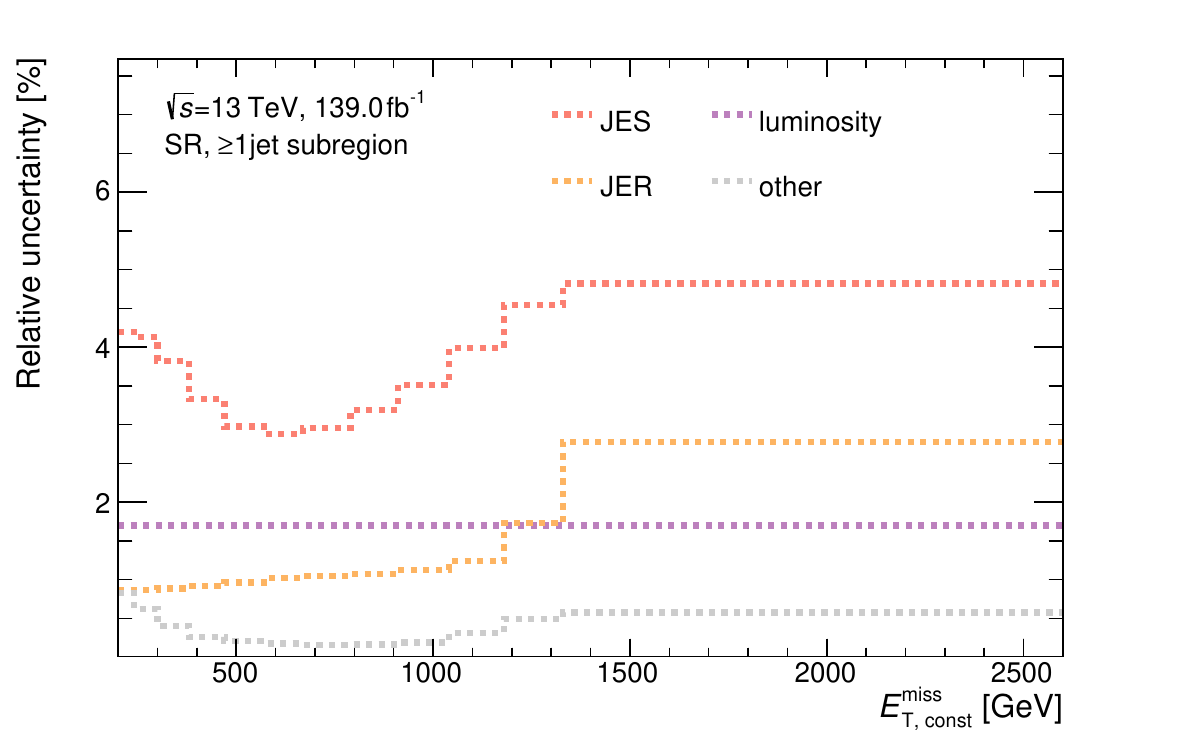}}\\
	\subfloat[]{\includegraphics[width=0.49\textwidth]{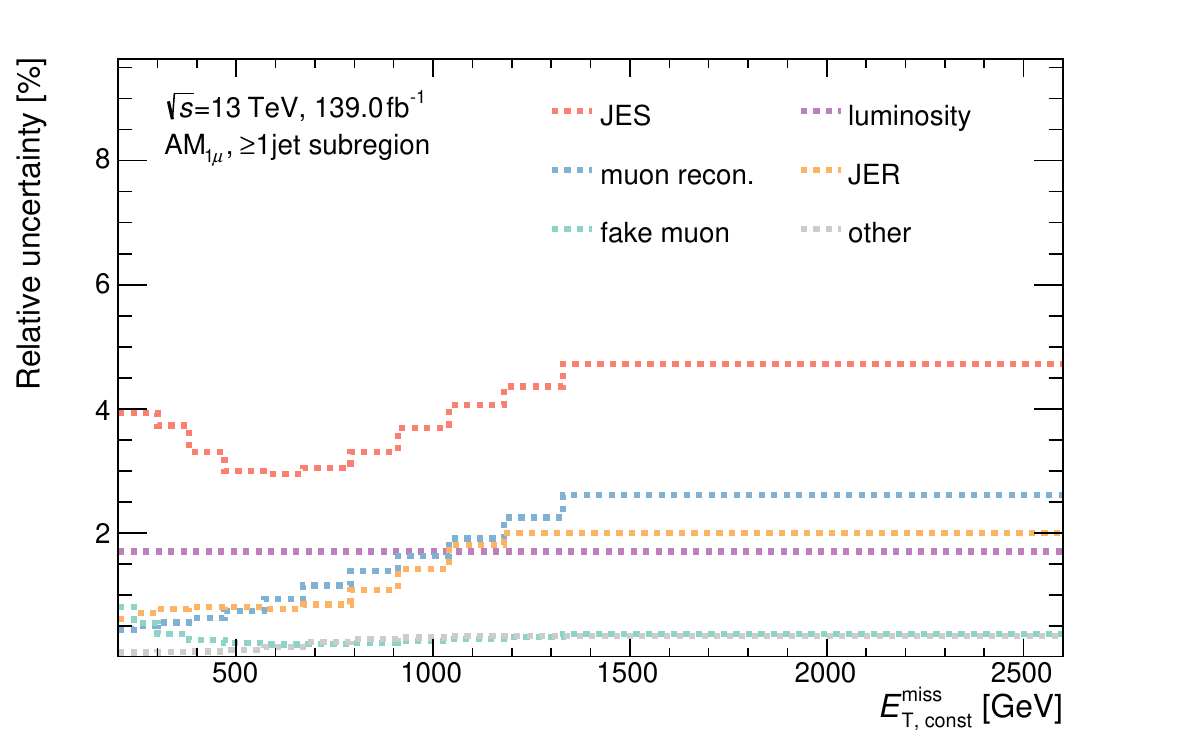}}
	\subfloat[]{\includegraphics[width=0.49\textwidth]{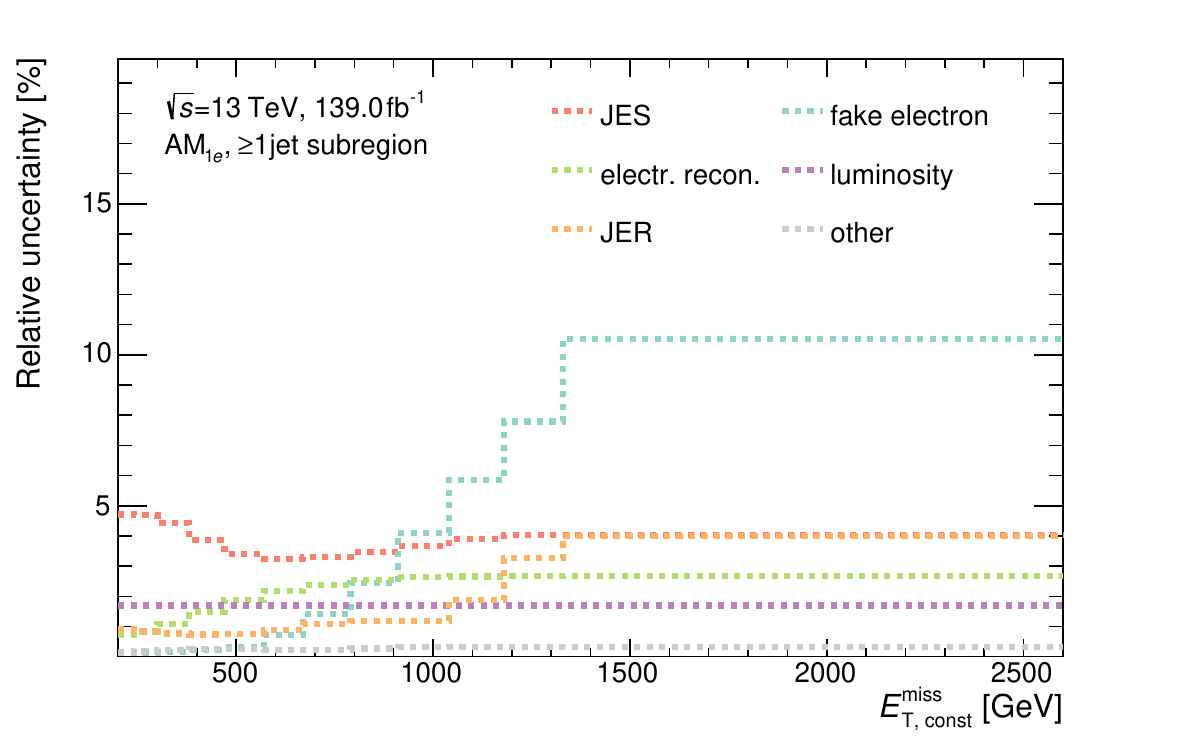}}\\
	\subfloat[]{\includegraphics[width=0.49\textwidth]{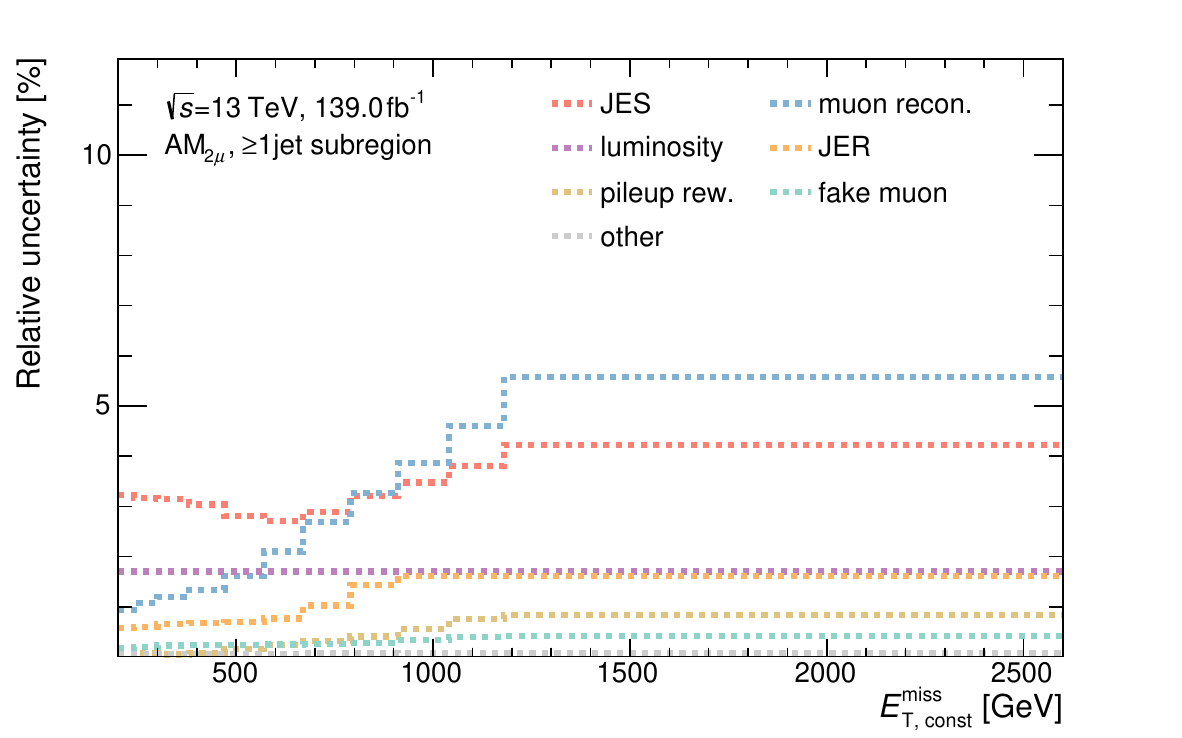}}
	\subfloat[]{\includegraphics[width=0.49\textwidth]{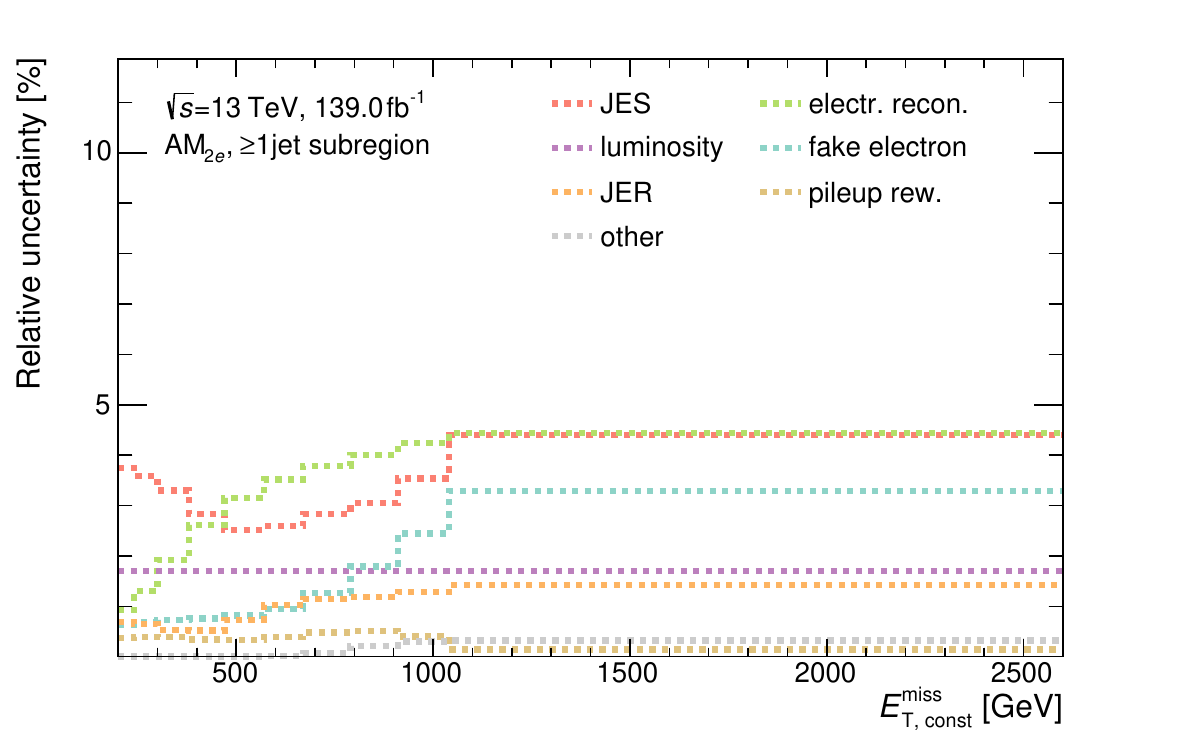}}\\
\end{myfigure}

\begin{myfigure}{		
		Experimental systematic uncertainties in detector-level representation for the five measurement regions in the \VBF subregion.
	}{fig:metJets_expSystematics_VBF}
	\subfloat[]{\includegraphics[width=0.49\textwidth]{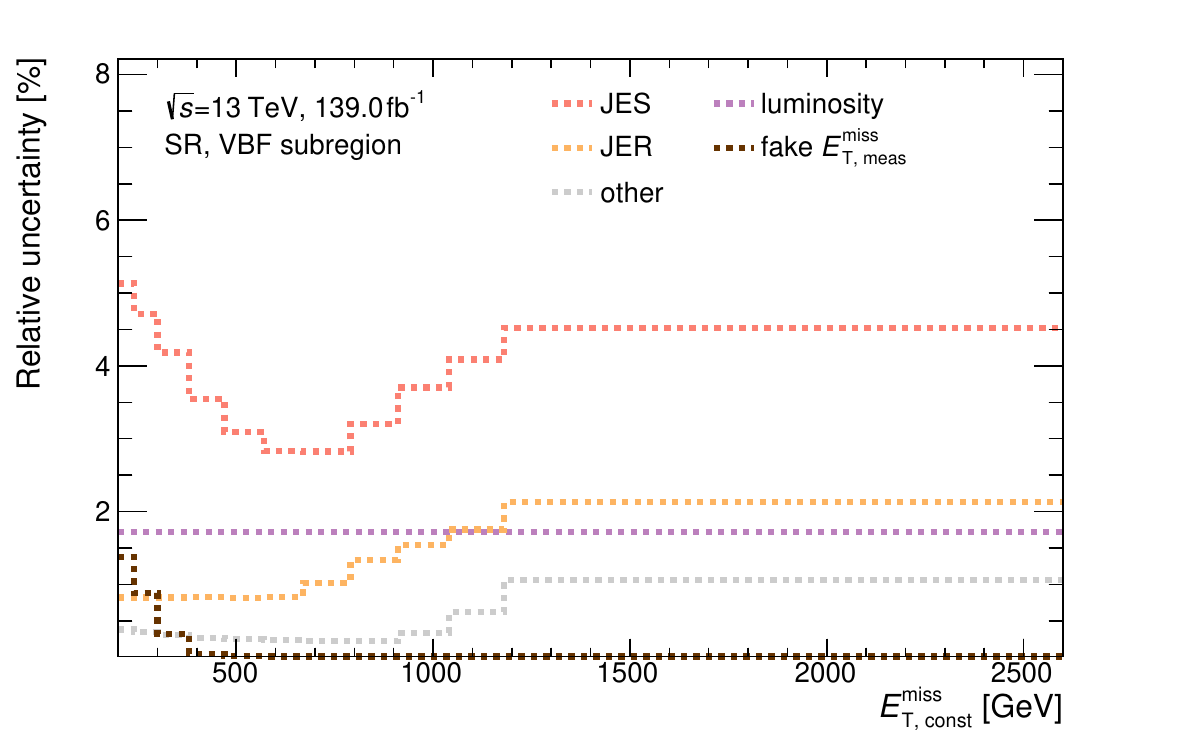}}\\
	\subfloat[]{\includegraphics[width=0.49\textwidth]{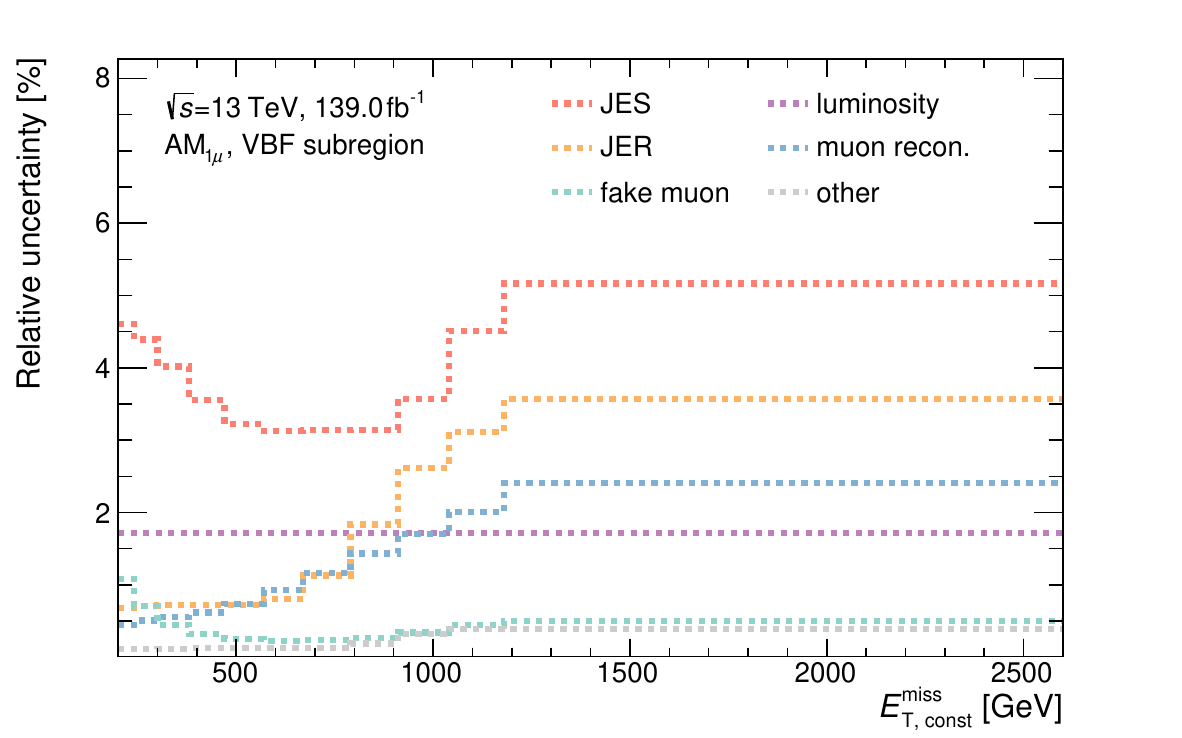}}
	\subfloat[]{\includegraphics[width=0.49\textwidth]{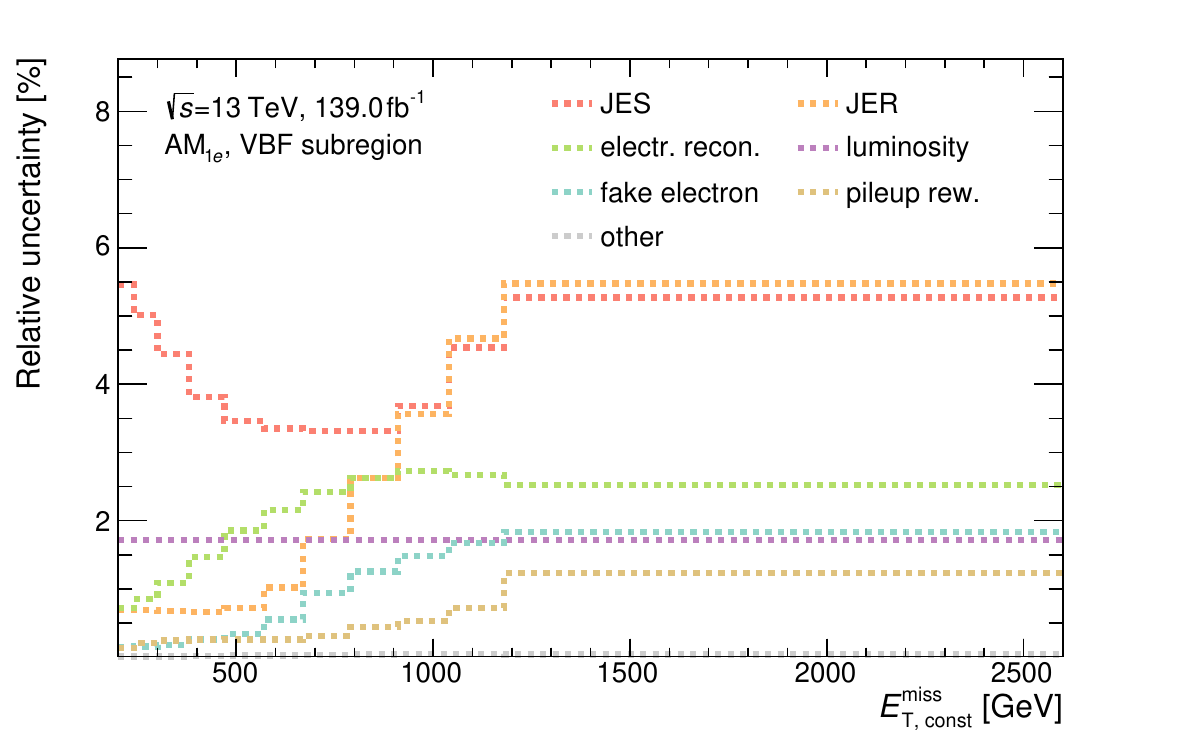}}\\
	\subfloat[]{\includegraphics[width=0.49\textwidth]{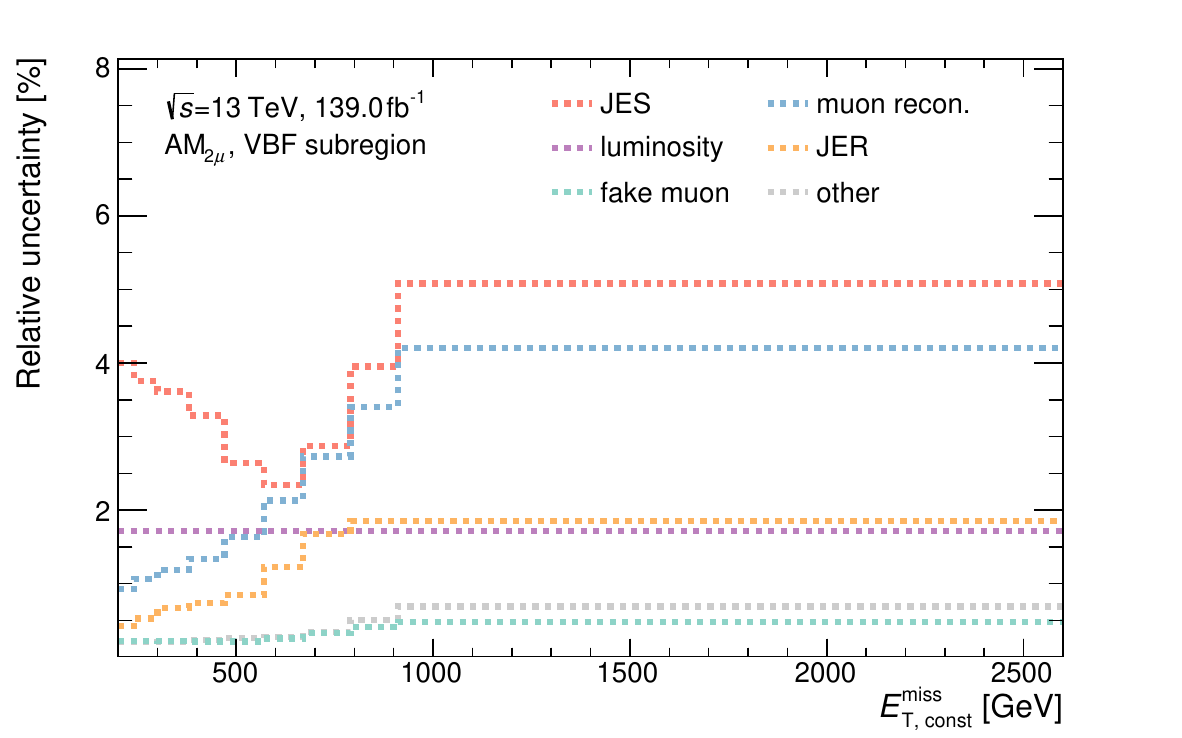}}
	\subfloat[]{\includegraphics[width=0.49\textwidth]{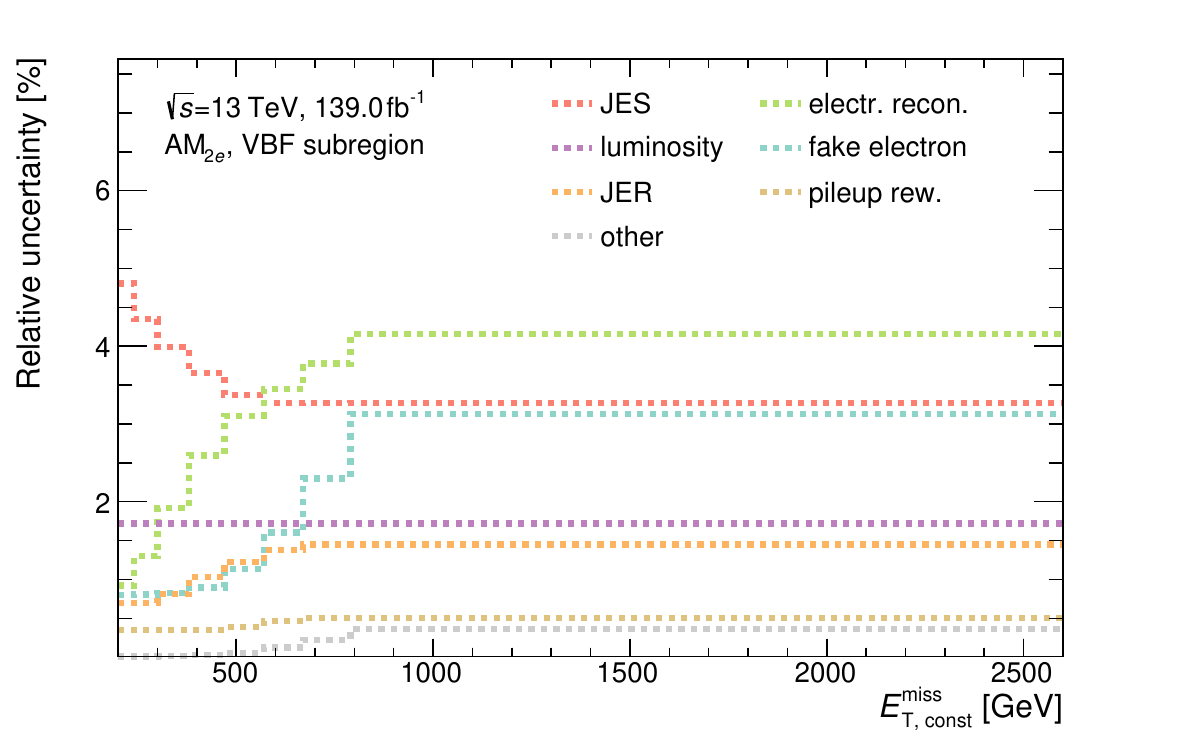}}\\
\end{myfigure}

The uncertainties in general increase with \METconst, where detector resolution and statistics for estimating the uncertainties degrade.
The uncertainty related to the jet energy scale also increases at low \METconst because of the higher uncertainty on the \JES estimate for low-momenta jets due to pileup~\cite{ATLAS:2020cli} and unlike detector responses to jets of different flavours, \ie quark- or gluon-initiated jets.
The uncertainty related to the estimate of fake \METmeas increases towards small \METconst, where the relative contribution of events with fake \METmeas to the total yield is larger.
The luminosity uncertainty impacts the yields by a constant factor.

\subsection{Theoretical systematic uncertainties}
\label{sec:metJets_theoSystUnc}

Theoretical systematic uncertainties to the simulation of \Vjets and electroweak \Vjj processes (\cf\secref{sec:metJets_MC_Vjets}) arise from uncertainties due to the used \PDF sets and scale variations~\cite{Lindert:2017olm,Lindert:2022ejn}.

\PDF uncertainties are estimated by taking the envelope of the statistical uncertainty from \NNPDFNNLO~\cite{NNPDF:2014otw} and the difference when weighting events according to the alternative \PDF sets \CTFTNNLO~\cite{Dulat:2015mca} and \MMHT~\cite{Harland-Lang:2014zoa}.

The normalisation impact from scale variations is assessed by independently taking one half, one or two times the nominal value for normalisation and factorisation scale. The envelope of the distributions is taken as uncertainty. Additionally, up- and downward variations of the strong coupling constant \alphas are taken into account.
Uncertainties of the shape of distributions stemming from scale variations are estimated as a function of the transverse momentum of the vector boson, $\pT^V$, as well as \mjj~\cite{Lindert:2017olm}.
The scale variations are treated as fully correlated across the different \Vjets processes and regions, assuming that the \Vjets processes are quite similar to each other for $\METconst\approx\pT^V\gg m_V$ where $m_V$ denotes the mass of the vector boson~$V$.
The impact of residual non-correlation effects is incorporated into differential scaling factors with respect to the \Zjets process according to the recommendation in \refcite{Lindert:2017olm}. Varying the resummation scale by factors one half and two gives an estimate of the impact of parton showering on the requirements on \SdPhi in the \VBF subregion and on \DeltaPhiMin.

Scale uncertainties can reach up to \SI{30}{\%} of the \Vjets yield at particle level, \PDF uncertainties are typically below \SI{5}{\%}.

\bigskip
For top, diboson and triboson processes (\cf\secref{sec:metJets_MC_DiTriboson}), theoretical uncertainties from choice of \PDF set and scale normalisation are estimated as described before for \Vjets. \PDF sets at \NLO are used for top processes to match the corresponding precision. Renormalisation and factorisation scale lead to an uncertainty of up to \SI{40}{\%} of the process's yield at particle level, the uncertainty from the strong coupling constant can exceed \SI{100}{\%}. \PDF uncertainties are typically below \SI{10}{\%}.

For single-top $tW$ processes, the difference between using the diagram removal rather than subtraction scheme for the interference between \ttbar and $tW$ processes~\cite{Frixione:2008yi} is taken as an additional uncertainty. This uncertainty amounts to less than \SI{2}{\%} of the total yield in most regions at particle level. In \OneLJetsAMs, where contributions from top processes are most dominant, it reaches up to \SI{11}{\%}.

\section{Detector-level comparisons}
\label{sec:metJets_detLevelResults}

After defining the selection of the \METjets measurement in the previous sections, comparisons between measured data and simulated \SM predictions in detector-level representation can be made.
They are shown in \figref{fig:metJets_detLevelResults} for a selection of the five regions in the two subregions as a function of \METconst.
The top panels give the total yield of data (black dots) and \SM simulation (blue crosses) with their respective statistical uncertainties.
The quadrature sum of experimental systematic and statistical uncertainties is displayed as red shaded areas.
Theoretical uncertainties are not propagated to detector level and are only shown in particle-level representation in \secref{sec:detCorr_partLevelResults}.
In the top panels, the contributions to the simulated yield from the different \SM processes are marked by filled areas.
In the bottom panels, the ratio of data to \SM simulation is shown.

\begin{myfigure}{
		Total yield and ratio of data to simulation at the detector level for the five measurement regions in the (left) \Mono and (right) \VBF subregion.
		Black dots (blue crosses) denote the measured data (total yields in \SM simulation) with their statistical uncertainty.
		The red shaded areas correspond to the total experimental uncertainty.
		In the top panels, the filled areas correspond to the different contributing processes.
	}{fig:metJets_detLevelResults}
	\subfloat[]{\includegraphics[width=0.49\textwidth]{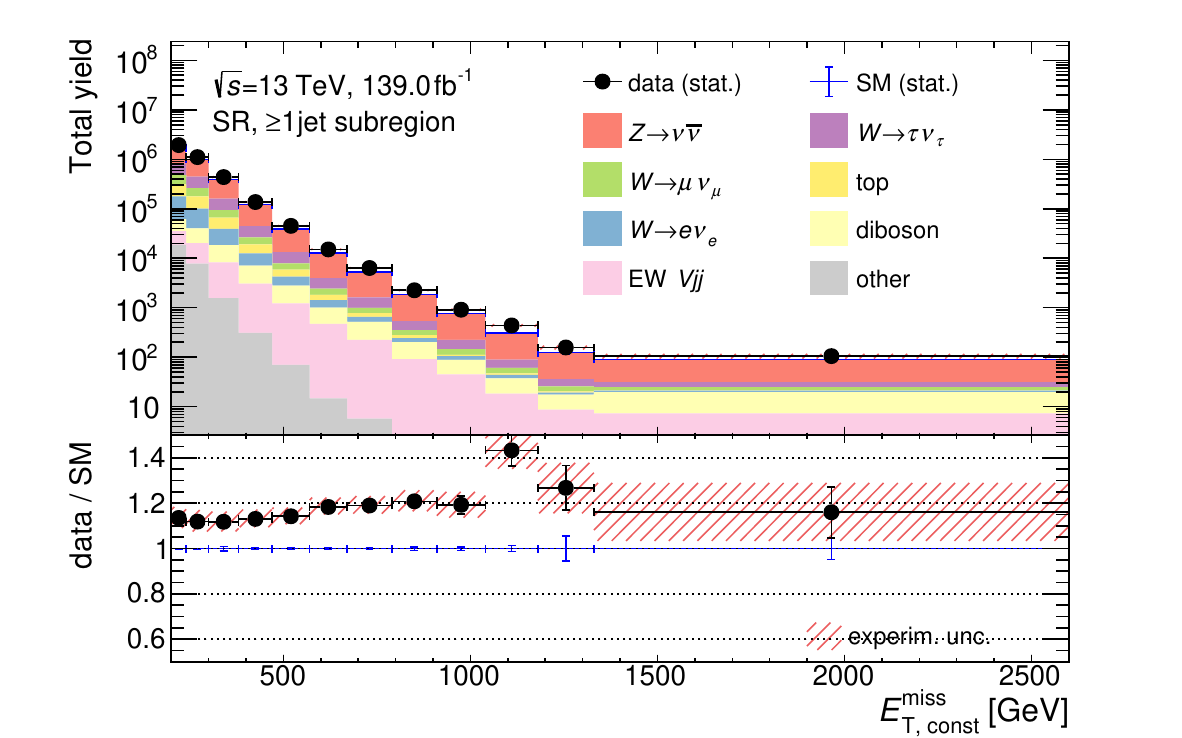}}
	\subfloat[]{\includegraphics[width=0.49\textwidth]{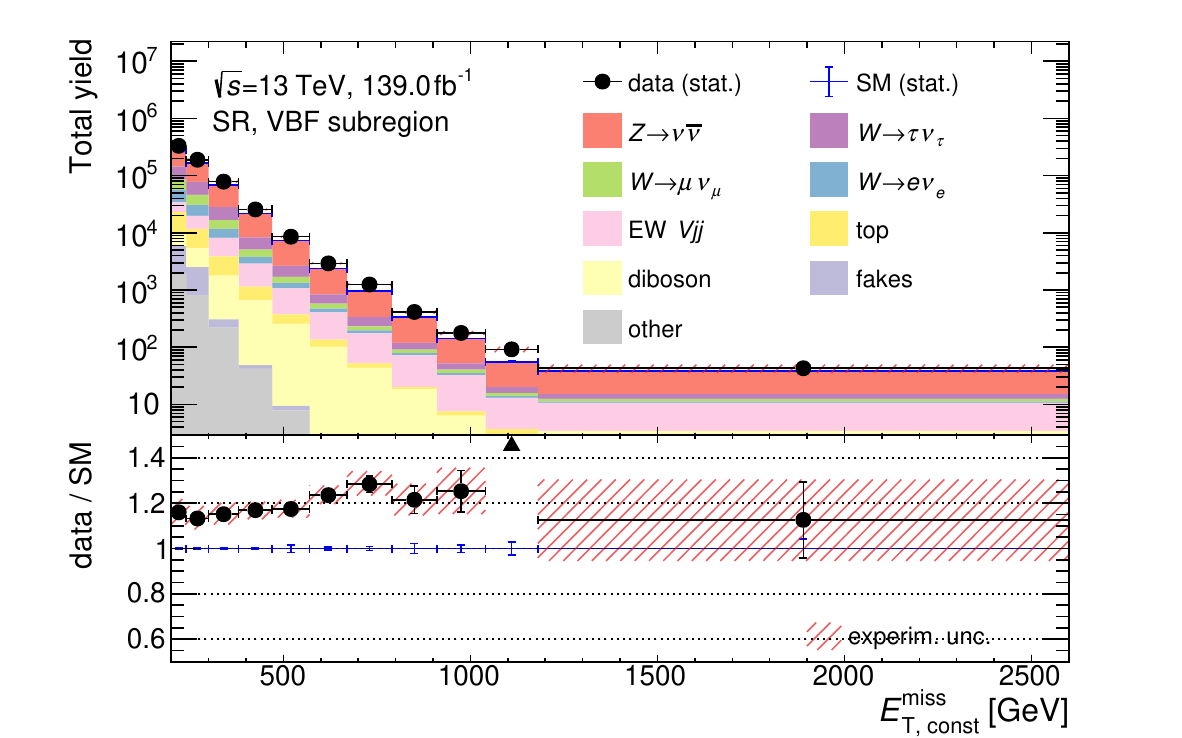}}\\
	\subfloat[]{\includegraphics[width=0.49\textwidth]{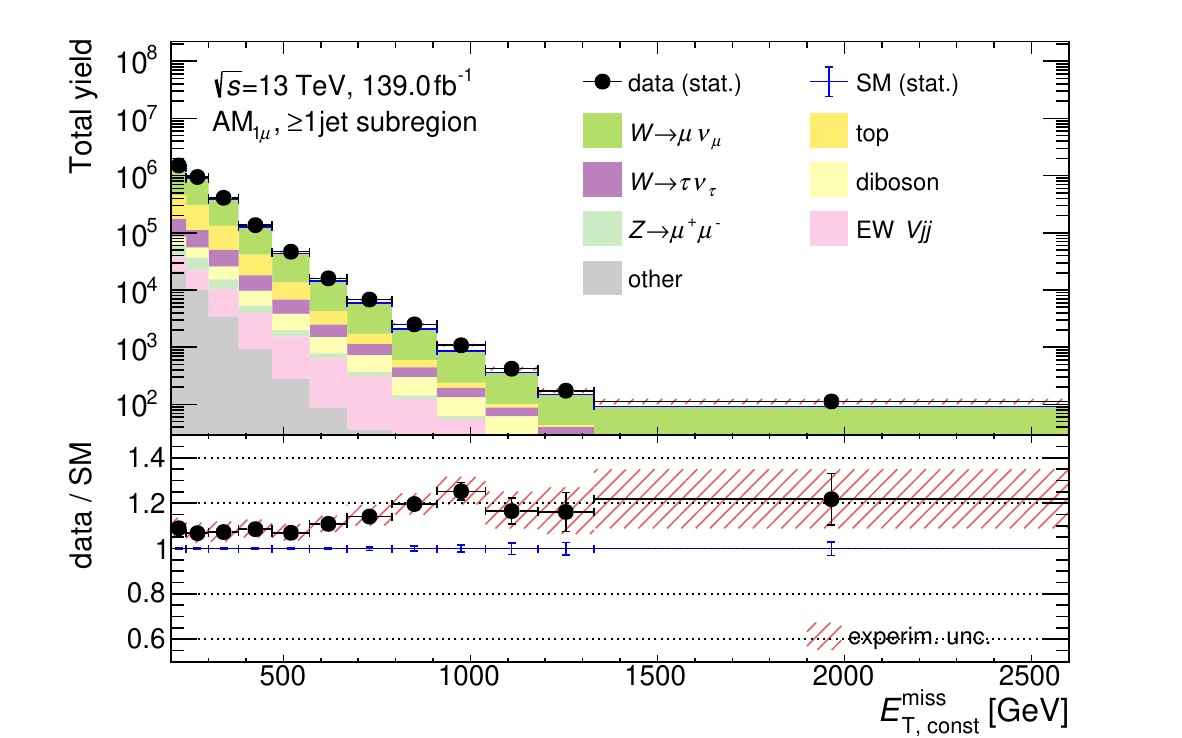}}
	\subfloat[]{\includegraphics[width=0.49\textwidth]{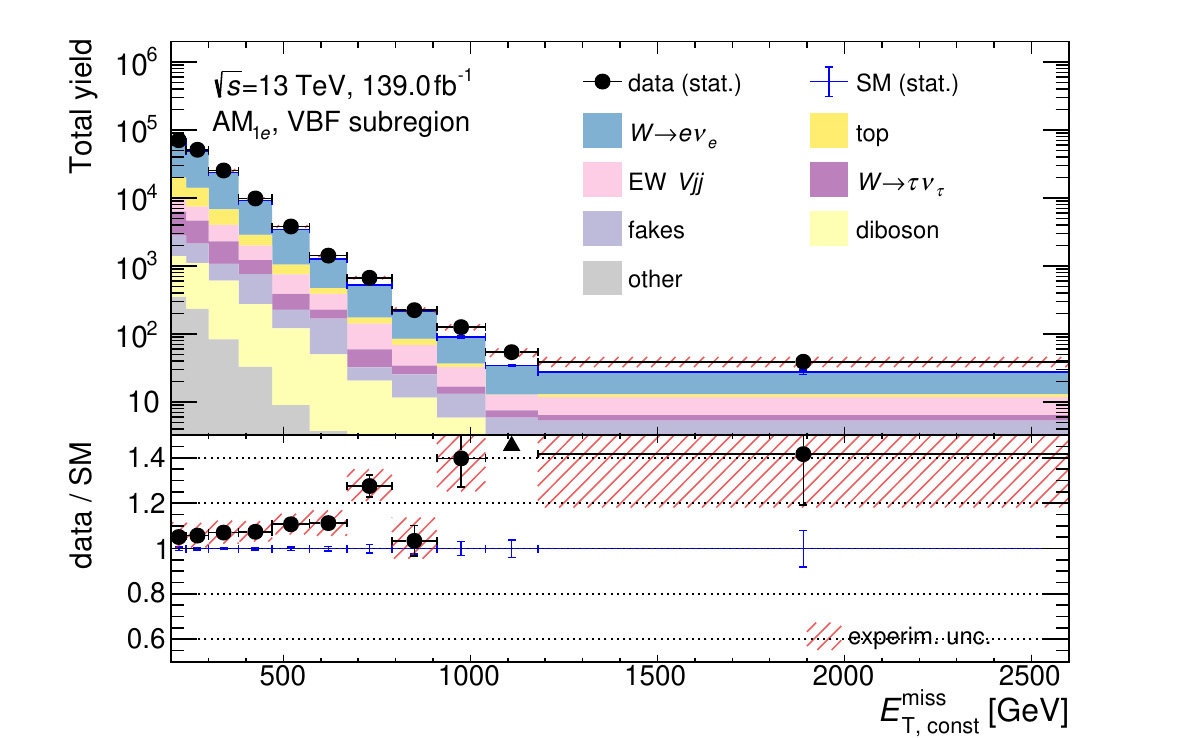}}\\
	\subfloat[]{\includegraphics[width=0.49\textwidth]{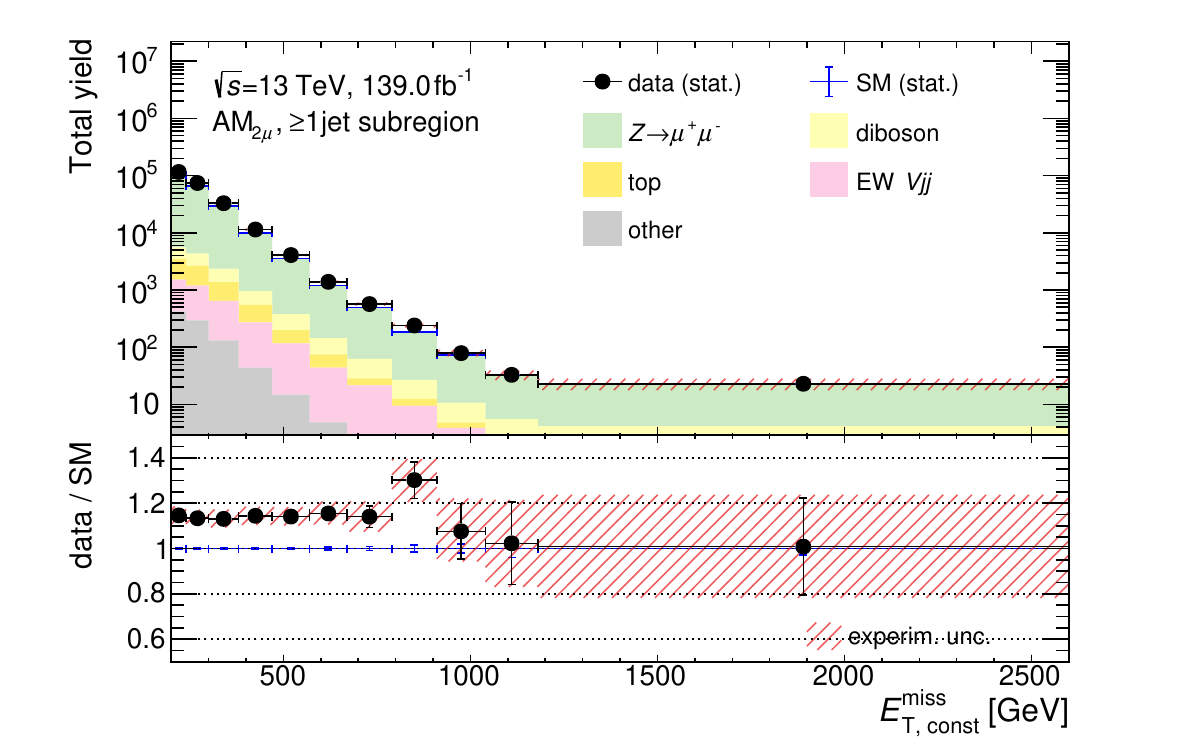}}
	\subfloat[]{\includegraphics[width=0.49\textwidth]{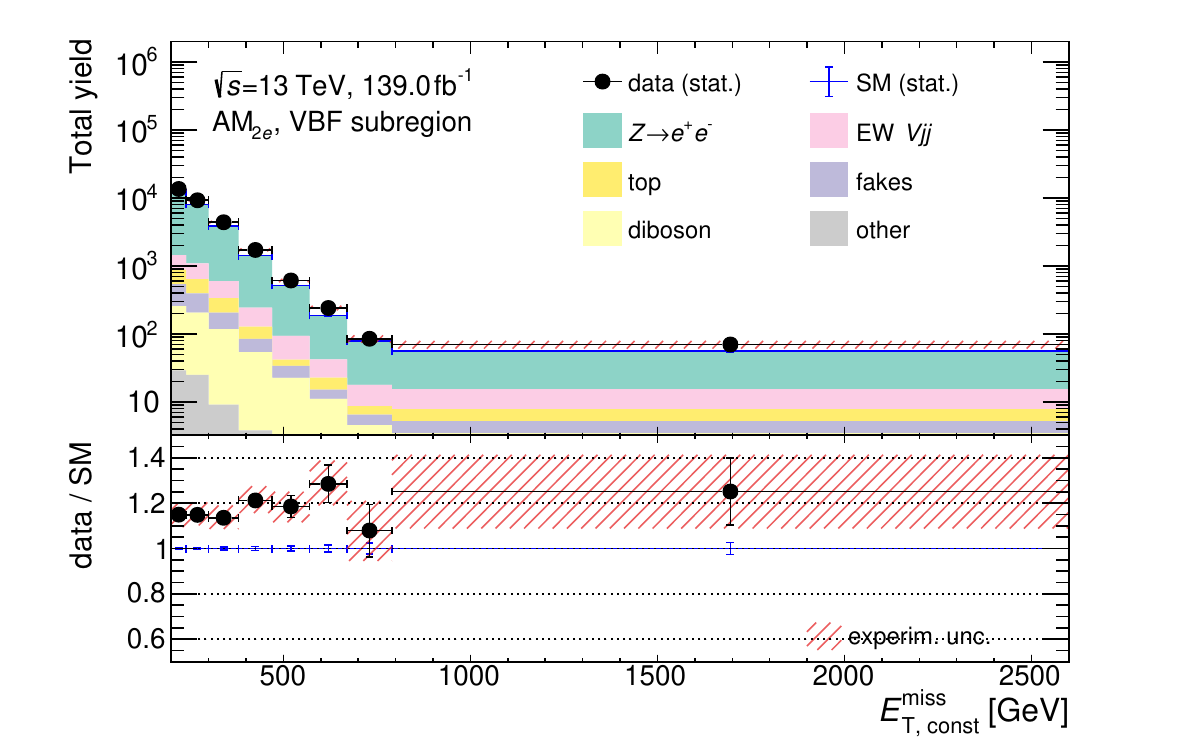}}\\
\end{myfigure}

\bigskip
Steeply falling spectra as a function of \METconst can be observed.
This illustrates that as a consequence of the parton distribution functions (see \secref{sec:MCEG_hardProcess}) in general the probability for a parton to have a specific momentum decreases the larger the momentum becomes.
Therefore, collisions involving larger parton momenta -- and consequently larger \METconst -- are rarer.

The composition of the \SM contributions is as expected from the analysis strategy described in \secref{sec:metJets_strategy}.
The dominant \SM contribution in the signal regions in the \Mono and \VBF subregion come from \Znunujets events.
The main subdominant \SM contributions arise from $W\to\ell\nu$ processes, either as \Wjets or with the $W$ boson originating from the decay of a top quark.
Minor contributions come from diboson and electroweak~(\EW) \Vjj processes.

In \OneLJetsAMs, the dominant \SM contributions stem from $W\to\ell\nu$ processes either as \Wjets or with the $W$ boson originating from the decay of a top quark.
Minor contributions stem from diboson and electroweak \Vjj processes.
In \OneEJetsAM, despite the stricter kinematic cuts compared to \OneMuJetsAM to suppress fake contributions, events with fake electrons give a non-negligible contribution.
In \OneMuJetsAM, \Zmumujets events where one of the muons is outside the detector acceptance contribute.

\TwoLJetsAMs are dominated by \Zlljets processes.
Minor contributions come from top quark decays involving two leptons, diboson and electroweak \Vjj processes.
In \TwoEJetsAM, similar to \OneEJetsAM, events with fake electrons give a non-negligible contribution.

In general, the relative fraction of the major contributions is constant across the subregions and as a function of \METconst.
The only exception is the contribution from electroweak \Vjj processes, which is dominated by \VBF production and therefore enhanced in the \VBF subregion (see the right plots in \figref{fig:metJets_detLevelResults}).
This is the case in particular when \METconst is large.
The statistics are about an order of magnitude smaller in the \VBF than in the \Mono subregion due to the stringent fiducial requirements in the former (\cf\secref{sec:metJets_subregions}).

\bigskip
Comparing simulated \SM prediction and data, the shape of the distributions is described generally well by the simulation, giving approximately constant ratios of data to simulation.
The normalisation is underestimated by $10-\SI{20}{\%}$ in all regions.
This underestimation is smaller in \OneLJetsAMs because they have a larger contribution of \ttbar processes:
the generated cross section for \ttbar processes tends to be overestimated~\cite{ATLAS:2022aof} while the \Vjets cross section tends to be underestimated~\cite{Aad:2019hga}.
The normalisation difference observed in \figref{fig:metJets_detLevelResults} is not covered by statistical or experimental systematic uncertainties.

Beyond this normalisation difference, there is a significant rise of the data over the prediction at $\METconst=\SI{1100}{GeV}$ in the signal regions (\cf\subfigsref{fig:metJets_detLevelResults}{a}{b}).
This could originate from extreme fluctuations within the uncertainties, unaccounted systematic effects or non-\SM contributions.
Whether these differences between measured data and \SM prediction can be accounted for everywhere simultaneously given the uncertainties, in particular when including the theoretical systematic uncertainties, is investigated in \secref{sec:interpretation_SM}.

There are also generally larger differences between data and prediction at large \METconst in the \VBF subregion.
These can be attributed to statistical fluctuations due to the stringent fiducial requirements (\cf\secref{sec:metJets_subregions}) and consequently smaller statistics in this extreme phase space.

\bigskip
In the next chapter, the distributions are corrected for detector effects.
This provides them in the particle-level representation and allows determining the actual differential cross sections of the investigated phase spaces including theoretical systematic uncertainties.
\Chapter{Approaching the particle level}{Correction for detector effects}{%
	Wo bleibt Ihr Einsatz, wo bleibt der Wille?\\
	Lasern Sie die Welt und dann weg mit der Brille.%
}{Christian Hartmann \textit{et al.}}{Deichkind:2015dsg}
\label{sec:metJets_detectorCorrection}

In the previous chapter, an analysis strategy and selection for the \METjets final state was described.
Results at that stage are in detector-level representation.
In this chapter, they shall be brought into particle-level representation.
This allows comparing the measured data to \SM and \BSM predictions without needing to simulate the detector response.
This is particularly useful when updated \SM predictions become available or a new \BSM model shall be investigated, as discussed in \chapref{sec:analysisPreservation}.

\secref{sec:detCorr_general} details how the correction for detector effects in general is performed.
This is applied to the phase space selected by the \METjets measurement in \secref{sec:detCorr_metJets}.

\section{Correction for detector effects}
\label{sec:detCorr_general}

In the following, the general procedure for correcting detector effects is discussed.
Types of detector effects that have to be taken into account are described in \secref{sec:detCorr_detEffects}.
Features related to the \textit{response matrix}, the matrix capturing the detector response function, are pointed out in \secref{sec:detCorr_responseMatrix}.
Technical details on the method that can be employed to remove detector effects using the response matrix, called \textit{unfolding}, are given in \secref{sec:detCorr_unfoldingMethod}.

\subsection{Types of detector effects}
\label{sec:detCorr_detEffects}
Events generally can be categorised for whether they pass the detector-level selection and for whether they pass the particle-level selection.
\textit{Particle level} thereby refers to quantities of detector-stable particles originating from Monte-Carlo generation (\cf\secref{sec:level_interpretation}).
Events that pass particle-level as well as detector-level selection are called \textit{matched}.
Detector effects impacting which events are selected at detector level comprise~\cite{Blobel:2002pu}:
\begin{itemize}
	\itembf{Efficiency}
	Events can pass the particle-level selection but fail the detector-level selection.
	This can happen when a physical object in an event, \eg an electron, is failed to be reconstructed.
	The \textit{efficiency} corresponds to the fraction of events that pass the particle-level selection that also pass the detector-level selection and are therefore matched.
	The efficiency depends on the true value $x$ of the observable.

	\itembf{Fiducial fraction and backgrounds}
	Events can pass the detector-level selection but fail the particle-level selection.
	This can happen when a physical object in the event is erroneously reconstructed as an object of another type, \eg a jet is reconstructed as an electron (\cf\secref{sec:metJets_expSystUnc}).
	The \textit{fiducial fraction} corresponds to the fraction of events that pass the detector-level selection that also pass the particle-level selection and are therefore matched.
	Events that are not matched are not part of the targetted particle-level phase space and are therefore considered to be background.

	\itembf{Transformations}
	The detector generally does not measure the true value $x$ of an observable but instead a related quantity $y$, \eg caused by a systematic shift.

	\itembf{Resolutions}
	The resolution of physical observables, \eg transverse momentum or rapidity, is limited in the detector.
	This smears the measured quantity $y$.
	Contrary to transformations, which capture deterministic effects, resolutions give a statistical relation between $x$ and $y$.
\end{itemize}

All of these detector effects can affect every single event, but it is a posteriori impossible to know to which amount a specific event was influenced.
Contrary to calibrations which adjust directly the properties of physics objects, detector corrections are therefore applied statistically to ensembles of events.

Let $f(x)$ be a distribution in an observable $x$ of events passing the particle-level selection, \ie in particle-level representation.
Let further $g(y)$ be the corresponding distribution in detector-level representation, \ie a distribution related to $f$ in a related observable $y$ of events passing the detector-level selection.
The relation between $f(x)$ and $g(y)$ is then given by
\begin{equation}
	\label{eq:detCorr_folding}
	\int A(y,x)f(x)\mathrm{dx}+b(y) = g(y).
\end{equation}
$A(y,x)$ is the response function and encodes the detector effects of efficiency, transformations and resolutions.
Fiducial-fraction effects are either incorporated in the response function as well or separately treated as a distribution of backgrounds $b(y)$.
Transferring $g(y)$ into $f(x)$ is called \textit{correcting for detector effects}.

For discrete measurements, \ie finite event counts as a function of intervals of an observable (\textit{bins}), \eqref{eq:detCorr_folding} can be transformed into the matrix equation~\cite{Schmitt:2016orm}
\begin{equation}
	\label{eq:detCorr_folding_matrix}
	A\vv x+\vv b=\vv y.
\end{equation}
The distribution $f(x)$ is discretised into vector~$\vv x$ of size $M_x$ giving the event counts in the bins. Similarly, $g(y)$ and $b(y)$ become vectors~$\vv y$ and~$\vv b$, respectively, of size $M_y$.
$A$ is the $M_x\times M_y$ \textit{response matrix} with entries $A_{ij}$ which give the probability to find an event produced in bin $i$ measured in bin $j$.

\subsection{Features of the response matrix}
\label{sec:detCorr_responseMatrix}

Particle simulation provides the particle-level as well as the detector-level information for each event and allows capturing efficiency, transformation and resolution effects.
The response matrix $A$ is therefore assessed, most commonly, in simulation.

The following quantities help to visualise features of the response matrix, the first two relating the probabilities to pass the selections at particle and reconstruction level, the second two relating the migration between bins:
\begin{itemize}
	\item The \textit{efficiency}, as already introduced in the previous section, is the fraction of events passing the particle-level selection in bin~$i$ that are also matched.
	It can be calculated from the response matrix as $\varepsilon_i=\sum_j A_{ij}$.
	A small efficiency indicates that the detector-level event selection is too stringent or the object identification or event reconstruction deficient.
	This causes many events that pass the particle-level selection to fail the detector-level selection.
	\item The \textit{fiducial fraction}, as already introduced in the previous section, is the fraction of events passing the detector-level selection in bin~$i$ that are also matched.
	A small fiducial fraction indicates that the detector-level event selection is not stringent enough.
	This allows many events to pass the detector-level selection, although they are not in the targetted particle-level phase space.
	\item The \textit{purity} of the migration, $p_i=\frac{A_{ii}}{\sum_jA_{ij}}=\frac{A_{ii}}{\varepsilon_i}$, is the fraction of events with respect to the total efficiency~$\varepsilon_i$ that are produced in bin~$i$ and reconstructed in the same bin.
	A small purity indicates that due to transformations and resolutions the particle- and detector-level values of the investigated observable of an event differ significantly with respect to the choice of bin widths for $\vv x$.
	\item The \textit{stability}, in opposition to the previous bullet, is the fraction of events with respect to the fiducial fraction that are reconstructed in bin $j$ and produced in the same bin.
	A small stability indicates that due to transformations and resolutions the particle- and detector-level values of the investigated observable of an event differ significantly with respect to the choice of bin widths for $\vv y$.
\end{itemize}

\subsection{Unfolding}
\label{sec:detCorr_unfoldingMethod}

The procedure to solve \eqref{eq:detCorr_folding_matrix} for the vector of interest $\vv x$ is called \textit{unfolding}.
\eqref{eq:detCorr_folding_matrix} obtains the characteristic of an inverse problem if $M_x=M_y$ and $\vv y$ is redefined as $\vv y \to\vv y'\coloneqq\vv y-\vv b$.
The statistical nature of the experiment, however, implies that the observed event counts follow a Poisson distribution and in general do not match perfectly the expectation~\cite{Blobel:2002pu}.
These statistical fluctuations to $\vv y'$ give at best rise to enormous statistical uncertainties if \eqref{eq:detCorr_folding_matrix} is solved by matrix inversion~\cite{Schmitt:2016orm,DAgostini:2010hil}.
At worst, the response matrix $A$ is not invertible at all due to the statistical fluctuations.

A more physical approach to obtain $\vv x$ from \eqref{eq:detCorr_folding_matrix} is therefore to impose that $\vv x$ has to be sufficiently smooth by a procedure called \textit{regularisation}~\cite{DAgostini:2010hil}.
Regularisation still makes use of the redefinition $\vv y'\coloneqq\vv y-\vv b$.
$M_x=M_y$ is not strictly required.
By regularisation, the impact of statistical fluctuations in the measured distribution $\vv y'$ on the detector-corrected result $\vv x$ is reduced.
Still, the number of events in each bin should be sizeable such that the degrees of freedom in the detector response can be accounted for.
Complex, composite observables, such as the missing transverse energy \MET, typically need event counts larger than 20 in each bin for a stable algorithm performance.
The binning of the distributions to be unfolded was chosen accordingly in the \METjets measurement.

One approach for regularisation is iterative Bayesian unfolding~\cite{Shepp:1982mlr,DAgostini:1994fjx}.
This is the approach adopted for the \METjets measurement.
Hereby, the unfolding result $x_j^{n+1}$ of iteration $n+1$ is obtained from the previous iteration according to~\cite{Schmitt:2016orm}
\begin{equation*}
	x_j^{n+1}=x_j^n\sum_{i=1}^{M_x}\frac{A_{ij}}{\varepsilon_j}\frac{y'_i}{\sum_{k=1}^{M_y} A_{ik}x_k^n}.
\end{equation*}
This corresponds to iteratively improving the unfolding result by reweighting the input distribution $\vv y'$ (\textit{prior}) to the unfolded data $\sum_{k=1}^{M_y} A_{ik}x_k^n$ (\textit{posterior}) of the previous iteration.
With each step, the bias from the choice of the original prior is reduced and the statistical uncertainty increased.
For infinitely many steps, the result of the approach using matrix inversion is recovered if the problem is not ill-posed.
For an intermediate number of steps, the true distribution $\vv x$ is obtained in good approximation.

Alternative methods for regularisation are for example Tikhonov regularisation~\cite{Tikhonov:1963reg}, Iterative Dynamically-Stabilised unfolding~\cite{Malaescu:2009dm} and Gaussian-Process unfolding~\cite{Bozson:2018asz}.

\bigskip
Different tests can be performed which investigate the success and stability of the unfolding procedure.
Important tests are detailed in the following.

\subsubsection{Closure test}

A simple test to ensure the employed unfolding method is stable is the \textit{closure test}.
Simulated detector-level distributions are used as pseudodata.
These are unfolded using the nominal response matrix and compared to the generated distributions at particle-level.
If both were also used for constructing the response matrix, it should \textit{close} trivially, \ie the distributions should always be in perfect correspondence.

\subsubsection{Stress test}

The response matrix is commonly assessed in simulation, which might exhibit considerable differences to data.
Discrepancies in normalisation in the unfolded observables are readily resolved by iterative unfolding methods~\cite{Schmitt:2016orm}.
Discrepancies in the shape of distributions of the unfolded observables can lead to a bias of the result.
Similarly, information is lost as the unfolding is only performed in one observable.
Discrepancies in the shape of distributions in variables that are not unfolded (\textit{hidden variables}) can also lead to a bias of the result.

These biases are assessed by a \textit{stress test}:
Generated events at particle level are assigned weights according to smooth functions such that the reweighted distribution for the simulated events at detector level matches the data.
The reweighted distribution at detector level is then used as pseudodata and unfolded using the un-reweighted response matrix.
The difference between the unfolded, reweighted distribution and the particle-level, reweighted distribution for generated events quantifies the stability of the unfolding procedure against shape differences.

\subsubsection{Signal-injection test}

Contributions from \BSM models could alter the measured distributions at detector level with respect to the \SM prediction at detector level.
They are, however, not accounted for in the response matrix.
Unfolding the measured distributions at detector level erroneously assuming no \BSM contributions might therefore not correctly reproduce the $\SM+\BSM$ distributions at particle level.

This is assessed in a \textit{signal-injection test}:
\SM distributions at detector level are overlaid with example \BSM distributions at detector level prior to unfolding and treated as pseudodata.
The overlaid distributions are then unfolded using the nominal response matrix and compared to overlaying the \BSM distributions at particle level to the \SM distributions at particle level.

\section{Application to the \METjets phase space}
\label{sec:detCorr_metJets}

The general procedure to correct a measurement for detector effects was explained in the previous section.
In the following, this is applied to the phase space selected by the \METjets measurement as described in \chapref{sec:metJets}.

The calculation and features of the response matrix are described in \secref{sec:detCorr_metJets_responseMatrix}.
\secref{sec:detCorr_metJets_unfoldingMethod} gives technical details on the employed unfolding method.
Systematic uncertainties related to this method are discussed in \secref{sec:detCorr_unfUncertainties}.
The results of the\linebreak \METjets measurement in particle-level representation are given in \secref{sec:detCorr_partLevelResults}.

\subsection{Response matrix}
\label{sec:detCorr_metJets_responseMatrix}

The \METjets phase space selected in \chapref{sec:metJets} is represented as event counts as a function of discrete intervals of \METconst.
The detector response can therefore be encoded in the form of a response matrix.
The response matrix is assessed as follows:
Events for the dominant \SM contributions are generated with Monte-Carlo event generators, as outlined in \secref{sec:metJets_MC}.
This gives events in the inclusive final state (\cf\chapref{sec:level_interpretation}).
Event counts in particle-level representation are obtained by applying the selection criteria described in \chapref{sec:metJets} to the objects in the inclusive final state following the definitions given in \secref{sec:objReco_particleLevel}.
At the same time, detector simulation is run on the inclusive final state to obtain detector signals.
The selection criteria described in \chapref{sec:metJets} are applied to these detector signals to obtain event counts in detector-level representation.
For each event it is therefore known whether it passes the particle-level selection, the detector-level selection or both.
From these events the response matrix is populated.
Contributions of events completely failing the particle- or detector-level selection are included in the response matrix in specific bins to allow a more stable unfolding procedure.

With this approach, most contributions passing the detector-level selection but failing the particle-level selection can be included in the definition of the response matrix as part of the fiducial fraction.
Only fake contributions (\cf\secref{sec:metJets_expSystUnc}) give a background that cannot be modelled well in simulation because of their rarity and close relation to the detector workings.
Fake contributions can therefore not be accounted for with the above approach in the response matrix.
These background contributions are subtracted in \eqref{eq:detCorr_folding_matrix}.

\bigskip
\figref{fig:detCorr_responseMatrix} shows an example response matrix for the signal region in the \Mono subregion.
The bin widths are chosen such that the number of events is larger than 20 or the purity is larger than \SI{60}{\%}.
The former is a requirement of the unfolding method (\cf\secref{sec:detCorr_unfoldingMethod}).
The latter ensures that the number of events migrating between bins is small, allowing for a more stable unfolding procedure.
Most events are reconstructed such that they contribute to the same bin at detector level they are produced in at particle level.
This leads to a prominent diagonal in the response matrix.
Off-diagonal elements are subdominant, with a tendency towards underestimating the particle-level \METconst during the reconstruction.

\begin{myfigure}{%
		Response matrix for the signal region in \METconst in the \Mono subregion.
		Given is the fraction of events generated at a specific particle-level \METconst that is reconstructed at another specific detector-level \METconst.
		The fraction missing from \SI{100}{\%} in a row (column) corresponds to events failing the detector-level (particle-level) selection.
	}{fig:detCorr_responseMatrix}
	\includegraphics[width=0.8\textwidth]{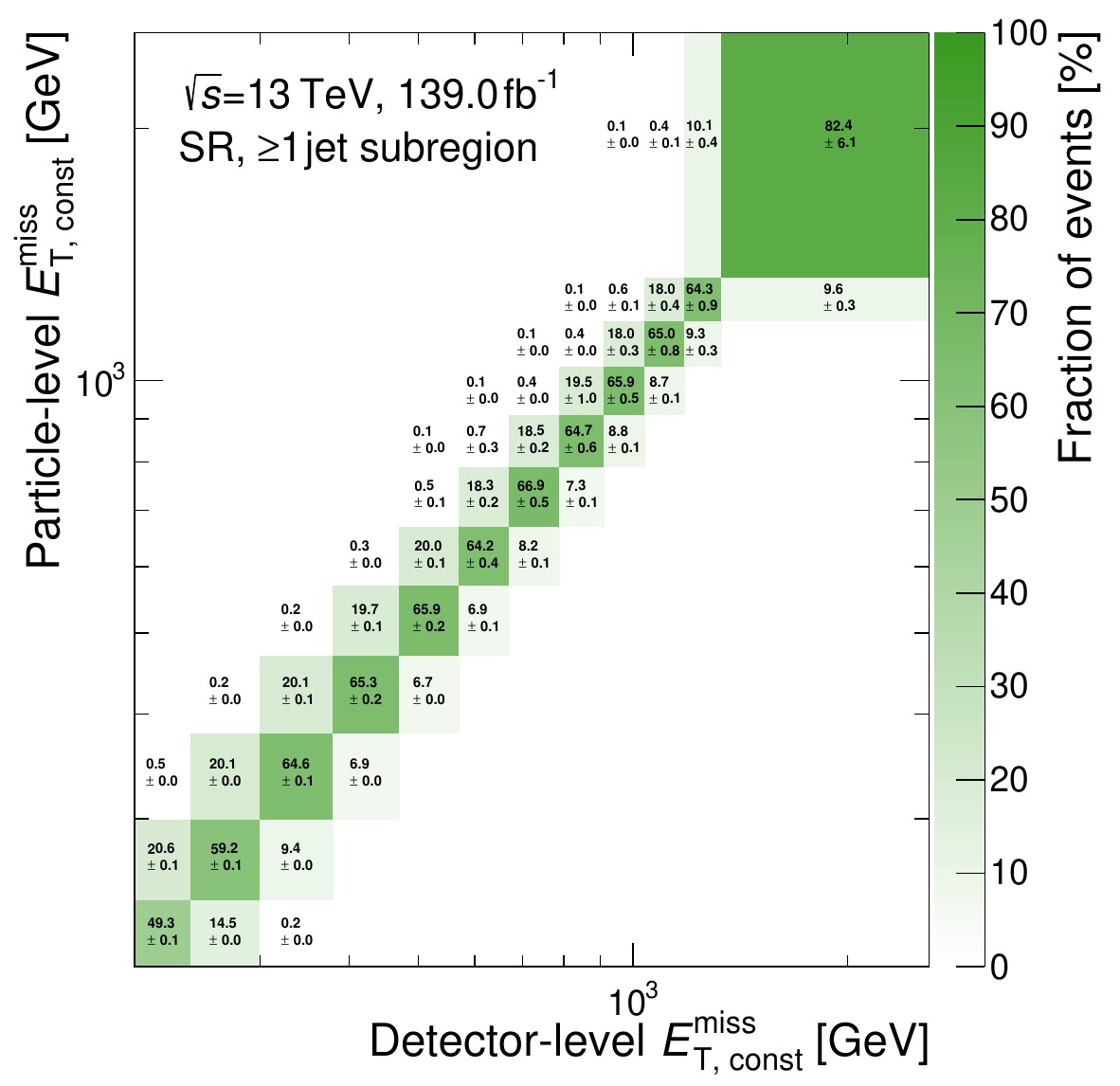}
\end{myfigure}

\bigskip
Specific features of the response matrix as introduced in \secref{sec:detCorr_responseMatrix} are shown as examples for the \Mono subregion in \figref{fig:detCorr_unfoldingDiagnostics}.

\subfigref{fig:detCorr_unfoldingDiagnostics}{a} gives the efficiency as a function of \METconst.
The efficiency is largest for the signal region because it does not require the reconstruction of a lepton.
The smallest efficiency can be observed in \OneEJetsAM because a stringent selection based on \METmeas and \mT is imposed in addition to requiring a tight identification (\cf\secref{sec:metJets_regions}).

\begin{myfigure}{%
		(a) Efficiency, (b) fiducial fraction, (c) purity and (d) stability in the \Mono subregion for the five different measurement regions.
	}{fig:detCorr_unfoldingDiagnostics}
	\subfloat[]{\includegraphics[width=0.49\textwidth]{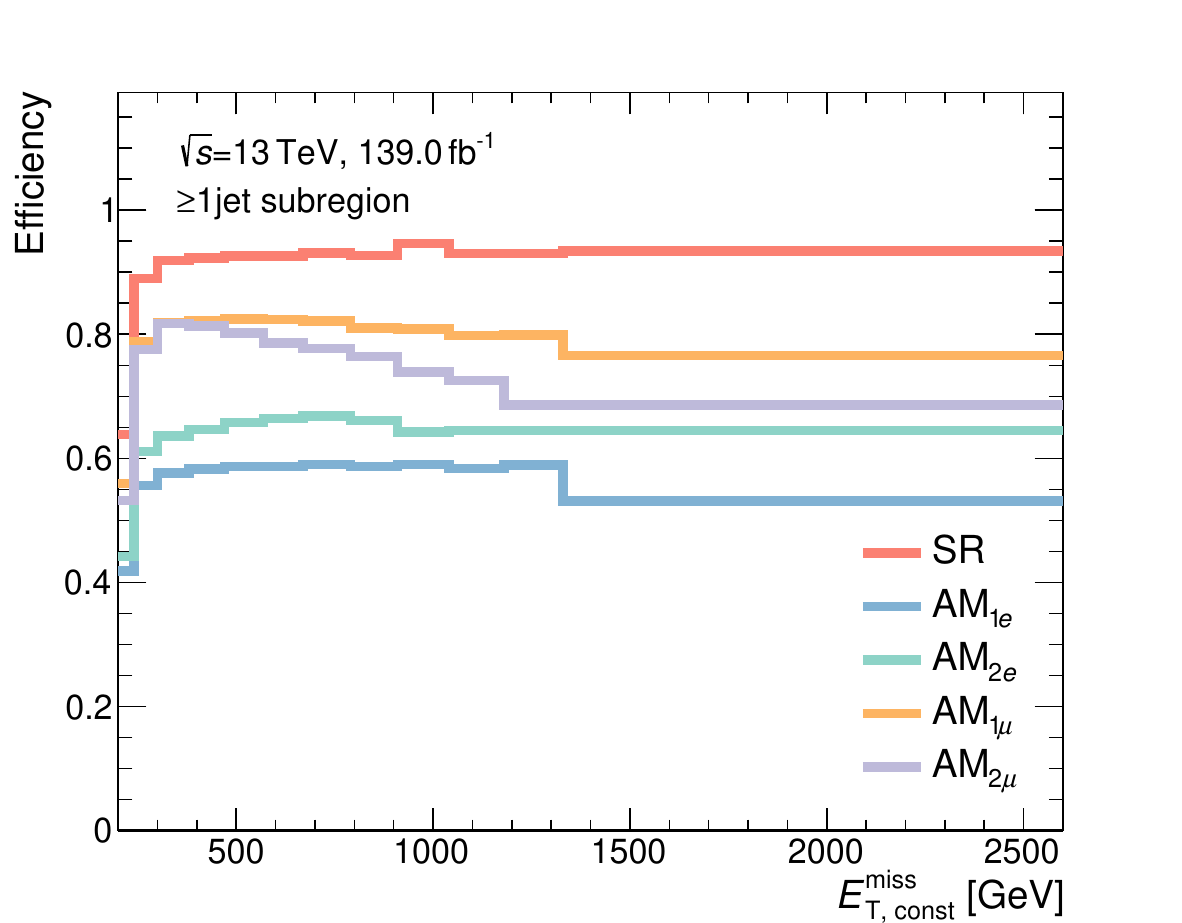}}
	\subfloat[]{\includegraphics[width=0.49\textwidth]{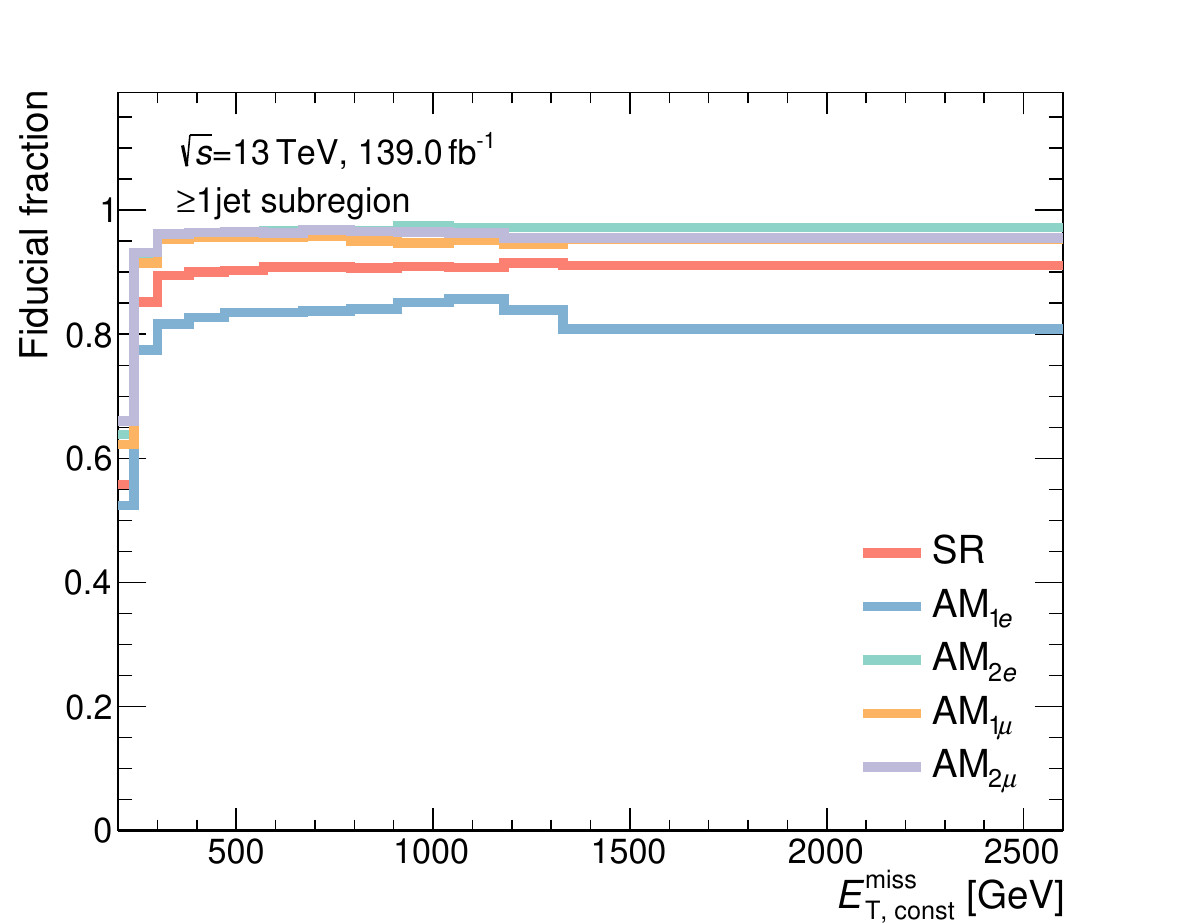}}\\
	\subfloat[]{\includegraphics[width=0.49\textwidth]{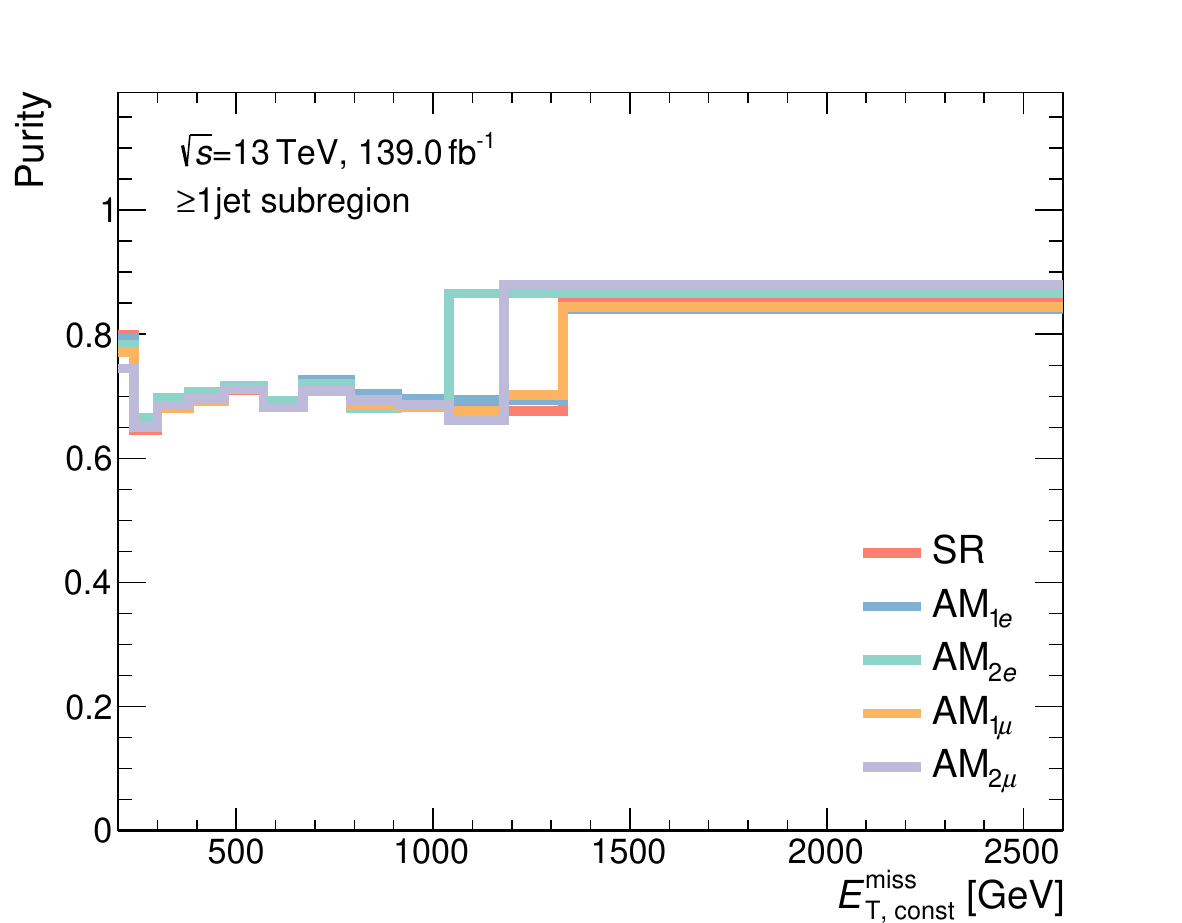}}
	\subfloat[]{\includegraphics[width=0.49\textwidth]{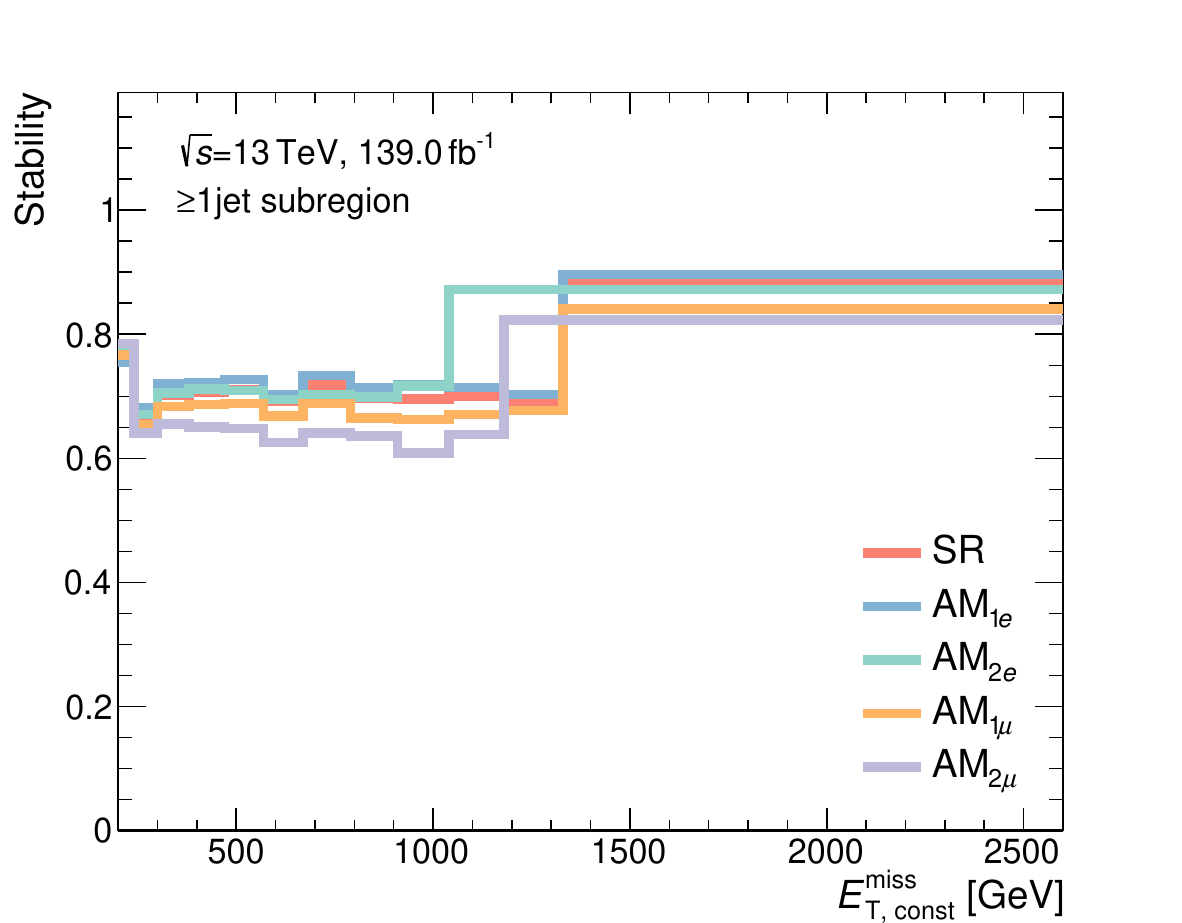}}\\
\end{myfigure}

\subfigref{fig:detCorr_unfoldingDiagnostics}{a} gives the fiducial fraction as a function of \METconst.
The fiducial fraction is approximately \SI{95}{\%} for almost all regions. The signal region exhibits a smaller fiducial fraction because of \Wenujets events that fail the particle-level selection because of an in-acceptance electron but pass the detector-level selection if the electron is not reconstructed. \OneEJetsAM has the smallest fiducial fraction due to migration out of the selected phase space in \mT.

Purity and stability are similar for all regions (\cf\subfigsref{fig:detCorr_unfoldingDiagnostics}{c}{d}).
They are larger than \SI{60}{\%} because the bin width in each distribution in the \METjets measurement is chosen such that the purity is larger than \SI{60}{\%} and the number of events is larger than 20, as mentioned above.
The transition from the former being the dominating condition to the latter being the dominant condition causes the discontinuity at large \METconst.

In general, none of the observed quantities depend strongly on \METconst, apart from border effects where events migrate into and out of the investigated phase space.

\subsection{Unfolding method}
\label{sec:detCorr_metJets_unfoldingMethod}

The \METjets measurement uses iterative Bayesian unfolding~\cite{Shepp:1982mlr,DAgostini:1994fjx} for regularising the result.
The simulation prediction at particle level is employed as the original prior.
Two unfolding iterations are used.
This poses a compromise between increasing the statistical uncertainty with each step and reducing the bias from the choice of original prior.

Every distribution is unfolded individually.
Correlations are taken into account by the bootstrap method (\cf\secref{sec:metJets_statUnc}).
After the unfolding, the event counts are transformed into differential cross sections by dividing by the luminosity (\cf\secref{sec:experiment_luminosity}) and the bin widths.

With this setup, the closure test closes perfectly.
In the stress test, there is small non-closure.
This non-closure is used as an uncertainty on the unfolding method and discussed in \secref{sec:detCorr_unfUncertainties}.

Signal-injection tests were performed among others using the \DMP model mentioned in \secref{sec:DM_models}.
Even when the yield from the pseudodata exceeds the \SM prediction by an order of magnitude, the difference between $\SM+\BSM$ distributions at particle level and unfolded pseudodata does not exceed \SI{10}{\%}.
This means that the unfolding method is also suitable for hypothetically unfolding data including all but the largest \BSM contributions with only mild biases.

\subsection{Unfolding uncertainties}
\label{sec:detCorr_unfUncertainties}

Uncertainties that are already present at detector level have to be propagated to particle level.
Statistical uncertainties are accounted for by unfolding the bootstraps in addition to the nominal distributions, as mentioned in \secref{sec:metJets_statUnc}. For experimental systematic uncertainties, the systematic variations as pointed out in \secref{sec:metJets_expSystUnc} are unfolded. For theoretical systematic uncertainties the uncertainties mentioned in \secref{sec:metJets_theoSystUnc} are assessed directly at particle level.
In addition, the unfolding procedure comes with uncertainties on its own, which are described in detail in the following.

\subsubsection{Stress-test uncertainty}

The non-closure in the stress test is taken as an uncertainty on the unfolding procedure.
The non-closure also takes into account effects of not perfectly reproducing the particle-level distributions in the unfolded distributions even of generated events.
This can for example originate from the limited number of used unfolding iterations, but is a very small effect.

\begin{myfigure}{
		Stress-test uncertainties for the five measurement regions in the (left) \Mono and (right) \VBF subregion when reweighting as a function of different variables.
		The total uncertainty from the sum in quadrature is shown as a solid black line.
	}{fig:detCorr_unfoldingUncertainties}
	\subfloat[]{\includegraphics[width=0.49\textwidth]{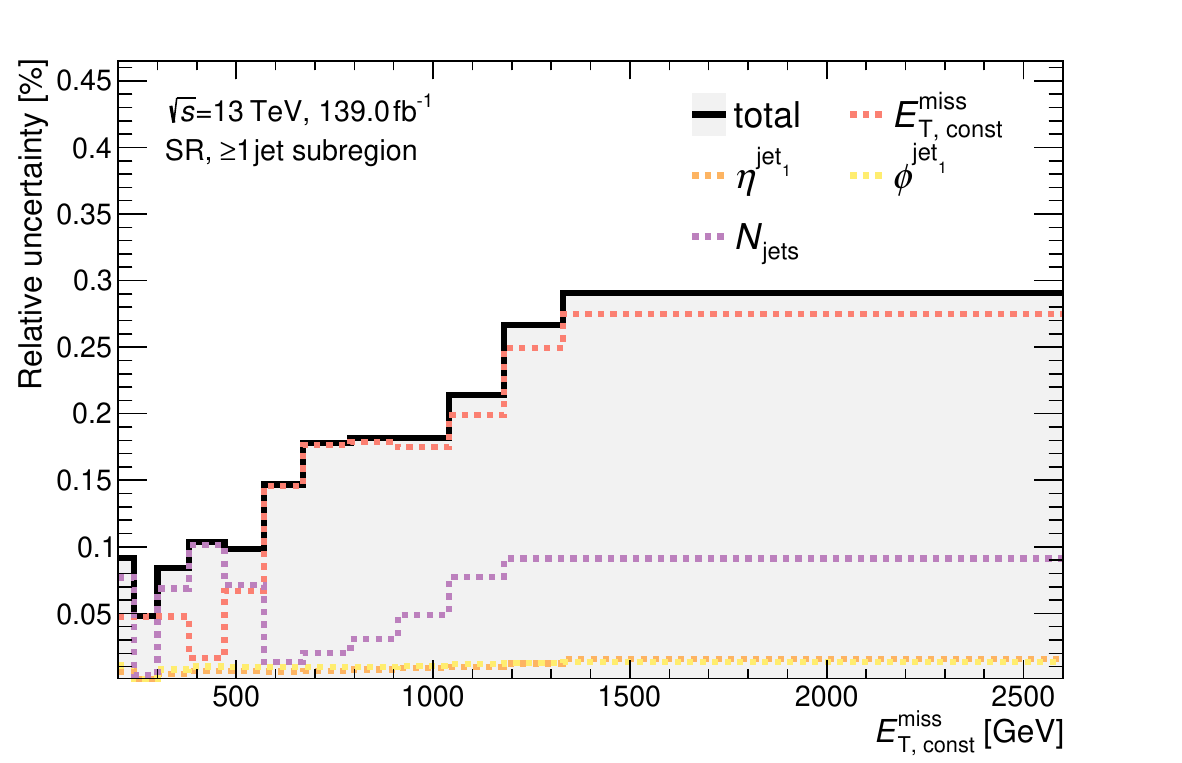}}
	\subfloat[]{\includegraphics[width=0.49\textwidth]{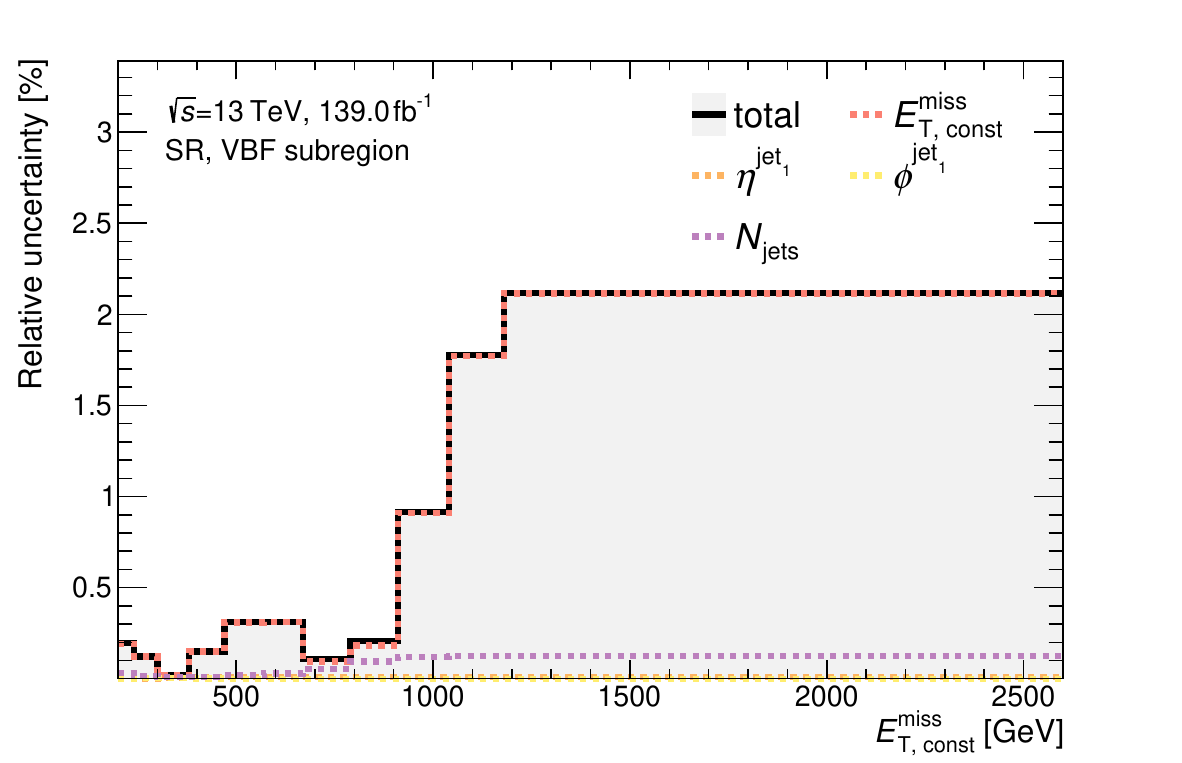}}\\
	\subfloat[]{\includegraphics[width=0.49\textwidth]{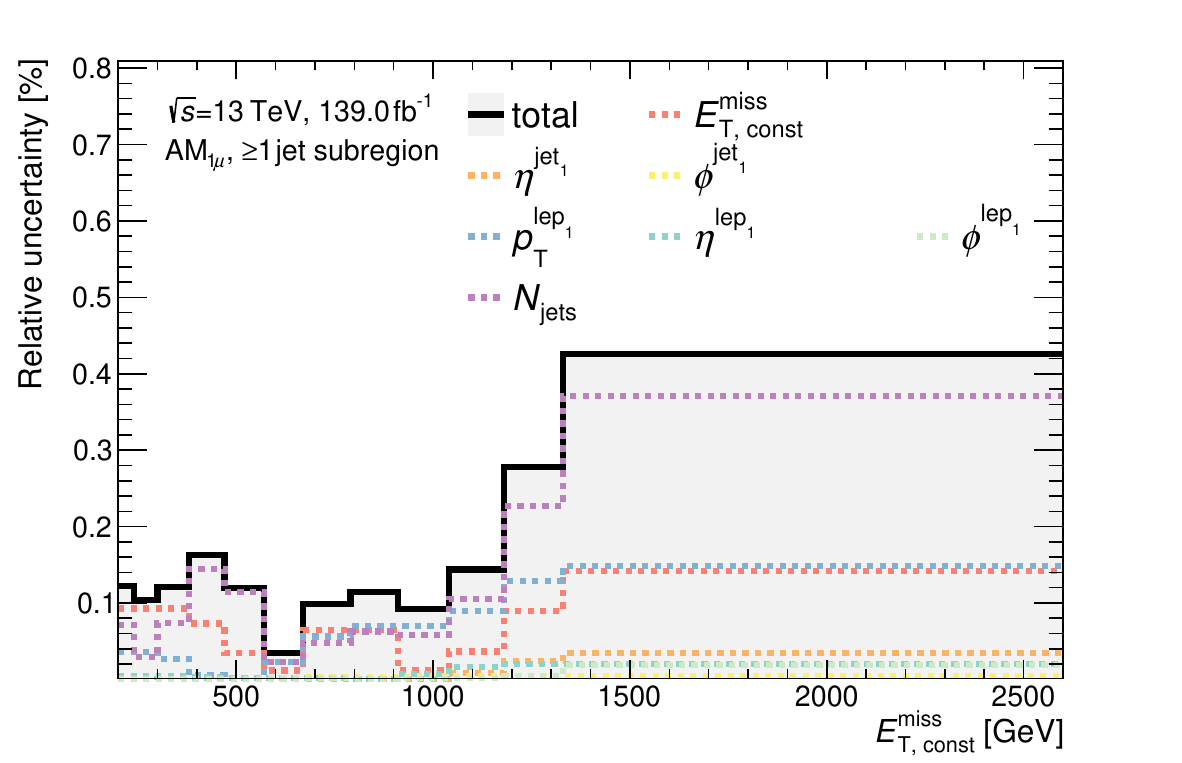}}
	\subfloat[]{\includegraphics[width=0.49\textwidth]{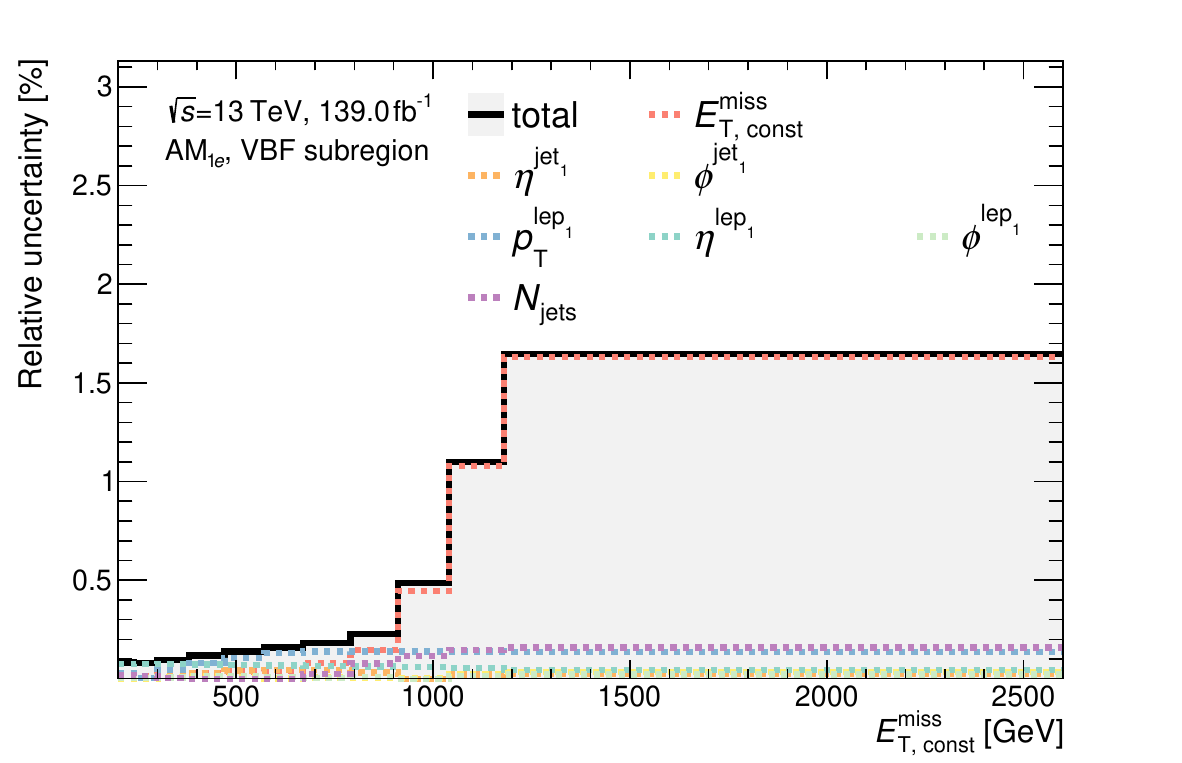}}\\
	\subfloat[]{\includegraphics[width=0.49\textwidth]{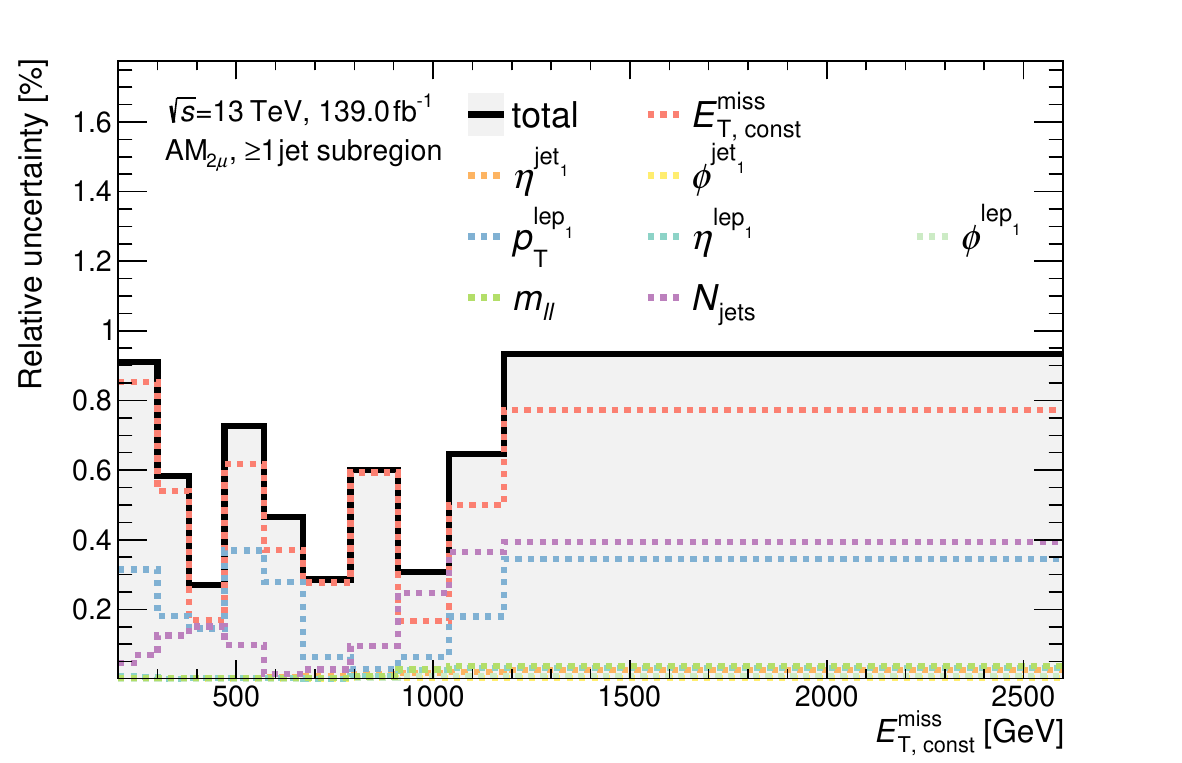}}
	\subfloat[]{\includegraphics[width=0.49\textwidth]{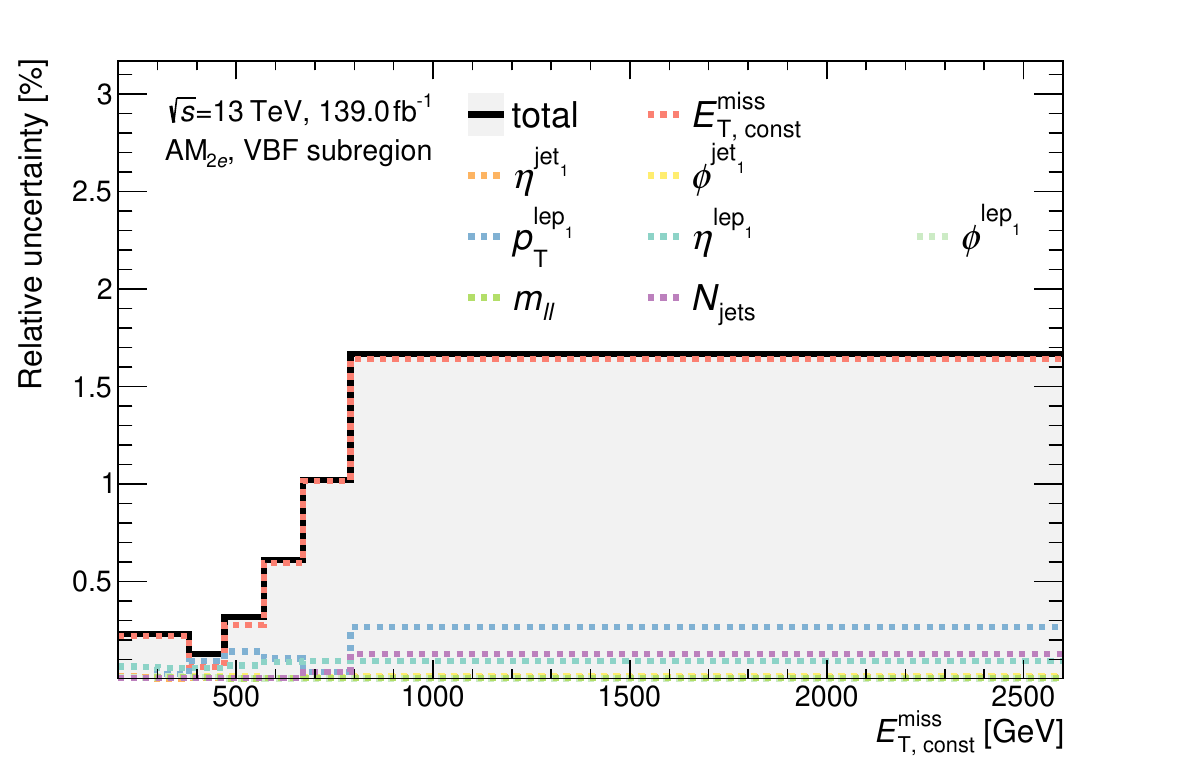}}\\
\end{myfigure}

\bigskip
The distribution of the stress-test uncertainty in a selection of the five measurement regions and two subregions is shown in \figref{fig:detCorr_unfoldingUncertainties}.
The uncertainty when reweighting as a function of a specific variable is given by a dashed coloured line.
The total uncertainty is obtained by summing the individual uncertainties in quadrature and marked by a solid black line.
The discontinuities in the uncertainty distributions give an impression of the statistical uncertainty of the method.

The most important uncertainty originates from reweighting as a function of the nominal unfolding variable, \METconst (red).
The resulting uncertainty in general is small but increases where the description of the data in simulation is poor.
This happens where statistics are smaller and fluctuations have a larger impact, \ie at large \METconst, in the two-lepton auxiliary measurements as well as in the \VBF subregion.
Here the uncertainty can reach up to \SI{2.5}{\%}.

Leading-jet kinematics (\eta, \phi), leading-lepton kinematics (\pT, \eta, \phi), the number of jets \Njets as well as the invariant mass of the dilepton system \mll where applicable are taken into account as hidden variables.
The uncertainty is not estimated for the leading-jet \pT because the strong correlation with \METconst would lead to double-counting with the uncertainty from reweighting as a function of \METconst.

In general, uncertainties related to the hidden variables are small, below \SI{0.4}{\%}.
They increase towards large \METconst, similar to the uncertainty when reweighting as a function of \METconst.
The dominant contributions stem from the modelling of \Njets (purple) as it impacts the event topology, as well as leading-lepton \pT (blue) and \eta (blue-green) if a lepton is in the event.
The latter arise because these variables influence the assignment of events to analysis regions.

\subsubsection{Uncertainty from sample composition}

In the \METjets measurement, an inclusive phase space is unfolded, consisting of a mixture of different \SM processes.
In \TwoEJetsAM and \TwoMuJetsAM, there is a clear dominating process.
In the \METjetsSR and \OneLJetsAM, however, there are important subdominant contributions from \Wlnujets and top processes.
Generally, these can exhibit different reconstruction efficiencies and overall distributions than the dominant \SM process.
This has an impact on the response matrix and therefore unfolding result.
The impact can be problematic if the relative yield of the contributions is mismodelled. The effect of alternative phase-space compositions is assessed by varying the yields of the most important subleading contributions by up to \SI{10}{\%}. The resulting differences of less than \SI{1}{\%} are treated as a systematic uncertainty.

Different \MCEG configurations predict different distributions for \SM processes. The impact of exchanging the used \MCEG configurations is assessed, in particular with respect to the splitting between hard process and parton shower for \Sherpa \Znunujets as well as the diagram removal and subtraction scheme for top processes. The differences are found to be negligible.

\subsubsection{Uncertainty from boundary effects}

Migration of events from bins below the selected phase space in \METconst into the selection can affect the unfolding results.
This effect is assessed by extending the phase space with additional bins in \METconst and comparing the unfolded results.
An impact of less than $\SI{1}{\%}$ is observed in the small \METconst region and assigned as an uncertainty.

\bigskip
Overall, the uncertainties of the \METjets measurement related to the unfolding are small compared to other uncertainties stemming from experiment, limited statistics and theory.

\subsection{Particle-level results}
\label{sec:detCorr_partLevelResults}

The measurement results at particle level are shown in \figref{fig:detCorr_partLevelResults} for a selection of the five regions in the two subregions.
The remaining measurement results can be found in \appref{app:detCorr_partLevelResults}.
The top panels give the differential cross section of data (black dots) and generated \SM prediction (blue crosses) with their respective statistical uncertainties as well as the contributions to the generated \SM cross section from the different processes (filled areas).
The quadrature sum of experimental (theoretical) systematic and statistical uncertainties is displayed as a red (blue) shaded area.
Unfolding uncertainties are counted towards the experimental uncertainties.
The bottom panels show the ratio of data to generated \SM prediction.

\begin{myfigure}{
		Differential cross section and ratio of data to generated \SM prediction at the particle level for the five measurement regions in the (left) \Mono and (right) \VBF subregion.
		Black dots (blue crosses) denote the measured data (yield from generated \SM prediction) with their statistical uncertainty.
		The red (blue) shaded areas correspond to the total experimental (theoretical) uncertainty. In the top panels, the filled areas correspond to the different contributing processes.
	}{fig:detCorr_partLevelResults}
	\subfloat[]{\includegraphics[width=0.49\textwidth]{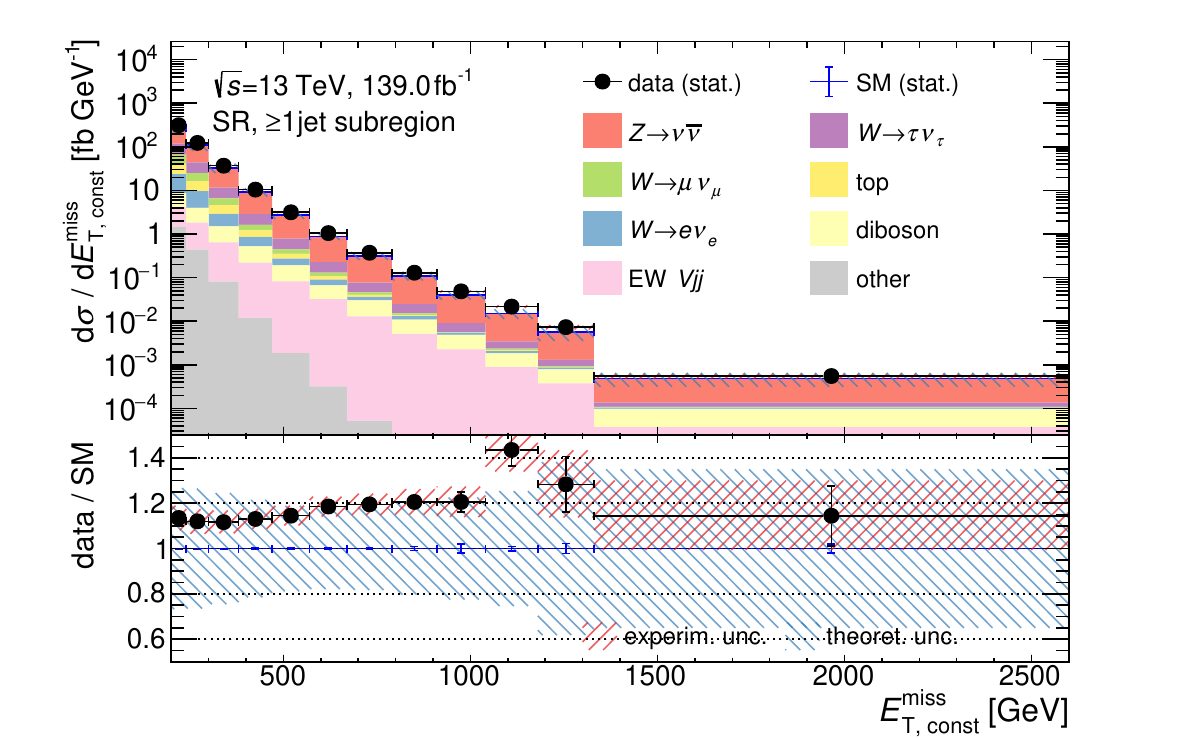}}
	\subfloat[]{\includegraphics[width=0.49\textwidth]{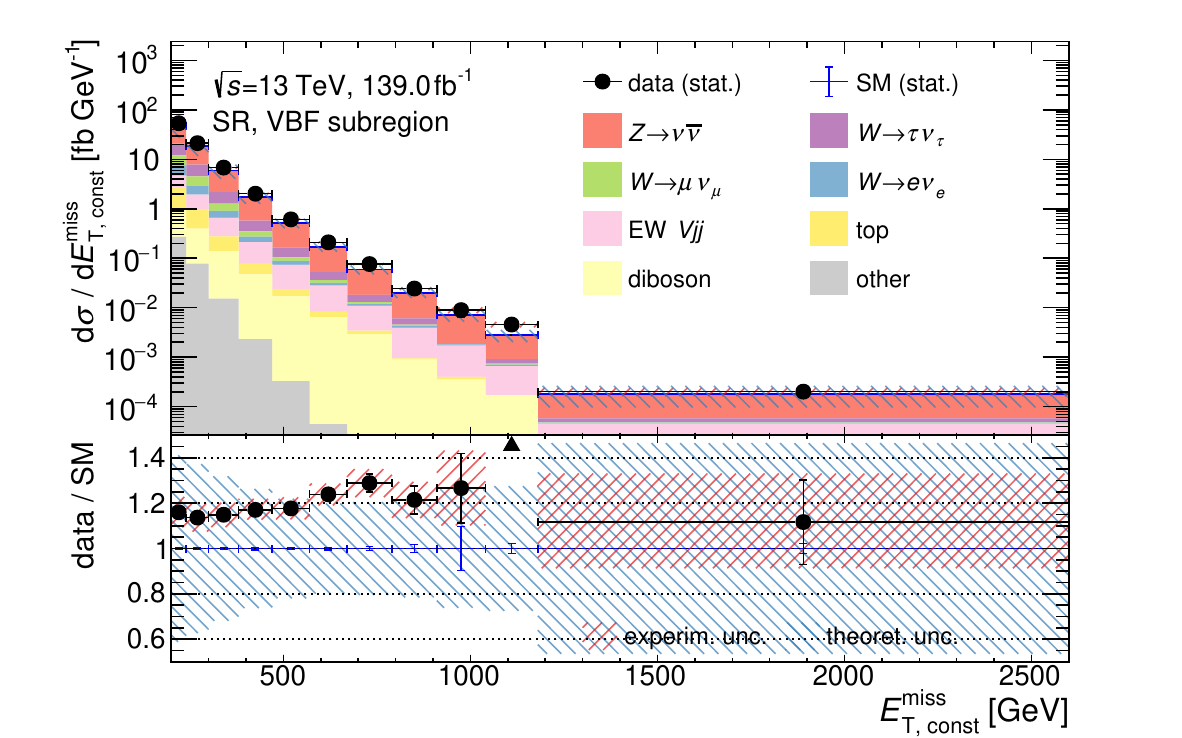}}\\
	\subfloat[]{\includegraphics[width=0.49\textwidth]{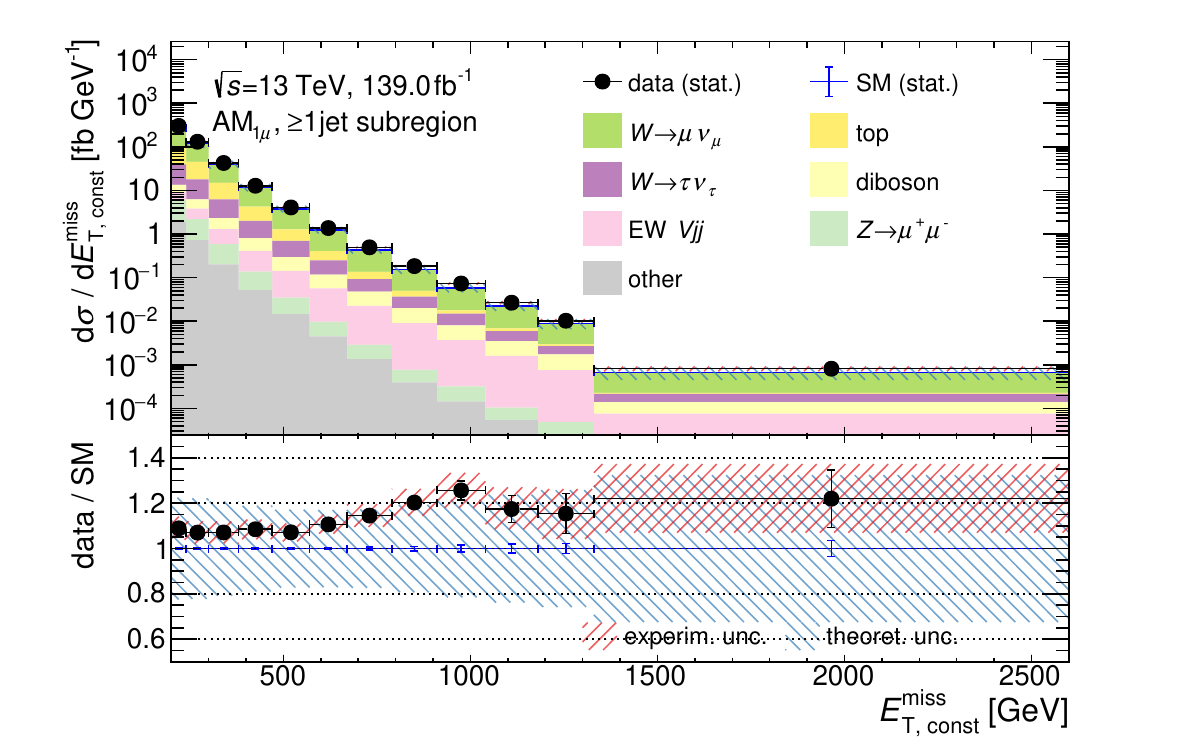}}
	\subfloat[]{\includegraphics[width=0.49\textwidth]{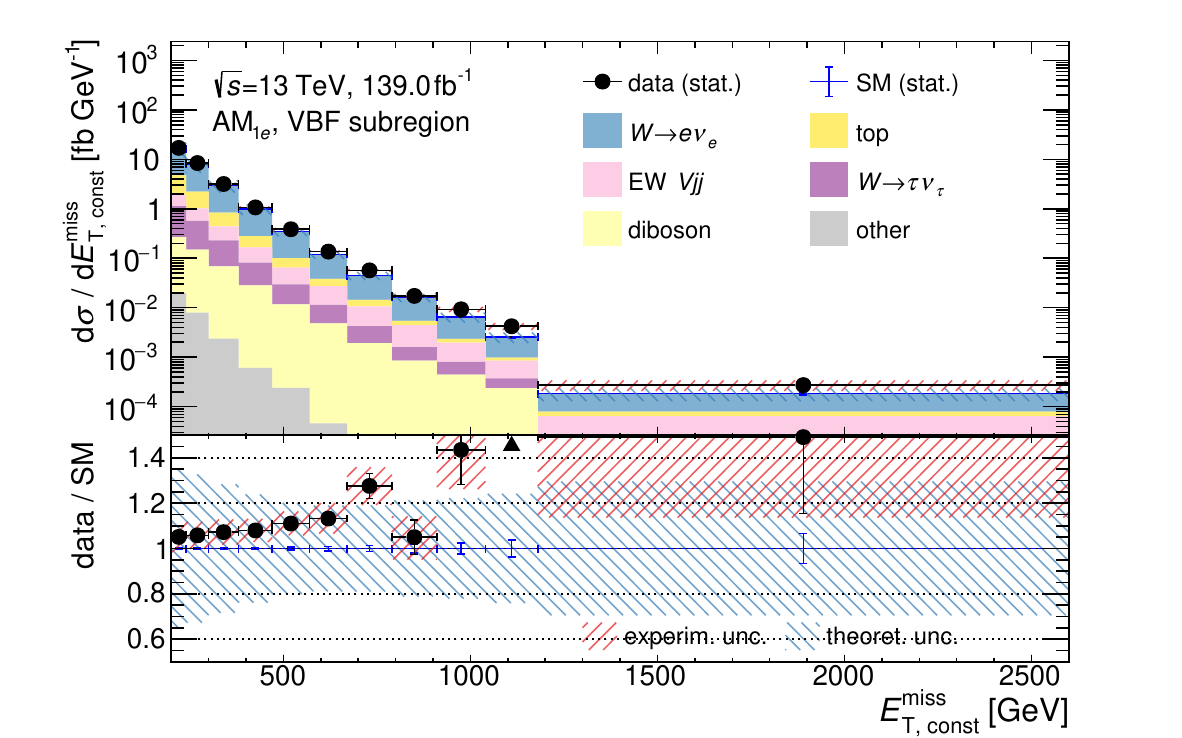}}\\
	\subfloat[]{\includegraphics[width=0.49\textwidth]{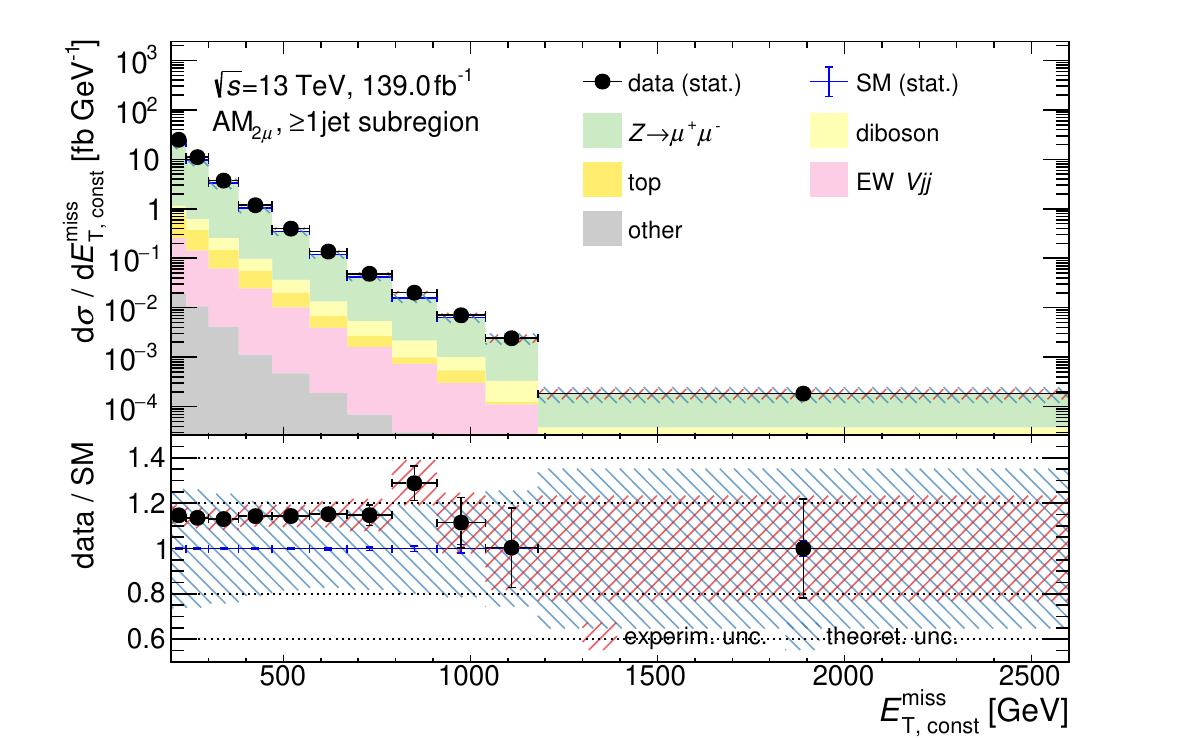}}
	\subfloat[]{\includegraphics[width=0.49\textwidth]{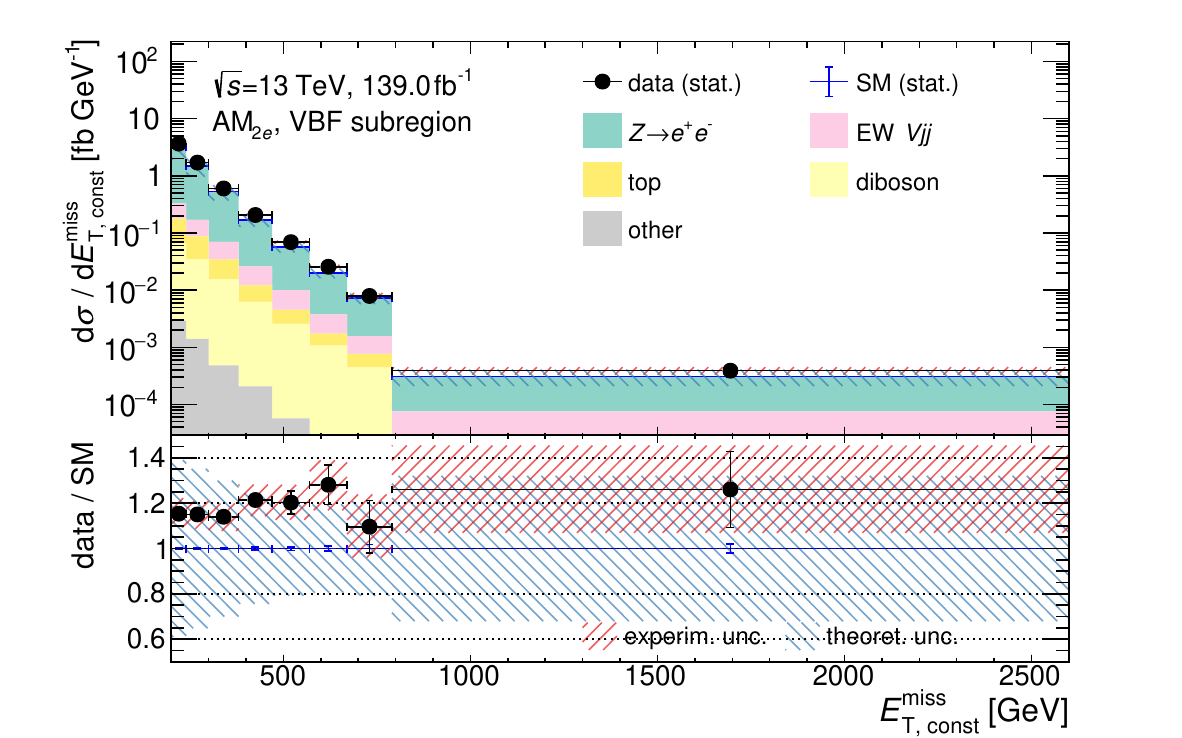}}\\
\end{myfigure}

The particle-level distributions from \figref{sec:detCorr_partLevelResults} are very similar to the detector-level distributions from \figref{sec:metJets_detLevelResults} in shape, normalisation and composition.
This is a consequence of the dominant diagonal in the response matrix and correspondingly almost constant efficiency and fiducial fraction as a function of \METconst.
No fake contributions are present in the particle-level representation because they have been removed as background when correcting for detector effects (\cf\secref{sec:detCorr_metJets_responseMatrix}).

As in detector-level representation (\cf\secref{sec:metJets_detLevelResults}), the \SM prediction gives a good description of the shape of the measured data but underestimates the normalisation by $10-\SI{20}{\%}$ in all regions.
The discrepancy is, however, mostly covered by the theoretical systematic uncertainties, predominantly by the scale uncertainties on the \Vjets estimate (\cf\secref{sec:metJets_theoSystUnc}).
Beyond this normalisation difference, there is a significant rise of the data over the prediction at $\METconst=\SI{1100}{GeV}$ in the signal regions (\cf\subfigsref{fig:detCorr_partLevelResults}{a}{b}), as previously seen in the detector-level representation.
There are also generally larger differences between data and prediction at large \METconst in the \VBF subregion due to statistical fluctuations in this extreme phase space.
Whether differences between measured data and \SM prediction can be accounted for everywhere simultaneously given the systematic uncertainties is investigated in \secref{sec:interpretation_SM}.

\bigskip
In \figref{fig:detCorr_Rmiss}, examples for the \Rmiss distributions, as introduced in \eqref{eq:metJets_Rmiss}, are shown in particle-level representation.
The top panels give \Rmiss in data (black dots) and generated \SM prediction (blue crosses) with their respective statistical uncertainties.
The quadrature sum of experimental (theoretical) systematic and statistical uncertainties is displayed as a red (blue) shaded area.

It can be observed that the signal region has a larger cross section than the auxiliary measurements requiring an electron or two leptons, \ie $\Rmiss>1$.
\Rmiss is larger than one for \TwoLJetsAMs because of the less restrictive signal-region selection and because of the branching fraction of $Z$-boson decays:
\Zll for a specific charged lepton~$\ell$ has a branching fraction of \SI{3.4}{\%}, \Znunu of \SI{20.0}{\%}~\cite{Zyla:2020zbs}.
\Zjets processes contribute approximately \SI{90}{\%} in \TwoLJetsAMs and approximately \SI{50}{\%} in the signal region (\cf\figref{fig:metJets_detLevelResults}).
In \OneLJetsAMs the dominant \SM process is \Wlnujets compared to \Znunujets in the signal region.
Here, \Rmiss can be close to 1.
\Rmiss is larger than 1 in \OneEJetsAM because of the more stringent phase-space selection in \OneEJetsAM (\cf\secref{sec:metJets_regions}).
In all regions, \Rmiss decreases with increasing \METconst, indicating a more steeply falling spectrum in the signal region.

In the bottom panels, the ratio of data to generated \SM prediction is shown.
The ratio does not depend significantly on \METconst.
For \TwoLJetsAMs, it is compatible with unity, while for \OneLJetsAMs an offset of approximately \SI{10}{\%} can be observed.
This is a consequence of the changing process composition in the Standard Model:
In \OneLJetsAMs, top processes give contributions up to \SI{25}{\%} while their contribution in all other regions is smaller than \SI{10}{\%}.
An incorrect prediction of the cross section for top processes then influences \Rmiss more in \OneLJetsAMs.
The significant rise of the data over the prediction at $\METconst=\SI{1100}{GeV}$ that was visible in the differential cross sections in the signal regions (\cf\subfigsref{fig:detCorr_partLevelResults}{a}{b}) is reduced in \Rmiss.

\begin{myfigure}{
		Ratio of the fiducial cross section in the signal region to the auxiliary measurements in the (left) \Mono and (right) \VBF subregion, calculated according to \eqref{eq:metJets_Rmiss}.
		Black dots (blue crosses) denote the measured data (yield from generated \SM prediction) with their statistical uncertainty.
		The red (blue) shaded areas correspond to the total experimental (theoretical) uncertainty.
		The bottom panel shows the ratio of data to the generated \SM prediction.
	}{fig:detCorr_Rmiss}
	\subfloat[]{\includegraphics[width=0.49\textwidth]{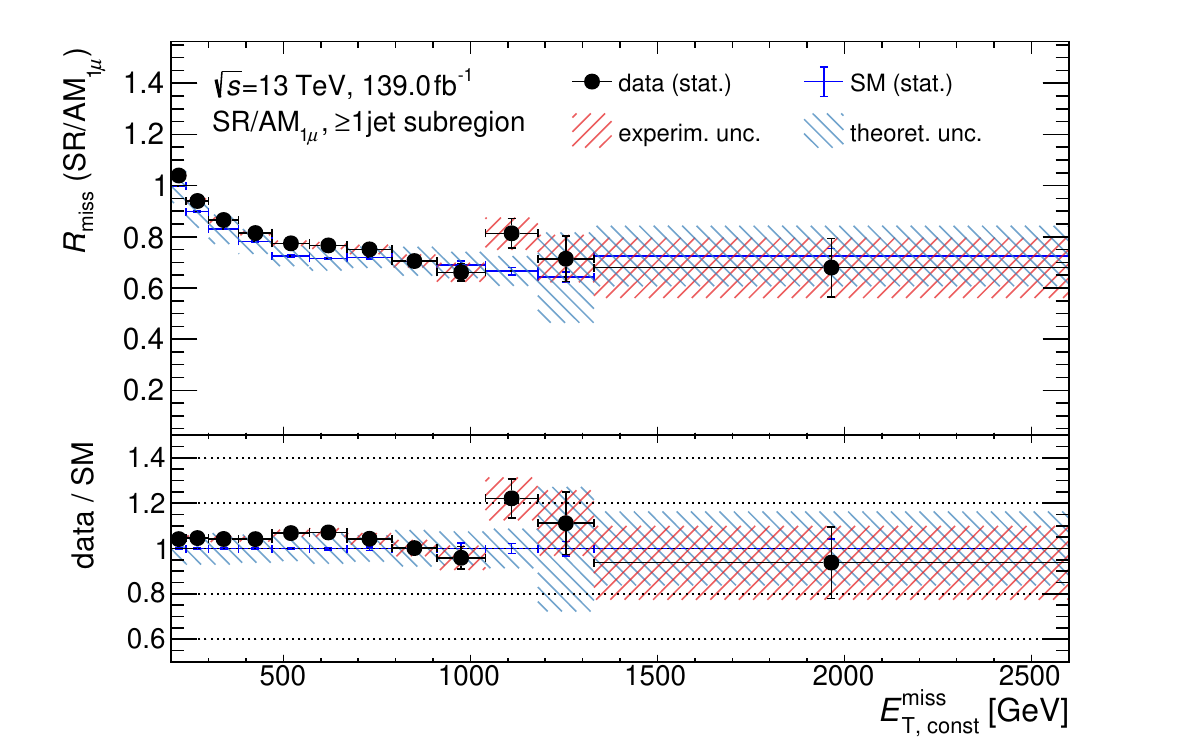}}
	\subfloat[]{\includegraphics[width=0.49\textwidth]{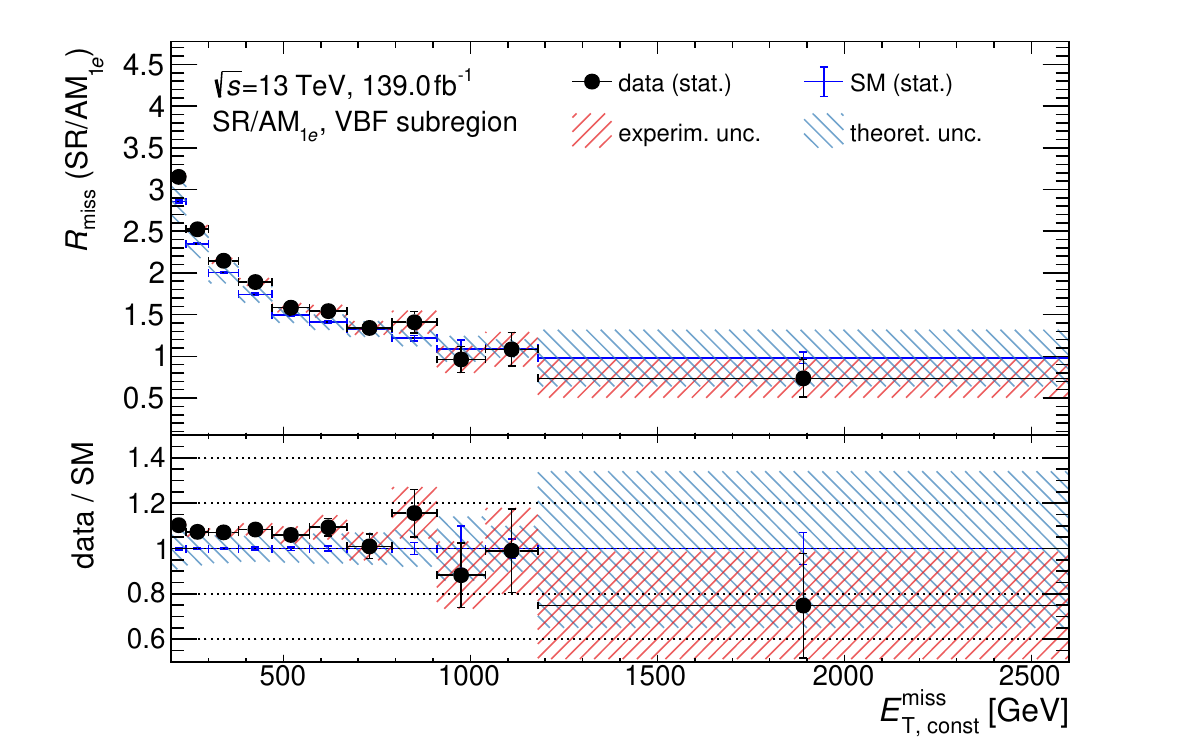}}\\
	\subfloat[]{\includegraphics[width=0.49\textwidth]{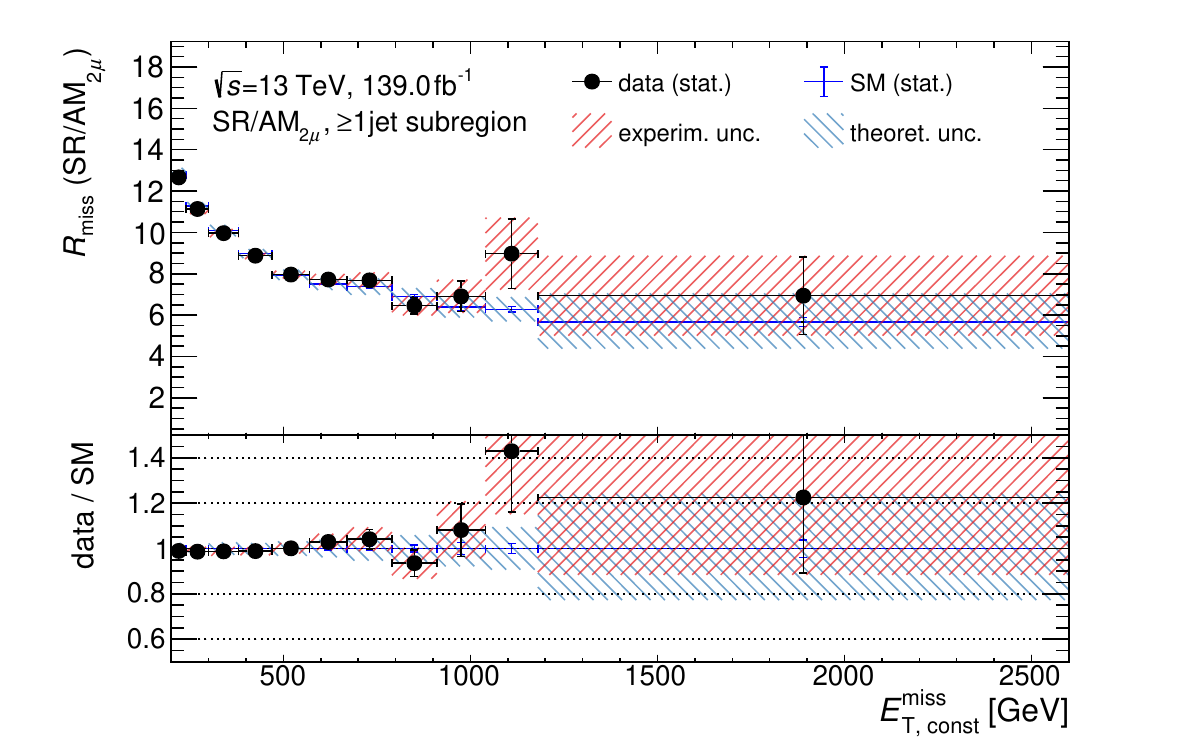}}
	\subfloat[]{\includegraphics[width=0.49\textwidth]{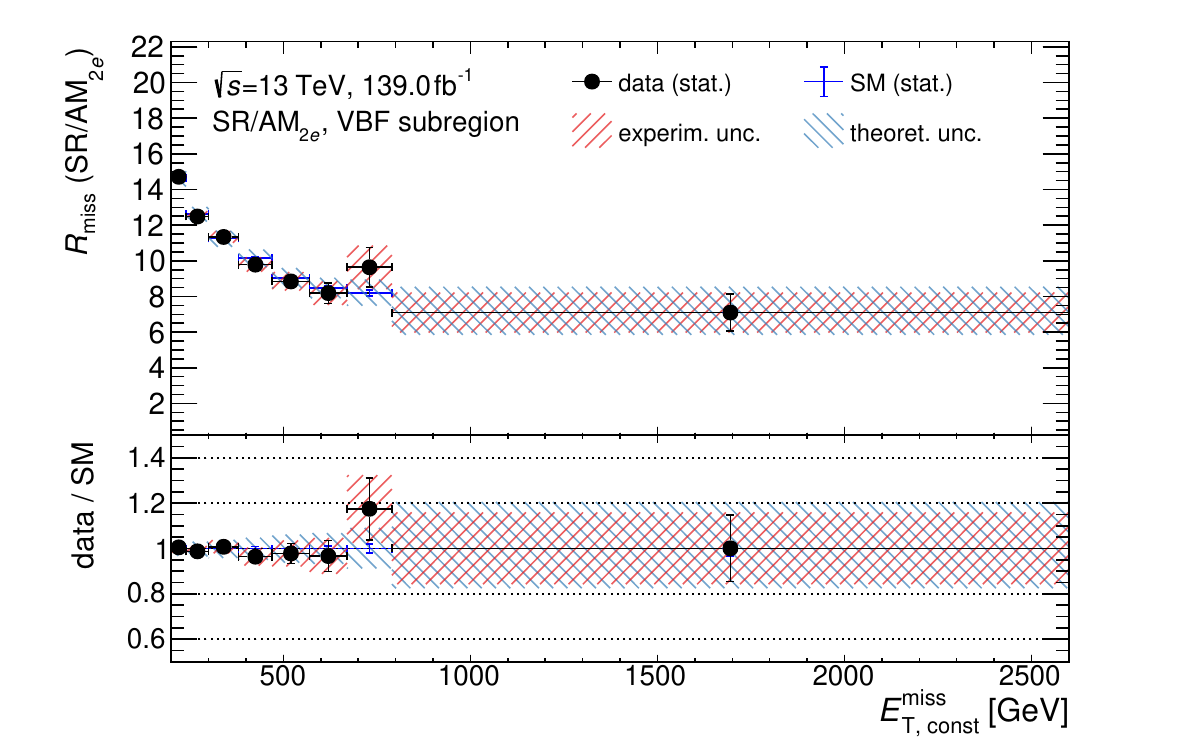}}\\
\end{myfigure}

In \Rmiss theoretical and experimental uncertainties that appear in both numerator and denominator are partly cancelled.
This concerns for example uncertainties related to jet energy scale or initial-state radiation.
This cancelling of systematic uncertainties that are correlated between auxiliary measurements and signal region leads to overall significantly smaller experimental and theoretical systematic uncertainties on \Rmiss compared to the bare differential cross sections.

\bigskip
The results of the \METjets measurement are now available in particle-level representation.
In the next chapter, the contributions a model for Dark Matter, the \THDMa, gives to the \METjets phase is discussed.
For this discussion, generated \THDMa predictions at particle level are employed.
It is not necessary to use a detector simulation, \ie consider the \THDMa contributions at detector level, because the \METjets measurement has been corrected for detector effects in this chapter.
\ChapterGeneral{Gauging the \BSM contributions}{2HDM$+\mathit{a}$ in the \METjets phase space}{\THDMa in the \METjets phase space}{%
	Mach mit. Du hast deine Pflicht zu erfüllen,\\
	deinen Beitrag zu leisten [...].\\
	Mach mit. Reih dich ein.%
}{Kolja Podkowik}{AntilopenGang:2015amm}
\label{sec:2HDMa_metJetsMeasurement}


The \THDMa is an excellently motivated model to investigate Dark Matter at colliders, as pointed out in \secref{sec:2HDMa}.
The production of the Dark-Matter particle $\chi$ leads to \METmeas in events as Dark Matter can leave the detector without interacting.
In this chapter, the contributions of the \THDMa to the \METjets phase space are studied before the results of the \METjets measurement are statistically interpreted with respect to the \THDMa in the next chapter.

In the whole chapter, \THDMa events are generated with the setup described in \secref{sec:interpretation_2HDMa_MC}.
All considerations are performed at particle level because the measurement has been corrected for detector events in the previous chapter.
Unless explicitly stated otherwise, the parameter settings according to \tabref{tab:LHCDMWG_params} are used.

The \THDMa processes that give important contributions to the \METjets final state are described in \secref{sec:interpretation_2HDMa_contributions}.
Two planes of \THDMa parameters are considered, as motivated in \secref{sec:2HDMa_parameterPlanes}.
How the cross sections of the important \THDMa processes change in the two parameter planes is discussed in \secref{sec:interpretation_2HDMa_xs}.
\secref{sec:interpretation_2HDMa_contribs} shows how the \THDMa processes contribute to the phase space selected by the \METjets measurement.

\section{\THDMa processes for the \METjets final state}
\label{sec:interpretation_2HDMa_contributions}

Different \THDMa processes contribute to the \METjets final state.
The nominal\linebreak \THDMa process targetted by the \METjets measurement is the production of a \DM pair in association with at least one jet from a quark or gluon, \xxqg.
For systematisation, all other processes are broadly categorised as \textit{other loop-induced} and \textit{other tree-level} processes by whether they involve loops or not.
The nominal process \xxqg includes loop-induced as well as tree-level processes.
All three categories -- \xxqg, loop induced and tree level -- are described in detail in the following.
In all these processes, \METmeas is produced if the pseudoscalar $a$ or $A$ decays to \DM particles.
The cross sections of these processes vary considerably as a function of the different \THDMa parameters and are discussed in detail in \secref{sec:interpretation_2HDMa_xs}.

\subsubsection{\xxqg}
Processes as depicted in \subfigref{fig:2HDMa_METjets_diagrams}{a} are the nominal final state targetted by the \METjets measurement.
In addition to the diagram considered in \subfigref{fig:metJets_Feynman_metJets}{a}, also decays of the pseudoscalar $A$ to \DM pairs contribute.
The jet required by the \METjets selection is often caused by initial-state radiation (\ISR).
The produced \METmeas in these processes has a strong correlation with the transverse momentum of the initial-state radiation, which is usually small.
Therefore, predominantly only small amounts of \METmeas are produced.

\begin{myfigure}{
		Feynman diagrams contributing to the \METjets final state in the \THDMa.	
	}{fig:2HDMa_METjets_diagrams}
	\subfloat[]{\includegraphics[width=0.32\textwidth]{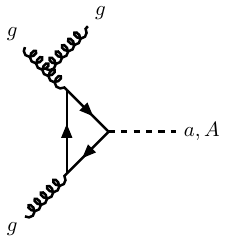}}
	\subfloat[]{\includegraphics[width=0.32\textwidth]{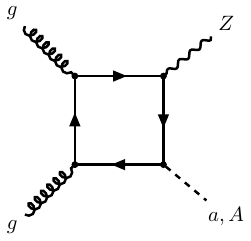}}
	\subfloat[]{\includegraphics[width=0.32\textwidth]{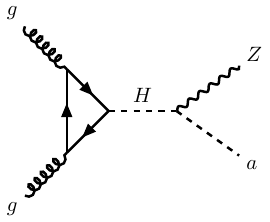}}\\
	
	\subfloat[]{\includegraphics[width=0.32\textwidth]{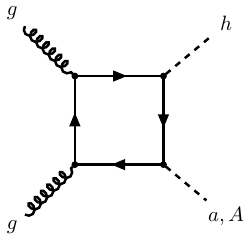}}
	\subfloat[]{\includegraphics[width=0.32\textwidth]{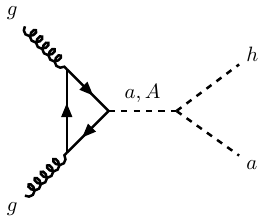}}
	\subfloat[]{\includegraphics[width=0.32\textwidth]{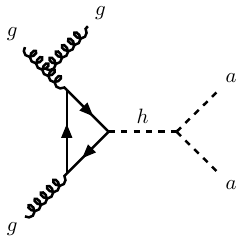}}\\
	
	\subfloat[]{\includegraphics[width=0.32\textwidth, valign=c]{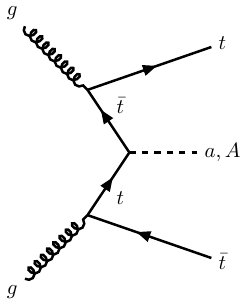}}
	\hspace{10pt}
	\subfloat[]{
		\includegraphics[width=0.32\textwidth,valign=c]{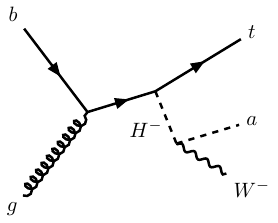}
		\vphantom{\includegraphics[width=0.29\textwidth,valign=c]{figures/Feynman_diagrams/2HDMa/gg_to_tta/diagram.pdf}} 
	}
\end{myfigure}

\subsubsection{Other loop-induced processes}
The different loop-induced contributions to the \METjets final state are

\begin{itemize}
	\itembf{\xxZ}
	Box diagrams (\subfigref{fig:2HDMa_METjets_diagrams}{b}) as well as $s$-channel production of scalars~$H$ (\subfigref{fig:2HDMa_METjets_diagrams}{c}) are possible.
	In the latter case, decays $H\to AZ$ are suppressed because either $H$ or $A$ has to be virtual due to the requirement \mAeqmHeqmHpm.
	There is at least one jet in \xxZ events, as required by the \METjets selection, if the $Z$ boson decays hadronically or there is initial-state radiation.
	
	\itembf{\xxh}
	Box diagrams (\subfigref{fig:2HDMa_METjets_diagrams}{d}) as well as $s$-channel production of pseudoscalars $A$ (\subfigref{fig:2HDMa_METjets_diagrams}{e}) are possible.
	The diagrams are similar to \xxZ production.
	There is at least one jet in \xxh events, as required by the \METjets selection, if the light scalar $h$ decays hadronically.
	In the case of $s$-channel production of the pseudoscalar $a$ and its subsequent decay to another pseudoscalar~$a$ and a scalar~$h$, one pseudoscalar $a$ in the process has to be virtual and the diagram is consequently suppressed.
	The process does nonetheless give important contributions if \ma is small.
	
	\itembf{\xxxxqg}
	This is the production of two \DM pairs, for example from invisible decays of the light scalar $h$ via two possibly virtual pseudoscalars $a$ as shown in \subfigref{fig:2HDMa_METjets_diagrams}{f}.
	The jet required by the \METjets selection is often caused by initial-state radiation.
\end{itemize}

\subsubsection{Other tree-level processes}
Various processes at tree-level can produce events with one or two \DM pairs in association with one or two Higgs bosons $h$, top quarks or vector bosons.
These can be selected by the \METjets measurement.
Examples of these processes are given in \subfigsref{fig:2HDMa_METjets_diagrams}{g}{h}.

\section{\THDMa cross sections}
\label{sec:interpretation_2HDMa_xs}

The contribution any \THDMa process makes to the phase space selected by the\linebreak \METjets measurement is correlated with the cross section of the process.
Investigating the \THDMa cross sections in the targetted parameter planes gives therefore insights into the sensitivity of the \METjets measurement to the \THDMa.
The two targetted parameter planes, \mamA and \matanB, as motivated in \secref{sec:2HDMa_parameterPlanes}, are discussed separately in the following.
A requirement of $\METmeas>\SI{150}{GeV}$ is placed in both cases to reduce the cross section not selected by the \METjets measurement selection.

\subsection{\mamA plane}

\figref{fig:interp_2HDMa_crossSections_mamA} gives the cross sections for the considered \THDMa processes in the \mamA plane.
In this plane, $\tanB=1$ and the coupling of uncharged \BSM bosons to top quarks is larger than to bottom quarks (see \secref{sec:2HDMa_parameterPlanes}).
The mass of the pseudoscalar $a$ ($A$) is varied between \SI{100}{GeV} and \SI{800}{GeV} (\SI{200}{GeV} and \SI{2000}{GeV}).
All other model parameters are chosen as given in \tabref{tab:LHCDMWG_params}.
The lower bounds on the masses of the pseudoscalars are selected because for too small masses of the \BSM bosons the decay branching-ratios of the \SM Higgs boson $h$ are changed significantly.
This alters the phenomenology of the model and can be in conflict with other measurements of the Higgs boson properties.
There are dedicated searches for these Higgs boson decays~\cite{ATLAS:2021ykg}.
The upper bounds on the masses are chosen because the sensitivity of the \METjets measurement decreases with increasing mass of the \BSM bosons (see \secref{sec:interpretation_2HDMa}).
In addition, the ratio of mass to width of \BSM bosons can be larger than \SI{20}{\%} at very large masses, \eg $\mA>\SI{1900}{GeV}$.
In this region, the narrow-width approximation (\cf\secref{sec:MCEG_hadronisation}) may be violated and predictions can become more unreliable.

\begin{myfigure}{
		Cross sections $\METmeas>\SI{150}{GeV}$ in the \mamA plane for (a) \xxqg, (b) other loop-induced and (c) other tree-level \THDMa processes.
		The solid black lines mark the indicated cross sections in fb.%
		\THDMaLines{white}
	}{fig:interp_2HDMa_crossSections_mamA}
		\subfloat[\xxqg]{\includegraphics[width=0.65\textwidth]{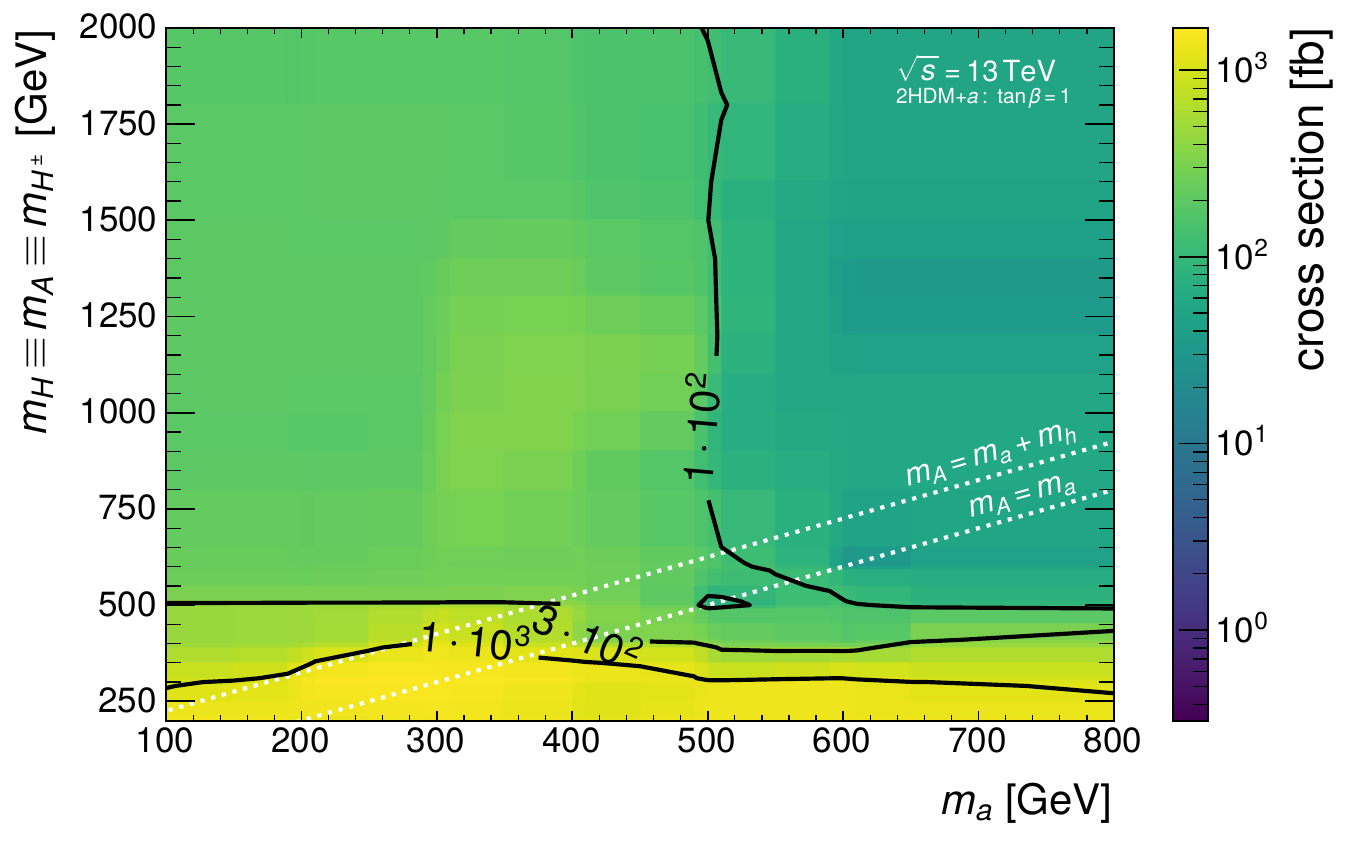}}\\
		\subfloat[other loop-induced processes]{\includegraphics[width=0.65\textwidth]{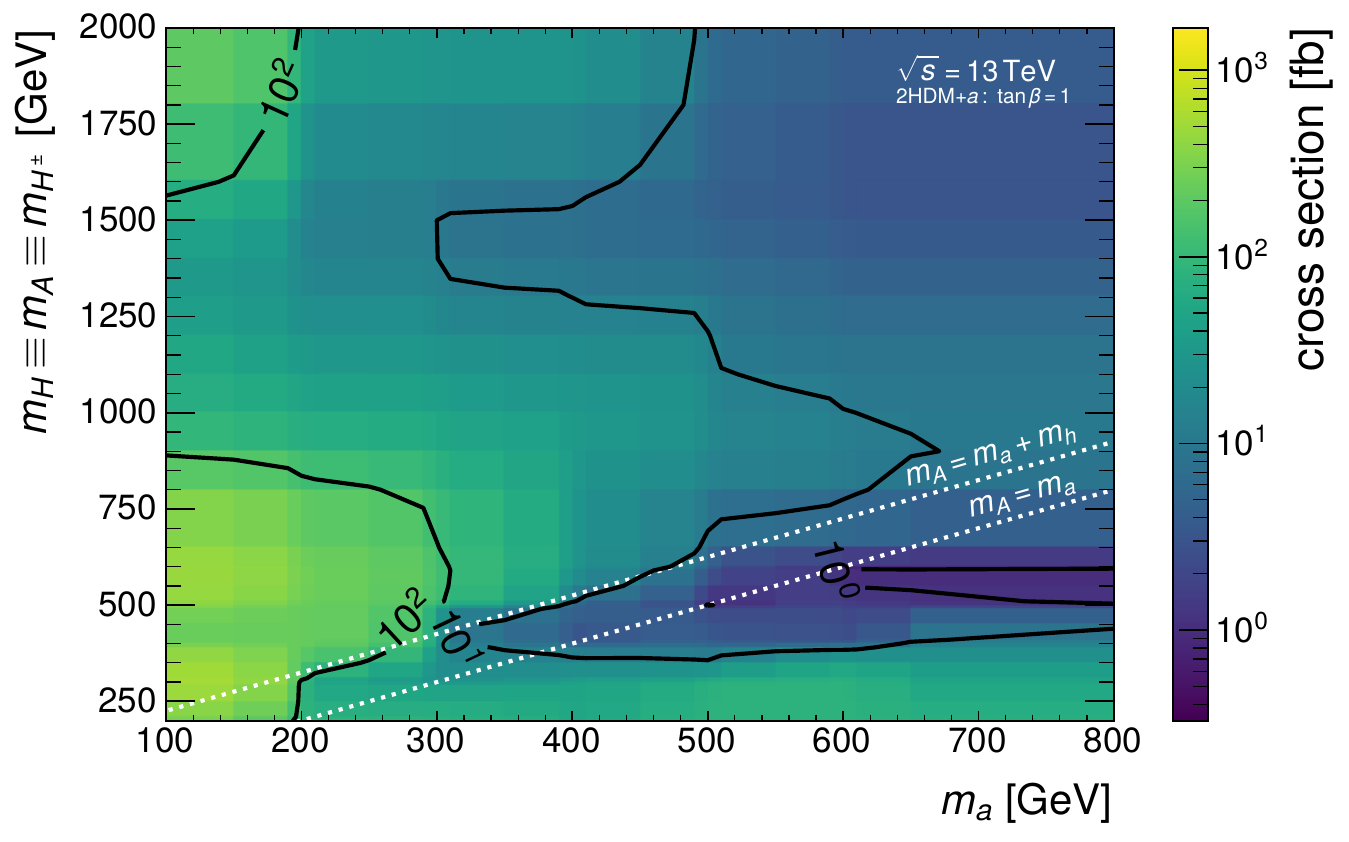}}\\
		\subfloat[other tree-level processes]{\includegraphics[width=0.65\textwidth]{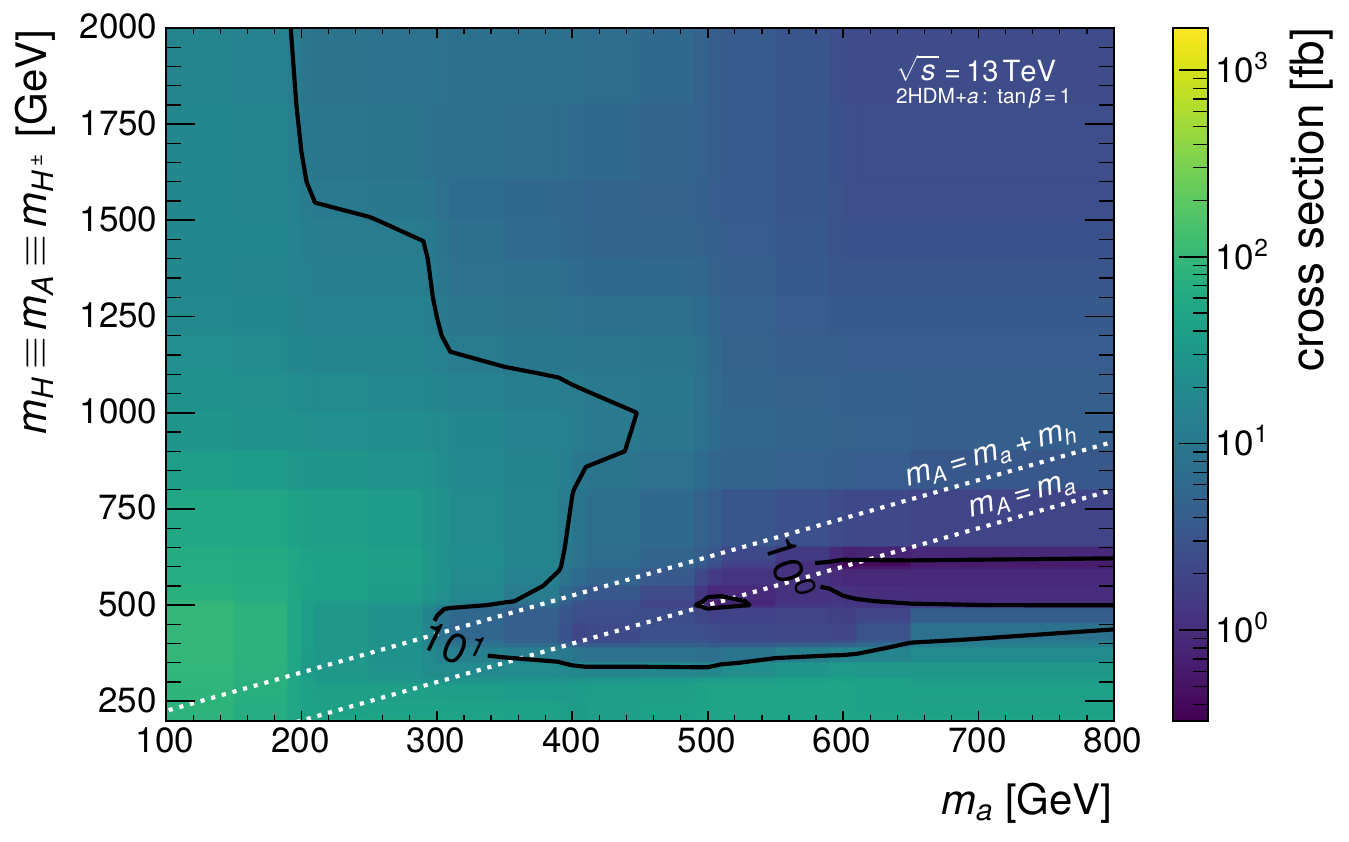}}
\end{myfigure}

\bigskip
In \subfigref{fig:interp_2HDMa_crossSections_mamA}{a}, the cross section for \xxqg processes is shown.
The cross section is larger than \SI{100}{fb} independent of \mAeqmHeqmHpm as long as $\ma<\SI{500}{GeV}$ because of the production of the pseudoscalar $a$ in association with jets and its subsequent decay into \DM particles.
The cross section decreases exponentially once the threshold for producing the pseudoscalar~$a$ resonantly from \ttbar fusion at $\ma=2m_t\approx\SI{350}{GeV}$ is passed.
Similarly, the cross section is larger than \SI{100}{fb} independent of \ma as long as $\mA<\SI{500}{GeV}$ because of the production of the pseudoscalar $A$ in association with jets and its subsequent decay into \DM particles.
The cross section decreases exponentially once the threshold for producing the pseudoscalar~$A$ resonantly from \ttbar fusion at $\mA=2m_t\approx\SI{350}{GeV}$ is passed.
In general, the cross section is therefore dominated by processes involving the pseudoscalar ($a$ or $A$) which has the smaller mass at the considered parameter point.
At $\ma=\mA$, the cross section is reduced by more than three orders of magnitude due to destructive interference between processes involving the pseudoscalar~$a$ and those involving the pseudoscalar~$A$~\cite{Bauer:2017ota}.

At \sinPLow, the coupling to top quarks of the pseudoscalar $A$ is larger than the coupling to top quarks of the pseudoscalar $a$ (\cf\eqsref{eq:2HDMa_Gamma_a_ff}{eq:2HDMa_Gamma_A_ff}).
This means that the cross section for loop-induced production of pseudoscalars $A$ is larger than for pseudoscalars $a$ if they have the same mass.
In consequence, the cross section for the total \xxqg process is larger at small \mA than small \ma and can exceed \SI{1000}{fb}.
When the decays $A\to\ttbar$ or $A\to ah$ become kinematically open, the cross section for \xxqg decreases.

\bigskip
In \subfigref{fig:interp_2HDMa_crossSections_mamA}{b}, the summed cross sections for loop-induced processes that do not come from the nominal \xxqg signature, \ie\xxZ, \xxh and \xxxxqg, are shown.
There are three main contributions leading to large cross sections:
\begin{itemize}
	\item At small \ma, box diagrams like those shown in \subfigsref{fig:2HDMa_METjets_diagrams}{b}{d} with the pseudoscalar $a$ in the final state contribute.
	Likewise, at small \mA box diagrams with the pseudoscalar $A$ in the final state contribute.
	The cross section for processes involving the scalar boson~$h$ is about an order of magnitude larger in this region than for processes involving the vector boson~$Z$.
	For very small \ma or very small \mA, \xxxxqg processes contribute up to \SI{30}{\%} of the cross section.
	At large \ma and \mA, the contribution from \xxxxqg processes is negligible.
	
	\item Above the kinematic limit of $\mA=\ma+\mh$, the cross section for the $s$-channel process $A\to ah$ as shown in \subfigref{fig:2HDMa_METjets_diagrams}{e} becomes large.
	Likewise, above the kinematic limit of $\mH=\ma+m_Z$ the cross section for the $s$-channel process $H\to aZ$ as shown in \subfigref{fig:2HDMa_METjets_diagrams}{c} becomes large.
	In both cases, the cross section decreases as a function of $\mH\equiv\mA$.
	When both processes are kinematically open, their cross sections are similar in size in this region.
	
	\item When the mass difference between the pseudoscalars $a$ and $A$ is large, Higgs radiation processes $a\to ah$ as depicted in \subfigref{fig:2HDMa_METjets_diagrams}{e} become important as a consequence of \eqref{eq:2HDMa_g_haa}.
	This leads to large cross sections at $\ma\approx\SI{100}{GeV}$ and $\mA\approx\SI{2000}{GeV}$.
	The interplay of $A\to ah$ and $a\to ah$ processes has a minimum of the cross section at $\mA\approx\SI{1300}{GeV}$.
\end{itemize}
In general, the cross section for \xxh processes dominates over the cross sections for \xxZ and \xxxxqg processes.
Destructive interference between diagrams involving the pseudoscalar~$a$ and those involving the pseudoscalar~$A$ reduces the cross section by approximately \SI{40}{\%} (\SI{80}{\%}) for processes with a scalar boson~$h$ (vector boson $Z$) in the final state if $\ma=\mA$.

\bigskip
In \subfigref{fig:interp_2HDMa_crossSections_mamA}{c}, the summed cross sections for tree-level processes that do not come from the nominal \xxqg signature are shown.

At small \ma, $t$-channel diagrams with the pseudoscalar $a$ in the final state contribute, \eg $\xx+\ttbar$ as shown in \subfigref{fig:2HDMa_METjets_diagrams}{g}.
Likewise, at small \mA $t$-channel diagrams with the pseudoscalar $A$ in the final state contribute.
At $\ma=\mA<2m_t$, the production channel involving pseudoscalars~$A$ dominates over the one involving pseudoscalars~$a$ because at \sinPLow the coupling to top quarks of the pseudoscalar $A$ is larger than the coupling to top quarks of pseudoscalar $a$ (\cf\eqsref{eq:2HDMa_Gamma_a_ff}{eq:2HDMa_Gamma_A_ff}).
For the same reason, the branching fraction \brf{A\to\xx} decreases more than the branching fraction \brf{a\to\xx} as soon as the decay $a\to\ttbar$ ($A\to\ttbar$) opens up at $\ma>2m_t$ ($\mA>2m_t$).
At this point, the cross section times branching fractions contributing to $\xx+\ttbar$ become comparable.
Destructive interference between diagrams involving the pseudoscalar~$a$ and those involving the pseudoscalar~$A$ reduces the cross section by approximately \SI{40}{\%} if $\ma=\mA$.

Above the kinematic limit of $\mHpm=\ma+m_{W^\pm}$, the cross section for the $s$-channel process $\Hpm\to aW^\pm$ as shown in \subfigref{fig:2HDMa_METjets_diagrams}{h} becomes large.
The cross section for this process decreases as a function of \mHpm.

\bigskip
In general, the cross section for \xxqg processes is the largest and can exceed \SI{1000}{fb}.
The next-largest cross sections originate from other loop-induced processes and can exceed \SI{100}{fb}.
Other tree-level processes have the smallest cross section and do not exceed \SI{100}{fb}.

\subsection{\matanB plane}

Similar to the previous section, \figref{fig:interp_2HDMa_crossSections_matanB} gives the cross sections for the considered \THDMa processes in the \matanB plane (see also \secref{sec:2HDMa_parameterPlanes}).
In this plane, $\mAeqmHeqmHpm=\SI{600}{GeV}$.
For all process types, the cross section at $\ma=\mA=\SI{600}{GeV}$ is reduced due to destructive interference~\cite{Bauer:2017ota} as already observed in the \mamA plane.
The mass of the pseudoscalar $a$ is varied between \SI{100}{GeV} and \SI{800}{GeV} as in the \mamA plane.
The parameter \tanB is varied between 0.5 and 40.
All other model parameters are set as given in \tabref{tab:LHCDMWG_params}.
The bounds on \tanB are chosen because for too extreme values of \tanB electroweak precision constraints exclude the \THDMa~\cite{Haber:1992py,Gerard:2007kn,Haber:2010bw,Haller:2018nnx}.
In addition, the ratio of mass to width of \BSM bosons can be larger than \SI{20}{\%} at extreme values of \tanB, \eg $\tanB<0.6$ and $\tanB>50$ for the pseudoscalar~$A$.

\begin{myfigure}{
		Cross sections $\METmeas>\SI{150}{GeV}$ in the \matanB plane for (a) \xxqg, (b) other loop-induced and (c) other tree-level \THDMa processes.
		The solid black lines mark the indicated cross sections in fb.%
		\THDMaLines{white}
	}{fig:interp_2HDMa_crossSections_matanB}
	\subfloat[\xxqg]{\includegraphics[width=0.65\textwidth]{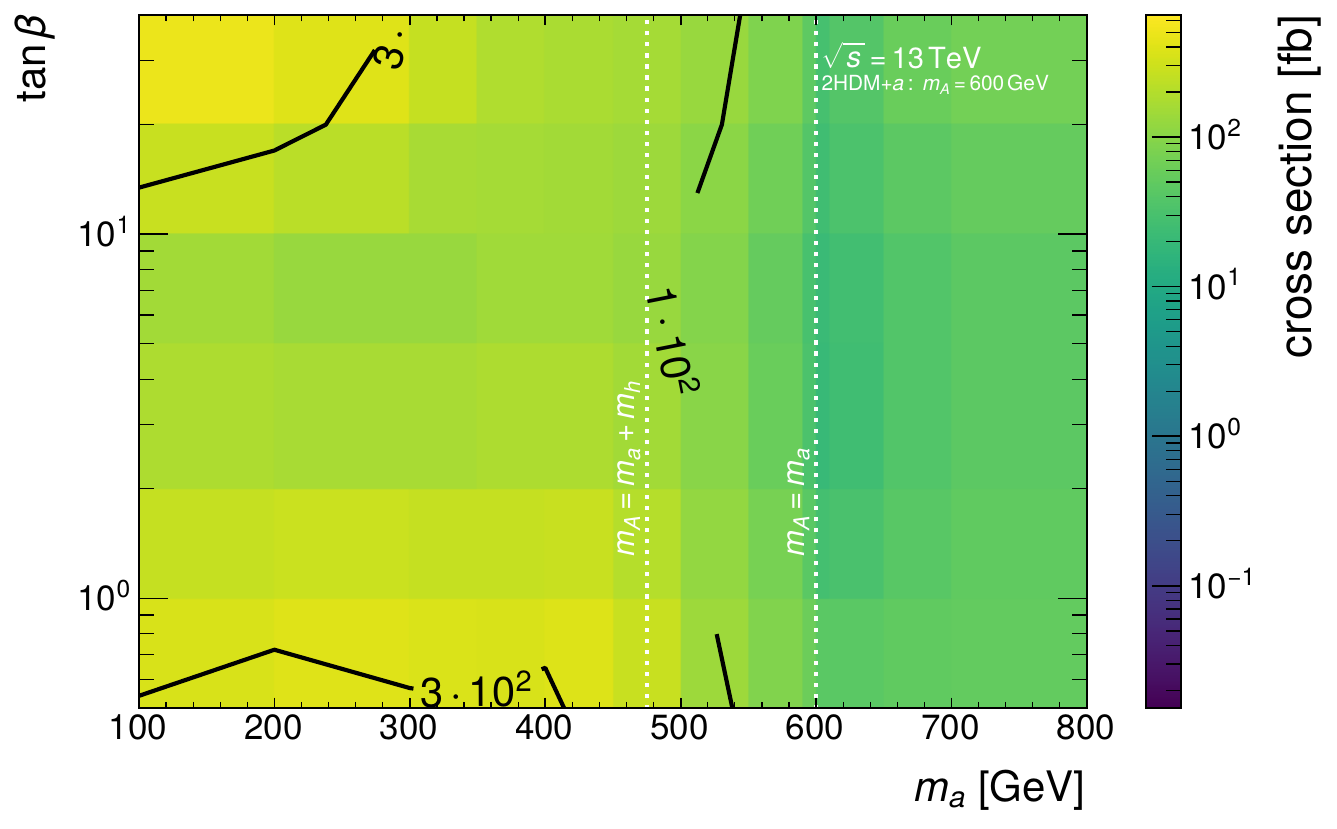}}\\
	\subfloat[other loop-induced processes]{\includegraphics[width=0.65\textwidth]{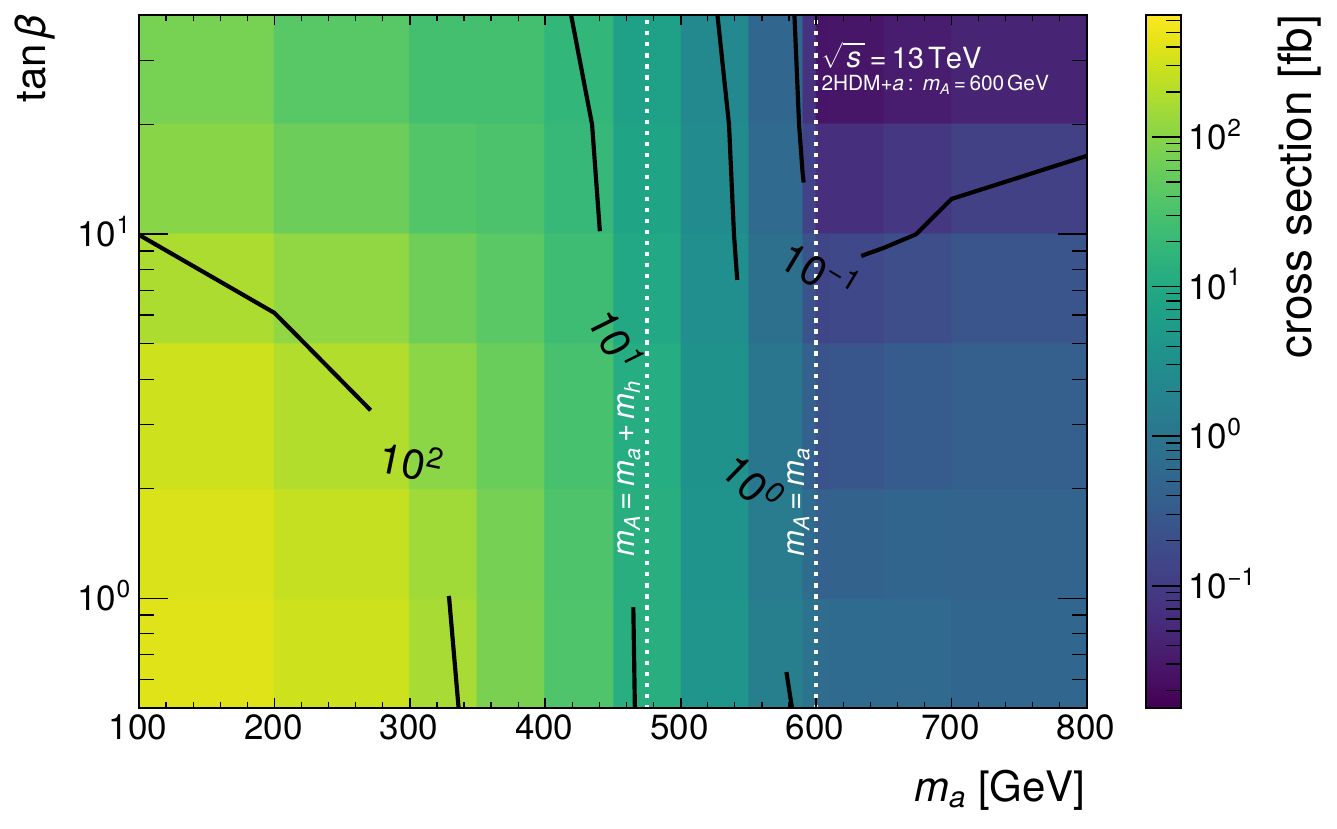}}\\
	\subfloat[other tree-level processes]{\includegraphics[width=0.65\textwidth]{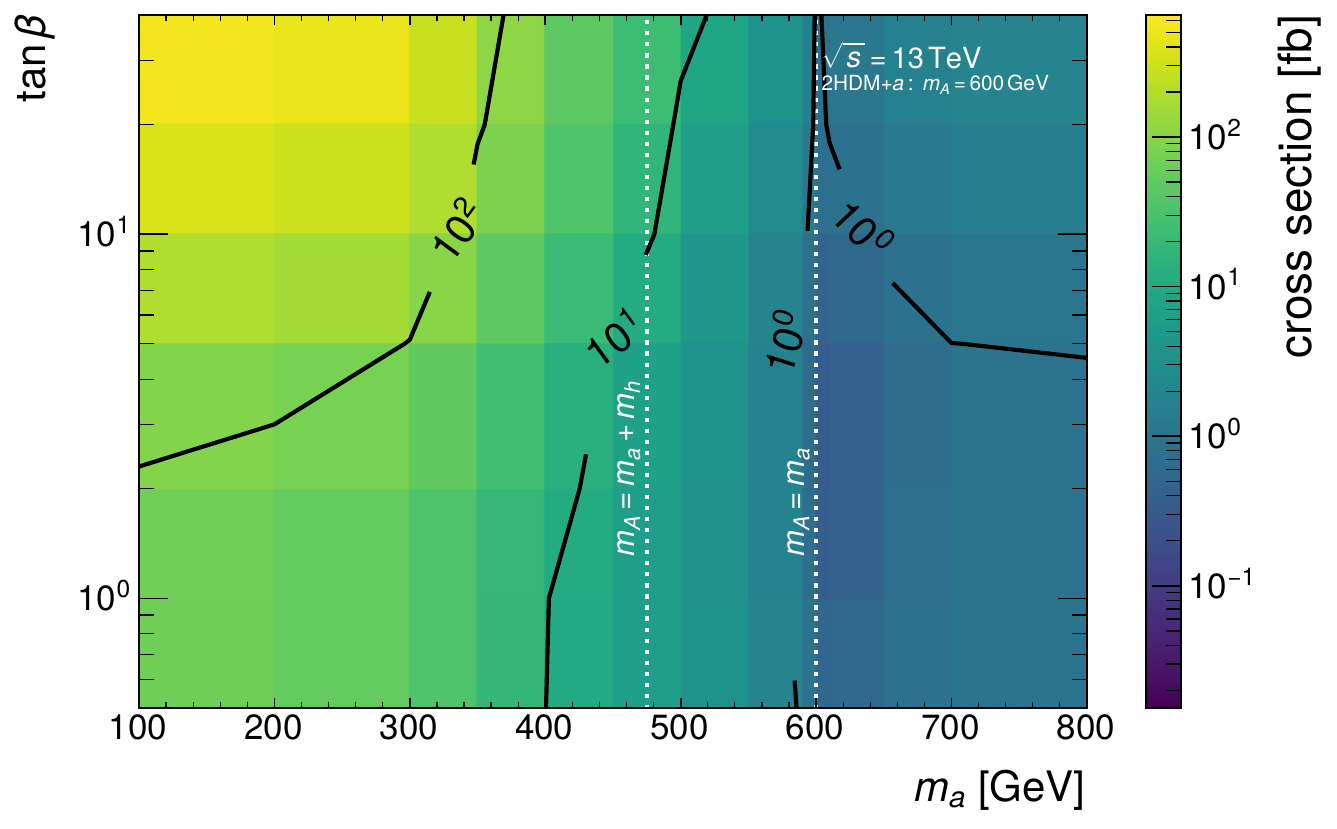}}
\end{myfigure}

\bigskip
In \subfigref{fig:interp_2HDMa_crossSections_matanB}{a}, the cross section for \xxqg processes is shown.
At small \ma and small \tanB, the cross section is large because the signature can be produced in gluon--gluon fusion (\cf\subfigref{fig:2HDMa_METjets_diagrams}{a}).
The cross section decreases exponentially with increasing \ma because the production of the pseudoscalar $a$ becomes less probable.
The cross section decreases moderately with increasing \tanB because top quarks contribute to the loop-induced production of the pseudoscalars~$a$ and $A$.
The coupling of the pseudoscalars to top quarks decreases with increasing \tanB (\cf\eqsref{eq:2HDMa_Gamma_a_ff}{eq:2HDMa_Gamma_A_ff}).
A second island of large cross section can be found at large \tanB because the coupling of the pseudoscalars to bottom quarks increases with increasing \tanB.
Here, the $s$-channel production is induced by bottom quarks from the proton \PDF.

\bigskip
In \subfigref{fig:interp_2HDMa_crossSections_matanB}{b}, the summed cross sections for loop-induced processes that do not come from the nominal \xxqg signature, \ie\xxZ, \xxh and \xxxxqg, are shown.
The cross section decreases with increasing \tanB because the coupling of the \BSM bosons~$a$, $A$ and $H$ to top quarks decreases.
This results in a decrease of loop-induced production and an increase in bottom-quark induced, tree-level production.
The latter can be seen in \subfigref{fig:interp_2HDMa_crossSections_matanB}{c} where the cross section increases with increasing \tanB.
In general, $\tanB\gg1$ is needed to obtain similar cross sections from loop-induced and tree-level processes because of the Yukawa coupling of the uncharged \BSM bosons to charged fermions and $m_t\gg m_b$.

In both figures, the cross section decreases with increasing \ma because the production of the pseudoscalar $a$ becomes less probable.
At $\ma>\mA=\mH$, processes involving the pseudoscalar $A$ or scalar~$H$ lead to cross sections independent of \ma, as was also seen in \figref{fig:interp_2HDMa_crossSections_mamA}{c}.
Contributions involve for example box diagrams for loop-induced processes and $t$-channel diagrams for tree-level processes. 

\section{\THDMa contributions to the \METjets measurement}
\label{sec:interpretation_2HDMa_contribs}

Following these considerations, it can now be investigated how the \THDMa concretely contributes to the phase space selected by the \METjets measurement.
As mentioned before, this can be studied at particle level because the \METjets measurement has been corrected for detector effects in \chapref{sec:metJets_detectorCorrection}.
The \THDMa contributions to a selection of the five measurement regions and two subregions are shown in \figref{fig:interp_signalBreakdown}.
As an example, the model point of $\ma=\SI{250}{GeV}$, $\mA=\SI{600}{GeV}$, $\tanB=1$ and all other parameters as defined in \tabref{tab:LHCDMWG_params} is used.
The selected yields at particle level are shown.
The top panels give differential cross section of Standard-Model (blue crosses) and \THDMa (green crosses) generation with their respective statistical uncertainties.
The quadrature sum of \SM (\THDMa) systematic and statistical uncertainties is displayed as a blue (green) shaded area.
In the top panels, the filled areas mark the contributions from the different \THDMa processes.
In the bottom panels, the filled areas mark the ratio of the different \THDMa processes to the total \THDMa contribution.
The blue crosses in the bottom panel denote the ratio of total \THDMa cross section to the \SM prediction.

\begin{myfigure}{
		Differential cross section at the particle level as a function of \METconst for the five measurement regions in the (left) \Mono and (right)~\VBF subregion.
		Blue (green) crosses denote the \SM (\THDMa) yield with their statistical uncertainty.
		The blue (green) shaded areas correspond to the total \SM (\THDMa) uncertainty.
		In the top panels, the filled areas correspond to the differential cross sections of the different contributing \THDMa processes.
		In the bottom panels, they give the fraction of the process from the total \THDMa contribution.
		The blue crosses in the bottom panels correspond to the ratio of the total \THDMa cross section to the \SM prediction, using the logarithmic scale drawn on the right.
	}{fig:interp_signalBreakdown}
	\subfloat[]{\includegraphics[width=0.49\textwidth]{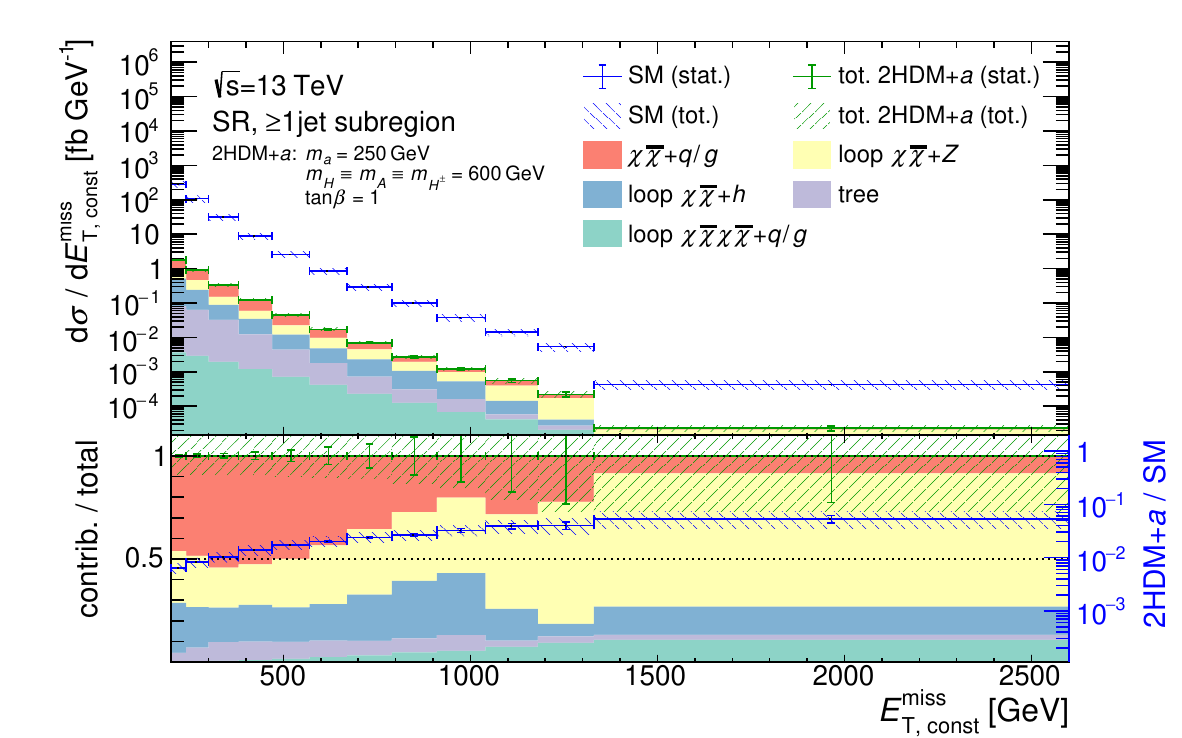}}
	\subfloat[]{\includegraphics[width=0.49\textwidth]{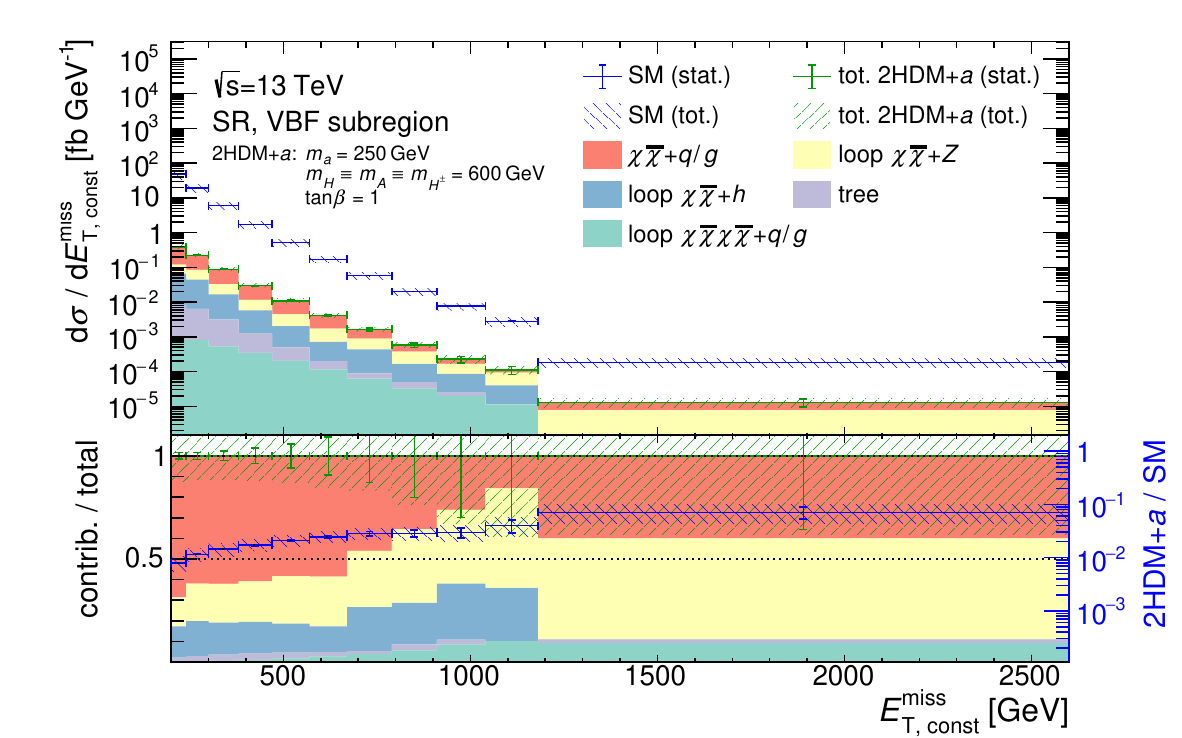}}\\
	\subfloat[]{\includegraphics[width=0.49\textwidth]{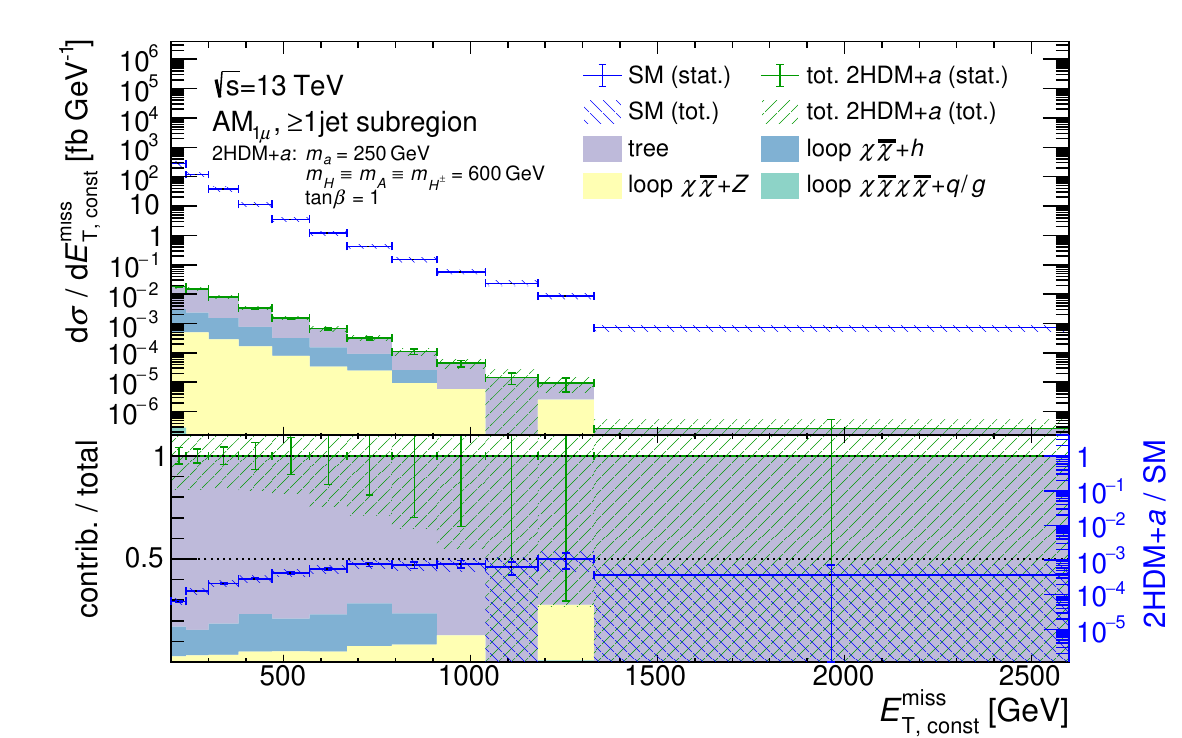}}
	\subfloat[]{\includegraphics[width=0.49\textwidth]{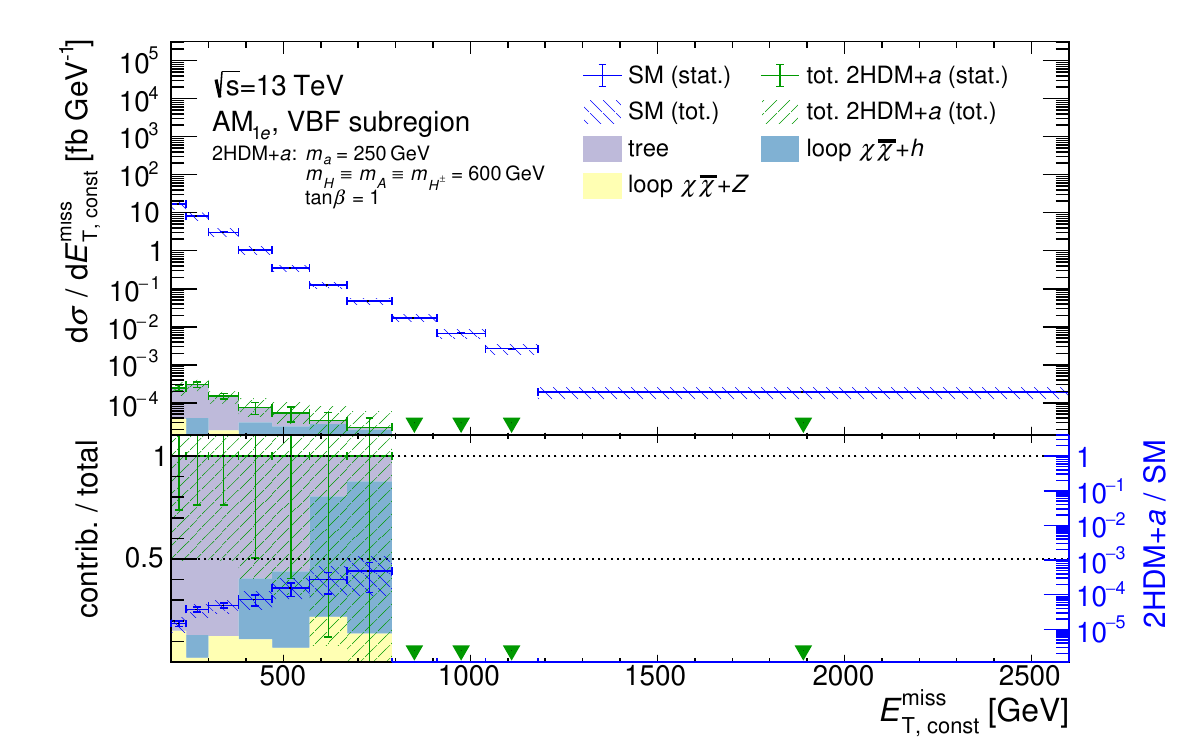}}\\
	\subfloat[]{\includegraphics[width=0.49\textwidth]{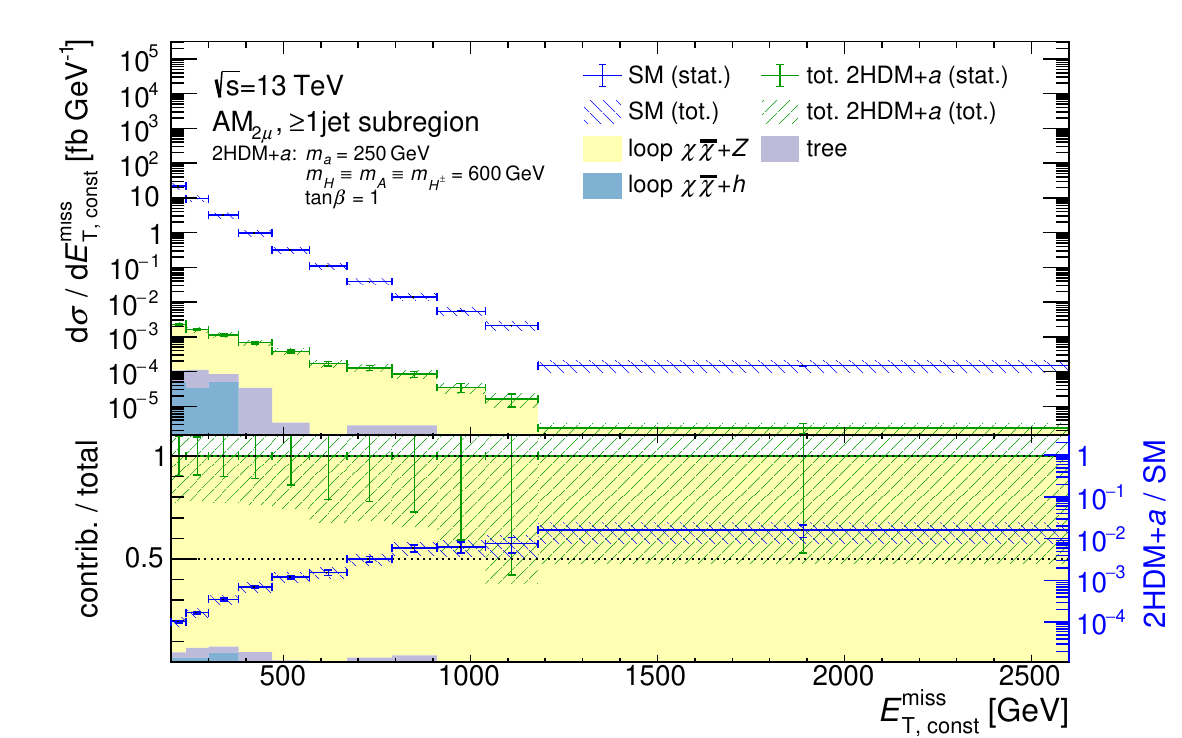}}
	\subfloat[]{\includegraphics[width=0.49\textwidth]{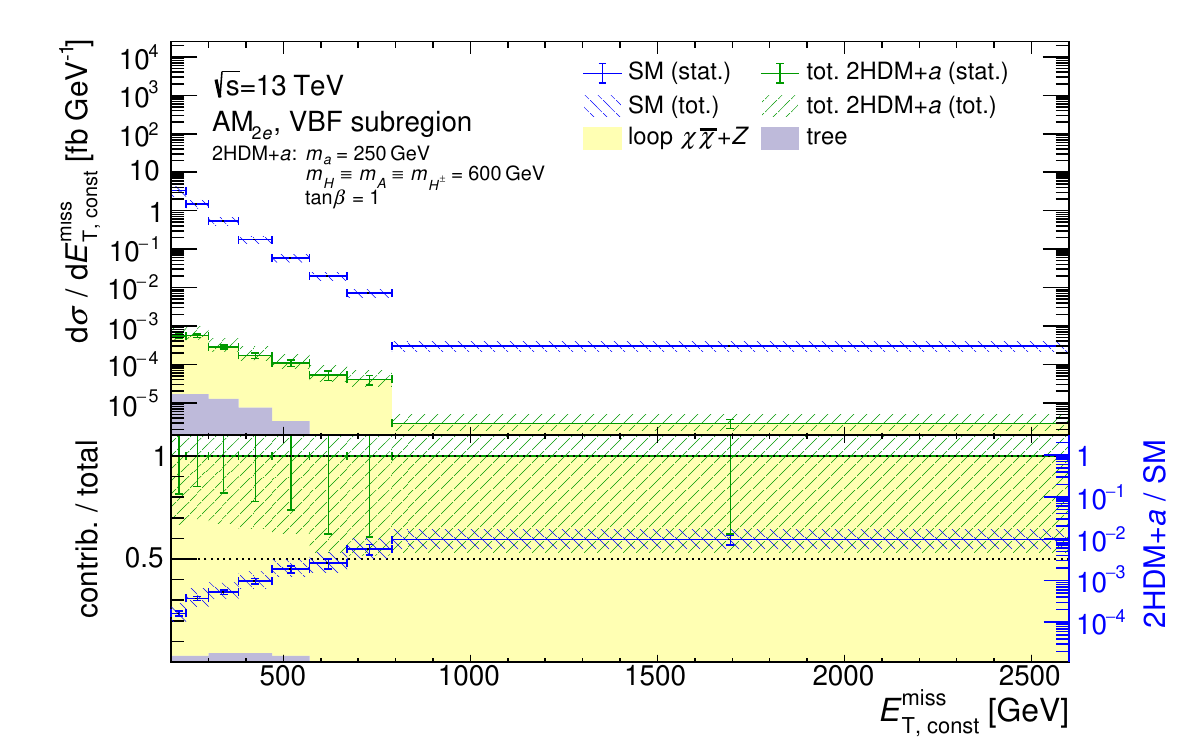}}\\
\end{myfigure}

\bigskip
In the signal region (\subfigsref{fig:interp_signalBreakdown}{a}{b}), $\METmeas\equiv\METconst$.
As $\tanB=1$, the Yukawa coupling of uncharged \BSM bosons to top quarks is larger than their coupling to bottom quarks.
In consequence, loop-induced processes dominate the sensitivity as they involve top quarks without having them in the final state (see \secref{sec:interpretation_2HDMa_contributions}).

\xxqg processes (red) exhibit mostly small values of $\METmeas\equiv\METconst$ because \METmeas often originates from one of the pseudoscalars recoiling against a small momentum \ISR jet.
There is a peak of the ratio of the \xxqg contribution to the total \THDMa cross section at $\METmeas\approx\SI{350}{GeV}$ because production of the pseudoscalar~$a$ in \ttbar fusion followed by invisible decays $a\to\xx$ gives rise to $\METmeas>2m_t$.

Loop-induced \xxZ processes (yellow) are selected by the signal region if the $Z$ boson decays hadronically, \Zqq, or if it decays to neutrinos, \Znunu.
In the latter case, the process has to be accompanied by initial-state radiation to have the required jet in the event.
This case leads to particularly large values of \METmeas as both the $Z$ and the pseudoscalar boson decay invisibly.

Loop-induced \xxh processes (blue) are selected by the signal region if the scalar~$h$ decays to bottom quarks, $h\to\bbbar$.
Loop-induced \xxxxqg processes (blue-green) give only a minor contribution in the signal region.
The same is true for tree-level processes~(purple), \eg the production of a \DM pair in association with hadronically decaying top quarks.

In total, the ratio of \THDMa to \SM prediction increases with \METconst because in particular the production of $\xx+Z(\to\nunubar)$ events in the \THDMa leads more often to events with large amounts of \METconst than the Standard Model.
The \THDMa contribution displays a gentle rise from approximately \SI{1}{\%} of the \SM yield at small \METconst to approximately \SI{5}{\%} at large \METconst.
The \THDMa contribution does not display a peaked shape.

In general, the systematic uncertainties for the \THDMa (\cf\secref{sec:interpretation_2HDMa_MC}) are independent of \METconst and amount to approximately \SI{10}{\%}.
It depends therefore on the available statistics, which decrease with \METconst, whether the systematic or statistical uncertainties dominate the total uncertainty.

The main difference between the \Mono and \VBF subregions (\subfigref{fig:interp_signalBreakdown}{a}~and~b, respectively) is that the \VBF subregion has an order of magnitude less statistics than the \Mono subregion.
This holds for \SM as well as \THDMa contributions.
The ratio of total \THDMa to \SM yields is marginally larger in the \VBF subregion.
The reason for this is that larger masses and therefore energy scales are involved in the production of \BSM bosons than in the production of \SM bosons.
This leads to generally larger transverse momenta of jets and can cause larger angular separation between jets in the event, mimicking the \VBF signature.

\bigskip
In \OneMuJetsAM in the \Mono subregion (\cf\subfigref{fig:interp_signalBreakdown}{c}), the dominant \THDMa contribution stems from tree-level processes~(purple).
This can for example be the production of a \DM pair in association with a leptonically decaying top quark or $W$ boson.
Minor additions are made by $\xx+h(\to WW^*)$ (blue) and $\xx+Z(\to\ell^+\ell^-)$ where one of the leptons is outside the fiducial volume (yellow).
This applies to \OneEJetsAM in the \Mono subregion, which is not shown, as well.

\bigskip
In \OneEJetsAM in the \VBF subregion (\cf\subfigref{fig:interp_signalBreakdown}{d}), the dominant \THDMa contribution at small \METconst also stems from tree-level processes~(purple) as was the case for the \Mono subregion.
At large \METconst, it is, however, dominated by production of pseudoscalars $A$ and their subsequent decays to a light pseudoscalar $a$ and a light scalar $h$ (blue).
Minor additions are made by $\xx+Z(\to\ell^+\ell^-)$ where one of the leptons is outside the fiducial volume (yellow).

The \VBF subregion has about two orders of magnitude less statistics for the \THDMa contributions than the \Mono subregion at small \METconst, contributing in total only about $10^{-5}$ of the \SM yield.
This is a larger difference than in the signal region because the tree-level contributions often include additional hadronically decaying top quarks or $W$ bosons.
The jets from these decays are in general not as well separated in \eta as required for the \VBF selection.
Due to the \xxh processes, the total \THDMa contribution at large \METconst is about \SI{0.1}{\%} of the \SM yield.

These smaller statistics have two striking consequences.
On the one hand, the region at large \METconst is not populated at all in the sample of generated \MC events.
On the other hand, the estimate for the parton-showering uncertainty is subject to large statistical uncertainties and dominates all other systematic \THDMa uncertainties.
It has a size similar to the nominal statistical uncertainty on the signal.

This applies to \OneMuJetsAM in the \VBF subregion, which is not shown, as well.

\bigskip
In \TwoLJetsAMs (\subfigsref{fig:interp_signalBreakdown}{e}{f}, respectively), the dominant \THDMa contribution by far comes from $\xx+Z(\to\ell^+\ell^-)$ processes (yellow) accompanied by initial-state radiation to have the required jet in the event.
As the momentum of the leptons is used in the calculation of \METconst, these processes lead to particularly large \METconst.
That is why the \THDMa contribution increases from \SI{0.01}{\%} at small \METconst to more than \SI{1}{\%} at large \METconst.
\Chapter[1]{Evaluating the results}{Statistical interpretation}{%
	Linkin Park}{Lost In The Echo~\cite{LinkinPark:2012lie}}{verse 2, line 2}
\label{sec:interpretation}





The results of the \METjets measurement obtained in the previous chapters can be interpreted with regard to the consistency of the measured data with the prediction according to the Standard Model or \BSM models. In principle, this can be performed either with the detector-level results from \secref{sec:metJets_detLevelResults} or the particle-level results from \secref{sec:detCorr_partLevelResults}.
In this chapter, the interpretation is performed at particle level because it removes the necessity to simulate the detector for the (B)\SM prediction, as discussed in \chapref{sec:analysisPreservation}.

\secref{sec:interpretation_framework} describes the statistical framework for the interpretation.
The results are interpreted with respect to the consistency of the measured data with the \SM prediction in \secref{sec:interpretation_SM} and with the prediction of the \THDMa in \secref{sec:interpretation_2HDMa}.

\section{Statistical framework}
\label{sec:interpretation_framework}

\subsection{Likelihood function}
\label{sec:interpretation_likelihood}

The \textit{likelihood} of observing the data $\vv x$ given a prediction $\vv\pi$ of a hypothesis is taken as
\begin{equation}
	\label{eq:metJets_likelihood}
	\lh\left(\vv{x}~|~\vv\pi, \vv\theta\right)
	\coloneqq
	\frac{1}{\sqrt{(2\pi)^{k}~\det\Cov}}
	\cdot
	e^{-\frac{1}{2}\chiSqYields\left(\vv{x}, \vv\pi, \vv\theta\right)}
	\cdot
	\prod_i
	\frac{1}{\sqrt{2\pi}}e^{-\frac{1}{2}\theta_i^2}.
\end{equation}
The full derivation of \eqref{eq:metJets_likelihood} is given in \appref{app:interpretation_likelihood}.
The covariance matrix \Cov is determined from statistical uncertainties and correlations according to \eqref{eq:metJets_covariance}.
The correlations have to be taken into account because the \Mono and \VBF subregions of the \METjets measurement are not orthogonal (\cf\secref{sec:metJets_subregions}).

The likelihood in \eqref{eq:metJets_likelihood} represents the product of $k$ independent measurements, \ie yields in bins, distributed according to normal distributions.
The product is rotated to incorporate correlations between the measurements as off-diagonal elements in the covariance matrix.
Normal instead of Poisson distributions can be assumed for the event counts because as a consequence of the unfolding the binning was chosen such that the event count in each bin is sufficiently high: for 20 or more events the relative error by assuming normal instead of Poisson distributions is less than \SI{9}{\%}~\cite{Rich:2009enp}.
The last factor is a product of probability density functions (\pdf{}s) for \textit{nuisance parameters} $\theta_i$ modelling the systematic uncertainties.
Nuisance parameters are any constrained parameters that are not the primary target of the investigation.
They are assumed to be distributed according standard normal distributions.

\chiSqYields in \eqref{eq:metJets_likelihood} is defined as
%
\begin{equation}
	\label{eq:interpretation_chiSqYields}
	\chiSqYields\left(\vv{x}, \vv\pi, \vv\theta\right)\coloneqq
	\left(\vv{x} - \vv{\pi}+\sum_i \theta_i \cdot \vv{\epsilon}_i \right)^\text{T}
	\Cov^{-1}
	\left(\vv{x} - \vv{\pi} + \sum_i \theta_i \cdot{\vv\epsilon}_i \right),
\end{equation}
where $\vv\epsilon_i$ is the absolute uncertainty amplitude associated with a nuisance parameter~$\theta_i$.
Systematic uncertainties are smoothed and symmetrised to decrease the impact of statistical fluctuations.
Depending on the performed statistical test, the prediction $\vv\pi$ can be set to different hypotheses.
This is discussed in the next paragraph.

\subsection{\SM predictions}
\label{sec:interpretation_SM_predictions}

The measured data are denoted $\vv x$ in \eqref{eq:interpretation_chiSqYields}.
In the interpretation with respect to the Standard Model (\cf\secref{sec:interpretation_SM}), the prediction~$\vv\pi$ is identical to the \SM prediction, $\vv\pi\equiv\vvpiSM$.
In the interpretation with respected to a \BSM model (\eg\secref{sec:interpretation_2HDMa}), the prediction additionally gets a component of the \BSM signal: $\vv\pi=f(\vv s)+\vvpiSM$, where $f$ denotes a general function of the \BSM signal~$\vv s$.

There are different approaches to define the \SM prediction~\vvpiSM, which are discussed in the following.
The yields of \SM process $i$ in the corresponding bins obtained in Monte-Carlo (\MC) generation shall be denoted $\vv b_i$.
One prediction for the Standard Model that can be used is
\begin{equation}
	\label{eq:interpretation_SM_pred_fixedNorm}
	\vvpiSM=\sum_i \vv b_i\eqqcolon \vv b.
\end{equation}
Hereby, $\vv b$ are the summed yields of all relevant \SM processes.
Interferences within one contribution $\vv b_i$ are taken into account.
This \SM prediction is useful because the prediction is taken from \MC generation right away and therefore a direct statement of whether the \MC generation agrees with the measured data can be made.
This approach is often taken in measurements of \SM processes and called "fixed normalisation" in the following.

\bigskip
In the above approach, the normalisation is associated to potentially mismodelled quantities, \eg the jet energy scale.
An alternative approach is to assign a cross-section parameter to the most important \SM contributions.
For this, additional normalisation parameters $\mu_i$ for \SM process $i$ with yield $\vv b_i$ are introduced.
The prediction becomes
\begin{equation}
	\label{eq:interpretation_SM_pred_floatNorm}
	\vvpiSM=\sum_i \mu_i \vv b_i.
\end{equation}
In this case, the shape of the yields for \SM processes is still taken from \MC generation, while the normalisation is now estimated using the observed data in a simultaneous fit.
Normalisation parameters are not constrained here, in contrast to the nuisance parameters mentioned earlier.

The approach in \eqref{eq:interpretation_SM_pred_floatNorm} can be understood as measuring the cross section of processes simultaneously to comparing the shape of the distributions between measured data and \MC generation.
This prediction is called "floating normalisation" in the following.
This approach is used among others in searches for \BSM physics~\cite{Baak:2014wma} because modelling the physics accurately in the extreme phase spaces selected in these searches is difficult.
The accuracy of the prediction is improved by determining cross-section parameters simultaneously in the data of related phase spaces.

\subsection{Hypothesis test}

In hypothesis tests it can be decided whether a hypothesis $H_0$ has to be rejected in favour of an alternative hypothesis~$H_1$.
The hypothesis $H_0$ which is rejected or failed to be rejected is called \textit{null hypothesis}.
For the hypothesis test, let
\begin{equation}
	\label{eq:interpretation_q}
	q\coloneqq
	-2\ln\frac{
		\lh_0\left(\vv{x}~|~\hat{\vv\pi}_0, \hat{\vv\theta}_0\right)
	}{
		\lh_1\left(\vv{x}~|~\hat{\vv\pi}_1, \hat{\vv\theta}_1\right)
	}
\end{equation}
be a \textit{test statistic} based on a likelihood ratio, where $\lh_i$ is the likelihood under hypothesis~$H_i$.
In this choice of test statistic, $\hat{\vv{\pi}}$ and $\hat{\vv\theta}_i$ are the \textit{pulled} values of the prediction~$\vv\pi$ and nuisance parameters $\vv\theta$.
These are the values that maximise the likelihood~$\lh_i$:
\begin{equation*}
	\hat{\vv\pi}_i, \hat{\vv\theta}_i\coloneqq\underset{\vv\pi, \vv\theta}{\arg\max}\,\lh_i\left(\vv x | \vv\pi_i, \vv\theta_i\right),
\end{equation*}
determined in a fit.
In the fit setup used in this chapter, the prediction~$\vv\pi$ only has fit parameters that can differ before and after the fit if the floating-normalisation prediction according to \eqref{eq:interpretation_SM_pred_floatNorm} is used.
In the fixed-normalisation approach, all parameters in the prediction~$\vv\pi$ are fixed.
In both cases, all other variability of the data~$\vv x$ and prediction~$\vv\pi$ is accounted for by the nuisance parameters~$\vv\theta$.

\subsection{Pulls and constraints}
\label{sec:interpretation_pullsAndConstraints}

Normal distributions constraining the nuisance parameters are assumed, as was mentioned in \secref{sec:interpretation_likelihood}.
The status before maximising the likelihood with respect to the nuisance parameters $\vv\theta$ and prediction~$\vv\pi$, if applicable, is called \textit{pre-fit}.
Pre-fit, the nuisance parameters are normalised such that they have a zero mean and unit standard deviation, \ie $\theta_0\coloneqq\theta_\textnormal{pre-fit}=0$ and $\sigma_{\theta_0}\coloneqq\sigma_{\theta_\textnormal{pre-fit}}=1$.

In the \METjets measurement, nuisance parameters for the theoretical uncertainties change the values for the \SM prediction~$\vvpiSM$.
Nuisance parameters for the experimental uncertainties change the values of the measured data~$\vv x$ as the measurement was corrected for detector effects.

The \textit{relative uncertainty} $u_{j, \vv y}^i$ for a nuisance parameter $i$ of the yield $y_j\in\left\{x_j, \pi_j\right\}$ in bin $j$ is
\begin{equation*}
	u_{j, \vv y}^i
	= \frac{\sigma_{\theta^i}\cdot\varepsilon^i_j}{y_j}.
\end{equation*}

\bigskip
After the fit (\textit{post-fit}), the nuisance parameters for systematic uncertainties can be pulled, \ie take non-zero values $\hat{\theta}\coloneqq\theta_\textnormal{post-fit}\neq0$.
This changes the yield according to the size and direction of the pull as well as the size of the uncertainty.
Pulling nuisance parameters allows better agreement between data and prediction to be achieved.
Pulling a nuisance parameter is penalised in the fit because of the normal distribution term for nuisance parameters in \eqref{eq:metJets_likelihood}.

Further, the allowed range for one standard deviation of the nuisance parameter can be \textit{constrained} post-fit, \ie take non-unit values $\sigma_{\hat\theta}\coloneqq\sigma_{\theta_\textnormal{post-fit}}\neq1$.
This effectively changes the size of the corresponding systematic uncertainty.

\subsection{Test statistics for the interpretations}

The likelihoods that are used in the test statistic given in \eqref{eq:interpretation_q} depend on the hypotheses that shall be tested.
These are described separately for the two interpretations covered in this chapter, Standard Model and \THDMa.

\subsubsection{Interpretation with respect to the Standard Model}

For the interpretation with respect to the Standard Model, a test of the so-called \textit{goodness of fit}~\cite{Cousins2013GeneralizationOC} is performed.
This test investigates how well the generated \SM prediction describes the measured data within their respective uncertainties.

The null hypothesis is the \SM prediction \vvpiSM to be true.
The alternative hypothesis is a \textit{saturated model} that sets the expected and observed yields to identical values, $\vv x\equiv\vv\pi$.
The test statistic becomes:

\begin{equation}
	\begin{aligned}
		\label{eq:interpretation_qSM}
		\qSM
		\coloneqq&
		-2\ln\frac{\lh_{\SM}}{\lh_\text{sat}}
		=
		-2\ln\frac{
			\lh\left(\vv{x}~|~\hatvvpiSM, \hat{\vv\theta}_{\SM}\right)
		}{
			\lh\left(\vv{x}~|~\vv{x}, 0\right)
		}\\
		=&
		\chiSqYields\left(\vv{x}, \hatvvpiSM, \hat{\vv\theta}_{\SM}\right)+\sum_i\hat{\theta}_{i,\SM}^2
	.
	\end{aligned}
\end{equation}

The saturated model thereby helps to obtain a correct normalisation in the statistical test.
The generated \SM prediction and the measured data are varied within their respective uncertainties, which is expressed by the nuisance parameters $\vv\theta$.

The test statistic \qSM is distributed according to a chi-square distribution, $\chi^2_{k-n}$.
The number of degrees of freedom are equal to the difference between number of considered bins~$k$ and number of free-floating parameters~$n$ in the prediction, if any.
As criterion for the goodness of fit, the \textit{reduced \chiSq}$\coloneqq \frac{\qSMobs}{k-n}$ for an observed value of the test statistic \qSMobs is used.

\subsubsection{Interpretation with respect to the \THDMa}
\label{sec:interpretation_LR_2HDMa}

\textit{Exclusion limits} on a \BSM model represent the parameter space in which the model is severely disfavoured.
They can be derived by taking the null hypothesis $H_0$ to be the joint yield of \BSM processes $\vv s$ and \SM prediction \vvpiSM,
\begin{equation}
	\label{eq:interpretation_piBSM}
	\vv\pi=\vv s+\vvpiSM.
\end{equation}
The hypothesis of only the \SM prediction being true becomes the alternative hypothesis~$H_1$, with likelihood $\lh_{\SM}$ as used in \eqref{eq:interpretation_qSM}.
As a test statistic,
\begin{equation}
	\label{eq:metJets_qBSM}
	\qBSM
	\coloneqq
	-2\ln\frac{\lh_{\BSM}}{\lh_{\SM}}
	=
	-2\ln\frac{
		\lh\left(\vv{x}|\vv s+\hatvvpiSM,\hat{\vv{\theta}}_{\BSM}\right)
	}{
		\lh\left(\vv{x}|\hatvvpiSM, \hat{\vv{\theta}}_{\SM}\right)
	}
\end{equation}
is employed in the interpretation with respect to the \THDMa.

The frequency of observing the data if the prediction of hypothesis $H_0$ is to be rejected against that of $H_1$ is described by the \textit{\pValue}
\begin{equation}
	\label{eq:interpretation_pValue}
	p\left(\qObs, \pdf\left(q\right)\right):=1-\int_{q_\text{obs}}^{\infty}\pdf\left(q\right)\,\mathrm{d}q.
\end{equation}
The \pValue quantifies the probability to find a value of the test statistic $q$ that is more extreme than the observed value $q_\text{obs}$.
The \pValue of a likelihood ratio like \eqref{eq:interpretation_q} has the highest probability at a given significance level $\alpha>p(\qObs,\pdf\left(q\right))$ to correctly reject $H_0$ if $H_1$ is true~\cite{Neyman:1933wgr}.

The \pValue for the signal-plus-background hypothesis given the background-only hypothesis is then $\pValueCLsb\coloneqq p\left(\qBSMobs,\pdf\left(\qBSM^{\vv s+\vvpiSM}\right)\right)$ where \qBSMobs is the observed val\-ue of the test statistic \qBSM.
The probability density function of \qBSM is determined assuming $H_0$ to be true, \ie $\vv\pi=\vv s+\vvpiSM$.
For this, alternative distributions of the prediction (\textit{toys}) are produced by varying the prediction $\vv\pi\rightarrow\vv\pi'$ within the uncertainties and calculating \qBSM for $\vv x\equiv\vv\pi'$.
The phase space of \BSM model parameters in which $H_0$ is significantly disfavoured compared to $H_1$ given the measured data, \ie $\pValueCLsb<\alpha$, is then called \textit{excluded} at confidence level $1-\alpha$.

Using \pValueCLsb for setting exclusion limits has the disadvantage that \BSM models can also be excluded at confidence level $1-\alpha$ in two undesirable cases:
on the one hand, when the signal-plus-background and background-only hypotheses are indistinguishable, $\vv s+\hatvvpiSM\approx\hatvvpiSM$; on the other hand when the likelihood for the background-only hypothesis is small, independent of the signal-plus-background hypothesis, $\lh_{\SM}\approx0$.
In both cases, \pValueCLsb allows for an exclusion, although it has to be questioned whether the performed measurement is suitable for the statement.

The \CLs technique~\cite{Read:2000ru,Read:2002hq} is therefore employed to be conservative.
It defines
\begin{equation}
	\label{eq:interpretation_CLsb}
	\begin{aligned}
		\CLsb&\coloneqq1-p\left(\qBSMobs,\pdf\left(\qBSM^{\vv s+\vvpiSM}\right)\right)=1-\pValueCLsb\\
		\CLb&\coloneqq1-p\left(\qBSMobs,\pdf\left(\qBSM^{\vvpiSM}\right)\right).
	\end{aligned}
\end{equation}

For the latter, the probability density function of \qBSM is determined assuming $H_1$ to be true, \ie $\vv\pi=\vvpiSM$.
With these definitions, a refined confidence level is constructed as
\begin{equation}
	\label{eq:metJets_CLs}
	\CLs\coloneqq\frac{\CLsb}{\CLb}.
\end{equation}
Exclusion limits at \SI{95}{\%} confidence are set on the \THDMa phase space if $\CLs<0.05$.

The \CLs technique is conservative because, if $H_0$ and $H_1$ are indistinguishable by the measurements, $\CLsb\approx\CLb$ and $\CLs\approx1$.
Alternatively, if the observed value of the test statistic is so small that the estimation of the \SM-only hypothesis has to be questioned, $\CLb\approx0$ and $\CLs\gg0$ independent of \CLsb.
In both cases, \CLs prevents discarding $H_0$ in favour of $H_1$, while \CLsb would incorrectly allow for an exclusion.

\bigskip
So called \textit{observed} exclusion limits are set by using the measured data $\vv x$.
\textit{Expected} exclusion limits are derived following the procedure described in \refcite{Cranmer:2014lly}:
A background-only fit of the data $\vv x$ assuming the background-only hypothesis, $\vv\pi=\vvpiSM$, is performed.
From this, a new "pre-fit" condition is defined:
The data is set to the post-fit \SM yield of the background-only fit, $\vv x=\hatvvpiSM$.
All fit parameters take as pre-fit values the post-fit values they assumed in the background-only fit, $\vv{\theta}_\mathrm{pre-fit}=\hat{\vv{\theta}}_{\SM}$.
Most importantly, the constraints of nuisance parameters are adjusted, $\vv\sigma_{\theta, \textnormal{pre-fit}}=\vv\sigma_{\hat\theta_{\SM}}$.
With this new pre-fit setup, exclusion limits for the signal-plus-background hypothesis are derived according to \eqref{eq:metJets_qBSM}.

\section{Interpretation with respect to the Standard Model}
\label{sec:interpretation_SM}

After the theoretical considerations regarding the statistical framework in the previous section,  the consistency of the measured data with the generated \SM prediction within the uncertainties can be assessed in this section.
Different input quantities and \SM predictions \vvpiSM are probed in \secsref{sec:interpretation_SM_fixedNorm_diffXS}{sec:interpretation_SM_floatNorm_diffXS}{sec:interpretation_SM_fixedNorm_Rmiss}.
A summary is provided in \secref{sec:interpretation_SM_summary}.

\tabref{tab:interpretation_SMpredictions} gives an overview of the employed \SM predictions.
For the \SM predictions in the whole chapter, an improved setup for \MC generation of \SM processes compared to the one in \chapsrefAnd{sec:metJets}{sec:metJets_detectorCorrection} is used.
The improved setup is described in \secref{sec:interpretation_SM_MC}.
Theoretical systematic uncertainties are adjusted correspondingly.
The simple swap of \MC setup highlights one of the advantages of correcting for detector effects mentioned in \secref{sec:analysisPreservation}: the ability to incorporate improvements in the theory predictions when they become available.

\begin{table}[b]
	\centering
	\begin{tabular}{ccc}
		\toprule
		name & \SM prediction \vvpiSM & normalisation parameters\\
		\midrule
		fixed normalisation & $\sum_i \vv b_i$ & none\\
		floating normalisation & $\sum_i \mu_i \vv b_i$ & \muVjets, \muTop (all others fixed to 1)\\
		\bottomrule
	\end{tabular}
	\caption{Settings used for the fixed- and floating-normalisation \SM predictions.}
	\label{tab:interpretation_SMpredictions}
\end{table}

\newcommand{\numExpSysts}{253\xspace}
\newcommand{\numThSysts}{75\xspace}

\newcommand{\fixedDiffXSNDists}{10\xspace}
\newcommand{\fixedDiffXSChiTwo}{173.4\xspace}
\newcommand{\fixedDiffXSNdF}{107\xspace}
\newcommand{\fixedDiffXSChiTwoNdF}{1.62\xspace}
\newcommand{\fixedDiffXSPVal}{\ensuremath{5.2\cdot10^{-5}}\xspace}
\newcommand{\fixedDiffXSMinPVal}{0.026\xspace} 

\newcommand{\floatDiffXSNDists}{10\xspace}
\newcommand{\floatDiffXSChiTwo}{170.3\xspace}
\newcommand{\floatDiffXSNbins}{107\xspace}
\newcommand{\floatDiffXSNdF}{105\xspace}
\newcommand{\floatDiffXSChiTwoNdF}{1.59\xspace}
\newcommand{\floatDiffXSPVal}{\ensuremath{5.7\cdot10^{-5}}\xspace}
\newcommand{\floatDiffXSMinPVal}{0.017\xspace} 

\newcommand{\fixedRmissNDists}{8\xspace}
\newcommand{\fixedRmissChiTwo}{114.9\xspace}
\newcommand{\fixedRmissNdF}{84\xspace}
\newcommand{\fixedRmissPVal}{\ensuremath{1.4\cdot10^{-2}}\xspace}
\newcommand{\fixedRmissMinPVal}{0.015\xspace} 

\subsection{Nominal Standard-Model fit}
\label{sec:interpretation_SM_fixedNorm_diffXS}

The differential cross sections ($\mathrm{d}\sigma/\mathrm{d}\METconst$) in all five measurement regions and both subregions binned in \METconst serve as the investigated quantity for data~$\vv x$ and prediction~\vvpiSM in the goodness-of-fit test.
The \SM prediction with fixed normalisation according to \eqref{eq:interpretation_SM_pred_fixedNorm} is used.
This prediction exploits all available information from measured data and \SM prediction and uses the \SM prediction directly as it is generated.

In total, \fixedDiffXSNDists distributions are included in the fit, amounting to \fixedDiffXSNdF bins.
This is also given in the first row of \tabref{tab:interpretation_SM_fitResults}.
\numExpSysts nuisance parameters for systematic uncertainties on the measured data and \numThSysts on the theory predictions are taken into account.

\begin{mytable}{
		Results for the \SM fit in the different approaches.
		The number of degrees of freedom corresponds to the difference between number of bins~$k$ and number of floating parameters~$n$.
	}{tab:interpretation_SM_fitResults}{cccccccc}
		norm. of prediction & quantity & num. distributions & num. deg. of freedom & test statistic & red. \chiSq\\
		&&& $k-n$ & \qSMobs & \qSMobs$/(k-n)$\\
		\midrule
		fixed & \dSigmaDMET & \fixedDiffXSNDists & \fixedDiffXSNdF & \fixedDiffXSChiTwo & \fixedDiffXSChiTwoNdF\\
		floating & \dSigmaDMET & \floatDiffXSNDists & \floatDiffXSNdF & \floatDiffXSChiTwo & \floatDiffXSChiTwoNdF\\
		fixed & \Rmiss & \fixedRmissNDists & \fixedRmissNdF & \fixedRmissChiTwo & \fixedRmissChiTwoNdF\\
\end{mytable}

\subsubsection{Differential cross sections}
The differential cross sections pre- and post-fit are shown in \figref{fig:interp_SM_distributions_postfit_fixedNorm_diffXS}.
The pre-fit values for the data are identical to those shown in \figref{fig:detCorr_partLevelResults}.
The pre-fit values for the \SM prediction are comparable despite the improved \SM prediction (\cf\secref{sec:interpretation_SM_MC}).
The bottom panels show the ratio to the pre-fit \SM prediction.
The pre-fit allowed range for the measured data (generated \SM values) within one standard deviation is shown as a red (blue) shaded band.
The post-fit values with their total uncertainties for the measured data (generated \SM values) are marked by black dots (blue crosses).
Nuisance parameters for the theoretical (experimental) uncertainties change the values for the \SM prediction~$\vvpiSM$ (data~$\vv x$), as mentioned in \secref{sec:interpretation_pullsAndConstraints}.

\newcommand{\postfitDistCaption}[6]{
		#1 and ratio to pre-fit \SM generation at the particle level for #2.
		The #3-normalisation \SM prediction#4 is employed.
		#5%
		The red (blue) shaded areas correspond to the pre-fit allowed range for the measured data (generated \SM values) within one standard deviation.
		Black dots (blue crosses) denote the post-fit measured data (values in \SM generation) with their total uncertainty.
		#6%
		For a fit with perfect agreement, the blue crosses and black dots would match in the top and bottom panels.
}
\begin{myfigure}{
		\postfitDistCaption{Differential cross section}{%
			the five measurement regions as a function of \METconst in the (left) \Mono and (right)~\VBF subregion%
		}
		{fixed}{}{}{}
	}{fig:interp_SM_distributions_postfit_fixedNorm_diffXS}
	\subfloat[]{\includegraphics[width=0.48\textwidth]{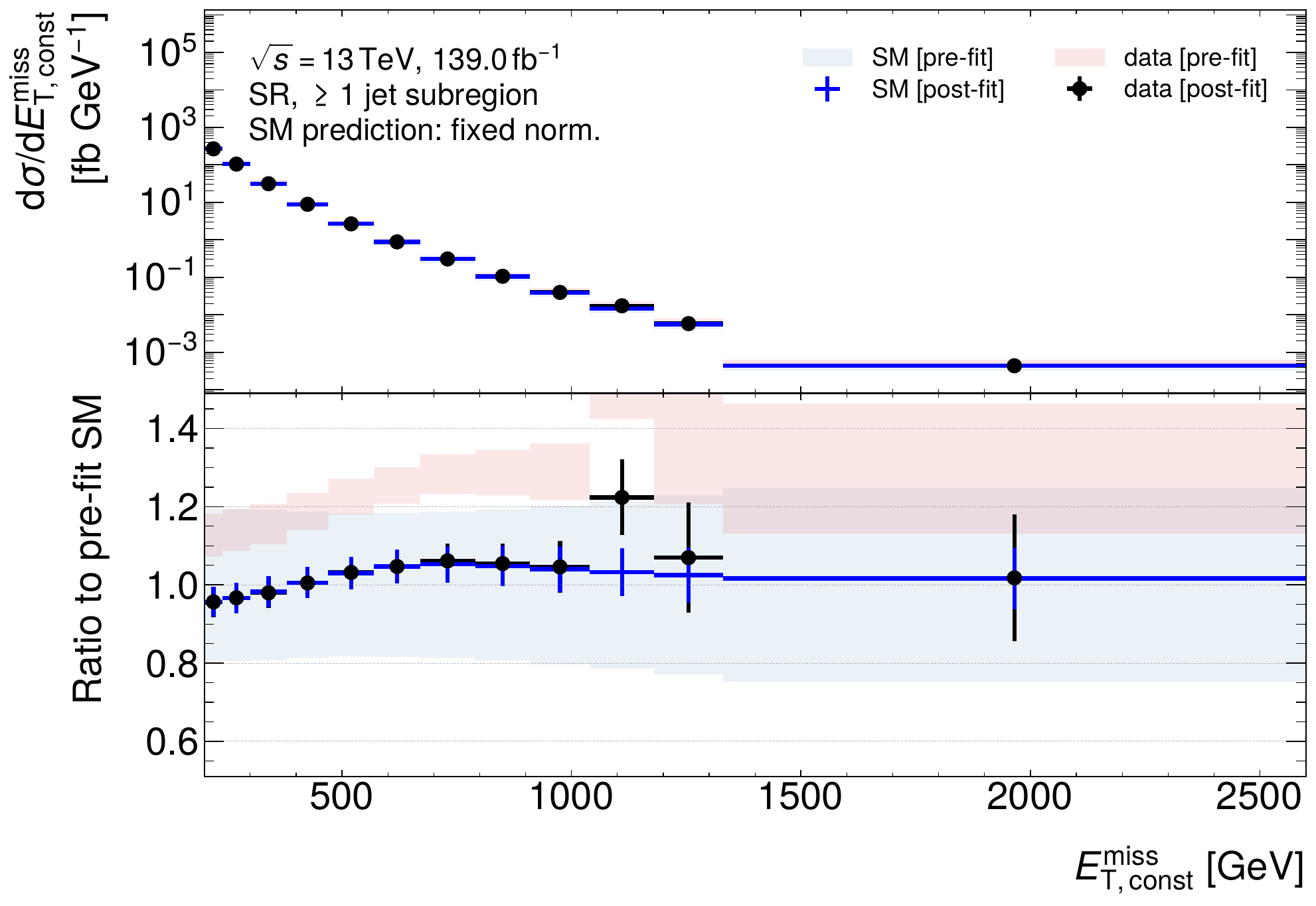}}
	\hspace{10pt}
	\subfloat[]{\includegraphics[width=0.48\textwidth]{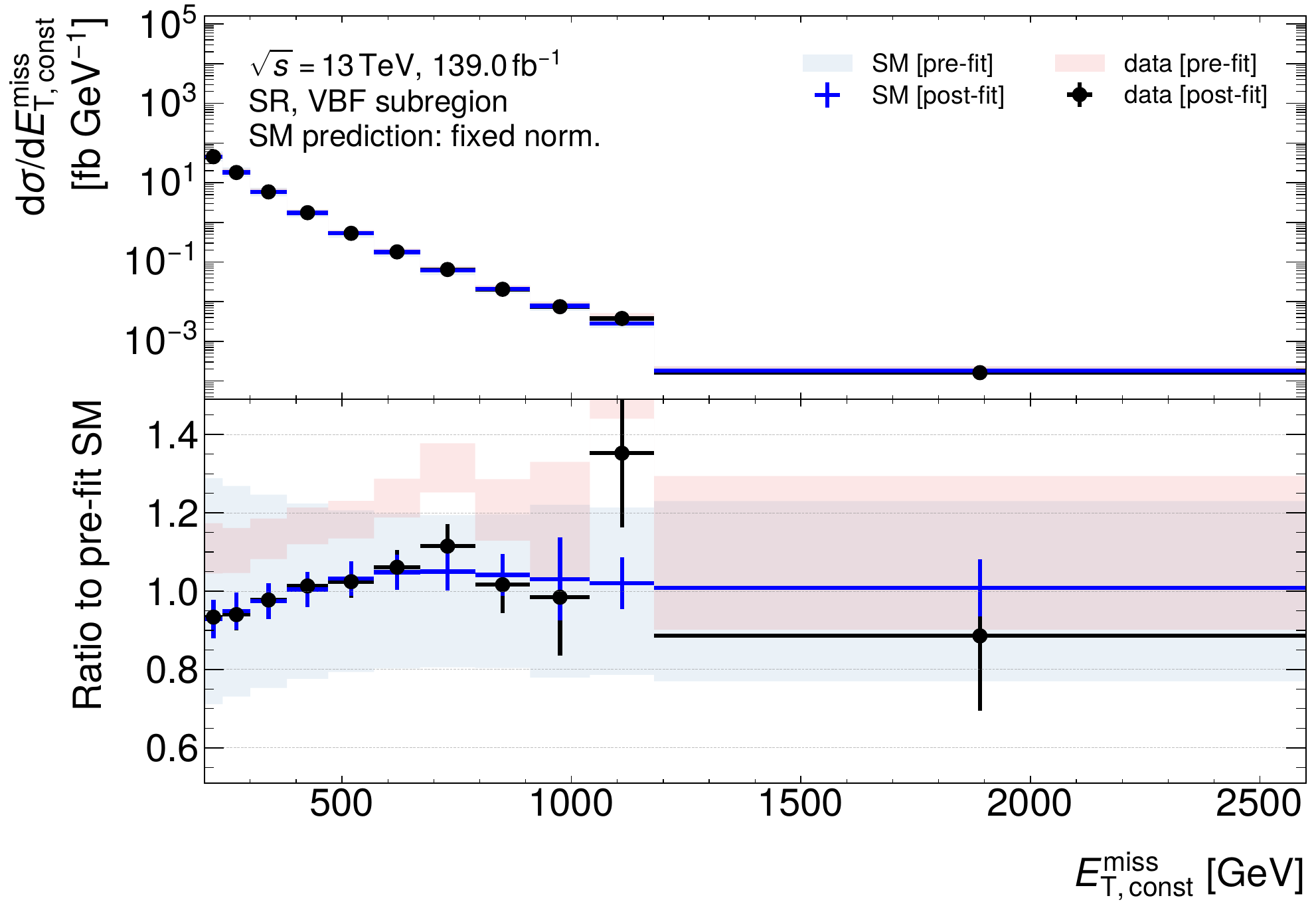}}\\
	\subfloat[]{\includegraphics[width=0.48\textwidth]{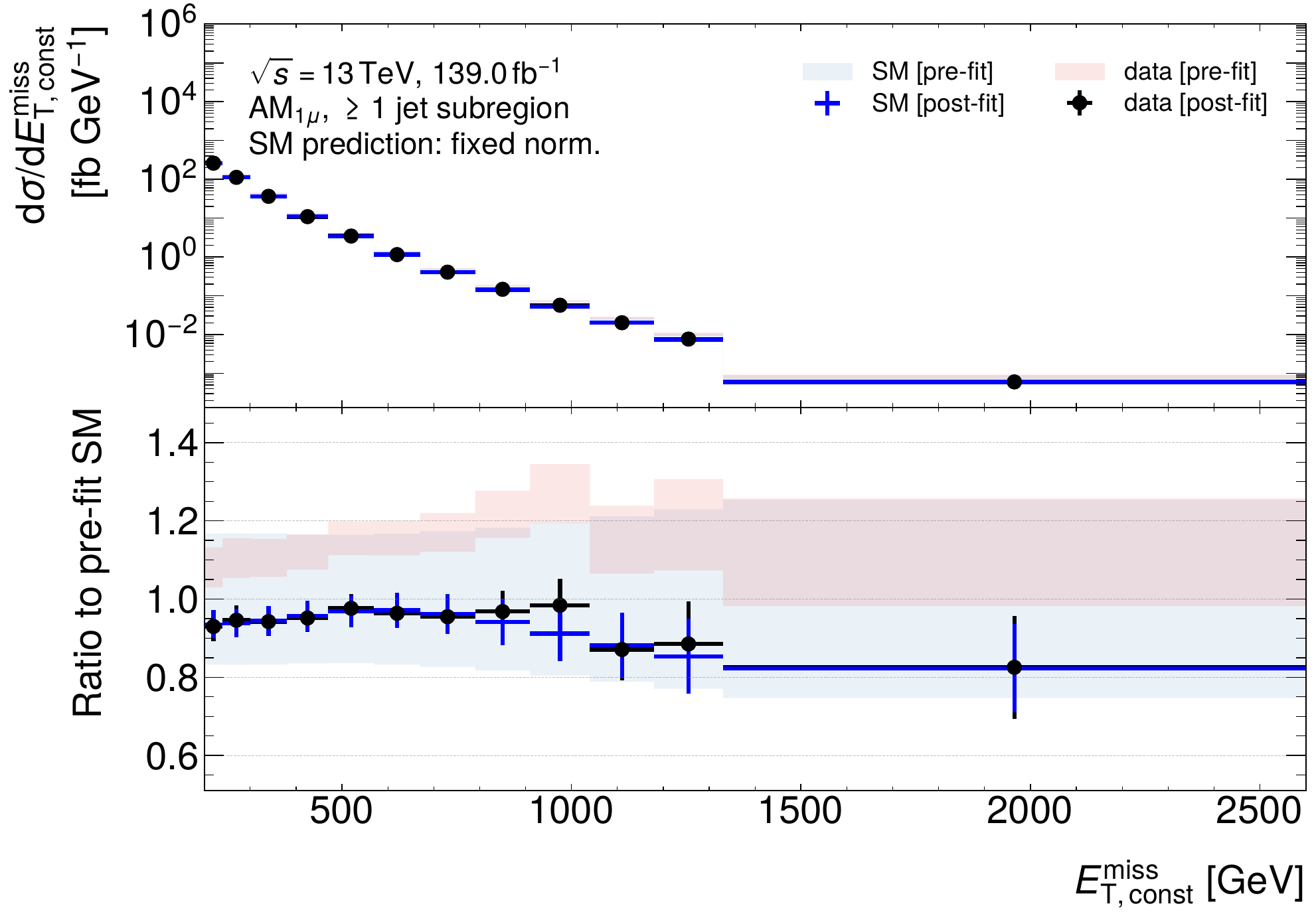}}
	\hspace{10pt}
	\subfloat[]{\includegraphics[width=0.48\textwidth]{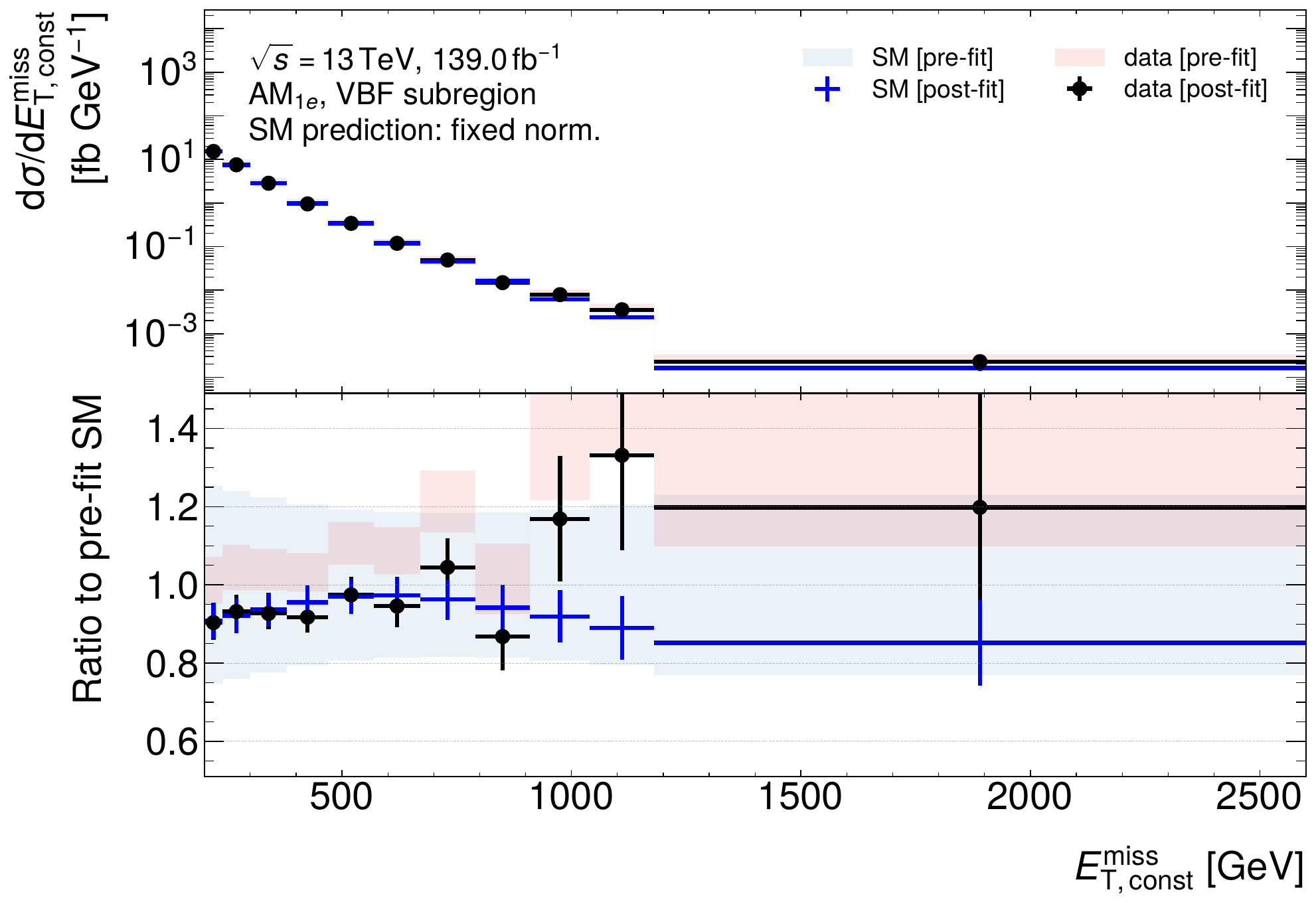}}\\
	\subfloat[]{\includegraphics[width=0.48\textwidth]{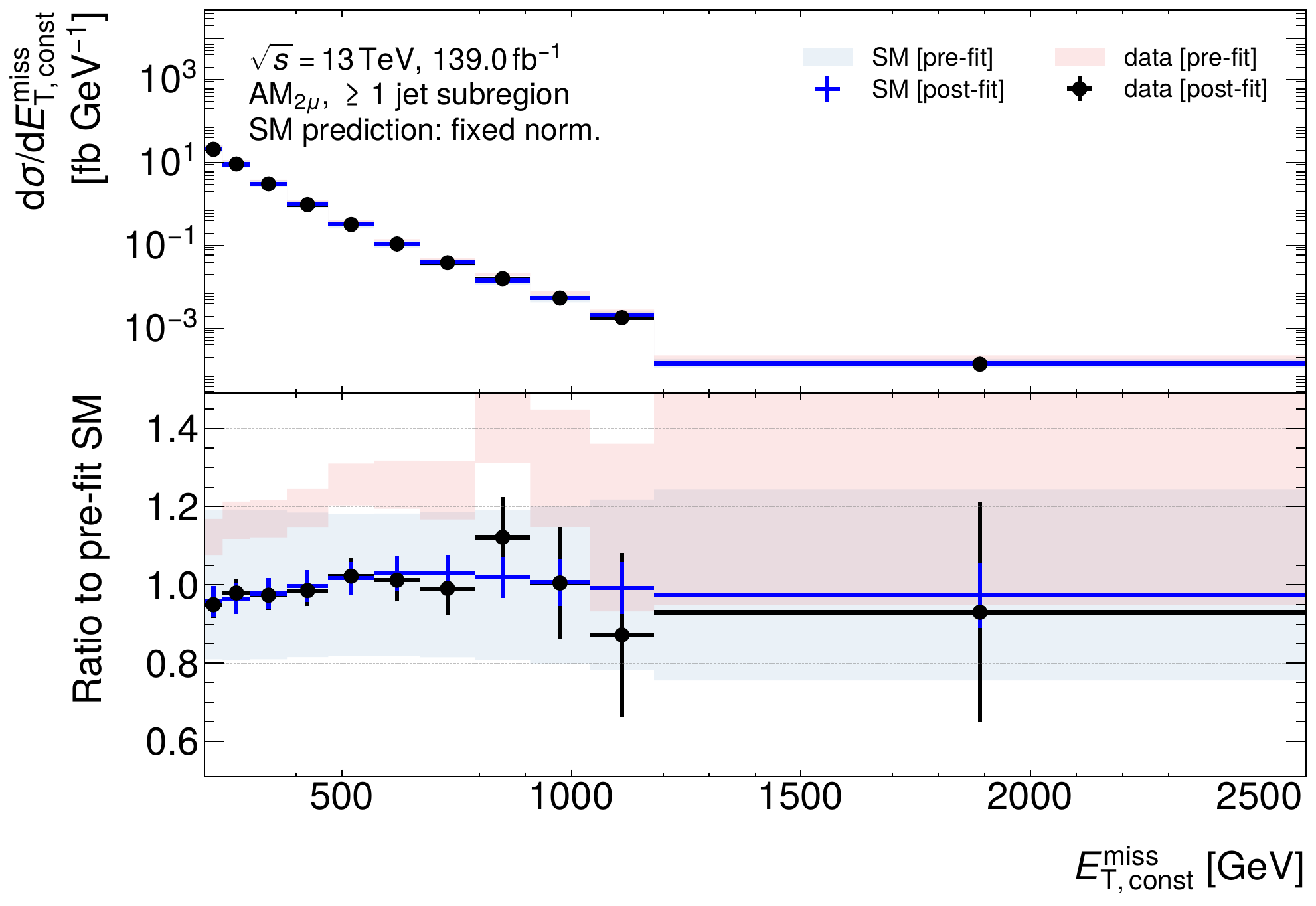}}
	\hspace{10pt}
	\subfloat[]{\includegraphics[width=0.48\textwidth]{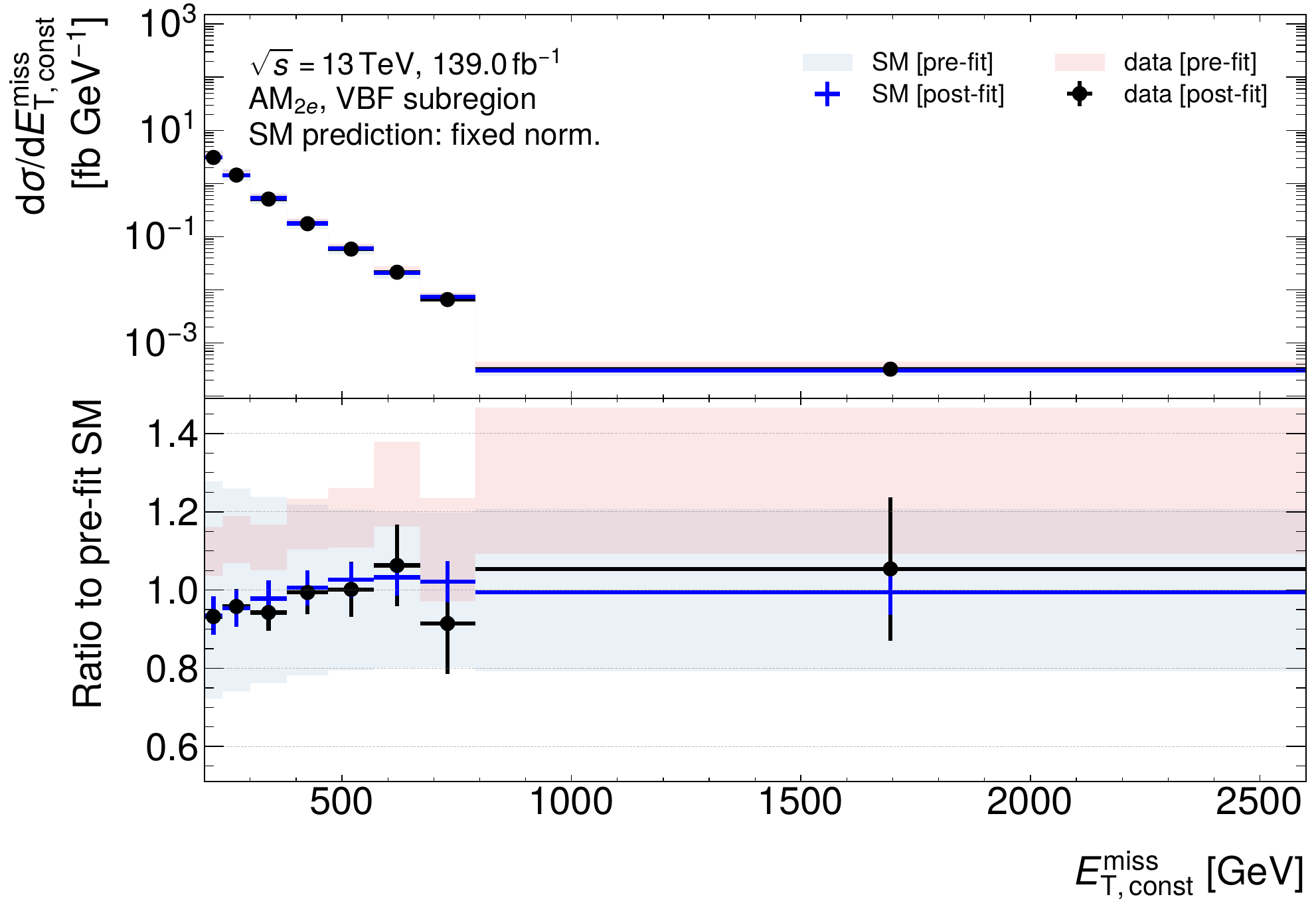}}\\
\end{myfigure}

Ideally, already the pre-fit values for the measured data and \SM prediction would agree.
This would correspond to the data values being located at 1 in the ratio panel.
However, as already discussed in \secsref{sec:metJets_detLevelResults}{sec:detCorr_partLevelResults}, there is a large discrepancy between the measured data and \SM prediction before the fit:
the normalisation of the measured data is underestimated in the \SM prediction by up to \SI{20}{\%} and there is also a difference in the shape of the distributions.
In consequence, nuisance parameters for systematic uncertainties are pulled in the fit to improve the agreement.

For a fit with perfect agreement, the post-fit \SM values (blue crosses) and measured data (black dots) would match in the top and bottom panels.
Indeed, post-fit \SM prediction and measured values agree within uncertainties because the nuisance parameters cover the pre-fit disagreement.
The most important exceptions are the bins at $\METconst\approx\SI{1100}{GeV}$ in the signal region of the \Mono and \VBF subregions where a discrepancy between measured data and \SM prediction remains.
Other discrepancies can be found at large \METconst in the \VBF subregion, in particular in \OneEJetsAM.
This can be attributed to statistical fluctuations at large \METconst in the \VBF subregion due to the stringent fiducial requirements and consequently smaller statistics (\cf\secref{sec:metJets_subregions}).
All of these discrepancies were already present in the pre-fit detector-level (\cf\figref{fig:metJets_detLevelResults}) and particle-level (\cf\figref{fig:detCorr_partLevelResults}) representation.
The considered nuisance parameters are not able to cover the discrepancies completely.

The shift between pre- and post-fit values for the measured data is larger than the shift for the \SM prediction because considerably more experimental than theoretical uncertainties are considered.
Their post-fit pulls impact the post-fit data values.

\subsubsection{Post-fit systematic uncertainties}

\figref{fig:interp_SM_systematics_postfit} shows the post-fit systematic uncertainties for the five measurement regions as a function of \METconst in the two subregions.
In the top panels, the relative uncertainties are given.
In the bottom panels, the ratio to the pre-fit uncertainties are shown.
The post-fit systematic uncertainties on the differential cross sections are analogous to post-fit impacts of nuisance parameters in a signal-strength measurement.

\bigskip
Theoretical systematics, particularly the uncertainties related to renormalisation and factorisation scale as well as \PDF{}s, dominate in both subregions in the signal region and one-lepton auxiliary measurements.
They are the largest uncertainties at large \METconst and often at small \METconst.
They increase in size with \METconst because the phase space for \SM predictions has to be extrapolated from the well-known low-energetic to the high-energetic regime.
The uncertainty can become as large as \SI{15}{\%}.

Uncertainties related to the renormalisation and factorisation scale are reduced considerably in size compared to their pre-fit values of up to \SI{30}{\%} (\cf\secref{sec:metJets_theoSystUnc}).
\PDF uncertainties are reduced less.

\bigskip
The dominant experimental uncertainties are jet energy scale (\JES) and\linebreak resolu\-tion~(\JER), as introduced in \secref{sec:objReco_jets_reconstruction}.
They can even be dominant at small \METconst, where the \JES estimate is impaired due to pileup and unlike detector responses to jets of different flavours, \ie quark- or gluon-initiated jets, \cf\secref{sec:metJets_expSystUnc}.
Apart from that, the most important experimental uncertainties are related to the reconstruction of electrons and muons, respectively, in the auxiliary measurements.
They can even be dominant in \TwoLJetsAMs at large \METconst when statistics are small.

The size of experimental uncertainties is generally only marginally decreased post- compared to pre-fit.
The most important exception are \JES uncertainties at small \METconst that are reduced by up to \SI{40}{\%}.

\begin{myfigure}{
		Relative post-fit systematic uncertainties for the five measurement regions as a function of \METconst in the (left) \Mono and (right)~\VBF subregion after constraining them in the fit.
		The bottom panels show the ratio of the absolute post- to the pre-fit uncertainties.
	}{fig:interp_SM_systematics_postfit}
	\subfloat[]{\includegraphics[width=0.48\textwidth]{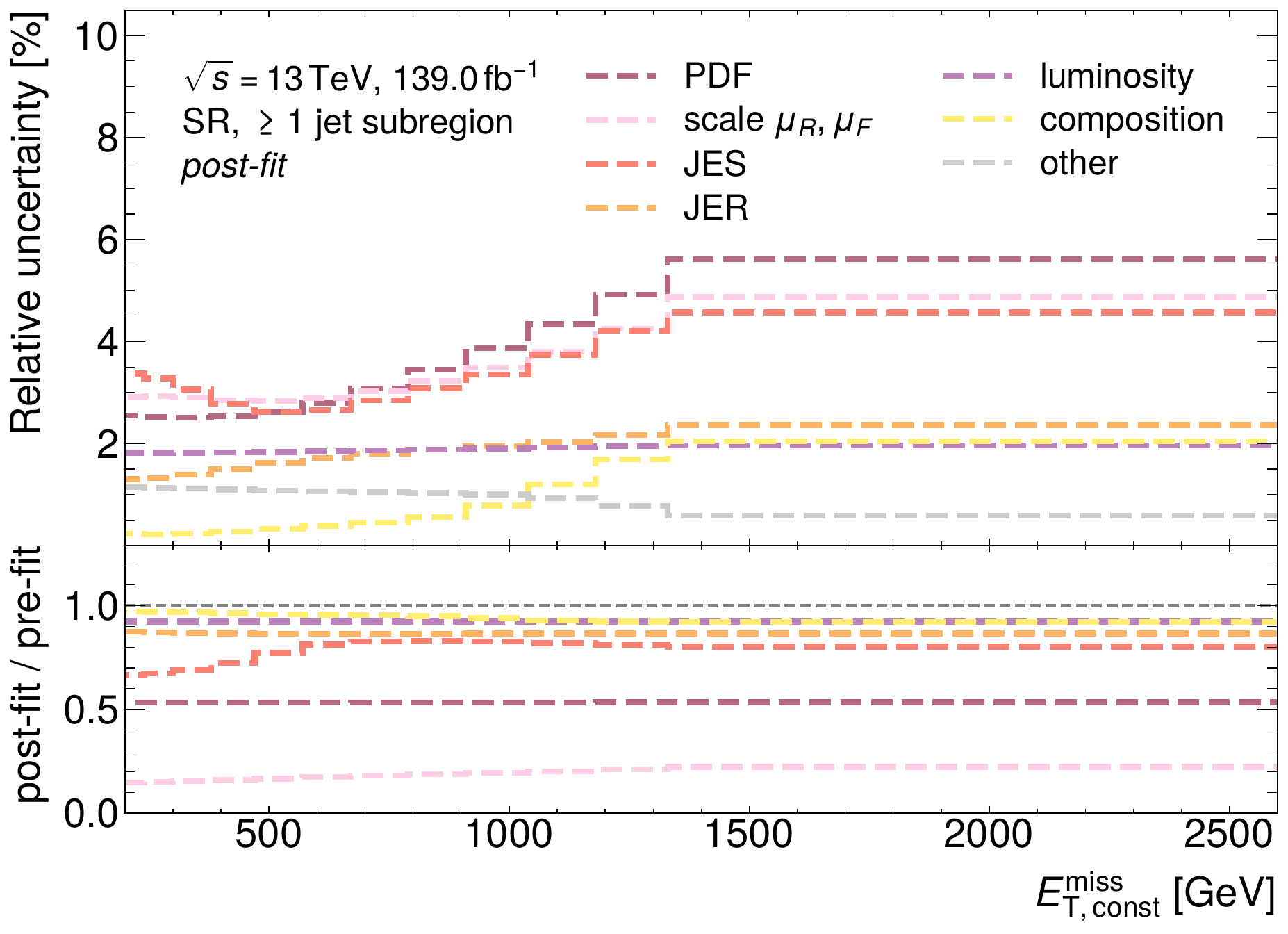}}
	\hspace{10pt}
	\subfloat[]{\includegraphics[width=0.48\textwidth]{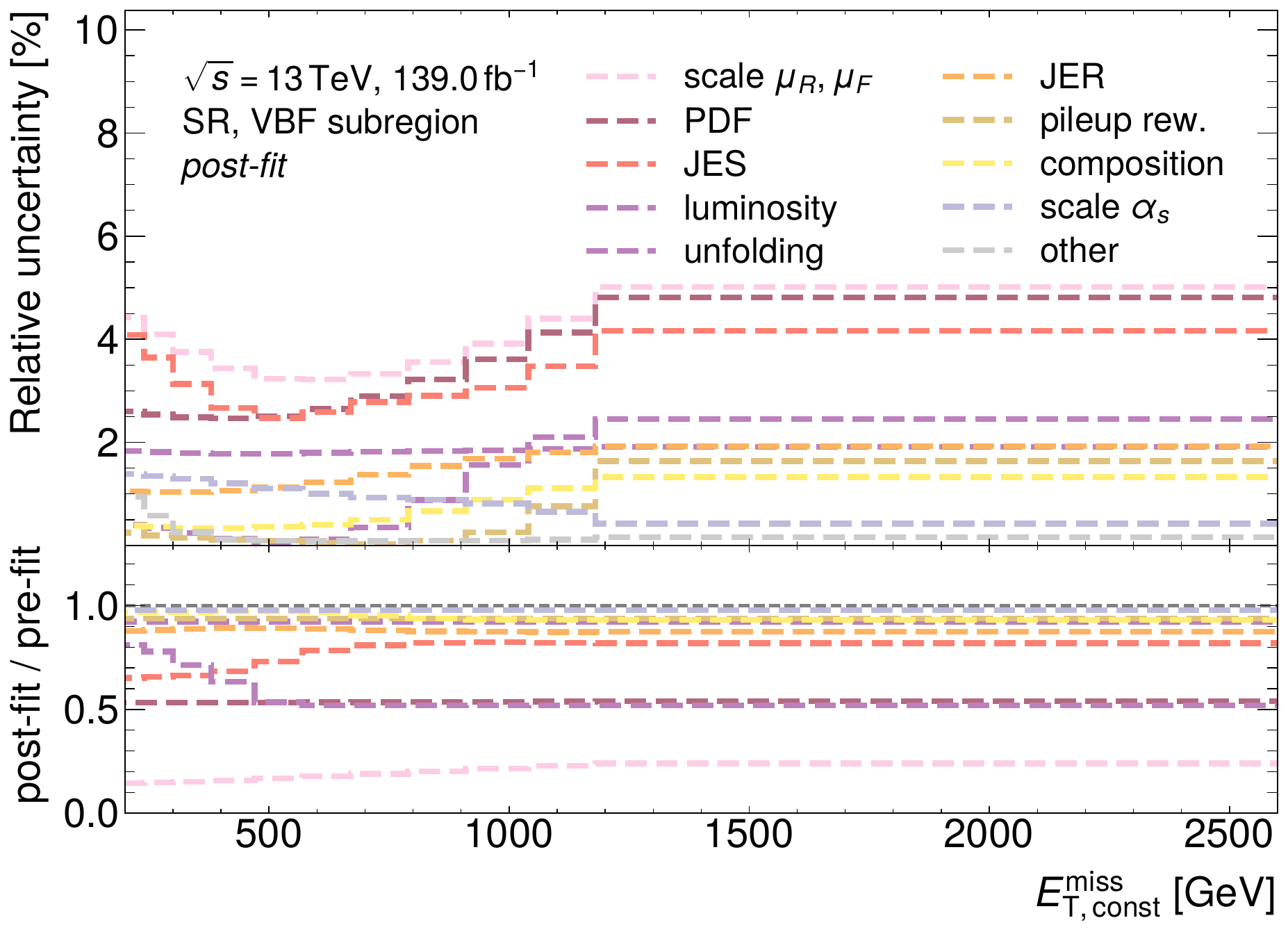}}\\
	\subfloat[]{\includegraphics[width=0.48\textwidth]{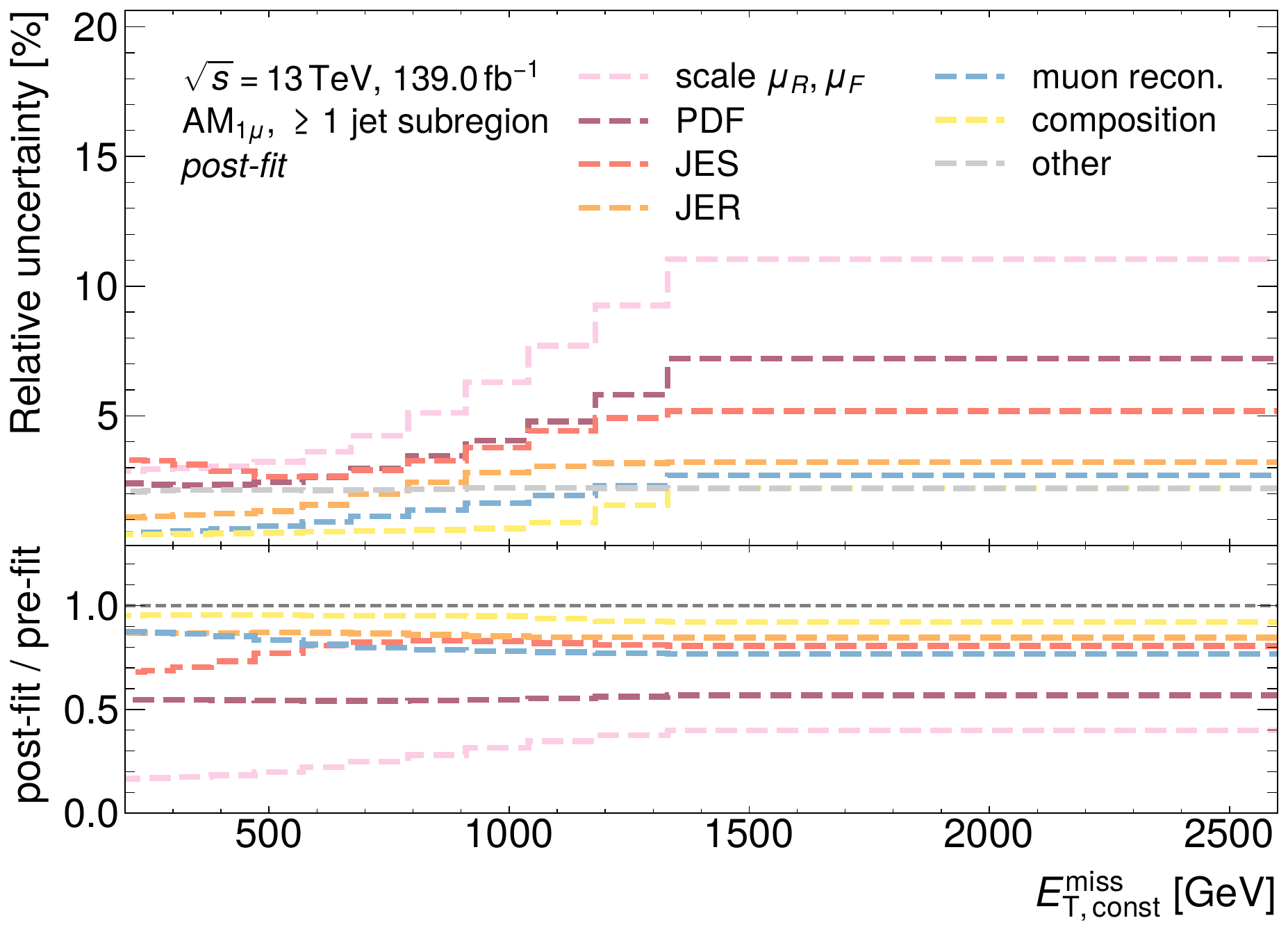}}
	\hspace{10pt}
	\subfloat[]{\includegraphics[width=0.48\textwidth]{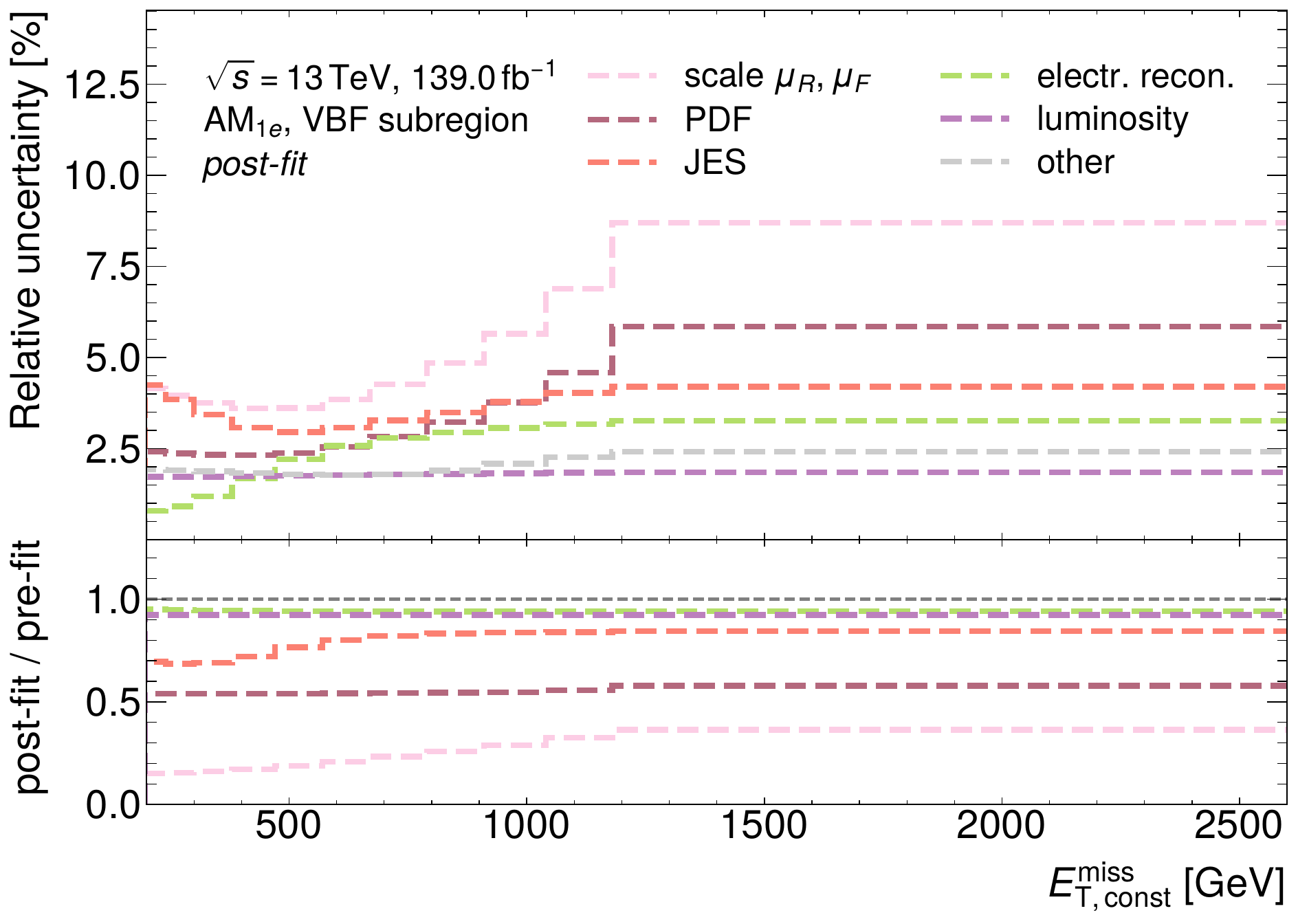}}\\
	\subfloat[]{\includegraphics[width=0.48\textwidth]{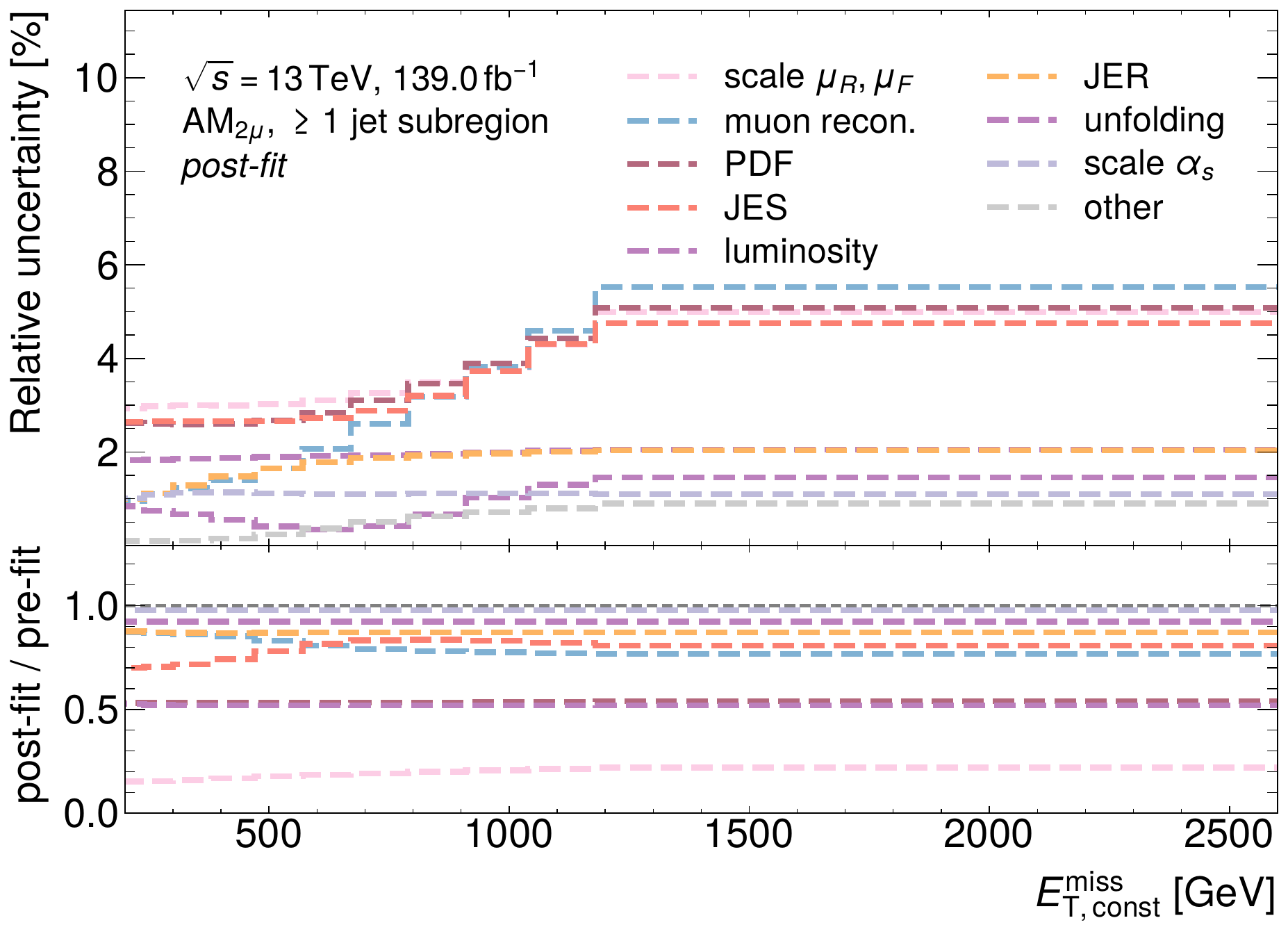}}
	\hspace{10pt}
	\subfloat[]{\includegraphics[width=0.48\textwidth]{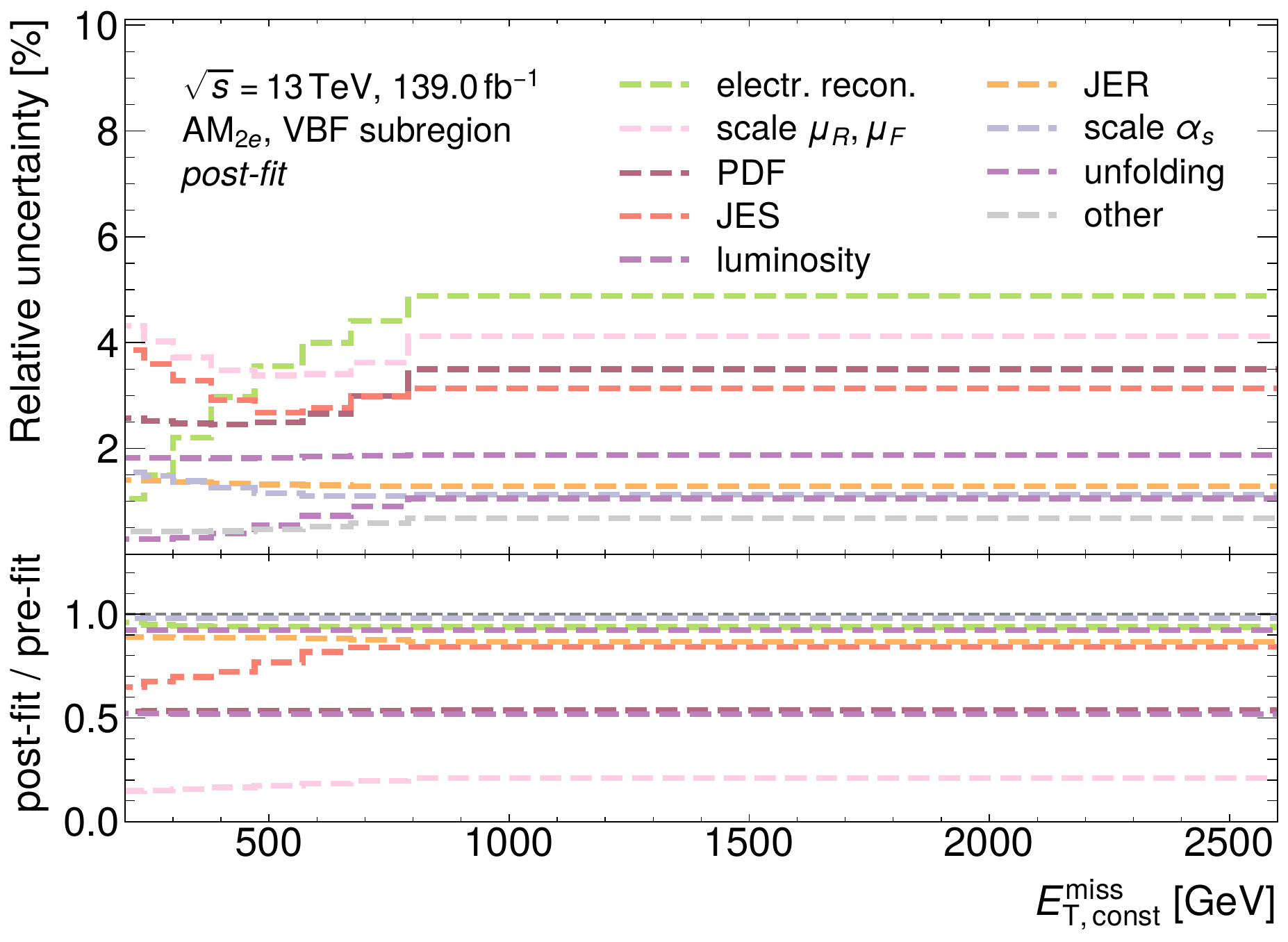}}\\
\end{myfigure}

\subsubsection{Pulls and constraints of nuisance parameters}

The effect of the pulls and constraints of the nuisance parameters was discernible in the previous paragraphs.
This is studied in more detail in the following.
\figref{fig:interp_SM_NPrankings_fixedNorm_diffXS} shows the post-fit pull $\hat\theta$ of the systematic uncertainties when maximising the likelihood in \eqref{eq:metJets_likelihood} for the data given the \SM prediction with fixed normalisation as black dots.
The post-fit constraints for the pulls $\sigma_{\hat\theta}$ are drawn as their errors.
The differential cross sections ($\mathrm{d}\sigma/\mathrm{d}\METconst$) are used as data $\vv x$ and prediction~\vvpiSM.

\newcommand{\NPrankingText}[3]{
	Post-fit pull \ensuremath{\hat{\theta}} of the nuisance parameters and median pre- as well as post-fit relative uncertainty \ensuremath{u_{j, \vv y}} for yield $\vv y$.
	The #1 serve as data and prediction and the #2-normalisation predictions #3are used.
	Nuisance parameters towards the top have larger values in the respective category.
	The bands for deviations of one and two standard deviations are marked in yellow and green, respectively.
}
\begin{myfigure}{%
		\NPrankingText{differential cross sections}{fixed}{}
	}{fig:interp_SM_NPrankings_fixedNorm_diffXS}{}
		\includegraphics[height=231pt]{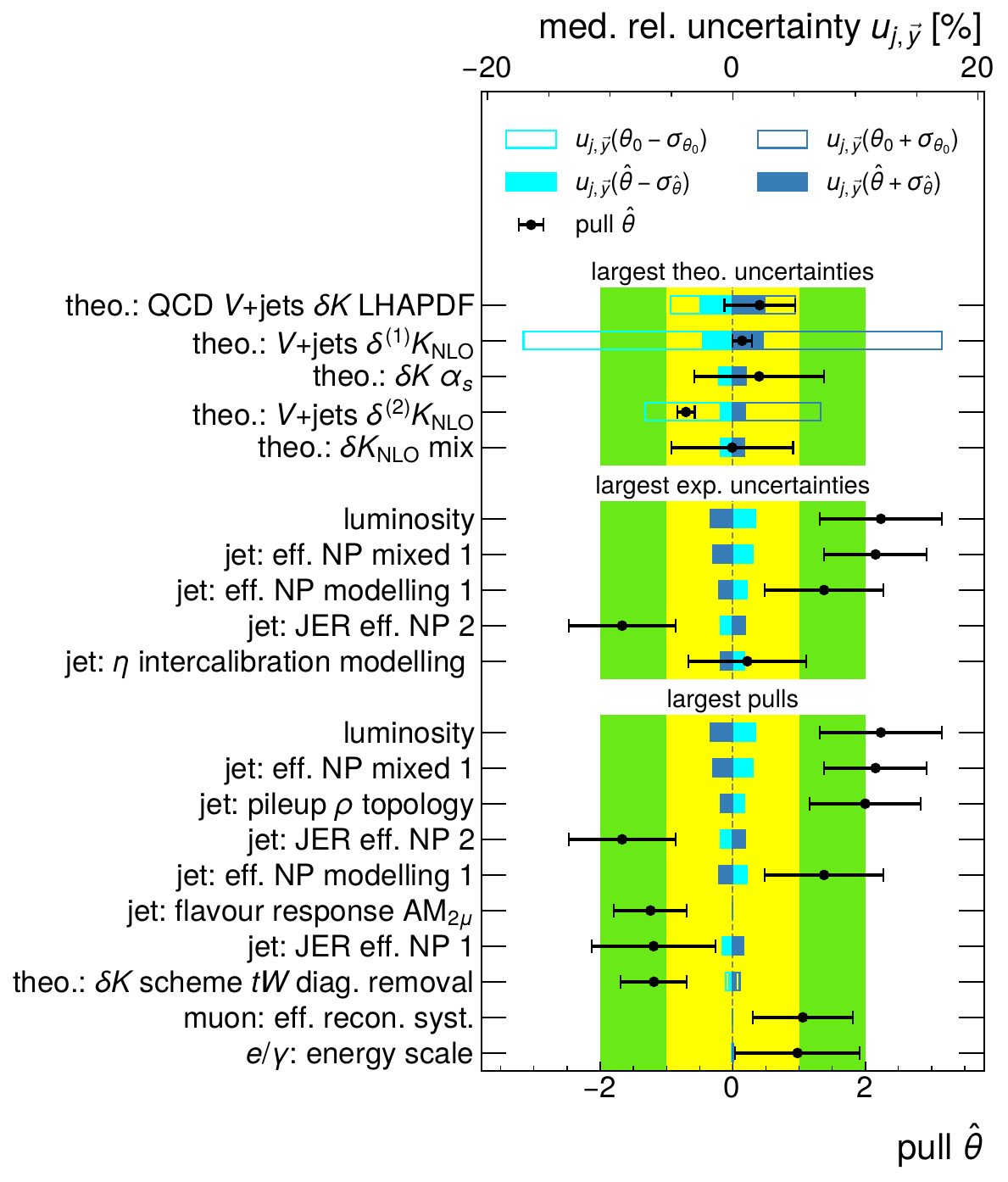}\rule{88pt}{0pt}
\end{myfigure}

The nominal result of the \METjets measurement are the differential cross sections and \Rmiss distributions, not the extraction of a single \SM parameter like the total \Vjets cross section.
As such, there is no single parameter on which the pre- and post-fit impact of each nuisance parameter could be shown.
The differential distributions that pose the nominal result are, however, significantly influenced by the pre- and post-fit size of the uncertainties.
For this reason, the relative uncertainties $u_{j, \vv y}$ of the yield $\vv y$ corresponding to a systematic uncertainty when varying the respective nuisance parameter by one standard deviation down (turquoise) or up (blue) are given in \figref{fig:interp_SM_NPrankings_fixedNorm_diffXS}.
Shown is the median of the relative uncertainties in all bins $j$.
The relative uncertainties pre-fit (post-fit) are displayed by open (closed) boxes.

In \figref{fig:interp_SM_NPrankings_fixedNorm_diffXS}, the five nuisance parameters for theoretical and experimental uncertainties leading to the largest post-fit relative uncertainty are given as well as the ten nuisance parameters with the largest absolute pulls.
Nuisance parameters towards the top have larger values in the respective category.

\bigskip
The theoretical uncertainties that have the largest median are all related to \Vjets modelling.
Note that their size in \figref{fig:interp_SM_NPrankings_fixedNorm_diffXS} is only partly comparable to the post-fit uncertainties in \figref{fig:interp_SM_systematics_postfit}:
the latter shows the quadrature sum over different uncertainty components and even over different \SM processes if the uncertainties originate from a related source, \eg \PDF{}s.

Uncertainties that are particularly large pre-fit, \eg $\delta K$ LHA\PDF and $\delta^{(1,2)} K_{\NLO}$~\cite{Lindert:2017olm}, cannot vary the yield post-fit as much as pre-fit without causing a significantly worse agreement between data and prediction.
They are therefore strongly constrained in the fit, as shown in \figref{fig:interp_SM_NPrankings_fixedNorm_diffXS}.
The other theoretical uncertainties are not constrained as much.

The experimental uncertainty that is the largest in the median is related to the luminosity estimate~\cite{ATLAS-CONF-2019-021}.
This uncertainty has the same size in every bin and can therefore amend normalisation differences between the measured data and the generated \SM prediction in all regions simultaneously.
This has the same effect as a cross section correction or a normalisation parameter on the total yield, apart from the constraint to be distributed according to a standard normal distribution.
The nuisance parameter for the luminosity uncertainty has a post-fit pull of 2.2 standard deviations which effectively decreases the data normalisation by approximately \SI{3.7}{\%}.
The next-largest experimental uncertainties all relate to jet reconstruction~\cite{ATLAS:2020cli}.
Jets are the primary selection requirement apart from \METconst in the \METjets measurement, and they are accordingly important for the selection of all regions.

The nuisance parameters with the largest post-fit pull are also related to the luminosity estimate and jet reconstruction, as discussed above.
Other nuisance parameters that have a large pull are related to top-quark modelling as well as muon and electron reconstruction.
All of them change the shape as well as the normalisation of the distributions.
The corresponding uncertainties, however, are considerably smaller than the aforementioned uncertainties.

The ensembles of pulls and constraints as a whole are studied in \appref{interpretation_SM_ensemble}.

\subsubsection{Test statistic and goodness of fit}

The test statistic takes a value of \fixedDiffXSChiTwo after this fit, resulting in a reduced~\chiSq obtained from the \SM fit of \fixedDiffXSChiTwoNdF given the \fixedDiffXSNdF degrees of freedom.
The results are summarised in the first row of \tabref{tab:interpretation_SM_fitResults}.
A conversion of the test statistic into a \pValue is discussed in \appref{app:interpretation_pValue}.

A reduced~\chiSq of 1 would correspond to perfect agreement between measured data and generated \SM prediction~\cite{Bevingtion:2003dre}.
Deviations from perfect agreement can be caused by statistical fluctuations, misrepresentation of the Standard Model by the setup used for generating \MC events or \BSM contributions in the data.

\subsubsection{Next steps}
Pre-fit, a difference in shape but also in normalisation of up to \SI{20}{\%} was observed.
Nonetheless, good post-fit agreement between the measured data and generated \SM prediction as well as a small reduced~\chiSq was achieved by pulling nuisance parameters, a few marginally more than two standard deviations.
In the following, it is investigated to which degree the pre-fit normalisation difference causes deviations from perfect post-fit agreement and whether the discrepancy between measured data and generated \SM prediction is consistent across all regions.
Both cases would point towards a generally imperfect representation of the Standard Model by the setup employed for \MC generation.
For this, two alternative fit approaches are studied:
\begin{itemize}
	\item In \secref{sec:interpretation_SM_floatNorm_diffXS}, a floating-normalisation approach according to \eqref{eq:interpretation_SM_pred_floatNorm} is used.
	This prediction allows for additional parameters that can account for pre-fit mismodelling in the normalisation of the most important \SM contributions.
	This unveils whether the cross section is mismodelled by the setup employed for \MC generation.

	\item In \secref{sec:interpretation_SM_fixedNorm_Rmiss}, the \SM prediction with fixed normalisation without normalisation parameters as discussed so far is used.
	The \Rmiss distributions, however, serve as the investigated quantity.
	In \Rmiss, discrepancies that appear in signal region as well as auxiliary measurements cancel.
	This could for example be a mismodelling of vector-boson \pT, which would show in all regions.
	Removing these duplicate discrepancies before the fit unveils whether there is a consistent mismodelling of the dominant \SM process across the regions in the setup employed for \MC generation.
\end{itemize}


\subsection{Floating instead of fixed normalisation}
\label{sec:interpretation_SM_floatNorm_diffXS}

\newcommand{\muVjetsNoRS}{\ensuremath{0.93\pm0.05}\xspace}
\newcommand{\muTopNoRS}{\ensuremath{0.88\pm0.16}\xspace}

\newcommand{\muVjetsRSmatchExp}{\ensuremath{\hatmuVjets=1.1691\pm0.0027}\xspace}
\newcommand{\muTopRSmatchExp}{\ensuremath{\hatmuTop=0.899\pm0.029}\xspace}

In the nominal fit, the pre-fit discrepancy between measured data and generated \SM prediction is removed by pulling nuisance parameters.
It is unclear, however, whether the uncertainties associated to the nuisance parameters properly represent the mismodelling of the underlying quantity.
This is verified in the following fit using a floating-normalisation \SM prediction according to \eqref{eq:interpretation_SM_pred_floatNorm}.
The fit thereby probes the impact of the normalisation on the post-fit agreement between data and \SM prediction.

The floating-normalisation prediction introduces parameters that scale the normalisation of the most important \SM contributions.
This approach describes the cross-section mismodelling of a process with its own parameters instead of associating it to a mismodelled quantity, \eg the jet energy scale.
The parameters are left unconstrained before the fit such that the normalisation of the corresponding \SM contributions can be suited to the fit optimum.
This does, however, ignore the theoretical computations that lead to the uncertainty estimates, in particular for the process cross sections.

The most important \SM contributions to the five measurement regions stem from \Zjets, \Wjets and top processes.
The systematic uncertainties for the modelling of \Zjets and \Wjets are closely correlated~\cite{Lindert:2017olm,Lindert:2022ejn}.
Therefore, a common normalisation parameter \muVjets for all \Vjets processes is used.
In addition, a normalisation parameter \muTop is used for all top processes.
No other process is assigned a variable normalisation parameter.
The setup is summarised in the second row in \tabref{tab:interpretation_SMpredictions}.

As in the nominal fit, the differential cross sections in all five measurement regions and both subregions binned in \METconst serve as the investigated quantity for data~$\vv x$ and prediction~\vvpiSM in the goodness-of-fit test.
In total, \fixedDiffXSNDists distributions are included in the fit, amounting to \fixedDiffXSNdF bins.
This is also given in the second row of \tabref{tab:interpretation_SM_fitResults}.

\bigskip
In the nominal \SM fit, the pre-fit disagreement between data and \SM prediction is exclusively solved by pulling nuisance parameters.
These pulls change not only the shape, but also the normalisation of the data and prediction.
This is illustrated in detail in \appref{app:interpretation_SM_NP_systExcl}.
In consequence, a balance has to be found for each nuisance parameter:
A too large pull on the nuisance parameter might give a good agreement in shape but a wrong estimate of the normalisation.
A too small pull on the nuisance parameter might give a good agreement in the normalisation but a wrong estimate of the shape.
Either would cause a discrepancy between the post-fit data and \SM prediction.
The only nuisance parameter for which this is not the case is the nuisance parameter for the luminosity estimate.
This nuisance parameter changes the yield completely constant as a function of \METconst.
In consequence, the nuisance parameter gets pulled by more than two standard deviations to give more freedom to improve the shape agreement by the other nuisance parameters (\cf\figref{fig:interp_SM_NPrankings_fixedNorm_diffXS}).

In the fit with a floating normalisation, the normalisation parameters take values of
\begin{align*}
	\hatmuVjets&=\muVjetsNoRS,\\
	\hatmuTop&=\muTopNoRS.
\end{align*} 
This means the normalisation of the prediction for \Vjets and top processes is decreased.
The fractions these processes contribute to the total \SM prediction are not constant as a function of \METconst.
Therefore, the total \SM prediction is not changed completely constant as a function of \METconst.
This is similar to most nuisance parameters, which also change the normalisation as well as the shape of the distribution.
The difference between the normalisation parameters and the nuisance parameters is that the normalisation parameters can be pulled without being penalised by the last term in \eqref{eq:interpretation_qSM}.
The luminosity uncertainty, in contrast to normalisation and other nuisance parameters, does change the \SM prediction completely constant as a function of \METconst.
A balance between normalisation parameters smaller than 1 and pulling the nuisance parameter for the luminosity is needed.
This balance provides more freedom to the other nuisance parameters to give a good shape agreement between data and prediction.

In \appref{app:interpretation_SM_NP_normSysts}, a study is performed in which the luminosity uncertainty and the normalisation component from other nuisance parameters are removed.
Here, the pre-fit normalisation difference between data and prediction must be amended by the normalisation parameters in the fit.
The normalisation parameters take values of \muVjetsRSmatchExp and \muTopRSmatchExp.
This corresponds to an effective increase of the \SM normalisation because \Vjets is the main \SM contribution.

Regarding the nominal post-fit normalisation parameters, it can further be noted that the relative uncertainty for \hatmuTop is almost \SI{20}{\%}.
This means that the measured phase spaces are not well suited to constrain the normalisation of top processes.
On the one hand, this is caused by the comparably small statistics in the regions with the largest top contributions, \OneLJetsAMs.
On the other hand, the normalisations of the distributions are not only adjusted by the normalisation parameters but also by the nuisance parameters.
In the study with removed normalisation components for the nuisance parameters in \appref{app:interpretation_SM_floatNormStudies}, the uncertainties on both normalisation parameters are significantly smaller.
Here, the normalisation of the data and \SM prediction has to be adjusted with the normalisation parameters.

\bigskip
The differential cross sections pre-fit and post-fit are shown for two examples in \figref{fig:interp_SM_distributions_postfit_floatNorm_diffXS}.
The bottom panels show the ratio to the pre-fit \SM prediction.
The pre-fit distributions are identical to those with the fixed-normalisation approach already given in \figref{fig:interp_SM_distributions_postfit_fixedNorm_diffXS}.
No large differences between post-fit fixed- and floating-normalisation distributions are visible (\cf\figsref{fig:interp_SM_distributions_postfit_fixedNorm_diffXS}{fig:interp_SM_distributions_postfit_floatNorm_diffXS}).

\begin{myfigure}{
		\postfitDistCaption{Differential cross section}{%
			(a) the signal region in the \Mono and (b) \OneEJetsAM in the \VBF subregion as a function of \METconst%
		}{floating}{}{}{}
	}{fig:interp_SM_distributions_postfit_floatNorm_diffXS}
	\subfloat[]{\includegraphics[width=0.48\textwidth]{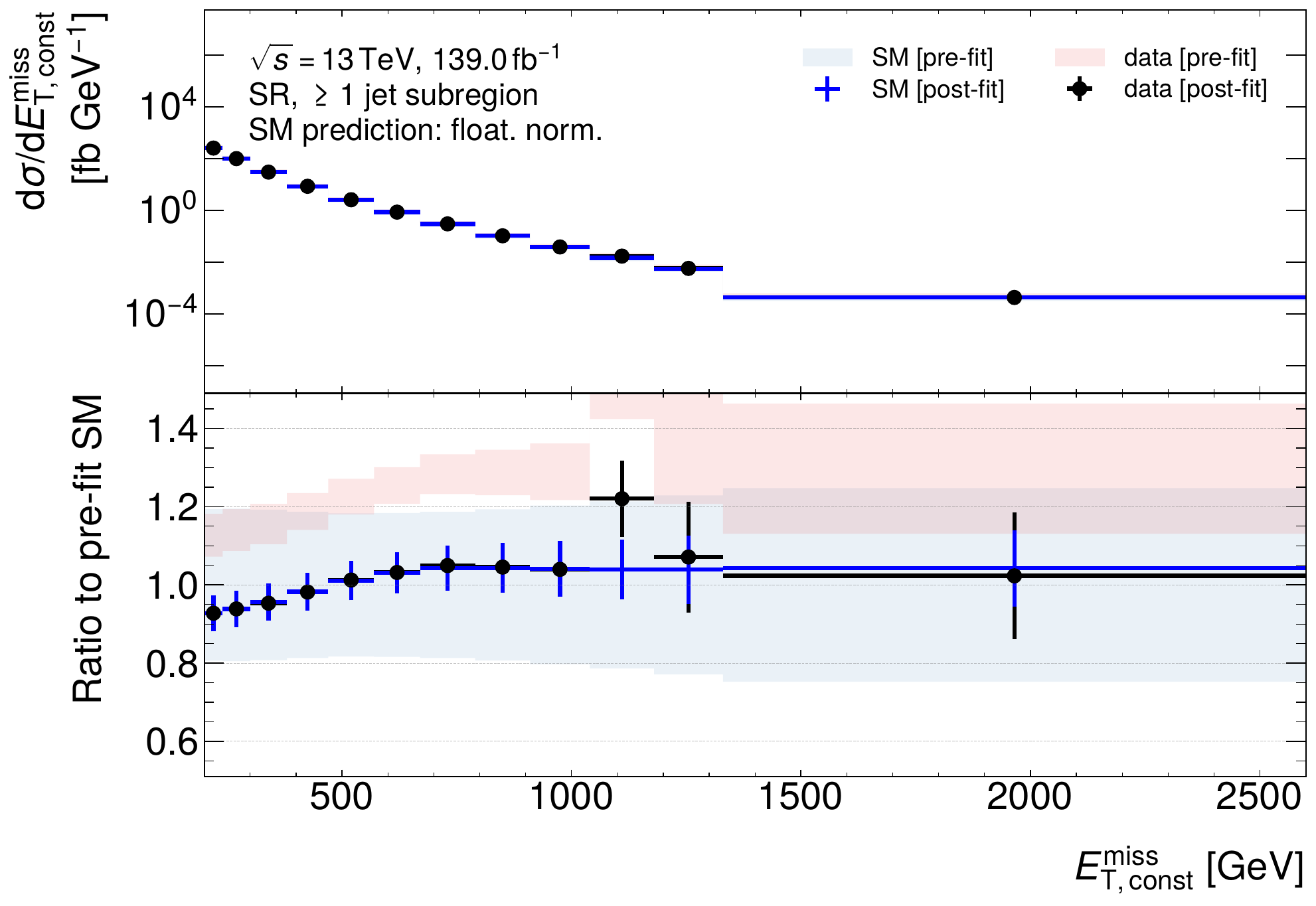}}
	\hspace{10pt}
	\subfloat[]{\includegraphics[width=0.48\textwidth]{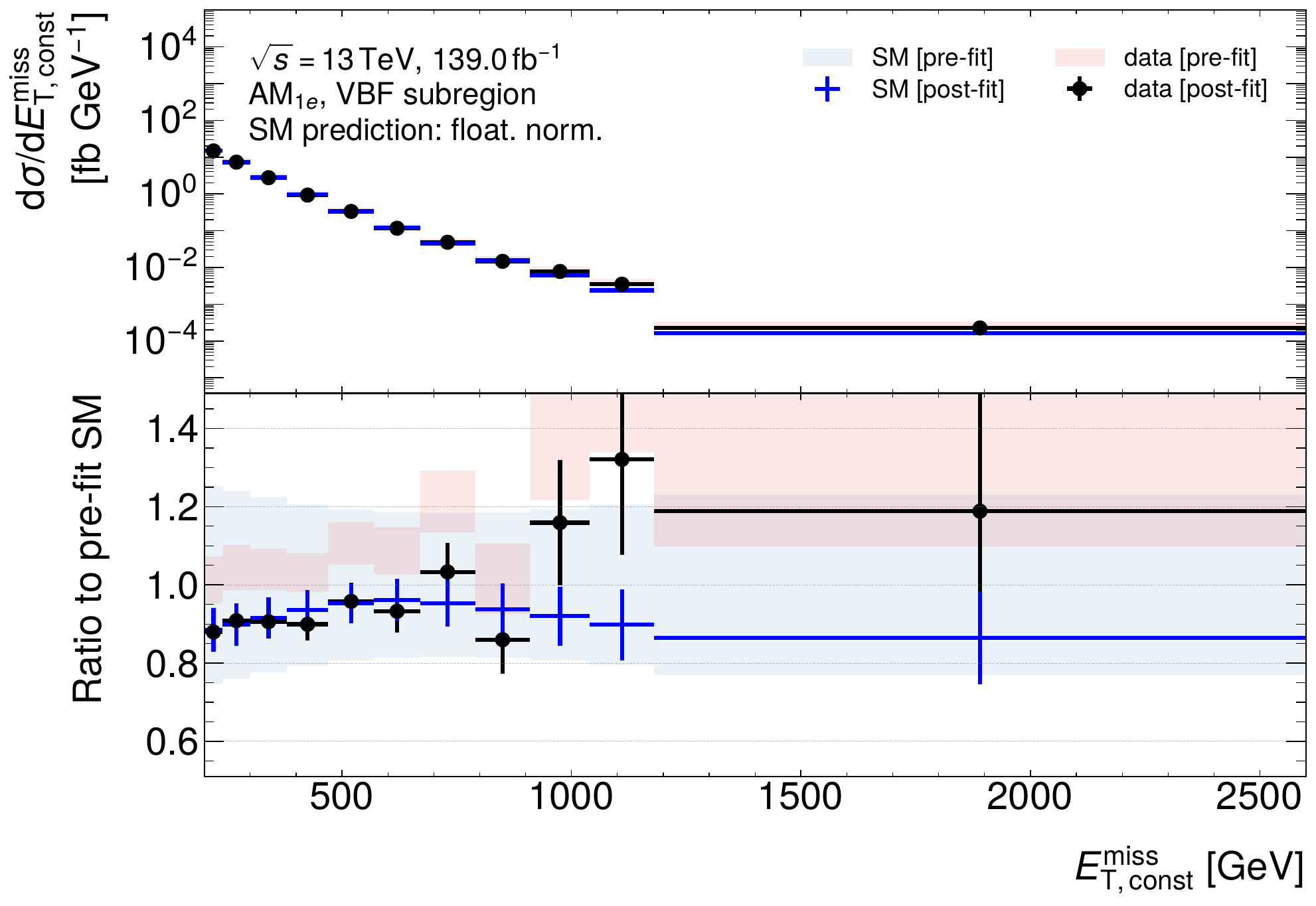}}\\
\end{myfigure}

\bigskip
\figref{fig:interp_SM_NPrankings_floatNorm_diffXS} shows the post-fit pull $\hat\theta$ and constraint for the pull $\sigma_{\hat\theta}$ of the systematic uncertainties.
Given are also the relative uncertainties $u_{j, \vv y}$ of the yield $\vv y$ corresponding to a systematic uncertainty.
In the top panel of \figref{fig:interp_SM_NPrankings_floatNorm_diffXS}, the post-fit values of the normalisation parameters $\mu$ are shown for the same fit.

\begin{myfigure}{%
		\NPrankingText{differential cross sections}{floating}{}
	}{fig:interp_SM_NPrankings_floatNorm_diffXS}
		\includegraphics[height=280pt]{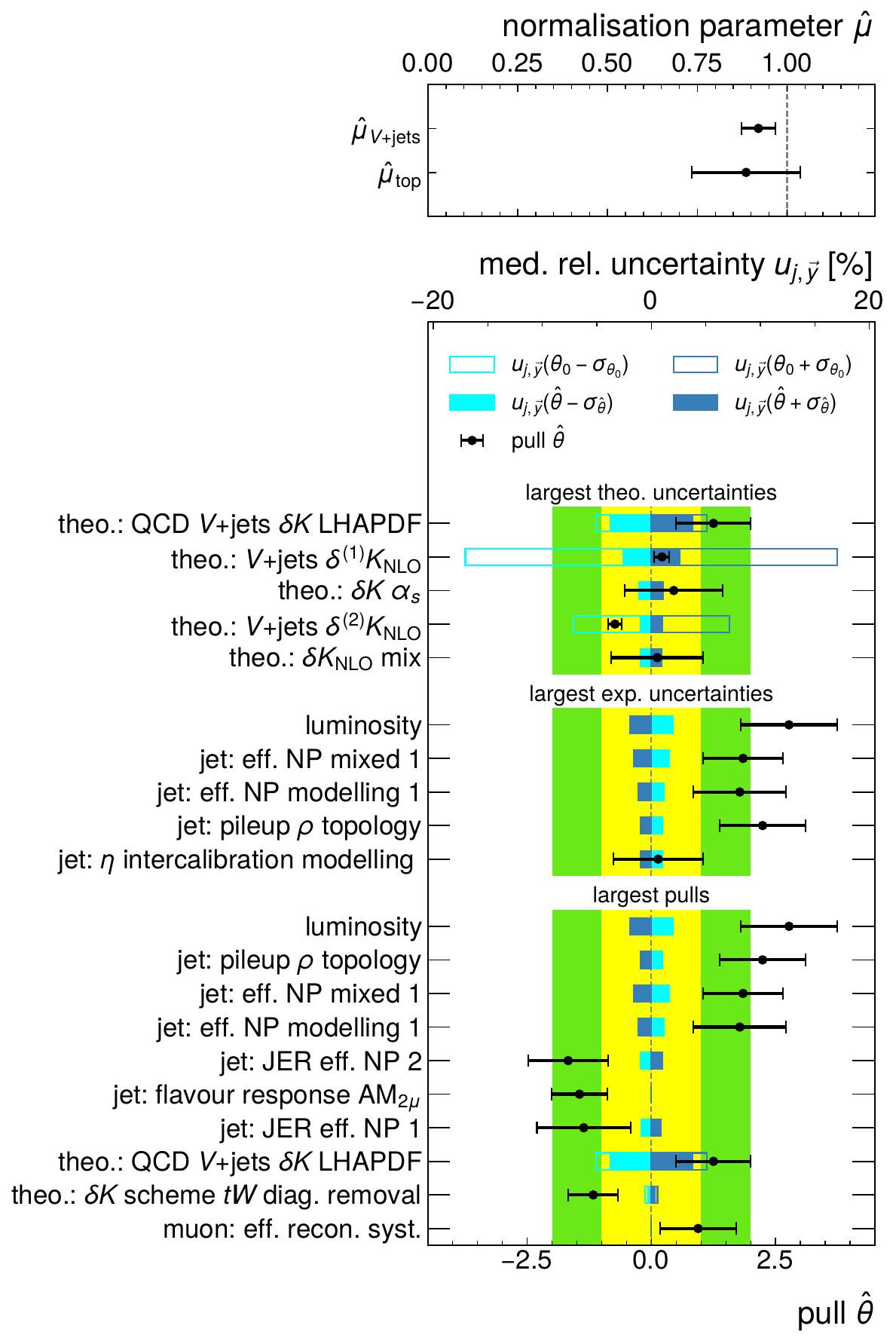}\rule{88pt}{0pt}
\end{myfigure}

The pulls and constraints differ only mildly from those using the \SM prediction with fixed normalisation in \figref{fig:interp_SM_NPrankings_fixedNorm_diffXS}.
The pulls and constraints for the nuisance parameters for the largest theoretical and experimental uncertainties are very similar.
The order of the nuisance parameters with the largest pulls changes a little but the pulls and constraints are still very similar.
The pull of the nuisance parameter for the luminosity uncertainty is increased to 2.8 standard deviations.
This decreases the data normalisation in the floating-normalisation compared to the fixed-normalisation approach.
This is required because the normalisation parameters $\hatmuVjets<1$ and $\hatmuTop<1$ analogously decrease the normalisation of the \SM prediction.

\bigskip
The fit results are summarised in the second row of \tabref{tab:interpretation_SM_fitResults}.
The test statistic takes a value of \floatDiffXSChiTwo.
The fit has \floatDiffXSNdF degrees of freedom as it uses \floatDiffXSNbins bins and two
pre-fit unconstrained normalisation parameters.
The resulting reduced~\chiSq is \floatDiffXSChiTwoNdF.
This is marginally smaller than the reduced~\chiSq for the fixed-normalisation prediction with differential cross sections as input that was discussed in \secref{sec:interpretation_SM_fixedNorm_diffXS}.

In conclusion, fixed- and floating-normalisation predictions give very similar fit results with respect to pulls and constraints of nuisance parameters as well as observed value of the test statistic and reduced~\chiSq.
Introducing unconstrained normalisation parameters for the most important \SM contributions does not improve the agreement between measured data and generated \SM prediction considerably.
The deviation from perfect post-fit agreement between data and \SM prediction observed in \secref{sec:interpretation_SM_fixedNorm_diffXS} is therefore not primarily a consequence of normalisation mismodelling of the dominant \SM processes in the setup employed for \MC generation.

\subsection[\Rmiss distributions instead of differential cross sections]{\RmissTitle distributions instead of differential cross sections}
\label{sec:interpretation_SM_fixedNorm_Rmiss}

As in \secref{sec:interpretation_SM_fixedNorm_diffXS}, the \SM prediction with fixed normalisation according to \eqref{eq:interpretation_SM_pred_fixedNorm} is used in this fit approach.
Contrary to before, the \Rmiss distributions are used as the investigated quantity for data~$\vv x$ and prediction~\vvpiSM in the goodness-of-fit test.
In \Rmiss, mismodellings that are common between signal region and auxiliary measurements cancel (\cf\eqref{eq:metJets_Rmiss}).
In consequence, a normalisation discrepancy that is present in the signal region as well as in the auxiliary measurements, \eg because it is inherent to all \Vjets calculations, does not impair the fit result.
Contrary to the floating-normalisation prediction studied in the previous section, in this approach not only common normalisation mismodellings can be accounted for but also common shape mismodellings.

In total, \fixedRmissNDists distributions are included in the fit, amounting to \fixedRmissNdF bins.
This is also given in the last row of \tabref{tab:interpretation_SM_fitResults}.
Fewer distributions are used in this approach than in the approaches before because meaningful \Rmiss distributions can only be calculated for the auxiliary measurements.
As the \SR yields are in the numerator of \Rmiss, calculating \Rmiss with the \SR yields also in the denominator trivially yields unit values.

\bigskip
The \Rmiss distributions pre- and post-fit are shown for two examples in \figref{fig:interp_SM_distributions_postfit_fixedNorm_Rmiss}.
The bottom panels give the ratio to the pre-fit \SM prediction.

\begin{myfigure}{
		\postfitDistCaption{\Rmiss distribution}{%
			(a) \TwoMuJetsAM in the \Mono and (b) \OneEJetsAM in the \VBF subregion as a function of \METconst%
		}{fixed}{}{}{}
	}{fig:interp_SM_distributions_postfit_fixedNorm_Rmiss}
	\subfloat[]{\includegraphics[width=0.48\textwidth]{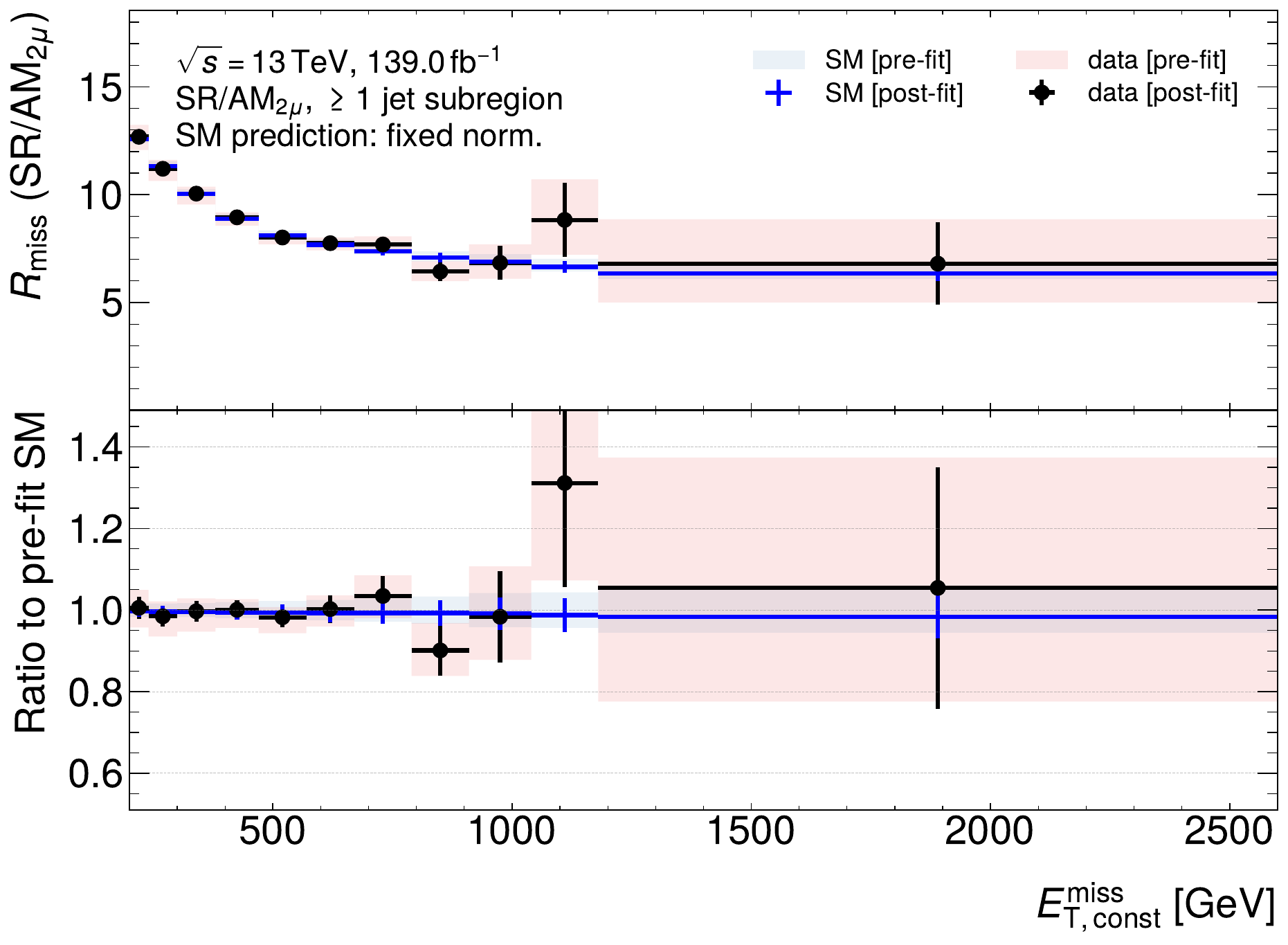}}
	\hspace{10pt}
	\subfloat[]{\includegraphics[width=0.48\textwidth]{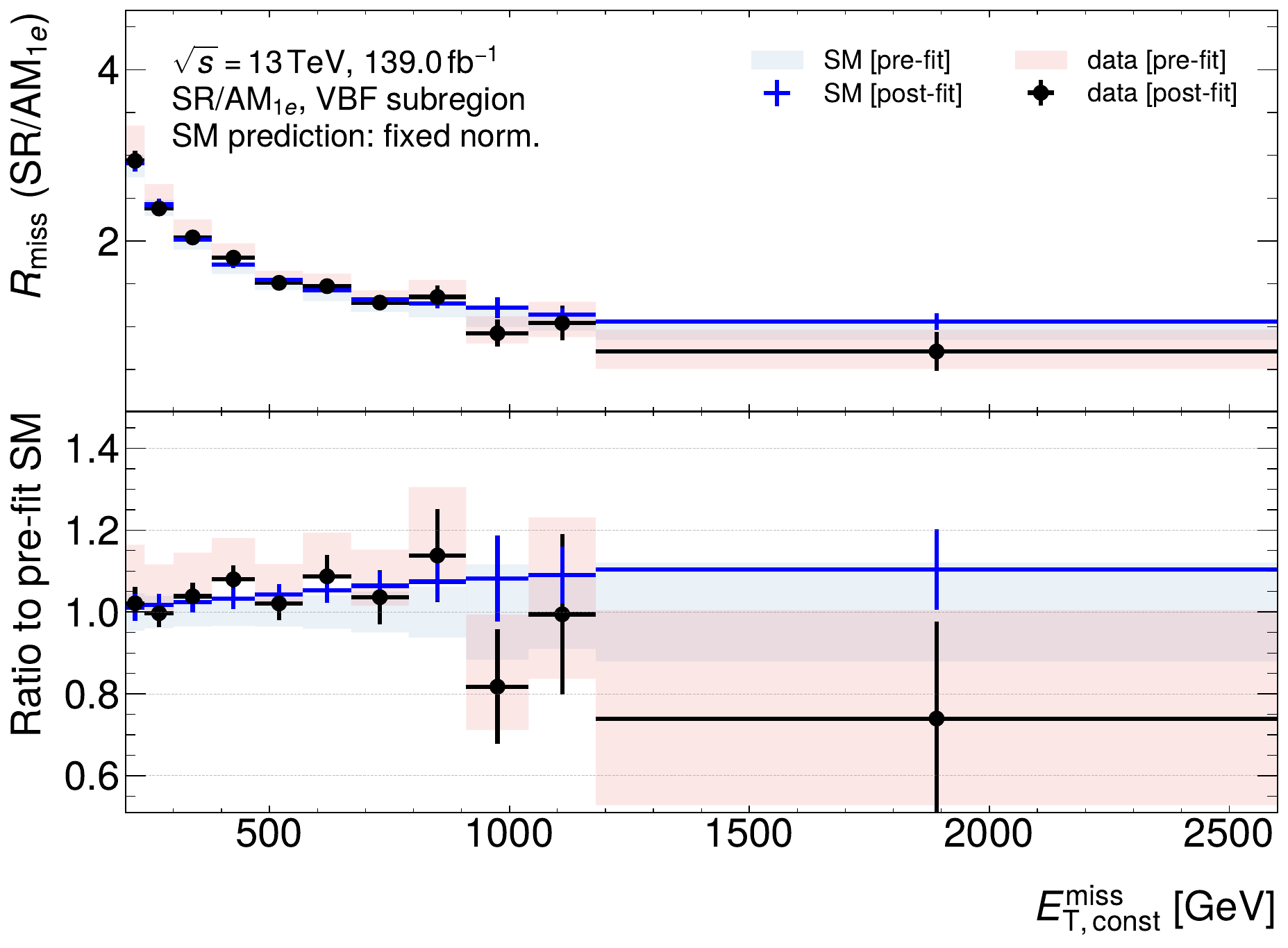}}\\
\end{myfigure}

Pre- and post-fit, the agreement between the measured data and generated \SM prediction is considerably better than when using the fixed-normalisation prediction (\cf\figsref{fig:interp_SM_distributions_postfit_fixedNorm_diffXS}{fig:interp_SM_distributions_postfit_floatNorm_diffXS}).
This is because mismodellings that are common for all regions cancel.
Many experimental and theoretical systematic uncertainties cancel as well which is why the total size of the experimental and theoretical uncertainties are also reduced in \Rmiss.
This holds for pre- as well as post-fit distributions.

\bigskip
\figref{fig:interp_SM_NPrankings_fixedNorm_Rmiss} shows the post-fit pull $\hat\theta$ and constraint for the pull $\sigma_{\hat\theta}$ of the systematic uncertainties.
Given are also the relative uncertainties $u_{j, \vv y}$ of the yield $\vv y$ corresponding to a systematic uncertainty.

\begin{myfigure}{%
		\NPrankingText{\Rmiss distributions}{fixed}{}
	}{fig:interp_SM_NPrankings_fixedNorm_Rmiss}
		\includegraphics[height=231pt]{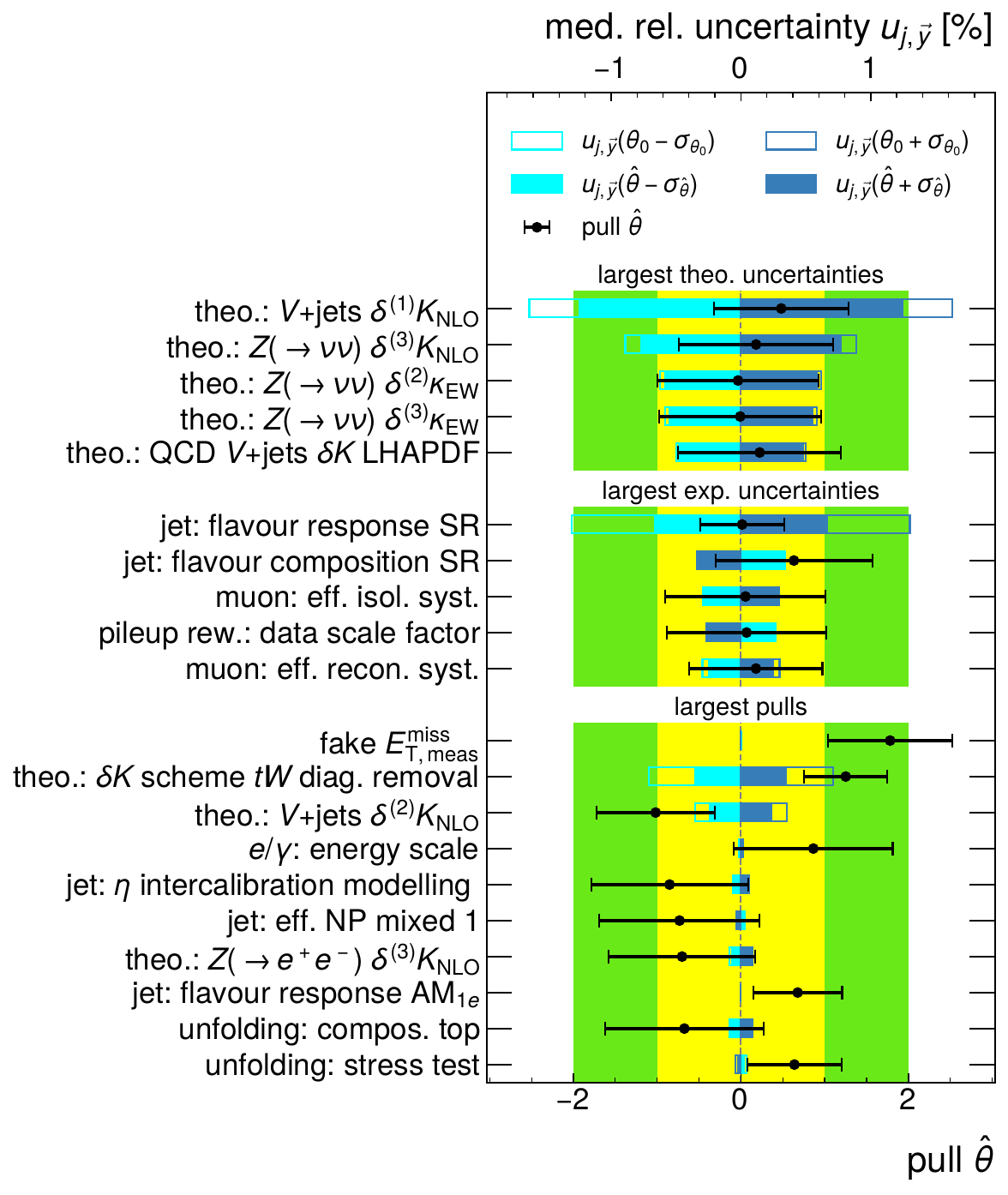}\rule{88pt}{0pt}
\end{myfigure}

The theoretical uncertainties that are the largest in the median are related to \Vjets processes, as was the case for differential cross sections as input quantity.
The pre-fit uncertainties, however, are an order of magnitude smaller than when using differential cross sections because their contributions cancel in the \Rmiss ratio.
In consequence, they are only mildly constrained.

The experimental uncertainty that is the largest in the median is related to the response of the detector to jets of different flavours, \ie quark- or gluon-initiated jets (\cf\secref{sec:objReco_jets_jetCalibration}), in the signal region~(\SR).
The nuisance parameter for this uncertainty is constrained significantly in the fit.
The next-largest experimental uncertainties are either also related to jets, to muon reconstruction~\cite{ATLAS:2020auj} or to pileup.
The corresponding nuisance parameters are, however, constrained less because their pre-fit size is considerably smaller.

The largest pulls are also smaller in general because the discrepancy between measured data and generated \SM prediction is smaller in \Rmiss than in the differential cross sections.
No nuisance parameter is pulled more than two standard deviations and only two are pulled more than one standard deviation in \figref{fig:interp_SM_NPrankings_fixedNorm_Rmiss}.
The nuisance parameters with the largest pulls correspond to a mixture of very different uncertainties, among others fake \METmeas, top-quark modelling, \Vjets modelling, the energy scale for electrons and photons as well as jet reconstruction.

\bigskip
The test statistic takes a value of \fixedRmissChiTwo in this fit approach, resulting in a reduced~\chiSq of \fixedRmissChiTwoNdF given the \fixedRmissNdF degrees of freedom.
The results are summarised in the third row of \tabref{tab:interpretation_SM_fitResults}.
This is a bit smaller than the values for the reduced \chiSq obtained for the other two fit approaches.
It proves that measured data and generated \SM prediction are in part subject to common sources of systematic errors across the regions.
This is foremost a mismodelling of the vector-boson \pT.
Their agreement improves when these common sources are cancelled in the \Rmiss ratio.

Unfortunately, \Rmiss is not a good quantity for all \BSM interpretations:
if the \BSM model contributes to signal region as well as auxiliary measurements, these contributions cancel and the \BSM model is indistinguishable from the Standard Model.
This effect is small for the \THDMa but can be large for other \BSM models.
Examples for this are models with vector-like quarks~\cite{Buchkremer:2013bha,Fuks:2016ftf}, which predominantly increase the production of $W$ and $Z$ bosons.

\subsection{Summary}
\label{sec:interpretation_SM_summary}

Three different fit approaches were used to probe the agreement of the generated \SM prediction with the measured data at particle level.
The fit results are summarised in \tabref{tab:interpretation_SM_fitResults}.

In the nominal approach, a \SM prediction with fixed normalisation is used, and the differential cross sections serve as data~$\vv x$ and prediction~\vvpiSM. 
This approach makes use of all available information from measured data and generated \SM prediction.
The setup gives an acceptable reduced~\chiSq of \fixedDiffXSChiTwoNdF.
There are remaining post-fit discrepancies between data and generated \SM prediction.

In a second approach, a floating-normalisation \SM prediction with normalisation parameters for the two largest \SM contributions is used, and the differential cross sections serve as data~$\vv x$ and prediction~\vvpiSM.
This approach allows compensating normalisation discrepancies between measured data and generated \SM prediction with the normalisation parameters.
This prediction discards information from theoretical calculations of the cross section of the normalised \SM contributions, however.
The fit setup gives results very similar to the nominal approach with respect to the pulls and constraints of nuisance parameters.
The reduced~\chiSq of \floatDiffXSChiTwoNdF is marginally smaller than in the nominal approach.
This means that the remaining discrepancy between measured data and generated \SM prediction does not originate primarily from differences in the normalisation of pre-fit distributions.

In a third approach, a \SM prediction with fixed normalisation is used, and the \Rmiss distributions serve as data~$\vv x$ and prediction~\vvpiSM. 
This approach maximally exploits the available correlations between the regions, \eg for \Vjets predictions and their uncertainties.
The resulting reduced~\chiSq of \fixedRmissChiTwoNdF is a bit smaller than for the other two approaches.
It proves that measured data and generated \SM prediction are subject to common sources of systematic errors across the regions.
Regarding the \SM prediction, this is foremost a mismodelling of the vector-boson \pT.
The post-fit agreement can be improved if these mismodellings are cancelled in the \Rmiss ratio but of course not by introducing normalisation parameters.

\section{Interpretation with respect to the \THDMa}
\label{sec:interpretation_2HDMa}

It can now be explored in which phase spaces \BSM models are disfavoured as a consequence of the measurement results.
As pointed out in \secref{sec:2HDMa}, the \THDMa is an excellently motivated model to investigate Dark Matter at colliders.
The model gives rise to significant contributions to the phase space selected by the \METjets measurement, as was seen in \chapref{sec:2HDMa_metJetsMeasurement}.
This can be quantised with the test statistic \qBSM from \eqref{eq:metJets_qBSM} based on the likelihood ratio between signal-plus-background and background-only likelihood.
However, also other test statistics are possible.
Exclusion limits with the profiled likelihood-ratio comparing the signal-plus-background likelihood to the global likelihood maximum are derived in \appref{app:interp_2HDMa_muOpt}.
They do not differ significantly from the exclusion limits derived in this section.
This means the exclusion limits for the \THDMa from the \METjets measurement do not depend significantly on the choice of test statistic.

In the whole section, \THDMa events are generated with the setup described in \secref{sec:interpretation_2HDMa_MC}.
\figref{fig:interp_toy_distributions} shows the probability density function $\pdf\left(\qBSM^{\vv s+\vvpiSM}\right)$ of \qBSM if the \THDMa was realised at the model point of $\ma=\SI{250}{GeV}$, $\mA=\SI{600}{GeV}$, $\tanB=1$ in green.
The fixed-normalisation prediction with differential cross sections as input quantities are used.
For illustration, the probability density functions are given for different multiples of the nominal signal cross section $\sigmaSig=\SI{0.64}{pb}$.
This demonstrates how a larger signal cross section would influence the distribution of the test statistic and its observed value.
The probability density function of \qBSM if the Standard Model is true is shown in blue.
The observed value of the test statistic given the data, \qBSMobs, is marked by a dashed black line.
\CLsb and \CLb correspond to the area under the curves to the right of \qBSMobs for the \THDMa and \SM hypothesis, respectively.

\newcommand{\myToyFigure}[3]{
	\begin{myfigure}{
			Probability density function of the test statistic $#1$ if the Standard Model~(blue) or \THDMa at the model point of $\ma=\SI{250}{GeV}$, $\mA=\SI{600}{GeV}$, $\tanB=1$ (green) was realised.
			The observed value of the test statistic \qBSMobs is marked by a dashed black line.
			The values for different multiples of the nominal signal cross section \sigmaSig are given.
		}{#2}
		\subfloat[\sigmaSig]{\includegraphics[width=0.3\textwidth]{figures/interpretation/2HDMa/toy_distributions/toys_combined_1.0_#3.pdf}}
		\hspace{10pt}
		\subfloat[$3\cdot\sigmaSig$]{\includegraphics[width=0.3\textwidth]{figures/interpretation/2HDMa/toy_distributions/toys_combined_3.0_#3.pdf}}
		\hspace{10pt}
		\subfloat[$6\cdot\sigmaSig$]{\includegraphics[width=0.3\textwidth]{figures/interpretation/2HDMa/toy_distributions/toys_combined_6.0_#3.pdf}}
	\end{myfigure}
}
\myToyFigure{\qBSM^{\vv s+\vvpiSM}}{fig:interp_toy_distributions}{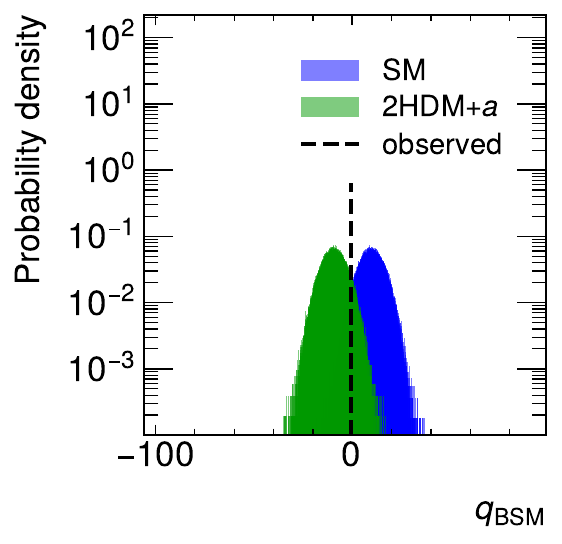}

If the signal cross section is small, the probability density functions form narrow, largely overlapping distributions.
The separation of the distribution increases with increasing signal cross section as the expected yields from the signal-plus-background and background-only hypotheses become distinguishable.
The distributions become wider with increasing signal cross section because the uncertainty of the signal grows proportionally.

The \CLs method is employed to exclude a model point and exclusion limits are set at \SI{95}{\%} confidence level (\cf\secref{sec:interpretation_LR_2HDMa}).
This means that a model point is excluded if the area to the right of \qBSMobs under the curve of the \THDMa hypothesis (\CLsb) is less than \SI{5}{\%} of the area to the right of \qBSMobs under the curve of the \SM hypothesis.
For the model point used in \figref{fig:interp_toy_distributions}, a signal with 3.2 times the nominal cross section is excluded.

\bigskip
In the following, the procedure is repeated successively for the two planes of \THDMa parameters that are particularly interesting to study, as identified in \secref{sec:2HDMa_parameterPlanes}.
On the one hand, the \mamA plane at $\tanB=1$ is investigated.
This plane is interesting because different decay channels of the \BSM bosons become kinematically open when the masses \ma and \mAeqmHeqmHpm are varied.
On the other hand, the \matanB plane at $\mAeqmHeqmHpm=\SI{600}{GeV}$ is investigated.
This plane is interesting as the coupling of the uncharged \BSM bosons to top quarks decreases and the one to bottom quarks increases with increasing \tanB (see \eg\eqref{eq:2HDMa_sigma_a}).
Therefore, different production channels become important.
All parameters that are not explicitly varied are chosen according to \tabref{tab:LHCDMWG_params}.

For each investigated parameter point, the probability density functions and \qBSMobs are determined.
From this, it can be decided if the signal can be excluded at \SI{95}{\%} confidence level.

\bigskip
In the following, the results of the \METjets measurement are interpreted with respect to their compatibility with the \THDMa.
The three fitting strategies investigated in the previous section vary in their interpretation of the original discrepancy between data and \SM prediction.
Nonetheless, all three strategies give a similarly good description of the data.
This description of the data by the prediction in the fit is the paramount handle defining the results of the \BSM exclusion limits.
Therefore, all three different fitting strategies discussed in the previous section are also studied in this section.

\subsection{\mamA plane}
\label{sec:interpretation_2HDMa_mamA_limits}

\figref{fig:interp_2HDMa_exclusion_mamA} shows the exclusion limits at \SI{95}{\%} confidence level from the \METjets measurement in the \mamA plane for the three different fit approaches.
All five measurement regions and both subregions binned in \METconst are used.

The observed and expected exclusion limits are marked by solid and dashed black lines, respectively.
The phase space to the left of the lines is excluded.
The expected exclusion limits are derived following the procedure outlined in \secref{sec:interpretation_LR_2HDMa}.
The uncertainty of the expected exclusion limits at \SI{95}{\%} confidence level is marked by coloured bands.
The central \SI{68}{\%} of the probability density function (\SI{\pm1}{\sigma}) for the expected test statistic \qBSM are marked by a green band.
The central \SI{95}{\%} of the probability density function (\SI{\pm2}{\sigma}) for the expected test statistic \qBSM are marked by a yellow band.

Dashed grey lines indicate where $\mA=\ma+\mh$ and $\mA=\ma$, respectively.

\begin{myfigure}{
		Expected (dashed lines) and observed (solid lines) exclusion limits at \SI{95}{\%} confidence level from the \METjets measurement in the \mamA plane for the three different fit approaches.
		The excluded parameter space is to the left of the lines.
		The green (yellow) band indicates the region of one (two) standard deviations from the expected exclusion limit.
		\THDMaLines{grey}
	}{fig:interp_2HDMa_exclusion_mamA}
	\subfloat[fixed normalisation, \dSigmaDMET]{\includegraphics[width=0.48\textwidth]{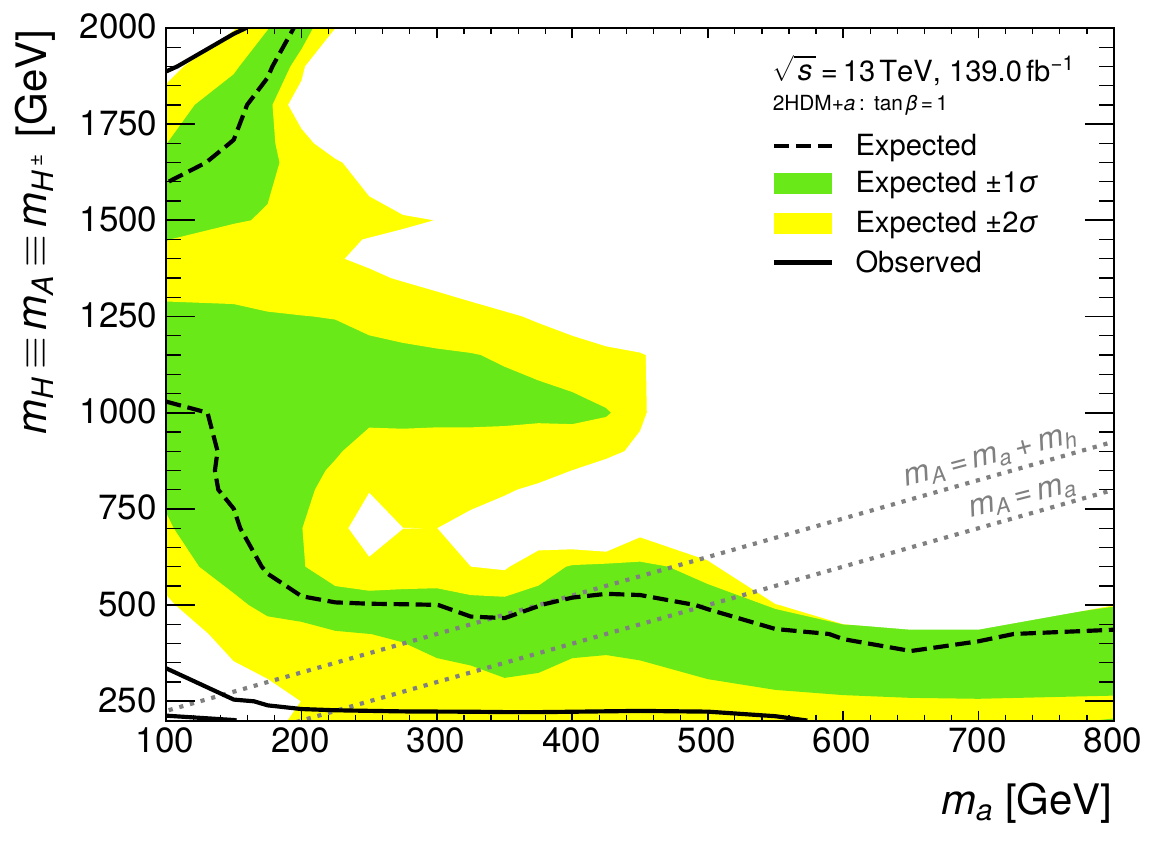}}\\
	\subfloat[floating normalisation, \dSigmaDMET]{\includegraphics[width=0.48\textwidth]{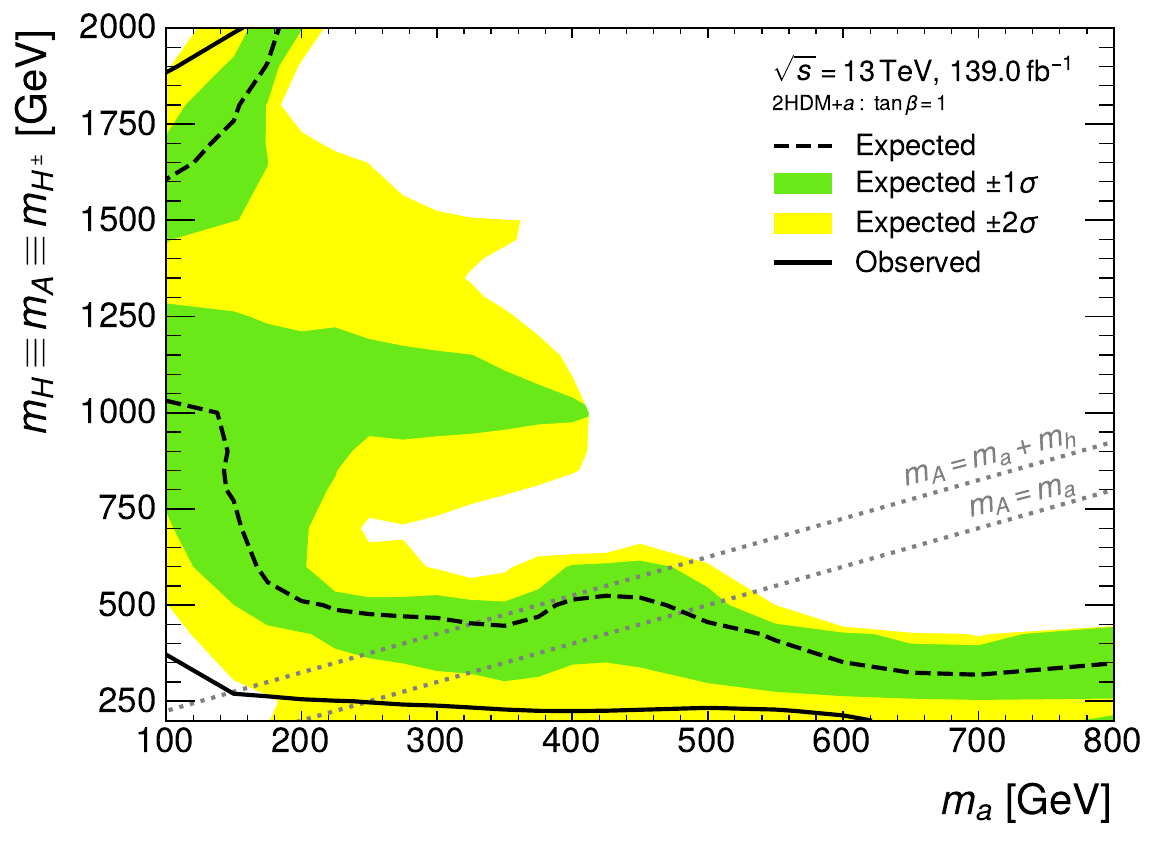}}
	\hspace{10pt}
	\subfloat[fixed normalisation, \Rmiss]{\includegraphics[width=0.48\textwidth]{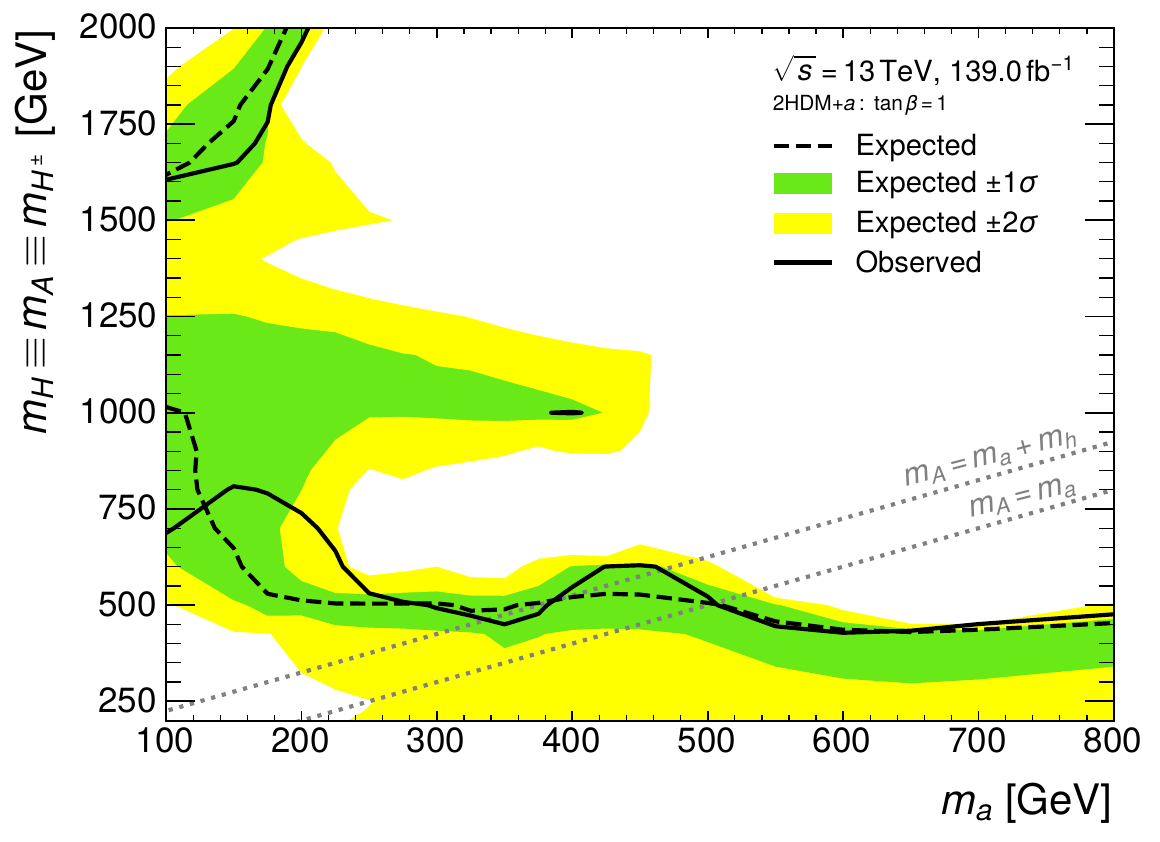}}
\end{myfigure}

\bigskip
In \subfigref{fig:interp_2HDMa_exclusion_mamA}{a}, the exclusion limits are shown when the differential cross sections ($\mathrm{d}\sigma/\mathrm{d}\METconst$) serve as data~$\vv x$ and prediction~$\vv\pi$ and the fixed-normalisation prediction according to \eqref{eq:interpretation_SM_pred_fixedNorm} is used.
This corresponds to the \SM fit discussed in \secref{sec:interpretation_SM_fixedNorm_diffXS}.

At small \mA, the expected exclusion is independent of \ma.
The region $\mA<\SI{500}{GeV}$ is expected to be excluded, with marginally stronger limits when $\mA>\ma$ but the decay channel $A\to ah$ is not yet kinematically open.
This exclusion closely follows the region of large cross section for \xxqg processes (\cf\subfigref{fig:interp_2HDMa_crossSections_mamA}{a}).

At small \ma and small \mA, the region $\ma<\SI{150}{GeV}, \mA<\SI{1000}{GeV}$ is expected to be excluded.
In this region, all considered \THDMa process types have large cross sections~(\cf\figref{fig:interp_2HDMa_crossSections_mamA}).

At small \ma and large \mA, the region $\ma<\SI{200}{GeV}, \mA>\SI{1600}{GeV}$ is expected to be excluded.
This exclusion closely follows the region of large cross section for \xxh processes at large \mA (\cf\subfigref{fig:interp_2HDMa_crossSections_mamA}{b}).

The uncertainty bands for the \SI{\pm1}{\sigma} (\SI{\pm2}{\sigma}) expected exclusion limits approximately follow these borders, with an offset of about \SI{100}{GeV} (\SI{200}{GeV}).
They deviate most notably at $\mA\approx\SI{1000}{GeV}$ where the bands reach up to $\ma=\SI{430}{GeV}$ (\SI{460}{GeV}).
In this region, resonant production of the heavy \BSM bosons and subsequent decays $A\to ah$, $H\to aZ$ and $\Hpm\to aW^\pm$ are all kinematically open.
At the same time, for $\ma>\SI{350}{GeV}\approx2m_t$ the pseudoscalar $a$ can be resonantly produced in \xxqg processes.
This region is currently just out of reach of the nominal expected limits.

The observed exclusion is considerably weaker than the expected exclusion.
This is a consequence of the broad signal shape (\cf\secref{sec:interpretation_2HDMa_contribs}) as well as the large post-fit pulls and residual discrepancy between \SM prediction and measured data (\cf\secref{sec:interpretation_SM}).
The post-fit pulls and residual discrepancy, \eg at $\METconst\approx\SI{1100}{GeV}$ in the signal regions, can be smaller than for the \SM-only hypothesis if specific \THDMa signals are included in the prediction.
This is demonstrated in \figref{fig:interp_2HDMa_NPrankings_fixedNorm_diffXS_signals}:
Given are the post-fit pull $\hat\theta$ and constraint for the pull $\sigma_{\hat\theta}$ of the systematic uncertainties when a \THDMa signal at $\ma=\SI{100}{GeV}, \mA=\SI{250}{GeV}$ or $\ma=\SI{200}{GeV}, \mA=\SI{600}{GeV}$ is included in the prediction.
The relative uncertainties $u_{j, \vv y}$ of the yield $\vv y$ corresponding to a systematic uncertainty (blue and turquoise) are shown as well.
The \THDMa signal used in the fit for \subfigref{fig:interp_2HDMa_NPrankings_fixedNorm_diffXS_signals}{a} is excluded, the \THDMa signal used in the fit for \subfigref{fig:interp_2HDMa_NPrankings_fixedNorm_diffXS_signals}{b} is not excluded.
In \figref{fig:interp_SM_NPrankings_fixedNorm_diffXS}, the pulls and uncertainties for a similar fit, but without a \THDMa signal, were shown.
Comparing the three figures, it can be observed that the pulls on parameters are in general larger if the excluded \THDMa signal ($\ma=\SI{100}{GeV}, \mA=\SI{250}{GeV}$) is considered.

This effect is quantified in \tabref{tab:interpretation_2HDMa_fixedNorm_diffXS_totalChi2}.
Given are the total contributions~$\sum_i\hat{\theta}_i^2$ from the pulled nuisance parameters to the test statistic (see also \eqref{eq:interpretation_qSM}).
The contributions are shown for the case without a \THDMa signal, that was already discussed in \secref{sec:interpretation_SM_fixedNorm_diffXS}, as well as for the two \THDMa signals in \figref{fig:interp_2HDMa_NPrankings_fixedNorm_diffXS_signals}.
If the \THDMa signal at $\ma=\SI{100}{GeV}, \mA=\SI{250}{GeV}$ is included in the prediction, larger pulls on the nuisance parameters are needed to remove the discrepancy between data and prediction.
This increases the total contribution of the pulls to the test statistic by approximately \SI{8}{\%} with respect to the no-signal case.
This increase is just enough to exclude the signal hypothesis at \SI{95}{\%} confidence level.
If the \THDMa signal at $\ma=\SI{200}{GeV}, \mA=\SI{600}{GeV}$ is included in the prediction, the obtained fit optimum uses smaller pulls.
The total contribution of the pulls to the test statistic is smaller than for the excluded signal ($\ma=\SI{100}{GeV}, \mA=\SI{250}{GeV}$) and even smaller than for the no-signal case.
Consequently, the signal hypothesis cannot be excluded at \SI{95}{\%} confidence level.

\begin{myfigure}{%
		\NPrankingText{differential cross sections}{fixed}{including the given signals }
	}{fig:interp_2HDMa_NPrankings_fixedNorm_diffXS_signals}{}
		\subfloat[$\ma=\SI{100}{GeV}, \mA=\SI{250}{GeV}$]{\includegraphics[height=230pt]{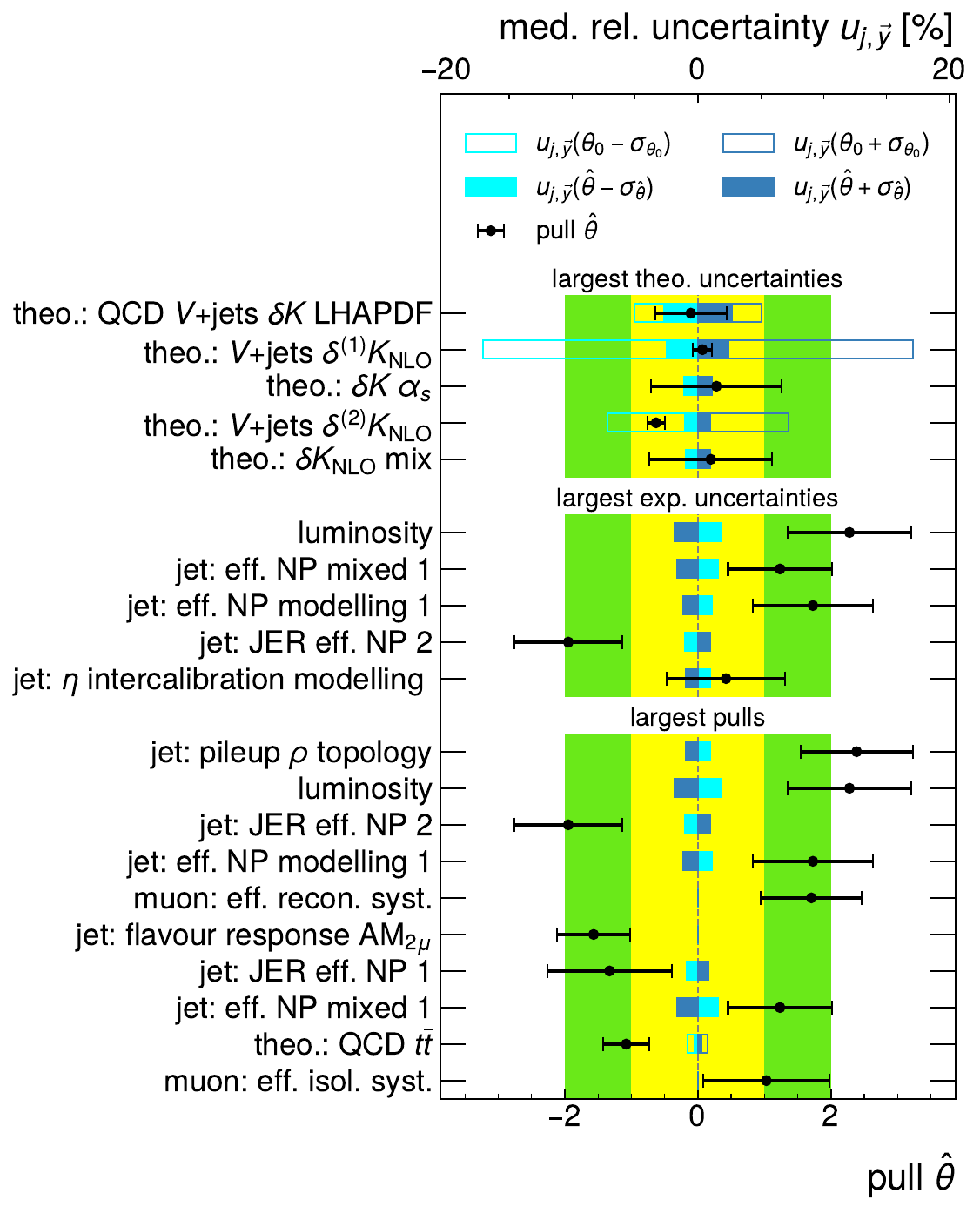}}
		\hspace{10pt}
		\subfloat[$\ma=\SI{200}{GeV}, \mA=\SI{600}{GeV}$]{\includegraphics[height=230pt]{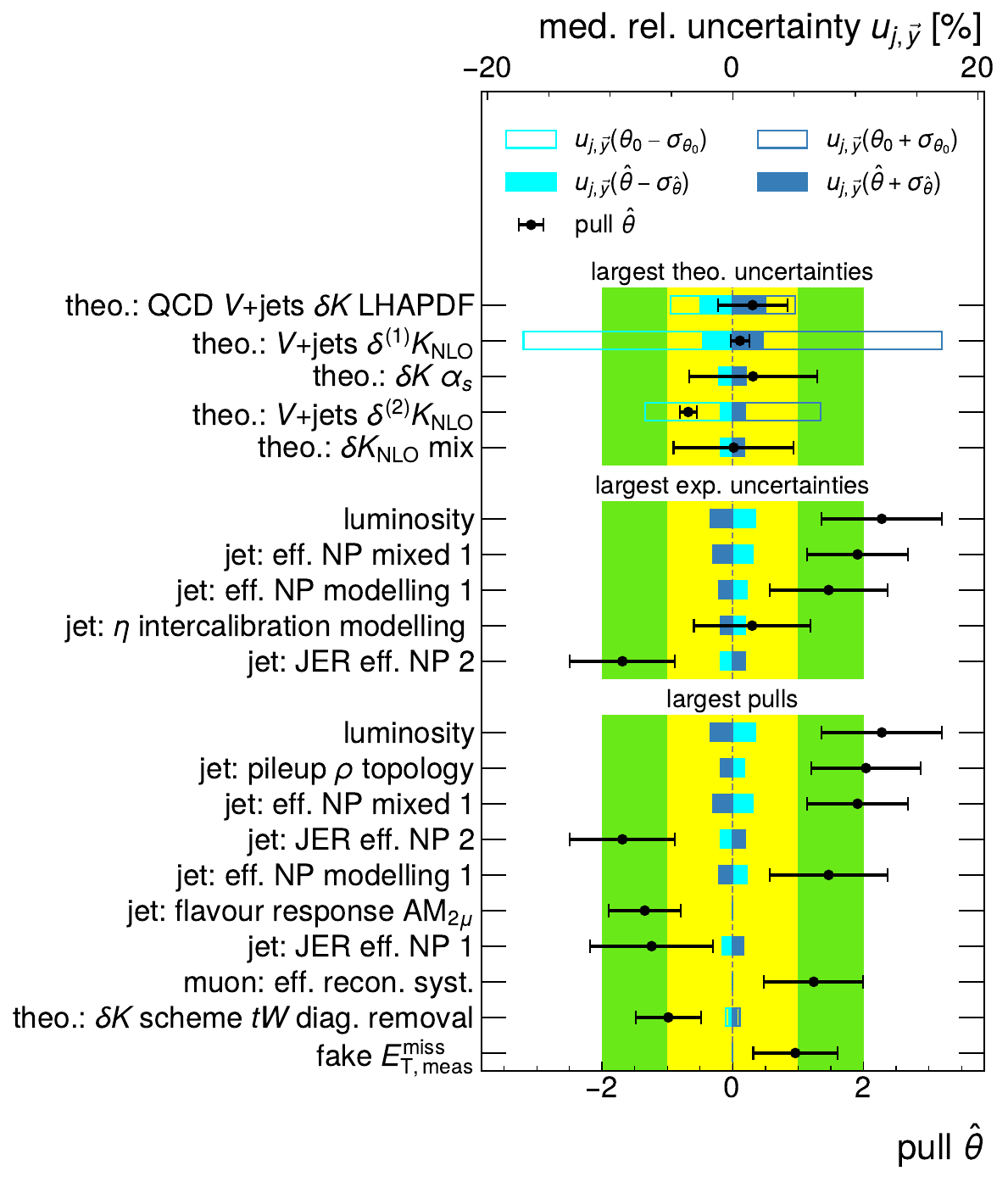}}\\
\end{myfigure}

\begin{mytable}{
		Total contribution from the pulled nuisance parameters to the test statistic for three different signal hypotheses.
		The differential cross sections are used for data and prediction.
	}{tab:interpretation_2HDMa_fixedNorm_diffXS_totalChi2}{cccc}
	& no signal & $\ma=\SI{100}{GeV}, \mA=\SI{250}{GeV}$ & $\ma=\SI{200}{GeV}, \mA=\SI{600}{GeV}$\\
	\midrule
	$\sum_i\hat{\theta}_i^2$ & 39.3 & 42.6 & 37.9\\
\end{mytable}

The observed exclusion limits in \subfigref{fig:interp_2HDMa_exclusion_mamA}{a} are those limits for the \THDMa that can be placed despite the large post-fit pulls and residual discrepancy between data and \SM prediction.
masses up to $\ma=\SI{570}{GeV}$ and up to $\mA=\SI{340}{GeV}$ can be excluded at \SI{95}{\%} confidence level in the region of small \mA, as well as masses up to $\ma=\SI{160}{GeV}$ and larger than $\mA=\SI{1900}{GeV}$ at large \mA.

\bigskip
In \subfigref{fig:interp_2HDMa_exclusion_mamA}{b}, the exclusion limits are shown when the differential cross sections serve as data~$\vv x$ and prediction~$\vv\pi$ and the floating-normalisation prediction according to \eqref{eq:interpretation_SM_pred_floatNorm} is used.
This corresponds to the \SM fit discussed in \secref{sec:interpretation_SM_floatNorm_diffXS}.

As in the \SM fit, the results for this fit approach are very similar to the one using the \SM prediction with fixed normalisation (\cf\subfigref{fig:interp_2HDMa_exclusion_mamA}{a}).
The expected exclusion limits and their uncertainty bands are almost identical because the post-fit \SM prediction does not differ significantly for the two approaches.
The observed exclusion limits are marginally stronger for the \SM prediction with fixed normalisation:
Masses up to $\ma=\SI{620}{GeV}$ and up to $\mA=\SI{370}{GeV}$ can be excluded at \SI{95}{\%} confidence level in the region of small \mA.
For large \mA, the exclusion for the \SM prediction with fixed and floating normalisation are identical.
This is because in this region the \THDMa contribution originates almost exclusively from \xxh processes, which give a more pronounced rise in yield with \METconst than the \SM contributions.
This shape difference can be accounted for equally bad by the \SM prediction with fixed and floating normalisation parameters, leading to the \THDMa exclusion.

\bigskip
In \subfigref{fig:interp_2HDMa_exclusion_mamA}{c}, the exclusion limits are shown when the \Rmiss distributions serve as data~$\vv x$ and prediction~$\vv\pi$ and the fixed-normalisation prediction according to \eqref{eq:interpretation_SM_pred_fixedNorm} is used.
This corresponds to the \SM fit discussed in \secref{sec:interpretation_SM_fixedNorm_Rmiss}.

The expected exclusion is almost identical to the one when using differential cross sections as input quantity (\cf\subfigsref{fig:interp_2HDMa_exclusion_mamA}{a}{b}).
This is because, for expected exclusion limits, the signal-plus-background hypothesis differs from the expected data in both cases in a similar way.
It does therefore not matter whether the \THDMa contributions, which are dominant in the signal region, are divided by the yields in the auxiliary measurements (\Rmiss) or not ($\mathrm{d}\sigma/\mathrm{d}\METconst$).

For the observed limits, $\mA<\SI{425}{GeV}$ is excluded independent of \ma.
At $\ma=\SI{150}{GeV}$ (\SI{450}{GeV}) masses $\mA<\limitmamAmAmin$ (\SI{600}{GeV}) are excluded.
At large \mA, masses larger than $\mA=\limitmamAmAmax$ as well as masses up to $\ma=\SI{200}{GeV}$ are excluded.
The observed exclusion limits are sometimes weaker and sometimes stronger than expected.
The reason for this is that in \Rmiss, the residual discrepancy between post-fit \SM prediction and data fluctuates around 0.
This means that the observed limits can be weaker as well as stronger than the expected ones, depending on the shape of the signal contributions.
In most cases, the observed exclusion limits are within \SI{68}{\%} of the expected exclusion limits.

In general, the pulls on nuisance parameters when including a signal hypothesis are larger than when no signal is included.
This is demonstrated in \figref{fig:interp_2HDMa_NPrankings_fixedNorm_Rmiss_signals}:
Given are the post-fit pull $\hat\theta$ and constraint for the pull $\sigma_{\hat\theta}$ of the systematic uncertainties when a \THDMa signal at $\ma=\SI{100}{GeV}, \mA=\SI{250}{GeV}$ or $\ma=\SI{200}{GeV}, \mA=\SI{600}{GeV}$ is included in the prediction.
The relative uncertainties $u_{j, \vv y}$ of the yield $\vv y$ corresponding to a systematic uncertainty (blue and turquoise) are shown as well.
Both \THDMa signals used in the fits are excluded (\cf\subfigref{fig:interp_2HDMa_exclusion_mamA}{c}).
This is in contrast to the case of differential cross sections shown in \subfigref{fig:interp_2HDMa_exclusion_mamA}{a} where the signal at $\ma=\SI{200}{GeV}, \mA=\SI{600}{GeV}$ could not be excluded.
In \figref{fig:interp_SM_NPrankings_fixedNorm_Rmiss}, the pulls and uncertainties for a similar fit, but without a \THDMa signal, were shown.
Comparing the three figures, it can be observed that the pulls on parameters are in general larger if either \THDMa signal is considered.
In the no-signal case, all nuisance parameters were pulled by less than two standard deviations.
The nuisance parameter for the fake \METmeas uncertainty is pulled by \SI{2.1}{standard} deviations for both considered signal hypotheses.

The effect of the nuisance parameters is further quantified in \tabref{tab:interpretation_2HDMa_fixedNorm_Rmiss_totalChi2}.
Given are the total contributions~$\sum_i\hat{\theta}_i^2$ from the pulled nuisance parameters to the test statistic, \eg according to \eqref{eq:interpretation_qSM} in the no-signal case.
The contributions are shown for the case without a \THDMa signal, that was already discussed in \secref{sec:interpretation_SM_fixedNorm_Rmiss}, as well as for the two \THDMa signals in \figref{fig:interp_2HDMa_NPrankings_fixedNorm_Rmiss_signals}.
If any \THDMa signal is included in the prediction, larger pulls on the nuisance parameters are needed to remove the discrepancy between data and prediction.
This increases the total contribution of the pulls to the test statistic by approximately \SI{50}{\%} (\SI{8}{\%}) for the signal at $\ma=\SI{100}{GeV}, \mA=\SI{250}{GeV}$ ($\ma=\SI{200}{GeV}, \mA=\SI{600}{GeV}$) with respect to the no-signal case.
This increase is enough to exclude both signal hypotheses at \SI{95}{\%} confidence level.
The signal at $\ma=\SI{200}{GeV}, \mA=\SI{600}{GeV}$ could not be excluded at \SI{95}{\%} confidence level if differential cross sections were used as input quantity.

\begin{myfigure}{%
		\NPrankingText{\Rmiss distributions}{fixed}{including the given signals }
	}{fig:interp_2HDMa_NPrankings_fixedNorm_Rmiss_signals}{}
		\subfloat[$\ma=\SI{100}{GeV}, \mA=\SI{250}{GeV}$]{\includegraphics[height=230pt]{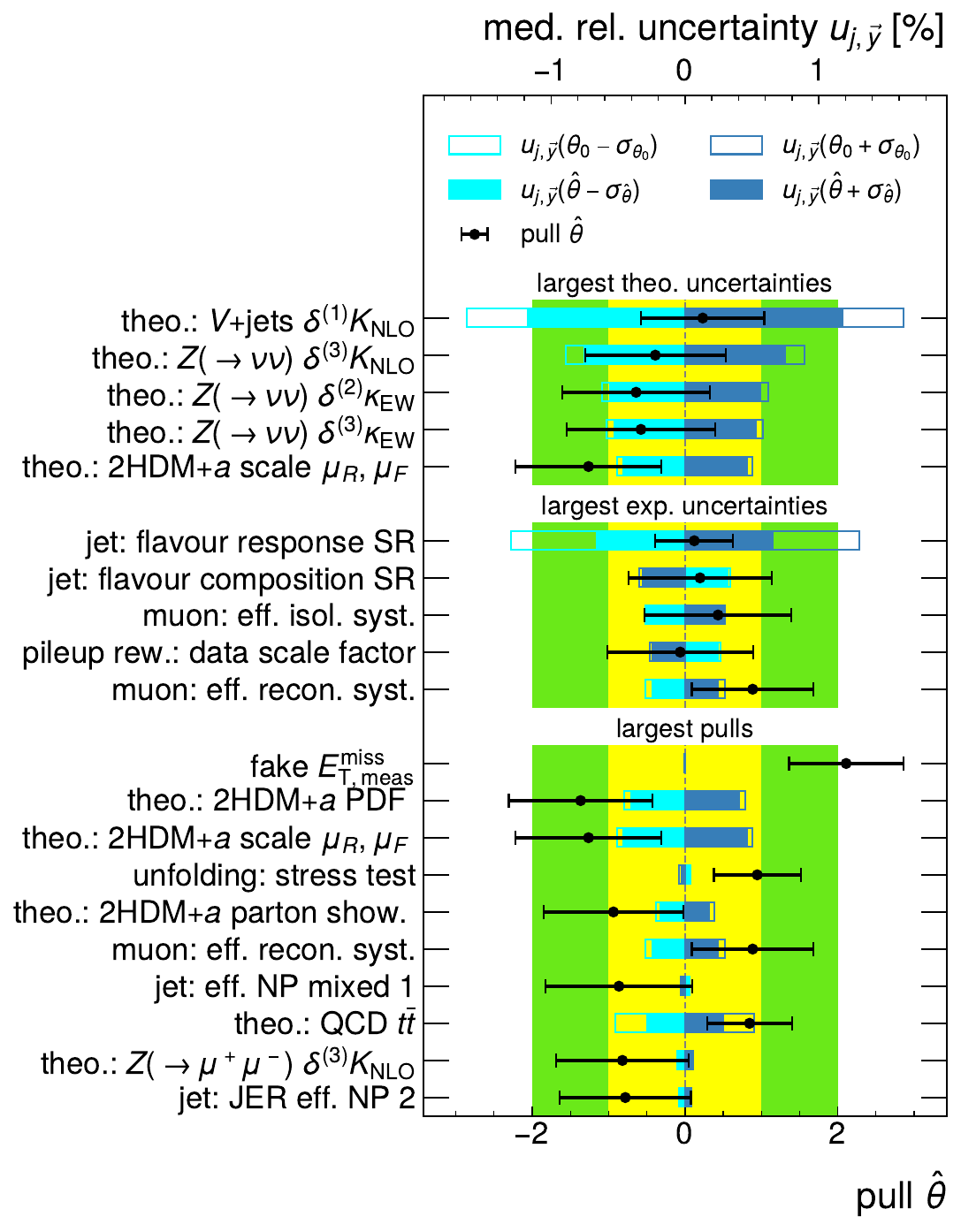}}
		\hspace{10pt}
		\subfloat[$\ma=\SI{200}{GeV}, \mA=\SI{600}{GeV}$]{\includegraphics[height=230pt]{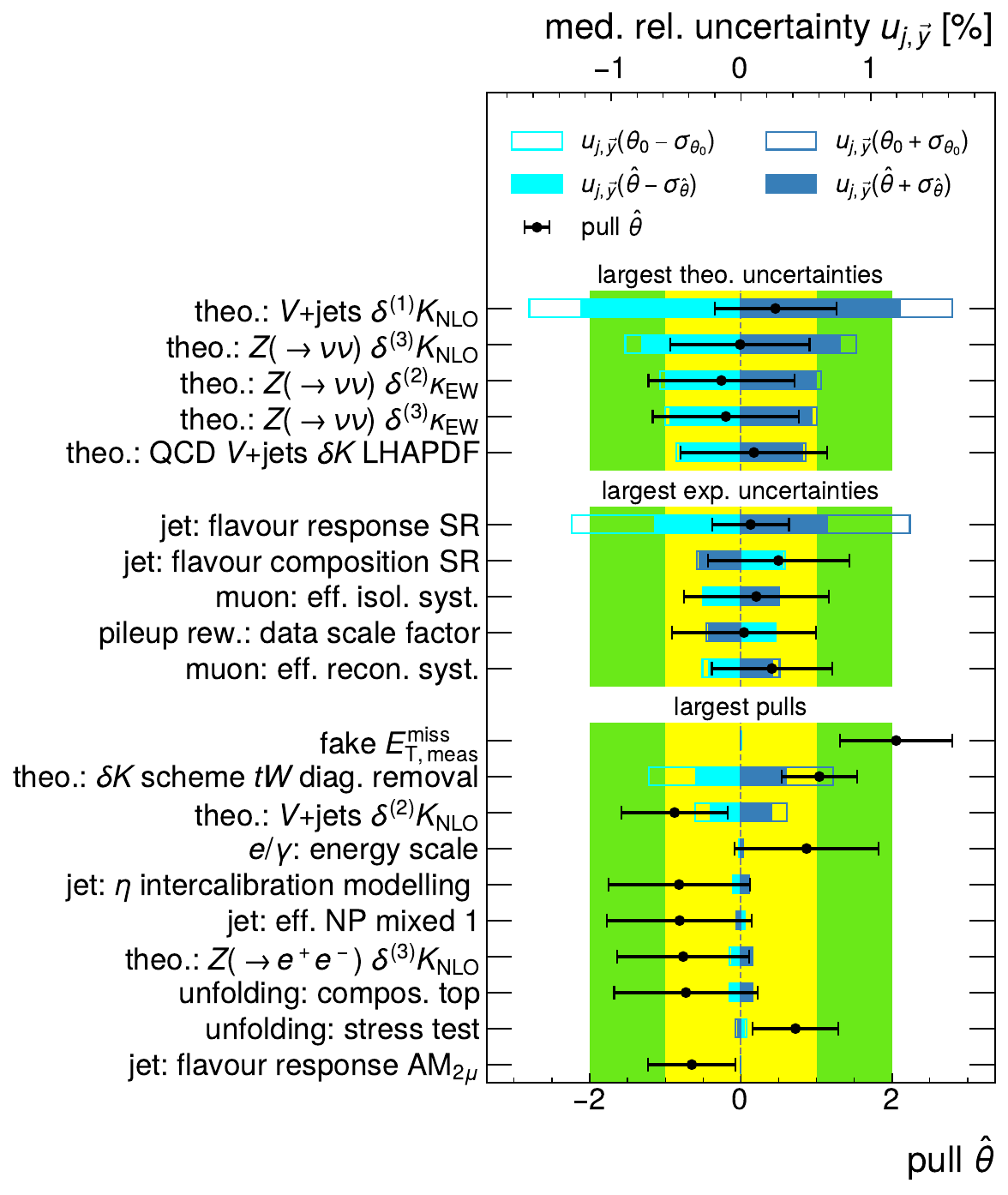}}\\
\end{myfigure}

\begin{mytable}{
		Total contribution from the pulled nuisance parameters to the test statistic for three different signal hypotheses.
		The \Rmiss distributions are used for data and prediction.
	}{tab:interpretation_2HDMa_fixedNorm_Rmiss_totalChi2}{cccc}
	& no signal & $\ma=\SI{100}{GeV}, \mA=\SI{250}{GeV}$ & $\ma=\SI{200}{GeV}, \mA=\SI{600}{GeV}$\\
	\midrule
	$\sum_i\hat{\theta}_i^2$ & 15.4 & 22.7 & 16.5\\
\end{mytable}

\bigskip
In summary, all three fit approaches yield similar results.
The fixed- and floating-normalisation fitting strategies using the differential cross sections give almost identical results, as was the case in the interpretation with respect to the Standard Model (\cf\secref{sec:interpretation_SM}).
Similarly, the fixed-normalisation fit approach using the \Rmiss distributions shows marginally superior observed exclusion limits because of the generally smaller post-fit pulls and residual discrepancy between measured data and \SM prediction.

\subsection{\matanB plane}

\figref{fig:interp_2HDMa_exclusion_matb} shows the exclusion limits at \SI{95}{\%} confidence level from the \METjets measurement in the \matanB plane.
The phase spaces at smaller masses and more extreme \tanB are excluded.

\begin{myfigure}{
		Expected (dashed lines) and observed (solid lines) exclusion limits at \SI{95}{\%} confidence level from the \METjets measurement in the \matanB plane for the three different fit approaches.
		The excluded parameter space is to the left or below of the lines.
		The green (yellow) band indicates the region of one (two) standard deviations from the expected exclusion limit.
		\THDMaLines{grey}
	}{fig:interp_2HDMa_exclusion_matb}
	\subfloat[fixed normalisation, \dSigmaDMET]{\includegraphics[width=0.48\textwidth]{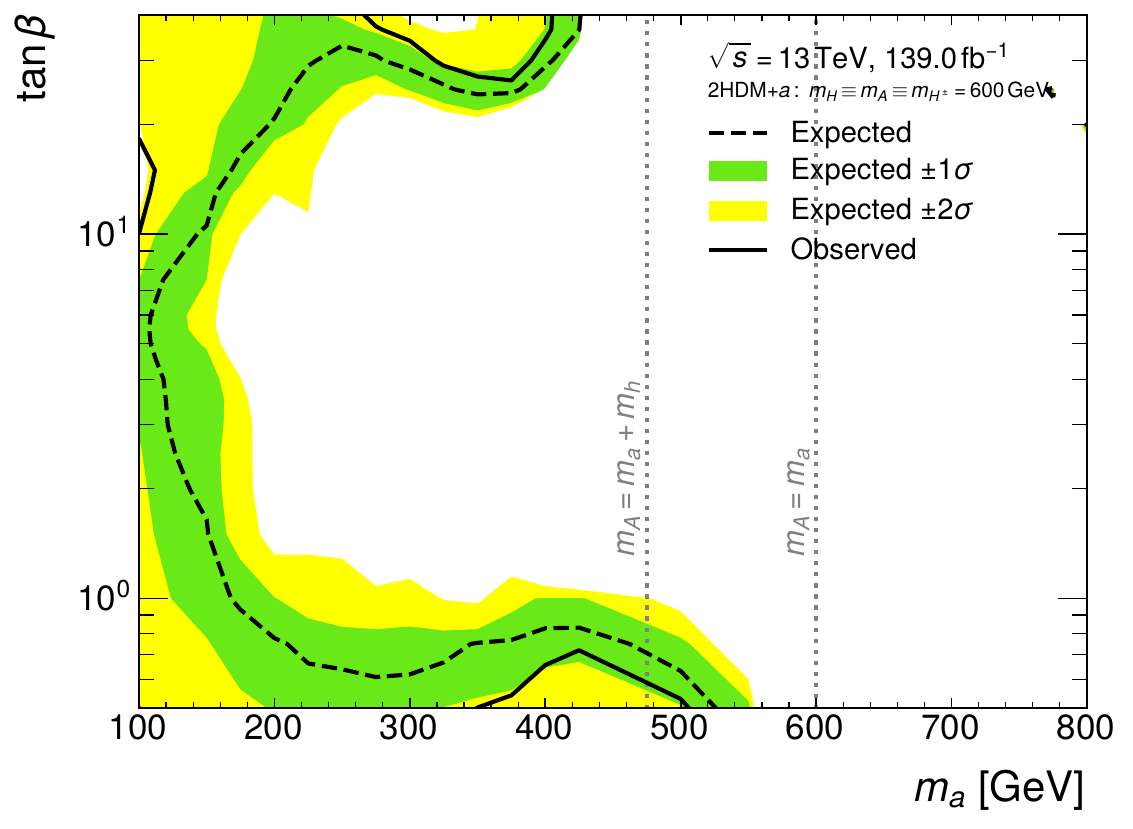}}\\
	\subfloat[floating normalisation, \dSigmaDMET]{\includegraphics[width=0.48\textwidth]{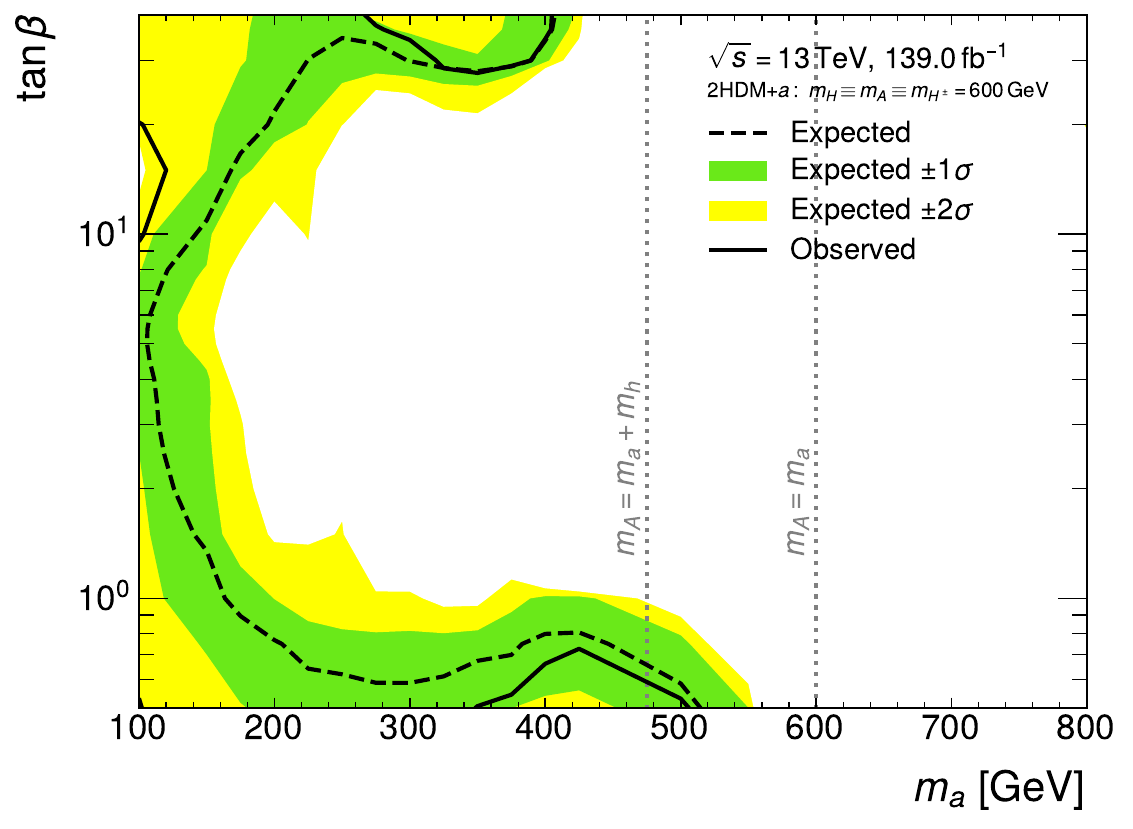}}
	\hspace{10pt}
	\subfloat[fixed normalisation, \Rmiss]{\includegraphics[width=0.48\textwidth]{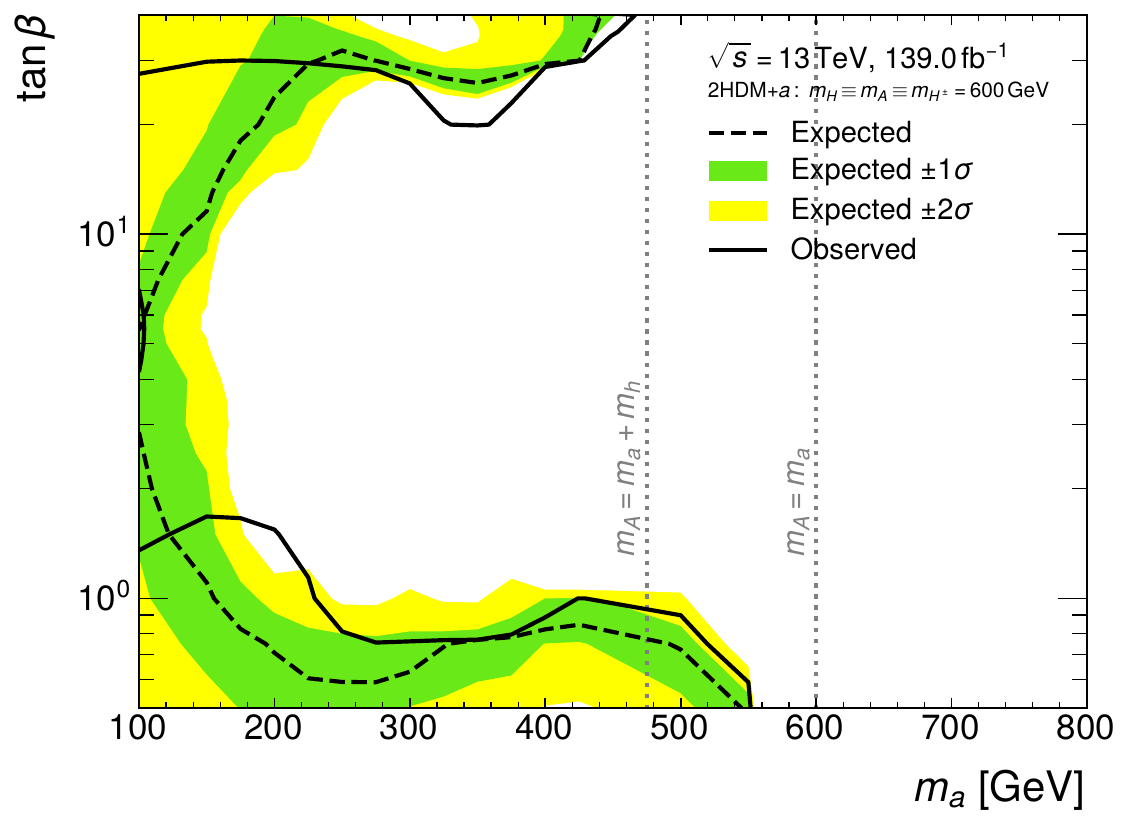}}
\end{myfigure}

\bigskip
In \subfigref{fig:interp_2HDMa_exclusion_matb}{a}, the exclusion limits are shown when the differential cross sections ($\mathrm{d}\sigma/\mathrm{d}\METconst$) serve as data~$\vv x$ and prediction~$\vv\pi$ and the fixed-normalisation prediction according to \eqref{eq:interpretation_SM_pred_fixedNorm} is used.
This corresponds to the \SM fit discussed in \secref{sec:interpretation_SM_fixedNorm_diffXS}.
There are three regions of expected exclusion:
\begin{itemize}
	\item At small \tanB, masses of the pseudoscalar $a$ are expected to be excluded up to $\ma=\SI{530}{GeV}$ if $\tanB<0.8$.
	This exclusion closely follows the region of large cross section at small \tanB for \xxqg processes (\cf\subfigref{fig:interp_2HDMa_crossSections_matanB}{a}) because of top-quark induced production of the pseudoscalars $a$ and $A$ in association with jets.
	The exclusion limits are marginally stronger at $\ma>\SI{350}{GeV}\approx2m_t$ because here the pseudoscalar $a$ can be resonantly produced from top quarks.
	
	\item At large \tanB, masses up to $\ma=\SI{440}{GeV}$ of the pseudoscalar $a$ are expected to be excluded  if $\tanB>30$.
	This exclusion closely follows the region of large cross section at large \tanB for \xxqg (\cf\subfigref{fig:interp_2HDMa_crossSections_matanB}{a}) and other tree-level processes (\cf\subfigref{fig:interp_2HDMa_crossSections_matanB}{c}) because of bottom-quark induced production of \BSM bosons.
	The exclusion limits are stronger at $\ma>\SI{300}{GeV}=\mH/2$ because the decay channel $H\to aa$ kinematically closes and the branching fraction for $H\to aZ$ increases in consequence.
	The \METjets measurement is more sensitive to the latter because the high-momentum jet required by the event selection can be obtained from \Zqq decays instead of initial-state radiation.
	The sensitivity is reduced when $\mH=\ma+m_Z$ is reached and this decay channel closes kinematically.
	
	\item At small \ma, masses of the pseudoscalar~$a$ are expected to be excluded up to $\ma=\SI{150}{GeV}$ independent of \tanB.
	This exclusion originates from a mixture of all considered \THDMa processes (\cf\figref{fig:interp_2HDMa_crossSections_matanB}).
	The sensitivity at small and large \tanB overlaps at intermediate \tanB if the mass of the pseudoscalar $a$ is small enough.
\end{itemize}

The uncertainty bands for the \SI{\pm1}{\sigma} and \SI{\pm2}{\sigma} expected exclusion limits approximately follow these borders.

As in the \mamA plane, the observed exclusion limits are significantly weaker.
At small \tanB, masses between \SI{340}{GeV} and \SI{500}{GeV} of the pseudoscalar~$a$ can be excluded if $\tanB<0.7$.
At large \tanB, masses between \SI{250}{GeV} and \SI{400}{GeV} of the pseudoscalar~$a$ can be excluded if $\tanB>30$.
At small \ma, values $10<\tanB<20$ can be excluded for $\ma<\SI{120}{GeV}$.

\bigskip
In \subfigref{fig:interp_2HDMa_exclusion_matb}{b}, the exclusion limits are shown when the differential cross sections serve as data~$\vv x$ and prediction~$\vv\pi$ and the floating-normalisation prediction according to \eqref{eq:interpretation_SM_pred_floatNorm} is used.
This corresponds to the \SM fit discussed in \secref{sec:interpretation_SM_floatNorm_diffXS}.

As in the \mamA plane, the exclusion limits are very similar to those using the fixed-normalisation prediction (\cf\subfigref{fig:interp_2HDMa_exclusion_matb}{a}).
	
\bigskip
In \subfigref{fig:interp_2HDMa_exclusion_matb}{c}, the exclusion limits are shown when the \Rmiss distributions serve as data~$\vv x$ and prediction~$\vv\pi$ and the fixed-normalisation prediction according to \eqref{eq:interpretation_SM_pred_fixedNorm} is used.
This corresponds to the \SM fit discussed in \secref{sec:interpretation_SM_fixedNorm_Rmiss}.

As in the \mamA plane, the expected exclusion limits are very similar to those using the fixed-normalisation prediction (\cf\subfigsref{fig:interp_2HDMa_exclusion_matb}{a}{b}).
The observed exclusion limits are stronger than when using the fixed-normalisation prediction.
At small \tanB, values of \tanB up to 1.0 can be excluded if $\ma<\limitmatbma$.
If $\ma\approx\SI{160}{GeV}$, values of \tanB up to \limitmatbtbmin can be excluded.
At large \tanB, values of \tanB larger than 30 can be excluded if $\ma<\SI{460}{GeV}$.
If $\ma\approx\SI{350}{GeV}$, values of \tanB larger than \limitmatbtbmax can be excluded.
At small \ma, values $4<\tanB<7$ can be excluded for $\ma<\SI{110}{GeV}$.

\bigskip
In summary, all three fit approaches yield similar results, as was the case for the \mamA plane.
The fit approach using the \Rmiss distributions again shows marginally superior observed exclusion limits.

\section{Conclusion}

In this chapter, the results of the \METjets measurement at particle-level were interpreted with respect to their agreement with \SM and \THDMa predictions.

The pre-fit agreement between measured differential cross sections and generated \SM predictions is small.
There is good agreement after performing a statistical fit to optimise nuisance parameters.
The impact of introducing normalisation parameters was tested.
Similar post-fit agreement was found.
Better pre- as well as post-fit agreement is achieved between measured data and \SM predictions for \Rmiss distributions.
This implies that there is a common mismodelling of the vector-boson \pT.
This mismodelling is cancelled in the \Rmiss ratio but of course cannot be fixed by normalisation parameters.

The \THDMa would give significant contributions to the phase space assessed by the \METjets measurement through a variety of processes.
This can be exploited to set exclusion limits on the \THDMa parameter space.
The measurement results are expected to give strong exclusion limits.
The observed exclusion limits using differential cross sections are considerably weaker than expected due to the large post-fit pulls and residual difference between the \SM prediction and the measured data.
The observed exclusion limits using \Rmiss distributions are superior because common mismodellings are cancelled.
Because of this, the post-fit pulls and residual difference between \SM prediction and measured data is smaller.
No significant difference in the derived exclusion limits depending on the usage of normalisation parameters or choice of test statistic is found.

\Chapter[1]{Exploring existing sensitivity}{The \Contur approach}{%
	Muse}{Explorers~\cite{Muse:2012exp}}{verse 1, lines 1-3}
\label{sec:Contur}


There has been no observation of \BSM physics at the \LHC to date.
Investigating the phase space that has been explored helps to identify gaps in the coverage and decide which \BSM models can be disfavoured.
In \chapsref{sec:metJets}{sec:interpretation} it was shown, on the example of the \METjets measurement, how a \textit{specific} phase space can be selected, corrected for detector effects and interpreted with respect to a chosen \BSM model.
Hundreds of measurements in analogue phase spaces have been performed at the \LHC alone.
In principle, the chosen model should be explored in each of these phase spaces to maximise the insights into the model as well as gain the maximum physics impact from each measurement.
At the same time, there is a multitude of different \BSM models worth investigating.
Not every single one of these models can be studied in every \LHC measurement, of course.
In particular, the main focus of a measurement should stay on the experimental details and not be diverted into numerous interpretations.
This is the problem the \Contur toolkit~\cite{contur_zenodo,Buckley:2021neu} addresses.
The toolkit uses hundreds of existing measurements to investigate the phase space of \BSM models that has been covered at colliders.
One focus of the toolkit thereby is to allow easy switching between models such that new \BSM models can be studied with little extra effort.

The method used by the \Contur toolkit is described in depth in \secref{sec:contur_method}.
In \secref{sec:contur_2HDMa}, \Contur is employed to investigate the parameter space of the \BSM model already used to interpret the \METjets measurement in \secref{sec:interpretation_2HDMa}, the \THDMa.

\vfill 

\section{The \Contur method}
\label{sec:contur_method}

\secref{sec:contur_idea} explains the general idea of the \Contur method.
The technical details are given in \secref{sec:contur_techDetails}.
The procedure of the statistical interpretation is described in \secref{sec:Contur_statistics}.
Limitations of the \Contur method are discussed in \secref{sec:Contur_Limitiations}.

\subsection{General idea}
\label{sec:contur_idea}

The broad concept behind the \Contur toolkit is that the Standard Model is well known and experimentally well studied.
Any theory adding effects beyond the Standard Model inevitably also results in a change to the \SM Lagrangian, as was seen for the \THDMa in \secref{sec:2HDMa}.
Modifications to the \SM Lagrangian, however, result in deviations from the \SM expectations in observed quantities.

The \Contur toolkit makes use of a repository of hundreds of measurements at colliders.
Given the variety of phase spaces probed by the large amount of measurements, many \BSM models give rise to significant deviations in at least some of the observed quantities.
The measurements have, however, been shown to be in general in good agreement with the \SM expectation.
In consequence, the repository can be used by \Contur to constrain the parameter space for a chosen \BSM model.

With this approach, the goal of the \Contur toolkit is to perform the widest possible analysis of a model coverage given existing measurements.
These not only include \LHC experiments but are continuously being extended to other present and past experiments.

The \Contur toolkit relies on well-established and -maintained external tools and interfaces.
This ensures that little effort is needed to switch parts of the \Contur workflow, \eg the examined \BSM model or employed Monte-Carlo event generator, and provide state-of-the-art results.

\subsection{Technical implementation}
\label{sec:contur_techDetails}

\begin{myfigure}{
		Schematic illustrating the workflow for obtaining exclusion limits with \Contur.
		The blue (red) arrows indicate steps taken for data (theory prediction).
		The framed boxes give the name of tools that are employed for these steps.
	}{fig:Contur_workflow}
	\includegraphics[width=\textwidth]{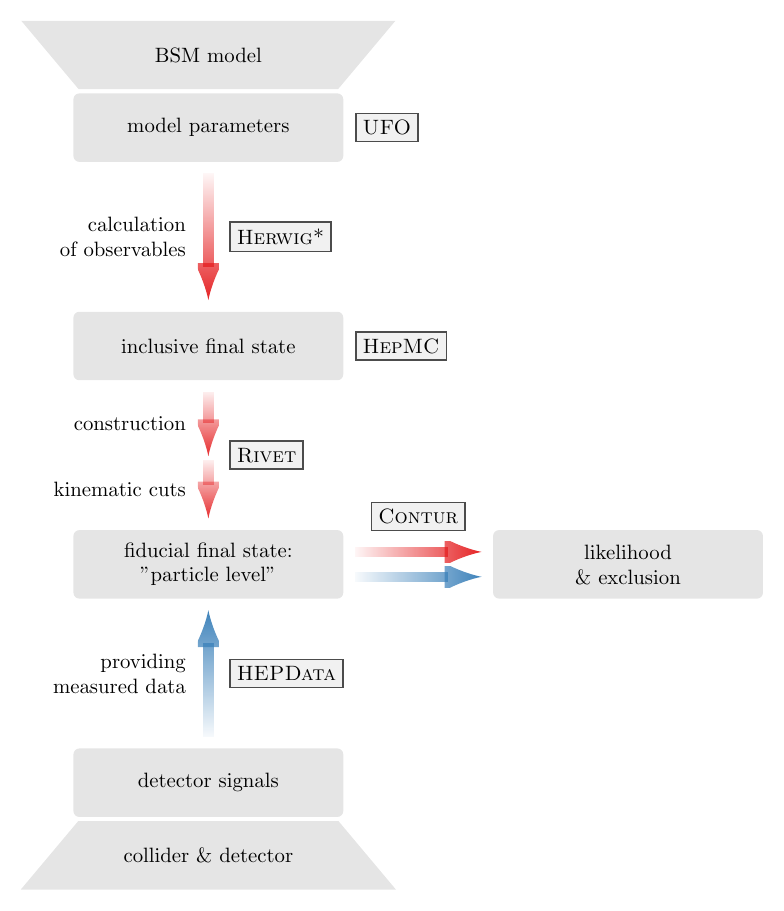}
\end{myfigure}

\figref{fig:Contur_workflow} is an adaption of \figref{fig:detectorCorrection} that gives an overview of the workflow for obtaining exclusion limits with \Contur.

First, a \BSM model that shall be investigated is chosen (top).
For this model, a specific set of parameters is studied.
Events are generated using a Monte-Carlo event generator (\MCEG) and provide observables in the inclusive final state.
High-level objects, like jets, can be constructed in these events and kinematic cuts applied to obtain a prediction in particle-level representation (\cf\secref{sec:level_interpretation}).
This forms the \BSM-prediction input for the statistical interpretation with \Contur.
As in the interpretations in \chapref{sec:interpretation}, these predictions are compared to measured data.

For the data, detector signals were measured with detectors at colliders (bottom).
In measurements, these signals were reconstructed and kinematic cuts applied.
Measurements providing their data at particle level form the nominal data input for the statistical interpretation with \Contur.
In principle, \Contur can also make use of data provided in detector-level representation.
For this, the theoretical prediction at particle level has to be smeared~\cite{Brooijmans:2020yij_RivetSmear}.
In this thesis, however, the focus shall be exclusively on particle-level inputs.

Both inputs, \BSM prediction and measured data, are used to calculate a likelihood function.
If the likelihood for the data given the prediction is small, the used set of parameters for the \BSM model can be excluded.

\bigskip
For the technical implementation of this workflow, \Contur builds upon the tremendous progress that has been made in recent years with regard to standardising approaches and unifying interfaces.
This standardisation is in line with the points raised in \chapref{sec:analysisPreservation}.
From top to bottom in \figref{fig:Contur_workflow}, the standardised tools employed by \Contur are:
\begin{itemize}
	\itembf{\UFO}
	Implementations of \BSM processes are by now routinely provided in the Universal \FeynRules Output (\UFO) format~\cite{Degrande:2011ua}.
	This common format simplifies switching between \BSM models, allowing to obtain exclusion limits for different \BSM models in quick succession.
	\itembf{\Herwig}
	The \UFO format can be read by \MCEGs like \Herwig~\cite{Bahr:2008pv,Bellm:2015jjp,Bellm:2019zci} and \MadGraph~\cite{Alwall:2014hca,Frederix:2018nkq} and used for \MC event generation in a straightforward way.
	\itembf{\HepMC}
	Events generated with a \MCEG can be saved in the common \HepMC format~\cite{Dobbs:2001ck,Buckley:2019xhk}.
	\itembf{\Rivet}
	\HepMC events can serve as an input for \Rivet~\cite{Bierlich:2019rhm}.
	\Rivet is a tool that facilitates comparisons between theoretical calculations and measurements at colliders.
	It serves as a library for software \textit{routines} which define the requirements imposed on a measured fiducial cross section.
	Concretely, they filter generated events that enter a fiducial region defined by selection cuts and project their physical observables into bins corresponding to those employed by the measurements.
	At the time of writing, the \Rivet library encompasses more than a thousand measurements, among others from various HERA, \LEP, \LHC, KEK and Tevatron experiments~\cite{Rivet:2022ric}.

	\Rivet is designed to work at particle level.
	As such, comparisons between measured data and \MC predictions can mainly be made in a meaningful way for measurements that are corrected for detector effects as described in \secref{sec:metJets_detectorCorrection}.
	Recently, however, it was demonstrated that the functionality of \Rivet to fold detector efficiencies and resolutions into the particle-level predictions achieves results compatible to those at detector level~\cite{Brooijmans:2020yij_RivetSmear}.
	This greatly extends the potential of this tool, and with it of \Contur, in the future.
	\itembf{\HEPData}
	Experiments commonly publish their results on \HEPData~\cite{Maguire:2017ypu} in an accessible format.
	The published results ideally include measured and predicted events as a function of the studied observables, a detailed breakdown of the corresponding uncertainties as well as bin-by-bin correlations.
	Often, the entry in the \HEPData database is published alongside a \Rivet routine to make the actually measured data and employed analysis logic, respectively, available.
	An example for the published results is given in \figref{fig:contur_Rivet_SM}.
\end{itemize}

\begin{myfigure}{
		Differential cross section at particle level as a function of the transverse momentum of the \ttbar system in a measurement of all-hadronic decays of highly energetic top quarks at $\sqrt{s}=\SI{13}{TeV}$ with the ATLAS detector~\cite{Aaboud:2018eqg}.
		The black (blue) crosses give the measured data (generated \SM prediction using \POWHEGBOX~\cite{Alioli:2010xd} and \Pythia~\cite{Sjostrand:2014zea}) with their respective uncertainties.
		The bottom panel shows the ratio of generated \SM prediction to measured data.
		The \pValue for the agreement of data and prediction is 0.87.
		The figure uses the \Rivet and \HEPData record of the analysis according to the technical setup described in \secref{sec:contur_methodology}.
	}{fig:contur_Rivet_SM}
	\includegraphics[width=0.7\textwidth]{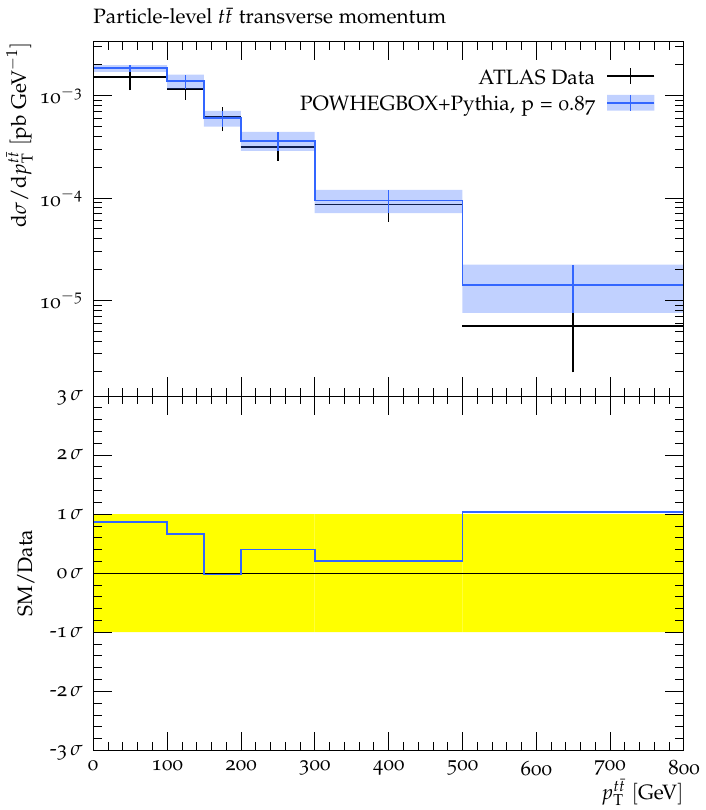}
\end{myfigure}

The \Contur toolkit forms the end of this long chain of tools and formats. 
It uses the available \Rivet routines to determine whether the model point that was generated can be statistically excluded given the data published on \HEPData.
Moreover, it provides the means for effortless steering of the whole chain and iteratively repeating the process.
As such, not only a single parameter point but whole parameter ranges of the \BSM model in question are explored.


\subsection{Statistical interpretation}
\label{sec:Contur_statistics}

For the statistical interpretation of the measured data and \BSM prediction in \Contur, an approach similar to the one outlined for the \METjets measurement in \secref{sec:interpretation_LR_2HDMa} is taken.
This section first discusses the choice of likelihood and test statistic.
Afterwards, it is described how different likelihoods are combined to obtain the exclusion limits for the \BSM model under consideration.

\subsubsection{Likelihood and test statistic}
The likelihood \lh is defined as in \eqref{eq:metJets_likelihood}.
This incorporates the various systematic uncertainties reported by the measurement as nuisance parameters as well as the statistical uncertainties for data, \SM~prediction and signal.

The total predicted yield is
\begin{equation}
	\label{eq:Contur_pred}
	\vv\pi=\mu\cdot\vv s+\vv b.
\end{equation}
This prediction~$\vv\pi$ is the sum of the \SM yield $\vvpiSM=\vv b$ and the \BSM yield $\vv s$.
In contrast to \eqref{eq:interpretation_piBSM}, this prediction includes a \textit{signal-strength} parameter $\mu$.
This is analogous to taking multiples of the nominal signal cross section as done in \secref{sec:interpretation_2HDMa}.

The nominal prediction according to the \BSM model, \ie $\mu=1$ and consequently $\vv\pi=\vv s+\vv b$, is used as the null hypothesis $H_0$.
This is identical to the null hypothesis used in \secref{sec:interpretation_LR_2HDMa}.
In the alternative hypothesis $H_1$, the signal strength parameter $\mu$ takes the value that maximises (\textit{profiles}) the likelihood, $\vv\pi=\hat{\mu}\cdot\vv s+\vv b$.
The test statistic is then the so-called \textit{profiled likelihood ratio} (\PLR):
\begin{equation}
	\label{eq:Contur_testStatistic}
	\qPLR\coloneqq
	-2\ln\frac{\lh_{\BSM}}{\lh_{\BSM, \text{opt}}}
	=
	-2\ln\frac{
		\lh\left(\vv{x}|\vv s+\vv b,\hat{\vv{\theta}}_{\mu=1}\right)
	}{
		\lh\left(\vv{x}|\hat{\mu}\cdot\vv s+\vv b, \hat{\vv{\theta}}_{\hat{\mu}}\right)
	}.
\end{equation}
It is the ratio of the likelihoods for the null hypothesis~$H_0$ and alternative hypothesis~$H_1$.
Hereby, $\hat{\vv\theta}_{\hat{\vv\mu}}$ are the values of the nuisance parameters $\vv\theta$ that maximise the likelihood simultaneously with $\hat{\mu}$.
The statistical method employed by the \Contur toolkit therefore compares the signal-plus-background likelihood to the global likelihood maximum.
The interpretation of the \METjets measurement with respect to the \THDMa in \secref{sec:interpretation_2HDMa} compared the signal-plus-background to the background-only likelihood.
Both approaches yield similar results, as shown in \appref{app:interp_2HDMa_muOpt}.

In practise, the nuisance parameters in \eqref{eq:Contur_testStatistic} are not explicitly profiled but treated as a contribution to the covariance matrix.
This results in identical values for the test statistic~\cite{Fogli:2002pt} as the nuisance parameters are assumed to be distributed according to a normal distribution.

\bigskip
In the large-sample limit, \qPLR can be approximated by~\cite{Wald:1943tsh,Cowan:2010js}
\begin{equation}
	\label{eq:Contur_Wald}
	\qPLR\approx \frac{\left(1-\hat{\mu}\right)^2}{\sigma^2}.
\end{equation}
Hereby, $\hat{\mu}$ is the profiled signal strength and is distributed according to a normal distribution with standard deviation $\sigma$.
The probability density function (\pdf) of the approximate test statistic in \eqref{eq:Contur_Wald} is a noncentral chi-square distribution.
Again in the large-sample limit, this probability density function can be approximated by a central chi-square distribution ($\chi^2_n$) with degrees of freedom $n$ equal to the difference in degrees of freedom between null and alternative hypothesis~\cite{Wilks:1938dza}:
\begin{equation*}
	\label{eq:Contur_Wilks}
	\pdf(\qPLR)\approx\frac{1}{\sqrt{2\pi\qPLR}}e^{-\frac{1}{2}\qPLR}\approx\chi^2_{n=1}.
\end{equation*}
The alternative hypothesis is fixed in one parameter less than the null hypothesis, \ie in the signal strength $\mu$.
Consequently, $n=1$.

Working in the large sample limit is justified because almost all measurements used by \Contur are unfolded.
Unfolding requires large numbers of events in all bins to allow for a stable algorithm performance, as discussed in \secref{sec:detCorr_unfoldingMethod}.
Often 20 or more events are used even in the tails of distributions, reducing the error from the large-sample approximation.

Using the profiled likelihood ratio in \eqref{eq:Contur_testStatistic} as test statistic has the advantage that its probability density function is analytically calculable in the large-sample approximation.
In contrast, the test statistic in \eqref{eq:interpretation_q} used for the interpretation with respect to the \THDMa of the \METjets measurement in \chapref{sec:interpretation} needs toys.
This is more precise also at small event numbers but more computationally expensive.

Identically to the approach outlined in \secref{sec:interpretation_LR_2HDMa}, \CLs is used as the final discriminant to set exclusion limits.
This ensures that the derived exclusion limits are conservative.

\subsubsection{Combination of likelihoods}
With the approach described above, the confidence in excluding the given \BSM model point for a single distribution from a chosen measurement can be obtained.
The exclusion power from the statistical analysis can be improved by combining statistically independent distributions.
Correlations across measurements or even across distributions within one measurement, however, are rarely determined and published.
Consequently, these correlations cannot be taken into account by \Contur.
Treating all distributions as uncorrelated would allow multiply-counting events and overestimating the sensitivity.
Likelihoods are therefore combined in the sophisticated procedure outlined below.

Different \textit{analysis pools} are defined in three steps which are known to be completely orthogonal by splitting
\begin{enumerate}
	\item different experiments, \eg CMS, ATLAS and LHCb.
	\item different centre-of-mass energies, \eg $\sqrt{s}=7, 8$ and \SI{13}{TeV}.
	\item different final states, \eg by photon multiplicity, lepton type or presence of jets.
\end{enumerate}
For each pool, only the single most sensitive distribution is considered and the rest discarded.
This is a statistically conservative approach because discarding less significant deviations means reducing the sensitivity to the \BSM model.

The likelihood in \eqref{eq:metJets_likelihood} uses bin-by-bin correlations across individual distributions in its calculation.
If the correlations are not available for a distribution, all bins of this distribution are considered to be individual distributions with unit bin number and assigned to their common analysis pool.
This prevents double counting even within an individual distribution.

\bigskip
Different analysis pools are then combined to improve the statistical statement by multiplying their likelihoods.
This corresponds to summing the test statistics according to \eqref{eq:Contur_testStatistic} for all pools $p$.
In short,
\begin{equation*}
	q_{\PLR, \text{obs}}^{\vv \pi,\text{tot}}\coloneqq
	\sum_{p\in\mathrm{pools}}\max_{d\in p}\,q_{\PLR, \text{obs}}^{\vv\pi}
\end{equation*}
is obtained for the total observed test statistic.
Hereby, $q_{\PLR, \text{obs}}^{\vv\pi}$ is the observed value of the test statistic for a prediction $\vv\pi$ in distribution~$d$ of pool $p$.
When employing the \CLs technique~\cite{Read:2000ru,Read:2002hq} to be conservative, the prediction~$\vv\pi$ in the numerator of the profiled likelihood ratio in \eqref{eq:Contur_testStatistic} is $\vv\pi=\vv s+\vv b$ for \CLsb and $\vv\pi=\vv b$ for \CLb, respectively.
Analogous to \eqref{eq:interpretation_CLsb}, this results in
\begin{alignat*}{2}
	\CLsb&\coloneqq1-p\left(\qPLRobsSB,\pdf\left(\qPLRSB\right)\right)&&=1-p\left(\qPLRobsSB, \chi^2_1\right)\\
	\CLb&\coloneqq1-p\left(\qPLRobsB,\pdf\left(\qPLRB\right)\right)&&=1-p\left(\qPLRB, \chi^2_1\right).
\end{alignat*}
Using the definition of the \pValue in \eqref{eq:interpretation_pValue} yields
\begin{equation*}
	\CLs\coloneqq
	\frac{\CLsb}{\CLb} =
	\frac{
		\int_{q_{\PLR, \text{obs}}^{\vv s+\vv b,\text{tot}}}^\infty
		\chi^2_1\,\mathrm{d}\chi^2_1
	}{
		\int_{q_{\PLR, \text{obs}}^{\vv b,\text{tot}}}^\infty
		\chi^2_1\,\mathrm{d}\chi^2_1
	}.
\end{equation*}
Exclusion limits are set where $\CLs<0.05$.

\subsection{Limitations}
\label{sec:Contur_Limitiations}

Given the approach and tools described before, the \Contur method proves very powerful.
The method also comes with a number of limitations, however, that can be divided into two categories and are detailed in the following.

\subsubsection{Limitations to usable measurements}
Measurements can only be used by \Contur if they are provided in a model-independent way.
That is why measurements for which a \HEPData entry and a \Rivet routine are published are curated by the \Contur authors for usage with the \Contur toolkit.
Important exclusion criteria shall be given in the following.

Measurements often rely upon data-driven methods for backgrounds that are not well modelled in simulation, like the \QCD multijet background in the \METjets measurement (\cf\secref{sec:metJets_expSystUnc}).
Hereby, the number of events in data in a \textit{control region}, a phase space not primarily targetted by the measurement, is used to extrapolate the number of events for this background into the target \textit{signal region}.
A \BSM model contributing to the control region would lead to a larger background estimate in the signal region.
If these backgrounds are subtracted prior to the publication, but the \BSM model is not taken into account in the estimation, the published data underestimates the joint yields from Standard Model and \BSM model in the signal region.
This would lead to an overestimated exclusion for the \BSM model.
Measurements for which this data-driven background subtraction prior to the publication is large are therefore excluded from being used by \Contur.

Some \Rivet routines make use of particle properties that are hard to determine experimentally at colliders.
An example for this is that in the \Rivet routine for \refcite{ATLAS:2019bsc} a \MET observable is constructed by selecting neutrinos in the events.
The observable is not, as is better practise, constructed as the negative vector-sum of the transverse momentum of visible particles (\cf\secref{sec:objReco_MET}).
\BSM models giving rise to the production of detector-invisible particles that are not neutrinos do therefore erroneously not contribute to this \MET observable.
\Rivet routines such as this have to be excluded for most \BSM models.

Another case that proves problematic is if kinematic requirements imposed at analysis level are not included in the definition of the fiducial phase space of the \Rivet routine.
This in particular arises for measurements rejecting events with $b$-tagged jets, \eg\refcite{CMS:2016ipg,CMS:2017gbl}.

Generally speaking, care has to be taken when evaluating a new \BSM model with \Contur and all measurements contributing sensitivity to the model be rigorously scrutinised.

\subsubsection{Limitations to statistical statement}
The second set of limitations for the \Contur toolkit arise for its statistical statement:
\begin{itemize}
	\item The test statistic employed by the \Contur toolkit implements a hypothesis test whether the null hypothesis that the \BSM model is true has to be discarded in favour of the alternative hypothesis that the Standard Model is true (\cf\secref{sec:Contur_statistics}).
	The \Contur toolkit can therefore only be used to set exclusion limits on \BSM models, not to infer any discoveries.
	
	\item Correlations are taken into account only across bins of the same distribution and naturally solely if they are provided with the \HEPData entry.
	Out of all distributions within one pool, the one giving the largest individual deviation is used, as explained in \secref{sec:Contur_statistics}.
	Correlations across distributions or across measurements are not taken into account.
	
	\item The nominal prediction $\vv\pi$ in the test statistic used by \Contur is
	\begin{equation}
		\label{eq:contur_nominalPrediction}
		\vv\pi=\vv s+\vv b,
	\end{equation}
	called "\SM-background".
	Hereby, $\vv s$ is the generated signal prediction and $\vv b$ the generated \SM prediction.
	At the time of writing, out of 158 measurements available to \Contur, only 43 supply the \SM prediction $\vv b$.
	First steps are being taken by the \Contur authors to provide the missing \SM predictions.
	In the meantime, the alternative prediction
	\begin{equation}
		\label{eq:contur_SMeqData}
		\vv\pi=\vv s+\vv x,
	\end{equation}
	called "data-background", can be used for setting exclusion limits with the \Contur toolkit.
	This prediction assumes that the measured data does not have \BSM contributions and the \SM prediction would have been equal to the measured data~$\vv x$.
	The approach is justified considering the good agreement of the measurements included in the repository with \SM predictions.
	It is also the method employed by many \BSM searches in data-driven control regions.
	The approximation is only valid in case the (missing) \SM uncertainties are smaller than the data uncertainties because otherwise the uncertainties would be reduced and the statistical statement artificially augmented.
	
	In principle \Contur would be insensitive to multiple statistically insignificant deviations adding up to a statistically significant one when assuming \eqref{eq:contur_SMeqData} and might therefore impose falsely strong limits.
	However, all analyses for which \Contur has to make this assumption are also unfolded to particle level.
	As mentioned in \secref{sec:Contur_statistics}, these analyses in general require a large number of events in all bins to allow for a stable unfolding procedure.
	This means that statistics are principally high, weakening the exclusion limit but also reducing the impact of outlying events.
	If nonetheless extraordinary caution wants to be taken in interpreting the results of \Contur, these alternative limits could be considered expected limits.
	They would in that case only outline regions where measurements are sensitive and deviations from the \SM disfavoured.
\end{itemize}

\section{Exclusion limits for the \THDMa}
\label{sec:contur_2HDMa}

In this section, the \Contur toolkit is used to gain more information about the \THDMa parameter space that is excluded by existing \LHC measurements.

\subsection{Methodology}
\label{sec:contur_methodology}

The implementation of the \THDMa physics in a corresponding \UFO format is used~\cite{Haisch:2017ufo}.
As in the previous chapters, the recommendations for the \THDMa parameters given by the \LHCDMWG~\cite{LHCDarkMatterWorkingGroup:2018ufk} detailed in \tabref{tab:LHCDMWG_params} are used unless stated otherwise.

Events are generated with \toolVersion{\Herwig}{7.2.2} as described in \secref{sec:MC_2HDMa_Contur}.
Measurements used by \Contur are typically defined inclusive in final state given a certain lepton, jet and photon multiplicity.
They are usually not searching for narrow resonances and thus not likely sensitive to differences in shapes of distributions.
For simplicity, interference effects with the Standard Model are therefore neglected in this study, although they can sizeably change the shape of distributions for $A/H\to\ttbar$ decays in the considered mass range $\SI{200}{GeV}<m_{A/H}<\SI{2000}{GeV}$~\cite{Gaemers:1984sj,Dicus:1994bm,Djouadi:2019cbm,ATLAS:2017snw,CMS:2019pzc}.

Objects are constructed from generated events by \toolVersion{\Rivet}{3.1.5}.
The tool also applies analysis selection-criteria given by the different \Rivet routines in the repository.

\toolVersion{\Contur}{2.4.1} is used for the statistical analysis.
All default measurements in particle-level representation at 7, 8 and \SI{13}{TeV}~\cite{Khachatryan:2016poo,Aaboud:2016ftt,Aad:2014tca,Aaboud:2017rwm,Aad:2015rka,Aaboud:2019lxo,Aad:2021ebo,Aad:2012awa,Aaboud:2016btc,Aaboud:2019jcc,Aaboud:2017hbk,Aad:2019hga,Aad:2020gfi,Aaboud:2017hox,Aad:2014dvb,Aaij:2012mda,Aad:2015auj,Sirunyan:2019bzr,Chatrchyan:2013vbb,Sirunyan:2018cpw,Aad:2013ysa,Aad:2013izg,Aad:2016sau,Aad:2013vka,Aaboud:2017soa,Aad:2014qxa,Aaboud:2017fye,Aad:2013gaa,Aaboud:2017vol,Chatrchyan:2013mwa,Aad:2013zba,Aad:2012tba,Aaboud:2017lxm,Aad:2016xcr,ATLAS:2021mbt,Aaboud:2017kff,Aaboud:2017skj,ATLAS:2012ar,Aad:2014lwa,ATLAS:2018nci,Aad:2013iua,Aad:2016zzw,Sirunyan:2018owv,ATLAS:2019zci,Chatrchyan:2014gia,ATLAS:2019rqw,Aad:2014vwa,Chatrchyan:2013qza,Sirunyan:2018xdh,Aaboud:2017wsi,Aaboud:2019aii,Sirunyan:2017skj,Aaboud:2017vqt,CMS:2021lxi,Aaboud:2017dvo,Chatrchyan:2012bja,Aad:2015nda,Aaboud:2017qwh,Aad:2020zcn,Aad:2014pua,Sirunyan:2021lwi,Khachatryan:2016mlc,Aad:2020fch,Chatrchyan:2012dk,Aad:2016mok,Aaboud:2016itf,Aad:2013tea,Aad:2014rma,Khachatryan:2016wdh,Aaij:2018imy,ATLAS:2012mec,Aad:2016wpd,Aaboud:2019nkz,Aad:2019hzw,Aad:2021dse,Aaboud:2018hip,Sirunyan:2020jtq,Khachatryan:2016iob,Aad:2014dta,Aad:2020sle,Chatrchyan:2013zja,Khachatryan:2014zya,Aad:2019gpq,Aad:2014qja,Sirunyan:2018wem,Sirunyan:2017yar,Khachatryan:2016mnb,Aaboud:2018eki,Aaboud:2018uzf,Aad:2014xca,Sirunyan:2019hqb,Aaboud:2017fha,Aad:2015hna,Aad:2015eia,Khachatryan:2016nbe,Aad:2019ntk,Sirunyan:2018ptc,Khachatryan:2016gxp,Aad:2015mbv,Aaboud:2017buf,AbellanBeteta:2016ugk,Sirunyan:2018hde,Aaij:2013nxa,Khachatryan:2015rja,Aaboud:2019nmv,Sirunyan:2019rfa,CMS:2019eih,Aaboud:2018eqg,ATLAS:2020ccu} are taken into account.
This includes in total 109 measurements at the \LHC from the ATLAS, CMS and LHCb experiments.
An overview of the used analysis pools and measurements is given in \appref{app:contur_app_pools}.
Measurements sensitive to the \METjets final state are excluded because this final state was extensively studied in \secref{sec:interpretation_2HDMa}.

Exclusion limits are placed with this setup according to the approach described in \secref{sec:Contur_statistics}.
An example for the contribution of the \THDMa to a measurement is shown in \figref{fig:contur_Rivet_2HDMa}.

\begin{myfigure}{
		Differential cross section at particle level as a function of the transverse momentum of the \ttbar system in a measurement of all-hadronic decays of highly energetic top quarks at $\sqrt{s}=\SI{13}{TeV}$~\cite{Aaboud:2018eqg}.
		The black crosses give the measured data, the blue crosses the generated \SM prediction.
		The green crosses give the generated \THDMa prediction using the "\SM-background" approach at $\ma=\SI{100}{GeV}$, $\mAeqmHeqmHpm\approx\SI{400}{GeV}$ and $\tanB=1$.
		The bottom panel shows the ratio to the measured data.
		The \pValue for the agreement of data and \THDMa prediction is 0.61.
	}{fig:contur_Rivet_2HDMa}
	\includegraphics[width=0.7\textwidth]{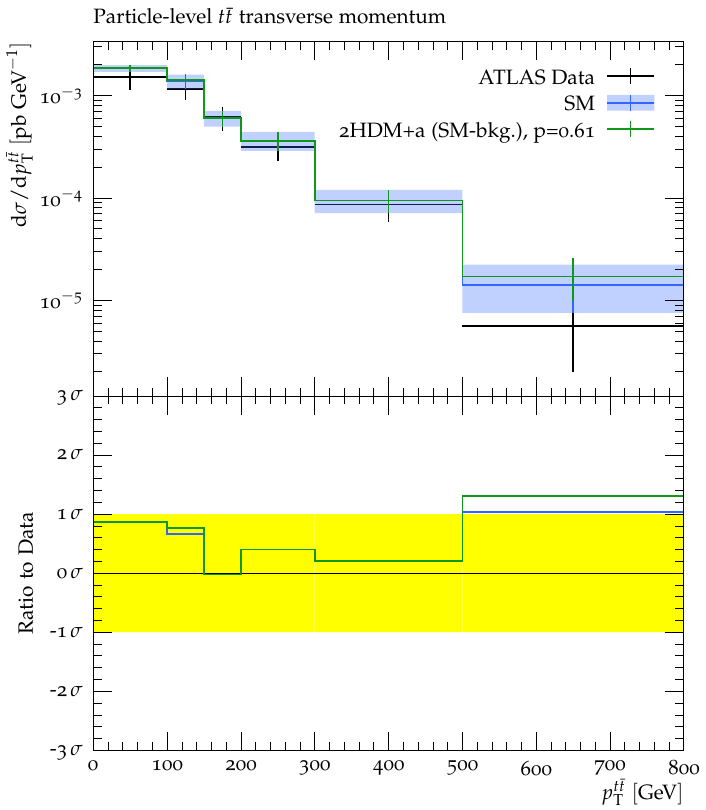}
\end{myfigure}

\subsection{Benchmark scenarios}
In this section the benchmark scans in the \mamA and \matanB plane (see\linebreak \secref{sec:2HDMa_parameterPlanes}) that were already investigated in \secref{sec:interpretation_2HDMa} are studied with \Contur.
There are stringent constraints from ATLAS searches in these planes~\cite{ATLAS:2021fjm,ATLAS:2022rxn}, among others in the \METZll and \METhbb final states.
These searches published their results in detector-level representation and are therefore not considered in this study with the \Contur toolkit.
Measurements of the same final states in particle-level representation, however, are taken into account if available.
A comparison between the exclusion limits obtained in this chapter with the \Contur toolkit and from a summary of ATLAS searches is given in \secref{sec:comparison_2HDMa}.

\subsubsection{\mamA plane}
The exclusion limits from \Contur in the \mamA plane are given in \figref{fig:Contur_benchmarks_mamA}.
As before, $\tanB=1$ is used in this plane and all other model parameters are chosen as given in \tabref{tab:LHCDMWG_params}.
In particular, $\mX=\SI{10}{GeV}$ and \mAeqmHeqmHpm.
Red and black lines give the exclusion limit assuming the prediction according to \eqsref{eq:contur_nominalPrediction}{eq:contur_SMeqData}, respectively.
The region below and to the left of the solid (dashed) lines can be excluded at \SI{95}{\%} (\SI{68}{\%}) confidence level.

Coloured tiles mark the analysis pool that contributes the most to the exclusion limit using the "\SM-background" prediction at the corresponding parameter point.
A legend matching colour to analysis pool is given below the figure.
In all investigated planes, the important physics processes and correspondingly the sensitive analysis pools do not change significantly if the "data-background" instead of "\SM-background" prediction is used.
Only the single most-sensitive analysis pool for a specific tile can differ.

Dashed white lines indicate the phase space where $\mA=\ma+\mh$ and $\mA=\ma$, respectively.
At very large masses, \eg $\mA>\SI{1900}{GeV}$, the ratio of mass to width of \BSM bosons can be larger than \SI{20}{\%}.
In this region, the narrow-width approximation (\cf\secref{sec:MCEG_hadronisation}) may be violated and predictions employing it can become more unreliable.

\begin{myConturFigure}{\mamA}{fig:Contur_benchmarks_mamA}{mA_sinP0.35}
        \swatch{darkgoldenrod}~ATLAS $\gamma$+\MET{} &
		\swatch{purple}~ATLAS $\ell^\pm\ell^\pm$+\MET{} &
		\swatch{mediumseagreen}~ATLAS $\ell^+\ell^-\gamma$ &
		\swatch{darkorange}~ATLAS $\mu^+\mu^-$+jet \\
		\swatch{blue}~ATLAS $\ell$+\MET{}+jet &
		\swatch{powderblue}~CMS $\ell$+\MET{}+jet &
		\swatch{orangered}~ATLAS $e^+e^-$+jet &
		\swatch{yellow}~ATLAS $\gamma$ \\
		\swatch{turquoise}~ATLAS $\ell_1\ell_2$+\MET{}+jet &
		\swatch{snow}~ATLAS hadronic $t\bar{t}$ &
		\swatch{magenta}~ATLAS 4$\ell$ &
		\swatch{wheat}~CMS hadronic $t\bar{t}$ \\
		\swatch{greenyellow}~ATLAS $\ell^+\ell^-$+\MET{} &
\end{myConturFigure}
\FloatBarrier 

\bigskip
The region $\mA<\SI{650}{GeV}$ (\SI{1000}{GeV}) independent of \ma can be excluded from the existing \LHC measurements at \SI{95}{\%} (\SI{68}{\%}) confidence level using the "data-background" prediction in \eqref{eq:contur_SMeqData}.
Using the "\SM-background" prediction in \eqref{eq:contur_nominalPrediction}, the region $\mA<\SI{350}{GeV}$ independent of \ma and $\mA<\SI{750}{GeV}$ at small \ma can be excluded at \SI{68}{\%} confidence level.
Only a small region $\SI{250}{GeV}<\mA<\SI{450}{GeV}$ and $\ma<\SI{130}{GeV}$ can be excluded at \SI{95}{\%} confidence level.

The large difference in the exclusion limit between the two predictions arises in part due to mismodelling of the \SM prediction in \MC generation as seen before for the\linebreak \METjets measurement in \chapsref{sec:metJets}{sec:interpretation}.
However, it is mostly caused by the fact that for the majority of measurements in the repository no \SM prediction was published alongside the measured data.
These measurements have to be neglected from the statistical analysis when deriving the exclusion limits with the "\SM-background" prediction.
This reinforces a point raised in \chapref{sec:analysisPreservation}: the need to not only publish the measured data but also the used theory prediction in the process of analysis preservation.

Numerous measurements contribute to the exclusion limits, as can be seen from the coloured tiles in \figref{fig:Contur_benchmarks_mamA}.
Cross sections and branching fractions helpful for understanding these exclusion limits are given in \figref{fig:contur_xsBR_mamA}.
The sensitivity in \figref{fig:Contur_benchmarks_mamA} arises as follows:
\begin{itemize}
	\item At small \ma and small \mA, an ATLAS measurement of the $\ell^+\ell^-$+\MET final state at \SI{7}{TeV}~(light-green,~\refcite{Aad:2012awa}) using \SI{4.6}{\ifb} dominates the sensitivity.
	Here, the $s$-channel production of the scalar $H$ and its subsequent decay into the pseudoscalar~$a$ and a $Z$ boson, $pp\to H\to aZ$, has a sizeable cross section times branching fraction.
	\MET is produced by invisible decays of the pseudoscalar, $a\to\xx$, and the two leptons by a leptonic decay of the $Z$ boson, \Zll.
	This process was already discussed extensively in \secref{sec:interpretation_2HDMa_contributions}.
	A Feynman diagram is shown in \subfigref{fig:2HDMa_METjets_diagrams}{c}.
	
	\item At small \ma and intermediate \mA, ATLAS measurements of the $\ell^+\ell^-+$jet final state (orange and red, \eg\refcite{Aad:2015auj}) at 7, 8 and \SI{13}{TeV} dominate.
	They are sensitive to the same processes as the ATLAS measurement of $\ell^+\ell^-+\MET$, \ie $pp\to aZ(\to\ell^+\ell^-)$.
	The jet in the selected final state either originates from $a\to\qqbar$ decays or from initial-state radiation if the pseudoscalar~$a$ decays invisibly to \DM particles, $a\to\xx$.
	The $\ell^+\ell^-+$jet measurements are less constraining in the small \mA region than the $\ell^+\ell^-+\MET$ measurement because they do not exploit the \MET signature.
	They do, however, make use of larger centre-of-mass energies and integrated luminosities which extends their sensitivity to larger values of \mA.

	\item At large \ma and not too large \mA, a mixture of ATLAS and CMS measurements of final states involving one or multiple leptons and often \MET dominate.
	Examples are the measurements in the ATLAS $\ell_1\ell_2$+\MET+jet pool (turquoise,~\refscite{Aad:2021dse,ATLAS:2019hau}) for measurements of unlike dileptons, \MET and jets.
	They are sensitive to the various \THDMa processes giving rise to the production of two $W$ bosons:
	There is the associated production of charged scalars \Hpm with either $W$ bosons, $pp\to W^\mp\Hpm$, or top quarks, $pp\to t\Hpm$, with subsequent decays of $\Hpm\to tb$.
	Furthermore, there is the resonant production of scalars $H$ as well as pseudoscalars $A$ and their subsequent decay to top-quark pairs, $pp\to H/A\to\ttbar$.
	
	\item In the whole parameter plane, \eg at large \mA independent of \ma but also at small \mA, ATLAS and CMS measurement of hadronic \ttbar decays (grey and light-yellow, \eg\refcite{Aaboud:2018eqg} and~\refcite{CMS:2019fak}, respectively) make important contributions.
	They are sensitive to the resonant production of either pseudoscalar $a$ or $A$ and their subsequent decay to top pairs, $pp\to a/A\to\ttbar$.
\end{itemize}

\subsubsection{\matanB plane}
The exclusion limits from \Contur in the \matanB plane are given in \figref{fig:Contur_benchmarks_matanB}.
As before, $\mAeqmHeqmHpm=\SI{600}{GeV}$ is used in this plane and all other model parameters are chosen as given in \tabref{tab:LHCDMWG_params}.
In particular, $\mX=\SI{10}{GeV}$.
At extreme values of \tanB, \eg $\tanB<0.6$ and $\tanB>50$ for the pseudoscalar $A$, the ratio of mass to width of \BSM bosons can be larger than \SI{20}{\%}.
In this region, the narrow-width approximation (\cf\secref{sec:MCEG_hadronisation}) may be violated and predictions employing it can become more unreliable.

\begin{myConturFigure}{\matanB}{fig:Contur_benchmarks_matanB}{tanB_sinP0.35}
        \swatch{blue}~ATLAS $\ell$+\MET{}+jet &
		\swatch{purple}~ATLAS $\ell^\pm\ell^\pm$+\MET{} &
		\swatch{mediumseagreen}~ATLAS $\ell^+\ell^-\gamma$ &
		\swatch{darkorange}~ATLAS $\mu^+\mu^-$+jet \\
		\swatch{turquoise}~ATLAS $\ell_1\ell_2$+\MET{}+jet &
		\swatch{powderblue}~CMS $\ell$+\MET{}+jet &
		\swatch{orangered}~ATLAS $e^+e^-$+jet &
		\swatch{yellow}~ATLAS $\gamma$ \\
		\swatch{greenyellow}~ATLAS $\ell^+\ell^-$+\MET{} &
		\swatch{snow}~ATLAS hadronic $t\bar{t}$ &
		\swatch{magenta}~ATLAS 4$\ell$ &
		\swatch{wheat}~CMS hadronic $t\bar{t}$ \\
		\swatch{darkolivegreen}~ATLAS high-mass Drell-Yan $\ell\ell$ &
\end{myConturFigure}

\bigskip
The exclusion limits principally decompose into two regions: one for $\tanB<2$ and one for $\tanB>10$.
The reason is rooted in \eqsref{eq:2HDMa_Lagrangian}{eq:2HDMa_xi}: while the coupling of $a$, $A$ and $H$ to bottom quarks and taus is proportional to \tanB, their coupling to top quarks is inversely proportional to it.
This means that different production and decay channels become important at small and large \tanB.
At intermediate \tanB, neither the couplings proportional to \tanB nor those inversely proportional to it are large enough to provide significant sensitivity to the model.

For small \tanB, the region $\tanB<1.3$ ($\tanB<0.5$) at small \ma and $\tanB<0.9$ ($\tanB<0.4$) at large \ma can be excluded at \SI{95}{\%} confidence level using the "data-background" ("\SM-background") prediction.
For large \tanB, the region $\tanB>13$ ($\tanB>60$) at small \ma and $\tanB>40$ ($\tanB>60$) at large \ma can be excluded at \SI{95}{\%} confidence level using the "data-background" ("\SM-background") prediction.

Numerous measurements contribute to the exclusion limits, as can be seen from the coloured tiles in \figref{fig:Contur_benchmarks_matanB}.
Cross sections and branching fractions helpful for understanding these exclusion limits are given in \figref{fig:contur_xsBR_matb}.
The sensitivity in \figref{fig:Contur_benchmarks_matanB} arises as follows:
\begin{itemize}
	\item For small \tanB, the sensitivity is driven by the various ATLAS and CMS measurements of leptons, \MET and jets (blue, turquoise and light-purple, \eg\refscite{ATLAS:2015mip,Aad:2021dse,CMS:2018tdx}).
	These measurements select final states with $W$ bosons or top quarks.
	They are sensitive to the resonant production of \BSM scalars and pseudoscalars and their subsequent decay to top-quark pairs, $pp\to H'\to\ttbar$ with $H'\in\left\{a,A,H\right\}$.
	The decay width for these processes is $\width[H'\to\ttbar]\propto\cotBSq$ according to \eqsref{eq:2HDMa_xi}{eq:2HDMa_Gamma_a_ff}{eq:2HDMa_Gamma_A_ff}{eq:2HDMa_Gamma_H_ff}, increasing the cross section times branching fraction at small \tanB.	

	\item For large \tanB, the sensitivity is driven by ATLAS and CMS measurements of hadronic \ttbar production (grey and light-yellow, \eg\refscite{Aaboud:2018eqg,CMS:2019eih}).
	They would select the resonant- or pair-production of \BSM scalars and pseudoscalars, $pp\to H'$ and $pp\to H'H'$, respectively, and their subsequent decay to bottom-quark or tau pairs, $H'\to ff'$ with $f\in\left\{b,\tau\right\}$.
	The decay width for these processes is $\width[H'\to ff']\propto\tanBSq$ according to \eqsref{eq:2HDMa_xi}{eq:2HDMa_Gamma_a_ff}{eq:2HDMa_Gamma_A_ff}{eq:2HDMa_Gamma_H_ff}, increasing the branching fraction at large \tanB.
	In addition, the scalar $H$ can decay to pairs of the pseudoscalar $a$, $H\to aa$.
	The width for this process is $\width[H\to aa]\propto\cot^2\left(2\beta\right)$ according to \eqsref{eq:2HDMa_Gamma_H_aa}{eq:2HDMa_g_Haa}, also increasing the branching fraction at large \tanB.
	All of these processes lead to final states with multiple bottom quarks or taus.

	It can be expected that measurements targetting bottom quarks or taus would be very sensitive in this \tanB region.
	The \Contur repository, however, does only include measurements of final states targetting exclusively bottom quarks at $\sqrt{s}\leq\SI{8}{TeV}$ with an integrated luminosity of less than \SI{12}{\ifb}.
	It does not include any measurements targetting taus.
	For this reason, measurements selecting events with hadronic decays of top-quark pairs (grey and light-yellow)  at \SI{13}{TeV} with an integrated luminosity larger than \SI{35}{\ifb} give the dominant contribution to the \Contur exclusion.
	These measurements select events with high-momentum jets with considerable substructure, which in this case originate from final states involving multiple bottom quarks or taus.
	
	\item For all \tanB, the measurements mentioned above of leptons, \MET and jets final states as well as of hadronic \ttbar production would select processes of top and bottom quarks being produced in association with charged scalars \Hpm and their subsequent decay also to top and bottom quarks, $pp\to tb\Hpm \to \ttbar\bbbar$.
	The cross section times branching fraction for these processes has factors proportional to \tanB as well as inversely proportional to it according to \eqref{eq:2HDMa_Lagrangian}.
	It is therefore less sensitive to \tanB variations.
	
	\item For intermediate \tanB and small \ma, the sensitivity is driven by ATLAS measurements of the $\ell^+\ell^-$+\MET and $\ell^+\ell^-+$jet final states (light-green, orange and red, \eg\refscite{Aad:2012awa,Aad:2015auj}).
	These measurements were already extensively discussed for the exclusion limits in the \mamA plane.
	They are sensitive to the $s$-channel production of the scalar $H$ and its subsequent decay into the pseudoscalar $a$ and a $Z$ boson, $pp\to H\to aZ$.
	The width of this decay is independent of \tanB according to \eqref{eq:2HDMa_Gamma_H_aZ}, allowing for a large branching fraction when the decay width of the scalar~$H$ to top quarks, bottom quarks and taus is small.
	This independence of \tanB was already observed in \secref{sec:interpretation_2HDMa_xs}.
\end{itemize}

In summary, \Contur sets stringent exclusion limits in both planes, \mamA and \matanB.
This in particular holds when assuming the "data-background" prediction.
In \secref{sec:comparison_2HDMa}, it is investigated how these exclusion limits compare to those derived from the \METjets measurement in \secref{sec:interpretation_2HDMa} and to existing exclusion limits from searches.

\subsection{Uncharted realms}
One advantage of the approach taken in this chapter is that broadened parameter ranges can be investigated easily.
It is phenomenologically the most interesting to additionally study parameter ranges in which new decay channels of the \BSM bosons become important and in consequence the measurements that are the most sensitive change.
On the one hand, this means to relax the assumption of degenerate heavy \BSM bosons, \mAeqmHeqmHpm.
In this case, decays of the heavy \BSM bosons into one another become kinematically allowed.
On the other hand, this means to vary the \DM mass because then different decay channels for the pseudoscalars $a$ and $A$ become important.

Both approaches, non-degenerate \BSM-boson masses and varied \DM mass, are investigated in the following.

\subsubsection{Non-degenerate masses}

The assumption \mAeqmHeqmHpm is often made when studying the \THDMa.
The mass degeneracy simplifies the available parameter space to a region that is favoured by electroweak precision constraints~\cite{Haller:2018nnx} and gives results that are adaptable to supersymmetric theories~\cite{Djouadi:2005gj} (\cf\secref{sec:2HDM_parameters}).
The electroweak constraints are, however, relaxed in light of the tension between \SM prediction and a recent measurement of the mass of the $W$ boson~\cite{CDF:2022hxs}:
the difference can be explained if the masses of the heavy \BSM bosons $A$, $H$ and \Hpm are not degenerate~\cite{Arcadi:2022dmt,Arcadi:2022lpp}.

The exclusion limits provided by existing \LHC measurements using the \Contur meth\-od are therefore studied as an example for the case $m_A\neq \mH\equiv\mHpm$.
The case $m_H\neq\mA\equiv\mHpm$ is investigated in \refcite{Butterworth:2020vnb}.
Other cases are left for future work.

\bigskip
\figref{fig:Contur_mHmHc_mA} shows the exclusion limits from existing \LHC measurements determined with the \Contur method for two values of \ma, $\ma=\SI{100}{GeV}$ and $\ma=\SI{500}{GeV}$, in the \mHmA plane with $\mH\equiv\mHpm$.
The different mass of the pseudoscalar $a$ means that different decay channels of the heavy \BSM bosons $A$, $H$ and \Hpm have the highest branching fraction.
All other parameters are kept at the previously used benchmark parameters given in \tabref{tab:LHCDMWG_params}.
In particular, $\mX=\SI{10}{GeV}$.

\begin{myConturFigureGeneral}{\mHmA}{}{%
		Dashed white lines indicate the phase space where $\mA=\mH=\mHpm$, $\mA=\ma+\mh$ and $\mA=\ma$, respectively, when inside the shown plane.
	}{fig:Contur_mHmHc_mA}{
		\subfloat[$\ma=\SI{100}{GeV}$]{\includegraphics[width=0.74\textwidth]{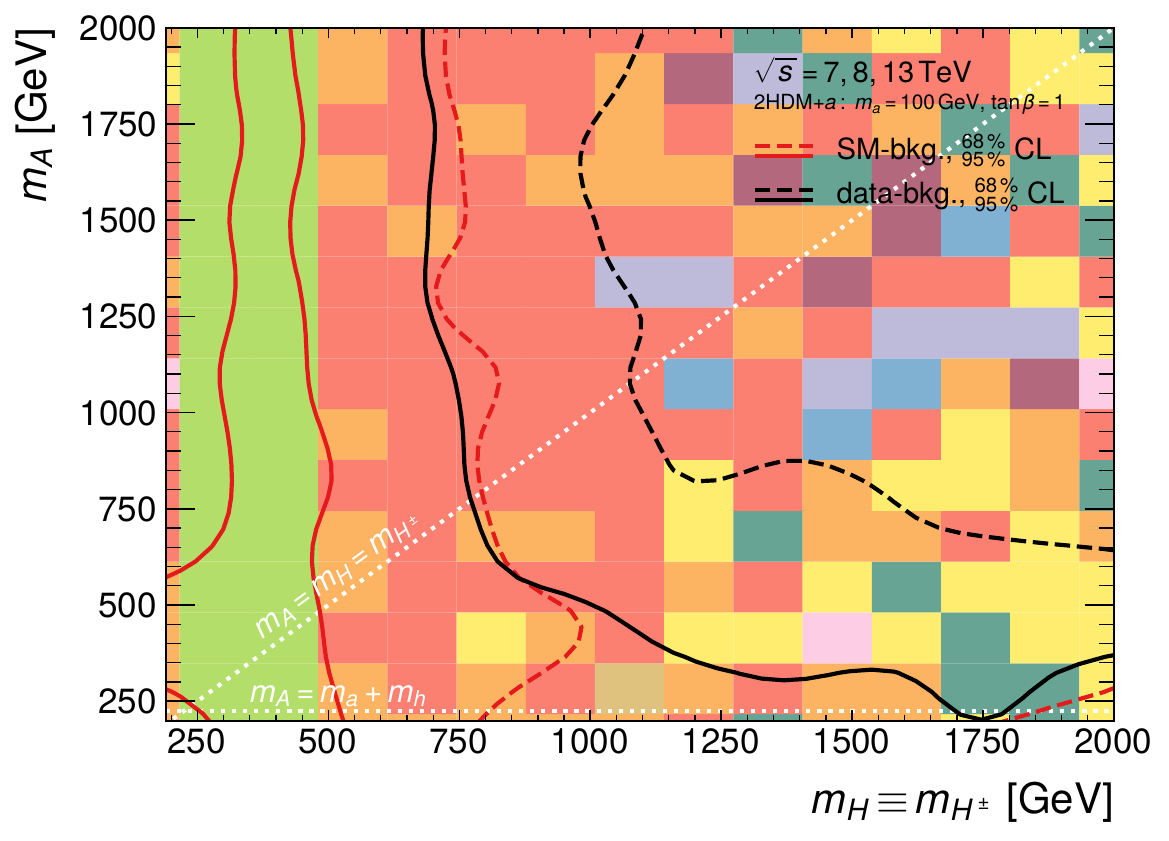}}\\
		\vspace{10pt}
		\subfloat[$\ma=\SI{500}{GeV}$]{\includegraphics[width=0.74\textwidth]{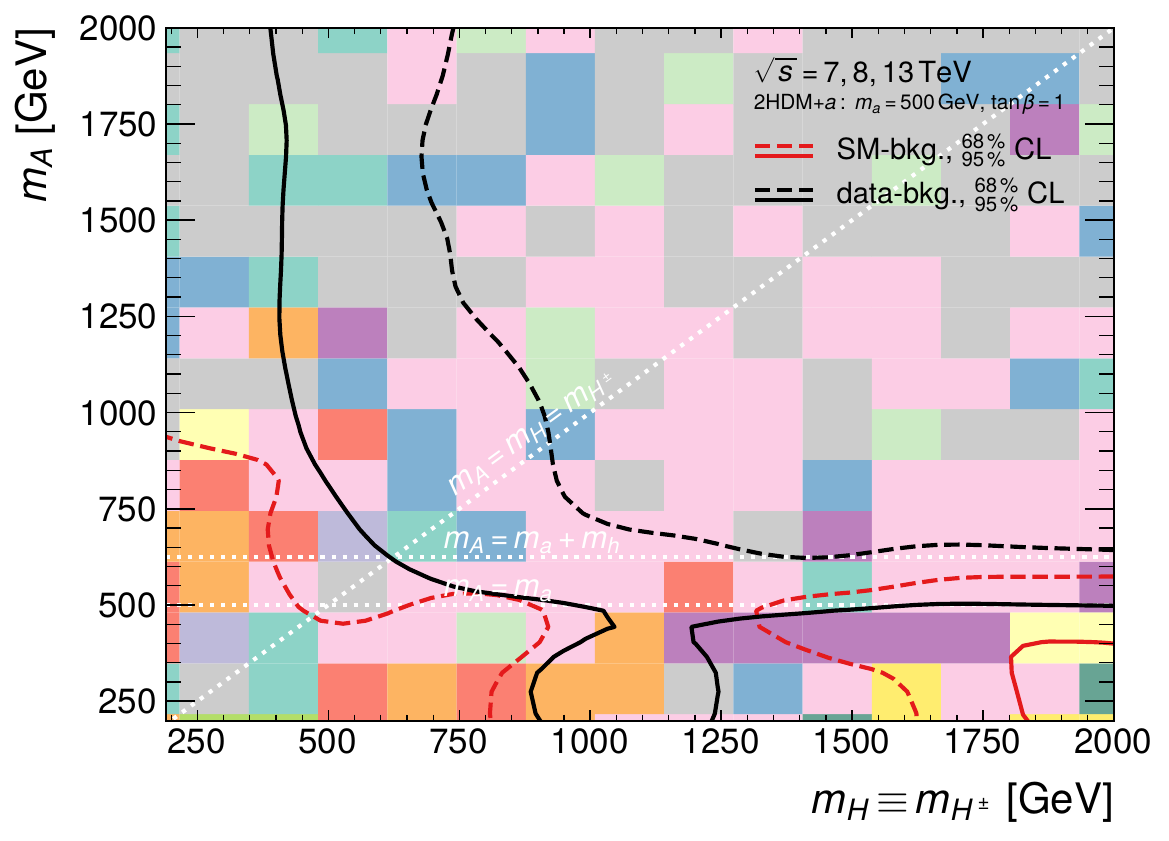}}\\
	}
        \swatch{darkgoldenrod}~ATLAS $\gamma$+\MET{} &
		\swatch{darkolivegreen}~ATLAS high-mass Drell-Yan $\ell\ell$ &
		\swatch{mediumseagreen}~ATLAS $\ell^+\ell^-\gamma$ &
		\swatch{darkorange}~ATLAS $\mu^+\mu^-$+jet \\
		\swatch{blue}~ATLAS $\ell$+\MET{}+jet &
		\swatch{purple}~ATLAS $\ell^\pm\ell^\pm$+\MET{} &
		\swatch{crimson}~ATLAS 3$\ell$ &
		\swatch{yellow}~ATLAS $\gamma$ \\
		\swatch{turquoise}~ATLAS $\ell_1\ell_2$+\MET{}+jet &
		\swatch{powderblue}~CMS $\ell$+\MET{}+jet &
		\swatch{orangered}~ATLAS $e^+e^-$+jet &
		\swatch{wheat}~CMS hadronic $t\bar{t}$ \\
		\swatch{greenyellow}~ATLAS $\ell^+\ell^-$+\MET{} &
		\swatch{snow}~ATLAS hadronic $t\bar{t}$ &
		\swatch{magenta}~ATLAS 4$\ell$ &
\end{myConturFigureGeneral}

The phase space below and to the left of the black (red) lines can be excluded assuming  "data-background" ("\SM-background") prediction.
In \subfigref{fig:Contur_mHmHc_mA}{a}, the exclusion at \SI{95}{\%} confidence level using the "\SM-background" prediction is completely driven by an ATLAS $\ell^+\ell^-+\MET$ measurement~(light-green,~\refcite{Aad:2012awa}) and exclusion limits at this confidence can only be set between the solid red lines.
In \subfigref{fig:Contur_mHmHc_mA}{b}, the exclusion at \SI{95}{\%} confidence level using the "\SM-background" prediction is only below and to the right of the solid red line.

\bigskip
Cross sections and branching fractions helpful for understanding these exclusion limits are given in \figsref{fig:contur_xsBR_mHmA_ma100}{fig:contur_xsBR_mHmA_ma500}.
Two distinct sets of final states contribute to the sensitivity for $\ma=\SI{100}{GeV}$ in \subfigref{fig:Contur_mHmHc_mA}{a}, as can be seen from the coloured tiles:
\begin{itemize}
	\item At small \mHeqmHpm, the sensitivity is driven by ATLAS measurements of the\linebreak $\ell^+\ell^-$+\MET and $\ell^+\ell^-+$jet final states (light-green, orange and red, \eg\refscite{Aad:2012awa,Aad:2015auj}).
	These measurements were already extensively discussed for the exclusion limits in the \mamA plane.
	They are sensitive to the $s$-channel production of the scalar $H$ and its subsequent decay into the pseudoscalar $a$ and a $Z$ boson, $pp\to H\to aZ$.
	This channel is independent of the mass of the pseudoscalar $A$.
	
	\item At large \mHeqmHpm, a significant contribution to the sensitivity comes from ATLAS diphoton (yellow,~\refcite{Aad:2014lwa}) and $\gamma+\MET$ (blue-green,~\refcite{Aad:2016sau}) measurements.
	They are sensitive to the production of a pseudoscalar $a$ in association with a light scalar $h$ from decays of resonantly produced pseudoscalars $A$, $pp\to A\to ah$.
	The latter gives a considerable contribution because, if \mHeqmHpm is significantly larger than the mass of all other Higgs bosons, \eqsref{eq:2HDMa_Gamma_A_ah}{eq:2HDMa_g_Aah} yield
	\begin{equation*}
		\width[A\to ah]\propto\mH^4.
	\end{equation*}
	Therefore, $A\to ah$ becomes the dominant decay mode of the pseudoscalar~$A$.
	The process was already discussed extensively in \secref{sec:interpretation_2HDMa_contributions}.
	A Feynman diagram is shown in \subfigref{fig:2HDMa_METjets_diagrams}{e}.
	Processes $a\to ah$ are negligible compared to $A\to ah$.
	\MET is produced by invisible decays of the pseudoscalar, $a\to\xx$, and the two photons by corresponding decays of the light scalar, $h\to\gamma\gamma$.
\end{itemize}

\bigskip
At larger \ma, $\ma=\SI{500}{GeV}$ (\cf\subfigref{fig:Contur_mHmHc_mA}{b}), the measurements contributing significantly to the sensitivity change drastically.
There are in general three regions of exclusion:
\begin{itemize}
	\item At small \mHeqmHpm, the various ATLAS and CMS measurements of leptons, \MET and jets (blue, turquoise and light-purple, \eg\refscite{ATLAS:2015mip,Aad:2021dse,CMS:2018tdx}) and hadronic \ttbar decays (grey and light-yellow, \eg\refscite{Aaboud:2018eqg,CMS:2019eih}) dominate.
	They are sensitive to $\Hpm\to tb$ and $H\to\ttbar$ decays.
	Additional sensitivity comes from ATLAS measurements of the $\ell^+\ell^-+$jet (orange and red, \eg\refscite{Aad:2015auj}) and four-lepton (rose,~\eg\refcite{Aad:2021ebo}) final states.
	These would select $A\to HZ(\to\ell^+\ell^-)$ decays if the mass difference between pseudoscalar $A$ and scalar $H$ is large enough.
	The scalars~$H$ subsequently dominantly decay to top-quark pairs, $H\to\ttbar$.
	
	\item At small \mA and small \mHeqmHpm, the same measurements are sensitive.
	They would additionally select events from the production of the pseudoscalar $A$ and its subsequent decays to top quarks, $pp\to A\to\ttbar$.
	
	\item At small \mA and large \mHeqmHpm, ATLAS measurements of same-sign dileptons and \MET (dark-purple,~\refcite{Aaboud:2019nmv}) and four-lepton final states (rose,~\eg\refcite{Aad:2021ebo}) dominate.
	They are sensitive to the resonant production of scalars $H$ and their subsequent decays to pseudoscalars $A$ and $Z$ bosons, $H\to AZ(\to\ell^+\ell^-)$.
	The pseudoscalars~$A$ subsequently dominantly decay to top-quark pairs, $A\to\ttbar$.
\end{itemize}

At small \mA, there is therefore sensitivity when \mHeqmHpm is particularly small and when it is particularly large.
The sensitivities for these two distinct regions do not overlap with the current measurements, leading to the sensitivity gap at $\mHeqmHpm\approx\SI{1000}{GeV}$.

\bigskip
In summary, there is significant sensitivity to the \THDMa also off the often investigated benchmark of degenerate masses for the heavy \BSM bosons.
The exclusion limits at \SI{95}{\%} confidence level using the "data-background" prediction for the masses of the heavy \BSM bosons typically reach about \SI{500}{GeV}.
This is generally not significantly weaker than the exclusion limits in the \mamA plane in \figref{fig:Contur_benchmarks_mamA}.
Similar measurements are sensitive but different processes, in particular cascade decays $H\to AZ$ and $A\to HZ$, become important.
In conclusion, \LHC measurements targetting \mAeqmHeqmHpm would not miss important signatures of the model.

\subsubsection{Varying the \DM mass}

The goal of searches for Dark Matter is to find a particle explaining the astrophysical observations discussed in \secref{sec:DM_evidence}.
A value of $\DMRD=0.12$ for the relic density of Dark Matter today is measured~\cite{Planck:2018vyg}.
The value of $\mX=\SI{10}{GeV}$ from the \LHCDMWG benchmark used in the studies so far is disfavoured by $\DMRD=0.12$~\cite{LHCDarkMatterWorkingGroup:2018ufk}.
Larger values of the \DM mass \mX are required to match the constraint.
It is therefore investigated in the following how the exclusion limits from \LHC measurements and their contributing signatures change if the \DM mass \mX is adjusted to more suitable values.

\figref{fig:Contur_mX} shows the exclusion limits from existing \LHC measurements determined with the \Contur method in the \mamX plane.
The mass of the pseudoscalar $a$ (Dark Matter \mX) is varied between \SI{100}{GeV} and \SI{800}{GeV} (\SI{1}{GeV} and \SI{800}{GeV}).
This means that one can have $\mX>\frac{\ma}{2}$ and the decay channel $a\to\xx$ is kinematically closed.
All other parameters are kept at the previously used benchmark parameters given in \tabref{tab:LHCDMWG_params}.
In particular, $\mAeqmHeqmHpm=\SI{600}{GeV}$ and $\tanB=1$.

Solid blue lines indicate where $\DMRD=0.12$ is fulfilled~\cite{LHCDarkMatterWorkingGroup:2018ufk} in \figref{fig:Contur_mX}.
The regions below the line at $\mX=\SI{50}{GeV}$ and between the lines at $\mX=\SI{70}{GeV}$ and $\mX=\SI{110}{GeV}$ at $\ma=\SI{100}{GeV}$ have $\DMRD>0.12$.
This region is strongly disfavoured because here too much Dark Matter would be produced in the early universe.
Another annihilation mechanism for Dark Matter is therefore required to obtain the \DM relic density observed today.
In the rest of the plane, $\DMRD<0.12$.
This is mildly disfavoured because here not enough Dark Matter would be produced in the early universe.
The particle~$\chi$ in the \THDMa could therefore not be the only source of Dark Matter.

\begin{myConturFigureGeneral}{\mamX}{
		Note that the whole plane is excluded at \SI{68}{\%} and \SI{95}{\%} confidence level using the "data-background" prediction and there is no exclusion at all at \SI{95}{\%} confidence level using the "SM-background" prediction.
	}{%
		Dashed white lines indicate the phase space where $\mA=\ma+\mh$, $\mA=\ma$ and $\ma=2\mX$, respectively.
		Solid blue lines indicate where the \DM relic density is $\DMRD=0.12$.
	}{fig:Contur_mX}{
	\includegraphics[width=\textwidth]{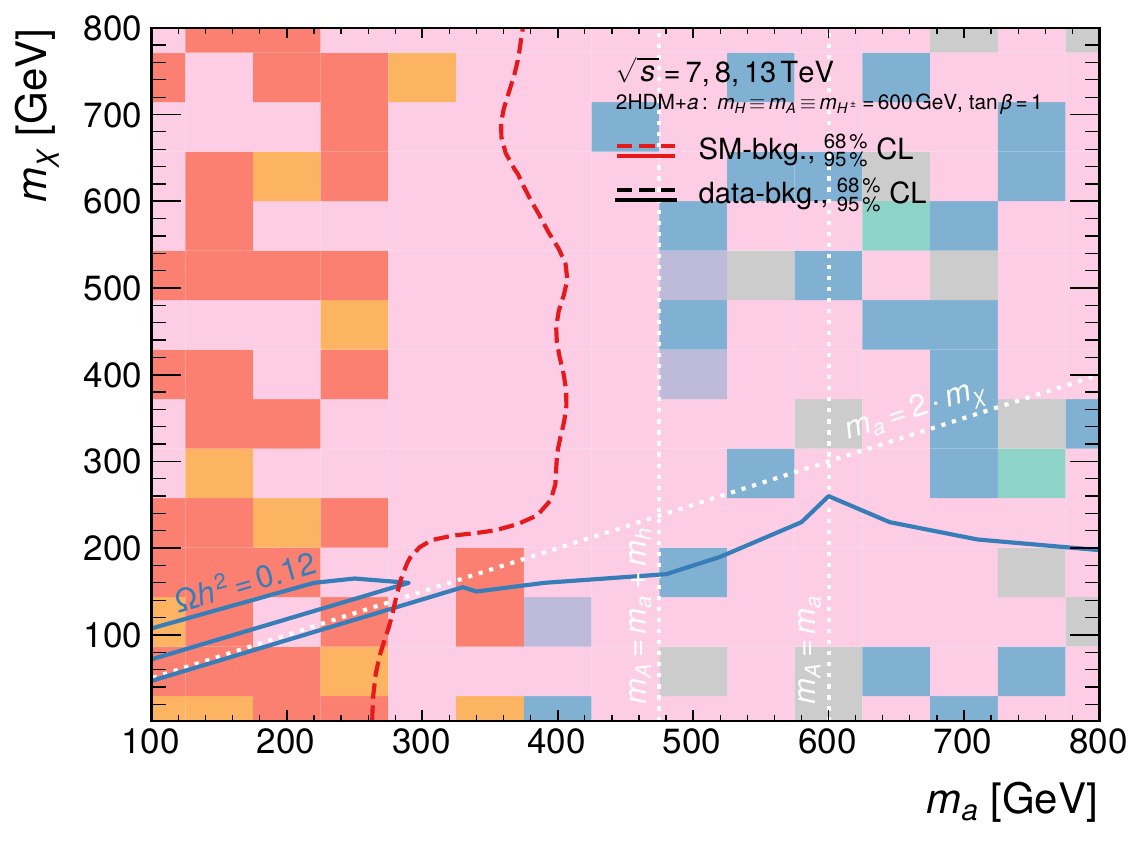}
	}
        \swatch{blue}~ATLAS $\ell$+\MET{}+jet &
		\swatch{powderblue}~CMS $\ell$+\MET{}+jet &
		\swatch{orangered}~ATLAS $e^+e^-$+jet &
		\swatch{darkorange}~ATLAS $\mu^+\mu^-$+jet \\
		\swatch{turquoise}~ATLAS $\ell_1\ell_2$+\MET{}+jet &
		\swatch{snow}~ATLAS hadronic $t\bar{t}$ &
		\swatch{magenta}~ATLAS 4$\ell$ &
\end{myConturFigureGeneral}


\bigskip
The whole plane is excluded at \SI{68}{\%} and \SI{95}{\%} confidence level using the "data-back\-ground" prediction in \figref{fig:Contur_mX}.
There is no exclusion at all at \SI{95}{\%} confidence level using the "SM-background" prediction.
At \SI{95}{\%} confidence level using the "SM-background" prediction, the phase space $\ma<\SI{270}{GeV}$ can be excluded approximately independent of \mX because the dominant sensitivity originates from measurements not relying on \MET from $a\to\xx$ decays.
Larger values of \ma are excluded for $\mX>\SI{200}{GeV}$.
Cross sections and branching fractions helpful for understanding these exclusion limits are given in \figref{fig:contur_xsBR_mamX}.
The origin of the sensitivity in \figref{fig:Contur_mX} is discussed in the following.

At small \ma, ATLAS measurements of the $\ell^+\ell^-+$jet final state (orange and red, \eg\refscite{Aad:2015auj}) dominate.
These measurements were already extensively discussed for the exclusion limits in the \mamA plane.
They would be sensitive to $H\to aZ$ decays where the $Z$ boson decays to lepton pairs and the pseudoscalar $a$ to quark pairs or gluons.

At large \ma, various ATLAS and CMS measurements of leptons, \MET and jets (blue, turquoise and light-purple, \eg\refscite{ATLAS:2015mip,Aad:2021dse,CMS:2018tdx}) as well as hadronic \ttbar decays (grey, \eg\refcite{Aaboud:2018eqg}) are sensitive.
These measurements were already extensively discussed for the exclusion limits in the \mamA plane.
They are sensitive to the various \THDMa processes giving rise to the production of two $W$ bosons from $pp\to A/H\to\ttbar$ as well as $pp\to W^\mp\Hpm$ and $pp\to t\Hpm$ with subsequent decays of $\Hpm\to tb$.
Also the production of pseudoscalars $a$ and their decays to top quarks, $pp\to a\to\ttbar$, gives significant contributions if $2m_t<\ma<2\mX$.

In the whole plane, ATLAS measurements of four-lepton final states (rose,\linebreak \eg\refcite{Aad:2021ebo}) are sensitive.
They would select $H\to aZ$ events in which the $Z$ boson decays to lepton pairs and the pseudoscalar $a$ to \ttbar if $2m_t<\ma$ ($\tau^+\tau^-$ if $2m_t>\ma$).
The \ttbar decays in particular lead to the stronger exclusion at \SI{68}{\%} confidence level using the "\SM-background" prediction (dashed red line) at $\ma=\SI{350}{GeV}$ and $\mX>\SI{200}{GeV}$ because decays $a\to\xx$ are kinematically closed.

\bigskip
In summary, the exclusion limits are mostly independent of \mX because the available measurements that dominate the sensitivity do not rely on large \MET from $a/A\to\xx$ decays.
From this, two conclusions can be drawn.
On the one hand, final states not involving the production of the actual \DM candidate are at least as important for constraining the \THDMa as \MET-based final states.
On the other hand, stronger \MET-based measurements are needed as input for \Contur which would lead to more powerful constraints in the phase space $\ma>2\mX$.

Regarding the \DM relic density, any combination of masses of the pseudoscalar $a$ and \DM particle $\chi$ is excluded at \SI{95}{\%} confidence level by existing \LHC measurements using the "data-background" prediction.
Using the "\SM-background" prediction, none is excluded at \SI{95}{\%} confidence level.
At \SI{68}{\%} confidence level, only the phase space $\ma>\SI{280}{GeV}$, $\mX>\SI{130}{GeV}$ is still allowed where the value $\DMRD=0.12$ can be obtained.
This is only at $\mAeqmHeqmHpm=\SI{600}{GeV}$ and $\tanB=1$.
It remains to be explored in future works how this changes with different parameter choices.

\section{Conclusion}
There is a tremendous amount of data measured at the \LHC and of analyses preserved as \Rivet routines.
The \Contur toolkit allows making use of these to investigate which parts of the parameter space for a given model are already excluded.

Strong exclusion limits can be obtained when the \Contur toolkit is applied to the \THDMa.
It is investigated in the next chapter how the exclusion limits in the \mamA and \matanB plane compare to those from the \METjets measurement obtained in \secref{sec:interpretation_2HDMa} and from a combination of ATLAS searches~\cite{ATLAS:2021fjm,ATLAS:2022rxn}.

Abstaining from the common assumption \mAeqmHeqmHpm in the \mHmA plane does not reveal clear gaps in coverage.
Insofar, this assumption presumably does not lead to important phase spaces of the \THDMa being missed.

Increasing the \DM mass to meet constraints on the \DM relic density does not completely change the picture either.
It does, however, indicate a generally lower importance of \MET-based final states.

\bigskip
The sensitivity to the \THDMa is mostly dominated by two kinds of measurements:
on the one hand, of the $\ell^+\ell^-+$jet finale state which is sensitive to $pp\to H\to aZ$ processes; on the other hand, of top-quark final states which are sensitive to $a/A/H\to\ttbar$ decays and processes involving the charged boson \Hpm.
Both points mean that final states not involving the production of the actual \DM candidate are at least as important for constraining the \THDMa as \MET-based final states.
These final states should be taken into consideration when trying to impose the strongest limits on the \THDMa.

The fact that rarely measurements requiring large \MET dominate the \Contur sensitivity points to a current weakness of the \Contur toolkit:
at the moment, there are very few analyses available to the repository that require large \MET.
This becomes particularly apparent regarding dilepton final states.
Measurements of $\ell^+\ell^-+$jet dominate the \Contur sensitivity at large parts of the parameter space because they are sensitive to $pp\to aZ$ processes.
Pseudoscalars $a$ decay dominantly to \DM particles in most of the phase space.
This gives rise to large \MET which is, however, not exploited by $\ell^+\ell^-+$jet measurements.
In consequence, significantly larger sensitivity to the \THDMa with the \Contur toolkit is expected once stronger $\ell^+\ell^-+$\MET measurements become available to the repository.

A similar argument holds for photon measurements which are sensitive to the increase in the production of \SM-like Higgs bosons by $pp\to ah(\to\gamma\gamma)$ processes.
They do not exploit \MET coming from $a\to\xx$ decays and larger sensitivity to the \THDMa with the \Contur toolkit is expected once stronger $\gamma\gamma+$\MET measurements become available to the repository.

\bigskip
Regarding the two different approaches for estimating the background in \Contur, the "data-background" leads to significantly stronger exclusion limits than the "\SM-back\-ground".
This is mostly a consequence of \SM predictions not being available to the \Contur toolkit.
This stresses the importance of publishing not only the data, but also the \SM prediction of an analysis on \HEPData.

\bigskip
The broad physics results obtained with the \Contur toolkit in this chapter suggest that analysis design should benefit more routinely from \Contur.
\Contur could, for example, be used to check automatically which parts of the parameter space of a model are already excluded when designing new searches.
If a foreseen search is tuned or abandoned because it does not add significant new information, considerable human and computational resources would be liberated for investigating more promising final states or less constrained models.

It has to be taken into consideration that to some degree \Contur works only in retrospect, however:
if no measurements of a final state are available to \Contur, these final states are at first sight indistinguishable from unimportant final states in the \Contur results.
Albeit if the prospective sensitivity of a new search for a \BSM model does not exceed the existing limits set by the various \LHC measurements already available, the need for that particular search should indeed be heavily questioned.

\Chapter[1]{Taking stock}{Comparison to related works}{%
	Peter Fox \& Cold Steel}{Die Affen steigen auf den Thron~\cite{PeterFox:2009ast}}{verse 2, lines 5-7}
\label{sec:comparison}





The \METjets final state and the \THDMa have both been investigated before.
In this chapter, the results of \chapsref{sec:metJets}{sec:Contur} are compared to existing, related works.
\secref{sec:metJets_relatedWork} focuses on existing investigations of the \METjets final state.
Existing investigations of the \THDMa at the \LHC are discussed in \secref{sec:relatedWork_2HDMa}.

\section{Works related to the \METjets measurement}
\label{sec:metJets_relatedWork}

The \METjets final state was targetted by \LHC analyses before because it arises in many \BSM models striving to explain Dark Matter.
Similarly, the most important \SM process contributing to the \METjets final state, \Zjets, was investigated in numerous manners.
It is impossible to give a just comparison of the measurement presented in \chapsref{sec:metJets}{sec:interpretation} to all these works.
The focus in this section shall therefore only be on the most important related works at \SI{13}{TeV} that require large \MET.
An overview of these is given in \tabref{tab:relatedWork_metJets_overview}.

\begin{mytable}{%
		Overview of the most important other works targetting the \METjets final state at \SI{13}{TeV}.
		Measuring event counts and cross sections is essentially equivalent.
	}{tab:relatedWork_metJets_overview}{cccccc}
		reference & experiment & subregion & representation & luminosity $[\ifb]$ & quantity\\
		\midrule
		this work & ATLAS & \Mono, \VBF & particle level & 139 & cross sections, \Rmiss\\
		\cite{ATLAS:2017txd} & ATLAS & \Mono, \VBF & particle level & 3.2 & \Rmiss (only for $\AM_X\in\left\{\TwoEJetsAM, \TwoMuJetsAM\right\}$)\\
		\cite{CMS:2020hkd} & CMS & \Mono & particle level & 35.9 & cross sections\\
		\cite{ATLAS:2021kxv} & ATLAS & \Mono & detector level & 139 & event counts\\
		\cite{CMS:2021far} & CMS & \Mono & detector level & 139 & event counts\\
		\cite{ATLAS:2022yvh} & ATLAS & \VBF & detector level & 139 & event counts\\
		\cite{CMS:2022qva} & CMS & \VBF & detector level & 137 & event counts, \Rmiss\\
\end{mytable}

\bigskip
A previous measurement of the \METjets final state at particle level used \SI{3.2}{\ifb} of ATLAS data~\cite{ATLAS:2017txd}.
It focused exclusively on measuring \Rmiss.
Only fiducial phase spaces corresponding to \TwoLJetsAMs were considered in the denominator of \Rmiss (\cf\eqref{eq:metJets_Rmiss}).
The work presented in this thesis uses a larger integrated luminosity of \SI{139}{\ifb}, considers additionally the phase spaces with exactly one muon or electron (\OneLJetsAMs, respectively) and investigates not only \Rmiss but also the bare differential cross sections.
Both measurements have subregions specifically targetted at \Mono and \VBF processes.

Measurements of the differential cross sections for the $Z$ boson decaying into two neutrinos were performed in \SI{35.9}{\ifb} of CMS data~\cite{CMS:2020hkd}.
They focused on exclusively measuring cross sections.
The work presented in this thesis uses a larger integrated luminosity of \SI{139}{\ifb}, investigates not only differential cross sections but also \Rmiss and has subregions specifically targetted at vector-boson fusion (\VBF) processes.

A search for new phenomena in the \METjets final state was performed in \SI{139}{\ifb} of ATLAS data~\cite{ATLAS:2021kxv}.
This search used an inclusive jet selection, \ie targetted the \Mono subregion.
Contrary to the work presented in this thesis, it additionally exploited an auxiliary measurement enriched in events with top quarks to constrain the background arising from these events in the signal region.
The search results were only provided at detector level, rendering reinterpretation more difficult, as discussed in \secref{sec:analysisPreservation}.

New particles were searched in the \METjets final state in \SI{137}{\ifb} of CMS data~\cite{CMS:2021far}.
The search used an inclusive jet selection, \ie targetted the \Mono subregion.
Using ma\-chine-learning techniques, events in the \METjets final state were separated into categories consistent with hadronic decays of $Z$ or $W$ bosons, or narrow jets from initial-state radiation. Primarily the latter category is comparable to the approach taken in this thesis.
Contrary to the work presented in this thesis, it additionally exploited an auxiliary measurement enriched in $\gamma$\plusJets events.
These events share many kinematic properties with the other \Vjets processes contributing to the signal and other auxiliary measurements.
Comparisons between data and prediction were only performed at detector level.

A search for the decay of the Higgs boson into invisible particles was performed in \SI{139}{\ifb} of ATLAS data~\cite{ATLAS:2022yvh}.
The search targetted the \VBF production of the Higgs boson.
Contrary to the work presented in this thesis, it imposed stricter requirements on the event selection to select \VBF events more exclusively.
Among others, maximally four jets were allowed in the event selection and further observables were exploited to assess the compatibility of jets with final-state radiation.
Comparisons between data and prediction were only performed at detector level.

A search for invisible decays of the Higgs boson was performed in \SI{101}{\ifb} of CMS data and combined with previous results using \SI{36}{\ifb}~\cite{CMS:2022qva}.
The search targetted the \VBF production of the Higgs boson.
Results were reported as event counts as well as \Rmiss distributions.
Contrary to the work presented in this thesis, the search additionally exploited an auxiliary measurement enriched in $\gamma$\plusJets events.
The search further imposed a requirement on the maximum azimuthal angle between the two leading jets to select \VBF events more exclusively.
Comparisons between data and prediction were only performed at detector level.


\bigskip
As such, the work presented in this thesis and in the publication of the ATLAS collaboration currently in preparation~\cite{ATLAS:2023mjt} is the first time the \METjets final state is investigated in such a large dataset of proton--proton collisions at $\sqrt{s}=\SI{13}{TeV}$ at particle level.
Contrary to many of the other analyses listed above, it has subregions specifically targetted at \Mono as well as \VBF processes and presents results as differential cross sections as well as \Rmiss distributions.

\section{Works related to the \THDMa results}
\label{sec:relatedWork_2HDMa}

No significant indication for the \THDMa has been found to date~\cite{ATLAS:2021fjm,ATLAS:2022rxn}.
Summarised exclusion contours on the \THDMa have therefore been derived by the ATLAS collaboration~\cite{ATLAS:2021fjm,ATLAS:2022rxn} and are shown in \figref{fig:2HDMa_combination} in the \mamA and \matanB plane.
At time of writing, the CMS collaboration did not publish summarised exclusion contours for the \THDMa.
In \figref{fig:2HDMa_combination}, unless a parameter is actively varied, the parameter settings from \tabref{tab:LHCDMWG_params} are applied.
In particular, \mAeqmHeqmHpm is used.
The colour of the shaded excluded parameter space indicates the constraining ATLAS search.
The hatched grey area marks the phase space where the width of the heavy pseudoscalar Higgs boson $A$ is larger than \SI{20}{\%} of its width and the assumed narrow-width approximation (\cf\secref{sec:MCEG_hadronisation}) can be violated.

\begin{myfigure}{
		Expected (dashed lines) and  observed (solid lines) exclusion limits by the ATLAS Experiment at \SI{95}{\%} confidence level in the (a) \mamA and (b) \matanB plane.
		The colour indicates the ATLAS search setting the exclusion limits.
		The hatched grey area marks the phase space where the width of the heavy pseudoscalar Higgs boson $A$ is larger than \SI{20}{\%} of its width.
		Figures taken from \refcite{ATLAS:2022rxn}.
	}{fig:2HDMa_combination}
	\subfloat[]{\includegraphics[width=0.49\textwidth]{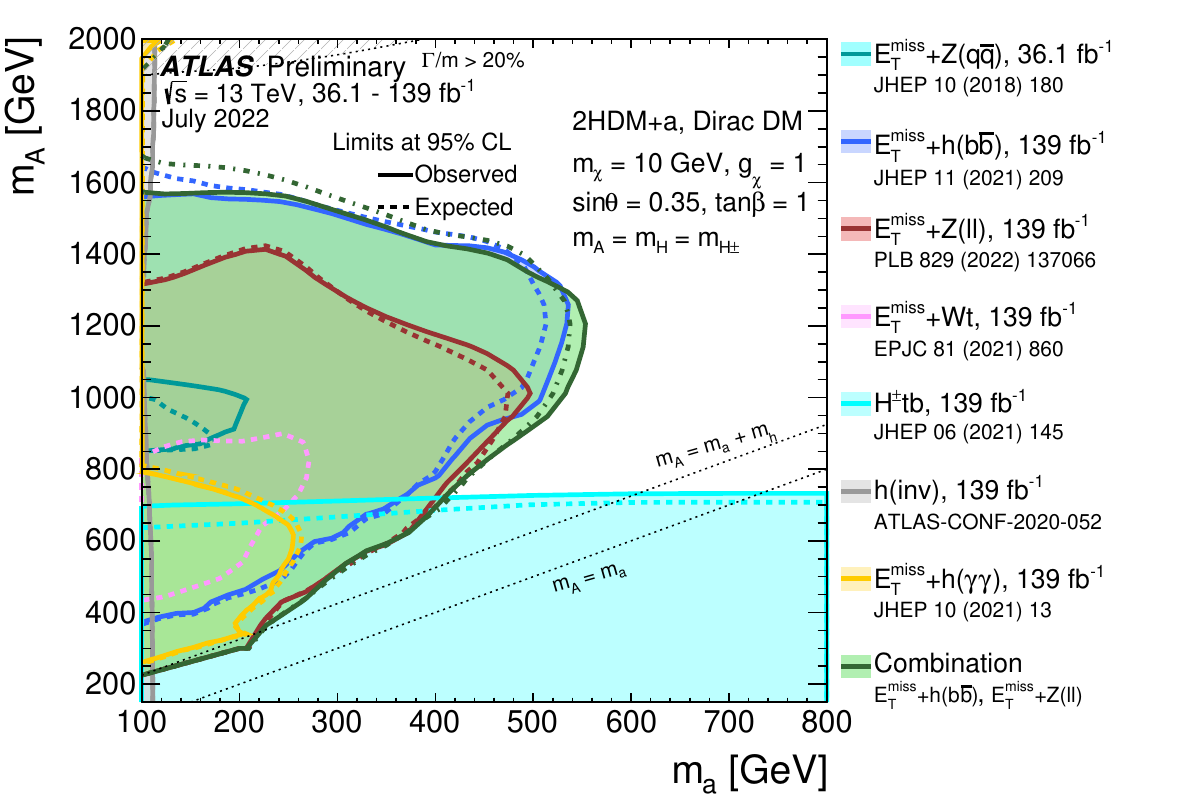}}
	\subfloat[]{\includegraphics[width=0.49\textwidth]{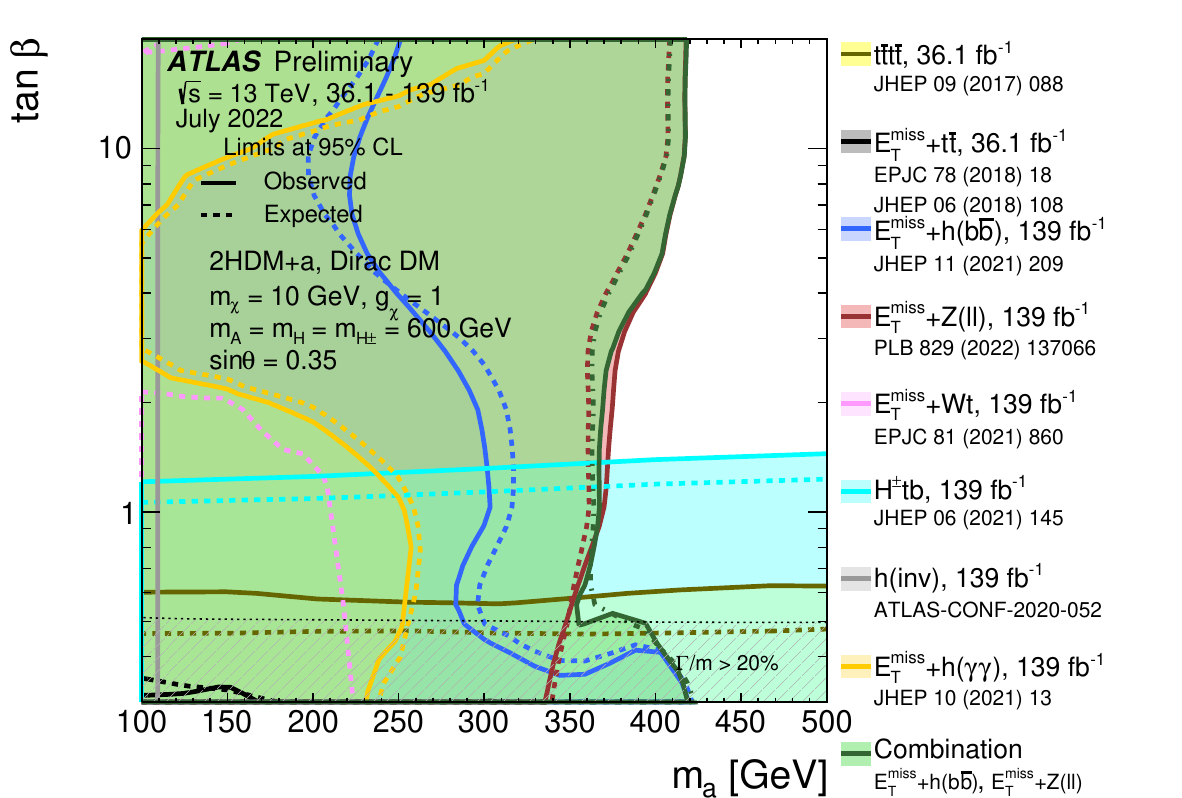}}
\end{myfigure}

\subsubsection{\mamA plane}
In the \mamA plane, there is significant exclusion from ATLAS searches for Dark Matter in events with large \MET and a scalar $h$ decaying to either \bbbar (blue, \refcite{ATLAS:2021qwk}) or $\gamma\gamma$ (yellow, \refcite{ATLAS:2021jbf}).
Their sensitivity is largely due to the resonant production of the heavy pseudoscalar $A$, decaying into the light scalar boson $h$ and the light pseudoscalar $a$ which subsequently decays into Dark Matter, $A\to a(\to\xx)h$, as depicted in \subfigref{fig:2HDMa_METjets_diagrams}{e}.

At small \mAeqmHeqmHpm, the \METhgamgam search reaches the kinematic limit of $\mA=\ma+\mh$.
This is not the case for the \METhbb search because it uses a \MET trigger and therefore imposes a requirement of $\MET>\SI{150}{GeV}$~\cite{ATLAS:2021qwk}, while the \METhgamgam search uses a diphoton trigger and can relax the requirement to $\MET>\SI{90}{GeV}$~\cite{ATLAS:2021jbf}.

At large \mAeqmHeqmHpm, the sensitivity is limited because the production cross section decreases as a function of \mH.
The main sensitivity extends to $\mA<\SI{1600}{GeV}$ and $\ma<\SI{500}{GeV}$ for \METhbb but only $\mA<\SI{800}{GeV}$ and $\ma<\SI{250}{GeV}$ for \METhgamgam.
This difference arises as a consequence of the $h$ branching fractions, which are approximately \SI{58}{\%} for $h\to\bbbar$ and \SI{0.23}{\%} for $h\to\gamma\gamma$~\cite{LHCHiggsCrossSectionWorkingGroup:2016ypw}.

When the mass difference between the pseudoscalar bosons $a$ and $A$ is large, Higgs radiation processes $a\to ah$, as also depicted in \subfigref{fig:2HDMa_METjets_diagrams}{e}, become important as a consequence of \eqref{eq:2HDMa_g_haa}.
This leads to additional sensitivity at $\ma\approx\SI{100}{GeV}$ and $\mA\approx\SI{2000}{GeV}$.
The interplay of $A\to ah$ and $a\to ah$ processes causes a minimum of the sensitivity at $\mA\approx\SI{1700}{GeV}$.

\bigskip
Significant exclusion power also comes from ATLAS searches for Dark Matter in events with large \MET and a vector boson decaying either to two leptons (red, \refcite{ATLAS:2021gcn}) or two quarks (blue-green, \refcite{ATLAS:2018nda}).
Their sensitivity is largely due to resonant production of the heavy scalar $H$ decaying into a $Z$ boson and the light pseudoscalar $a$ which subsequently decays into Dark Matter, $H\to a(\to\xx)Z$, as depicted in \subfigref{fig:2HDMa_METjets_diagrams}{c}.

For the \METZll search, a single-lepton trigger is used, allowing the selection of events with $\MET>\SI{90}{GeV}$ and the exclusion contour to almost reach the kinematic limit of $\mA=\ma+m_Z$.
At large \mAeqmHeqmHpm, the sensitivity is limited because the production cross section decreases as a function of \mA.
No process corresponding to Higgs radiation is available for $Z$ bosons, so the sensitivity does not reach as high as for the \METhbb search.

The \METZqq search exhibits much weaker exclusion limits than the \METZll search because it is subject to larger multijet backgrounds.

\bigskip
Another important contribution to the exclusion comes from an ATLAS search for Dark Matter produced in association with a top quark and an energetic $W$ boson~\cite{ATLAS:2022znu}.
The \METtW search is sensitive to the production of the charged scalar \Hpm and its subsequent decay to a $W$ boson and an invisibly decaying pseudoscalar $a$, $pp\to\Hpm\to W^\pm a(\to\xx)$.
The \METtW search exhibits an exclusion shape similar to that of the \METZll search.
The exclusion region is smaller due to a smaller cross section for the targetted processes as well as smaller acceptance of the analysis.

\bigskip
Complementary to the above-mentioned exclusion contours are those of a search for charged Higgs bosons (light blue, \refcite{ATLAS:2021upq}) and a search for invisible decays of the \SM Higgs boson (grey, \refcite{ATLAS-CONF-2020-052}).

The former is sensitive to the associated production of a charged scalar \Hpm with a top and bottom quark and subsequent decays of the charged scalar to another pair of top and bottom quarks, $pp\to\Hpm(\to tb)tb$.
This production channel is independent of the mass of the light pseudoscalar $a$, apart from changes to the branching fraction of \Hpm, \eg concerning the decays $\Hpm\to tb$ and $\Hpm\to aW^\pm$.
Exclusion bounds are set up to $\mHpm\equiv\mA=\SI{700}{GeV}$, limited by the decrease in production cross section for \Hpm.

The search for invisible decays of the \SM Higgs boson is sensitive to decays of the \SM-like, light scalar $h$ to two light pseudoscalars $a$ and their subsequent decays into \DM particles, $h\to aa\to\xx\xx$, as depicted in \subfigref{fig:2HDMa_METjets_diagrams}{f}.
In this process, any number of light pseudoscalars $a$ can be virtual.
This search gives rise to an exclusion contour that is mostly independent of \mA, excluding the region $\ma<\SI{110}{GeV}$.

\subsubsection{\matanB plane}
In the \matanB plane, there is again significant exclusion power from the \METZll, \METtW, \METhbb and \METhgamgam searches.
The latter two are strongly influenced by the interplay of loop- and \bbbar-induced production modes of the pseudoscalars $a$ and $A$ as well as the scalar $H$.
The loop contributions decrease proportional to $\cotB^2$ with increasing \tanB, while the \bbbar contributions increase proportional to $\tanB^2$ (see \eg\eqref{eq:2HDMa_sigma_a}).
This leaves a sensitivity gap at $3<\tanB<10$.
With decreasing \tanB, the exclusion power of \METhbb increases rapidly, apart from a reduced sensitivity at $\ma\approx\SI{350}{GeV}$ when decays of the light pseudoscalar $a$ to \ttbar pairs become kinematically allowed.

The exclusion from the \METtW search is limited to small \tanB because the production cross section of \Hpm decreases with \tanB.
The exclusion from the \METZll search is largely independent of \tanB.

Overall, at small and large \tanB, masses of the light pseudoscalar of $\ma<\SI{420}{GeV}$ are excluded.
At intermediate \tanB, marginally weaker exclusion limits of $\ma<\SI{370}{GeV}$ are observed.

\bigskip
The search for a charged Higgs boson \Hpm exhibits complementary exclusion contours, as was the case in the \mamA plane.
The allowed parameter space is limited to $\tanB<1.1$, with only minimal dependence on \ma.

Similarly, the search for invisible decays of the \SM Higgs boson excludes the region $\ma<\SI{110}{GeV}$ independent of \tanB.

\bigskip
The search for final states with two top-quark pairs (ochre, \refcite{ATLAS:2017oes}) is sensitive to the associated production of a top-quark pair with $a$, $A$ or $H$ and subsequent decay of the \BSM boson to a second top-quark pair.
The constraints from this search limit the allowed parameter space independent of \ma to $\tanB<0.6$.

Likewise, searches for final states with \MET and one top-quark pair (black, \refscite{ATLAS:2017hoo,ATLAS:2017eoo}) are sensitive to associated production of a top-quark pair with a pseudoscalar $a$ or $A$ and subsequent decay of the pseudoscalar to a \DM pair (\cf\subfigref{fig:2HDMa_METjets_diagrams}{g}).
They constrain the allowed parameter space for very small \tanB and very small \ma.

\bigskip
All in all, there is still considerable phase space in the \THDMa that is not excluded.
In particular, no exclusion limits have been set on the \THDMa in the \METjets final state to date despite the many measurements targetting this final state.

\subsection{Comparison of all exclusion limits}
\label{sec:comparison_2HDMa}

It can now be investigated how the exclusion limits derived for the \THDMa in this work compare to one another and to those from the ATLAS combination.
This is done separately for the \mamA and \matanB planes in the following.

\subsubsection{\mamA plane}

\figref{fig:comparison_2HDMa_mamA} shows a summary of the exclusion limits in the \mamA plane at \SI{95}{\%} confidence level discussed in this thesis.

The exclusion limits from the \METjets measurement performed in \chapsref{sec:metJets}{sec:interpretation} are marked in red.
The dashed line represents the border of the phase space that is expected to be excluded using the \Rmiss distributions and the \SM prediction with fixed normalisation.
The solid line represents the border of the observed exclusion in the same setup.

The exclusion limits from existing \LHC measurements derived in \chapref{sec:Contur} using the \Contur toolkit are marked in blue.
The dashed line represents the border of the excluded phase space if the "data-background" prediction is employed.
The solid line represents the border of the same setup if the "\SM-background" prediction is employed.

The exclusion limits from the ATLAS combination~\cite{ATLAS:2021fjm,ATLAS:2022rxn} are marked in black.
The dashed line represents the border of the excluded phase space if all searches considered in the ATLAS combination are taken into account.
The solid line represents the border for the same setup if only searches selecting events with large \MET are taken into account.

\begin{myfigure}{
		Exclusion limits at \SI{95}{\%} confidence level in the \mamA plane for the complete ATLAS combination (dashed black line), the ATLAS combination taking only into account \MET-based signatures (solid black line), \Contur using the "data-background" prediction (dashed blue line), \Contur using the "\SM-background" prediction (solid blue line) as well as the \METjets measurement as expected (dashed red line) and observed (solid red line).
		The coloured region is excluded.
	}{fig:comparison_2HDMa_mamA}
	\includegraphics[width=\textwidth]{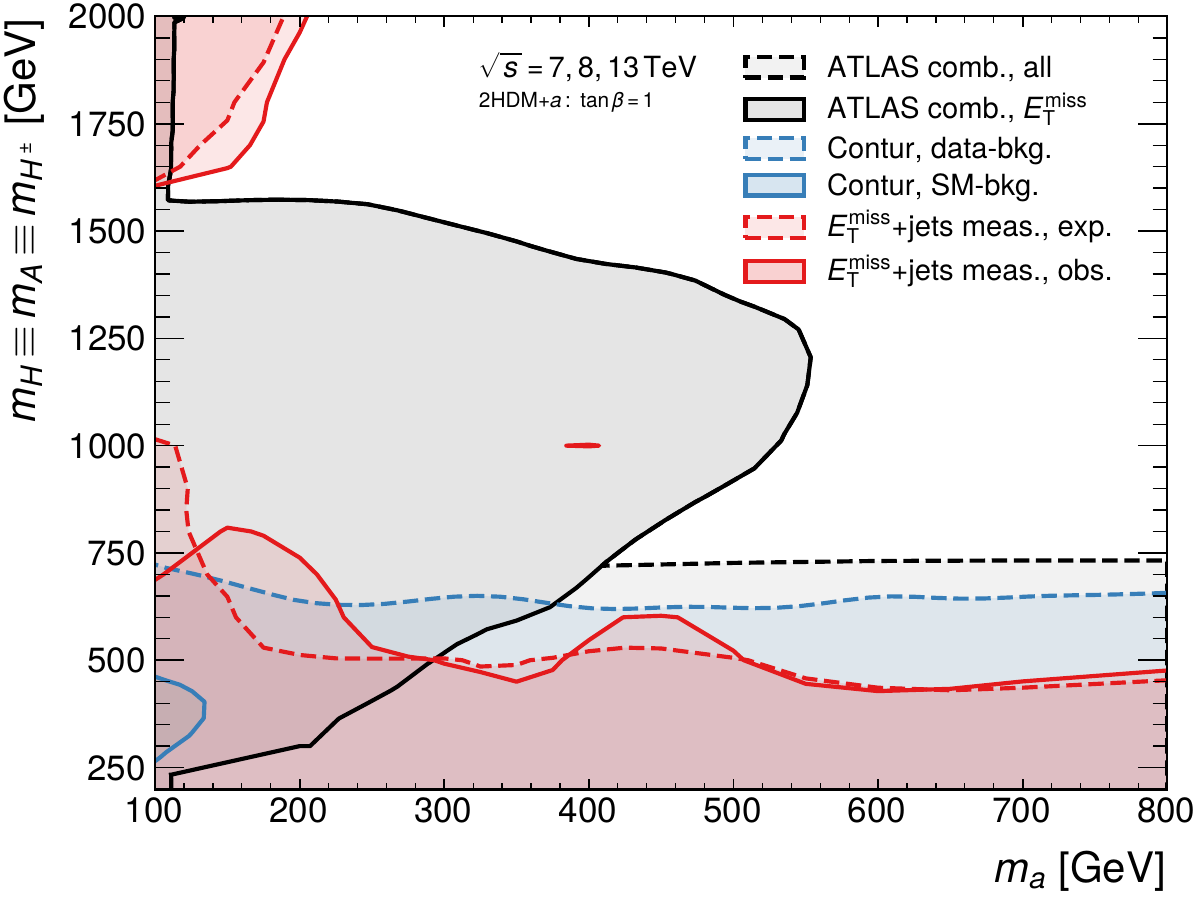}
\end{myfigure}

\bigskip
Comparing the exclusion limits from \METjets measurement and ATLAS combination, the former excludes considerable phase space that was unconstrained before.

At large \mA, the \METjets measurement extends the existing exclusion limits on the mass of the pseudoscalar~$a$ by up to \SI{80}{GeV}.
The process that is selected by the \METjets measurement in this phase is the production of a pseudoscalar~$a$ in association with a scalar~$h$ which decay to pairs of \DM particles and bottom quarks, respectively, $pp\to a(\to\xx)h(\to\bbbar)$.
Contrary to the ATLAS searches targetting this decay~\cite{ATLAS:2021qwk,ATLAS:2021jbf}, the \METjets measurement does not require the explicit reconstruction of photons, jets tagged as coming from bottom-quarks or scalars~$h$.
In addition, \Rmiss can be used in which major systematic uncertainties cancel.
	
At small \mA, the \METjets measurement sets exclusion limits in a phase space that has not been excluded to date using \MET-based final states.
Masses up to $\mA=\limitmamAmAmin$ independent of the mass of the pseudoscalar~$a$ can be excluded.
In principle, this phase space has already been excluded by an ATLAS search for charged Higgs bosons decaying to top and bottom quarks~\cite{ATLAS:2021upq}.
This does, however, not exploit the presence of large \MET in the final state.
The cross section for the production of charged Higgs bosons as well as the branching fraction of the charged Higgs boson decaying to top and bottom quarks depends on the exact parameter choice of the \THDMa, in particular with respect to \tanB, \sinP and masses of the \BSM bosons.
Likewise, the production of the \METjets final state depends on the exact parameter choice of the \THDMa.
It is therefore important to derive exclusion limits using orthogonal, \ie \MET- as well as non-\MET-based, final states.
This suggests the inclusion of an analysis targetting the \METjets final state in a future ATLAS combination.

\bigskip
Comparing the exclusion limits derived from the \METjets measurement and from the \Contur toolkit, the former excludes considerable phase space that is currently not excluded by other existing \LHC measurements when using the "\SM-background" prediction.
The \METjets measurement excludes only significant additional phase space at large masses of the pseudoscalar~$A$ if the "data-background" prediction is employed by \Contur.
This emphasises that the \Contur toolkit will provide much more stringent limits when more and better \SM predictions for measurements are available.

At the time of writing, only two measurements targetting the \METjets final state are available in the \Contur repository:
the \Rmiss measurement that was discussed in \secref{sec:metJets_relatedWork}~\cite{ATLAS:2017txd} as well as a search for supersymmetric particles~\cite{ATLAS:2016dwk}.
Both of them make only use of an integrated luminosity of \SI{3.2}{\ifb}.
Insofar, the \METjets measurement presented in \chapsref{sec:metJets}{sec:interpretation} will prove a valuable addition to the repository of measurements used by the \Contur toolkit as soon as its data and \Rivet routine are published.


\bigskip
Comparing the exclusion limits derived from existing \LHC measurements using the \Contur toolkit and from ATLAS searches, the latter outperform the former everywhere.
This is in large parts caused by the lack of measurements making use of the full integrated luminosity at \SI{13}{TeV} of \SI{139}{\ifb} in the \Contur repository.
Most notably, for the important $Z(\to\ell^+\ell^-)+\MET$ final state only a measurement at \SI{7}{TeV} using an integrated luminosity of \SI{4.6}{\ifb}~\cite{Aad:2012awa} is available.
The ATLAS combination exploits a search at \SI{13}{TeV} using an integrated luminosity of \SI{139}{\ifb} for this final state~\cite{ATLAS:2021gcn}.

The comparison also highlights some important final states that are not yet included in the ATLAS combination.
Most importantly, final states involving two $W$ bosons or two top quarks give significant sensitivity to the \THDMa for the \Contur toolkit at small \mA independent of the mass of the pseudoscalar~$a$.
Including analyses targetting these final states in the ATLAS combination could strengthen the exclusion limits.
These analyses could either be dedicated searches in these final states or directly existing ATLAS measurements.

Despite the exclusion limits from the \Contur toolkit being generally weaker than from the ATLAS combination, it is striking what amount of parameter space can be excluded by simply re-using existing \LHC measurements.
In addition, none of these measurements has been specifically targetted at \BSM models, contrary to the searches used by the ATLAS combination.
The excluded parameter space can be expected to be continuously expanded when new \LHC measurements become available and no sign of the \THDMa is found.
Integrating their sensitivity into the \Contur exclusion limits is straightforward and could also be performed automatically in the foreseeable future.
This is in stark contrast to the ATLAS combination which requires laudable but immense human effort.

\subsubsection{\matanB plane}

\figref{fig:comparison_2HDMa_matanB} shows a summary of the exclusion limits in the \matanB plane at \SI{95}{\%} confidence level discussed in this thesis.
Note that the displayed range of \tanB is chosen such that it meets the requirements of all three different sources of exclusion limits.
The shown range therefore differs from the one in \figsref{fig:interp_2HDMa_exclusion_matb}{fig:Contur_benchmarks_matanB}{fig:2HDMa_combination}\hspace{-3pt}b.

\begin{myfigure}{
		Exclusion limits at \SI{95}{\%} confidence level in the \matanB plane for the complete ATLAS combination (dashed black line), the ATLAS combination taking only into account \MET-based signatures (solid black line), \Contur using the "data-background" prediction (dashed blue line), \Contur using the "\SM-background" prediction (solid blue line) as well as the \METjets measurement as expected (dashed red line) and observed (solid red line).
		The coloured region is excluded.
	}{fig:comparison_2HDMa_matanB}
	\includegraphics[width=\textwidth]{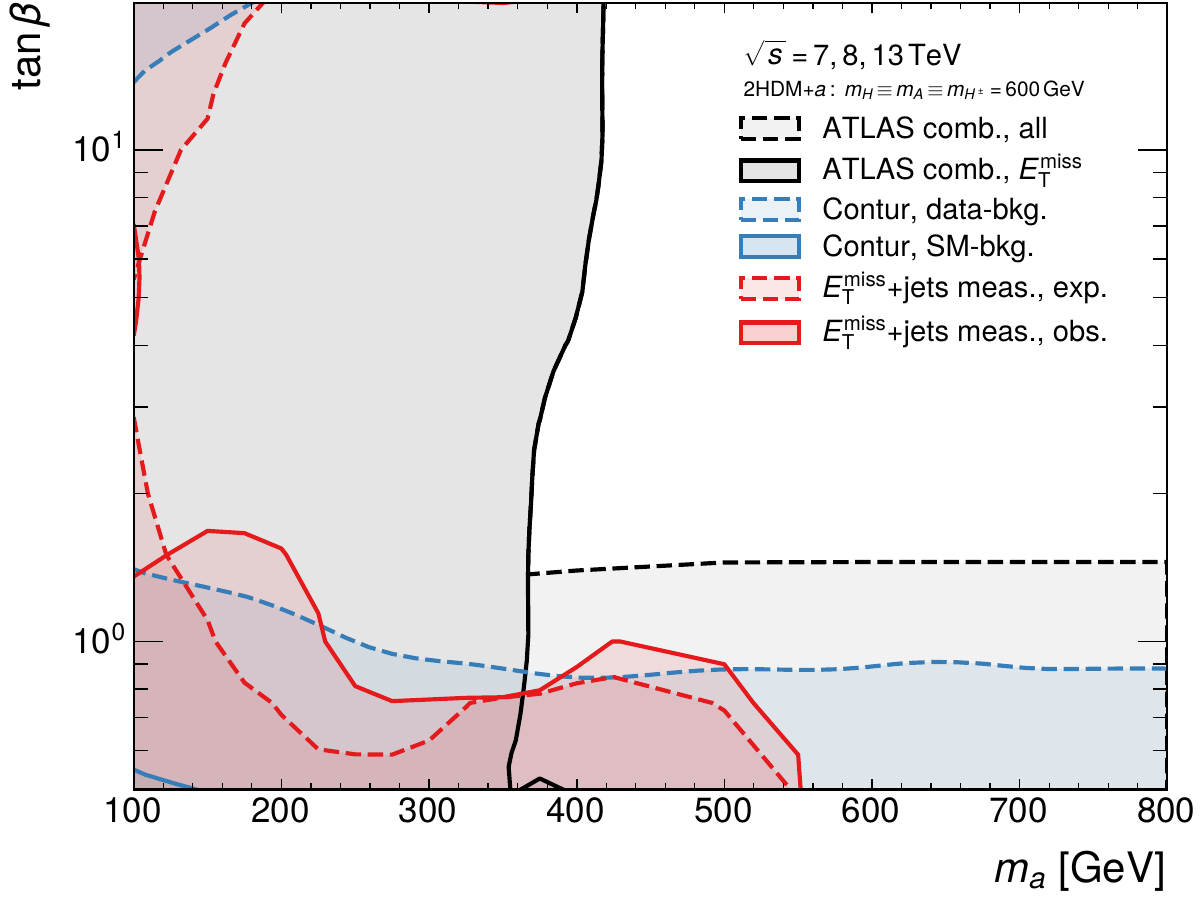}
\end{myfigure}

\bigskip
Comparing the exclusion limits from \METjets measurement and ATLAS combination, the former excludes a small amount of phase space that was unconstrained using \MET-based final states before.
At small \tanB, the \METjets measurement extends the existing exclusion limits on the mass of the pseudoscalar~$a$ by up to \SI{190}{GeV}.
At large \tanB, the ATLAS combination does not discuss as large values of \tanB as are available for the \METjets measurement.
A conclusion concerning the excluded region at large \tanB from the \METjets measurement (\cf\figref{fig:interp_2HDMa_exclusion_matb}) can therefore not be drawn.

\bigskip
Comparing the exclusion limits derived from the \METjets measurement and from the \Contur toolkit, the latter only excludes a small region in the displayed \tanB range when using the "\SM-background" prediction.
A larger amount of parameter space is excluded when using the "data-background" prediction.
In both cases, the \METjets measurement excludes significant phase space that is currently not excluded by other \LHC measurements.
The \METjets measurement will therefore prove a valuable addition to the repository of measurements used by the \Contur toolkit when it is published also with regard to the \matanB plane.

\bigskip
Comparing the exclusion limits derived from existing \LHC measurements using the \Contur toolkit and from ATLAS searches, the latter outperform the former everywhere.
Having no measurement with large integrated luminosity of the $Z(\to\ell^+\ell^-)+\MET$ final state available in the \Contur repository has its largest consequence in the \matanB plane:
in the ATLAS combination, the search for this final state~\cite{ATLAS:2021gcn} excludes the region $\ma<\SI{360}{GeV}$ for all \tanB.
The region of intermediate \tanB cannot be constrained by \Contur because an appropriate measurement is missing in the repository.
\Chapter[1]{Summing up}{Conclusions and outlook}{%
	Peter Fox}{Zukunft Pink~\cite{PeterFox:2022zpk}}{verse 2, line 16 - pre-hook, line 1}
\label{sec:conclusion}






The Standard Model of particle physics (\SM) gives an astonishingly accurate description of elementary particles and their interactions.
The model does not, however, provide a suitable candidate for Dark Matter $\chi$ -- a massive, stable and collision-free particle hinted at by various astrophysical observations.
The two-Higgs-doublet model with a pseudoscalar mediator to Dark Matter (\THDMa) is an extension of the Standard Model consistent at high energies that is able to explain Dark Matter.
Important parameters of this model are the masses of the pseudoscalars~$a$ and~$A$, \ma and \mA, respectively, as well as the ratio of the vacuum expectation values of the two Higgs doublets, \tanB.

The ATLAS and CMS Experiments are two general-purpose detectors at the Large Hadron Collider.
They provide large datasets of proton--proton collisions at the highest energies.
For comparison, physics processes can also be simulated with Monte-Carlo event-generators.
From both, measured and simulated detector signals, physics objects are reconstructed.
This enables straightforward comparisons between measured data and simulated prediction.
These allow measuring \SM properties and searching for hints of Dark Matter.

In this thesis, a measurement of the final state of large missing transverse energy (\MET) and at least one jet was performed.
This nominal final state is sensitive to \Znunujets events in the Standard Model but for example also \xxqg processes in the \THDMa.
Auxiliary measurements were constructed exploiting final states with different lepton multiplicities to constrain systematic uncertainties.
The \METjets measurement was performed in two subregions, in an inclusive subregion requiring only at least one jet and in a subregion targetting processes of vector-boson fusion.
The measurement is dominated by theoretical uncertainties.

The measurement was corrected for detector effects by removing contributions from mismeasurements as well as unfolding.
Following best scientific practise, this ensures the reusability of the data:
it removes the need to understand and simulate the detector response when the results are investigated by other researchers in the future.
This facilitates reinterpretation of the results as well as re-evaluation of the agreement between measured data and theory predictions whenever the latter are improved.
The \METjets measurement is the topic of an upcoming publication of the ATLAS collaboration~\cite{ATLAS:2023mjt}.

Interpreting the results of the \METjets measurement with respect to \SM predictions in a statistical fit showed good agreement between measured data and generated \SM prediction.
This agreement did not significantly depend on the choice of normalisation parameters in the fit model.
Improved agreement between data and prediction was, however, achieved by comparing the ratio of the yields in the nominal final state to auxiliary measurements.
In this ratio, common mismodellings, \eg of the vector-boson \pT, and many systematic uncertainties cancel.

Interpreting the results of the \METjets measurement with respect to \THDMa predictions showed complex behaviour for the \THDMa processes contributing to the selected final states.
Stringent exclusion limits on the \THDMa parameter space were set in two parameter planes.
In the \mamA plane at $\tanB=1$, masses of the pseudoscalar~$A$ up to \limitmamAmAmin and larger than \limitmamAmAmax are excluded.
In the \matanB plane at $\mAeqmHeqmHpm=\SI{600}{GeV}$, masses of the pseudoscalar~$a$ up to \limitmatbma and of $\tanB<\limitmatbtbmin$ or $\tanB>\limitmatbtbmax$ are excluded.
Further tests showed that these exclusion limits do not significantly depend on the usage of normalisation parameters in the fit model or choice of test statistic.

The \Contur toolkit exploits existing \LHC measurements, like the \METjets measurement, to set exclusion limits on models beyond the Standard Model.
Stringent exclusion limits were set by employing the \Contur toolkit on the \THDMa in the aforementioned two planes, \mamA and \matanB.
Parameter scenarios beyond these two were studied, among others varying the mass of the \DM particle.
This showed large sensitivity of \LHC measurements to the \THDMa even without Dark Matter in the final state when the $a\to\xx$ decay is kinematically closed.

The exclusion limits on the \THDMa derived with the results of the \METjets measurement can compete with those from a recent ATLAS combination~\cite{ATLAS:2021fjm,ATLAS:2022rxn}.
In the \mamA plane the existing exclusion limits are extended by up to \SI{80}{GeV} on the mass of the pseudoscalar~$a$.
In parts of the \mamA as well as \matanB plane, the \METjets measurement further sets the first exclusion limits using \MET-based final states to date.

The exclusion limits derived with the \Contur toolkit are in general weaker than those of the ATLAS combination.
This stresses the need for more and better theory predictions as well as the addition of more recent measurements to the \Contur repository.
After its publication, the \METjets measurement can be expected to add significantly to the \Contur sensitivity.

\bigskip
The biggest foreseen improvement for the \METjets measurement is the \LHC Run 3, that is already taking place at the time of writing, and beyond.
The \LHC Run 3 is planned to approximately double the available dataset for proton--proton collisions.
Statistical uncertainties are not the dominant source of uncertainty in the \METjets measurement presented in this thesis.
They do, however, contribute significantly at large \MET.
An increased dataset will further reduce the statistical uncertainties.
Alternatively, an increased dataset allows investigating smaller intervals of \MET.
This enables further studies on the shape agreement between data and prediction.

Insufficiently accurate \SM predictions have to be considered the main reason for the observed disagreement between measured data and prediction in this work.
A big improvement in this regard can be expected from better \SM predictions, in particular regarding the transverse momenta of vector-bosons and the total cross section.
These could be directly applied to the results from this work because the results are presented in a detector-independent, particle-level representation.

The \Contur toolkit has immense physics potential.
Recently it was demonstrated that smearing generated model predictions to mimic the detector response with \Rivet gives results comparable to those directly taken in detector-level representation~\cite{Brooijmans:2020yij_RivetSmear}.
This allows the \Contur toolkit to make use of analyses that were not corrected for detector effects.
Massive improvements of the sensitivity of the \Contur toolkit can be expected once more of these analyses provide \Rivet plug-ins.

Apart from that, a large strength of the \Contur toolkit lies in its potential automatisation.
On the one hand, model points newly generated by the \LHC collaborations could be automatically tested for exclusion from existing \LHC measurements.
On the other hand, the \Contur toolkit could provide exclusion limits on selected models that are automatically updated as soon as new measurements are added to the repository.

Regarding the \THDMa, it remains to be seen whether this model is the right one to explain the astrophysical observations of Dark Matter.
In the future, it would be worthwhile to search for the \THDMa, or set exclusion limits, not only in final states involving the \DM candidate.
Other final states, in particular those involving two $W$ bosons or top quarks, would also provide powerful handles.

Regarding \LHC analyses in general, searches in detector-level representation do not necessarily provide significantly stronger exclusion limits on new models than measurements in particle-level representation.
More analyses should therefore be corrected for detector effects in the future to ensure reusability.
In addition to that, theory predictions and correlations should be published more routinely alongside the nominal results of the measured data (see also \refcite{LHCReinterpretationForum:2020xtr}).
This would largely improve the statistical statement that can be made in reinterpretations.

\ChapterStarQuote[1]{\COTe statement}{CO$_2$e statement}{%
	Muse}{The 2nd Law: Unsustainable~\cite{Muse:2012uns}}{verse 2, line 4 - chorus, line 1}


Climate change is a severe threat to planetary health and human well-being~\cite{IPCC:2022iav}.
Emission of greenhouse gases is the main driver of this development~\cite{IPCC:2021psb}.
Assessing and reporting on the greenhouse gas emission of all activities, also in research, are two first steps in limiting the negative impacts on the world's climate~\cite{Banerjee:2023avd}.
This can raise awareness as well as help in identifying and ultimately mitigating the sources of greenhouse gas emission.
Regarding this work, these emissions should be put into perspective to the achieved gain in scientific knowledge.
The estimates help to judge the overall greenhouse-gas impact of the work, the main drivers of the emission as well as ultimately which activities are deemed acceptable or unacceptable.

Computations on the Worldwide \LHC Computing Grid~\cite{Bird:2005js} for this work caused emissions equivalent to the global warming potential of \SI{8.8}{t} CO$_2$ (\COTe), estimated with the methodology described in \refcite{Brueers:2023twm}.
This is the result of \SI{269}{years} single-core CPU usage on distributed computers around the world, tracked with \refcite{Kibana:2023kdb}, and takes into account the energy mix in the respective hosting countries of the computing centres using \toolVersion{\GreenAlgorithms}{2.2}~\cite{GreenAlg:2021gal}.

Computations on batch systems amounted to \SI{325}{years} single-core CPU usage, determined with the command \texttt{condor\_userprio -\mbox{-allusers}} on \HTCondor systems.
This corresponds to \SI{13.3}{\tCOTe} using \GreenAlgorithms for computer clusters located in Germany.

Local computing was found to cause \order{\SI{0.2}{\tCOTe}} under circumstances similar to this work~\cite{Brueers:2023twm}.

Far-distance travel relating to this work amounted to \SI{6,800}{km} (\SI{2,700}{km}) by means of planes (trains), corresponding to \SI{1.21}{\tCOTe} (\SI{0.08}{\tCOTe})~\cite{UBA:2020cfa,COTR:2023cot}.

In total, this work therefore caused emissions of \SI{5.5}{\tCOTe/year} for computing and \SI{0.34}{\tCOTe/year} for travels over a period of \SI{48}{months}.
This is to be compared to a "carbon budget" of \SI{1.1}{\tCOTe/year} per capita until 2050 if the global warming is to be contained below \SI{1.5}{\degreeCelsius} compared to $1850-1900$ at \SI{83}{\%} probability~\cite{Bloom:2022gux}.
Arguably, the per capita emissions in industrialised countries -- like the country this work was mostly performed in, Germany -- should be smaller to account for emissions in the past.

Unfortunately, the estimates derived in this chapter have to be understood as a lower limit on the emissions to this work as they do not take into account many other sources of greenhouse gases related to this work, \eg in the construction and operation of the ATLAS detector and the \LHC or from commuting.
Nonetheless, it can be seen that already from the usage of computing resources alone the annual carbon budget of the author would be exceeded manifoldly.
As a first step, increased attention should therefore be paid to sustainable power sources for computing facilities as well as improved software efficiency in future endeavours~\cite{Banerjee:2023avd}.

It is time to take action against human-made climate change.
Now.
\ChapterStarQuote[1]{Acknowledgements}{Acknowledgements}{%
	Muse}{Liberation~\cite{Muse:2022lib}}{chorus 1, lines 4-7}


Working on my PhD project and writing this thesis would not have been possible without the support and encouragement of a plenitude people.
It is about time I thank you all for everything.

My deepest gratitude goes to my -- first technical, then official -- supervisor, Priscilla Pani.
Thanks for your vision for this project, the extensive physics discussions along the way and the never-ending support.
Thanks for the inspiring brainstorming sessions for the project and the future, even or in particular when they went faster than I could keep up or when I was too stubborn to listen to your advice.
Thanks also for the criticism which taught me to think beyond only the next plot ahead, but instead of the big physics picture that is behind it.
This made this thesis considerably more solid and myself a better physicist.

My gratitude also goes to my first official supervisor and now second referee, David Berge.
Without you, I would have never embarked on this journey.
Thanks for the opportunity and for discussions in which I learned what is important about a thesis, and what is not.

Thanks also to the other referees, Christophe Grojean, Thomas Kuhr and Thorsten Kamps, for investing the work of reading another thesis of 200-ish pages just in the spirit of good scientific practise and collegiality.
This means a lot to me.

A big thank you to everyone in the \METjets measurement:
Jonathan Butterworth, Louie Corpe, Monica Dunford, Christian Gütschow, Aidan Kelly, Martin Klassen, Vasilis Konstantinides, Stephen Menary, Emily Nurse, Pavel Starovoitov, Matouš Vozák, Sebastian Weber and Yoran Yeh.
Working with you on the measurement was always fun and engaging, although this seemingly straightforward measurement turned out to be something of a beast at every corner.
Thanks for everything, I learned a lot!
At this point, a particular thanks to Stephen Menary -- without you this work would have been both, harder and easier.
Thanks for making me learn so much about statistics.

A heartfelt "thank you!" also to the whole team of the \Contur toolkit,
most importantly Andy Buckley, Jonathan Butterworth, Louie Corpe and David Yallup.
There is a ton of things I love about this project.
Thanks to you, the entry barrier is really low; one always feels invited and ideas welcome.

Very importantly -- thanks to everyone in Priscilla Pani's Young Investigator's Group:
Marawan Barakat, Ben Brüers, Baishali Dutta, Marianna Liberatore, Eleonora Loiacono, Álvaro Lopez Solis, Sana Tabbassum,
as well as Claudia Seitz' Young Investigator's Group.
It has been a blast meeting and working with you all!
We had more fun trips to restaurants, engaging discussions and enlightening late-afternoon coffees than I could ever list here.

Thanks to all the people at DESY I had the pleasure to meet.
It was an incredibly fun and engaging time.
I will always look back to it with one laughing and one crying eye.
At this point also thanks to all those who cannot be thanked enough considering this work:
the people operating the DESY computer clusters, the \LHC grid and the ATLAS detector.
You are doing a terrific job without which works like these would not be possible!

At this point also thanks to Martin Bauer, Ulrich Haisch and Felix Kahlhoefer -- your paper regarding the \THDMa proved to contain all the relevant information again and again and was a huge help during the whole project.

Thanks also to all those artists that allowed me to use their quotes in this work.
Your work accompanied me through the whole project, so it is all the more fitting to quote it in this thesis. 

A particularly big thank you also to those that read and commented on parts of the thesis in earlier or later stages:
Jonathan Butterworth, Ben Brüers, Álvaro Lopez Solis, Claudia Seitz, Tadej Novak and in particular Priscilla Pani and Thorsten Kuhl.
You have found so many unclear paragraphs, minor inconsistencies, or outright mistakes -- your input to this work was invaluable!

Last but not least -- thanks to my family and friends!
It was always a relief to be able to discuss thesis-related topics or, on the contrary, clear my mind of anything thesis-related.
You helped me push through to the very end.
I would not have been able to finish this without you.
Thanks.

\bookmarksetup{bold} 
\appendix
\part*{\textcolor{subtitlecolor}{Appendices}}
\addcontentsline{toc}{part}{Appendices}
\bookmarksetup{bold=false} 

\chapter{Triggers for the \METjets measurement}
\label{app:metJets_triggers}

\tabref{tab:metJets_triggers} gives an overview of the triggers used for selecting events in the \METjets measurement described in \chapsref{sec:metJets}{sec:interpretation}.
An "xe" ("e") followed by an integer denotes the threshold with respect to \METmeas (the electron \pT) that has to be fulfilled for an event to pass the trigger in GeV, \eg $\METmeas>\SI{70}{GeV}$ for HLT\_xe70.
For electron triggers, "lhloose", "lhmedium" and "lhtight" indicate a working point for electron identification; "ivarloose" a working point for electron isolation (\cf\secref{sec:objReco_elePhot}).

Signal region, \OneMuJetsAM and \TwoMuJetsAM use the same trigger requirements because the calculation of \METmeas in the high-level trigger in the ATLAS Experiment does not use information from the Muon spectrometer~\cite{ATLAS:2020atr}.
Therefore, all events with sufficient transverse momentum of the muon are already selected by the \METmeas triggers.

\begin{mytable}{Summary of the trigger requirements applied to the five regions of the \METjets measurement in the different years. See the references in the second column for further details.}{tab:metJets_triggers}{cccl}
	Region & Reference & Year & Trigger requirement\\
	\midrule
	\multirow{7}{70pt}{signal region, \OneMuJetsAM, \TwoMuJetsAM} & \multirow{7}{*}{\cite{ATLAS:2020atr}}
		& 2015
			& HLT\_xe70\\
		\cline{3-4}
		&& \multirow{2}{*}{2016}
			& HLT\_xe90\_mht\_L1XE50\\
			&&& HLT\_xe110\_mht\_L1XE50\\
		\cline{3-4}
		&& 2017
			& HLT\_xe110\_pufit\_L1XE55\\
		\cline{3-4}
		&& \multirow{3}{*}{2018}
			& HLT\_xe110\_pufit\_xe70\_L1XE50\\
			&&& HLT\_xe120\_pufit\_L1XE50\\
			&&& HLT\_xe110\_pufit\_xe65\_L1XE50\\
	\midrule
	\multirow{16}{*}{\OneEJetsAM, \TwoEJetsAM} & \multirow{16}{*}{\cite{ATLAS:2019dpa}}
		& \multirow{3}{*}{2015}
			& HLT\_e24\_lhmedium\_L1EM20VH\\
			&&& HLT\_e60\_lhmedium\\
			&&& HLT\_e120\_lhloose\\
		\cline{3-4}
		&& \multirow{6}{*}{2016}
			& HLT\_e24\_lhmedium\_nod0\_L1EM20VH\\
			&&& HLT\_e60\_lhmedium\\
			&&& HLT\_e26\_lhtight\_nod0\_ivarloose\\
			&&& HLT\_e60\_lhmedium\_nod0\\
			&&& HLT\_e140\_lhloose\_nod0\\
			&&& HLT\_e300\_etcut\\
		\cline{3-4}
		&& \multirow{4}{*}{2017}
			& HLT\_e26\_lhtight\_nod0\_ivarloose\\
			&&& HLT\_e60\_lhmedium\_nod0\\
			&&& HLT\_e140\_lhloose\_nod0\\
			&&& HLT\_e300\_etcut\\
		\cline{3-4}
		&& \multirow{4}{*}{2018}
			& HLT\_e26\_lhtight\_nod0\_ivarloose\\
			&&& HLT\_e60\_lhmedium\_nod0\\
			&&& HLT\_e140\_lhloose\_nod0\\
			&&& HLT\_e300\_etcut\\
\end{mytable}
\chapter{Particle-level results in the \METjets measurement}
\label{app:detCorr_partLevelResults}

\figref{fig:detCorr_app_partLevelResults} shows the measurement results at particle level not displayed in \secref{sec:detCorr_partLevelResults}.
The top panels give the differential cross section of data (black dots) and generated \SM prediction (blue crosses) with their respective statistical uncertainties as well as the contributions to the generated cross section from the different processes (filled areas).
The quadrature sum of experimental (theoretical) systematic and statistical uncertainties is displayed as a red (blue) shaded area.
Unfolding uncertainties are counted towards the experimental uncertainties.
The bottom panels show the ratio of data to generated \SM prediction.

\begin{myfigure}{
		Differential cross section and ratio of data to generated \SM prediction at the particle level in the (left) \Mono and (right) \VBF subregion.
		Black dots (blue crosses) denote the measured data (yield from generated \SM prediction) with their statistical uncertainty.
		The red (blue) shaded areas correspond to the total experimental (theoretical) uncertainty. In the top panels, the filled areas correspond to the different contributing processes.
	}{fig:detCorr_app_partLevelResults}
	\subfloat[]{\includegraphics[width=0.49\textwidth]{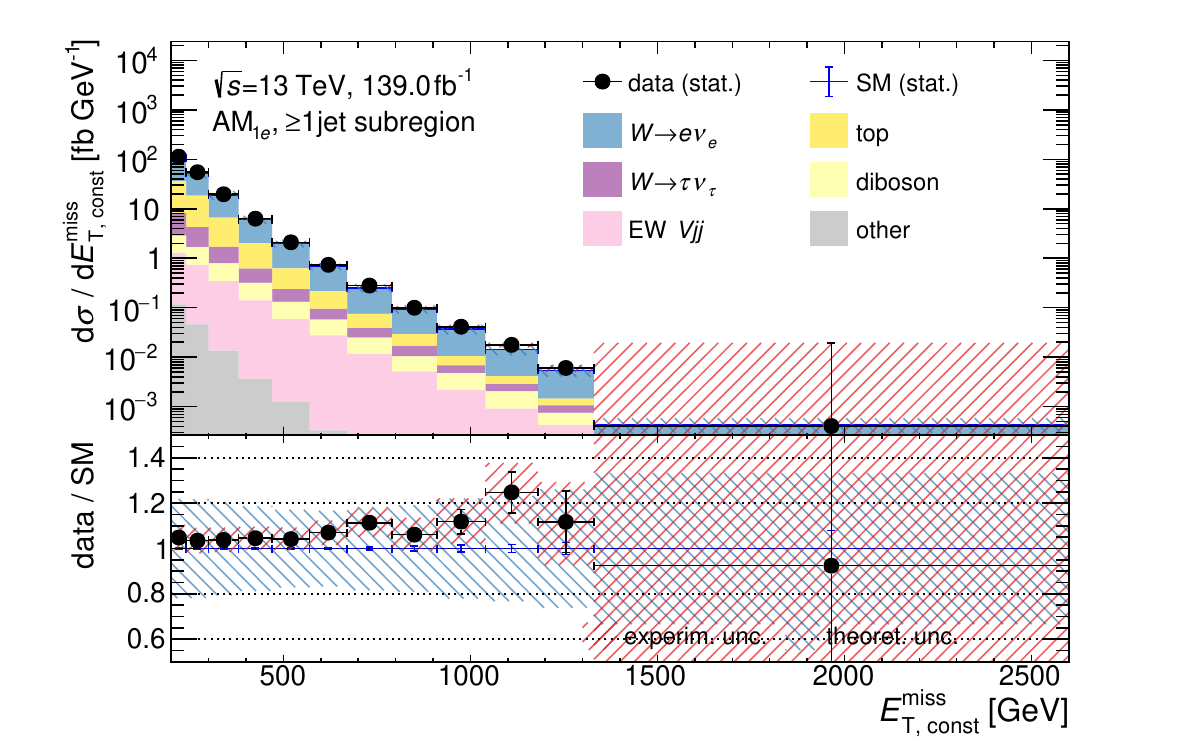}}
	\subfloat[]{\includegraphics[width=0.49\textwidth]{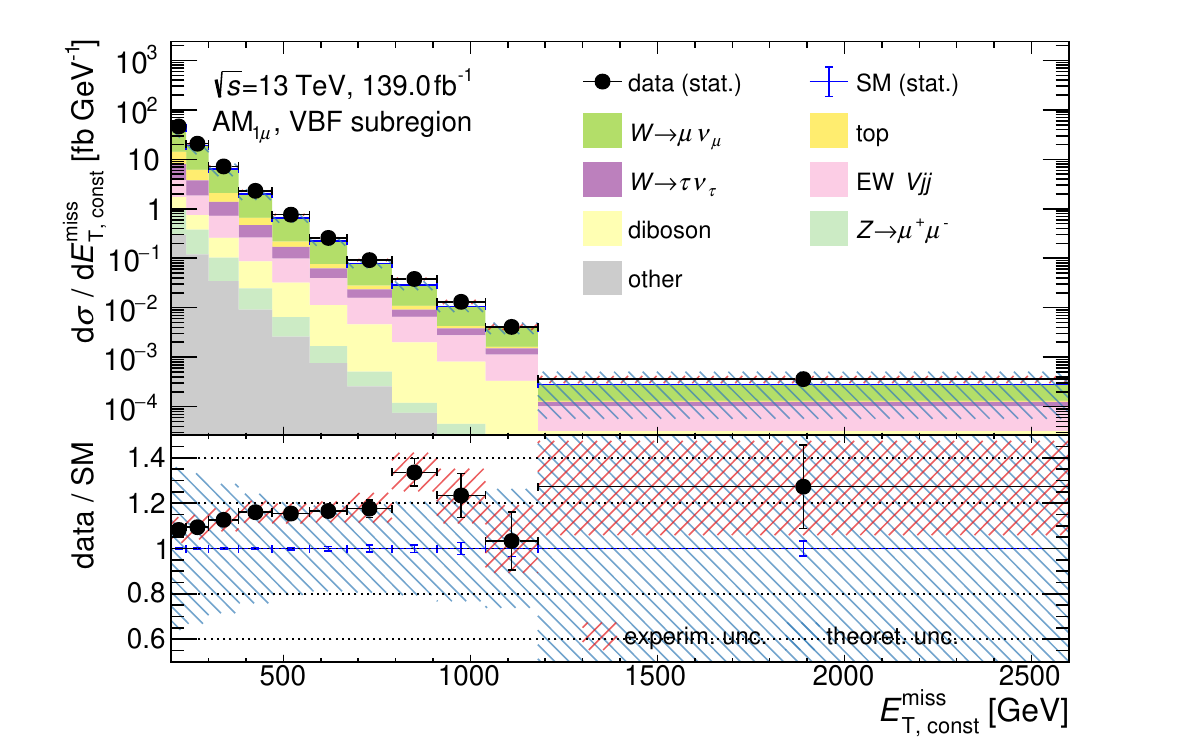}}\\
	\subfloat[]{\includegraphics[width=0.49\textwidth]{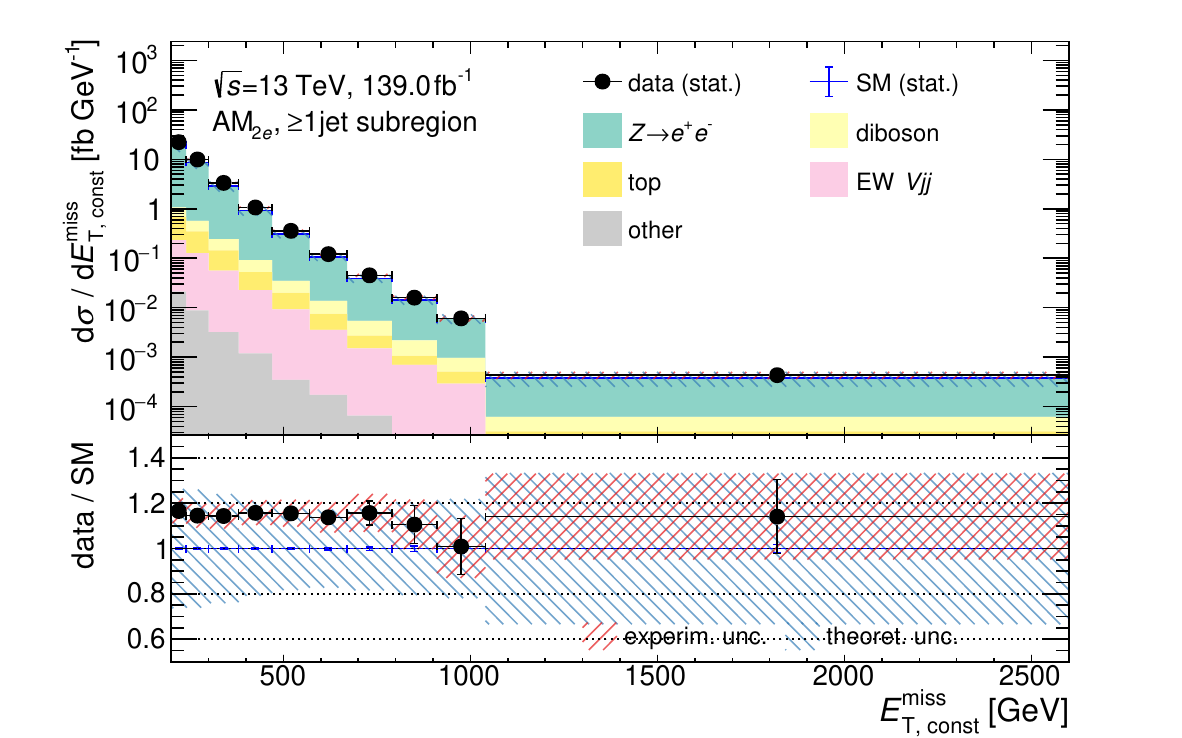}}
	\subfloat[]{\includegraphics[width=0.49\textwidth]{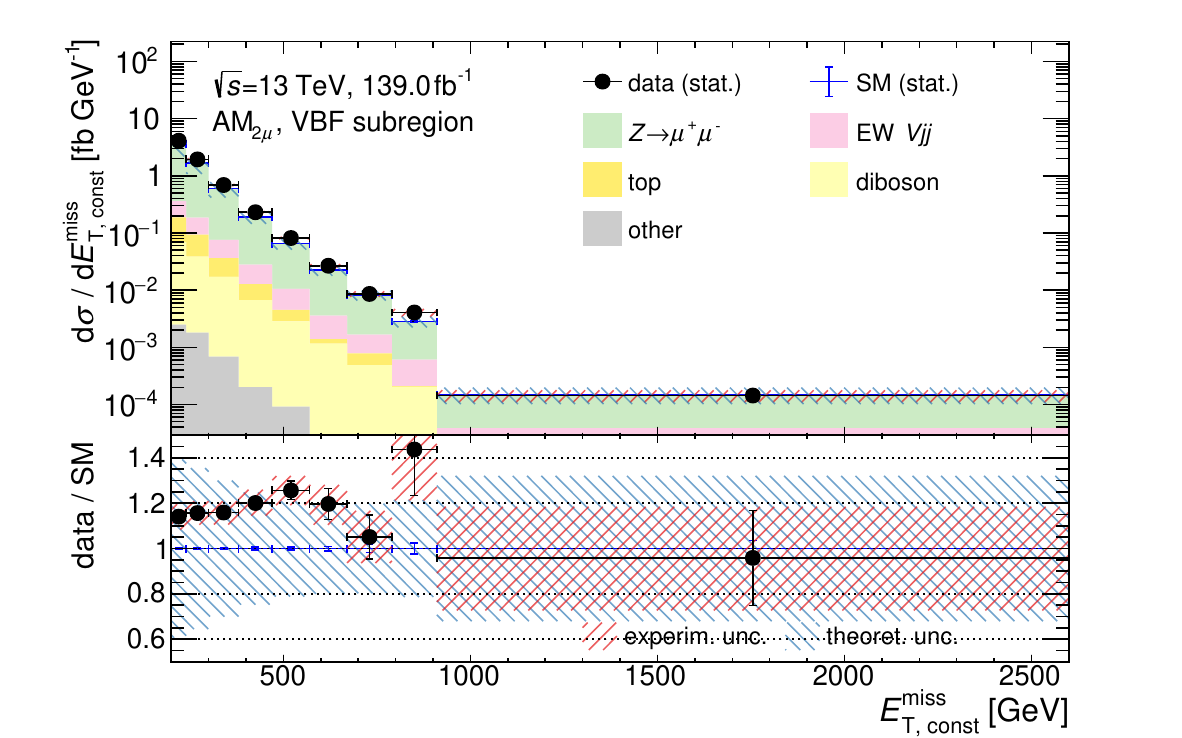}}\\
\end{myfigure}

The generated \SM prediction gives a good description of the shape of the measured data but underestimates the normalisation by $10-\SI{20}{\%}$ in all regions.
The discrepancy is smaller in \OneLJetsAMs because of the different process composition in the Standard Model (\cf\secref{sec:metJets_detLevelResults}).
The discrepancy in all regions is mostly covered by the theoretical systematic uncertainties.

\bigskip
In \figref{fig:detCorr_app_partLevelResults_Rmiss}, the ratio of the fiducial cross section in the signal region to auxiliary measurements not displayed in \secref{sec:detCorr_partLevelResults} is shown.
\Rmiss is calculated according to \eqref{eq:metJets_Rmiss}.
The top panels give \Rmiss in data (black dots) and generated \SM prediction (blue crosses) with their respective statistical uncertainties.
The quadrature sum of experimental (theoretical) systematic and statistical uncertainties is displayed as a red (blue) shaded area.
In the bottom panels, the ratio of data to generated \SM prediction is shown.

\begin{myfigure}{
		Ratio of the fiducial cross section in the signal region to the auxiliary measurements in the (left) \Mono and (right) \VBF subregion, calculated according to \eqref{eq:metJets_Rmiss}.
		Black dots (blue crosses) denote the measured data (yield from generated \SM prediction) with their statistical uncertainty.
		The red (blue) shaded areas correspond to the total experimental (theoretical) uncertainty.
		The bottom panel shows the ratio of data to generated \SM prediction.
	}{fig:detCorr_app_partLevelResults_Rmiss}
	\subfloat[]{\includegraphics[width=0.49\textwidth]{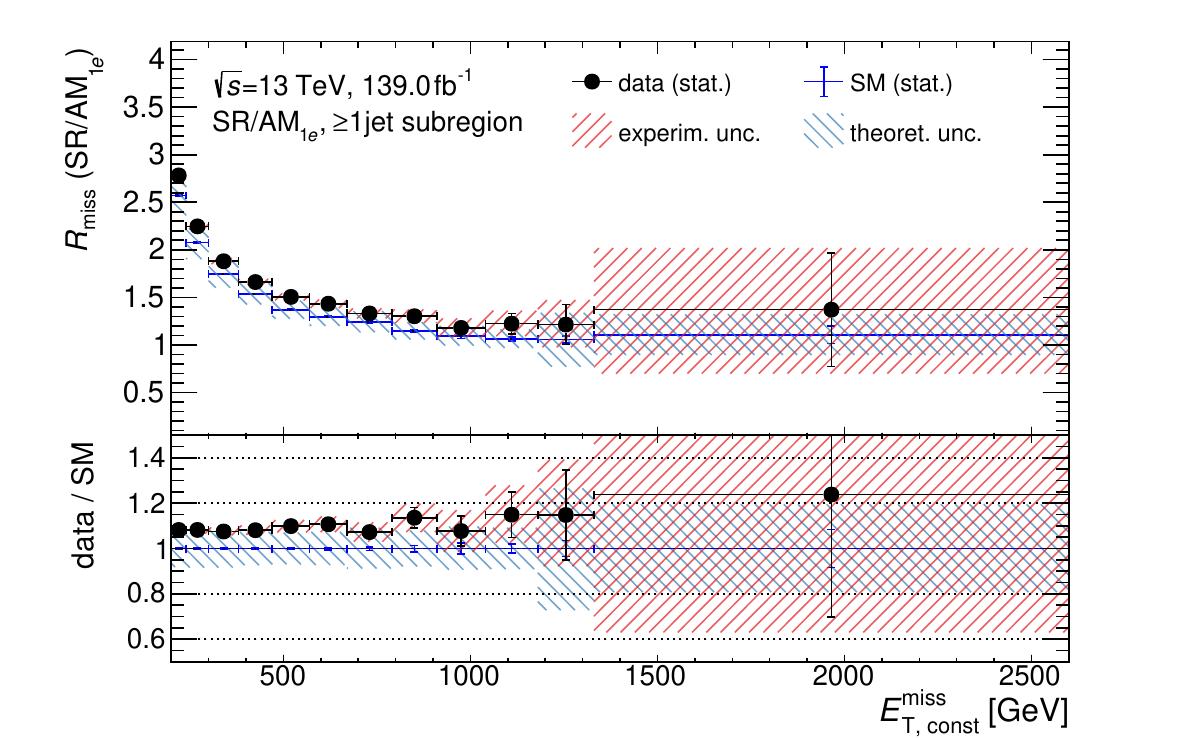}}
	\subfloat[]{\includegraphics[width=0.49\textwidth]{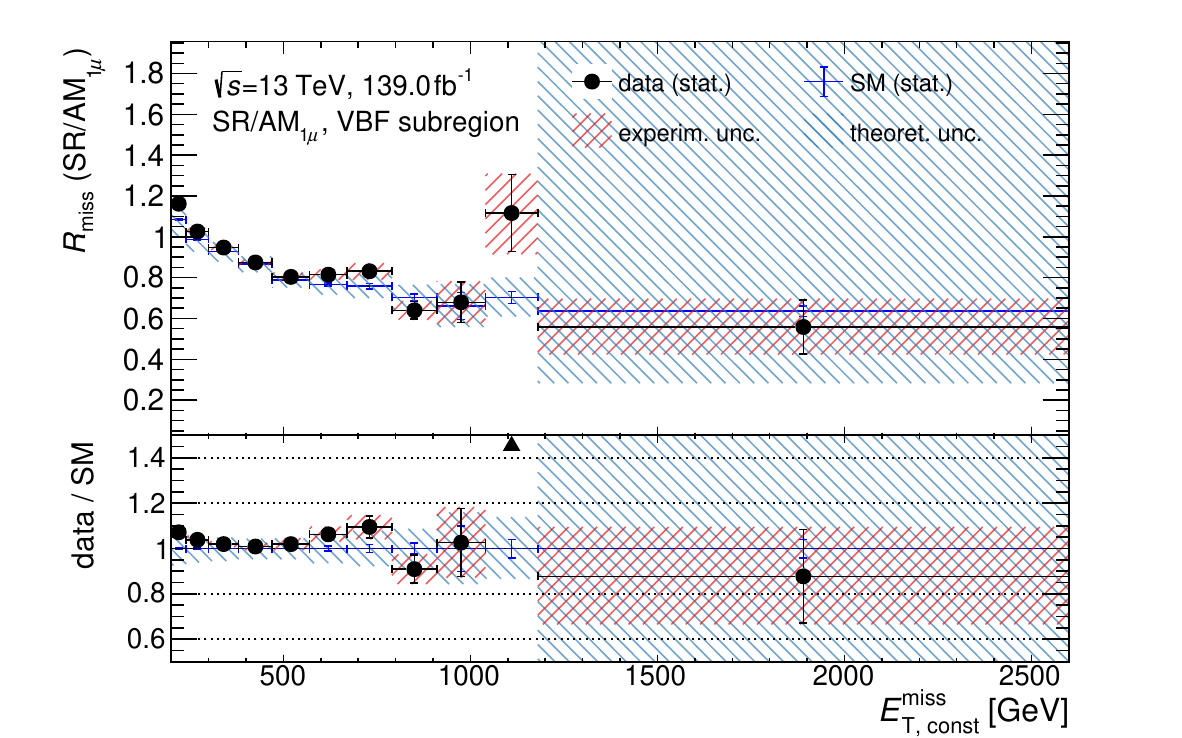}}\\
	\subfloat[]{\includegraphics[width=0.49\textwidth]{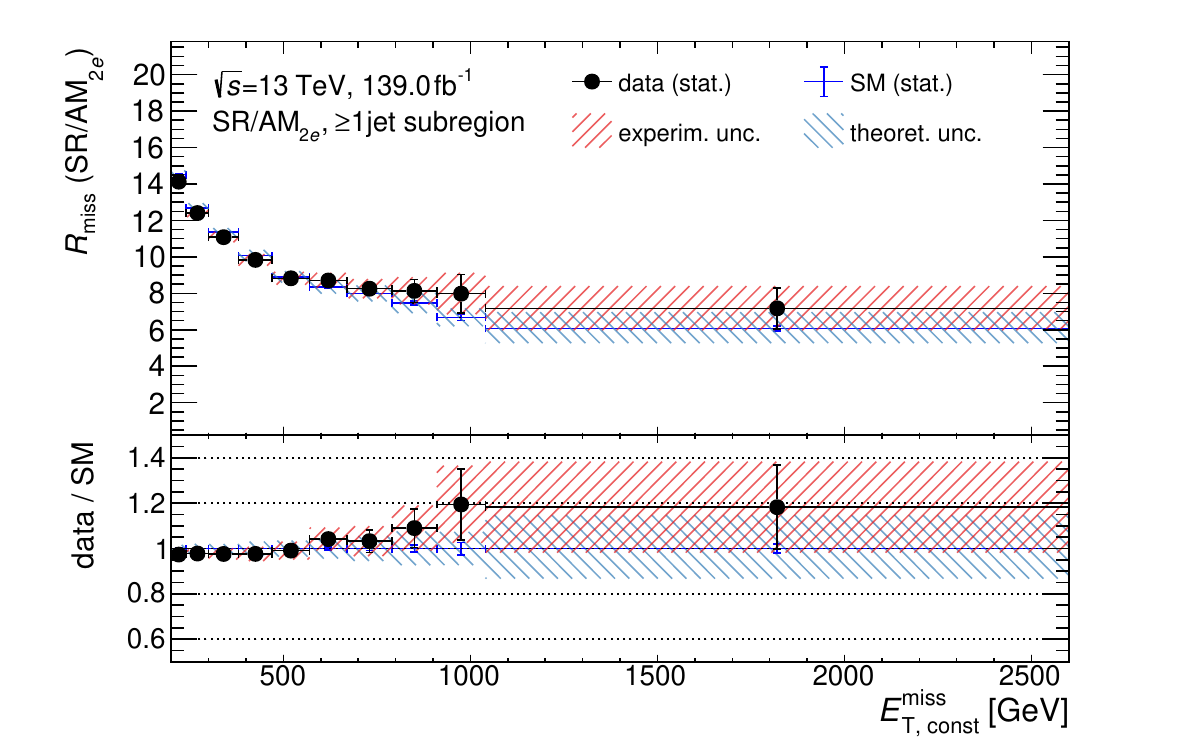}}
	\subfloat[]{\includegraphics[width=0.49\textwidth]{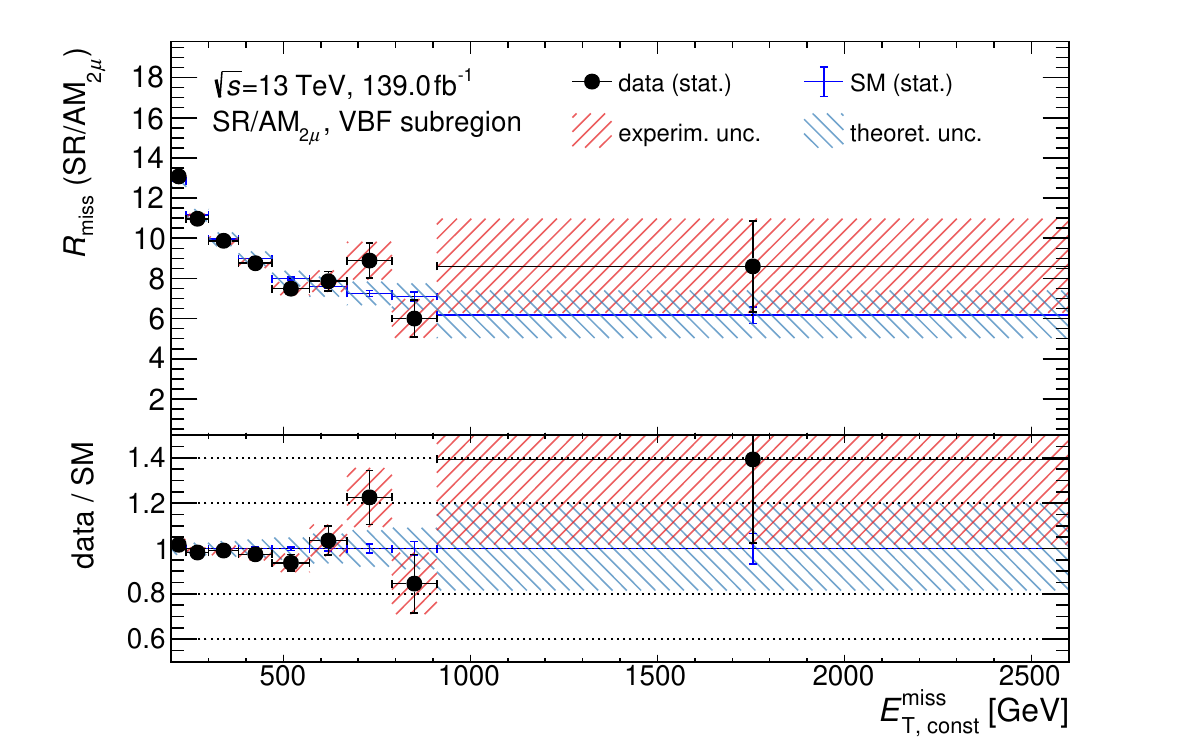}}\\
\end{myfigure}

It can be observed that the signal region has a larger cross section than the auxiliary measurements requiring an electron or two leptons, \ie $\Rmiss>1$.
\Rmiss is larger than one for \TwoLJetsAMs because of the branching fraction of $Z$ boson decays into two specific charged leptons (\SI{3.4}{\%}) and two neutrinos (\SI{20.0}{\%})~\cite{Zyla:2020zbs}.
In \OneLJetsAMs, the dominant \SM process is \Wlnujets compared to \Znunujets in the signal region.
Here, \Rmiss can be close to 1.
\Rmiss is larger than 1 in \OneEJetsAM because of the more stringent phase-space selection in \OneEJetsAM (\cf\secref{sec:metJets_regions}).
In all regions, \Rmiss decreases with increasing \METconst, indicating a more steeply falling spectrum in the signal region.

The ratio between data and generated \SM prediction (bottom panels) does not depend significantly on \METconst.
For \TwoLJetsAMs it is compatible with unity, while for \OneLJetsAMs an offset of approximately \SI{10}{\%} can be observed.
This is a consequence of the changing process composition in the Standard Model.
\chapter{Additional material on the \METjets interpretation}

\section{Derivation of the likelihood function}
\label{app:interpretation_likelihood}

The probability density function (\pdf) of a random variable $x$ in a counting experiment given the expectation $\mu$ is given by the Poisson distribution
\begin{equation*}
	\mathcal{L}\left(x|\mu\right)=\frac{\mu^xe^{-\mu}}{x!}.
\end{equation*}
Its \pdf is called \textit{likelihood} $\mathcal{L}$ and $x$ ($\mu$) are the observed (predicted) counts in a parameter interval.
In the large-sample limit, $\mu\to\infty$, it can be approximated by a normal distribution
\begin{equation*}
	\mathcal{L}\left(x|\mu\right)\xrightarrow[\mu\to\infty]{}\frac{1}{\sqrt{2\pi}~\sigma}\cdot e^{-\frac{1}{2}\left(\frac{x-\mu}{\sigma}\right)^2}
\end{equation*}
with standard deviation $\sigma=\sqrt{\mu}$.

The likelihood for $k$ independent measurements, \eg $k$ independent parameter intervals, is the product of all individual \pdf{}s:
\begin{align*}
	\mathcal{L}\left(\vec x|\vec\mu\right)
	&=\prod_i \frac{1}{\sqrt{2\pi}~\sigma_i}\cdot e^{-\frac{1}{2}\left(\frac{x_i-\mu_i}{\sigma_i}\right)^2}\\
	&=\frac{1}{\left(\sqrt{2\pi}\right)^k~\prod_i\sigma_i}\cdot e^{-\frac{1}{2}\sum_i\left(\frac{x_i-\mu_i}{\sigma_i}\right)^2}\\
	&=\frac{1}{\left(\sqrt{2\pi}\right)^k~\prod_i\sqrt{\sigma_i^2}}\cdot e^{-\frac{1}{2}\sum_i\left(\frac{x_i-\mu_i}{\sigma_i}\right)^2}\\
	&=\frac{1}{\left(\sqrt{2\pi}\right)^k~\sqrt{\prod_i\sigma_i^2}}\cdot e^{-\frac{1}{2}\sum_i\left(\frac{x_i-\mu_i}{\sigma_i}\right)^2}.
\end{align*}
The vectors of expected and observed counts in all considered measurements are denoted $\vv\mu$ and $\vv x$, respectively, with standard deviation $\sigma_i$ in measurement $i$.

Let $\sigma$ be the matrix of variances.
In the current approach, it is diagonal because all measurements are independent and correspondingly uncorrelated.
Then
\begin{equation*}
	\mathcal{L}\left(\vec x|\vec\mu\right)
	=\frac{1}{\left(\sqrt{2\pi}\right)^k~\sqrt{\det\sigma^2}}\cdot e^{-\frac{1}{2}(\vec x-\vec\mu)^\text{T}\left(\sigma^2\right)^{-1}(\vec x-\vec\mu)}.
\end{equation*}

Not all measurements, however, are uncorrelated in general.
In consequence, the basis is changed according to the transformation
\begin{alignat*}{2}
	\vec x &\rightarrow \vec x'&\coloneqq& U\vec x\\
	\vec{x}^\text{T} &\rightarrow \vec{x}^{\prime\text{T}}&=&\vec{x}^\text{T}U^\text{T}\\
	\vec\mu &\rightarrow \mu'&\coloneqq& U\vec\mu\\
	\vec\mu^\text{T} &\rightarrow \mu^{\prime\text{T}}&=&\vec\mu^\text{T}U^\text{T}\\
	\sigma^2 &\rightarrow \text{Cov}&\coloneqq& U\sigma^2U^\text{T}
\end{alignat*}
with transformation matrix $U$.
Hereby, $\sigma^2$ is transformed into the covariance matrix $\text{Cov}$, a matrix with off-diagonal elements which reflect both the degree of variance and the degree of correlation.
With this, the likelihood becomes
\begin{equation*}
	\mathcal{L}\left(\vec x'|\vec\mu'\right)
	=\frac{1}{\left(\sqrt{2\pi}\right)^k~\sqrt{\det\text{Cov}}}\cdot e^{-\frac{1}{2}(\vec x'-\vec\mu')^\text{T}\text{Cov}^{-1}(\vec x'-\vec\mu')}.
\end{equation*}
Theoretically speaking, this means a multidimensional normal distribution is modelled, which is stretched and rotated by a linear transformation.
The likelihood of the set of measurements is evaluated by undoing this transformation and comparing to normal distributions.

In a last step, the likelihood is modified by introducing systematic uncertainties as nuisance parameters $\vv\theta$.
They are constrained parameters that are not the primary target of the investigation.
They are expected to be distributed according to standard normal distributions with zero mean and unit standard deviation.
Multiplying the probabilities gives
\begin{equation*}
	\mathcal{L} \left(\vec{x}'~|~\vec\mu', \vec\theta\right)
	=
	\frac{1}{\sqrt{(2\pi)^k~\det\text{Cov}}}
	\cdot
	e^{-\frac{1}{2}\chiSqYields\left(\vv x', \vv\mu', \vv\theta\right)}
	\cdot
	\prod_i
	\frac{1}{\sqrt{2\pi}}e^{-\frac{1}{2}\theta_i^2},
\end{equation*}
where for simplification
\begin{equation*}
	\chiSqYields\left(\vv x', \vv\mu', \vv\theta\right)\coloneqq
	\left(\vv x' - \vv\mu'+\sum_i \theta_i \cdot \vv{\epsilon}_i \right)^\text{T}
	\Cov^{-1}
	\left(\vv x' - \vv \mu' + \sum_i \theta_i \cdot{\vv\epsilon}_i \right)
\end{equation*}
was defined.
$\vv\epsilon_i$ is the absolute uncertainty amplitude associated with a nuisance parameter~$\theta_i$.

\section{\pValue{}s in the interpretation with respect to the Standard Model}
\label{app:interpretation_pValue}

In \secref{sec:interpretation_SM}, the reduced~\chiSq is used as the criterion for the goodness of fit of the generated \SM prediction with respect to the measured data.
It is also possible to assess the goodness of fit using \pValue{}s.
This shall be performed in the following.

The \pValue for the \SM fit with an observed value of the test statistic \qSMobs is calculated as $p\left(\qSMobs,\pdf\left(\chi^2_{k-n}\right)\right)$, following the definition in \eqref{eq:interpretation_pValue}.
This calculation assumes that the test statistic~\qSM is distributed according to a $\chi^2_{k-n}$~distribution with $k-n$ degrees of freedom, where $k$ is the number of bins and $n$ the number of free-floating parameters in the prediction, respectively.
This \pValue uses the complete differential distributions in the calculation of \qSM as well as of the number of degrees of freedom.
It is comparable to the \pValue{}s used in other physics analyses only within certain limits because these often employ inclusive bins, \ie sum over many bins.
This summation, however, is not recommendable for the interpretation of \METjets measurement because the differential cross sections are steeply falling over six orders of magnitude and the results would therefore be completely dominated by the agreement at small \METconst in all regions.

The \pValue{}s obtained for the three different fit approaches discussed in \secref{sec:interpretation_SM} are given in \tabref{tab:interpretation_SM_fitResults_pVal}.
The nominal fit approach with a fixed-normalisation prediction and differential cross sections as input quantity that was discussed in \secref{sec:interpretation_SM_fixedNorm_diffXS} has a very small \pValue of \fixedDiffXSPVal.
This \pValue is dominated by the post-fit disagreement between measured data and prediction in the signal region and \Mono as well as \VBF subregions at $\SI{1040}{GeV}<\METconst<\SI{1180}{GeV}$ (\cf\subfigsref{fig:interp_SM_distributions_postfit_fixedNorm_diffXS}{a}{b}).

\begin{table}[b]
	\begin{adjustbox}{max width=\textwidth, center}
		\begin{tabular}{cccccccc}
			\toprule
			norm. of prediction & quantity & num. distributions & num. deg. of freedom & test statistic & \pValue\\
			&&& $k-n$ & \qSMobs\\
			\midrule
			fixed & \dSigmaDMET & \fixedDiffXSNDists & \fixedDiffXSNdF & \fixedDiffXSChiTwo & \fixedDiffXSPVal\\
			floating & \dSigmaDMET & \floatDiffXSNDists & \floatDiffXSNdF & \floatDiffXSChiTwo & \floatDiffXSPVal\\
			fixed & \Rmiss & \fixedRmissNDists & \fixedRmissNdF & \fixedRmissChiTwo & \fixedRmissPVal\\
			\bottomrule
		\end{tabular}
	\end{adjustbox}
	\caption{%
		Results for the \SM fit in the different approaches.
		The number of degrees of freedom corresponds to the difference between number of bins~$k$ and number of floating parameters~$n$.
	}
	\label{tab:interpretation_SM_fitResults_pVal}
\end{table}

In the fit approach using the floating-normalisation prediction with differential cross sections as input that was discussed in \secref{sec:interpretation_SM_floatNorm_diffXS}, the \pValue is \floatDiffXSPVal.
This is only marginally larger than before.
This insignificant improvement matches the similarities between the two fit approaches with respect to pulls and constraints of nuisance parameters as well as observed value of the test statistic and reduced~\chiSq.
This \pValue is also dominated by the post-fit disagreement between measured data and prediction in the signal region and \Mono as well as \VBF subregions at $\SI{1040}{GeV}<\METconst<\SI{1180}{GeV}$ (\cf\subfigref{fig:interp_SM_distributions_postfit_floatNorm_diffXS}{a}).

In the fit approach using the fixed-normalisation prediction with the \Rmiss distributions as input that was discussed in \secref{sec:interpretation_SM_fixedNorm_Rmiss}, the \pValue is \fixedRmissPVal.
This is significantly better than the \pValue{}s for the other two fit approaches.
This significant improvement matches the smaller reduced~\chiSq observed in \secref{sec:interpretation_SM_fixedNorm_Rmiss}.
The larger \pValue proves that in principle, measured data and generated \SM prediction are in acceptable agreement if common sources of systematic errors across the regions are cancelled.
The \pValue is dominated by the post-fit disagreement between measured data and prediction in the \Mono as well as \VBF subregions at $\SI{1040}{GeV}<\METconst<\SI{1180}{GeV}$  (see \eg\subfigref{fig:interp_SM_distributions_postfit_fixedNorm_Rmiss}{a}).
This is particularly the case in the \VBF subregion for $\Rmiss\left(\SR/\OneMuJetsAM\right)$ where an excess in the signal region meets a deficit in \OneMuJetsAM.

\section{Nuisance parameters in the interpretation with respect to the Standard Model}
\label{app:interpretation_SM_NPrankings}

The \SM fits discussed in \secref{sec:interpretation_SM} give very complex results.
More studies are conducted here, in particular with respect to the nuisance parameters, to improve the understanding of the fit results.

\subsection{Ensemble of pulls and constraints}
\label{interpretation_SM_ensemble}

Further quantities are defined here in addition to those introduced in \secref{sec:interpretation_pullsAndConstraints}.
Pre-fit, the \textit{relative deviation} $d_{j, \vv y}^i$ for a nuisance parameter $i$ changing the yield $y_j\in\left\{x_j, \pi_j\right\}$ in bin $j$ is
\begin{equation*}
	d_{j, \vv y}^i = \frac{\theta^i_0\cdot\varepsilon^i_j}{y_j}
	=0.
\end{equation*}

The total relative deviation $d_{j, \vv y}$ of the yield $y_j$ in a bin $j$ from all nuisance parameters associated to $\vv y$ is
\begin{equation}
	d_{j, \vv y} = \sum_{i}\frac{\theta^i_0\cdot\varepsilon^i_j}{y_j}
	=0.
\end{equation}
and post-fit it becomes
\begin{equation}
	\label{eq:interpretation_relDev}
	\hat d_{j, \vv y} = \sum_i\frac{\hat\theta^i\cdot\varepsilon^i_j}{y_j}.
\end{equation}

The pre-fit total relative uncertainty $u_{j, \vv y}$ of the yield $y_j$ in bin $j$ for uncorrelated uncertainties is obtained by adding the individual uncertainties in quadrature:
\begin{equation*}
	u_{j, \vv y}
	= \sum_i\left(\frac{\sigma_{\theta^i_0}\cdot\varepsilon^i_j}{y_j}\right)^2
	= \frac{\left(\sum_i\varepsilon^i_j\right)^2}{y_j}.
\end{equation*}
Post-fit it becomes
\begin{equation}
	\label{eq:interpretation_relUnc}
	\hat u_{j, \vv y} = \frac{\left(\sum_i\sigma_{\hat\theta^i}\cdot\varepsilon^i_j\right)^2}{y_j}.
\end{equation}

\bigskip
In \figref{fig:interp_SM_NPsummary_pulls} (\figref{fig:interp_SM_NPsummary_constraints}), the ensembles of pulls (constraints) as a whole for the three fit setups discussed in \secref{sec:interpretation_SM} are shown.

\begin{myfigure}{
		Post-fit (left) pull and (right) relative deviation per bin $j$ following \eqref{eq:interpretation_relDev} for the systematic uncertainties for the three different approaches.
		Blue (red) lines mark the post-fit quantity related to nuisance parameters of theoretical (experimental) uncertainties.
		The median is indicated by a black line.
		A green (yellow) line marks pulls of one (two) standard deviation(s) in the left figure.
	}{fig:interp_SM_NPsummary_pulls}
	\includegraphics[width=0.48\textwidth,valign=t]{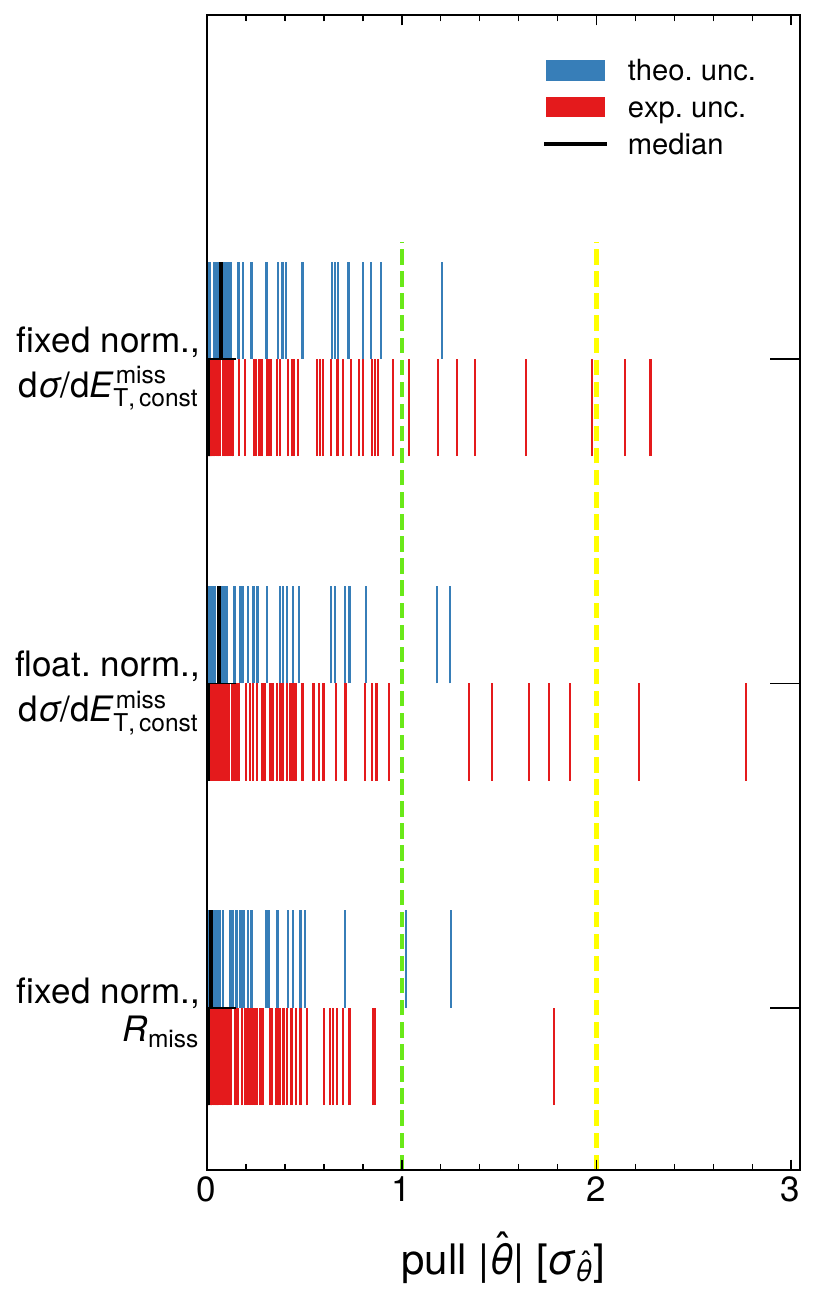}
	\hspace{10pt}
	\includegraphics[width=0.48\textwidth,valign=t]{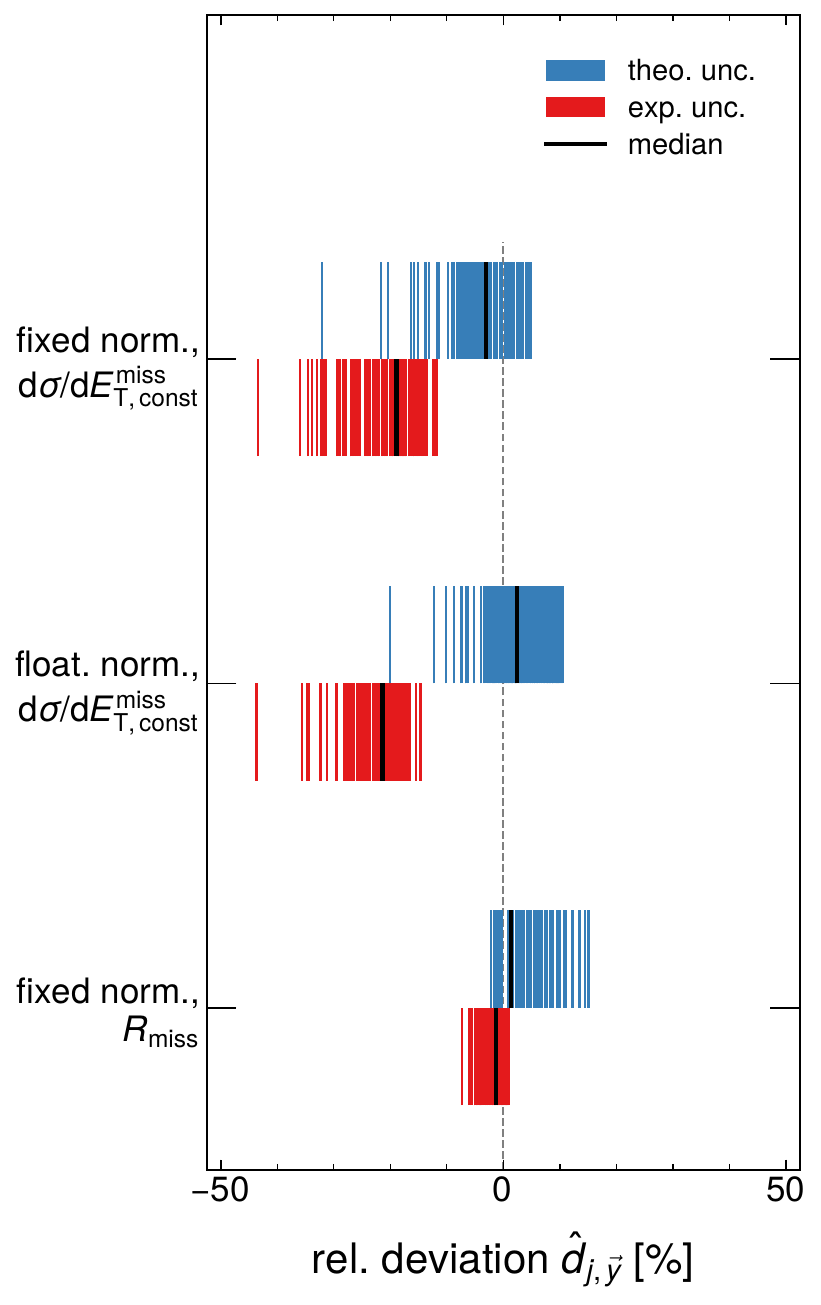}
\end{myfigure}

\subsubsection{Nominal \SM fit}

First, the ensemble of pulls and constraints when using the differential cross sections as input quantity and the fixed-normalisation prediction is investigated.
This corresponds to the fit discussed in \secref{sec:interpretation_SM_fixedNorm_diffXS}.

\bigskip
\subfigref{fig:interp_SM_NPsummary_pulls}{a} gives the post-fit pull $\hat{\theta}$ for each nuisance parameter.
The pre-fit pull $\theta_0$ of all nuisance parameters is 0.
The sign of a pull depends on the definition of "up" and "down" variation for a systematic uncertainty.
That is why only the absolute values of the pulls are investigated here.
Each nuisance parameter for an experimental (theoretical) uncertainty is marked by a red (blue) line.
The median of the pulls in either category is indicated by a black line.

In general, most nuisance parameters are pulled less than one standard deviation.
There is one nuisance parameter for a theoretical uncertainty, however, -- the difference between diagram subtraction and diagram removal scheme for the interference between \ttbar and $tW$ processes (\cf\secref{sec:metJets_theoSystUnc}) -- that is pulled by almost 1.3 standard deviations.

There are eight nuisance parameters for experimental uncertainties which are pulled more than one standard deviation.
Two of them -- the nuisance parameters for the uncertainty on the luminosity estimate and the uncertainty on the jet energy scale~-- are even pulled more than two standard deviations.
This is because they are large uncertainties present in all regions.
Therefore, pulling their nuisance parameters can reduce the discrepancy in normalisation between measured data and generated \SM prediction in all regions simultaneously.
Nonetheless, most nuisance parameters for experimental uncertainties are pulled very little and the median is less than the one for theoretical uncertainties.

\bigskip
\subfigref{fig:interp_SM_NPsummary_pulls}{b} gives the post-fit relative deviation per bin according to \eqref{eq:interpretation_relDev}.
The total relative deviation from experimental (theoretical) uncertainties for each bin is\linebreak marked by a red (blue) line.
The median over the bins in either category is indicated by a black line.

The nuisance parameters for the theoretical uncertainties change the yield by \SI{-30}{\%} to $+\SI{5}{\%}$ in the different bins.
The median of the deviations, however, is close to 0.
The nuisance parameters for the experimental uncertainties change the yield by \SI{-40}{\%} to \SI{-10}{\%} in the different bins, with a median of \SI{-20}{\%}.
This amends the pre-fit underestimation of the measured data by the \SM prediction.

\begin{myfigure}{
		Post-fit (left) constraint for the systematic uncertainties and (right) relative uncertainty per bin $j$ following \eqref{eq:interpretation_relUnc} for the systematic uncertainties before and after the fit for the three different approaches.
		Blue (red) lines mark the post-fit quantity related to nuisance parameters of theoretical (experimental) uncertainties.
		Light-blue (light-red) lines additionally mark the pre-fit quantity related to nuisance parameters of theoretical (experimental) uncertainties in the right figure.
		The median is indicated by a black line.
		A green line marks constraints of 1 in the left figure.
	}{fig:interp_SM_NPsummary_constraints}
	\includegraphics[width=0.48\textwidth,valign=t]{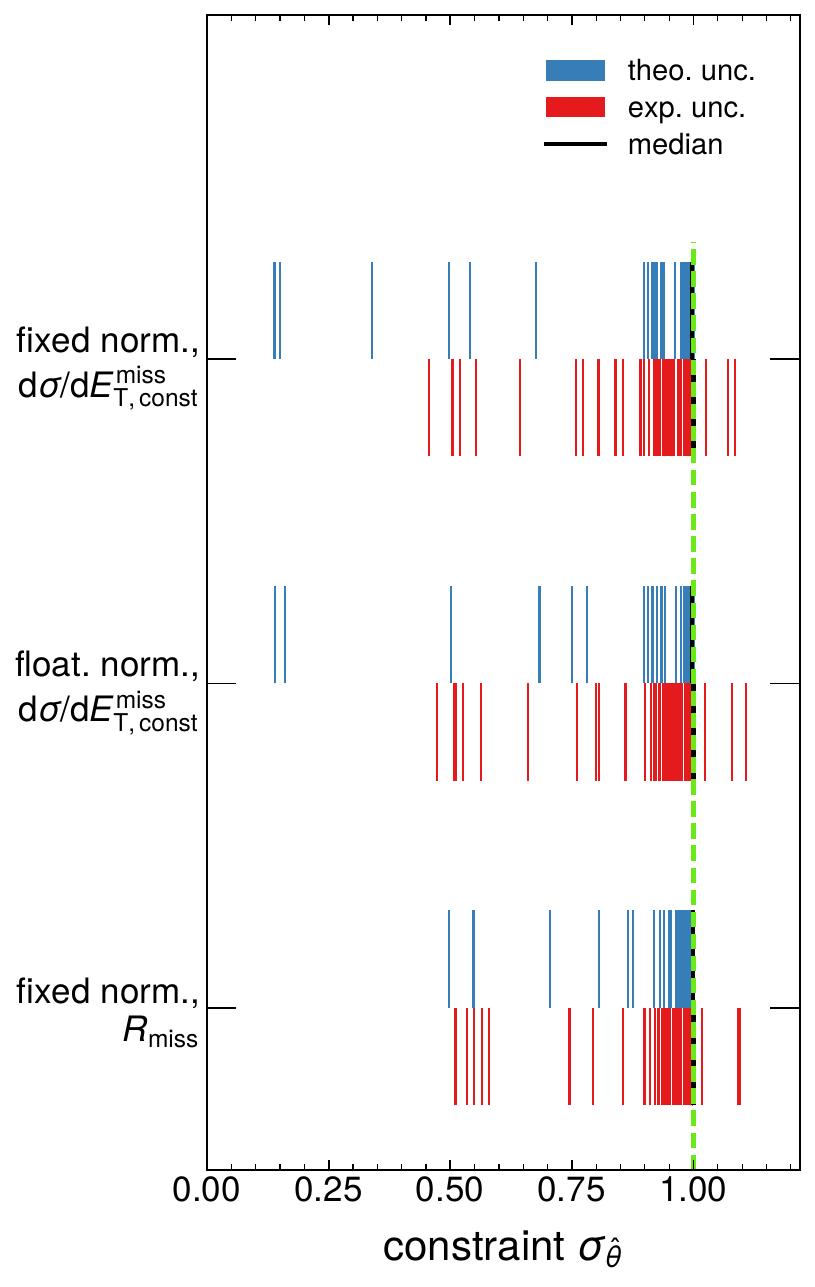}
	\hspace{10pt}
	\includegraphics[width=0.48\textwidth,valign=t]{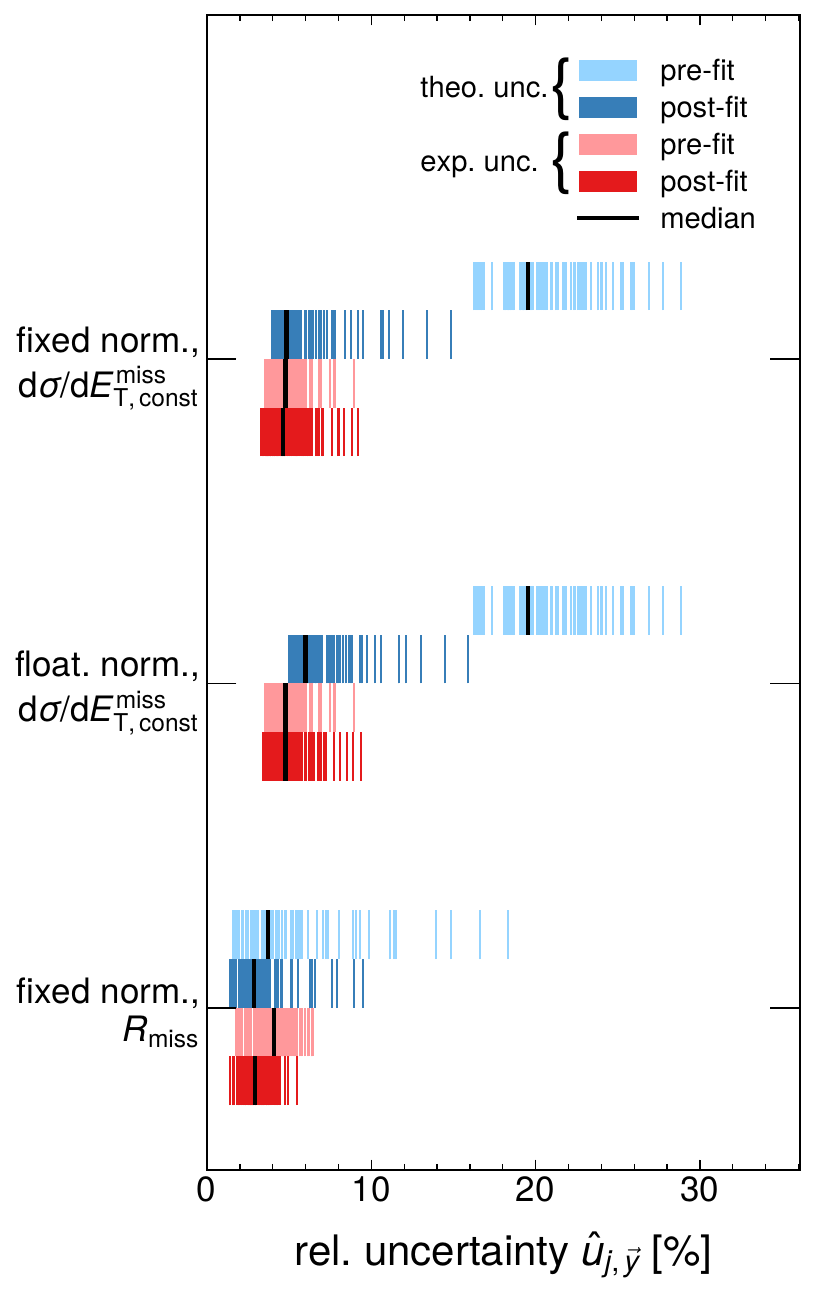}
\end{myfigure}

\bigskip
\subfigref{fig:interp_SM_NPsummary_constraints}{a} gives the post-fit constraint $\sigma_{\hat\theta}$ for each nuisance parameter.
The pre-fit constraint of all nuisance parameters is 1.
Each nuisance parameter for an experimental (theoretical) uncertainty is marked by a red (blue) line.
The median of the constraints in either category is indicated by a black line.

Theoretical uncertainties are very large pre-fit.
Post-fit, they are not allowed to vary the \SM prediction as much for a good agreement between data and prediction.
In consequence, the post-fit constraints for nuisance parameters of all theoretical uncertainties are smaller than 1.
Some nuisance parameters even have a post-fit constraint as small as 0.1.
They stem from nuisance parameters of large uncertainties related to the \Vjets normalisation as well as shape.
These are not allowed to change the normalisation nearly as much after the fit as before.
The median constraint for nuisance parameters of theoretical uncertainties, however, is close to 1.

Experimental uncertainties in general are smaller than theoretical uncertainties.\linebreak
Therefore, the post-fit constraints for nuisance parameters of experimental uncertainties vary between values smaller and greater than 1.
Similar to the nuisance parameters for theoretical uncertainties, the median constraint is close to 1.

\bigskip
\subfigref{fig:interp_SM_NPsummary_constraints}{b} gives the post-fit relative uncertainty per bin according to \eqref{eq:interpretation_relUnc}.
The total relative uncertainty from experimental (theoretical) uncertainties for each bin post-fit is marked by a red (blue) line.
A lighter shading marks the pre-fit values.
The median over the bins in either category is indicated by a black line.

The total theoretical uncertainty in each bin is smaller after the fit than before the fit:
pre-fit the total relative uncertainty is \SI{15}{\%} to \SI{30}{\%}, post-fit it is drastically reduced to \SI{4}{\%} to {15}{\%}.
The total experimental uncertainty is in general smaller after the fit than before the fit.
This is apparent for example from the smaller post-fit median uncertainty.
In a few bins, the total experimental uncertainty, however, becomes larger because of nuisance parameters with post-fit constraints larger than one.

The total experimental uncertainties are pre-fit considerably smaller than the total theoretical uncertainties.
Post-fit, they still have a smaller median but can be larger in individual cases.

\subsubsection{Floating instead of fixed normalisation}

The ensemble of pulls and constraints when using the differential cross sections as input quantity and the floating-normalisation prediction is investigated next.
This corresponds to the fit discussed in \secref{sec:interpretation_SM_floatNorm_diffXS}.

The ensemble of pulls and constraints are shown in the second rows of \figsref{fig:interp_SM_NPsummary_pulls}{fig:interp_SM_NPsummary_constraints}, respectively.
For the pulls, constraints and relative uncertainty per bin, no large difference to the fixed-normalisation approach before can be seen.
For the relative deviations per bin, the relative deviation for nuisance parameters of theoretical uncertainties is larger in all bins.
In the median, the relative deviation now increases the \SM prediction a little.
This is related to the normalisation parameters $\hatmuVjets$ and $\hatmuTop$, which both take values smaller than 1 and therefore give the nuisance parameters for theoretical uncertainties more freedom to increase the predicted yield (see also \secref{app:interpretation_SM_floatNormStudies}).
The median deviation for nuisance parameters of experimental uncertainties is marginally smaller than in the fixed-normalisation approach before.

\subsubsection{\RmissTitle distributions instead of differential cross sections}

Lastly, the ensemble of pulls and constraints when using the \Rmiss distributions as input quantity and the fixed-normalisation prediction is investigated.
This corresponds to the fit discussed in \secref{sec:interpretation_SM_fixedNorm_Rmiss}.
The ensemble of pulls and constraints are shown in the third rows of \figsref{fig:interp_SM_NPsummary_pulls}{fig:interp_SM_NPsummary_constraints}, respectively.

In \subfigref{fig:interp_SM_NPsummary_pulls}{a}, the pulls for nuisance parameters of theoretical uncertainties are in general reduced, and the median pull is close to 0.
This is because the pre-fit discrepancy between measured data and generated \SM prediction is smaller.
The largest pulls for nuisance parameters of theoretical uncertainties stay similar because also the size of most uncertainties is reduced.
For pulls of nuisance parameters of experimental uncertainties the picture is similar.
Here, only the uncertainty related to the fake estimate of \METmeas (\cf\secref{sec:metJets_expSystUnc}) is pulled by more than one standard deviation.
The pull for the nuisance parameter of this uncertainty is large because this uncertainty impacts the signal region and therefore all \Rmiss distributions.
A pull on the nuisance parameter can therefore correct discrepancies between data and prediction in all regions simultaneously.

In \subfigref{fig:interp_SM_NPsummary_pulls}{b}, the relative deviation per bin from pulling the nuisance parameters is significantly reduced compared to the approaches using the differential cross sections.
The reason for this is the reduced pre-fit discrepancy between data and prediction.
The deviations for nuisance parameters of both theoretical and experimental uncertainties are very close to 0 in the median.

In \subfigref{fig:interp_SM_NPsummary_constraints}{a}, the constraints when using the \Rmiss distributions are similar to the other approaches.
The median constraint is very close to 1.
The most extreme constraints for nuisance parameters of the theoretical uncertainties are closer to 1 than for the approaches using differential cross sections.
They were large uncertainties related to the \Vjets normalisation as well as shape.
These uncertainties are consistent across the regions and therefore largely cancelled in \Rmiss.
Due to their reduced pre-fit sizes, their post-fit constraints are smaller.

In \subfigref{fig:interp_SM_NPsummary_constraints}{b}, the pre-fit relative uncertainties per bin are reduced significantly for theoretical uncertainties and mildly for experimental uncertainties when using the \Rmiss distributions.
The median size of experimental and theoretical uncertainties is comparable, with a longer tail towards large values for theoretical uncertainties.
After the fit, the relative uncertainties per bin are reduced further but the relation between theoretical and experimental uncertainties remains similar.

\subsection{Nuisance parameters for the floating-normalisation prediction}
\label{app:interpretation_SM_floatNormStudies}

In this section, further studies regarding the nuisance parameters when using the floating-normalisation prediction and differential cross sections as input quantity (\cf~Sec\-tion~\ref{sec:interpretation_SM_floatNorm_diffXS}) are performed.

First, the change of the normalisation when pulling nuisance parameters is demonstrated in more detail.
Then an alternative approach for defining systematic uncertainties that results in more intuitive post-fit values of the normalisation parameters is studied.

\subsubsection{Illustration of normalisation change by pulling nuisance parameters}
\label{app:interpretation_SM_NP_systExcl}

\newcommand{\floatDiffXSSystExclNDists}{two\xspace}
\newcommand{\floatDiffXSSystExclChiTwo}{31.9\xspace}
\newcommand{\floatDiffXSSystExclNbins}{24\xspace}
\newcommand{\floatDiffXSSystExclNdF}{22\xspace}
\newcommand{\floatDiffXSSystExclPVal}{\ensuremath{7.9\cdot10^{-2}}\xspace}

\newcommand{\muVjetsNoRSSystExcl}{\ensuremath{\hatmuVjets=0.94\pm0.10}\xspace}
\newcommand{\muTopNoRSSystExcl}{\ensuremath{\hatmuTop=0.77\pm0.26}\xspace}

In \secref{sec:interpretation_SM_floatNorm_diffXS}, it was mentioned that pulling a nuisance parameter in the interpretation of the \METjets measurement changes the normalisation of the corresponding distribution.
This is apparent when looking at the data distributions in \figref{fig:interp_SM_distributions_postfit_floatNorm_diffXS} where the yield in all bins is lowered simultaneously as a result of pulls on many nuisance parameters for experimental uncertainties.
As a counter-intuitive consequence, both normalisation parameters therefore post-fit take values smaller than 1 when using the floating-normalisation \SM prediction although the generated \SM prediction underestimates the measured data pre-fit.

The normalisation change by pulling nuisance parameters can be illustrated further when performing a dedicated fit which gives similar post-fit values for the normalisation parameters but in which only few nuisance parameters are used.
This is done in the following.

\bigskip
The differential cross sections in the signal region and \OneEJetsAM in the \Mono subregion binned in \METconst serve as the investigated quantity in data $\vv x$ and prediction~\vvpiSM in the goodness-of-fit test.
The floating-normalisation \SM prediction with one normalisation parameter for \Vjets contributions, \muVjets, and one for top contributions, \muTop, is used.

In total, \floatDiffXSSystExclNDists distributions are included in the fit, amounting to \floatDiffXSSystExclNbins bins.
Only two nuisance parameters for the theoretical uncertainties and none for the experimental uncertainties are taken into account.
The remaining 73 theoretical and 254 experimental uncertainties are not assigned nuisance parameters and only contribute to the covariance matrix in \eqref{eq:metJets_likelihood}.

After the fit, the normalisation parameters take values of \muVjetsNoRSSystExcl and \muTopNoRSSystExcl.
This is compatible with the results of the full fit in \secref{sec:interpretation_SM_floatNorm_diffXS}.

The normalisation change by the systematic uncertainties can be observed in \figref{fig:interp_SM_distributions_postfit_floatNormSystExcl}.
The top panel shows the differential cross sections, the bottom panel the ratio to the pre-fit \SM prediction.
The pre-fit allowed range for the measured data (generated \SM values) within one standard deviation is shown as a red (blue) shaded band.
The post-fit values with their total uncertainties for the measured data are marked by black dots.
They are identical to the pre-fit values as no nuisance parameters for experimental systematic uncertainties are considered.

\begin{myfigure}{
	\postfitDistCaption{Differential cross section}{%
			(a) the signal region in the \Mono and (b) \OneEJetsAM in the \VBF subregion as a function of \METconst%
		}{floating}{}{%
			A simplified fit setup is used, see text for details.
		}{
			Dashed blue crosses mark the post-fit values in \SM generation with their total uncertainty when neglecting the pulls of nuisance parameters.
		}
	}{fig:interp_SM_distributions_postfit_floatNormSystExcl}
		\subfloat[]{\includegraphics[width=0.48\textwidth]{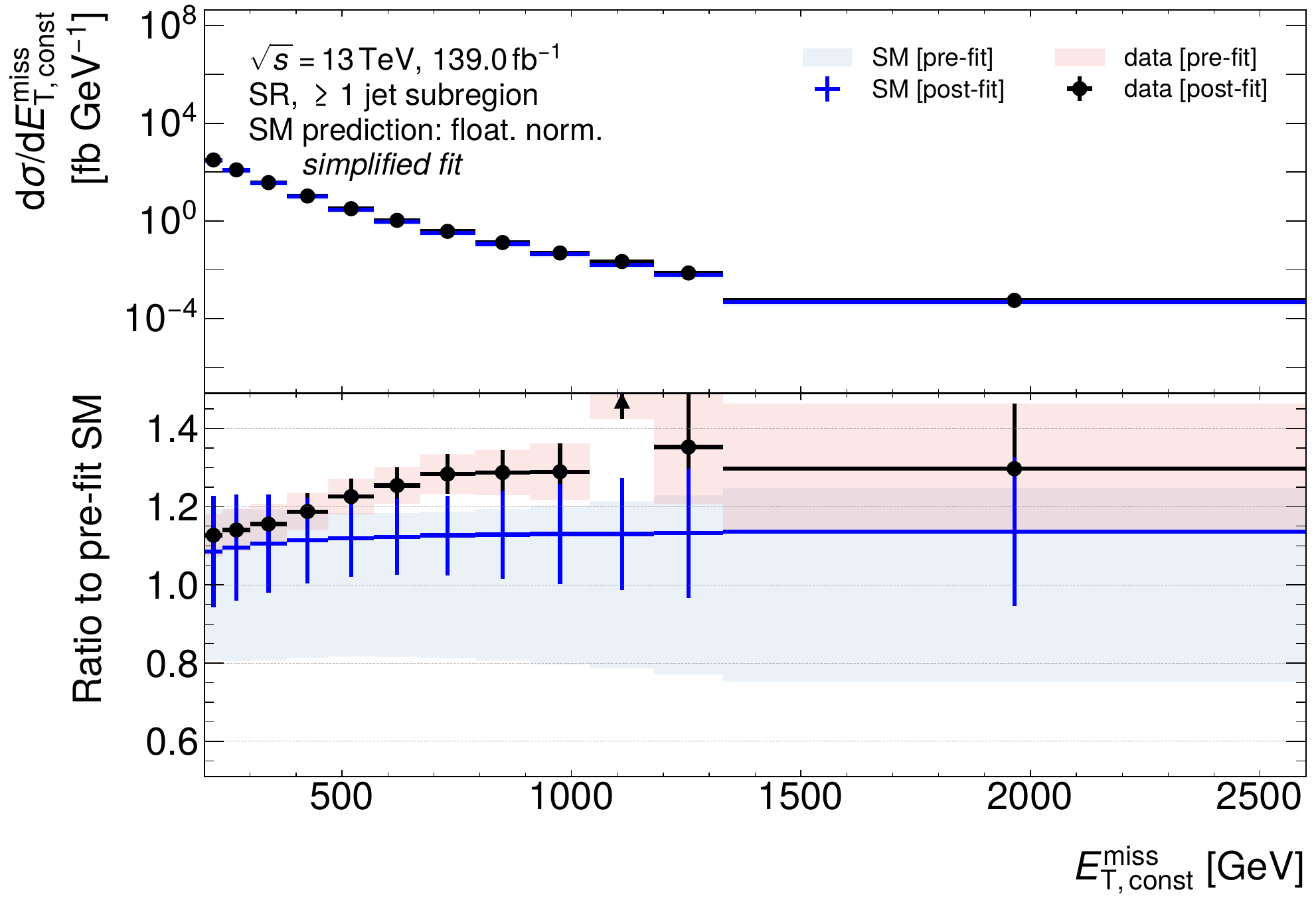}}
		\hspace{10pt}
		\subfloat[]{\includegraphics[width=0.48\textwidth]{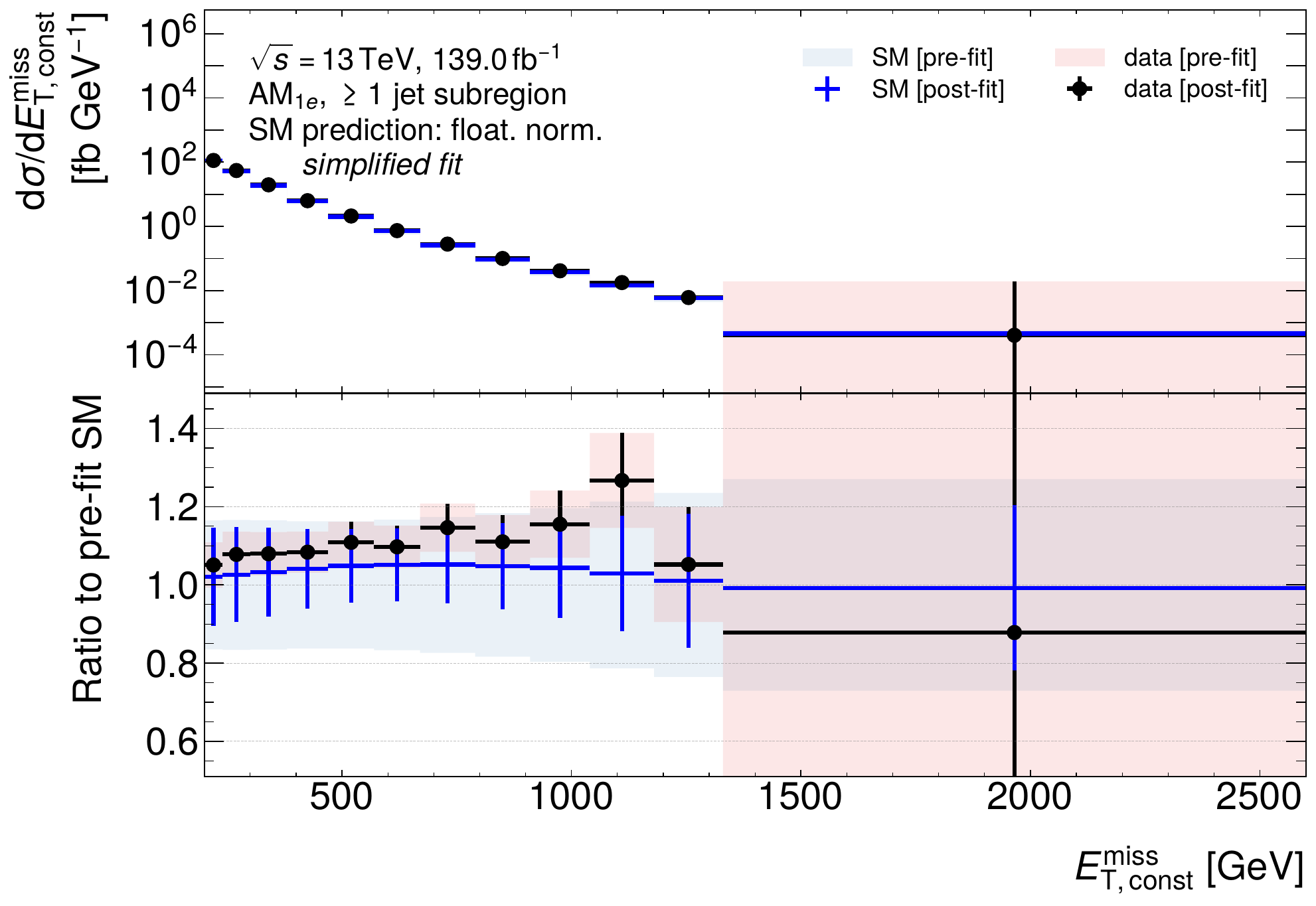}}\\
\end{myfigure}

The post-fit values with their total uncertainties for the generated \SM prediction when using the two normalisation parameters \muVjets as well as \muTop and nuisance parameters for the two largest theoretical uncertainties are marked by solid blue crosses.
The two nuisance parameters taken into account relate to the \Vjets normalisation and the difference between diagram-subtraction and -removal scheme for top processes (\cf\secref{sec:metJets_theoSystUnc}).

\figref{fig:interp_SM_distributions_postfit_floatNormSystExcl} additionally shows the post-fit values for the generated \SM prediction as dashed blue crosses for the same fit when applying only the post-fit normalisation parameters but neglecting the pulls of the nuisance parameters.

\bigskip
Pre-fit, the previously observed discrepancy between measured data and generated \SM prediction is visible, among others about \SI{12}{\%} in normalisation in the signal region (\subfigref{fig:interp_SM_distributions_postfit_floatNormSystExcl}{a}).
Post-fit when including no pulls (dashed blue crosses for the \SM prediction), the total normalisation of the \SM prediction is reduced by about \SI{6}{\%} in the signal region according to the post-fit normalisation parameters for \Vjets and top processes, which both take values smaller than one.
The ratio is not perfectly constant as a function of \METconst because the relative contribution of the processes to the total yield changes as a function of \METconst.

Post-fit already when including only two nuisance parameters (solid blue crosses for the \SM prediction), the normalisation of the \SM prediction is increased by about \SI{15}{\%} compared to the case without nuisance parameters.
This reduces the normalisation difference between measured data and generated \SM prediction post-fit to less than \SI{5}{\%}.

\subsubsection{Alternative approach: normalised systematic uncertainties}
\label{app:interpretation_SM_NP_normSysts}

The fit in \secref{sec:interpretation_SM_floatNorm_diffXS} results in post-fit values smaller than 1 for both normalisation parameters because pulling nuisance parameters changes the normalisation of the distributions, as discussed before.
More intuitive post-fit values for the normalisation parameters can be obtained when systematic uncertainties are redefined such that their nuisance parameters do not change the normalisation of distributions.
This is performed in the following.

In general, the same fit approach as described in \secref{sec:interpretation_SM_floatNorm_diffXS} is used.
The systematic uncertainties, however, are defined differently:
\begin{itemize}
	\item The change in normalisation that is caused by theoretical uncertainties corresponding to one of the normalisation factors, \ie\Vjets uncertainties for \muVjets and top uncertainties for \muTop, is in principle already accounted for by the normalisation factors themselves.
	These uncertainties are therefore renormalised such that pulling their nuisance parameter by one standard deviation causes the same difference in shape of the prediction, but the normalisation of the prediction is unaltered.
	The size of the uncertainties is in general reduced by this renormalisation.
	
	\item The experimental uncertainties are likewise estimated in simulation in the\linebreak \METjets measurement.
	The change in normalisation of the corresponding \SM contribution that is caused by experimental uncertainties is therefore in principle also already accounted for by the corresponding normalisation factor.
	These uncertainties are consequently renormalised identically to the theoretical uncertainties mentioned above.
	This neglects that the change in normalisation of the \SM contributions that are not normalised, \eg diboson processes, caused by the experimental uncertainties is not accounted for by the normalisation factors.
	This effect is small, however, as normalisation factors are used for the dominant \SM contributions.
	
	\item The uncertainty on the luminosity estimate causes a change in normalisation of the measured data that is constant as a function of \METconst.
	This is mostly redundant with the introduced normalisation parameters.
	The uncertainty on the luminosity estimate is therefore neglected completely in this study.
\end{itemize}

With this renormalisation approach, the post-fit normalisation parameters become \muVjetsRSmatchExp and \muTopRSmatchExp.
The overall normalisation of the \SM prediction is increased as expected as the normalisation parameter for the dominant \SM contribution in all regions, \hatmuVjets, is larger than 1.
The normalisation of top processes is decreased by $\hatmuTop<1$ because the discrepancy between the normalisation of data and \SM prediction is smaller in the regions that have the largest top contributions, \OneLJetsAMs (\cf\figref{fig:detCorr_partLevelResults}).
Both normalisation factors are considerably more constrained because most nuisance parameters for the systematic uncertainties cannot change the normalisation of the \SM prediction any more.

\bigskip
The differential cross sections pre-fit and post-fit are shown for two examples in \figref{fig:interp_SM_distributions_postfit_floatNorm_RSmatch}.
The bottom panels show the ratio to the pre-fit \SM prediction.

\begin{myfigure}{
		\postfitDistCaption{Differential cross section}{%
			(a) the signal region in the \Mono and (b) \OneEJetsAM in the \VBF subregion as a function of \METconst
		}{floating}{}{%
			A specialised fit setup with renormalised systematic uncertainties is used, see text for details.
		}{}
	}{fig:interp_SM_distributions_postfit_floatNorm_RSmatch}
	\subfloat[]{\includegraphics[width=0.48\textwidth]{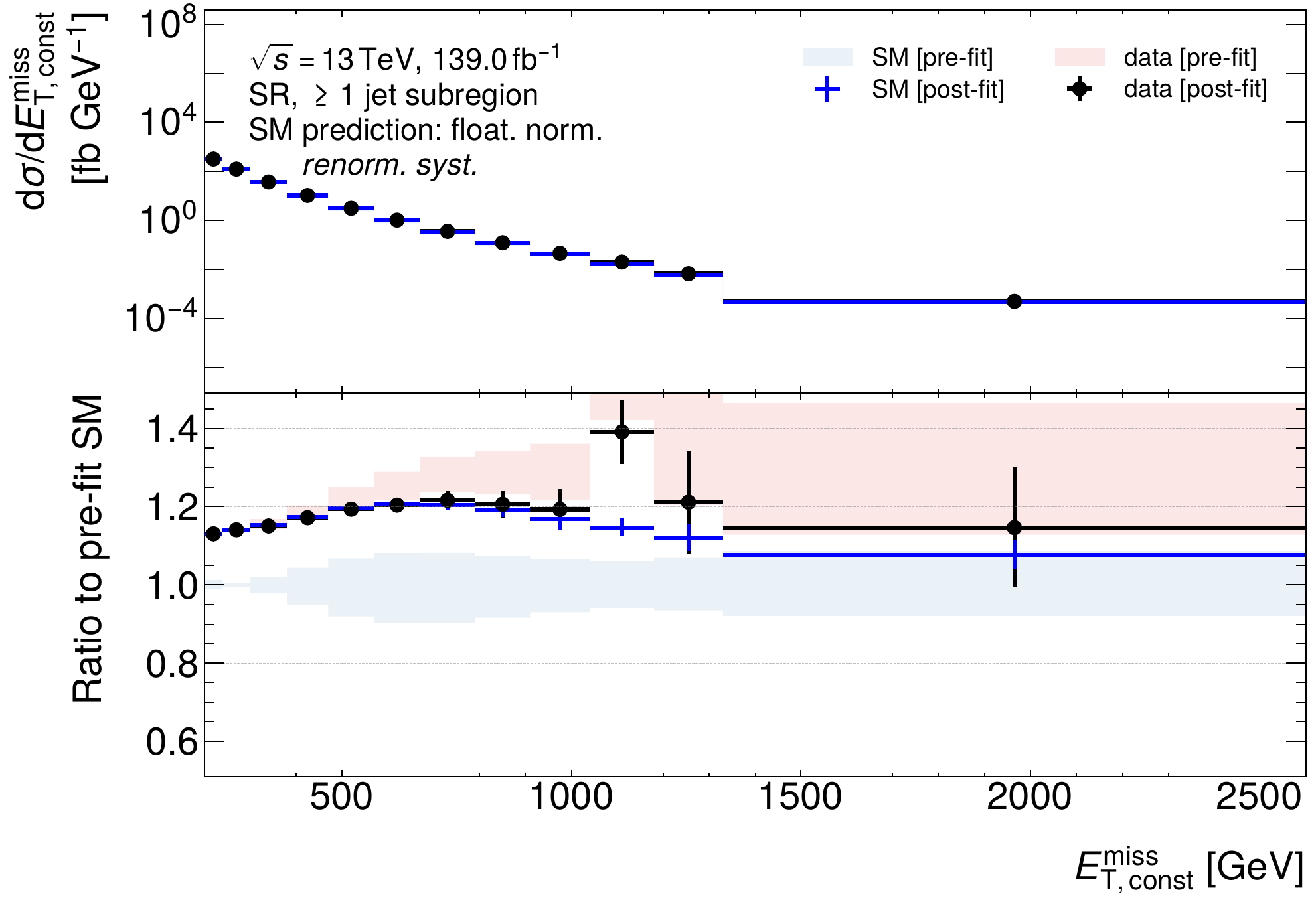}}
	\hspace{10pt}
	\subfloat[]{\includegraphics[width=0.48\textwidth]{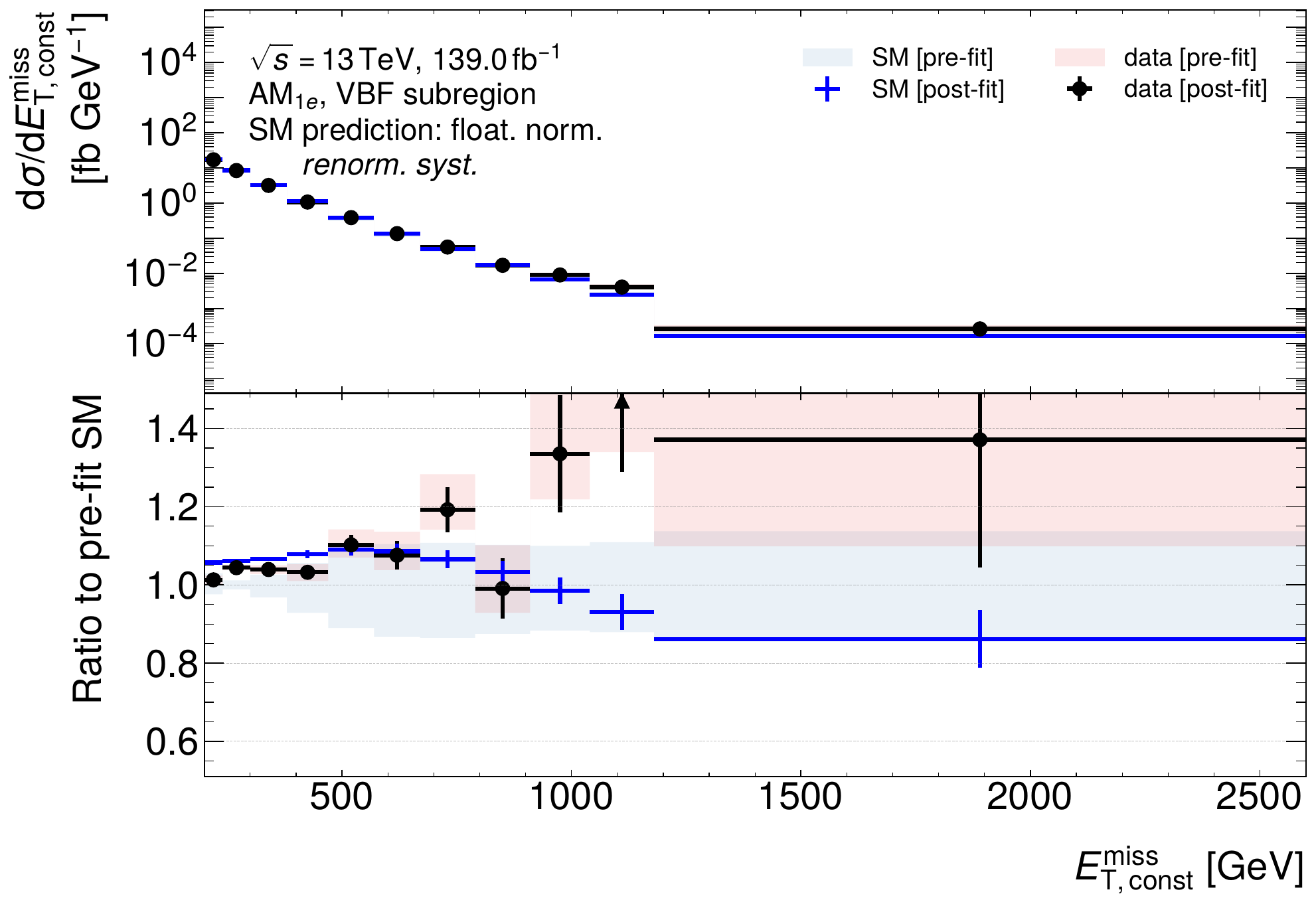}}\\
\end{myfigure}

The disagreement between measured data and generated \SM prediction is reduced considerably in the fit.
In particular in the signal region in the \Mono subregion (\subfigref{fig:interp_SM_distributions_postfit_floatNorm_RSmatch}{a}), the agreement becomes excellent apart from $\METconst\approx\SI{1100}{GeV}$.
The agreement is worse in \OneEJetsAM in the \VBF subregion (\subfigref{fig:interp_SM_distributions_postfit_floatNorm_RSmatch}{b}) because both this region and subregion have lower statistics and consequently larger statistical uncertainties than their counterparts, reducing the weight in the fit.
The systematic uncertainties pre-fit and post-fit are significantly reduced by the renormalisation procedure compared to the original approach (\cf\figref{fig:interp_SM_distributions_postfit_floatNorm_diffXS}), particularly at small \METconst.

\bigskip
In \figref{fig:interp_SM_NPrankings_renormSyst}, it can be seen how this approach changes the post-fit pulls~$\hat\theta$ and constraints~$\sigma_{\hat\theta}$ for the nuisance parameters compared to the original approach in \figref{fig:interp_SM_NPrankings_floatNorm_diffXS}.
Given are also the relative uncertainties $u_{j, \vv y}$ of the yield $\vv y$ corresponding to a systematic uncertainty when varying the respective nuisance parameter by one standard deviation down (turquoise) or up (blue).
Shown is the median of the relative uncertainties in all bins $j$.
The relative uncertainties pre-fit (post-fit) are displayed by open (closed) boxes.
In the top panel, the post-fit values of the normalisation parameters $\mu$ are shown.

The theoretical uncertainties that cause the largest median relative uncertainty are smaller than in the renormalised approach and get further reduced in the fit.
They exhibit larger pulls than before, however, because of their reduced pre-fit size.

\begin{myfigure}{%
		\NPrankingText{differential cross sections}{fixed}{with renormalised systematic uncertainties }
	}{fig:interp_SM_NPrankings_renormSyst}
	\includegraphics[height=280pt]{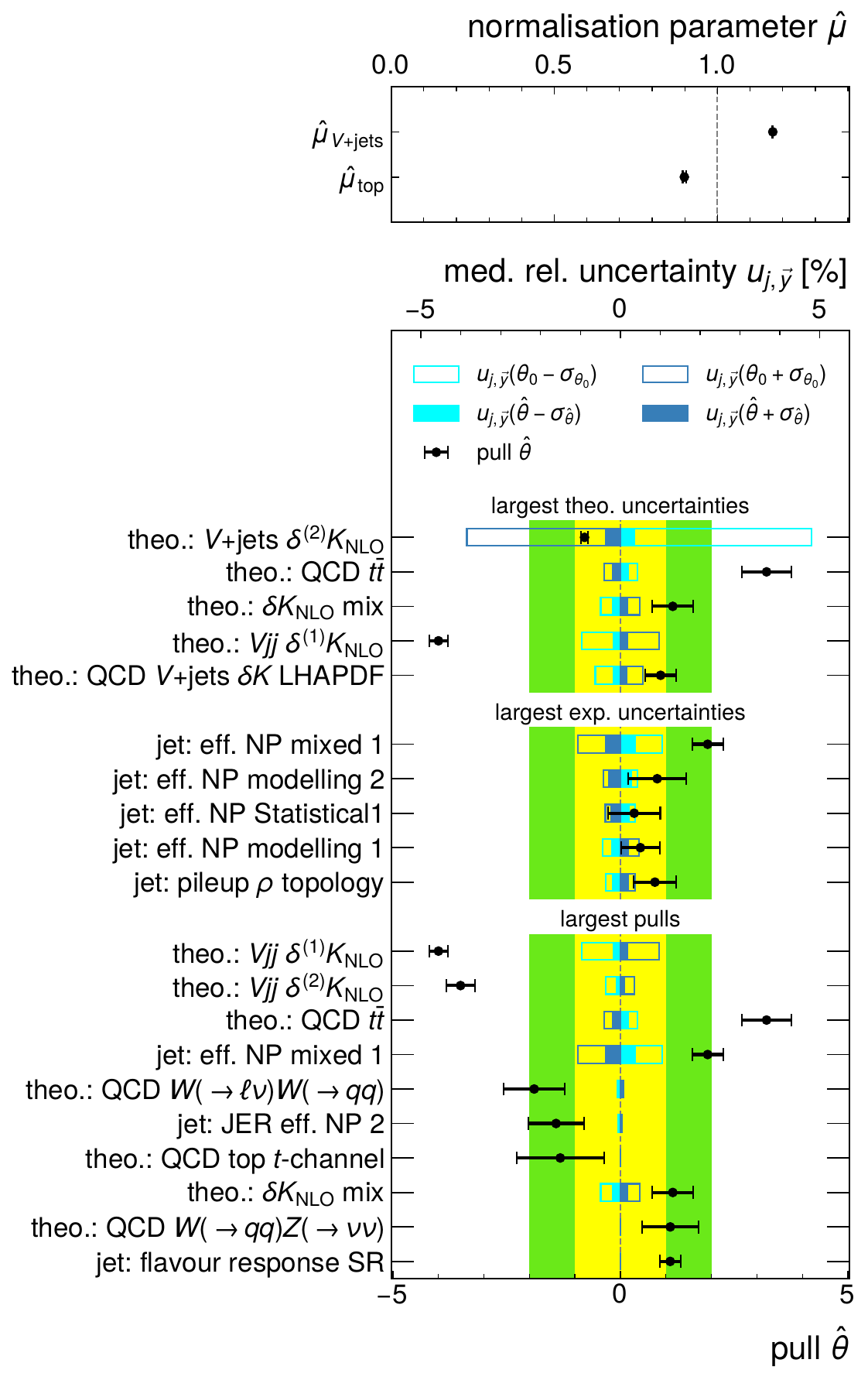}\rule{75pt}{0pt}
\end{myfigure}

For the largest experimental uncertainties, more similarities can be observed between the original and the renormalised approach:
many of the largest uncertainties are similar to before although their exact ordering changes.
They are all related to jet reconstruction.
For those uncertainties that are also among the largest experimental uncertainties in the original approach, the absolute value of the pull is in general reduced.
This is because in the original approach they are needed to change the normalisation as well as the shape of the distributions.
In the renormalised approach, they only change the shape.
The luminosity uncertainty is not among the largest experimental uncertainties any more because it is neglected in this fit approach.

The nuisance parameters with the largest pulls are now dominated by nuisance parameters of uncertainties that change the shape at large \METconst, where the shape discrepancy between measured data and generated \SM prediction is the largest.
An example for this are the \Vjj \alphas uncertainties (\Vjj $\delta^{(1,2)} K_{\NLO}$~\cite{Lindert:2022ejn}).

The size of the uncertainties is in general reduced by the renormalisation.
This means that to remove the same shape discrepancy between data and prediction, the nuisance parameter for an uncertainty has to be pulled more.
This results in pulls that are in general larger in the renormalised approach than in the original approach.

\bigskip
Overall, the renormalisation of appropriate systematic uncertainties has the desired effect of providing more intuitive post-fit values of the normalisation parameters while keeping sensible pulls and constraints for the nuisance parameters.
This renormalisation, however, is an ad-hoc procedure and therefore not taken as the nominal approach when using floating-normalisation predictions.

\section{\THDMa exclusion with alternative test statistic}
\label{app:interp_2HDMa_muOpt}

The test statistic used for the interpretation with respect to the \THDMa in \secref{sec:interpretation_2HDMa} takes the likelihood ratio of the signal-plus-background hypothesis to the background-only hypothesis (\cf\eqref{eq:metJets_qBSM}).
An alternative approach is to use a \textit{profiled} likelihood ratio comparing the signal-plus-background likelihood to the global likelihood maximum~\cite{Cranmer:2014lly}.
This is also the test statistic used by the \Contur toolkit (\cf\secref{sec:Contur_statistics}).
The test statistic has the advantage that its probability density function (\pdf) can be approximated by asymptotic formulae~\cite{ATLAS:2011tau,Cowan:2010js} and therefore no toys (\cf\secref{sec:interpretation_LR_2HDMa}) need to be produced to estimate the \pdf.

In this approach, a signal strength parameter $\mu$ is introduced into the likelihood that multiplies the process cross section by a constant value.
One possible test statistic then is
\begin{equation}
	\label{eq:metJets_qBSM_muOpt}
	\qBSM \coloneqq -2\ln
	\begin{cases}
		1,& \hat{\mu}>1\\
		\frac{\lh\left(\vv{x}|\vv s+\hatvvpiSM,\hat{\vv{\theta}}_{\mu=1}\right)}{\lh\left(\vv{x}|\hat{\mu}\cdot\vv s+\hatvvpiSM,\hat{\vv{\theta}}_{\hat\mu}\right)}, & 0<\hat{\mu}\leq1.\\
		\frac{\lh\left(\vv{x}|\vv s+\hatvvpiSM,\hat{\vv{\theta}}_{\mu=1}\right)}{\lh\left(\vv{x}|\hatvvpiSM,\hat{\vv{\theta}}_{\mu=0}\right)}, & \hat{\mu}\leq0
	\end{cases}
\end{equation}
Hereby, $\hat\mu$ is the signal strength that maximises the likelihood.
This is a modified version of \eqref{eq:Contur_testStatistic}.
The nominal case is $0<\hat{\mu}\leq1$.
The signal strength in the denominator is not allowed to exceed $\mu=1$ such that large fluctuations of the data \textit{over} the signal-plus-background prediction do not count as evidence against the signal-plus-background hypothesis.
Similarly, the signal strength in the denominator is not allowed to be smaller than $\mu=0$ such that large fluctuations of the data \textit{under} the \SM prediction do not count as evidence against the background-only hypothesis.

After adopting this test statistic, the procedure follows the one outlined in Sec-\linebreak{}tion~\ref{sec:interpretation_LR_2HDMa}.
In this example, no asymptotic formulae for approximating the probability density function are employed.

\bigskip
\figref{fig:interp_toy_distributions_muOpt} shows the probability density function $\pdf\left(\qBSM^{\vv s+\vvpiSM}\right)$ of the test statistic \qBSM in \eqref{eq:metJets_qBSM_muOpt} if the \THDMa was realised at the model point of $\ma=\SI{250}{GeV}$, $\mA=\SI{600}{GeV}$, $\tanB=1$ in green.
For illustration, the probability density functions are given for different multiples of the nominal signal cross section $\sigmaSig=\SI{0.64}{pb}$.
This corresponds to different signal strengths $\mu$.
The test demonstrates how a larger signal cross section would influence the distribution of the test statistic and its observed value.
The probability density function of \qBSM if the Standard Model is true is shown in blue.
The observed value of the test statistic given the data, \qBSMobs, is marked by a dashed black line.
\CLsb and \CLb correspond to the area under the curves to the right of \qBSMobs for the \THDMa and \SM hypothesis, respectively.

\myToyFigure{\qBSM^{\vv s+\vvpiSM}}{fig:interp_toy_distributions_muOpt}{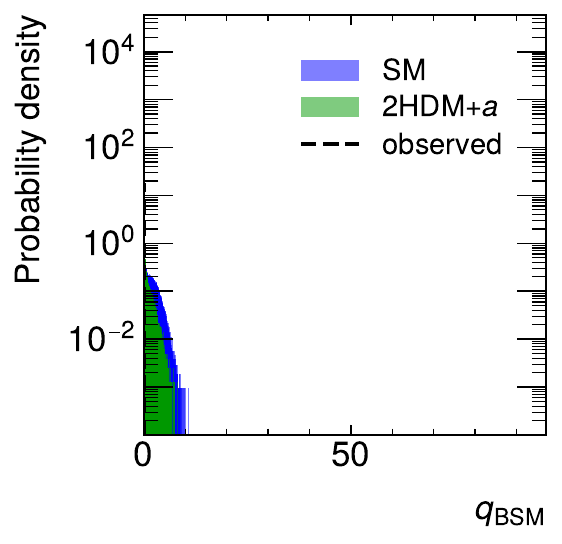}

If the signal cross section is small, the probability density functions form narrow, largely overlapping distributions.
The separation of the distribution increases with increasing signal cross section as the expected yields from the signal-plus-background and background-only hypotheses become distinguishable.
The distributions become wider with increasing signal cross section because the uncertainty of the signal grows proportionally.

Comparing to the probability density function for the nominal test statistic from \eqref{eq:metJets_qBSM} shown in~\figref{fig:interp_toy_distributions}, the most striking difference is that the new test statistic is bounded from below at $\qBSM=0$.
This is a consequence of comparing to the global likelihood maximum in the denominator.

The probability density function for \qBSM assuming the \SM prediction to be true~(blue) is very similar under both test statistics.
This is because, if the \SM prediction is true, $\hat\mu$ will in general be very close to 0.
At this point, both test statistics yield similar values.

For the model point used in \figref{fig:interp_toy_distributions_muOpt}, a signal with 3.3 times the nominal cross section is excluded at \SI{95}{\%} confidence level.
This is almost identical to the signal cross section that is excluded with the nominal test statistic (\cf\secref{sec:interpretation_2HDMa}).

\bigskip
\figref{fig:interp_2HDMa_exclusion_muOpt} shows the exclusion limits at \SI{95}{\%} confidence level from the \METjets measurement in the \mamA plane using the profiled likelihood ratio.
All five measurement regions and both subregions binned in \METconst are used.
The differential cross sections serve as data~$\vv x$ and prediction~$\vv\pi$ and the fixed-normalisation prediction according to \eqref{eq:interpretation_SM_pred_fixedNorm} is employed.
This corresponds to the \SM fit discussed in \secref{sec:interpretation_SM_fixedNorm_diffXS}.
The region to the left of the lines is excluded.

\begin{myfigure}{
		Expected (dashed lines) and observed (solid lines) exclusion limits at \SI{95}{\%} confidence level from the \METjets measurement in the \mamA plane using the profiled likelihood ratio.
		The excluded parameter space is to the left of the lines.
		The green (yellow) band indicates the region of one (two) standard deviations from the expected exclusion limit.%
		\THDMaLines{grey}
	}{fig:interp_2HDMa_exclusion_muOpt}
	\subfloat[\mamA plane]{\includegraphics[width=0.48\textwidth]{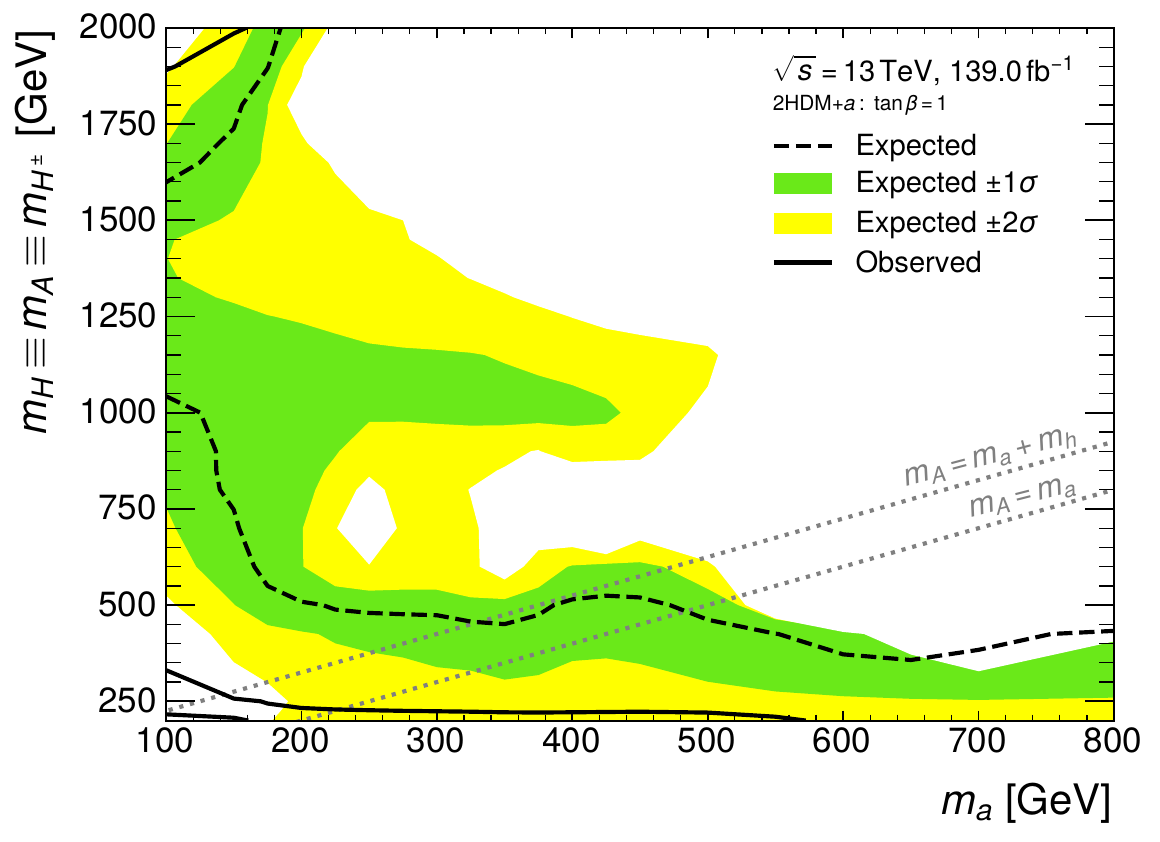}}\\
\end{myfigure}

The expected as well as observed exclusion limits assuming the profiled likelihood ratio are almost identical to those using the nominal test statistic (\cf\subfigref{fig:interp_2HDMa_exclusion_mamA}{a}).

\bigskip
In summary, the exclusion limits derived in \secref{sec:interpretation_2HDMa} change depending on whether the differential cross sections or \Rmiss distributions are used.
They do, however, not change significantly depending on whether \SM predictions with fixed or floating normalisation are used or which test statistic is employed.
\chapter{Additional material for the \Contur studies}
\label{app:contur_app}

\section{Details on \Contur pools and measurements}
\label{app:contur_app_pools}
\tabref{tab:contur_pools} gives details on the analysis pools and measurements used by \Contur.
The \LHC experiment for which the analysis pools is available is listed in the second column.
The third column gives a detailed description of the pool selection.
The last column gives references to the measurements that are considered in the analysis pool.
Measurements can appear for multiple analysis pools if they have regions that fulfil the criteria for either analysis pool.
This is for example the case for the measurement of the differential cross section for $W$ boson production in association with bottom-quark induced jets~\cite{Aad:2013vka} which considers an electron as well as a muon channel.
It has regions contributing to the $e+\MET+$jet analysis pool as well as regions contributing to the $\mu+\MET+$jet analysis pool.

\begin{mytable}{
		The different analysis pools used by \Contur, the experiments for which these pools are available, a detailed description of the pools and references to the contributing measurements.
	}{tab:contur_pools}{llp{0.5\linewidth}p{0.2\linewidth}}
		Pool name & Experiments & Description & References\\
		\midrule
    3$\ell$ & ATLAS, CMS & trileptons & \cite{Khachatryan:2016poo,Aaboud:2016ftt}\\
	4$\ell$ & ATLAS & four leptons & \cite{Aad:2012awa,Aad:2015rka,Aad:2021ebo,Aaboud:2017rwm,Aad:2014tca,Aaboud:2019lxo}\\
	$e^+e^-$+jet & ATLAS, CMS, LHCb & dielectrons at the $Z$ pole, plus optional jets & \cite{Sirunyan:2019bzr,Aad:2019hga,Aaboud:2017hbk,Aaij:2012mda,Aad:2014dvb,Aad:2015auj,Chatrchyan:2013vbb,Aaboud:2019jcc,Aaboud:2016btc,Aad:2013ysa,Sirunyan:2018cpw,Aaboud:2017hox,Aad:2020gfi}\\
	$e^+e^-\gamma$ & ATLAS & dielectrons plus photon(s) & \cite{Aad:2013izg,Aad:2016sau}\\
	$e$+\MET{}+jet & ATLAS, CMS & electron, missing transverse energy, plus optional jets (typically $W$, semi-leptonic $t\bar{t}$ analyses) & \cite{Aad:2014qxa,Aaboud:2017fye,Chatrchyan:2013vbb,Aaboud:2017soa,Aaboud:2016btc,Aad:2013vka}\\
	$e$+\MET{}+$\gamma$ & ATLAS & electron, missing transverse energy, plus photon & \cite{Aad:2013izg}\\
	$\gamma$ & ATLAS, CMS & inclusive (multi)photons & \cite{Aad:2013gaa,ATLAS:2012ar,Aad:2012tba,Aaboud:2017lxm,Aad:2013zba,Aaboud:2017vol,Aaboud:2017kff,ATLAS:2021mbt,Aaboud:2017skj,Aad:2016xcr,Chatrchyan:2013mwa,Aad:2014lwa}\\
	$\gamma$+\MET{} & ATLAS & photon plus missing transverse energy & \cite{ATLAS:2018nci,Aad:2016sau}\\
	high-mass Drell-Yan $\ell\ell$ & ATLAS, CMS & dileptons above the $Z$ pole & \cite{Aad:2013iua,Aad:2016zzw,ATLAS:2019zci,Sirunyan:2018owv}\\
	jets & ATLAS, CMS & inclusive hadronic final states & \cite{Sirunyan:2021lwi,Aaboud:2017qwh,Khachatryan:2016mlc,Chatrchyan:2012bja,Aad:2014vwa,Aad:2014rma,Aad:2015nda,Chatrchyan:2012dk,Sirunyan:2017skj,Aad:2013tea,Aad:2016mok,Chatrchyan:2014gia,Aaboud:2017wsi,Aad:2020zcn,Khachatryan:2016wdh,Aad:2020fch,Chatrchyan:2013qza,Aaboud:2017dvo,Aad:2014pua,CMS:2021lxi,Aaboud:2016itf,Aaboud:2017vqt,Sirunyan:2018xdh,Aaboud:2019aii,ATLAS:2019rqw}\\
	$e\mu+b$ & LHCb & top pairs via $e\mu+b$ & \cite{Aaij:2018imy}\\
	$\ell_1\ell_2$+\MET{} & ATLAS & $W^+W^-$ analyses in dileptons plus missing transverse energy channel & \cite{Aad:2016wpd,ATLAS:2012mec,Aaboud:2019nkz,Aaboud:2016ftt}\\
	$\ell_1\ell_2$+\MET{}+jet & ATLAS & unlike dileptons plus missing transverse energy and jets & \cite{Aad:2019hzw,Aad:2021dse,Aaboud:2019jcc}\\
	$\ell_1\ell_2$+\MET{}+$\gamma$ & ATLAS & fully leptonic $t\bar{t}$ plus photon & \cite{Aaboud:2018hip}\\
	$\ell^+\ell^-$+jet & ATLAS, CMS & $W^+W^-$ analyses in dileptons plus zero or one jet & \cite{Aad:2020sle,Khachatryan:2014zya,Sirunyan:2020jtq,Khachatryan:2016iob,Chatrchyan:2013zja,Aad:2014dta}\\
	$\ell^+\ell^-$+\MET{} & ATLAS & dileptons plus missing transverse energy & \cite{Aad:2012awa}\\
	$\ell^+\ell^-\gamma$ & ATLAS & dileptons plus photon(s) & \cite{Aad:2019gpq,Aad:2016sau}\\
	low-mass Drell-Yan $\ell\ell$ & ATLAS & dileptons below the $Z$ pole & \cite{Aad:2014qja}\\
	$\ell$+\MET{}+jet & ATLAS, CMS & lepton, missing transverse energy, plus optional jets (typically $W$, semi-leptonic $t\bar{t}$ analyses) & \cite{Sirunyan:2018ptc,Aaboud:2017fha,Aad:2015eia,Khachatryan:2016nbe,Aad:2019ntk,Sirunyan:2019hqb,Aad:2014xca,Aaboud:2018uzf,Aad:2015mbv,Sirunyan:2018wem,Khachatryan:2016gxp,Aaboud:2018eki,Aad:2015hna,Khachatryan:2016mnb,Sirunyan:2017yar}\\
	$\ell$+\MET{}+$\gamma$ & ATLAS & semileptonic $t\bar{t}$ plus photon & \cite{Aaboud:2018hip}\\
	\MET{}+jet & ATLAS & missing transverse energy plus jets & \cite{Aaboud:2017buf}\\
	$\mu$+jet & LHCb & muons (aimed at $W$) & \cite{AbellanBeteta:2016ugk}\\
	$\mu$+\MET{}+jet & ATLAS, CMS & muon, missing transverse energy, plus optional jets (typically $W$, semi-leptonic $t\bar{t}$ analyses) & \cite{Aad:2014qxa,Sirunyan:2018hde,Aaboud:2017fye,Aaboud:2017soa,Aaboud:2016btc,Aad:2013vka}\\
	$\mu$+\MET{}+$\gamma$ & ATLAS & muon, missing transverse energy, plus photon & \cite{Aad:2013izg}\\
	$\mu^+\mu^-$+jet & ATLAS, CMS, LHCb & dimuons at the $Z$ pole, plus optional jets & \cite{Sirunyan:2021lwi,Sirunyan:2019bzr,Aad:2020gfi,Aaij:2013nxa,Khachatryan:2016nbe,Aad:2014dvb,Aad:2015auj,Aaboud:2017hox,Aaboud:2019jcc,Aaboud:2016btc,Aad:2013ysa,Sirunyan:2018cpw,AbellanBeteta:2016ugk,Aaboud:2017hbk}\\
	$\mu^+\mu^-\gamma$ & ATLAS, CMS & dimuons plus photon & \cite{Aad:2013izg,Aad:2016sau,Khachatryan:2015rja}\\
	$\ell^\pm\ell^\pm$+\MET{} & ATLAS & two same-sign leptons ($e/\mu$) plus missing transverse energy and jets & \cite{Aaboud:2019nmv}\\
	hadronic $t\bar{t}$ & ATLAS, CMS & fully hadronic top events & \cite{ATLAS:2020ccu,Aaboud:2018eqg,CMS:2019eih,Sirunyan:2019rfa}\\
\end{mytable}

\section{Branching fractions and cross sections in the \THDMa}

In Figures~\ref{fig:contur_xsBR_mamA}~to~\ref{fig:contur_xsBR_mamX}, the main branching ratios and cross sections needed for understanding the sensitivity in the different planes in \secref{sec:contur_2HDMa} are given.
The values are obtained using the generation setup described in \secref{sec:MC_2HDMa_Contur}.
The parameters are set according to \tabref{tab:LHCDMWG_params} unless stated otherwise.
The cross section for producing \BSM bosons in the $s$-channel without any other particles is only given if the cross section times branching fraction for the decay of the \BSM boson to exclusively \SM particles is sizeable.
In practise, this means it is only given for \Hpm.

\newcommand{\myConturXSBRFigure}[7]
{%
	\begin{myfigure}{
			Auxiliary (a) branching fractions and (b) cross sections for the #1 plane#2 as a function of #3 for different values of #4.
			Only the most important branching ratios and cross sections are considered.
			The parameters that are not varied are most importantly #5.
		}{fig:contur_xsBR_#6}
		\newcommand{\xsbrWidth}{0.43\textwidth}
		\subfloat[branching fractions]{
			\begin{tabular}{cc}
				\includegraphics[width=\xsbrWidth]{figures/Contur/xsecBR/#7/BR_a.pdf}&
				\includegraphics[width=\xsbrWidth]{figures/Contur/xsecBR/#7/BR_A.pdf}\\
				\includegraphics[width=\xsbrWidth]{figures/Contur/xsecBR/#7/BR_H.pdf}&
				\includegraphics[width=\xsbrWidth]{figures/Contur/xsecBR/#7/BR_H+.pdf}
			\end{tabular}
		}\\
		\subfloat[cross sections]{
			\begin{tabular}{cc}
				\includegraphics[width=\xsbrWidth]{figures/Contur/xsecBR/#7/xs_a.pdf}&
				\includegraphics[width=\xsbrWidth]{figures/Contur/xsecBR/#7/xs_A.pdf}\\
				\includegraphics[width=\xsbrWidth]{figures/Contur/xsecBR/#7/xs_H.pdf}&
				\includegraphics[width=\xsbrWidth]{figures/Contur/xsecBR/#7/xs_H+.pdf}
			\end{tabular}
		}
	\end{myfigure}
}

\myConturXSBRFigure{\mamA}{}{\ma}{\mA}{$\tanB=1$ and $\mX=\SI{10}{GeV}$}{mamA}{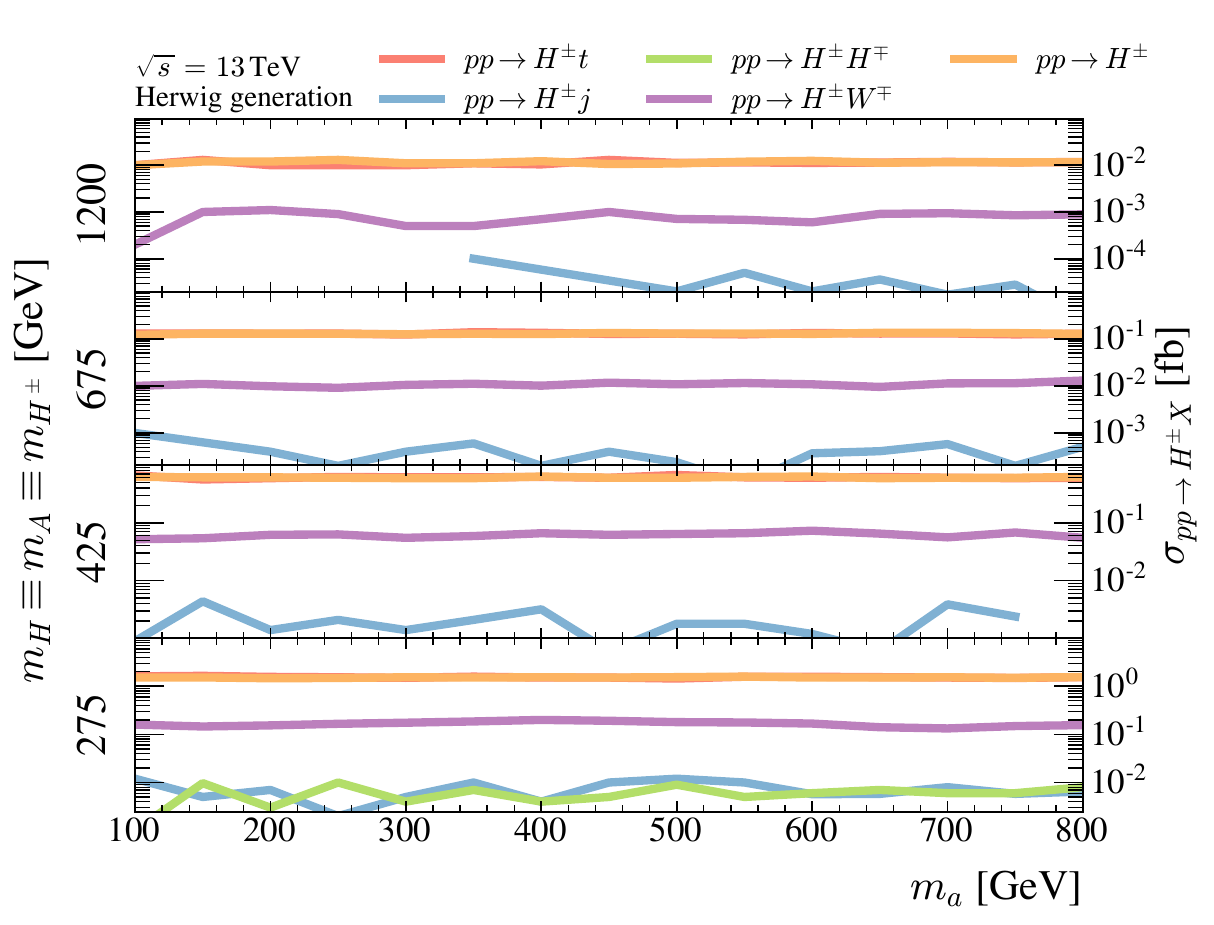}

\myConturXSBRFigure{\matanB}{}{\ma}{\tanB}{$\mAeqmHeqmHpm=\SI{600}{GeV}$ and $\mX=\SI{10}{GeV}$}{matb}{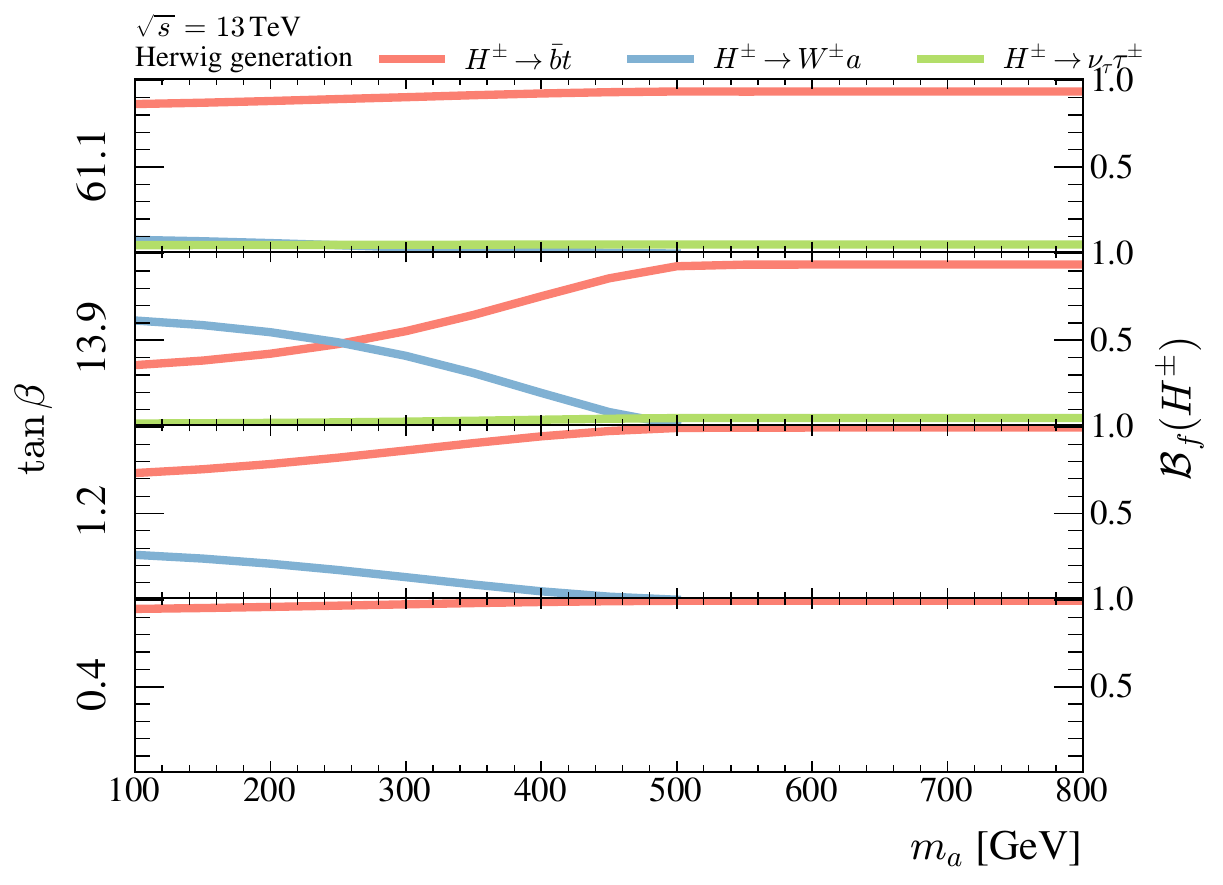}

\myConturXSBRFigure{\mHmA}{ at $\ma=\SI{100}{GeV}$}{\mHeqmHpm}{\mA}{$\tanB=1$ and $\mX=\SI{10}{GeV}$}{mHmA_ma100}{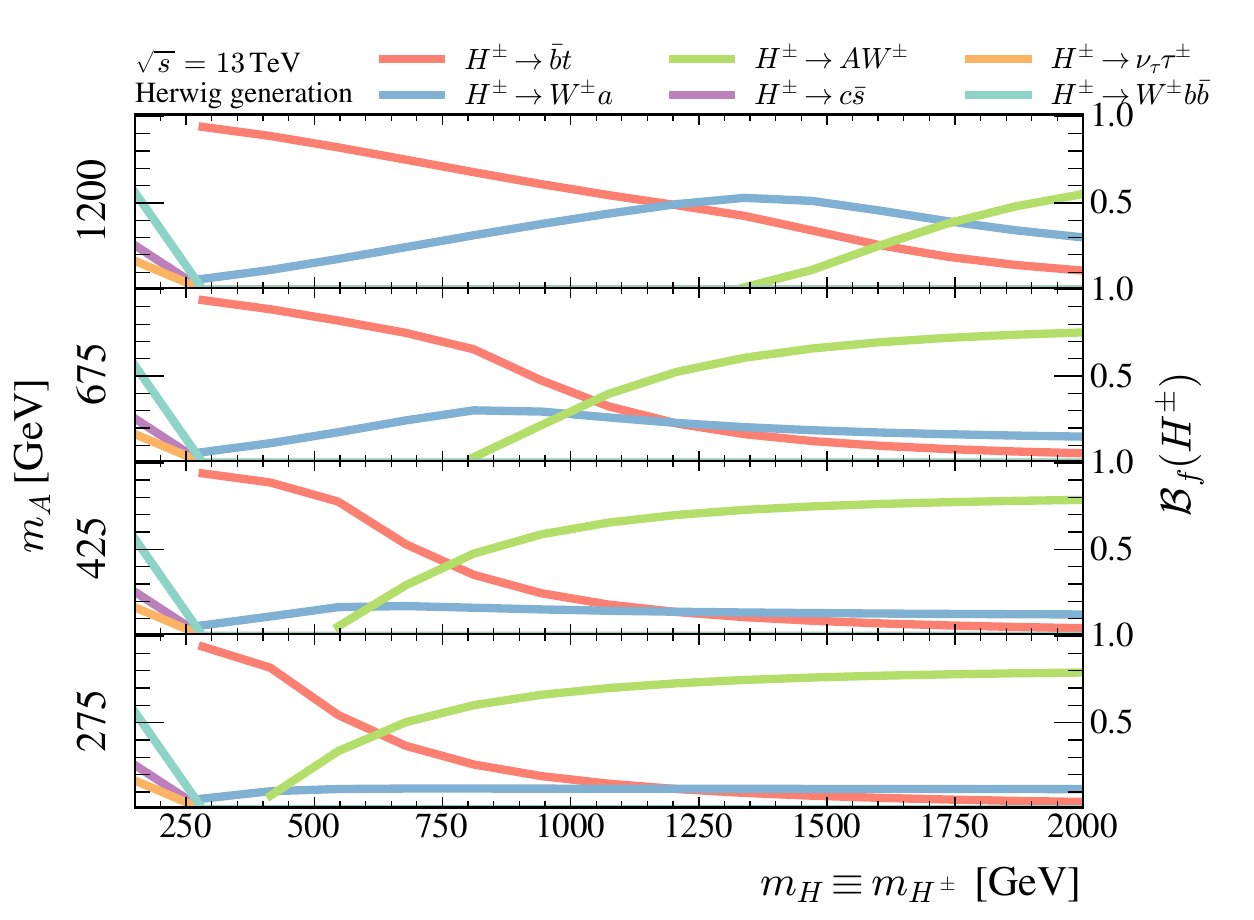}

\myConturXSBRFigure{\mHmA}{ at $\ma=\SI{500}{GeV}$}{\mHeqmHpm}{\mA}{$\tanB=1$ and $\mX=\SI{10}{GeV}$}{mHmA_ma500}{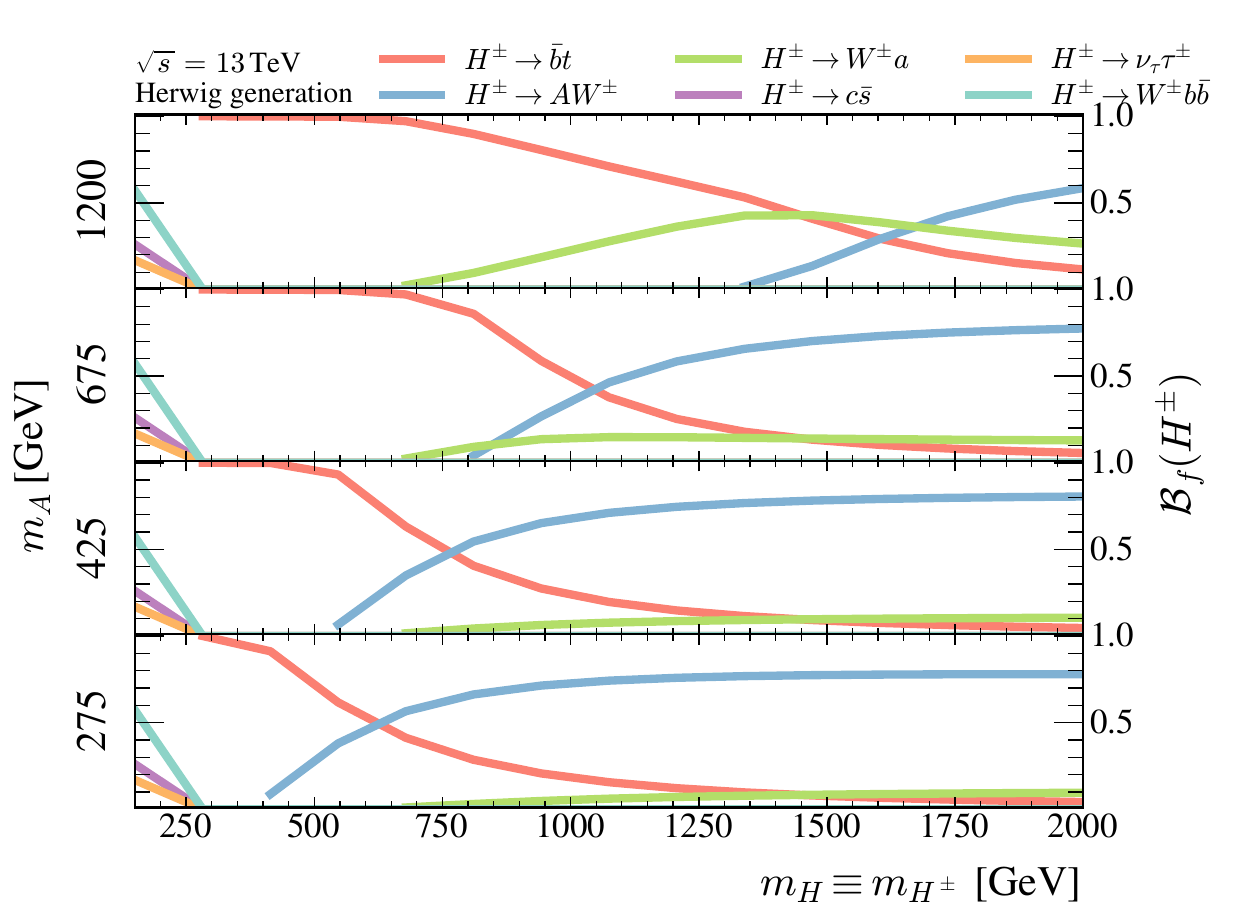}

\myConturXSBRFigure{\mamX}{}{\ma}{\mX}{$\mAeqmHeqmHpm=\SI{600}{GeV}$ and $\tanB=1$}{mamX}{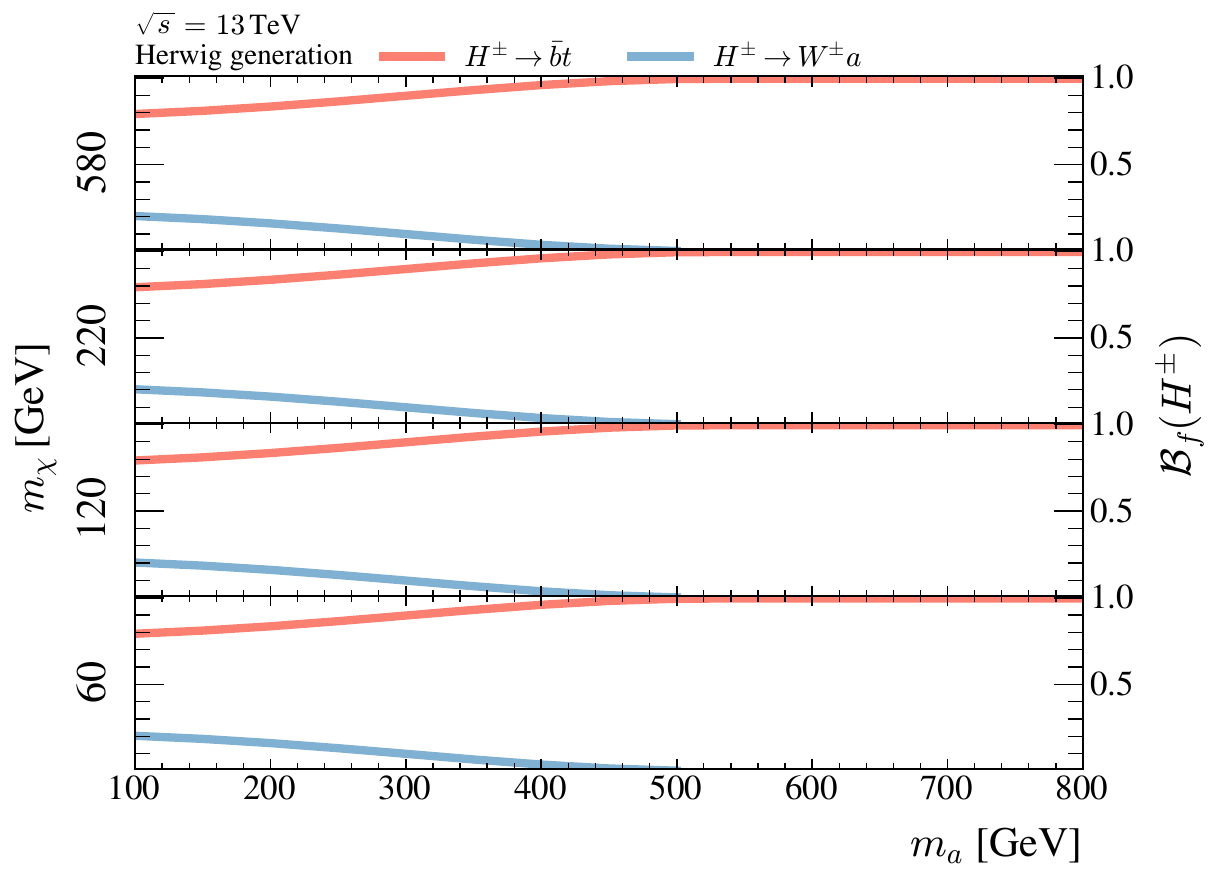}

\ChapterStarSimple{List of Acronyms}
\renewcommand{\glossarysection}[2][]{} 
\setlength{\glsdescwidth}{6.5cm} 
\hspace{40pt} 
\printglossary[type=\acronymtype, style=super2colleft, nogroupskip]
\printbibliography[heading=bibintoc]
\end{document}